\documentclass{ILCTDR}
\pdfoutput=1
\usepackage{savesym}
\savesymbol{iint}
\savesymbol{iiint}
\usepackage[maybess]{heppennames2} %symbols for physics processes (from CERN)
\usepackage{amsmath,amssymb}
\usepackage{dsfont}
\restoresymbol{}{iint}
\restoresymbol{}{iiint}

\newcommand{\mathbold}[1]{\mbox{\boldmath $#1$}}
\def\leqn#1{(\ref{#1})}
\providecommand{\ee}   {\mathrm{e^+e^-}}
\def\beqpeskin{\begin{equation}}
\def\eeqpeskin#1{\label{#1}\end{equation}}

\newif\ifversionthreeD

\newif\ifversionforidag
\versionforidagfalse %set this to true if making the version for the IDAG

\newif\ifversionforpac
\versionforpacfalse %set this to true if making the version for the PAC

\newif\ifversionforilcsc
\versionforilcscfalse %set this to true if making the version for the PAC

\newif\ifbibliographyend
\bibliographyendtrue %set this to true if you like to put the bibliography at the end

\ifbibliographyend %depending were the bibliography goes
        \usepackage[numbers,sort]{natbib}
	\usepackage{bibunits}	
\else
	\usepackage[numbers,sort,sectionbib]{natbib}
	\usepackage[sectionbib]{bibunits}
\fi

\definecolor{violet}{rgb}{0.1,0,0.45}

\usepackage[hypertexnames=false,pdfpagelayout=TwoColumnRight,bookmarksopen=true,bookmarksopenlevel=0]{hyperref}
\defaultbibliographystyle{kp}
\defaultbibliography{Bibliography/ilchiggsbibliography}
\makeatletter
\let\oldtb=\thebibliography
\renewcommand{\thebibliography}{%
 \renewcommand{\@mkboth}[2]{}%
  \oldtb}
\makeatother

\graphicspath{%
{Bibliography/}
{Chapter_Benchmarking/}
{Chapter_Calorimetry/}
{Chapter_ConceptOverview/}
{Chapter_Costs/}
{Chapter_ElectronicsDAQ/}
{Chapter_ForwardSystems/}
{Chapter_Introduction/}
{Chapter_Magnet/}
{Chapter_MDI/}
{Chapter_MuonSystem/}
{Chapter_SimReco/}
{Chapter_Summary/}
{Chapter_Tracker/}
{Chapter_Vertex/}
{ILCHIGGSMain/}
}

\newcommand{\sid}{{SiD}\xspace}

\newcommand{\fbinv}{\ensuremath{\mathrm{fb}^{-1}}\xspace}

\newcommand{\geant}{\textsc{Geant4}\xspace}
\newcommand{\slic}{\textsc{SLIC}\xspace}

\newcommand{\roots}{\ensuremath{\sqrt{s}}\xspace}

\newcommand{\xo}{\ensuremath{\mathrm{X}_{0}}\xspace}

\renewcommand{\epem}{\ensuremath{\mathrm{\Pep\Pem}}\xspace} %e+e-
\newcommand{\mpmm}{\ensuremath{\PGmp\PGmm}\xspace}  %mu+mu-
\newcommand{\bbbar}{\ensuremath{\mathrm{\PQb}\mathrm{\PAQb}}\xspace}
\newcommand{\ccbar}{\ensuremath{\mathrm{\PQc}\mathrm{\PAQc}}\xspace}
 
\newcommand{\qqbar}{\ensuremath{\mathrm{\PQq}\mathrm{\PAQq}}\xspace} 
 
\newcommand{\nuenuebar}{\ensuremath{\PGne\PAGne}\xspace} %nu_e nue_e
\newcommand{\gamgam}{\ensuremath{\upgamma\upgamma}\xspace} %gamma gamma
\newcommand{\smH}{\ensuremath{\PSh}\xspace}  % SM higgs
\newcommand{\nuenueH}{\ensuremath{\nuenuebar\smH}\xspace}  % nuenue H
\newcommand{\gghadrons}{\ensuremath{\upgamma\upgamma \rightarrow \mathrm{hadrons}}\xspace}

\newcommand{\micron}{\ensuremath{\upmu\mathrm{m}}\xspace}

\newcommand{\pT}{\ensuremath{p_\mathrm{T}}\xspace}

\begin{document}

%decide which table to use
\ifversionforidag
	\fancyfoot[RO,LE]{\thepage}
	\fancyfoot [LO]{\slshape IDAG Draft version 24/09/2012}
	\fancyfoot [RE]{\slshape IDAG Draft version 24/09/2012}
	\input{IDAGtitlepage.tex}  % the title page
	\newpage
\begin{center}
{\Large{\textsc{Authors}}}\\
\end{center}
\begin{center}
{
D.M.~Asner$^{1}$, T.~Barklow$^{2}$, C.~Calancha$^{3}$, K.~Fujii$^{3}$, N.~Graf$^{2}$, H.~E.~Haber$^{4}$, A.~Ishikawa$^{5}$, S.~Kanemura$^{6}$, S.~Kawada$^{7}$, M.~Kurata$^{8}$, A.~Miyamoto$^{3}$, H.~Neal$^{2}$, H.~Ono$^{9}$, C.~Potter$^{10}$, J.~Strube$^{11}$, T.~Suehara$^{5}$, T.~Tanabe$^{8}$, J.~Tian$^{3}$, K.~Tsumura$^{12}$, S.~Watanuki$^{5}$, G.~Weiglein$^{13}$, K.~Yagyu$^{14}$, H.~Yokoya$^{6}$
}
\end{center}

\noindent  $^{\;\;\:\:1}$\begin{minipage}[t]{0.95\textwidth}\textit{Pacific Northwest National Laboratory, Richland, USA}
\end{minipage}\\
\noindent  $^{\;\;\:\:2}$\begin{minipage}[t]{0.95\textwidth}\textit{SLAC National Accelerator Laboratory, Menlo Park, USA}
\end{minipage}\\
\noindent  $^{\;\;\:\:3}$\begin{minipage}[t]{0.95\textwidth}\textit{KEK, Tsukuba, Japan}
\end{minipage}\\
\noindent  $^{\;\:\:\:4}$\begin{minipage}[t]{0.95\textwidth}\textit{University of California, Santa Cruz, USA}
\end{minipage}\\
\noindent  $^{\;\;\:\:5}$\begin{minipage}[t]{0.95\textwidth}\textit{Tohoku Univesity, Sendai, Japan}
\end{minipage}\\
\noindent  $^{\;\;\:\:6}$\begin{minipage}[t]{0.95\textwidth}\textit{University of Toyama, Toyama, Japan}
\end{minipage}\\
\noindent  $^{\;\;\:\:7}$\begin{minipage}[t]{0.95\textwidth}\textit{Hiroshima University, Hiroshima, Japan}
\end{minipage}\\
\noindent  $^{\;\;\:\:8}$\begin{minipage}[t]{0.95\textwidth}\textit{University of Tokyo, Tokyo, Japan}
\end{minipage}\\
\noindent  $^{\;\;\:\:9}$\begin{minipage}[t]{0.95\textwidth}\textit{Nippon Dental University, Niigata, Japan}
\end{minipage}\\
\noindent  $^{\;\:\:\:10}$\begin{minipage}[t]{0.95\textwidth}\textit{University of Oregon, Eugene, USA}
\end{minipage}\\
\noindent  $^{\;\;\:\:11}$\begin{minipage}[t]{0.95\textwidth}\textit{CERN, Geneva, Switzerland}
\end{minipage}\\
\noindent  $^{\;\;\:\:12}$\begin{minipage}[t]{0.95\textwidth}\textit{University of Nagoya, Nagoya, Japan}
\end{minipage}\\
\noindent  $^{\;\;\:\:13}$\begin{minipage}[t]{0.95\textwidth}\textit{DESY, Hamburg, Germany}
\end{minipage}\\
\noindent  $^{\;\;\:\:14}$\begin{minipage}[t]{0.95\textwidth}\textit{National Central University, Zhongli, Taiwan}
\end{minipage}\\
\cleardoublepage
%list of main and chapter editors
\else
	\ifversionforpac
		\fancyfoot[RO,LE]{\thepage}
		\fancyfoot [LO]{\slshape PAC Draft version 10/12/2012}
		\fancyfoot [RE]{\slshape PAC Draft version 10/12/2012}
		\input{PACtitlepage.tex}  % the title page
		%list of main and chapter editors		
	\else
		\ifversionforilcsc
			\fancyfoot[RO,LE]{\thepage}
			\fancyfoot [LO]{\slshape Final Draft 21/02/2013}
			\fancyfoot [RE]{\slshape Final Draft 21/02/2013}
			\input{ILCSCtitlepage.tex}  % the title page
			%list of main and chapter editors		
		\else
		
			\begin{titlepage}
\vspace*{1cm}
%\begin{center}
%\includegraphics[width=0.7\textwidth]{SiDlogo.pdf}
%\end{center}

\begin{center}
\vspace*{1cm}
\Huge{\textsc{ILC Higgs White Paper}}
\vspace*{0.5cm}
\end{center}
%\begin{center}
%\Large{\textsc{Detailed Baseline Report}} 
%\end{center}
\vspace*{3cm}

%\begin{center}
%\Large{Version: \buildrevision} \\
%\Large{Last Change:\buildtimestamp } 
%\end{center}
%\vspace*{2cm}

\cleardoublepage

\end{titlepage}
  % the title page
			 %list of main and chapter editors
		\fi	
		
	\fi

\fi

\phantomsection % fix hyperlinking
\addcontentsline{toc}{chapter}{Table of Contents} % add table of contents entry to table of contents
\tableofcontents
%\makeatletter
%\input{ILCHiggs.toc}
%\makeatother

\pdfoutput=1
\chapter*{Introduction \label{sid:chapter_introduction}}
\addcontentsline{toc}{chapter}{Introduction}

Following an intense and successful R\&D phase, the ILC has now achieved a state of maturity and readiness, culminating recently with the publication 
of the Technical Design Report~\cite{Behnke:2013xla,Baer:2013cma,Adolphsen:2013jya,Adolphsen:2013kya,Behnke:2013lya}. Several important physics goals at the TeV energy scale have motivated this effort.  These include a detailed study of the properties of the recently discovered 
Standard Model-like Higgs boson, including  precision measurements of its couplings to fermions and bosons,  and an improved knowledge of the top quark 
and W, Z boson interactions to a high level of precision.  In all of these, the ILC will yield substantial improvements over LHC measurements and will 
have a qualitative advantage on signatures that have high backgrounds at LHC or are difficult to trigger on.  Moreover, the ILC provides a unique 
sensitivity in the search for signals of new physics beyond the Standard Model arising from the electroweak production of new particles (assuming these 
are kinematically accessible), as well as extending the probe of new interactions to higher mass scales via the precision measurements 
of  W, Z and two-fermion processes.  In this way, the ILC experiments will be sensitive to new phenomena such as supersymmetric partners of known 
particles, new heavy gauge bosons, extra spatial dimensions, and particles connected with alternative theories of electroweak symmetry breakingt~\cite{Baer:2013cma}.   Indeed, the 
ILC experiments will bring qualitatively new capabilities; detailed simulations with realistic detector designs show that the ILC can reach the precision goals needed~\cite{Behnke:2013lya}.   

The requirements of the ILC~\cite{Heuer:2003nn} include tunability between center-of-mass energies of 200 and 500 GeV, with rapid changes in energy over a limited range for threshold scans.   
The luminosity, which must exceed $10^{34}$ cm$^{-2}$s$^{-1}$ at 500 GeV, roughly scales proportionally with center-of-mass collision energy. Highly polarized electrons ($>80$\%) are specified, 
with polarized positrons desirable.  The TDR design~\cite{Adolphsen:2013jya,Adolphsen:2013kya} has met these specifications. R\&D has achieved 
the accelerating gradient goal of 35 MV/m in test stands and 31.5 MV/m in 
installed cryomodules with beam loading. Cavity fabrication to these specifications has been industrialized. The effects of the electron cloud in the positron damping ring have 
been studied experimentally, leading to proven techniques for its mitigation.  Fast kickers needed for damping ring beam injection and ejection have been developed.  The required 
small final focus spot size is being demonstrated in a test facility.  The final focus and interaction region, including the detector push-pull system, has been designed. Two 
detailed detector designs have been developed~\cite{Behnke:2013lya}, with R\&D supporting these designs.  Beam tests with highly granular calorimeters have demonstrated the calorimetry performance 
needed by using the particle flow technique.  Similarly, tracking R\&D has advanced for vertex detection based on thin CMOS monolithic pixel sensors, outer tracking with low-mass 
supported silicon microstrips, and advanced TPC technologies employing micropattern gas detectors or silicon sensors for readout.

Recently, the Japanese government has expressed a desire to host the ILC, and international negotiations are underway. In a staged approach, beginning at a center-of-mass energy 
of 250 GeV, a physics program would start with precision measurements of the Higgs branching ratios and properties. Raising the energy to 500 GeV would move to precision measurements 
of top quark properties well beyond those possible at the LHC.  Measurements of the top coupling to the Higgs and the Higgs self coupling would begin at 500 GeV.  Should there be 
accessible new particles such as supersymmetric partners of gauge bosons, Higgs bosons and leptons, the ILC with the power of polarized beams is the only place where they can be studied in full detail. If there are 
additional Higgs boson states (which are often difficult to observe at the LHC even if not too heavy), the ILC would be needed to measure their masses, quantum numbers, and 
couplings to Standard Model particles.  Extension of the ILC to 1 TeV is straightforward, with lengthened linac tunnels and additional cryomodules, building on the original 
ILC sources, damping rings, final focus and interaction regions, and beam dumps.

%%%%%%%%%%%%%%
%% Roman numerals
%%%%%%%%%%%%%%
\makeatletter
\newcommand{\rmnum}[1]{\romannumeral #1}
\newcommand{\Rmnum}[1]{\expandafter\@slowromancap\romannumeral #1@}
\makeatother
\newcommand\longdash{\mathrm{\phantom{a}---\phantom{a}}}

\newenvironment{Eqnarray}%
         {\arraycolsep 0.14em\begin{eqnarray}}{\end{eqnarray}}
\def\beqa{\begin{Eqnarray}}
\def\eeqa{\end{Eqnarray}}
\def\beq{\begin{equation}}
\def\eeq{\end{equation}}
\def\vev#1{\langle #1 \rangle}
\def\nn{\nonumber}
\def\half{\tfrac{1}{2}}
\def\quarter{\tfrac{1}{4}}
\def\sinb{\sin\beta}
\def\cosb{\cos\beta}
\def\tanb{\tan\beta}
\def\sinbma{s_{\beta-\alpha}}
\def\cosbma{c_{\beta-\alpha}}
\renewcommand\Re{{\mathrm Re}}
\renewcommand\Im{{\mathrm Im}}
\def\Eq#1{Eq.~(\ref{#1})}
\def\eq#1{eq.~(\ref{#1})}
\def\eqs#1#2{eqs.~(\ref{#1}) and (\ref{#2})}
\def\Eqs#1#2{Eqs.~(\ref{#1}) and (\ref{#2})}
\def\eqst#1#2{eqs.~(\ref{#1})--(\ref{#2})}
\def\ifmath#1{\relax\ifmmode #1\else $#1$\fi}
\def\lsub#1{\ifmath{_{\lower1.5pt\hbox{$\scriptstyle #1$}}}}
\def\lsup#1{^{\lower 6pt\hbox{$\scriptstyle#1$}}}
\def\ddel{\!\!\mathrel{\raise1.5ex\hbox{$\leftrightarrow$\kern-.85em
\lower1.7ex\hbox{$\partial$}}}}
\def\llsup#1{^{\lower 2pt\hbox{$\scriptstyle#1$}}}
\def\mstopa{M_{\widetilde t_1}}
\def\mstopb{M_{\widetilde t_2}}
\def\msusy{M_{\mathrm S}}
\def\msusyy{M_{\mathrm S}^2}
\def\hl{h}
\def\ha{A}
\def\hh{H}
\def\hp{H^+}
\def\hpm{H^\pm}
\def\mha{m_{\ha}}
\def\mhl{m_{\hl}}
\def\mhh{m_{\hh}}
\def\mhpm{m_{\hpm}}
\def\phaa{\phantom{a}}
\def\anti{\overline}
\def\wtil{\widetilde}
\def\ql{Q_L}
\def\ur{U_R}
\def\dr{D_R}
\def\mud{M_U}
\def\mdd{M_D}
\def\kpu{\kappa^U}
\def\rhu{\rho^U}
\def\kpd{\kappa^D}
\def\rhd{\rho^D}
\def\cbma{c_{\beta-\alpha}}
\def\cbmasq{c^2_{\beta-\alpha}}
\def\sbma{s_{\beta-\alpha}}
\def\sbmasq{s^2_{\beta-\alpha}}
\def\lsim{\mathrel{\raise.3ex\hbox{$<$\kern-.75em\lower1ex\hbox{$\sim$}}}}
\def\gsim{\mathrel{\raise.3ex\hbox{$>$\kern-.75em\lower1ex\hbox{$\sim$}}}}
\def\phm{\phantom{-}}
\def\mw{m_W}
\def\mz{m_Z}
\def\mt{m_t}

\chapter{Higgs Theory \label{sid:chapter_theory}}

\section{Introduction: the Higgs mechanism}

Quantum field theory has been enormously successful in describing the
behavior of fundamental point particles and their interactions in a framework
that is consistent with the principles of relativity and quantum mechanics.
Indeed, once these principles are invoked, quantum field theory appears to
be the only consistent framework for incorporating
interacting fundamental point particles.   If such a framework is
to be predictive (i.e., dependent only on a finite number of input parameters
that are provided by experimental measurements), then
the properties of such fundamental particles are highly constrained---only
spin 0, spin 1/2 and spin 1 are allowed~\cite{Weinberg:1995mt,Weinberg:1996kr}.
Moreover, if the spin 1 particles are self-interacting,
they must be described by a gauge theory.  It is remarkable that this is precisely the
spectrum of fundamental particles that have been observed in nature.

A gauge theory of fundamental self-interacting spin-1 gauge bosons
naively appears to require that gauge bosons should be massless, since
an explicit mass term for the gauge boson in the Lagrangian manifestly
violates the gauge symmetry.  However, due to the seminal work of
Brout, Englert~\cite{Englert:1964et} and
Higgs~\cite{Higgs:1964ia,Higgs:1964pj} and subsequent work by Guralnik, Hagen and
Kibble~\cite{Guralnik:1964eu,Guralnik:1965uza,Kibble:1967sv}, a
mass-generation mechanism for gauge bosons that is consistent with the
gauge symmetry was developed.  The simplest realization of this
mechanism was subsequently employed by Weinberg, when he incorporated
a self-interacting complex scalar doublet into a gauge theory of
electroweak interactions~\cite{Weinberg:1967tq}.  The neutral scalar
of the doublet acquires a vacuum expectation value (vev), which
spontaneously breaks the gauge symmetry and generates mass for the
$W^\pm$ and $Z$ bosons of electroweak theory while leaving the photon
massless.  Moreover, by coupling the complex scalar doublet to the
chiral fermions of the Standard Model (where no gauge-invariant mass
terms for the fermions are allowed prior to symmetry breaking), one
can also generate masses for all quarks and charged leptons.  In the
Glashow-Weinberg-Salam theory of the electroweak
interactions~\cite{Glashow:1961tr,Weinberg:1967tq,Salam:1968rm}, the
gauge bosons acquire mass via the Higgs mechanism by absorbing three
of the four degrees of freedom of the complex scalar doublet, which
provide for the longitudinal degrees of freedom of the $W^\pm$ and $Z$
bosons.  One physical scalar degree of freedom is left over---the
Higgs boson of the Standard Model.

There are other possible dynamics that can be used for achieving a
spontaneously broken gauge theory of the electroweak interactions (via
the Higgs mechanism) in which elementary scalar bosons are not
employed.  For example, it is possible to spontaneously break a gauge
theory by introducing a strongly interacting fermion pair that
condenses in the vacuum, in analogy with Cooper pairs of
superconductivity (for a nice review, see Ref.~\cite{King:1994yr}).
However, in the summer of 2012 a new scalar boson
was discovered at the LHC by the ATLAS and CMS
Collaborations~\cite{Aad:2012tfa,Chatrchyan:2012ufa}, whose
properties appear to be consistent(within the experimental error) with
those expected of the Standard Model Higgs boson~\cite{Aad:2013wqa,Aad:2013xqa,Chatrchyan:2013lba,Chatrchyan:2012jja}.  Consequently, it
appears that nature has chosen to realize the Higgs mechanism via
scalar fields that are either elementary or appear elementary at the
electroweak scale.  Although the scalar sector need not be a minimal
one, the data seems to favors the existence of one state of the scalar
sector whose properties resemble those of the Standard Model Higgs
bosons; any deviations from Standard Model behavior, if they
exist, are likely to be small.  Clearly, precision measurements of
the newly discovered scalar state will be critical for establishing
and testing the theory that governs the dynamics of electroweak
symmetry breaking.

\subsection{Vector boson mass generation and the unitarity of $VV\to VV$ scattering ($V=W$ or $Z$)}

Consider the theory of electroweak interactions without the attendant scalar sector.  If one attempts
to simply add an explicit mass term to the $W^\pm$ and $Z$ bosons, then the resulting theory would
be mathematically inconsistent.  One signal of this inconsistency would be revealed by using the
theory to compute the cross section for the scattering of longitudinally polarized gauge bosons,
$VV \rightarrow VV$ (where $V=W$ or $Z$) at tree-level.  Such a calculation would yield a
scattering amplitude whose energy dependence grows with the square of the center of mass energy,
a result that grossly violates unitarity.  Such a result would be in violation of one of the sacred
principles of quantum mechanics (which requires that the sum of probabilities can never exceed unity).

It is remarkable that this tree-level unitarity violation can be
mitigated by postulating the existence of an elementary scalar
particle that couples to $W^+ W^-$ and $ZZ$ with coupling strength
$gm_V^2/m_W$ (where $V=W$ or $Z$).  This new interaction introduces an
additional contribution to $VV \rightarrow VV$, which exactly cancels
the bad high energy behavior of the scattering amplitude, and leaves a
result that approaches a constant at infinite energy.  Thus, one can
reconstruct the Standard Model by imposing tree-level unitarity on all
scattering amplitudes of the
theory~\cite{LlewellynSmith:1973ey,Cornwall:1973tb,Cornwall:1974km}.
Thus, if the newly discovered scalar $h$ is to be interpreted as the
Higgs boson of the Standard Model, one should confirm that
\begin{equation} \label{hvv}
g_{hVV}=\frac{\sqrt{2}\,m_V^2}{v}\,,\qquad g_{hhVV}=\frac{\sqrt{2}\,m_V^2}{v^2}\,,
\end{equation}
where the Higgs vev, $v=174$~GeV, is related to the $W$ mass via $m_W=gv/\sqrt{2}$.

Suppose deviations from eq.~(\ref{hvv}) were uncovered by experimental Higgs studies.  Then, one would surmise
that the scalar Higgs sector is not minimal, and other scalar states play a role in achieving tree-level unitarity~\cite{Gunion:1990kf}.
Indeed, one can examine the case of an arbitrary scalar sector and derive unitarity sum rules that
replace eq.~(\ref{hvv}) in the more general Higgs model.   We shall impose one constraint on all extended Higgs
sectors under consideration---the scalar multiplets and vevs employed should satisfy the tree-level constraint that
\beq
\rho\equiv \frac{m_W^2}{m_Z^2\cos^2\theta_W}= 1\,,
\eeq
a result that is strongly suggested by precision electroweak
measurements~\cite{ErlerPDG,Baak:2012kk}.  For example, consider a CP-conserving
extended Higgs sector that has the
property that $\rho=1$ and no tree-level $ZW^\pm\phi^\mp$ couplings (where $\phi^\pm$ are physical charged scalars
that might appear in the scalar spectrum), then it follows that~\cite{Gunion:1990kf}
\beqa
      \sum_i  g^2_{h_iVV}&=&\frac{2m_V^4}{v^2}\,,\label{sumv}       \\
m_W^2 g_{h_i ZZ}&=&m_Z^2 g_{h_i WW}\,,\label{rel}
\eeqa
where the sum in eq.~(\ref{sumv}) is taken over all neutral CP-even scalars $h_i$.
In this case, it follows that $g_{h_i VV}\leq g_{hVV}$ for all $i$ (where $h$ is the Standard Model Higgs boson).
Models that contain only scalar singlets and doublets satisfy the requirements stated above and hence respect
the sum rule given in eq.~(\ref{sumv}) and the coupling relation given in eq.~(\ref{rel}).
However, it is possible to violate $g_{h_i VV}\leq g_{hVV}$
and $m_W^2 g_{h_i ZZ}=m_Z^2 g_{h_i WW}$ if tree-level $ZW^\pm\phi^\mp$
couplings are present.  Indeed, in this case, one can show that doubly charged
Higgs bosons must also occur in the model~\cite{Gunion:1990kf}..

\subsection{Chiral fermion mass generation and the unitarity of $VV \rightarrow f\bar{f}$ scattering}

In the Standard Model, left-handed fermions are electroweak doublets and right-handed fermions are
electroweak singlets.  A fermion mass term would combine a left-handed and right-handed fermion field, so
that gauge invariance does not allow for explicit fermion mass terms.  However, in the Standard Model, it
is possible to couple a left-handed and right-handed fermion field to the scalar doublets.   Such interactions
comprise the Yukawa couplings of the Standard Model.  When the scalar field acquires a vev, mass terms for
the quarks and charged leptons are generated.

One can repeat the previous analysis by again considering the theory of electroweak interactions
without the attendant scalar sector.  If one attempts
to simply add explicit mass terms to the quarks and charged leptons and the $W^\pm$ and $Z$ bosons, then the resulting theory would
again be mathematically inconsistent.  One signal of this inconsistency would be revealed by using the
theory to compute the cross section for the scattering of longitudinally polarized gauge bosons into a pair of top quarks,
$VV \rightarrow t\bar{t}$ at tree-level.
Such a calculation would yield a
scattering amplitude whose energy dependence grows with the the center of mass energy,
which violates tree-level unitarity.

Once again, the addition of an elementary particle that couples to $W^+ W^-$ and $ZZ$ with coupling strength $gm_V^2/m_W$
and couples to $t\bar{t}$ with coupling strength $gm_t/(2m_W)$ is sufficient to cure the bad high energy behavior.
Thus, if the newly discovered scalar $h$ is to be interpreted as the Higgs boson of the Standard Model,
one should confirm that
\beq \label{htt}
g_{hVV}g_{hff}=\frac{m^2_V m_f}{v^2}\,,
\eeq
for all quarks and charged leptons $f$ (in practice, $f=t,b, c$ and $\tau$ are the most relevant).   In models of extended
Higgs sectors, eq.~(\ref{htt}) would be replaced by a unitarity sum rule in which the right-hand side of eq.~(\ref{htt}) would
be the result of summing over multiple Higgs states in the model~\cite{Gunion:1990kf}.

\section{Theoretical structure of the Standard Model Higgs boson}

\subsection{Tree level Higgs boson couplings}

The Higgs sector of the Standard Model (SM), which takes the minimal form,
consists of
one isospin doublet scalar field $\Phi$ with the hypercharge
$Y=1$~\cite{Gunion:1989we}.
The most general SU(2)$\times$U(1)-invariant renormalizable Higgs
potential is give by
\begin{eqnarray}
V(\Phi)= \mu^2 |\Phi|^2 + \half\lambda |\Phi|^4\,.
\end{eqnarray}
The Higgs doublet field is parameterized as
\begin{eqnarray}
     \Phi = \left(
       \begin{array}{c} \omega^+ \\
                               v + (h + i z)/\sqrt{2} \end{array}\right),
\end{eqnarray}
where $\omega^\pm$ and $z$ represent the Nambu-Goldstone boson, $h$ is a
physical state, the Higgs boson, and  $v=174$ GeV is the vacuum expectation
value (vev) of the Higgs field.
The self-coupling constant $\lambda$ is positive to guarantee the stability of vacuum.
Assuming that $\mu^2 < 0$, the shape of the potential resembles the
Mexican hat, and the minimum of the scalar potential occurs
at $\langle \Phi \rangle = v$, where $\mu^2=-\lambda v^2$.
The SU(2)$\times$U(1)  electroweak symmetry is then broken down to
U(1)$_{\mathrm EM}$.  Expanding the scalar field around its vacuum
expectation value, the scalar potential immediately yields
the mass and the self-couplings of the Higgs boson $h$,
\begin{eqnarray}
  m_h^2 = 2\lambda v^2\,,  \qquad \lambda_{hhh} = 3\sqrt{2}\,\lambda
  v\,,\qquad \lambda_{hhhh}=3\lambda\,.
\end{eqnarray}
Hence, the Higgs mass and self-couplings are related by
\begin{eqnarray}
  \lambda_{hhh} = \frac{3m_h^2}{v\sqrt{2}}\,,\qquad \lambda_{hhhh}=\frac{3m_h^2}{2v^2}\,.
\end{eqnarray}
That is, the Higgs mass is directly related to the dynamics of the
Higgs sector.
In particular, the heavier the Higgs mass the stronger the strength of
the Higgs self-couplings.
Indeed, the observed Higgs mass of 126 GeV implies that
$\lambda_{hhhh}\simeq 0.787$, which implies that the Higgs
dynamics is weakly coupled.

The Higgs field couples to the weak gauge bosons (W and Z) via the
covariant derivative~\cite{Abers:1973qs},
$|{\cal D}_\mu \Phi |^2$, where ${\cal D}_\mu
= \partial_\mu + \half i g A_\mu^a \tau^a + \half i g' B_\mu Y$.  Here,
the $\tau^a$ are the usual Pauli matrices, the electric charge
operator is $Q=\half(\tau^3+Y)$, and $g$ and $g'$ are gauge coupling
constants for SU(2)$_T$ and U(1)$_Y$, respectively.  The masses of the
gauge bosons, which are generated by electroweak symmetry breaking via
the Higgs mechanism, are proportional to the neutral scalar field vev,
\begin{eqnarray}
 m_W^2 = \tfrac{1}{2} g^2 v^2, \hspace{1cm} m_Z^2 = \tfrac{1}{2} (g^2+g'^2) v^2.
\end{eqnarray}
Electroweak symmetry breaking also generates the $hWW$ and $hZZ$
couplings,
\beqa
 g(hWW) = \frac{1}{\sqrt{2}} g^2 v, \hspace{1cm} g(hZZ) = \frac{1}{\sqrt{2}} (g^2+g'^2) v.
\eeqa
Therefore, the gauge boson masses and the couplings to the Higgs boson
are related as noted previously in \eq{hvv}.
%\begin{eqnarray}
% g(hWW) = \frac{2 m_W^2}{v}, \hspace{1cm} g(hZZ) = \frac{2 m_Z^2}{v}.
%\end{eqnarray}

The Higgs field also couples to quarks and leptons via Yukawa
interactions.  For example, the coupling of the Higgs fields to the
three generations of quarks is given by
\beq \label{smyuk}
-\mathcal{L}_{\mathrm Yukawa}=Y_U^{0ij}(\overline U_L\lsup{\,0i} U_R^{0j}\Phi^{0\,\ast}
-\overline D_L\lsup{\,0i}
U_R^{0j}\Phi^-)+Y_D^{0ij}(\overline D_L\lsup{\,0i} D_R^{0j}
\Phi^0
+\overline U_L\lsup{\,0i} D_R^{0j}\Phi^{+})+{\mathrm h.c.}\,,
\eeq
where $i,j$ are generation labels, $U^{0}=(u^0,c^0,t^0)$ and
$D^{0}=(d^0,s^0,b^0)$ are the interaction-eigenstate quark fields,
and $Y^0_U$ and $Y^0_D$ are arbitrary
complex $3\times 3$ matrices (the sum over repeated indices is
implied).  In \eq{smyuk} we have introduced left and right-handed
quark fields via $Q^0_L\equiv P_LQ^0$ and $Q^0_R\equiv P_R Q^0$ where
$P_{R,L}\equiv\half(1\pm\gamma\lsub{5})$.
Setting the Goldstone boson fields to
zero and writing $\Phi^0=v+h^0/\sqrt{2}$, we identify the
quark mass matrices,
\beq
M_U^0\equiv vY^0_U\,,\qquad\quad
M_D^0\equiv vY^0_D\,.
\eeq
We now determine the quark mass eigenstate fields, $U=(u,c,t)$ and
$D=(d,s,b)$ by introducing the following unitary transformations,
\beq
U_L=  V_L^U U^0_L\,,\qquad U_R= V_R^U U^0_R\,,\qquad
D_L =  V_L^D D^0_L\,,\qquad D_R= V_R^D D^0_R\,,
\eeq
where $V_L^U$, $V_R^U$, $V_L^D$, and $V_R^D$ are unitary matrices chosen
such that
\beq
M_U\equiv V_L^{U} M^0_U V_R^{U\,\dagger}={\mathrm diag}(m_u\,,\,m_c\,,\,m_t)\,,\qquad
M_D\equiv V_L^{D} M^0_D V_R^{D\,\dagger}={\mathrm diag}(m_d\,,\,m_s\,,\,m_b)\,,
\eeq
such that the $m_i$ are the positive quark masses (this is the
\textit{singular value decomposition} of linear algebra).

Having diagonalized the quark mass matrices, the neutral Higgs Yukawa couplings
are automatically flavor-diagonal.   That is, if we define
\beq
Y_U\equiv  V_L^{U} Y^0_U V_R^{U\,\dagger}=M_D/v\,,\qquad\quad
Y_D\equiv  V_L^{D} Y^0_U V_D^{U\,\dagger}=M_U/v\,,
\eeq
then \eq{smyuk} can be rewritten in terms of quark mass eigenstates
as:
\beq \label{yukmass}
-\mathcal{L}_{\mathrm Yukawa}=\overline{U}_LY_U
U_R\Phi^{0\,\ast}-\overline{D}_LK^\dagger
Y_UU_R\Phi^-+\overline{U}_LKY_D D_R \Phi^++\overline{D}_L Y_D
D_R\Phi^0+{\mathrm h.c.}\,,
\eeq
where
\beq
K\equiv V_L^U V_L^{D\,\dagger}\,,
\eeq
is the Cabibbo-Kobayashi-Maskawa (CKM) matrix.
Hence the SM possesses no flavor-changing neutral currents (FCNCs) mediated
by neutral Higgs boson exchange at tree-level.
Note that independently  of the Higgs sector,
the quark couplings to $Z$ and $\gamma$ are automatically flavor diagonal.
Flavor dependence only enters the quark couplings to the $W^\pm$ via
the CKM matrix.

The Yukawa coupling of the Higgs doublets to the
leptons can be similarly treated  by
replacing $U\to N$, $D\to E$, $M_U\to 0$, $M_D\to M_E$ and
$K\to\mathds{1}$, where $N=(\nu_e,\nu_\mu,\nu_\tau)$, $E=(e,\mu,\tau)$
and $M_E$ is the diagonal charged lepton mass matrix.  In the present
treatment, the right-handed neutrino fields are absent, in which case the
neutrinos are exactly masses.  One can accommodate the very small
neutrino masses by  including the right-handed neutrino fields and
adding a SU(2)$\times$U(1)-invariant mass term $\overline{N}_L M_N
N_R+{\mathrm h.c.}$ to \eq{yukmass}.  Assuming that the eigenvalues of $M_N$ are much
larger than the scale of electroweak symmetry breaking, one obtains
three very light Majorana neutrino mass eigenstates due to the
seesaw mechanism~\cite{Minkowski:1977sc,GellMann:1980vs,Yanagida:1980xy,Mohapatra:1979ia,Mohapatra:1980yp}.
The very small neutrino masses have almost no impact on Higgs physics.
Consequently, we shall simply treat the neutrinos as massless in this chapter.

In the SM,  there is an universal relation between the
masses of the fundamental particles and their couplings to the Higgs boson,
\begin{eqnarray} \label{universal}
 \frac{g(hWW)}{\sqrt{2}m_W^2} = \frac{g(hZZ)}{\sqrt{2}m_Z^2} = \frac{y_c}{m_c} = \frac{y_\tau}{m_\tau}=\frac{y_b}{m_b}
 =\frac{y_t}{m_t}=\frac{\sqrt{2}\lambda(hhh)}{3m_h^2} = \cdot\cdot\cdot = \frac{1}{v}.
\end{eqnarray}
This is a unique feature of the SM with one Higgs doublet field.
By accurately measuring the mass and coupling to the Higgs boson independently for each particle,
one can test the mass generation mechanism of the SM by using this relation.
In Fig.~\ref{FIG:c-m},  the Standard Model relation is shown along with expected
precision from the full ILC program for the coupling determinations.
If the Higgs sector takes a non-minimal form, deviations from this
universal relation are expected.
%The deviation from the SM relation is the clear signature for
%non-minimal Higgs sectors.
Each non-minimal Higgs sector possesses a specific pattern of
deviations.  Thus, if the Higgs couplings can be measured with
sufficient precision, these measurements would provide a way to
distinguish among different models of extended Higgs sectors.
%The ILC is an ideal machine for such a purpose.
\begin{figure}[t]
\begin{center}
\includegraphics[width=90mm]{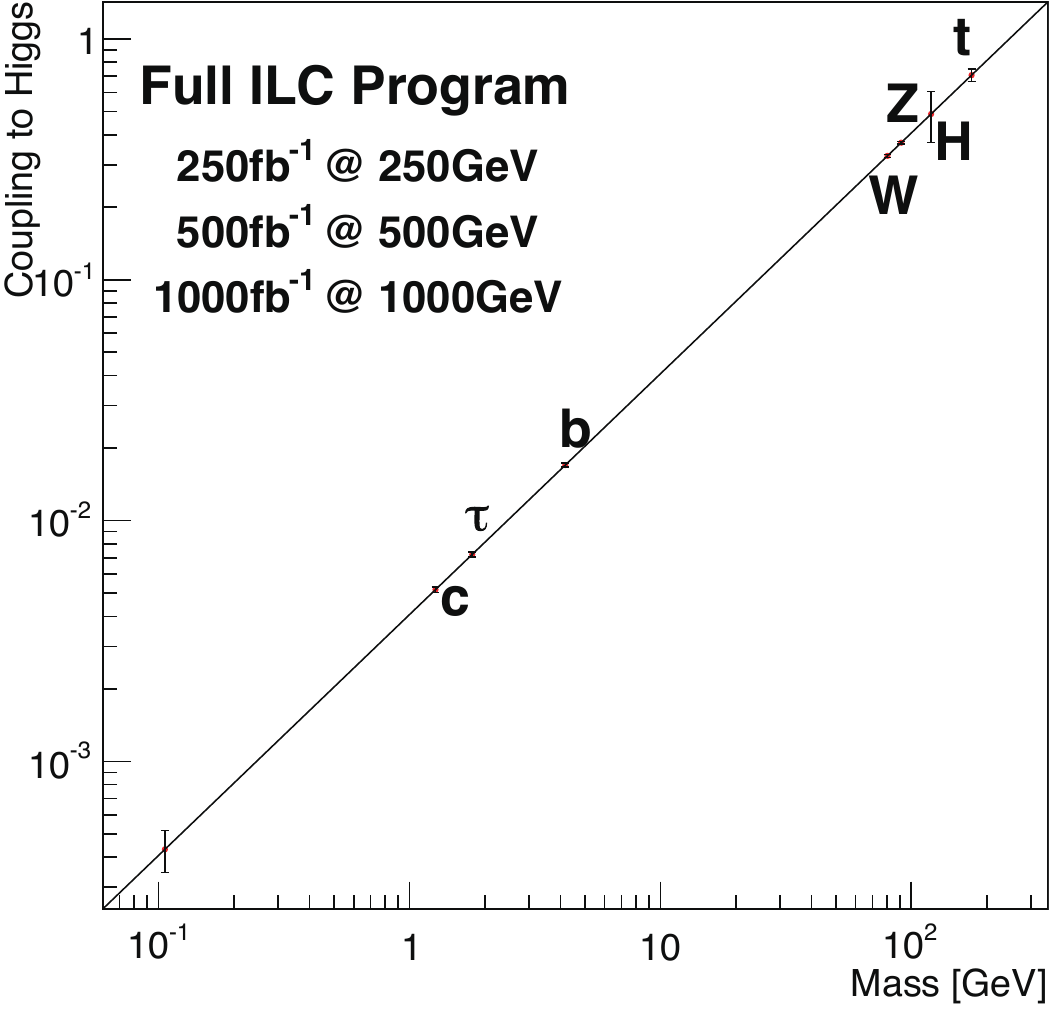}
\caption{ The Standard Model prediction
that the Higgs coupling to each particle is proportional to its mass.
Expected precision from the full ILC program for the coupling determination is also shown.}
\label{FIG:c-m}
\end{center}
\end{figure}

\subsection{Higgs couplings at one-loop}

The Higgs boson $h$ is a charge and color neutral state; hence, it does not couple to photons and gluons at the tree level.
However, beyond the tree level, the coupling $hgg$, $h\gamma\gamma$ and $h\gamma Z$ appear via the dimension-6 operators,
\begin{eqnarray}
       \frac{1}{\Lambda^2} |\Phi|^2 F_{\mu\nu} F^{\mu\nu}, \hspace{1cm}  \frac{1}{\Lambda^2} |\Phi|^2 G_{\mu\nu} G^{\mu\nu} .
\end{eqnarray}
In the SM, the effective $hgg$ coupling is induced at the one-loop
level via quark-loop diagrams, with the dominant contribution arising
from the top quark loop.  In contrast, the $h\gamma\gamma$ and
$h\gamma Z$ couplings are induced via the top loop diagram and the
W-loop diagram in the SM.  The leading contribution to the coupling of
$h\gamma\gamma$ is the $W^\pm$ boson loop, which is roughly 4.5
times larger in amplitude than the contribution of the top quark
loop.  Analytic expressions for the $h\to gg$ decay width and the
diphoton partial width are given by \cite{Ellis:1975ap,Shifman:1979eb}
\begin{eqnarray}
\Gamma(h\to gg) &=&
\frac{G_F \alpha_s^2 m_h^3}{512\sqrt{2}\pi^3} \left|N_c Q_t^2  A_{1/2}(\tau_t) \right |^2 ,   \\
\Gamma(h\to \gamma\gamma) &=&
\frac{G_F \alpha^2 m_h^3}{128\sqrt{2}\pi^3}\left|A_1(\tau_W)+ N_c Q_t^2  A_{1/2}(\tau_t) \right |^2 ,
\end{eqnarray}
where $G_F$ is the Fermi constant, $N_c=3$ is the number of color, $Q_t=+2/3$ is
the top quark electric charge in units of $e$, and $\tau_i\equiv
4m_i^2/m_h^2$
(for $i=t, W$).
Below the $WW$ threshold, the loop functions  for spin-1 ($W$ boson) and spin-1/2 (top quark)  particles are
given in the Appendix in Ref.~\cite{Carena:2012xa}.

In the limit that the particle running in the loop has a mass much heavier than the Higgs, we have
\begin{equation}
\label{eq:limit}
A_1 \rightarrow -7  \ , \qquad  N_c Q_t^2\, A_{1/2} \rightarrow \frac{4}3 N_c Q_t^2 \ .
\end{equation}
For a Higgs mass below the $WW$ threshold, the $W$ boson contribution is always dominant and monotonically decreasing from $A_1=-7$ for very small Higgs masses to $A_1\approx -12.4$ at the threshold,
while the top quark contribution is well-approximated by the asymptotic value of $(4/3)^2\approx 1.78$.
If we consider a Higgs mass at 126 GeV, the $W$ and top contributions are
\begin{equation}
m_h=126 \ \ {\mathrm GeV}: \ A_1=-8.32 \ , \quad  N_c Q_t^2 A_{1/2}=1.84\ .
\end{equation}
There have been many studies on the new physics loop
contributions to the $hgg$ as well as $h\gamma\gamma$ couplings.
Recently, Carena, Low and Wagner have performed a comprehensive study for the effects on
the diphoton width from adding new colorless charged particles of spin-0, spin-1/2, and spin-1,
which would interfere with the SM contributions~\cite{Carena:2012xa}.

In general, the contribution of a heavy particle in loop diagrams to
a decay amplitude
scales inversely as a positive power of the corresponding particle mass.
That is, in the infinite mass limit, the effects of the heavy particle loops
decouple, as a consequence of the decoupling theorem of Appelquist and
Carazzone~\cite{Appelquist:1974tg}.   However, the validity of the
decoupling theorem depends on the assumption that all couplings
are held fixed.  In cases where the
origin of the heavy particle mass is due to electroweak symmetry
breaking, the squared-mass of the boson or the mass of the
fermion is proportional to the vacuum expectation value as indicated
in \eq{universal}, and the constant of proportionality is the
corresponding Higgs coupling.  Thus in this case, one can only take the limit of large
mass by taking the corresponding Higgs coupling to be large.
As a result, the corresponding contribution of such particles in
loop diagrams do \textit{not} decouple.
% behaving positive power or logarithmic of the mass in the large mass
% limit.

For example, the loop contributions of weak gauge bosons
($W$ and $Z$) and chiral fermions such as the top quark to $h\to gg$ and
$h\to\gamma\gamma$ are examples where the corresponding
one-loop contributions to the decay amplitudes approach a constant in
the large mass limit.
Non-decoupling effects can also appear in radiative corrections to various observables.
As a dramatic example,
the one loop correction to the triple Higgs boson coupling $hhh$ is large because it receives a non-decoupling effect
proportional to the quartic power of the top quark mass after renormalization~\cite{Sirlin:1985ux,Kanemura:2004mg},
\begin{eqnarray}
    \lambda_{hhh}^{\mathrm ren} \simeq  \frac{3 m_h^2}{\sqrt{2}\,v} \left(1 - \frac{N_c m_t^4}{16 \pi^2} \right).
\end{eqnarray}

In theories that go beyond the Standard Model (BSM), new particles may exist that couple to the Higgs boson.
%One-loop effects can also appear in the loop contributions of new particles in models beyond the Standard Model.
For example, new bosonic loops yield positive contributions and fermionic loops yield negative
contributions to the $hhh$ coupling.
The loop induced couplings $hgg$, $h\gamma\gamma$, $hZ\gamma$ and the radiatively-corrected $hhh$ coupling
are particularly sensitive to
new particles in the loop when electroweak symmetry breaking provides the dominant contributions to the corresponding
new particle masses.  Thus, the couplings of the Higgs boson to SM fields can exhibit deviations from SM
predictions due to BSM loop effects even when the corresponding tree-level couplings are fixed to their SM values.

%Notice that these loop effect deviates coupling constant even when there is no mixing between the SM-like Higgs $h$
%and new particles.

The non-decoupling contribution of new particles can affect the
effective potential at finite temperatures.   For example, a new bosonic loop
contribution can make the electroweak phase transition sufficiently strongly
first order as required for a successful scenario of electroweak
baryogenesis~\cite{Cohen:1993nk,Morrissey:2012db}.  Such a
non-decoupling effect results in a large deviation in the $hhh$
coupling, so that one can test this scenario by measuring the $hhh$
coupling accurately.  In Ref.~\cite{Kanemura:2004ch}, the correlation
between the condition for a first order phase transition and the
deviation in the $hhh$ coupling is studied.  To test this scenario
of electroweak baryogenesis requires a determination of the $hhh$
coupling with a $10$--$20\%$ accuracy.  A
measurement of the $hhh$ coupling with the required precision can be achieved at
the ILC as shown in Section 5.6.

\subsection{Higgs decays}

The Higgs boson couples to all the particles of the SM. Therefore, there are many decay modes. In particular,
with the mass of about 126 GeV the Higgs boson decays
into $b\bar b$, $WW^\ast$,  $\tau^+\tau^-$, $gg$, $c\bar c$, $ZZ^\ast$,
$\gamma\gamma$ and $\gamma Z$,  $\mu\mu$,
where $\gamma\gamma$ and $\gamma Z$ are one-loop induced decay processes.

\begin{figure}[b]
\begin{center}
\includegraphics[width=90mm]{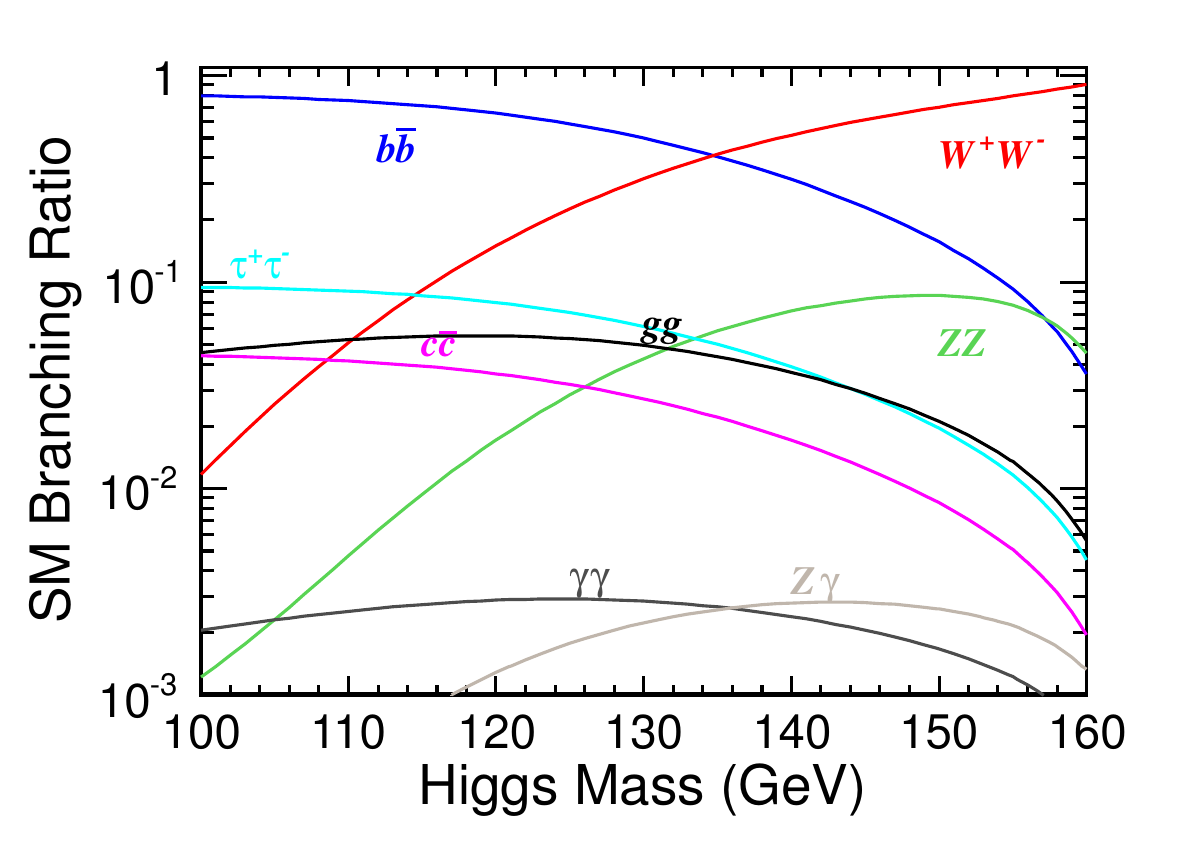}
\caption{Branching ratio of the Higgs boson in the SM as a function of the mass.\label{FIG:hbr_SM}}
\end{center}
\end{figure}

In Fig.~\ref{FIG:hbr_SM}, branching ratios for various decay modes are
shown as a function of the mass of the Higgs boson.  The decay
branching ratios strongly depend on the mass of the Higgs boson $m_h$.
In Tables~\ref{tab_bf_hff} and \ref{tab_bf_hvv}, the predicted values
of decay branching ratios of the Standard Model Higgs boson are listed
for $m_h= 125.0$, 125.3, 125.6, 125.9, 126.2 and 126.5
GeV~\cite{Heinemeyer:2013tqa}.  In Table~\ref{tab_bf_hvv} the
predicted values of the total decay width of the Higgs boson are also
listed.  It is quite interesting that with a Higgs mass of 126 GeV, a
large number of decay modes have similar sizes and are accessible to
experiments.  Indeed, the universal relation between the mass and the
coupling to the Higgs boson for each particle shown in
Fig.~\ref{FIG:c-m} can be well tested by measuring these branching
ratios as well as the total decay width accurately at the ILC.  For
example, the top Yukawa coupling and the triple Higgs boson coupling
are determined respectively by measuring the production cross sections
of top pair associated Higgs boson production and double Higgs boson
production mechanisms.

\begin{table}[t!]
\centering
\caption{The Standard Model values of branching ratios of fermionic decays of the Higgs boson
for each value of the Higgs boson mass $m_h$.
\label{tab_bf_hff}
\\}
\begin{tabular}{|c||c|c|c|c|c|}\hline
$m_h$ (GeV)   & $b\bar b$ & $\tau^+\tau^-$ & $\mu^+\mu^-$ & $c\bar c$ & $s \bar s$   \\
\hline
125.0  & 57.7 \% & 6.32 \% & 0.0219 \% & 2.91 \% & 0.0246 \%  \\
125.3  & 57.2 \% & 6.27 \% & 0.0218 \% & 2.89 \% & 0.0244 \%  \\
125.6  & 56.7 \% & 6.22 \% & 0.0216 \% & 2.86 \% & 0.0242 \%  \\
125.9  & 56.3 \% & 6.17 \% & 0.0214 \% & 2.84 \% & 0.0240 \%  \\
126.2  & 55.8 \% & 6.12 \% & 0.0212 \% & 2.81 \% & 0.0238 \%  \\
126.5  & 55.3 \% & 6.07 \% & 0.0211 \% & 2.79 \% & 0.0236 \%  \\
\hline
\end{tabular}
\end{table}
\begin{table}[t!]
\centering
\caption{The Standard Model values of branching ratios of bosonic decays of the Higgs boson for each
value of the Higgs boson mass $m_h$.
The predicted value of the total decay width of the Higgs boson is also listed for each value of $m_h$.
\label{tab_bf_hvv}
\\}
\begin{tabular}{|c||c|c|c|c|c|c|}\hline
$m_h$ (GeV)   & $gg$ & $\gamma\gamma$ & $Z\gamma$ & $W^+W^-$ & $ZZ$ & $\Gamma_H$  (MeV) \\
\hline
 125.0  & 8.57 \% & 0.228 \% & 0.154 \% & 21.5 \% & 2.64 \% & 4.07\\
 125.3  & 8.54 \% & 0.228 \% & 0.156 \% & 21.9 \% & 2.72 \% & 4.11\\
 125.6  & 8.52 \% & 0.228 \% & 0.158 \% & 22.4 \% & 2.79 \% & 4.15 \\
 125.9  & 8.49 \% & 0.228 \% & 0.162 \% & 22.9 \% & 2.87 \% & 4.20\\
 126.2  & 8.46 \% & 0.228 \% & 0.164 \% & 23.5 \% & 2.94 \% & 4.24\\
 126.5  & 8.42 \% & 0.228 \% & 0.167 \% & 24.0 \% & 3.02 \% & 4.29\\
\hline
\end{tabular}
\end{table}

\subsection{Higgs production at the ILC}

\begin{figure}[b!]
\begin{center}
\includegraphics[width=40mm]{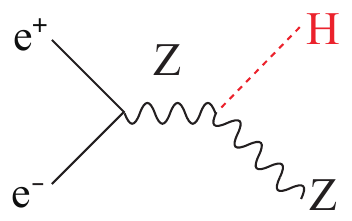}
\includegraphics[width=40mm]{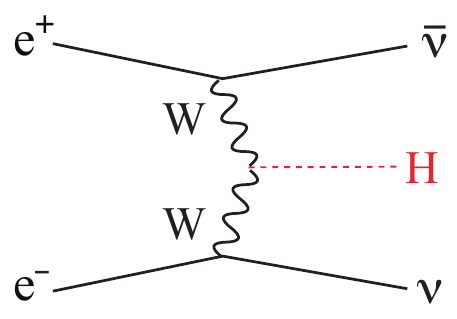}
\includegraphics[width=40mm]{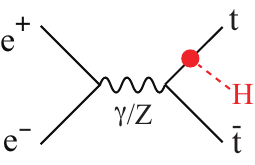}
\caption{Two important Higgs boson production processes at the ILC.
The Higgsstrahlung process (Left), the W-boson fusion process (Middle) and the top-quark association (Right). }
\label{FIG:Hpro_diags_SM}
\end{center}
\end{figure}

At the ILC, the SM Higgs boson $h$ is produced mainly via production mechanisms such as
the Higgsstrahlung process $e^+e^- \rightarrow Z^\ast  \rightarrow Z h$ (Fig.~\ref{FIG:Hpro_diags_SM} Left)
and the the weak boson fusion
processes $e^+e^- \rightarrow W^{+\ast} W^{-\ast} \nu \bar \nu \rightarrow  h \nu \bar \nu$
(Fig.~\ref{FIG:Hpro_diags_SM} (Middle))
and
$e^+e^- \rightarrow Z^{\ast} Z^{\ast} e^+e^- \rightarrow  h e^+e^-$.
The Higgsstrahlung process is an $s$-channel process so that it is maximal just above
the threshold of the process, whereas vector boson fusion is a
$t$-channel process which yields a cross section that grows logarithmically
with the center-of-mass energy.
 The Higgs boson is also produced in association with a fermion pair.
 The most important process of this type is Higgs production
in association with a top quark pair, whose typical diagram
 is shown in Fig.~\ref{FIG:Hpro_diags_SM} (Right).
The corresponding production cross sections at the ILC
are shown in Figs.~\ref{FIG:Higgs_xsec_SM} (Left) and (Right)
as a function of the collision energy by assuming
the initial electron (positron) beam polarization to be $-0.8$ ($+0.2$).

\begin{figure}[t!]
\begin{center}
%\begin{minipage}
\includegraphics[width=60mm]{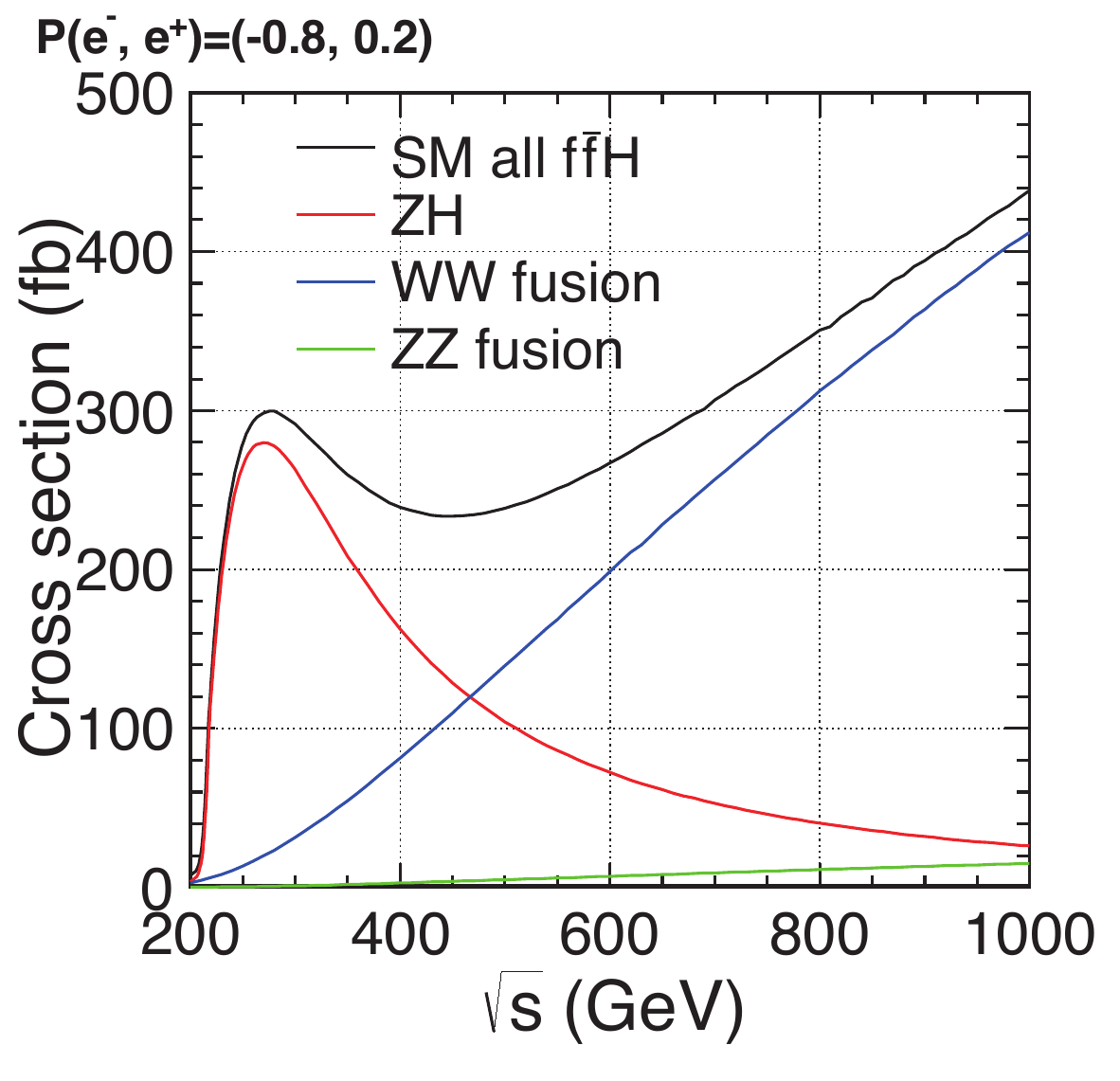}
%\end{minipage}
%\begin{minipage}
\includegraphics[width=85mm]{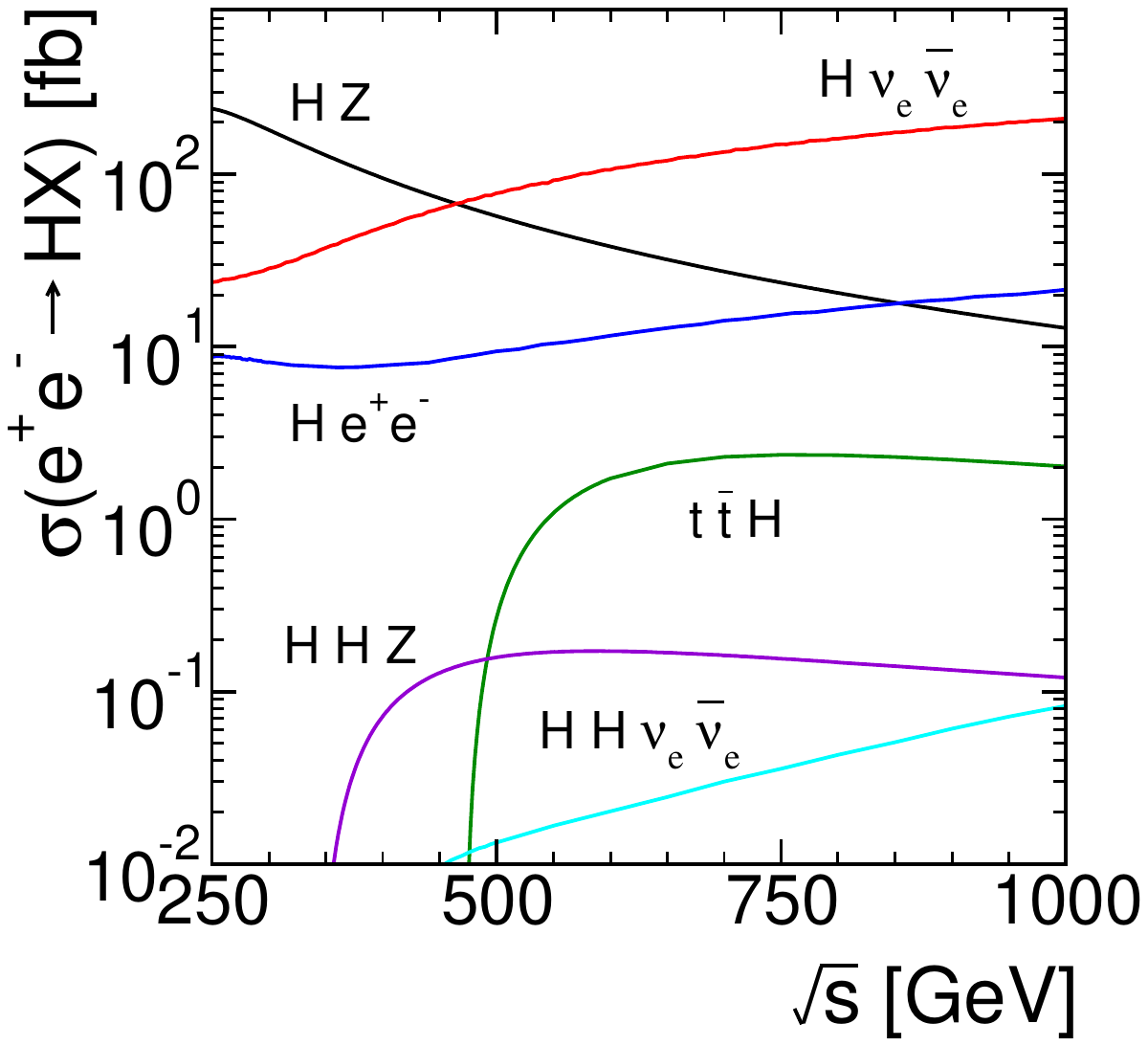}
%\end{minipage}
\caption{(Left)The production cross sections of the Higgs boson with the mass of 125 GeV
at the ILC as a function of the collision energy
$\sqrt{s}$. Polarization of the electron beam (80\%) and the positron beam (20\%) is
assumed.
(Right) The cross sections of the production processes $e^+e^- \to hZ$, $e^+e^- \to H \nu_e \bar \nu_e$,
$e^+e^- \to H e^+ e^-$, $e^+e^- \to t \bar t H$,
$e^+e^- \to HHZ$ and $e^+e^- \to HH \nu_e \bar \nu_e$ as a function of
the collision energy for the mass of 125 GeV.
No polarization is assumed for the initial electron and positron beams. \label{FIG:Higgs_xsec_SM}}
\end{center}
\end{figure}
\begin{figure}[h!]
\begin{center}
\includegraphics[width=50mm]{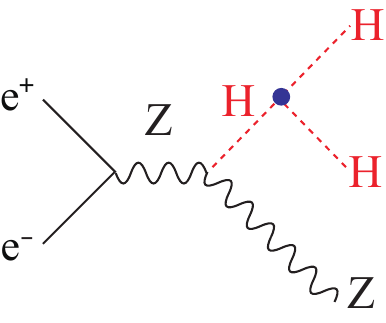}
\includegraphics[width=50mm]{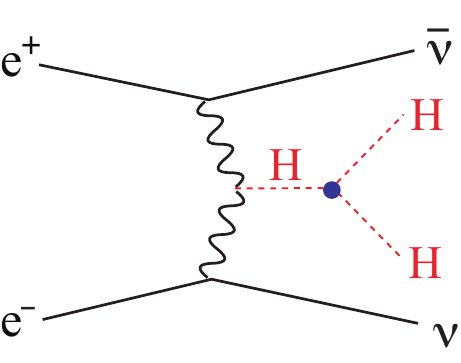}
\caption{Typical diagrams for double Higgs boson production via off-shell Higgsstrahlung (Left) and $W$-boson fusion (Right) processes. \label{FIG:HHpro_diags_SM}}
\end{center}
\end{figure}

The ILC operation will start with the $e^+e^-$ collision energy of 250
GeV (just above threshold for $hZ$ production), where
the Higgsstrahlung process is dominant and the contributions of
the fusion processes are small, as shown in Fig.~\ref{FIG:Higgs_xsec_SM} (Left) .
As the center-off-mass energy,$\sqrt{s}$  increases, the Higgsstrahlung
cross-section falls off as $1/s$.  Consequently,
the $W$-boson fusion mechanism is more significant at higher energies, and
its production cross section grows logarithmically and becomes
larger than that of the Higgsstrahlung cross section for $\sqrt{s} > 450$ GeV.
At $\sqrt{s} = 500$ GeV, both the Higgsstrahlung process and the W-boson fusion process are
important, and at $\sqrt{s}=1$ TeV the W-boson fusion is dominant.
The cross section of $e^+e^- \to t \bar t h$ is shown in
Fig.~\ref{FIG:Higgs_xsec_SM} (Right) .  The threshold of the production process
is roughly 480 GeV, so that the $t\bar{t}h$ cross section can be measured at the
ILC with the energy of 1 TeV.

Finally, the triple Higgs boson coupling can be determined from measuring the
double Higgs production mechanisms $e^+e^- \to Z hh$ and $e^+e^- \to \nu \bar \nu hh$
by extracting  the contribution of the Feynman diagram shown in Fig.~\ref{FIG:HHpro_diags_SM}.
The production cross section for the $Zhh$ process is typically of the order of 0.1 fb at the
collision energy just above the threshold at about 400 GeV
as shown in Fig.~\ref{FIG:Higgs_xsec_SM}(Right).
At the ILC with a center-of-mass energy of 500 GeV, the triple Higgs boson coupling can be measured
via this process.
On the other hand, at higher energies the cross section of the fusion process
$e^+e^- \to \nu \bar \nu hh$ becomes larger. This process becomes relevant
for the measurement of the triple Higgs boson coupling
at the energies around 1 TeV.

\subsection{Vacuum Stability}

The mass of the Higgs boson is proportional to the strength of the
Higgs
self-coupling $\lambda$ via $m_h^2 = 2 \lambda v^2$.
The magnitude of $\lambda$ at high energies can be predicted from the size of $\lambda$ at the electroweak scale
by using the renormalization group equation (RGE). The RGE for the coupling constant $\lambda$ is given by~\cite{Cabibbo:1979ay}
\begin{eqnarray}
16 \pi^2 \mu \frac{d}{d\mu} \lambda = 12(\lambda^2 +\lambda y_t^2-
y_t^4) -3\lambda(3g^2+g^{\prime\,2})+\tfrac{3}{4}\bigl[2g^4+(g^2+g^{\prime\,2})^2\bigr]+
\ldots,
 \end{eqnarray}
 where the $\ldots$ indicates terms proportional to the Yukawa
 couplings of the five light quarks and the charged leptons, which can
 be neglected in first approximation.  If the mass is large, $\lambda$
 is large and the $\beta$-function is positive. Then, $\lambda$ is
 larger for higher energies and blows up at some high energy point
 (the Landau pole), which can be below the Planck scale.  In contrast,
 when the mass is small and the $\beta$-function is negative due to
 the term proportional to the fourth power of the top-quark Yukawa
 coupling.  In this case, the coupling $\lambda$ decreases
as the energy scale increases and finally becomes negative.
If $\lambda$ is driven negative below the Planck scale (at which point
quantum gravitational effects would have to be taken into account), then we could
conclude that electroweak vacuum is not the global minimum,
since either a deeper scalar potential minimum exists or the
scalar potential is unbounded from below.  In either case, the
electroweak minimum would no longer be stable.
%When the mass of the Higgs boson
% was unknown, the allowed region of the mass was evaluated by using
 %these properties.
By assuming that the electroweak minimum is stable up to a given
 high energy scale $\Lambda$, below which the coupling $\lambda$ does
 not blow up nor is driven negative,
one can derive upper and the lower Higgs mass bounds
 as a function of $\Lambda$.

Given that the mass of the Higgs boson is
now known to be around 126 GeV, which corresponds to
$\lambda\sim 0.26$ at the electroweak scale,
it follows that the
$\beta$-function is negative. The recent RGE analysis in the NNLO
approximation~\cite{Degrassi:2012ry} shows that the scale $\Lambda$ where
$\lambda$ becomes negative is between $10^{7}$ GeV to $10^{15}$ GeV at the 3$\sigma$ level.
The main uncertainty comes from the top quark mass, $\alpha_s$, and the theoretical uncertainties in QCD corrections.
When the mass of the top quark is measured with the accuracy of about 30 MeV at the ILC, the cut-off scale of the
SM can be much better determined, as exhibited by Fig.~\ref{FIG:SM_VS}.
\begin{figure}[b!]
\includegraphics[width=55mm]{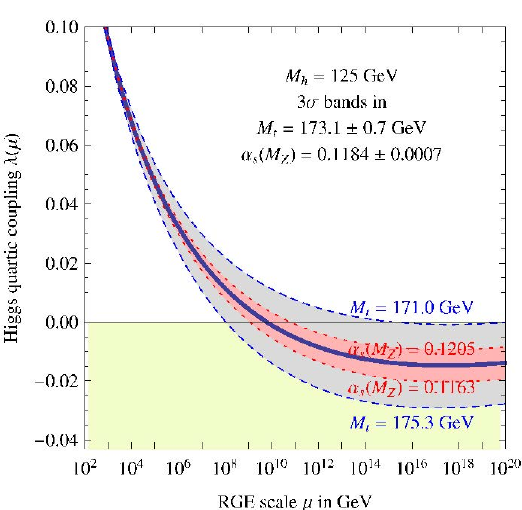}
\includegraphics[width=85mm]{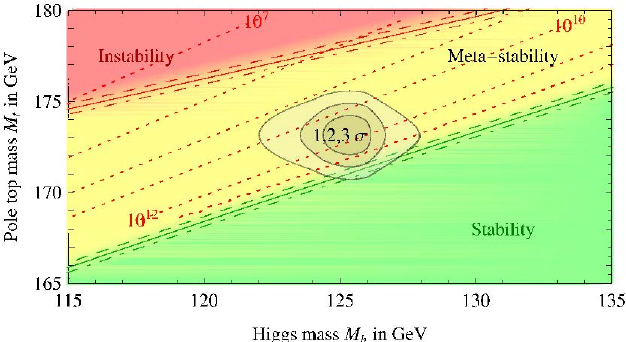}
\caption{
\textit{Left}: RG evolution of $\lambda$ varying $M_t$ and $\alpha_{\mathrm s}$ by $\pm 3\sigma$.
\textit{Right}: Regions of absolute stability, metastability and instability of the SM vacuum in the $M_t$--$M_h$
plane in the region of the preferred experimental range of $M_h$ and $M_t$ (the gray areas denote the allowed
region at 1, 2, and 3$\sigma$).
The three  boundaries lines correspond to $\alpha_s(M_Z)=0.1184\pm 0.0007$, and the grading of the
colors indicates the size of the theoretical error.
The dotted contour-lines show the instability scale $\Lambda$ in GeV assuming $\alpha_s(M_Z)=0.1184$.
}
\label{FIG:SM_VS}
\end{figure}

With a Standard Model Higgs mass of 126 GeV, the central value of
$\lambda$ is negative at the Planck scale.  Therefore, the electroweak
vacuum is not stable in the Standard Model unless new physics enters
below the Planck scale.  However, if we only require that the
electroweak vacuum is metastable, with a lifetime considerably longer
than the age of the Universe, then Fig.~\ref{FIG:SM_VS} indicates that
the Standard Model can be valid with no new physics required all the
way up to the Planck scale.

Finally, note that the bound from vacuum stability is largely relaxed when extended Higgs sectors
are considered where the lightest scalar behaves like the SM
Higgs boson\cite{Sher:1988mj,Kanemura:1999xf}.
For example, if we consider the scalar sector with two Higgs doublets, the cut off scale
where the vacuum stability is violated can be easily above the Planck scale.
Due to the loop contribution of extra scalar fields, the beta-function of the
quartic coupling constant of the SM-like Higgs boson is in general larger than that in the SM.
Therefore, the cutoff scale is higher than that of the SM.

\section{The two-Higgs-doublet model (2HDM)}
\label{tdhm}

Given that there are multiple generations of quarks and leptons, it is
reasonable to consider the possibility that the Higgs sector of
electroweak theory is also non-minimal.  The introduction of the
two-Higgs doublet extension of the Standard Model (2HDM)~\cite{Branco:2011iw}
was motivated for various reasons over the years.  It was initially introduced to
provide a possible new source of CP violation mediated by neutral
scalars~\cite{Lee:1973iz}.  Subsequently, the 2HDM was studied for phenomenological
reasons, as it provides for new scalar degrees of freedom including a
charged Higgs pair, a neutral CP-odd Higgs scalar in the case of a
CP-conserving scalar potential and neutral scalars of indefinite CP in
the case of a CP-violating scalar potential and/or vacuum~\cite{Gunion:1989we}.
These features yield new phenomenological signals for the production
and decay of fundamental spin-0 particles.

Today, the main motivation for the 2HDM is connected with models of
TeV-scale supersymmetry.  Such models provide the only natural
framework for weakly-coupled fundamental scalar particles (for further
details, see Section~\ref{alternate}).  In particular, the
minimal supersymmetric extension of the Standard Model (MSSM)
requires a Higgs sector with at least two Higgs doublet fields.
The MSSM Higgs sector is a 2HDM that is highly constrained
by supersymmetry.   The structure of the MSSM Higgs sector will be
explored further in Section~\ref{mssm}.

The most general version of the 2HDM, which contains all possible
renormalizable terms (mass terms and interactions) allowed by the
electroweak gauge invariance, is not phenomenologically viable due to
the presence of Higgs--quark Yukawa interaction terms that lead to
tree-level Higgs-mediated flavor changing neutral currents (FCNCs)~\cite{Glashow:1976nt,Paschos:1976ay}.
Such effects are absent in the MSSM Higgs sector due to the
constraints imposed by supersymmetry on the Yukawa interactions.  In
non-supersymmetric versions of the 2HDM, one can also naturally avoid
FCNCs by imposing certain simple discrete symmetries on the the scalar
and fermion fields, as discussed in Section~\ref{specialforms}.  These symmetries reduce the parameter freedom of
the 2HDM and automatically eliminate the dangerous FCNC interactions.
Nevertheless, it is instructive to examine the structure of the most
general 2HDM, as constrained versions of the 2HDM can then be examined
as limiting cases of the most general 2HDM.

\subsection{Model-independent treatment}
\label{modelind}

The scalar fields of the 2HDM are complex SU(2) doublet, hypercharge-one fields, $\Phi_1$ and $\Phi_2$,
where the corresponding vevs are $\langle\Phi_i\rangle=v_i$, and
$v^2\equiv |v_1|^2+|v_2|^2=(174~{\mathrm GeV})^2$ is fixed by the observed
$W$ mass, $m_W=gv/\sqrt{2}$.  The most general
renormalizable SU(2)$\times$U(1) gauge invariant scalar potential is given by
\beqa
\mathcal{V}&=& m_{11}^2 \Phi_1^\dagger \Phi_1+ m_{22}^2 \Phi_2^\dagger \Phi_2 -[m_{12}^2
\Phi_1^\dagger \Phi_2+{\mathrm h.c.}]
+\half \lambda_1(\Phi_1^\dagger \Phi_1)^2+\half \lambda_2(\Phi_2^\dagger \Phi_2)^2
+\lambda_3(\Phi_1^\dagger \Phi_1)(\Phi_2^\dagger \Phi_2)
\nn\\
&&\quad
+\lambda_4( \Phi_1^\dagger \Phi_2)(\Phi_2^\dagger \Phi_1)
 +\left\{\half \lambda_5 (\Phi_1^\dagger \Phi_2)^2 +\big[\lambda_6 (\Phi_1^\dagger
\Phi_1) +\lambda_7 (\Phi_2^\dagger \Phi_2)\big] \Phi_1^\dagger \Phi_2+{\mathrm
h.c.}\right\}\,.\label{genpot}
\eeqa

In the most general 2HDM, the fields $\Phi_1$ and $\Phi_2$ are indistinguishable.
Thus, it is always possible to define two orthonormal linear combinations of the two
doublet fields without modifying any prediction of the model.  Performing such a
redefinition of fields leads to a new scalar potential with the same form as \eq{genpot}
but with modified coefficients.  This implies that the coefficients that parameterize
the scalar potential in \eq{genpot} are not directly physical~\cite{Davidson:2005cw}.

To obtain a scalar potential that is more closely related to physical observables, one
can introduce the so-called \textit{Higgs basis} in which the redefined doublet fields (denoted below
by $H_1$ and $H_2$ have the property that $H_1$ has a non-zero vev whereas $H_2$ has a zero
vev~\cite{Branco:1999fs}. In particular, we define new Higgs doublet fields:
\beq \label{higgsbasispot}
H_1=\begin{pmatrix}H_1^+\\ H_1^0\end{pmatrix}\equiv \frac{v_1^* \Phi_1+v_2^*\Phi_2}{v}\,,
\qquad\quad H_2=\begin{pmatrix} H_2^+\\ H_2^0\end{pmatrix}\equiv\frac{-v_2 \Phi_1+v_1\Phi_2}{v}
 \,.
\eeq
It follows that $\vev{H_1^0}=v$ and $\vev{H_2^0}=0$.
The Higgs basis is uniquely defined
up to an overall rephasing, $H_2\to e^{i\chi} H_2$ (which does not alter the fact that
$\vev{H_2^0}=0$).  In the Higgs basis, the scalar potential is
given by~\cite{Branco:1999fs,Davidson:2005cw}:
\beqa \mathcal{V}&=& Y_1 H_1^\dagger H_1+ Y_2 H_2^\dagger H_2 +[Y_3
H_1^\dagger H_2+{\mathrm h.c.}]
+\half Z_1(H_1^\dagger H_1)^2+\half Z_2(H_2^\dagger H_2)^2
+Z_3(H_1^\dagger H_1)(H_2^\dagger H_2)
\nn\\
&&\quad
+Z_4( H_1^\dagger H_2)(H_2^\dagger H_1)
+\left\{\half Z_5 (H_1^\dagger H_2)^2 +\big[Z_6 (H_1^\dagger
H_1) +Z_7 (H_2^\dagger H_2)\big] H_1^\dagger H_2+{\mathrm
h.c.}\right\}\,,
\eeqa
where $Y_1$, $Y_2$ and $Z_1,\ldots,Z_4$ are real and uniquely defined,
whereas $Y_3$, $Z_5$, $Z_6$ and $Z_7$ are complex and transform under
the rephasing of $H_2$,
\vspace{-0.1in}
\beq \label{rephase}
[Y_3, Z_6, Z_7]\to e^{-i\chi}[Y_3, Z_6, Z_7] \quad{\mathrm and}\quad
Z_5\to  e^{-2i\chi} Z_5\,.
\eeq
After minimizing the scalar potential, $Y_1=-Z_1 v^2$
and $Y_3=-Z_6 v^2$.  This leaves 11 free parameters:
1 vev, 8 real parameters, $Y_2$, $Z_{1,2,3,4}$, $|Z_{5,6,7}|$, and two relative phases.

If $\Phi_1$ and $\Phi_2$ are indistinguishable fields, then
observables can only depend on combinations of Higgs basis
parameters that are independent of $\chi$.  Symmetries, such as
discrete symmetries or supersymmetry, can distinguish between
$\Phi_1$ and $\Phi_2$, which then singles out a specific physical basis for
the Higgs fields, and can yield additional observables such as
$\tan\beta\equiv |v_2|/|v_1|$ in the MSSM.

In the general 2HDM,
the physical charged Higgs boson is the charged component of the Higgs-basis doublet $H_2$, and its mass
is given by
\beq \label{chhiggsmass}
m_{H^\pm}^2=Y_{2}+Z_3 v^2\,.
\eeq
The three physical neutral Higgs boson mass-eigenstates
are determined by diagonalizing a $3\times 3$ real symmetric squared-mass
matrix that is defined in the Higgs basis~\cite{Branco:1999fs,Haber:2006ue}
\beq   \label{mtwo}
\mathcal{M}^2=2v^2\left( \begin{array}{ccc}
Z_1&\,\, \Re(Z_6) &\,\, -\Im(Z_6)\\
\Re(Z_6)  &\,\, \half (Z_{345}+Y_2/v^2) & \,\,
- \half \Im(Z_5)\\ -\Im(Z_6) &\,\, - \half \Im(Z_5) &\,\,
 \half (Z_{345}+Y_2/v^2)-\Re(Z_5)\end{array}\right),
\eeq
where $Z_{345}\equiv Z_3+Z_4+\Re(Z_5)$.The diagonalizing matrix is a $3\times 3$
real orthogonal matrix that depends on three angles:
$\theta_{12}$, $\theta_{13}$ and~$\theta_{23}$.
Under the rephasing $H_2\to e^{i\chi}H_2$~\cite{Haber:2006ue},
\beq
\theta_{12}\,,\, \theta_{13}~{\hbox{\text{are invariant, and}}}\quad
\theta_{23}\to  \theta_{23}-\chi\,.
\eeq
By convention, we choose
\beq \label{range}
-\half\pi\leq\theta_{12}\,,\,\theta_{13}<\half\pi\,.
\eeq

It is convenient to define
invariant combinations of  $\theta_{12}$ and $\theta_{13}$, denoted by $q_{k1}$ and $q_{k2}$
in Table~\ref{tabinv} below, where $k=1,2,3$ corresponds to the associated neutral Higgs mass eigenstate $h_k$~\cite{Haber:2006ue}.
\begin{table}[h!]
\centering
\caption{Invariant combinations of neutral Higgs mixing angles $\theta_{12}$ and $\theta_{13}$,
where $c_{ij}\equiv\cos\theta_{ij}$ and $s_{ij}\equiv\sin\theta_{ij}$.          \label{tabinv}
\\}
\begin{tabular}{|c||c|c|}\hline
$\phaa k\phaa $ &\phaa $q_{k1}\phaa $ & \phaa $q_{k2} \phaa $ \\
\hline
$1$ & $c_{12} c_{13}$ & $-s_{12}-ic_{12}s_{13}$ \\
$2$ & $s_{12} c_{13}$ & $c_{12}-is_{12}s_{13}$ \\
$3$ & $s_{13}$ & $ic_{13}$ \\ \hline
\end{tabular}
\end{table}

\noindent
The physical neutral Higgs states ($h_{1,2,3}$) are then given by:
\beq
h_k=\frac{1}{\sqrt{2}}\biggl\{q_{k1}^*\left(H_1^0-v\right)+q_{k2}^*H_2^0 e^{i\theta_{23}}+{\mathrm h.c.}\biggr\}\,.
\eeq
It is convenient to choose the mass ordering of the states such that $m_1<m_{2,3}$.  The mass ordering fixes the neutral
Higgs mixing angles $\theta_{12}$ and $\theta_{13}$.  Although the explicit formulae for the Higgs masses and mixing
angles are quite complicated, there are numerous relations among them which take on rather simple forms.  The
following results are noteworthy~\cite{Haber:2006ue,Haber:2010bw}:
\beqa
2Z_1 v^2&=&m_1^2 c_{12}^2 c_{13}^2+m_2^2 s_{12}^2 c_{13}^2 + m_3^2
s_{13}^2\,,\label{z1v} \\[5pt]
2\Re(Z_6\,e^{-i\theta_{23}})\,v^2 &=& c_{13}s_{12}c_{12}(m_2^2-m_1^2)\,,
\label{z6rv} \\[5pt]
2\Im(Z_6\,e^{-i\theta_{23}})\,v^2 &=& s_{13}c_{13}(c_{12}^2 m_1^2+s_{12}^2
m_2^2-m_3^2) \,, \label{z6iv}   \\[5pt]
2\Re(Z_5\,e^{-2i\theta_{23}})\,v^2 &=& m_1^2(s_{12}^2-c_{12}^2 s_{13}^2)+m_2^2(c_{12}^2-s_{12}^2 s_{13}^2)-m_3^2 c_{13}^2\,,
\label{z5rv} \\[5pt]
\Im(Z_5\,e^{-2i\theta_{23}})\,v^2 &=& s_{12}c_{12}s_{13}(m_2^2-m_1^2)\,. \label{z5iv}
\eeqa

%The corresponding neutral Higgs masses, denoted by
%$m_1$, $m_2$ and $m_3$, can be conveniently expressed in terms of $Z_1$, $Z_6$ and the neutral Higgs mixing angles,
%\beqa
%m_1^2 &=&
%\left[Z_1-\frac{s_{12}}{c_{12}c_{13}}\Re(Z_6\,e^{-i\theta_{23}})
%+\frac{s_{13}}{c_{13}}\Im(Z_6\,e^{-i\theta_{23}})\right]v^2\,,
%\label{m12}\\[5pt]
%m_2^2 &=&
%\left[Z_1+\frac{c_{12}}{s_{12}c_{13}}\Re(Z_6\,e^{-i\theta_{23}})
%+\frac{s_{13}}{c_{13}}\Im(Z_6\,e^{-i\theta_{23}})\right]v^2\,,
%\label{m22}\\[5pt]
%m_3^2 &=&
%\left[Z_1-\frac{c_{13}}{s_{13}}\Im(Z_6\,e^{-i\theta_{23}})\right]v^2\,.
%\label{m32}
%\eeqa
%Note that the physical masses are independent of a rephasing of the Higgs basis, $H_2\to e^{i\chi}H_2$, as expected.

If we also define the physical charged Higgs state by $H^\pm=e^{\pm i\theta_{23}}H_2^\pm$, then all the Higgs mass eigenstate
fields ($h_1$, $h_2$, $h_3$ and $H^\pm$) are invariant under the rephasing $H_2\to e^{i\chi}H_2$.
Thus, we have established a second well-defined basis of the general
2HDM, which corresponds to the
mass-eigenstate basis for the neutral Higgs bosons.

\subsection{Constraints on 2HDM scalar potential parameters}

The assumption of tree-level unitarity in the scattering of longitudinal gauge bosons
yields via the equivalence theorem upper bounds on the quartic couplings of the scalar
potential.  The bounds are rather simple when expressed in the Higgs basis.
For example, the following bounds obtained in Ref.~\cite{Haber:2010bw} are based on single
channel scattering processes,
\beqa
&& |Z_1|<4\pi\,,\qquad |Z_3|<8\pi\,,\qquad |Z_3+Z_4|<8\pi\,,\quad
|{\mathrm Re}(Z_5 e^{-2i\theta_{23}})|<2\pi\,,\nonumber \\
&& |{\mathrm Im}(Z_5 e^{-2i\theta_{23}})|<8\pi\,,\qquad
|{\mathrm Re}(Z_6 e^{-i\theta_{23}})|<2\pi \qquad   |{\mathrm Im}(Z_6 e^{-i\theta_{23}})|<\tfrac{8}{3}\pi\,.
\eeqa
There are no unitarity restrictions at tree-level on $Z_2$ and $Z_7$
as these quantities are absent from the neutral scalar mass matrix.
One can obtain somewhat improved tree-level bounds by considering
multi-channel scattering processes and analyzing the eigenvalues of
the corresponding amplitude matrices.  If the $|Z_i|$ are too large,
then the scalar sector becomes strongly coupled, and the tree-level
unitarity bounds become suspect.  Nevertheless, it is common practice
to consider weakly-coupled scalar sectors, in which case one should
not allow any of the $|Z_i|$ to become too large.  For practical
purposes, we shall assume that $|Z_i|\lesssim 2\pi$, in order to
maintain unitarity and perturbativity of tree-level amplitudes.

Additional constraints on the 2HDM scalar potential parameters arise from
the analysis of precision electroweak observables, which are sensitive to
Higgs bosons via loop corrections to Standard Model processes.
The $S$, $T$, and $U$ parameters, introduced by Peskin and
Takeuchi~\cite{Peskin:1991sw}, are independent ultraviolet-finite combinations of
radiative corrections to gauge boson two-point functions (the
so-called ``oblique'' corrections). The parameter $T$ is related to
the well known $\rho$-parameter of electroweak physics~\cite{Veltman:1977kh} by
$\rho - 1 = \alpha T$.  The oblique parameters can be expressed in
terms of the transverse part of the gauge boson two-point
functions.  For example,
\beq
\label{oblique}
\alpha\, T \equiv \frac{\Pi^{\mathrm new}_{WW}(0)}{m_W^2}
-\frac{\Pi^{\mathrm new}_{ZZ}(0)}{m_Z^2}\,,
\eeq
where $\alpha\equiv e^2/(4\pi)$ is the electromagnetic coupling defined
in the $\overline{\mathrm MS}$ scheme evaluated at $m_Z$.
The $\Pi_{V_a V_b}^{\mathrm new}$ are the new physics contributions
to the one-loop $V_a$---$V_b$ vacuum polarization functions (where $V=W$ or $Z$).
New physics contributions are defined as those that enter relative
to the Standard Model with the Higgs mass fixed to its observed value.
The definition of the two other oblique parameters $S$ and $U$ can be found
in Ref.~\cite{ErlerPDG}.

Explicit expressions for $S$, $T$ and $U$ in the general 2HDM have been written down
and the numerical contributions of the 2HDM states
(relative to that of the SM)  to the oblique parameters in the
2HDM have been studied in Refs.~\cite{Haber:2010bw,Froggatt:1991qw,Froggatt:1992wt,Grimus:2008nb}.
The general conclusion is that corrections
to $S$ and $U$ due to the contribution of the Higgs sector are small, consistent
with the present experimental limits.  However, the contributions
to $T$ may or may not be significant depending on the amount of custodial symmetry breaking
introduced by the 2HDM scalar potential.  Indeed, in Ref.~\cite{Haber:2010bw} it is shown that
a custodial symmetric scalar potential is one in which CP is conserved and in addition,
the following condition is satisfied,
\beq \label{basindcust}
\hspace{-0.65in} Z_4=\begin{cases}  \varepsilon_{56}|Z_5|\,, & \quad \text{for}~~Z_6\neq 0\,, \\[5pt]
 \varepsilon_{57}|Z_5|\,, & \quad \text{for}~~Z_7\neq 0\,, \\[5pt]
\pm |Z_5|\,, & \quad \text{for}~~Z_6=Z_7=0\,,
\end{cases}
 \eeq
where the two sign factors, $\varepsilon_{56}$ and $\varepsilon_{57}$ are defined by:
\beq  \label{signs}
\varepsilon_{56}=Z_5^* Z_6^2=\varepsilon_{56}|Z_5| |Z_6|^2\,,\quad\qquad
\varepsilon_{57}=Z_5^* Z_7^2=\varepsilon_{57}|Z_5| |Z_6|^2\,.
\eeq
Note that since the scalar potential is assumed to be CP conserving (otherwise the custodial
symmetry is violated), it follows that ${\mathrm Im}(Z_5^* Z_6^2)={\mathrm Im}(Z_5^* Z_7^2)=0$.
Hence $\varepsilon_{56}$ and $\varepsilon_{57}$ are real numbers of unit modulus.

A numerical study shows that the 2HDM contribution to $T$ is within the experimentally
measured bounds as long as there is not a significant mass splitting between the
charged Higgs boson and the heavy neutral Higgs bosons.  Such mass splittings would
require rather large values of some of the scalar quartic couplings (which would be
approaching their unitarity limits).

\subsection{Tree-level Higgs boson couplings--the general case}

The interactions of the Higgs bosons with the gauge bosons and the Higgs self-interactions,
when computed in the Higgs basis,  can be expressed in terms of the
parameters $Z_i$, $\theta_{12}$, $\theta_{13}$
and $\theta_{23}$~\cite{Haber:2006ue}.  In fact, the only combinations that
appear will be invariant with respect to the rephasing $H_2\to e^{i\chi} H_2$
(since observables cannot depend on the arbitrary angle $\chi$).
Indeed, the interaction terms will depend on the invariant quantities $q_{k1}$ and $q_{k2}$ defined
in Table~\ref{tabinv} and on invariant combinations of the $Z_i$ and $e^{-i\theta_{23}}$.

The interactions of the Higgs bosons and vector bosons of the Standard Model are given by:
\beqa
\mathcal{L}_{VVH}&=&\left(gm_W W_\mu^+W^{\mu\,-}+\frac{g}{2c_W} m_Z
Z_\mu Z^\mu\right)q_{k1} h_k
\,,\label{VVH}\\[8pt]
\mathcal{L}_{VVHH}&=&\left[\quarter g^2 W_\mu^+W^{\mu\,-}
+\frac{g^2}{8c_W^2}Z_\mu Z^\mu\right]h_k h_k \nonumber \\
&& +\left[\half g^2 W_\mu^+ W^{\mu\,-}+
e^2A_\mu A^\mu+\frac{g^2}{c_W^2}\left(\half
-s_W^2\right)^2Z_\mu Z^\mu +\frac{2ge}{c_W}\left(\half
-s_W^2\right)A_\mu Z^\mu\right]H^+H^-
\nn \\
&& +\biggl\{ \left(\half eg A^\mu W_\mu^+
-\frac{g^2s_W^2}{2c_W}Z^\mu W_\mu^+\right)
q_{k2}H^-h_k +{\mathrm h.c.}\biggr\}
\,,\label{VVHH} \\[8pt]
\mathcal{L}_{VHH}&=&\frac{g}{4c_W}\,\epsilon_{jk\ell}q_{\ell 1}
Z^\mu h_k\ddel_\mu h_j -\half g\biggl[iq_{k2}W_\mu^+
H^-\ddel\lsup{\,\mu} h_k
+{\mathrm h.c.}\biggr]\nonumber \\
&& +\left[ieA^\mu+\frac{ig}{c_W}\left(\half -s_W^2\right)
Z^\mu\right]H^+\ddel_\mu H^-\,, \label{VHH}
\eeqa
where $s_W\equiv \sin\theta_W$, $c_W\equiv\cos\theta_W$,
and the sum over pairs of repeated indices $j,k=1,2,3$ is implied.

The trilinear Higgs self-interactions are given by
\beqa
\mathcal{L}_{3h}&=&-\frac{v}{\sqrt{2}}\, h_j h_k h_\ell
\biggl[q_{j1}q_{k1}q_{\ell 1} Z_1
+q_{j2}q^*_{k2}q_{\ell 1}(Z_3+Z_4) + q_{j1} \Re(q_{k2}q_{\ell 2}Z_5\,
e^{-2i\theta_{23}}) \nonumber \\
&&
+3q_{j1}q_{k1}\Re\!\left(q_{\ell
2}Z_6\,e^{-i\theta_{23}}\right) +\Re(q_{j2}^*q_{k2}q_{\ell
2}Z_7\,e^{-i\theta_{23}})
\biggr]\nonumber \\
&&
-\sqrt{2}\,v\,h_k
H^+H^-\biggl[q_{k1}Z_3+\Re(q_{k2}\,e^{-i\theta_{23}}Z_7)\biggr]\,, \label{hhh}
\eeqa
where there is an implicit sum over repeated indices.
Note that the complex $Z_{5,6,7}$ are always paired with the correct
power of $e^{-i\theta_{23}}$ such that the corresponding product is
invariant under the rephasing of $H_2$.   Finally, for completeness, the
quadralinear Higgs self-interactions are exhibited,
\beqa
\mathcal{L}_{4h}&=&
-\tfrac{1}{8} h_j h_k h_l h_m \biggl[q_{j1}q_{k1}q_{\ell 1}q_{m1}Z_1
+q_{j2}q_{k2}q^*_{\ell 2}q^*_{m2}Z_2
+2q_{j1}q_{k1}q_{\ell 2}q^*_{m2}(Z_3+Z_4)\nonumber \\
&& +2q_{j1}q_{k1}\Re(q_{\ell 2}q_{m2}Z_5\,e^{-2i\theta_{23}})
+4q_{j1}q_{k1}q_{\ell 1}\Re(q_{m2}Z_6\,e^{-i\theta_{23}})
+4q_{j1}\Re(q_{k2}q_{\ell
  2}q^*_{m2}Z_7\,e^{-i\theta_{23}})\biggr]\nonumber\\
&& -\half h_j h_k H^+ H^-\biggl[q_{j2}q^*_{k2} Z_2 +
q_{j1}q_{k1}Z_3
+2q_{j1}\Re(q_{k2}Z_7\,e^{-i\theta_{23}})\biggr]  -\half Z_2 H^+H^- H^+ H^-.    \label{hhhh}
\eeqa
It is remarkable how compact the expressions are for the Higgs boson interactions
when written explicitly in terms of invariant quantities that can be directly
related to observables.

We next turn to the Higgs-fermion Yukawa couplings.  For simplicity, we focus
on the interaction of the Higgs bosons with three generations of quarks.  The
corresponding interactions with leptons are easily obtained from the latter by
the appropriate substitutions.   One starts out initially with a Lagrangian
expressed in terms of the scalar doublet fields $\Phi_i$ ($i=1,2$) and
interaction--eigenstate quark fields.  After electroweak symmetry breaking,
one can transform the scalar doublets into the Higgs basis fields $H_1$ and $H_2$.
At the same time,
one can identify the $3\times 3$ quark mass matrices.  By redefining the left
and right-handed quark fields appropriately, the quark
mass matrices are transformed into diagonal form, where the diagonal elements are real
and non-negative.  The resulting Higgs--quark Yukawa couplings are given by~\cite{Haber:2010bw}
\beqa
-\mathcal{L}_{\mathrm Y}&=&\overline U_L (\kappa^U H_1^{0\,\dagger}
+\rho^U H_2^{0\,\dagger})\ur
-\anti D_L K^\dagger(\kappa^U H_1^{-}+\rho^U H_2^{-})\ur \nonumber \\
&& +\anti U_L K (\kappa^{D\,\dagger}H_1^++\rho^{D\,\dagger}H_2^+)\dr
+\anti D_L (\kappa^{D\,\dagger}H_1^0+\rho^{D\,\dagger}H_2^0)\dr+{\mathrm h.c.},
\label{lyuk}
\eeqa
where $U=(u,c,t)$ and $D=(d,s,b)$ are the mass-eigenstate quark fields, $K$ is the
%Cabibbo-Kobayashi-Maskawa (CKM)   %defined earlier
CKM
mixing matrix and $\kappa$ and $\rho$ are $3\times 3$
Yukawa coupling matrices.  Note that $Q_{R,L}\equiv P_{R,L}Q$,
where $Q=U$ or~$D$ and $P_{R,L}\equiv\half(1\pm\gamma\lsub{5})$ are
the right and left handed projection operators, respectively.

By setting $H_1^0=v$ and $H_2^0=0$, one can relate
$\kappa^U$ and $\kappa^D$ to the diagonal
quark mass matrices $M_U$ and $M_D$,
respectively,
\beq \label{mumd}
M_U=v\kappa^U={\mathrm diag}(m_u\,,\,m_c\,,\,m_t)\,,\qquad
M_D=v\kappa^{D\,\dagger}={\mathrm
diag}(m_d\,,\,m_s\,,\,m_b) \,.
\eeq
However, the complex matrices $\rho^Q$ ($Q=U,D$) are unconstrained.   Moreover,
\beq \label{rhoq}
\rho^Q\to e^{i\chi}\rho^Q\,,
\eeq
under the rephasing $H_2\to e^{i\chi}H_2$.

The Yukawa coupling of the Higgs doublets to the
leptons can be similarly treated  by
replacing $U\to N$, $D\to E$, $M_U\to 0$, $M_D\to M_E$ and
$K\to\mathds{1}$, where $N=(\nu_e,\nu_\mu,\nu_\tau)$, $E=(e,\mu,\tau)$
and $M_E$ is the diagonal charged lepton mass matrix.

To obtain the physical Yukawa couplings of the Higgs boson, one must relate the
Higgs basis scalar fields to the Higgs mass-eigenstate fields.  This yields the
physical Higgs--quark Yukawa couplings,
\beqa
  -\mathcal{L}_Y &=& \frac{1}{\sqrt{2}}\,\overline D\sum_{k=1}^3
\biggl\{q_{k1}\frac{M_D}{v} +
q_{k2}\,[e^{i\theta_{23}}\rho^D]^\dagger P_R+
q^*_{k2}\,e^{i\theta_{23}}\rho^D P_L\biggr\}Dh_k \nonumber \\
&&  +\frac{1}{\sqrt{2}}\,\overline U \sum_{k=1}^3\biggl\{q_{k1}\frac{M_U}{v}+
q^*_{k2}\,e^{i\theta_{23}}\rho^U P_R+
q_{k2}\,[e^{i\theta_{23}}\rho^U]^\dagger P_L\biggr\}U h_k
\nonumber \\
&& +\biggl\{\overline U\bigl[K[e^{i\theta_{23}}\rho^D]^\dagger
P_R-[e^{i\theta_{23}}\rho^U]^\dagger KP_L\bigr] DH^+ +{\mathrm
h.c.}\biggr\}\,. \label{YUK}
\eeqa
The
combinations $e^{i\theta_{23}}\rho^U$ and $e^{i\theta_{23}}\rho^D$ that appear in the interactions above are invariant under
the rephasing of $H_2$.

Note that no $\tan\beta$ parameter appears above!  This is because $\tan\beta$ is
the absolute value of the ratio of the
two neutral Higgs vevs defined with respect to some arbitrary basis of the scalar doublets.
But, since the two Higgs doublet fields
are identical at this stage, there is no physical principle that
singles out a particular basis.  Indeed, physical observables cannot depend on the choice of basis.  Hence,
$\tan\beta$ is an unphysical parameter.  In contrast, all parameters that appear in \eq{YUK} are physical
and can be directly related to some observable.

It is convenient to rewrite the Higgs-fermion Yukawa couplings in terms of
the following two $3\times 3$ hermitian matrices that are invariant
with respect to the rephasing of $H_2$,
\beqa
\rho^Q_R &\equiv& \frac{v}{2}\,M^{-1/2}_Q
\biggl\{e^{i\theta_{23}}\rho^Q +
[e^{i\theta_{23}}\rho^Q]^\dagger\biggr\}M^{-1/2}_Q\,,
\qquad \text{for $Q=U,D$}\,,\nonumber \\[6pt]
\rho^Q_I &\equiv& \frac{v}{2i}M^{-1/2}_Q
\biggl\{e^{i\theta_{23}}\rho^Q -
[e^{i\theta_{23}}\rho^Q]^\dagger\biggr\}M^{-1/2}_Q\,,
\qquad \text{for $Q=U,D$}\,.\label{rhodefs}
\eeqa
%where $M_U$ and $M_D$ are the diagonal up and down-type
%fermion mass matrices [cf.~\eq{mumd}] and the matrix Yukawa
%coupling matrices are introduced in \eq{lyuk}.
Then, the Yukawa couplings take the following form:
\beqa
\!\!\!\!\!\!\!\!\!\!
-\mathcal{L}_Y &=& \frac{1}{v\sqrt{2}}\,\overline D\sum_{k=1}^3 M_D^{1/2}
\biggl\{q_{k1}\mathds{1}  + \Re(q_{k2})\rho^D_R+\Im(q_{k2})\rho^D_I+i\gamma\lsub{5}\bigl[\Im(q_{k2})\rho^D_R
-\Re(q_{k2})\rho^D_I\bigr] \biggr\} M_D^{1/2}Dh_k  \nonumber \\
&&  +\frac{1}{v\sqrt{2}}\,\overline U \sum_{k=1}^3 M_U^{1/2}\biggl\{q_{k1}\mathds{1}
+ \Re(q_{k2})\rho^U_R+\Im(q_{k2})\rho^U_I-i\gamma\lsub{5}\bigl[\Im(q_{k2})\rho^U_R
-\Re(q_{k2})\rho^U_I\bigr] \biggr\} M_U^{1/2}Uh_k \nonumber \\
&& +\frac{1}{v}\biggl\{\overline U\bigl[KM_D^{1/2}(\rho^D_R-i\rho^D_I)
M_D^{1/2}P_R-M_U^{1/2}(\rho^U_R-i\rho^U_I) M_U^{1/2}KP_L\bigr] DH^+ +{\mathrm
h.c.}\biggr\}, \label{YUK2}
\eeqa
where $\mathds{1}$ is the $3\times 3$ identity matrix.
The appearance of unconstrained complex $3\times 3$ Yukawa matrices
$\rho^Q_{R,I}$ in \eq{YUK2} indicates the presence of potential flavor-changing neutral Higgs--quark interactions.
If the off-diagonal elements of $\rho^Q_{R,I}$ are unsuppressed, they will generate tree-level Higgs-mediated FCNCs
that are incompatible with the strong suppression of FCNCs observed in nature.

\subsection{Tree-level Higgs boson couplings---the CP-conserving case}

It is instructive to consider the case of a CP-conserving Higgs scalar potential.  If CP is \textit{explicitly} conserved,
then there exists a basis for the scalar fields in which all the parameters of the scalar potential are simultaneously
real.  Such a basis (if it exists) is called a \text{real basis}.  If in addition the vacuum conserves CP, then
a real basis exists in which the vevs are simultaneously real.  In
this case, a
real Higgs basis exists that is unique up to a redefinition of $H_2\to -H_2$.
Thus, without loss of generality, we can adopt a convention in which
the sign of $Z_6$ or $Z_7$ is fixed to be either positive or negative.

Having chosen a real Higgs basis, one can diagonalize the neutral Higgs mass matrix given in
\eq{mtwo}.
One immediately finds two neutral CP-even scalars, $h$ and $H$ (with $m_h<m_H)$ with squared-masses,
\beq
m_{H,h}^2=\tfrac{1}{2}\biggl\{Y_2+\bigl(Z_{345}+2Z_1\bigr)v^2\pm
\sqrt{\bigl[Y_2+\bigl(Z_{345}-2Z_1\bigr)v^2\bigr]^2+16Z_6^2v^4}\,\biggr\}\,,
\eeq
where $Z_{345}\equiv Z_3+Z_4+Z_5$ (since $Z_5$ is real by assumption),
and a CP-odd scalar $A$, with squared-mass
\beq
m_A^2=Y_2+(Z_3+Z_4-Z_5)v^2\,.
\eeq
Only one neutral Higgs mixing angle $\theta_{12}$ is required, since
$\theta_{13}=0$ and  $e^{i\theta_{23}}={\mathrm sgn}~Z_6$.
It is conventional to
rotate from the Higgs basis to an arbitrary basis by an angle
$\beta$.  In this basis, the conventionally defined Higgs mixing angle
$\alpha$ is related to $\theta_{12}$ by,
\beq \label{alphacp}
\alpha=\beta-\theta_{12}-\tfrac{1}{2}\pi\,.
\eeq
The quantity $\beta-\alpha=\theta_{12}+\tfrac{1}{2}\pi$ is clearly
independent of the choice of basis used to define $\beta$.  In this
notation, we have
\beq \label{bma}
\cos\theta_{12}=\sin(\beta-\alpha)\,,\qquad\quad
\sin\theta_{12}=-\cos(\beta-\alpha)\, {\mathrm sgn}~Z_6\,,
\eeq
where $0\leq\beta-\alpha<\pi$ [in light of \eq{range}].
\Eqs{z1v}{z6rv} yield
\beqa
\cos^2(\beta-\alpha)&=&\frac{2Z_1 v^2-m_1^2}{m_2^2-m_1^2}\,,\label{c2exact}\\
\sin(\beta-\alpha)\cos(\beta-\alpha)&=&-\frac{2Z_6v^2}{m_2^2-m_1^2}
\label{scexact}\,.
\eeqa

It is convenient to
adopt a convention where $Z_6>0$ in which case we can take
$\theta_{23}=0$.  In this convention, \eq{scexact} implies that the
sign of $\cos(\beta-\alpha)$ is negative (since by assumption,
$0\leq\sin(\beta-\alpha)\leq 1$ and $m_2>m_1$).
The corresponding invariant combinations of neutral Higgs mixing
angles given in Table~\ref{tabinv} simplify as shown in
Table~\ref{tabinvcp} below.
\begin{table}[h!]
\centering
\caption{Invariant combinations of Higgs mixing angles in the
CP-conserving case, where $\cbma\equiv\cos(\beta-\alpha)$ and
$\sbma\equiv\sin(\beta-\alpha)$, in a convention where $Z_6>0$.  These are obtained from
Table~\ref{tabinv} by setting $\theta_{12}=\beta-\alpha-\half\pi$
and $\theta_{13}=0$.  \label{tabinvcp}
\\}
\begin{tabular}{|c||c|c|}\hline
$\phaa k\phaa $ &\phaa $q_{k1}\phaa $ & \phaa $q_{k2} \phaa $ \\
\hline
$1$ & $\phantom{-}\sbma$ & $\cbma$ \\
$2$ & $-\cbma$ & $\sbma$ \\
$3$ & $0$ & $i$ \\ \hline
\end{tabular}
\end{table}

\noindent
Using the results of Table~\ref{tabinvcp} and identifying
$h_1=h$, $h_2=-H$ and $h_3=A$ to match the standard conventions
of the CP-conserving 2HDM, we can obtain
from eqs.~(\ref{VVH})--(\ref{hhhh}) and \eq{YUK2}
the complete list of Higgs couplings in the CP-conserving case.
The properties of the three-point and
four-point Higgs boson-vector boson couplings are conveniently summarized
by listing the couplings that are proportional
to either $\sbma$ or $\cbma$, and the couplings
that are independent of $\beta-\alpha$~\cite{Gunion:1989we}:
\beq
\renewcommand{\arraycolsep}{1cm}
\let\us=\underline
\begin{array}{lll}
\us{\cos(\beta-\alpha)}&  \us{\sin(\beta-\alpha)} &
\us{\mathrm{angle-independent}} \\ [3pt]
\noalign{\vskip3pt}
       \hh W^+W^-&        \hl W^+W^- &  \qquad\longdash   \\
       \hh ZZ&            \hl ZZ  & \qquad\longdash  \\
       Z\ha\hl&          Z\ha\hh  & ZH^+H^-\,,\,\,\gamma H^+H^-\\
       W^\pm H^\mp\hl&  W^\pm H^\mp\hh & W^\pm H^\mp\ha \\
       ZW^\pm H^\mp\hl&  ZW^\pm H^\mp\hh & ZW^\pm H^\mp\ha \\
    \gamma W^\pm H^\mp\hl&  \gamma W^\pm H^\mp\hh & \gamma W^\pm H^\mp\ha \\
   \quad\longdash    &\quad\longdash  & VV\phi\phi\,,\,Z\gamma H^+
                                        H^-,\,\gamma\gamma H^+ H^-
\end{array}
\label{littletable}
\eeq
where $\phi\phi=\hl\hl$, $\hh\hh$, $AA$, or $H^+ H^-$ and $VV=W^+W^-$ or $ZZ$.
Note in particular that \textit{all} vertices
in the theory that contain at least
one vector boson and \textit{exactly one} non-minimal Higgs boson state
($\hh$, $\ha$ or $\hpm$) are proportional to $\cos(\beta-\alpha)$.
This can be understood as a consequence of unitarity sum rules which
must be satisfied by the tree-level amplitudes of the
theory \cite{Cornwall:1974km,Lee:1977eg,Weldon:1984wt,Gunion:1990kf}.

\subsection{The decoupling/alignment limit of the 2HDM}
\label{sec:decouplalign}

Many models of extended Higgs sectors possess a decoupling limit, in which
there exists one scalar whose properties coincide with those of the
Standard Model Higgs boson~\cite{Haber:1989xc}.
The decoupling limit of the 2HDM corresponds
to the limiting case in which the Higgs doublet $H_2$ (in the Higgs basis)
receives a very large mass and is therefore decoupled from the
theory.  This can be achieved by
assuming that $Y_2\gg v^2$ and $|Z_i|\lesssim\mathcal{O}(1)$ for all~$i$~\cite{Gunion:2002zf,Haber:2006ue}.
The effective low energy theory consists of a single Higgs doublet field
(namely, $H_1$), corresponding to the Higgs sector of the Standard Model.
The alignment limit of the 2HDM corresponds to the limiting case in
which the mixing of the two Higgs doublet fields $H_1$ and $H_2$ (in
the Higgs basis) is suppressed~\cite{Craig:2013hca}.  This can be achieved by assuming that
$|Y_3|\ll 1$ [which implies that $|Z_6|\ll 1$ via the scalar potential
minimum conditions given below \eq{rephase}].  In both the decoupling
and alignment limits, the neutral Higgs mass eigenstate is
approximately given by $\sqrt{2}\,\Re(H_1^0-v)$, and its couplings
approach those of the Standard Model (SM) Higgs boson.
In this section, we provide a general analysis of the
decoupling/alignment limit of the 2HDM following the work of Ref.~\cite{Haber:2013}.

It is convenient to order the neutral scalar masses such that $m_1\leq m_{2,3}$
and define the invariant Higgs mixing angles accordingly.
If we identify $h_1$ as the SM-like Higgs boson so that
\beq
\frac{g_{h_1VV}}{g_{h_{\mathrm SM}VV}}=c_{12} c_{13}\simeq 1\,,\qquad \text{where $V=W$ or $Z$}\,,
\eeq
then it follows that $s_{12}$, $s_{13}\ll 1$.  Thus, in the
decoupling/alignment limit, \eqs{z6rv}{z6iv} yield~\cite{Haber:2006ue}:
\beqa
s_{12}\equiv\sin\theta_{12}&\simeq &
\frac{2\,\Re(Z_6 e^{-i\theta_{23}})v^2}{m_2^2-m_1^2}\ll 1\,, \label{done}\\
s_{13}\equiv\sin\theta_{13}&\simeq&
-\frac{2\,\Im(Z_6 e^{-i\theta_{23}})v^2}{m_3^2-m_1^2}\ll 1\,.\label{dtwo}
\eeqa
In addition, eq.~(\ref{z5iv}) implies that one additional small quantity characterizes the
decoupling/alignment limit,
\beq \label{dthree}
\Im(Z_5 e^{-2i\theta_{23}})\simeq \frac{(m_2^2-m_1^2) s_{12}s_{13}}{v^2}
\simeq-\frac{2\,\Im(Z_6^2 e^{-2i\theta_{23}})v^2}{m_3^2-m_1^2}\ll 1\,.
\eeq
Note that in the decoupling/alignment limit, \eq{z5rv} yields
\beq
m_2^2-m_3^2\simeq 2\,\Re(Z_5 e^{-2i\theta_{23}})v^2\,.
\eeq

In the decoupling limit, $m_1^2\simeq 2Z_1 v^2\ll m_2^2$, $m_3^2$,
$m_{H^\pm}^2$, which guarantees that \eqst{done}{dthree} are satisfied.
In addition, $m_2^2-m_3^2\simeq
m_{H^\pm}^2-m_3^2=\mathcal{O}(v^2)$.
That is, the
mass splittings among the heavy Higgs states are of order $v^2/m_3$.
In the alignment limit, $|Z_6|\ll 1$ ensures that \eqst{done}{dthree} are satisfied.
We again find that $m_1^2\simeq 2Z_1 v^2$, but
with no requirement that $h_2$, $h_3$ and $H^\pm$ must be
significantly heavier than $h_1$.
The couplings of $h_1$ to the vector bosons and fermions and the Higgs
self-couplings in the approach to the decoupling/alignment limit are exhibited in
Table~\ref{tabdecouplings}.
\begin{table}[h!]
\centering
\caption{2HDM couplings of the SM-like Higgs boson $h\equiv h_1$ normalized to
those of the SM Higgs boson, in the
decoupling/alignment limit.   In the Higgs couplings to vector bosons,
$VV=W^+ W^-$ or $ZZ$.
In the Higgs self-couplings,
$Z_{6R}\equiv \Re(Z_6 e^{-i\theta_{23}})$ and $Z_{6I}\equiv \Im(Z_6 e^{-i\theta_{23}})$.
For the fermion couplings,
$D$ is a column vector of three down-type fermion fields
(either down-type quarks or charged leptons)
and $U$ is a column vector of three up-type quark fields.  The
$3\times 3$ hermitian matrices, $\rho^Q_R$ and $\rho^Q_I$ (where $Q=U$
or $D$) are defined in \eq{rhodefs}.  The normalization of the
pseudoscalar coupling of the Higgs boson $h$ to fermions is relative to
the corresponding scalar coupling to fermions.  In the third column, the first non-trivial correction to decoupling/alignment is exhibited.
\label{tabdecouplings} \\}
\begin{tabular}{|c||c|c|}\hline
Higgs interaction & 2HDM coupling & decoupling/alignment limit\\
\hline
$hVV$ & $c_{12} c_{13}$ & $1-\half s_{12}^2-\half s_{13}^2$
 \\[6pt]
$hhh$ & see eq.~(\ref{hhh}) &  $1-2(s_{12}Z_{6R}-s_{13}Z_{6I})/Z_1$
  \\[6pt]
$hhhh$ & see eq.~(\ref{hhhh})  &  $1-3(s_{12}Z_{6R}-s_{13}Z_{6I})/Z_1$
  \\[6pt]
$h\overline{D}D$ & $c_{12} c_{13}\mathds{1}-s_{12}\rho^D_R-c_{12}s_{13}\rho^D_I$ &
$\mathds{1}-s_{12}\rho^D_R-s_{13}\rho^D_I$
\\[6pt]
$ih\overline{D}\gamma\lsub{5}D$ &
$s_{12}\rho^D_I-c_{12}s_{13}\rho^D_R$ &
$s_{12}\rho^D_I-s_{13}\rho^D_R$
\\[6pt]
$h\overline{U}U$ & $c_{12} c_{13}\mathds{1}-s_{12}\rho^U_R-c_{12}s_{13}\rho^U_I$ &
$\mathds{1}-s_{12}\rho^U_R-s_{13}\rho^U_I$
 \\[6pt]
$ih\overline{U}\gamma\lsub{5}U$ &
$-s_{12}\rho^U_I+c_{12}s_{13}\rho^U_R$ &
$-s_{12}\rho^U_I+s_{13}\rho^U_R$
\\[6pt] \hline
\end{tabular}
\end{table}

If the scalar potential is CP-conserving, then in the
conventions established above, $\theta_{13}=\theta_{23}=0$ and $Z_6$
is real and positive.  In this
case eqs.~(\ref{dtwo}) and (\ref{dthree}) are automatically satisfied.
The decoupling/alignment limit is then
achieved when eq.~(\ref{done}) is satisfied.  Using \eq{scexact}, the
decoupling/alignment limit corresponds to~\cite{Gunion:2002zf}:
\beq \label{z6decoupling}
\cos(\beta-\alpha)\simeq -\frac{2Z_6 v^2}{m_H^2-m_h^2}\ll 1\,.
\eeq
In this limit, the neutral Higgs masses are given by,
\beq
m_h^2\simeq 2 Z_1 v^2\,,\qquad\quad m^2_{H,A}\simeq Y_2+(Z_3+Z_4\pm
Z_5)v^2\,.
\eeq

In the 2HDM with a CP-conserving scalar potential, the couplings of $h$ to the vector bosons and fermions and the Higgs
self-couplings in the approach to the decoupling/alignment limit are exhibited in
Table~\ref{tabdecouplingscp}.
Note that if the Yukawa coupling matrices $\rho^U$ and/or $\rho^D$ are complex,
then small CP-violating pseudoscalar couplings of the SM-like
Higgs boson to fermion pairs will be present, suppressed by a factor
of $\cbma$.
\begin{table}[h!]
\centering
\caption{2HDM couplings of the SM-like Higgs boson $h$ normalized to
those of the SM Higgs boson, in the
decoupling/alignment limit.  The $hH^+H^-$ coupling given below is normalized to
the SM $hhh$ coupling.
The scalar Higgs potential is taken to be CP-conserving.
For the fermion couplings,
$D$ is a column vector of three down-type fermion fields
(either down-type quarks or charged leptons)
and $U$ is a column vector of three up-type quark fields.  The
$3\times 3$ hermitian matrices, $\rho^Q_R$ and $\rho^Q_I$ (where $Q=U$
or $D$) are defined in \eq{rhodefs}.  The normalization of the
pseudoscalar coupling of the Higgs boson $h$ to fermions is relative to
the corresponding scalar coupling to fermions.  In the third column, the first non-trivial correction to decoupling/alignment is exhibited.
\label{tabdecouplingscp} \\}
\begin{tabular}{|c||c|c|}\hline
Higgs interaction & 2HDM coupling & decoupling/alignment limit \\
\hline
$hVV$ & $\sbma$ & $1-\half\cbmasq$ \\[6pt]
$hhh$ & see eq.~(\ref{hhh})
& $1+2(Z_6/Z_1)\cbma$  \\[6pt]
$hH^+H^-$ & see \eq{hhh} &
$\tfrac{1}{3}\left[(Z_3/Z_1)+(Z_7/Z_1)\cbma\right]$\\[6pt]
$hhhh$ & see eq.~(\ref{hhhh})
& $1+3(Z_6/Z_1)\cbma$  \\[6pt]
$h\overline{D}D$ & $\sbma\mathds{1}+\cbma\rho^D_R$
& $\mathds{1}+\cbma\rho^D_R$\\[6pt]
$ih\overline{D}\gamma\lsub{5}D$ &
$\cbma\rho^D_I$
& $\cbma\rho^D_I$\\[6pt]
$h\overline{U}U$ & $\sbma\mathds{1}+\cbma\rho^U_R$
& $\mathds{1}+\cbma\rho^U_R$ \\[6pt]
$ih\overline{U}\gamma\lsub{5}U$ &
$\cbma\rho^U_I$
& $\cbma\rho^U_I$\\[6pt] \hline
\end{tabular}
\end{table}

The 2HDM couplings of $H$ and $A$ in the decoupling/alignment limit
are also noteworthy.  The couplings to vector boson pairs and fermion
pairs are displayed in Table~\ref{tabheavyhiggs}.
The pattern of Higgs couplings noted in \eq{littletable} indicate
that all couplings that involve at least one vector boson and exactly
one of the non-minimal Higgs states ($H$, $A$ or $H^\pm$) is
suppressed by a factor of $\cbma$ in the decoupling/alignment limit.
\begin{table}[h!]
\centering
\caption{2HDM couplings of $H$ and $A$ normalized to
those of the SM Higgs boson, in the
decoupling/alignment limit.  The $Hhh$ coupling given below is
normalized to the SM $hhh$ coupling.
The scalar Higgs potential is taken to be
CP-conserving.  In the convention of $Z_6>0$, we identify $H\equiv -h_2$
and $A\equiv h_3$.
See caption to Table~\ref{tabdecouplingscp}.
\label{tabheavyhiggs} \\}
\begin{tabular}{|c||c|c|}\hline
Higgs interaction & 2HDM coupling & decoupling/alignment limit \\
\hline
$HW^+W^-$\,,\,$HZZ$& $\cbma$ & $\cbma$ \\[6pt]
$Hhh$ & see \eq{hhh} & $-Z_6/Z_1+[1-\tfrac{2}{3}(Z_{345}/Z_1)]\cbma$
\\[6pt]
$H\overline{D}D$ & $\cbma\mathds{1}-\sbma\rho^D_R$ &
$\cbma\mathds{1}-\rho^D_R$
\\[6pt]
$iH\overline{D}\gamma\lsub{5}D$ &
$\sbma\rho^D_I$ &
$\rho^D_I$
\\[6pt]
$H\overline{U}U$ & $\cbma\mathds{1}-\sbma\rho^U_R$ &
$\cbma\mathds{1}-\rho^U_R$
 \\[6pt]
$iH\overline{U}\gamma\lsub{5}U$ &
$-\sbma\rho^U_I$ &
$-\rho^U_I$
\\[6pt]
$AW^+W^-$\,,\,$AZZ$ & $0$ & $0$ \\[6pt]
$A\overline{D}D$ & $\rho^D_I$ &
$\rho^D_I$
\\[6pt]
$iA\overline{D}\gamma\lsub{5}D$ & $\rho^D_R$
&
$\rho^D_R$
\\[6pt]
$A\overline{U}U$ & $\rho^U_I$ &
$\rho^U_I$
 \\[6pt]
$iA\overline{U}\gamma\lsub{5}U$ &
$-\rho^U_R$ &
$-\rho^U_R$
\\[6pt] \hline
\end{tabular}
\end{table}

\clearpage
For completeness we note that it may be possible to identify the SM-like Higgs
boson with $h_2=-H$.  In this case,
we have $\cbma\simeq 1$ and $\sbma\ll 1$, in order to achieve a
SM-like $HVV$ coupling.  In this case, \eq{scexact} yields
\beq \label{Hsm}
\sbma\simeq -\frac{2Z_6 v^2}{m_H^2-m_h^2}\ll 1\,.
\eeq
This cannot be satisfied in the decoupling limit, since by assumption
we are identifying $H$ with the SM-like Higgs boson, with $m_H>m_h$.
However, \eq{Hsm} can be satisfied in the alignment limit where
$Z_6\ll 1$.
The corresponding neutral Higgs masses are:
\beq
m_H^2=2Z_1 v^2\,,\qquad\quad
m_{h,A}^2=Y_2+(Z_3+Z_4\pm Z_5)v^2\,,
\eeq
which requires that $2Z_1 v^2>Y_2+(Z_3+Z_4+Z_5)v^2$ (since
$m_h<m_H$).  In order for this interpretation to be viable, one must
check that the other Higgs states would have not been discovered at
LEP.  Although it is not yet possible to fully rule out this case,
we shall not consider it further here.

\subsection{Higgs production at the ILC}

In the CP-conserving 2HDM, the neutral Higgs bosons are produced via
Higgsstrahlung and fusion processes as in the SM. In  these
production mechanisms, the CP-even Higgs bosons are produced
via the coupling to gauge bosons.
Consequently, the production cross section of $h$ and $H$
via these processes are simply given by
\begin{eqnarray}
\sigma_{\mathrm 2HDM}(h) &=& \sigma_{\mathrm SM}(h) \sin^2(\beta-\alpha), \\
\sigma_{\mathrm 2HDM}(H) &=& \sigma_{\mathrm SM}(h)\cos^2(\beta-\alpha) .
\end{eqnarray}
In the decoupling regime where $\sin^2(\beta-\alpha) \simeq 1$ and $\cos^2(\beta-\alpha) \ll 1$,
the production cross section of $h$ is similar to that of the Higgs boson in the SM,
while that of $H$ is small.
The production of the CP-odd Higgs boson $A$ via Higgsstrahlung or
gauge boson fusion is highly suppressed since the $A$
does not couple to weak gauge boson pairs at tree-level.

\begin{figure}[h!]
\begin{center}
\includegraphics[width=50mm]{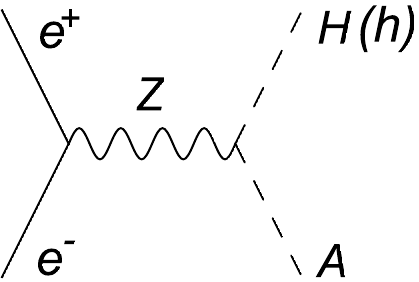}\hspace{1cm}
\includegraphics[width=45mm]{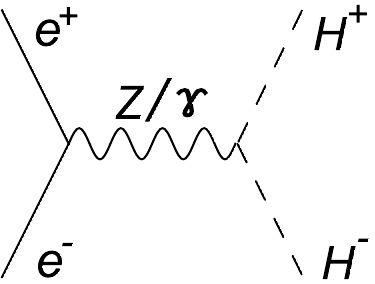}
\caption{Pair production diagrams for neutral and charged Higgs bosons. }
\label{FIG:HA}
\end{center}
\end{figure}
\begin{figure}[h!]
\begin{center}
\begin{minipage}{0.4\hsize}
\includegraphics[width=50mm,angle=-90]{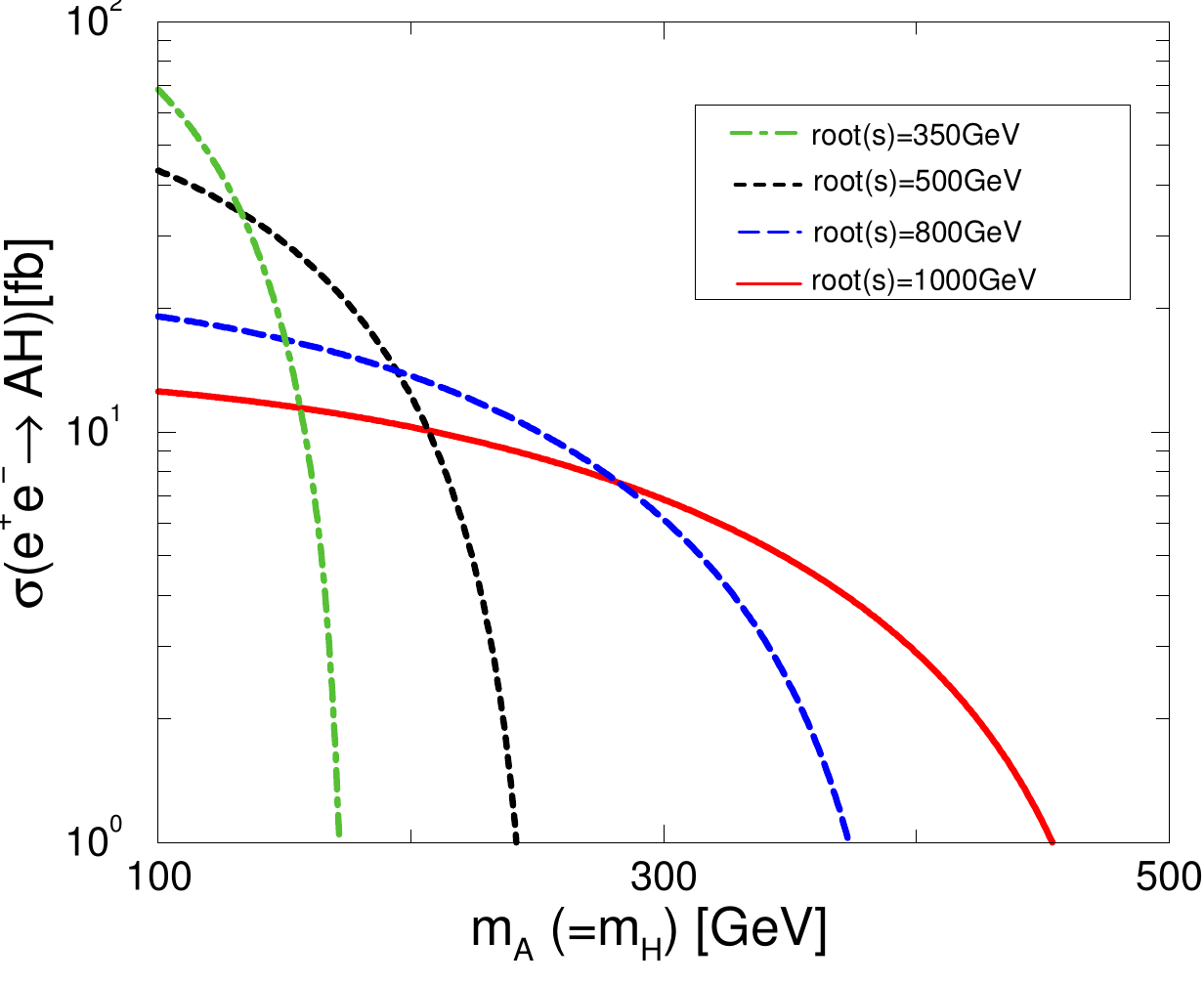}
\end{minipage}
\hspace{1cm}
\begin{minipage}{0.4\hsize}
\includegraphics[width=80mm]{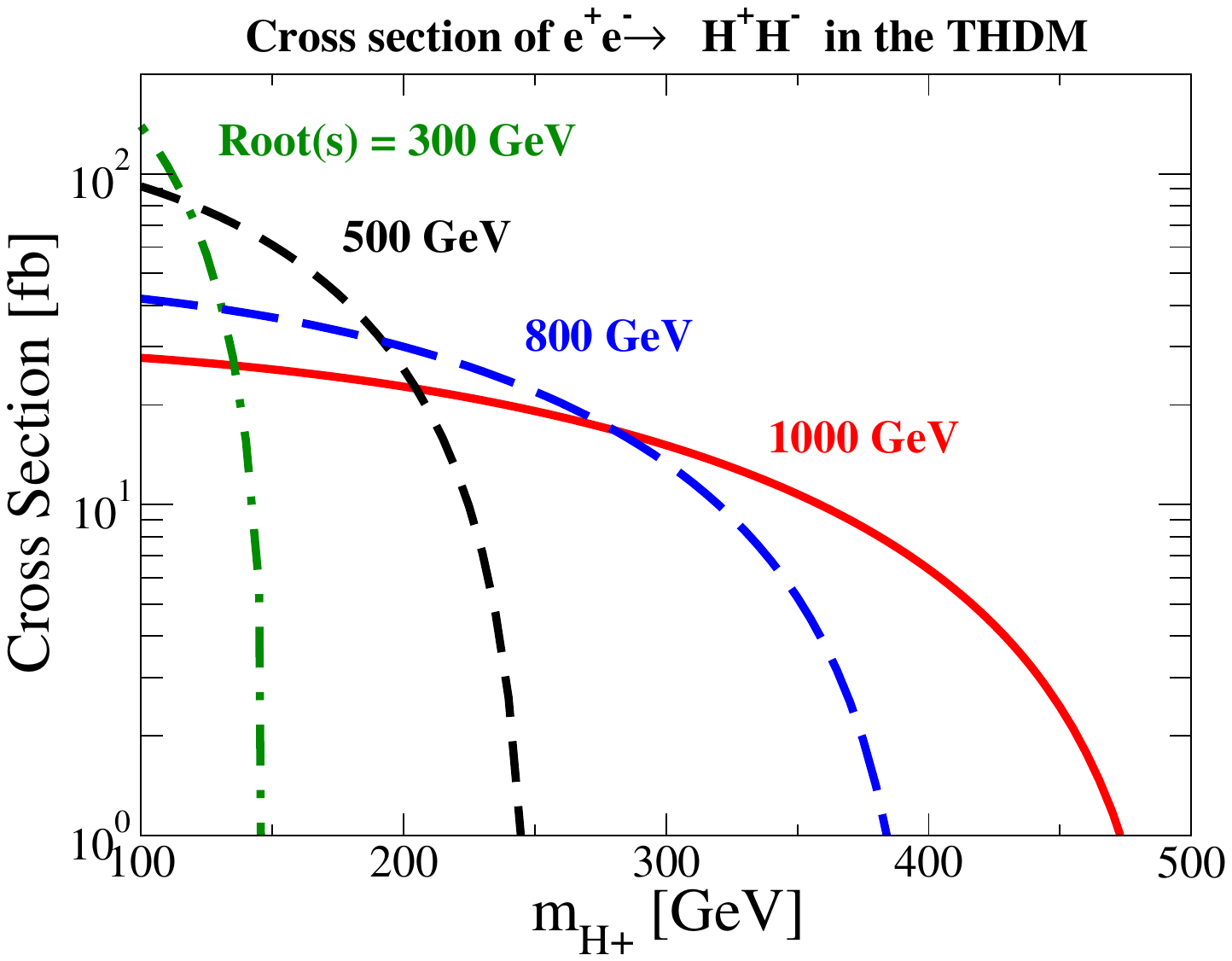}
\end{minipage}
\caption{Production cross sections of $e^+e^- \to H^+H^-$ and $e^+e^-\to AH$}
\label{FIG:HApair}
\end{center}
\end{figure}

In addition, $H$ (or $h$) and $A$ are pair produced via the couplings
$H A Z$ and $h A Z$, as shown in
Fig.~\ref{FIG:HA} (Left).  In light of \eq{littletable}, the corresponding cross-sections
are proportional to $\sin^2(\beta-\alpha)$ and $\cos^2(\beta-\alpha)$, respectively.
In the decoupling regime, where $\sin^2(\beta-\alpha) \simeq 1$, the $HA$ production
is maximal. In Fig.~\ref{FIG:HApair} (Left), the production cross section of $e^+e^- \to Z^\ast \to
HA$ is shown as a function of $m_A$, assuming $m_A=m_H$ for $\sqrt{s}=250$ and 500 GeV.
In these pair production mechanisms, the mass reach is kinematically limited by $\sqrt{s}/2$.
Beyond the threshold of pair production, single production processes
$e^+e^- \to f \bar f H$ ($f \bar f A$) could be used although the cross sections are
rather small due to the limited phase space.
%For a heavier charged Higgs boson as compared to the threshold of pair production,
%$\sqrt{s}/2$,

Charged Higgs bosons are produced in pairs via $e^+e^-\to H^+H^-$ as long as
it is kinematically allowed as illustrated in Fig.~\ref{FIG:HA} (Right).
 In Fig.~\ref{FIG:HApair} (Right), the production cross section
of $e^+e^- \to Z^\ast (\gamma) \to H^+H^-$ is shown
as a function of $m_{H^\pm}$  for $\sqrt{s}=300$, 500, 800 and 1000 GeV.
The associated production process, $e^+e^- \to H^\pm W^\mp$, is highly
suppressed in the
2HDM due to the absence of a $H^\pm W^\mp Z$ vertex at tree level.
Therefore, this process is especially sensitive to charged Higgs bosons
in extended Higgs sectors with exotic scalar fields in a
triplet or septet representation, where a tree-level $H^\pm W^\mp Z$
vertex is present.

If $m_{H^\pm}>\half\sqrt{s}$, then pair production of charged Higgs
bosons is not kinematically allowed.  In this case,
single charged Higgs boson production processes such as
$e^+e^- \to\tau \nu H^\pm$, $e^+e^- \to c s  H^\pm$ and $e^+e^- \to t b H^\pm$
can be studied, although the cross sections for these processes
are rather small, typically 0.1 fb or less.
%are not so large even just above the threshold of
%pair production.  Their production cross sections are typically 0.1 fb or less.

The production of multiple Higgs boson states can in principle
probe many of the Higgs self-coupling parameters and allow for a
(partial) reconstruction of the Higgs potential.  The
mechanisms for double Higgs and triple Higgs production
in high energy $e^+e^-$ collisions have been
considered in Refs.~\cite{Djouadi:1996ah,Djouadi:1999gv,Djouadi:1999rca,Ferrera:2007sp,Hodgkinson:2009uj,LopezVal:2009qy}.

\subsection{Special forms for the Higgs-fermion Yukawa interactions}
\label{specialforms}

In the most general 2HDM, there is no reason why the matrices
$\rho_R^Q$ and $\rho_I^Q$, which appear in the Higgs-fermion Yukawa interactions
[cf.~eq.~(\ref{YUK2})], should be approximately diagonal,
as required by the absence of large FCNCs (mediated by tree-level
neutral Higgs exchange) in the data.  Indeed in the general case, the diagonal structure
of $\rho_R^Q$ and $\rho_I^Q$ is not stable with respect to radiative
corrections, so that imposing such a condition
requires an artificial fine-tuning of parameters.
However, for special forms for the Higgs-fermion Yukawa interactions,
it turns out that the matrices $\rho_R^Q$ and $\rho_I^Q$ are
automatically diagonal, due to the presence of a symmetry that
guarantees that the diagonal structure is radiatively stable.  In this
case, tree-level Higgs mediated FCNCs are ''naturally'' absent (in the
same way that tree-level FCNCs mediated by $Z$-exchange are absent due
to the GIM mechanism~\cite{Glashow:1970gm}).

In a general extended Higgs model, tree-level Higgs mediated FCNCs are
absent if for some choice of basis of the scalar fields,
at most one Higgs multiplet is responsible for
providing mass for quarks or leptons of a given electric
charge~\cite{Glashow:1976nt,Paschos:1976ay}.  This Glashow-Weinberg-Pascos (GWP)
condition can be imposed by a symmetry
principle, which guarantees that the absence of  tree-level Higgs
mediated FCNCs is natural.  By an appropriate choice of symmetry transformation
laws for the fermions and the Higgs scalars, the resulting
Higgs-fermion Yukawa interactions take on the required form in some
basis.  The symmetry also restricts the form of the Higgs scalar
potential in the same basis.  These considerations were first applied in
the 2HDM in Refs.~\cite{Haber:1978jt} and \cite{Donoghue:1978cj}.

More generally, consider the Higgs--quark
Yukawa interactions of the 2HDM in the $\Phi_1$--$\Phi_2$ basis,
\beqa
-\mathcal{L}_{\mathrm Y}&=&\overline U_L \Phi_{a}^{0\,*}{{h^U_a}} \ur -\anti
D_L K^\dagger\Phi_{a}^- {{h^U_a}}\ur
+\overline U_L K\Phi_a^+{{h^{D\,\dagger}_{a}}} \dr
+\overline D_L\Phi_a^0 {{h^{D\,\dagger}_{a}}}\dr \nn \\
&&\qquad\quad +\overline N_L\Phi_a^+{{h^{L\,\dagger}_{a}}} E_R
+\overline E_L\Phi_a^0 {{h^{L\,\dagger}_{a}}}E_R +{\mathrm h.c.}\,,
\label{higgsql}
\eeqa
where we have made explicit both the couplings to the quarks and leptons.
%$U=(u,c,t)$, $D=(d,s,b)$, $N=(\nu_e,\nu_\mu,\nu_\tau)$,
%$E=(e,\mu,\tau)$, the
In \eq{higgsql},
$h^{U,D,L}$ are $3\times 3$ Yukawa coupling matrices and there is an
implicit sum over $a=1,2$.
The GWP condition can be implemented in four different ways~\cite{Hall:1981bc,Barger:1989fj,Aoki:2009ha}:
\begin{enumerate}
\item Type-\Rmnum{1} Yukawa couplings: $h_1^U=h_1^D=h_1^L=0$,
\item Type-\Rmnum{2} Yukawa couplings: $h_1^U=h_2^D=h_2^L=0$.
\item Type-X Yukawa couplings: $h_1^U=h_1^D=h_2^L=0$,
\item Type-Y Yukawa couplings: $h_1^U=h_2^D=h_1^L=0$.
\end{enumerate}
%Both the Type-\Rmnum{1} and Type-\Rmnum{2} conditions can be imposed
%with the following $\mathbb{Z}_2$ symmetries: $\Phi_1\to -\Phi_1$,
%$\Phi_2\to +\Phi_2$, $U_{L,R}\to +U_{L,R}$ and $D_{L,R}\to \pm
%D_{L,R}$, where the upper [lower] sign is taken in
%Type-\Rmnum{1} [Type-\Rmnum{2}].
The four types of Yukawa couplings can be implemented by a discrete
symmetry as shown in Table~\ref{Tab:type}.
%%%%%%%%%%%%%%%%%%%%%%%%%%%%%%%%%%%%%%%%%%%%%%%%%
\begin{table}[ht!]
 \begin{center}
 \caption{Four possible $\mathbb{Z}_2$ charge assignments that forbid
tree-level Higgs-mediated FCNC effects in the 2HDM.~\cite{Aoki:2009ha}.}
\label{Tab:type}
\begin{tabular}{|cl||c|c|c|c|c|c|}
\hline && $\Phi_1$ & $\Phi_2$ & $U_R^{}$ & $D_R^{}$ & $E_R^{}$ &
 $U_L$, $D_L$, $N_L$, $E_L$ \\  \hline
Type I  && $+$ & $-$ & $-$ & $-$ & $-$ & $+$ \\
Type II &(MSSM like)& $+$ & $-$ & $-$ & $+$ & $+$ & $+$ \\
Type X  &(lepton specific) & $+$ & $-$ & $-$ & $-$ & $+$ & $+$ \\
Type Y  &(flipped) & $+$ & $-$ & $-$ & $+$ & $-$ & $+$ \\
\hline
\end{tabular}
\end{center}
\end{table}
%%%%%%%%%%%%%%%%%%%%%%%%%%%%%%%%%%%%%%%%%%%%%

The imposition of the discrete symmetry also restricts the form of
the Higgs scalar potential given in \eq{genpot} by setting
$m_{12}^2=\lambda_6=\lambda_7=0$.  In this case, one can always
rephase $\Phi_1$ such that $\lambda_5$ is real, in which case the
scalar potential is CP-conserving.  Moreover, assuming that a
U(1)$_{\mathrm EM}$-conserving potential minimum exists, the
corresponding vacuum is CP-conserving, corresponding to real
vacuum expectation values, $v_i\equiv\vev{\Phi_i^0}$.  Thus, the
parameter
\beq \label{tanbeta}
\tan\beta\equiv\frac{v_2}{v_1}\,,
\eeq
is now meaningful since it refers to vacuum expectation values with respect to the basis
of scalar fields where the discrete symmetry has been imposed.   By
convention, we shall take $0\leq \beta\leq\half\pi$, in which case
$\tan\beta$ is non-negative.  This can be achieved by redefining
$\Phi_2\to -\Phi_2$ if $\tan\beta$ is negative.  However, such a redefinition would
also reverse the signs of $Z_6$ and $Z_7$.
Thus, by adopting the convention that $\tan\beta$ is non-negative, we
can no longer a choose the convention where, say, $Z_6>0$.  Indeed, in a convention
where $\tan\beta$ and $\sin(\beta-\alpha)$ are non-negative, both
$Z_6$ and $\cos(\beta-\alpha)$ can be of either sign [subject to the
constraint that $Z_6\cos(\beta-\alpha)<0$ due to \eq{scexact}].
%The special case of $\sin 2\beta=0$, corresponding to $\beta=0$ or $\half\pi$, is known
%as the inert 2HDM and will be treated separately in Section~\ref{inertsec}.

It is straightforward to evaluate the $\rho^Q_{R,I}$ for the
Type-\Rmnum{1} and Type-\Rmnum{2} Higgs-quark Yukawa couplings.
Using the corresponding results for the $\rho^Q_{R,I}$, the couplings
of $h$, $H$ and $A$ are easily obtained from Tables~\ref{tabdecouplingscp} and \ref{tabheavyhiggs}.
\begin{enumerate}
\item
Type-\Rmnum{1}: $\rho^D_R=\rho^U_R=\mathds{1}\cot\beta$\,,\qquad $\rho^D_I=\rho^U_I=0$.
\beqa
h\overline{D}D\,,\, h\overline{U}U:&&
\phantom{-}\frac{\cos\alpha}{\sin\beta}=\sbma+\cbma\cot\beta\,,\nn\\
H\overline{D}D\,,\, H\overline{U}U:&&
\phantom{-}\frac{\sin\alpha}{\sin\beta}=\cbma-\sbma\cot\beta\,,\nn\\
iA\overline{D}\gamma\lsub{5}D:&& \phm\cot\beta\,,\nn\\
iA\overline{U}\gamma\lsub{5}U:&& -\cot\beta\,.\label{typeoneff}
\eeqa
\item
Type-\Rmnum{2}:
 $\rho^D_R=-\mathds{1}\tan\beta$\,,\qquad $\rho^U_R=\mathds{1}\cot\beta$\,,\qquad $\rho^D_I=\rho^U_I=0$.
\beqa
h\overline{D}D:&& -\frac{\sin\alpha}{\cos\beta}=\sbma-\cbma\tan\beta\,,\nn\\
h\overline{U}U:&&
\phantom{-}\frac{\cos\alpha}{\sin\beta}=\sbma+\cbma\cot\beta\,,\nn \\
H\overline{D}D:&& \phantom{-}\frac{\cos\alpha}{\cos\beta}=\cbma+\sbma\tan\beta\,,\nn\\
H\overline{U}U:&&
\phantom{-}\frac{\sin\alpha}{\sin\beta}=\cbma-\sbma\cot\beta\,,\nn\\
iA\overline{D}\gamma\lsub{5}D:&& -\tan\beta\,,\nn\\
iA\overline{U}\gamma\lsub{5}U:&& -\cot\beta\,.\label{t2}
\eeqa
\end{enumerate}
Likewise, the charged Higgs Yukawa couplings to quarks are given by
\beqa
-\mathcal{L}_Y&\ni&
\phm\frac{1}{v}\cot\beta\biggl(\overline{U}\bigl[KM_DP_R-M_UKP_L\bigr]DH^+
  + {\mathrm h.c.}\biggr)\,,\qquad\qquad\quad \text{Type-I}\,,\label{chhiggsy1}\\
-\mathcal{L}_Y&\ni&
-\frac{1}{v}\biggl(\overline{U}\bigl[KM_DP_R\tan\beta-M_UKP_L\cot\beta\bigr]DH^+
  + {\mathrm h.c.}\biggr)\,,\qquad\quad\!\!\! \text{Type-II}\,,\label{chhiggsy2}
\eeqa
where
$M_{U,D}$ are the diagonal up-type and down-type $3\times 3$
quark mass matrices and $K$ is the CKM mixing matrix.
The Type-I [Type-II] neutral and charged
Higgs Yukawa coupling to quarks also apply to Type-X [Type-Y],
respectively.

Following the prescription below \eq{rhoq}, the charged
Higgs Yukawa coupling to leptons are obtained from the couplings to
quarks given above by
replacing $U\to N$, $D\to E$, $M_U\to 0$, $M_D\to M_E$ and
$K\to\mathds{1}$.   The Type-I [Type-II] neutral and charged
Higgs Yukawa coupling to leptons also apply to Type-Y [Type-X], respectively.
The neutral Higgs Yukawa couplings to quarks and leptons (relative
to the corresponding couplings of the SM Higgs boson) are
conveniently summarized
in Table~\ref{yukawa_tab} for the four possible implementations of
the GWP condition.
\begin{table}[hb!]
 \begin{center}
\caption{Higgs--fermion couplings in the 2HDM subject to the
$\mathbb{Z}_2$ symmetries given in Table~\ref{Tab:type}.  The
couplings listed below are normalized relative to the SM Higgs
couplings $h_{\mathrm SM}\overline{U}U$, $h_{\mathrm SM}\overline{D}D$,
and $h_{\mathrm SM}\overline{E}E$.}
\label{yukawa_tab}
{\renewcommand\arraystretch{1.5}
\begin{tabular}{|c||ccccccccc|}\hline
&$h\overline{U}U$
&$h\overline{D}D$&$h\overline{E}E$&$H\overline{U}U$&$H\overline{D}D$&$H\overline{E}E$&$iA\overline{U}\gamma\lsub{5}U$&$iA\overline{D}\gamma\lsub{5}D$
& $iA\overline{E}\gamma\lsub{5}E$\\
&$\xi_h^u$&$\xi_h^d$&$\xi_h^e$&$\xi_H^u$&$\xi_H^d$&$\xi_H^e$&$\xi_A^u$&$\xi_A^d$&$\xi_A^e$\\
\hline\hline
Type I &$\frac{\cos\alpha}{\sin\beta}$&$\phm\frac{\cos\alpha}{\sin\beta}$&$\phm\frac{\cos\alpha}{\sin\beta}$&$\frac{\sin\alpha}{\sin\beta}$&$\frac{\sin\alpha}{\sin\beta}$&$\frac{\sin\alpha}{\sin\beta}$&$-\cot\beta$&$\phm\cot\beta$&$\phm\cot\beta$\\\hline
Type II &$\frac{\cos\alpha}{\sin\beta}$&$-\frac{\sin\alpha}{\cos\beta}$&$-\frac{\sin\alpha}{\cos\beta}$&$\frac{\sin\alpha}{\sin\beta}$&$\frac{\cos\alpha}{\cos\beta}$&$\frac{\cos\alpha}{\cos\beta}$&$-\cot\beta$&$-\tan\beta$&$-\tan\beta$\\\hline
Type X &$\frac{\cos\alpha}{\sin\beta}$&$\phm\frac{\cos\alpha}{\sin\beta}$&$-\frac{\sin\alpha}{\cos\beta}$&$\frac{\sin\alpha}{\sin\beta}$&$\frac{\sin\alpha}{\sin\beta}$&$\frac{\cos\alpha}{\cos\beta}$&$-\cot\beta$&$\phm\cot\beta$&$-\tan\beta$\\\hline
Type Y &$\frac{\cos\alpha}{\sin\beta}$&$-\frac{\sin\alpha}{\cos\beta}$&$\phm\frac{\cos\alpha}{\sin\beta}$&$\frac{\sin\alpha}{\sin\beta}$&$\frac{\cos\alpha}{\cos\beta}$&$\frac{\sin\alpha}{\sin\beta}$&$-\cot\beta$&$-\tan\beta$&$\phm\cot\beta$\\\hline
\end{tabular}}
\end{center}
\end{table}
%%%%%%%%%%%%%%%%%%%%%%

In implementing the $\mathbb{Z}_2$ discrete symmetries given in
Table~\ref{Tab:type}, we noted above that the parameters of the scalar
Higgs potential are restricted such that
$m_{12}^2=\lambda_6=\lambda_7=0$ in the basis in which the discrete
symmetry is manifest.  However, these latter conditions can be slightly
relaxed by taking $m_{12}^2\neq 0$ (while maintaining
$\lambda_6=\lambda_7=0$).  In this case, it is convenient to introduce a
squared-mass parameter,
\beq \label{Mtwo}
M^2\equiv \frac{2m_{12}^2}{\sin 2\beta}=m_A^2+2\lambda_5 v^2\,.
\eeq
When $M^2\neq 0$, the discrete symmetry is softly broken by a
dimension-two term in the scalar potential (while it is respected by
all dimension-four terms of the Lagrangian).  In a 2HDM of this type,
Higgs--mediated FCNCs are still absent at tree-level, but can be
generated at the one-loop level.  Since the neutral Higgs Yukawa
couplings are suppressed by fermion masses, one can check that
sensible parameter regimes exist in which the
radiatively generated FCNCs in this model are sufficiently suppressed
so as not to be in conflict with experimental data.

The existence of a softly-broken $\mathbb{Z}_2$ symmetry
that imposes $\lambda_6=\lambda_7=0$ in some basis yields the following constraint
on the Higgs basis scalar potential parameters:
\beq
(Z_6+Z_7)(Z_2-Z_1)(Z_1+Z_2-2Z_{345})+(Z_6-Z_7)\left[(Z_2-Z_1)^2-4(Z_6+Z_7)^2\right]=0\,,
\eeq
where $Z_{345}\equiv Z_3+Z_4+Z_5$.
The parameter $\beta$ is also determined
%(by convention, $0\leq \beta\leq\half\pi$),
\beq
\tan 2\beta=\frac{2(Z_6+Z_7)}{Z_2-Z_1}\,.
\eeq
The case of $Z_1=Z_2$ and $Z_6=-Z_7$ must be treated separately.  In this case, a
$\mathbb{Z}_2$ symmetry governing the quartic terms of the scalar potential is automatically present, and the corresponding value
of $\beta$ is determined from the following quadratic equation,
\beq
(Z_1-Z_{345})\tan 2\beta+2Z_6(1-\tan^2 2\beta)=0\,.
\eeq
%This special case arises in the case of the MSSM Higgs sector, where $Z_7=-Z_6=%\quarter(g^2+g^{\prime\,2})\sin 2\beta\cos 2\beta$.

In the constrained 2HDMs considered in this subsection, there are a
number of benefits in allowing for a soft breaking of the
$\mathbb{Z}_2$ discrete symmetry, which permits a
nonzero $m_{12}^2$ in the basis where $\lambda_6=\lambda_7=0$.
First, this allows us to treat the MSSM Higgs sector, which employs
the Type-II Higgs--fermion Yukawa couplings as a consequence
of supersymmetry rather than a $\mathbb{Z}_2$ discrete symmetry.
Second, the 2HDM with a soft breaking of the $\mathbb{Z}_2$ discrete
symmetry possesses a decoupling limit (which corresponds to large
$m_{12}^2$).  If $m_{12}^2=0$, no decoupling limit exists since the
scalar potential minimum conditions imply that $Y_2\sim\mathcal{O}(Z_i v^2)$.
Thus, in this latter case, a SM-like Higgs boson emerges only in the
alignment limit.  Finally, taking $m_{12}^2\neq 0$ allows for
a new source of CP-violation in the Higgs sector.  One can check~\cite{Gunion:2002zf}
that the Higgs scalar potential is explicitly
CP-violating if $\Im[(m_{12}^2)^2\lambda_5^*]\neq 0$.
If the scalar potential is explicitly CP-conserving, then one can
rephase the scalar fields such that $m_{12}^2$ and $\lambda_5$ are
real.  In this case spontaneous CP-violation can occur if
$0<|m_{12}^2|<2\lambda_5 |v_1||v_2|$, in which case the minimum
of the scalar potential yields a relative phase
$\vev{\Phi_1^\dagger\Phi_2}= |v_1||v_2|e^{i\xi}$, where
$\cos\xi=m_{12}^2/(2\lambda_5 |v_1||v_2|)$.

%%%%%%%%%%%%%%%%%%%%%%%%%%%%%%%%%%%%%%%%

The decays of the Higgs bosons in the constrained
2HDM depend on the Type of Yukawa interactions.
When $\sin(\beta-\alpha)=1$,  the decay pattern of $h$ is
the same as those in the Standard Model at tree level.
When $\sin(\beta-\alpha)$ differs from from 1, the couplings
of $h$ will deviate from Standard Model expectations.  In particular,
the couplings of $h$
to down-type quarks and leptons will differ from those of the SM
with a pattern of deviations that
strongly depends on the Type of Yukawa Interactions.
The precision measurement of these coupling make it possible to
discriminate among the various Types of Yukawa interactions of the 2HDM.
%and further
%determine extended Higgs sectors. We discuss this in more details
%in the later subsection.

On the other hand,
the decay patterns of $H$, $A$, and $H^\pm$ can vary over a large range~\cite{Barger:1989fj,Aoki:2009ha,Su:2009fz,Logan:2009uf}.
Figure~\ref{FIG:br_200} shows the decay branching ratios of $H$, $A$ and $H^\pm$
as a function of $\tan\beta$ for Higgs boson masses of $200$ GeV
and $\sin(\beta-\alpha)=1$ for $m_H^{}=m_A^{}=m_{H^\pm}^{}=M=200$ GeV
[where $M$ is defined in \eq{Mtwo}], assuming Type I, II, X and Y Yukawa couplings.
For example, the amplitudes for the
fermionic and the $gg$ decays mode of $H$, $A$ and $H^\pm$
in the Type-I 2HDM are all proportional to
$\cot\beta$.  Consequently the corresponding branching ratios are
roughly independent of $\tan\beta$.
Note that for $\sin(\beta-\alpha)=1$, the
couplings of $H$ and $A$ to fermion pairs are equal in magnitude [cf.~\eq{typeoneff}].
The differences among
the $A$ and $H$ branching ratios can be attributed to
$\Gamma(A\to gg, \gamma\gamma)>\Gamma(H\to gg, \gamma\gamma)$, which arises due to different loop factors
for the CP-odd and CP-even scalar couplings to $gg$ and $\gamma\gamma$ via the dominant top-quark loop.
In principle, the partial widths $\Gamma(H,A\to\gamma\gamma)$ can also differ due to
the contributions of the $W$ and charged Higgs loops to the amplitude for
$H\to\gamma\gamma$ (the corresponding couplings to $A$ are absent due to the assumed
CP conservation).  However, in the limit of $\lambda_6=\lambda_7=\cos(\beta-\alpha)=0$,
the $W^+ W^- H$ coupling vanishes and
the $H^+ H^- H$ coupling takes a particularly simple form,
$g\lsub{H^+H^- H}=2(m_H^2-M^2)\cot 2\beta/v$~\cite{Gunion:2002zf},
which vanishes for the parameter choices employed in Figures~\ref{FIG:br_200} and \ref{FIG:br_400}.
\begin{figure}[h!]
\begin{center}
 \includegraphics[width = 35mm]{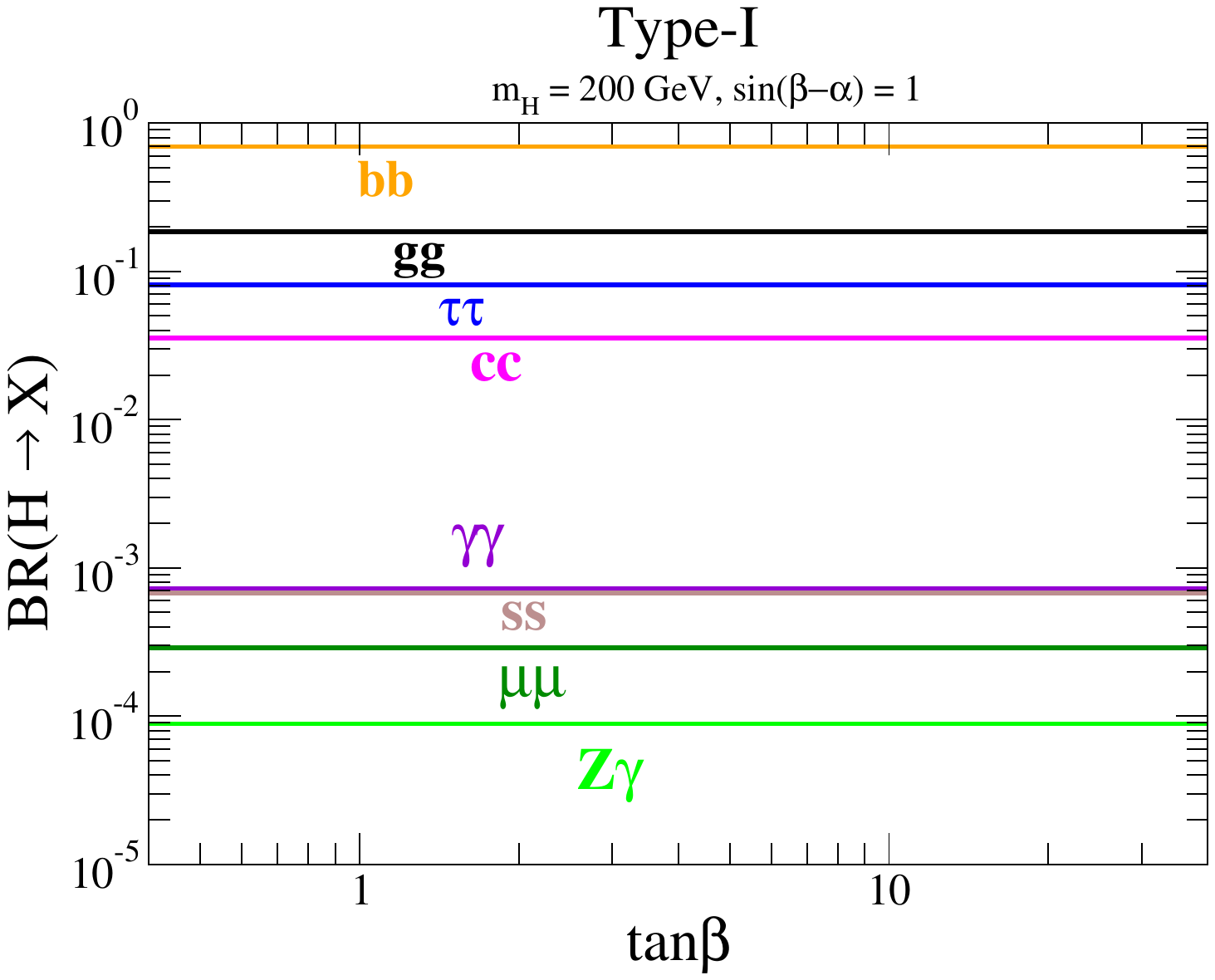}%\hspace{2mm}
 \includegraphics[width = 35mm]{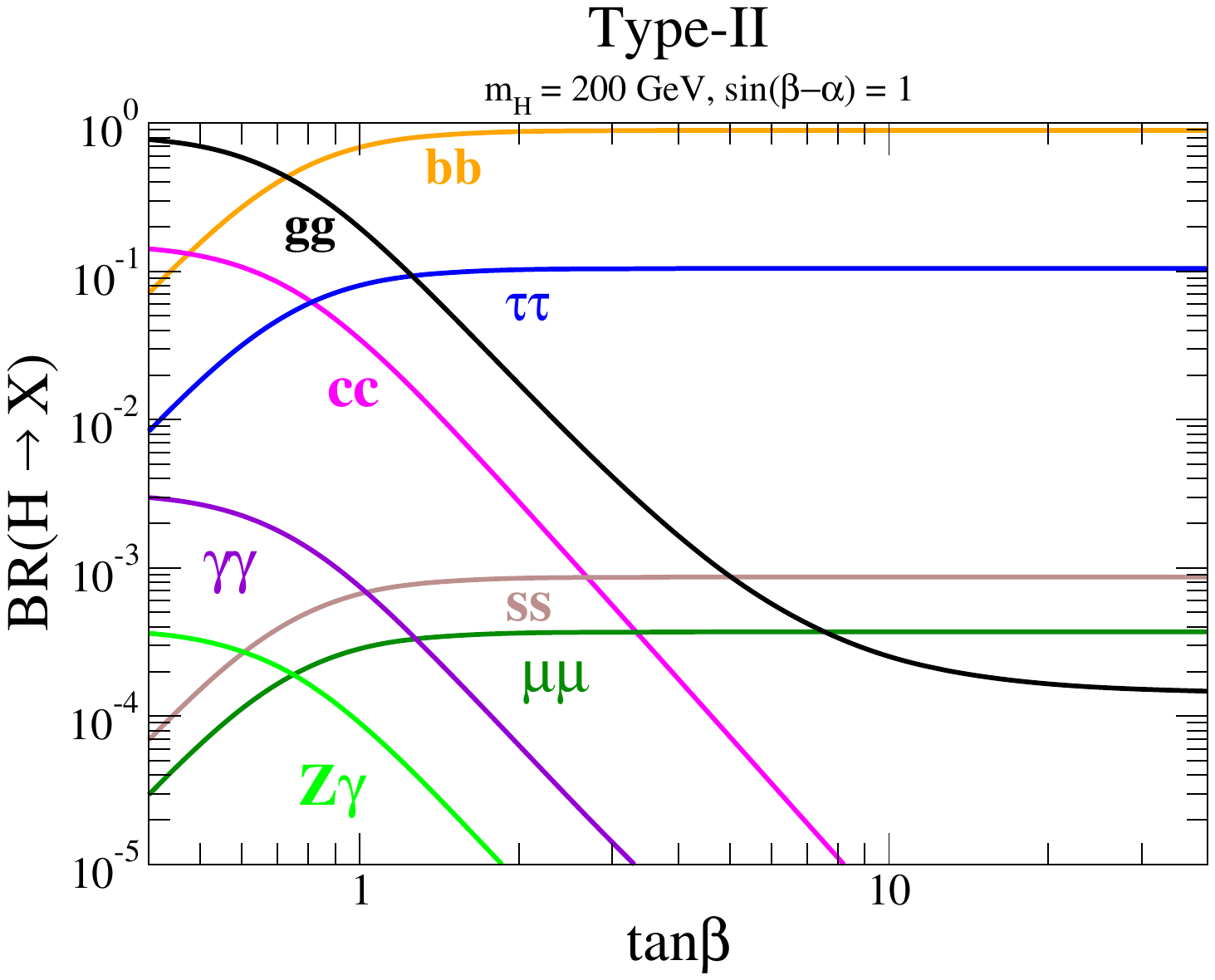}%\hspace{2mm}
 \includegraphics[width = 35mm]{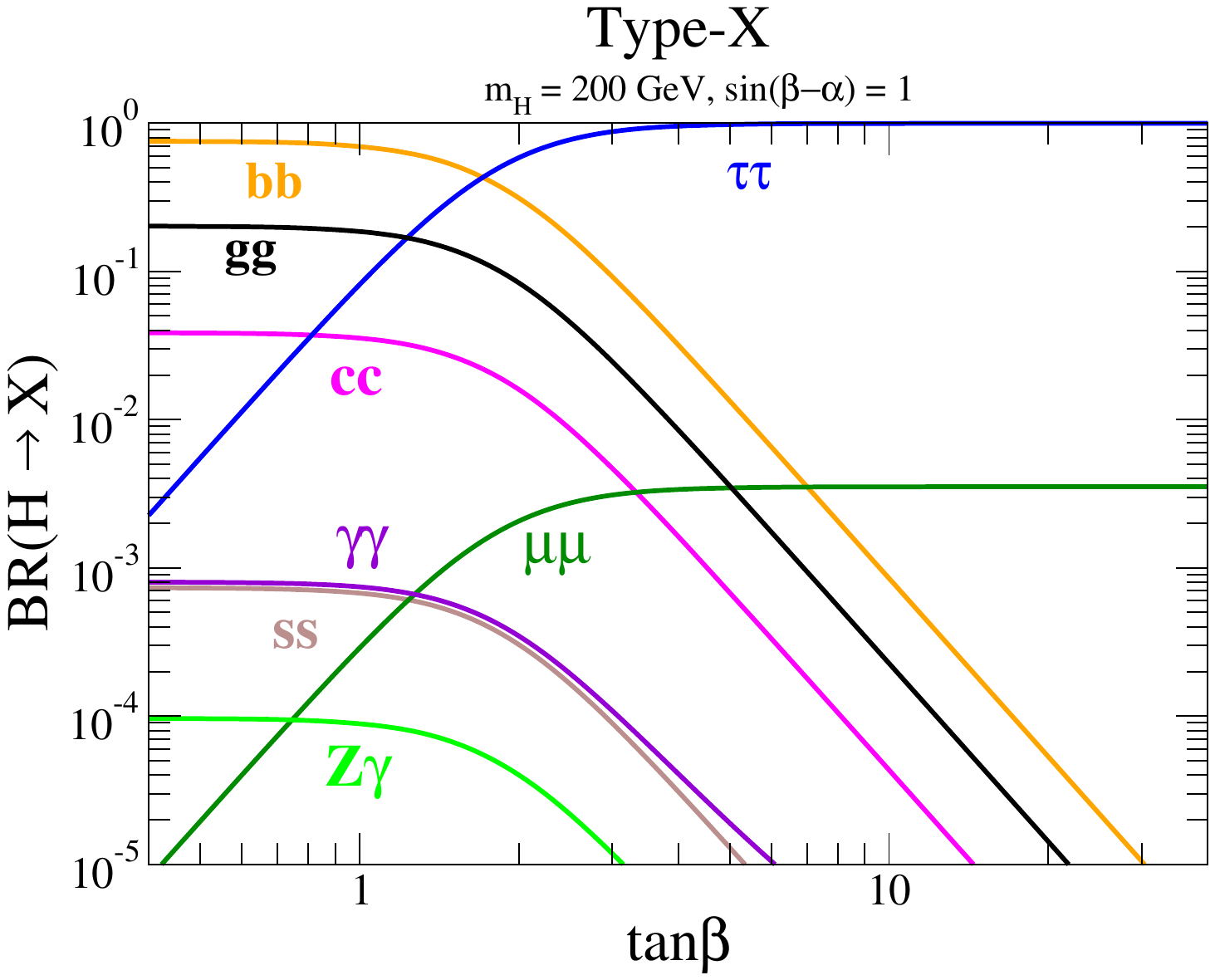}%\hspace{2mm}
 \includegraphics[width = 35mm]{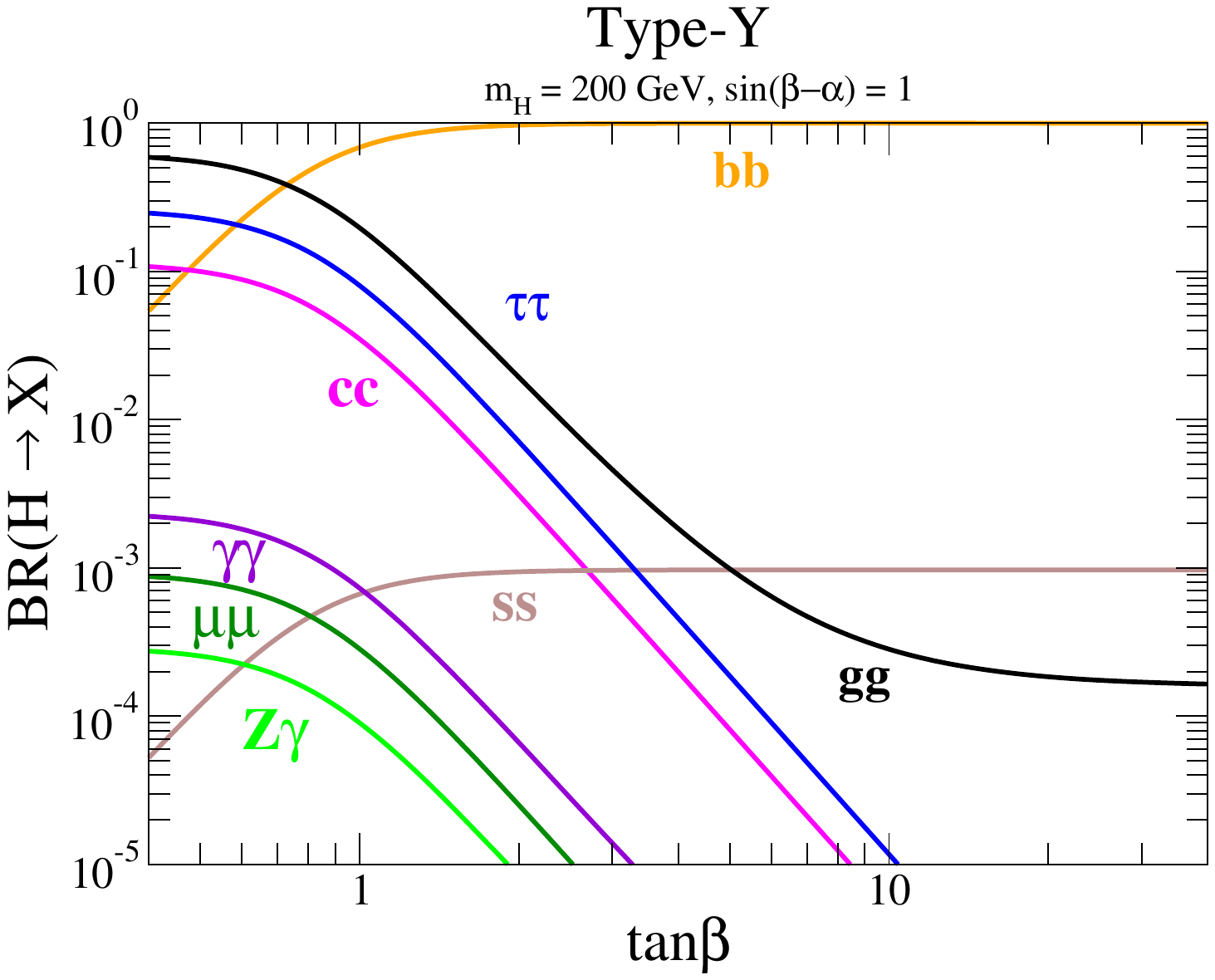}\\
\vspace{5mm}
  %\caption{XXX}
%\end{center}
%\end{figure}
%\begin{figure}[h]
%\begin{center}
 \includegraphics[width = 35mm]{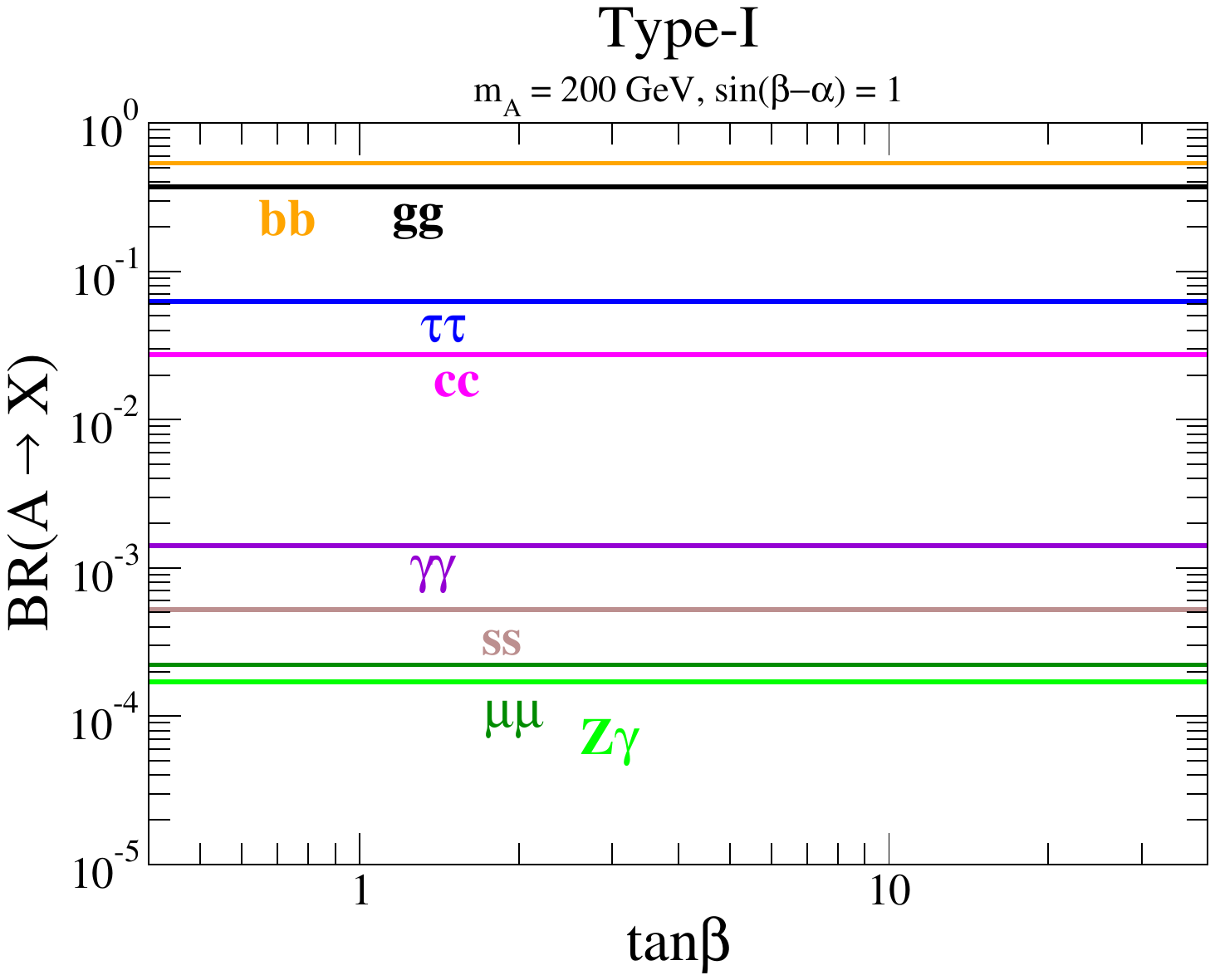}%\hspace{2mm}
 \includegraphics[width = 35mm]{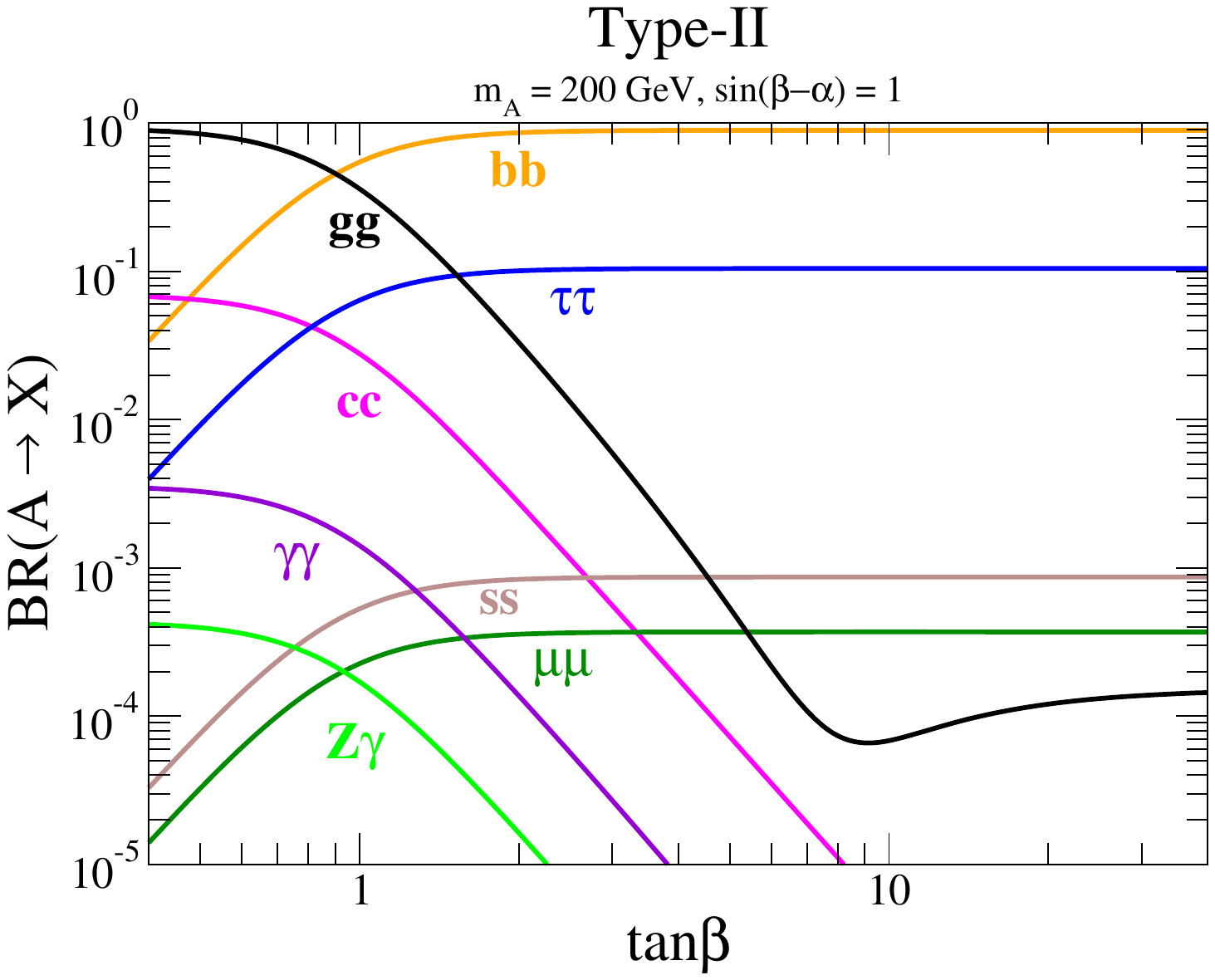}%\hspace{2mm}
 \includegraphics[width = 35mm]{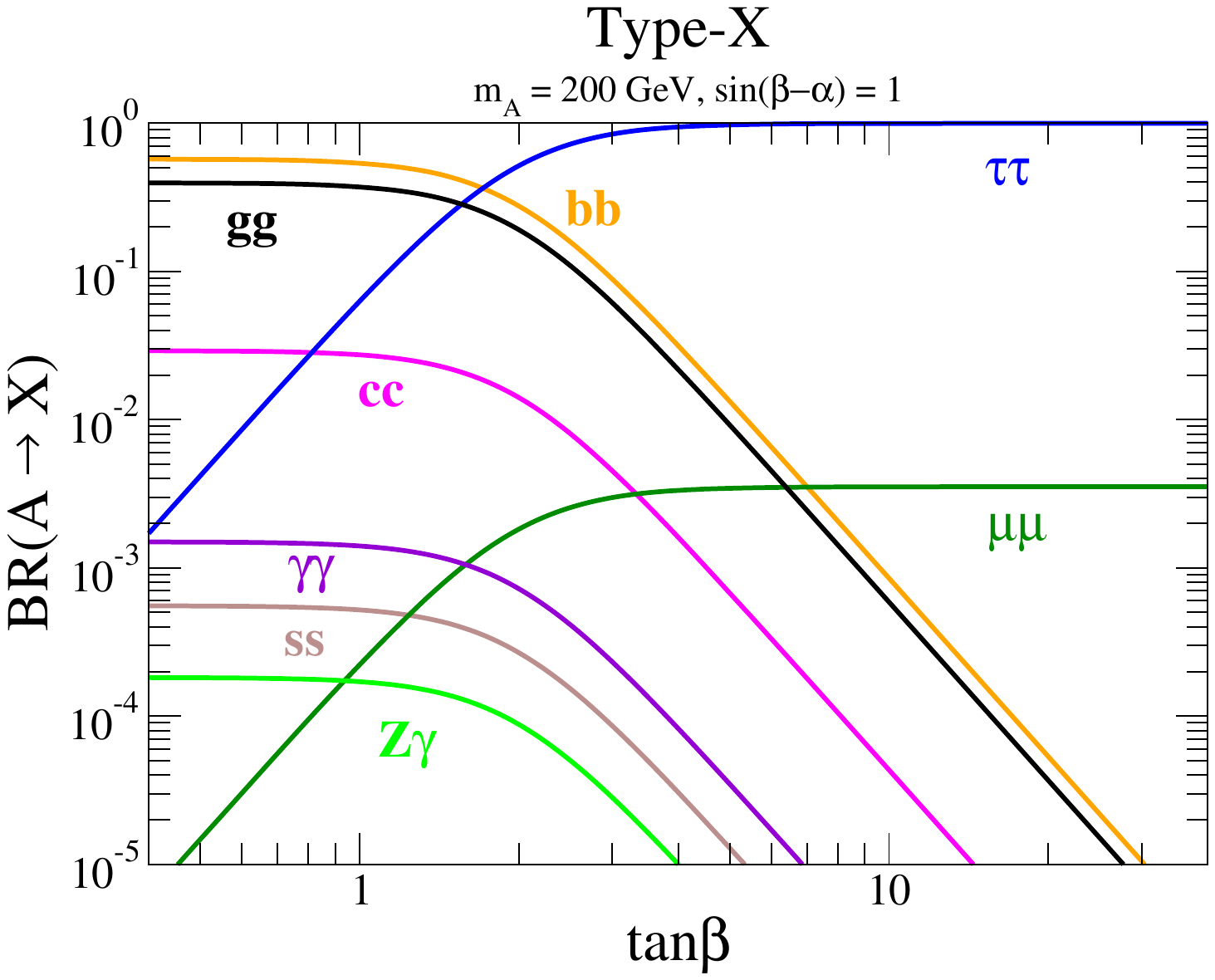}%\hspace{2mm}
 \includegraphics[width = 35mm]{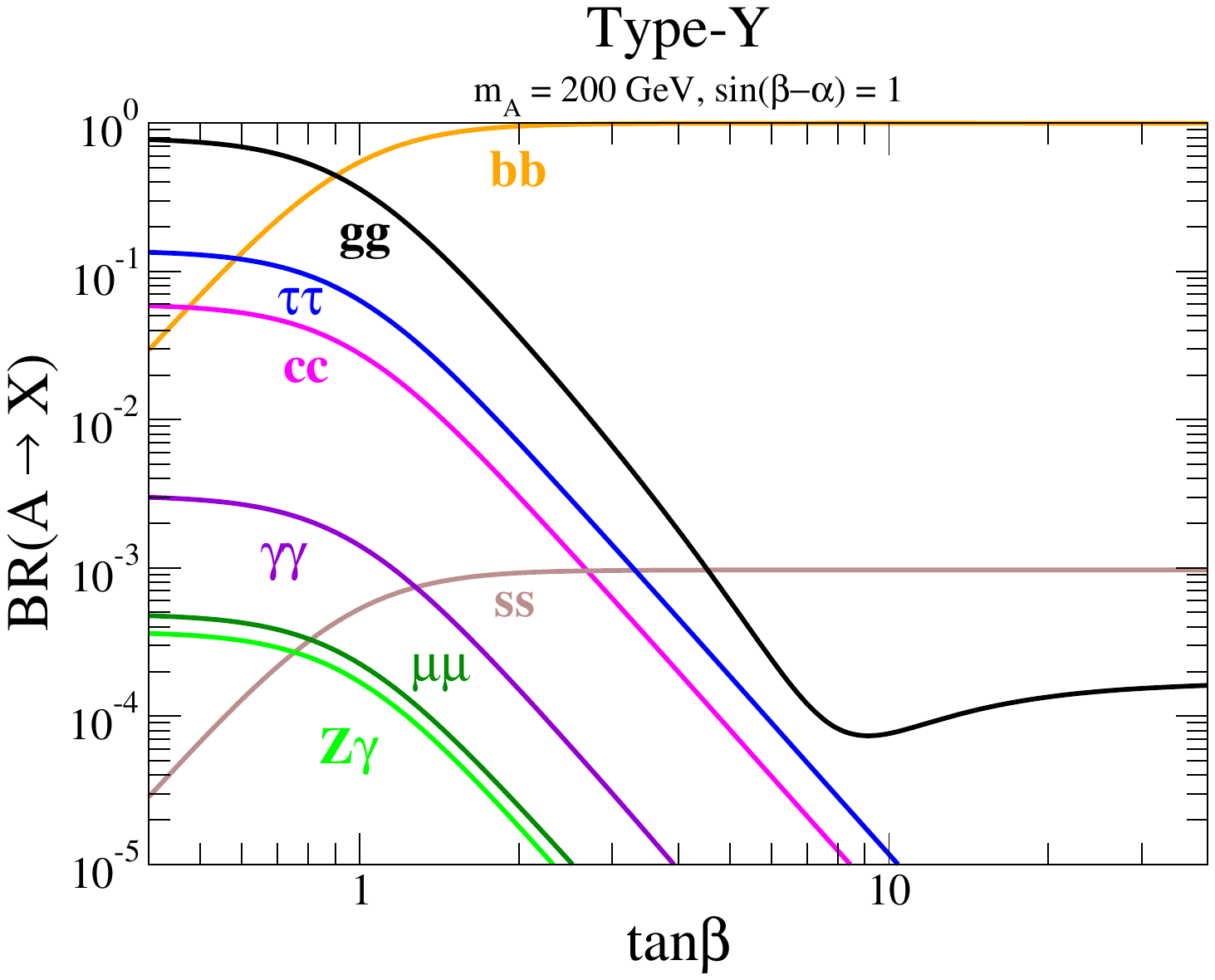}\\
\vspace{5mm}
  %\caption{XXX}
%\end{center}
%\end{figure}
%\begin{figure}[h]
%\begin{center}
 \includegraphics[width = 35mm]{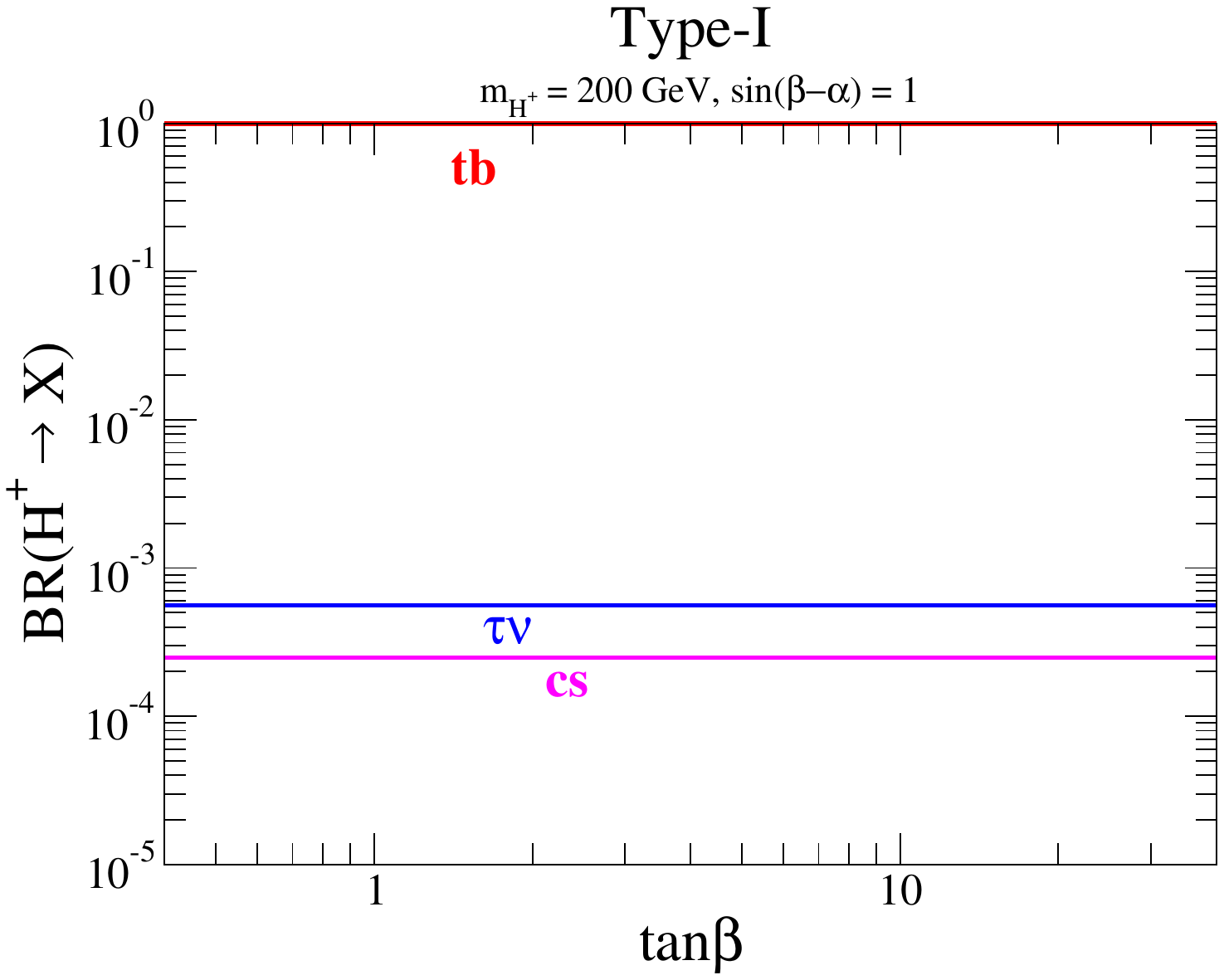}%\hspace{2mm}
 \includegraphics[width = 35mm]{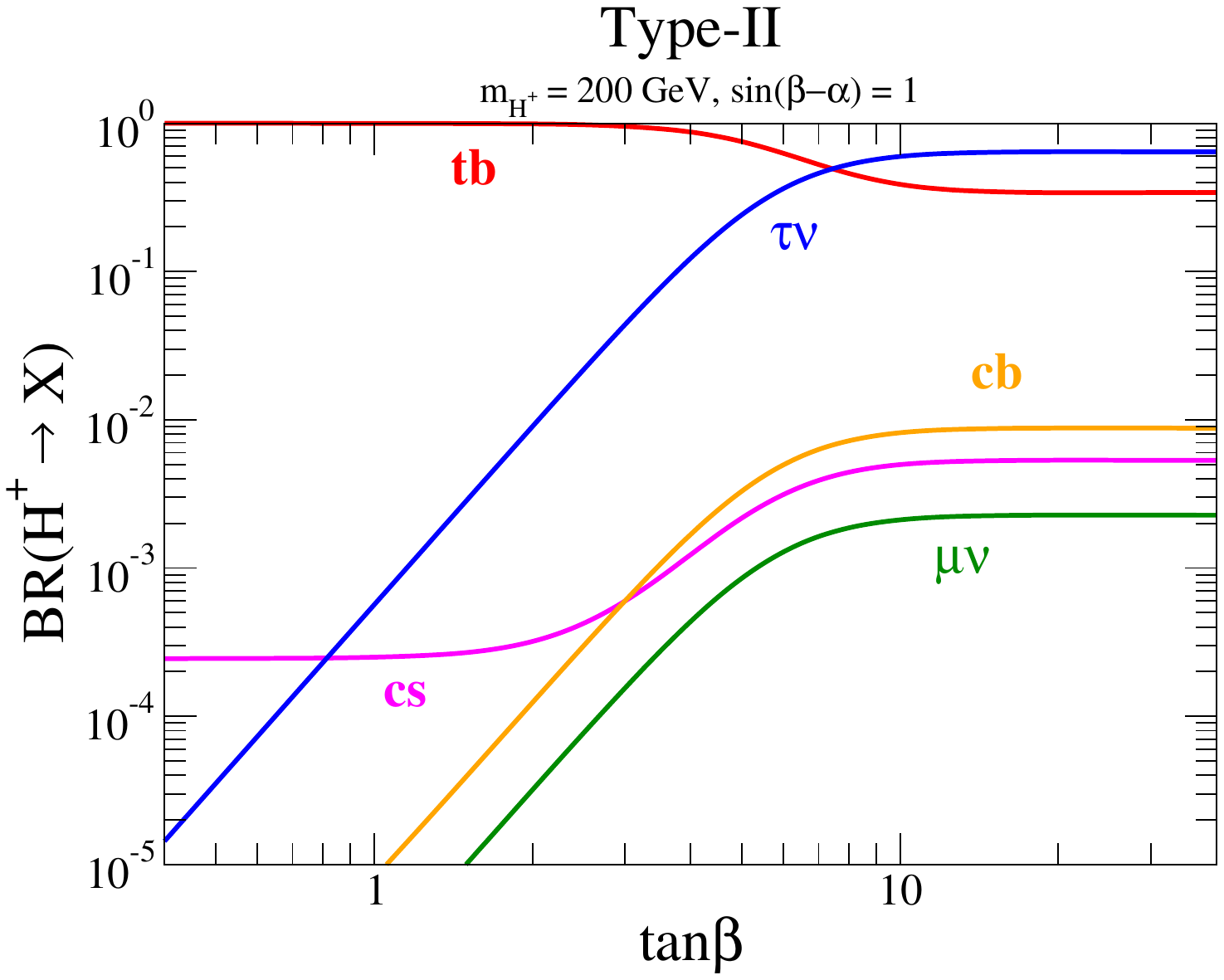}%\hspace{2mm}
 \includegraphics[width = 35mm]{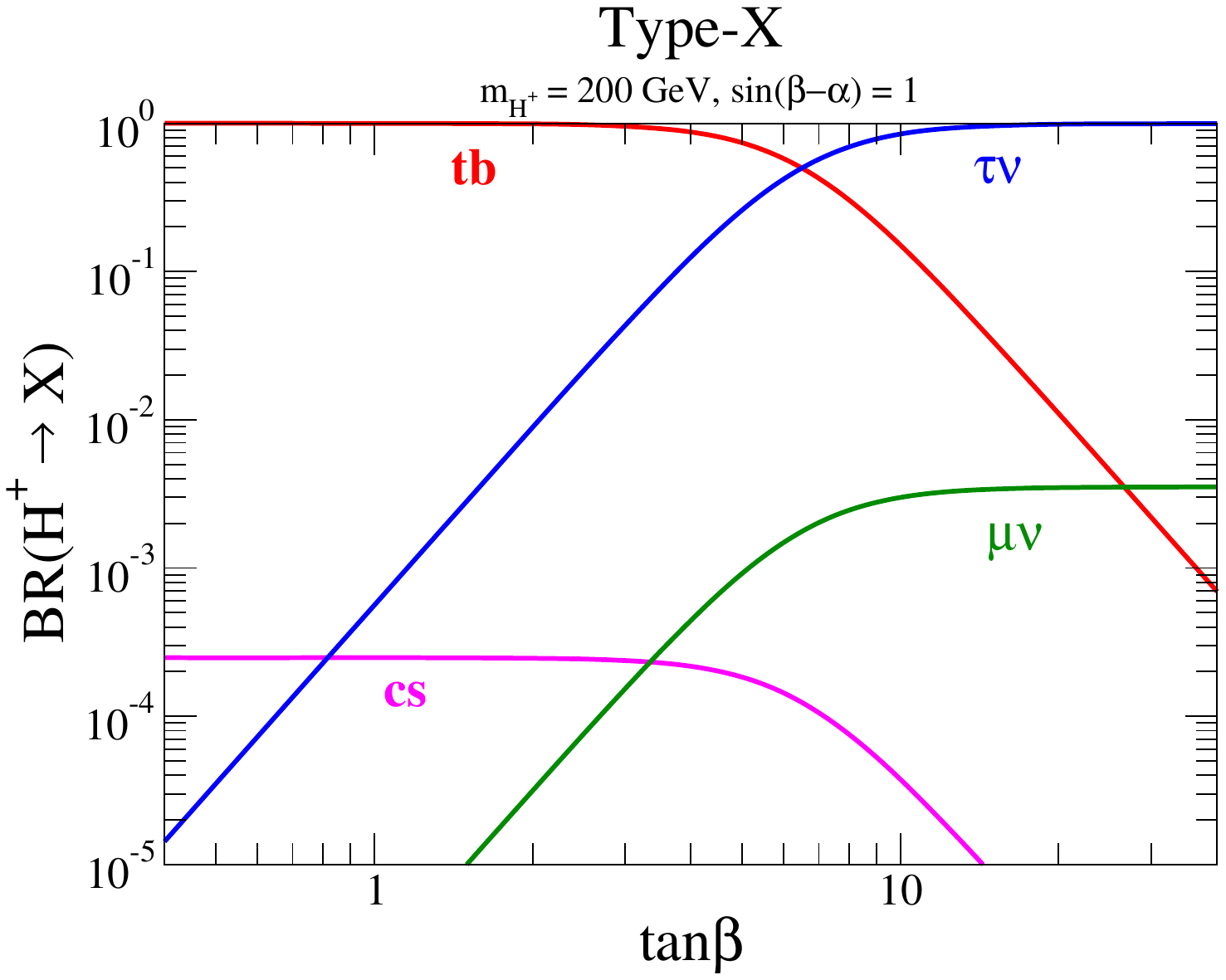}%\hspace{2mm}
 \includegraphics[width = 35mm]{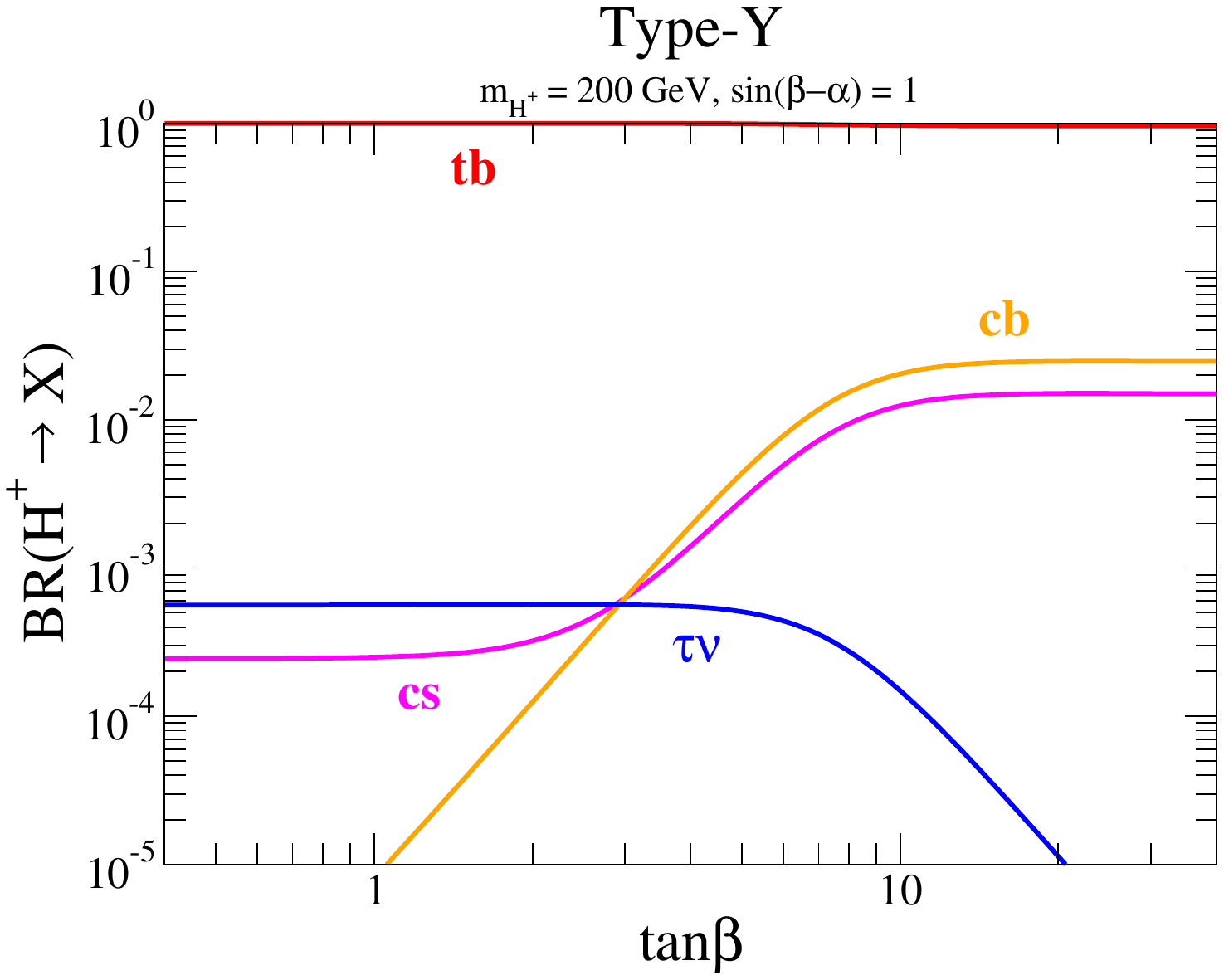}
\caption{Decay branching ratios of $H$, $A$ and $H^\pm$
in the four different Types of 2HDM as a function of $\tan\beta$
for $m_H^{}=m_A^{}=m_{H^\pm}^{}=M=200$ GeV.
The SM-like limit $\sin(\beta-\alpha) =1$ is taken.\label{FIG:br_200}}
\end{center}
\end{figure}
\begin{figure}[h!]
\begin{center}
 \includegraphics[width = 35mm]{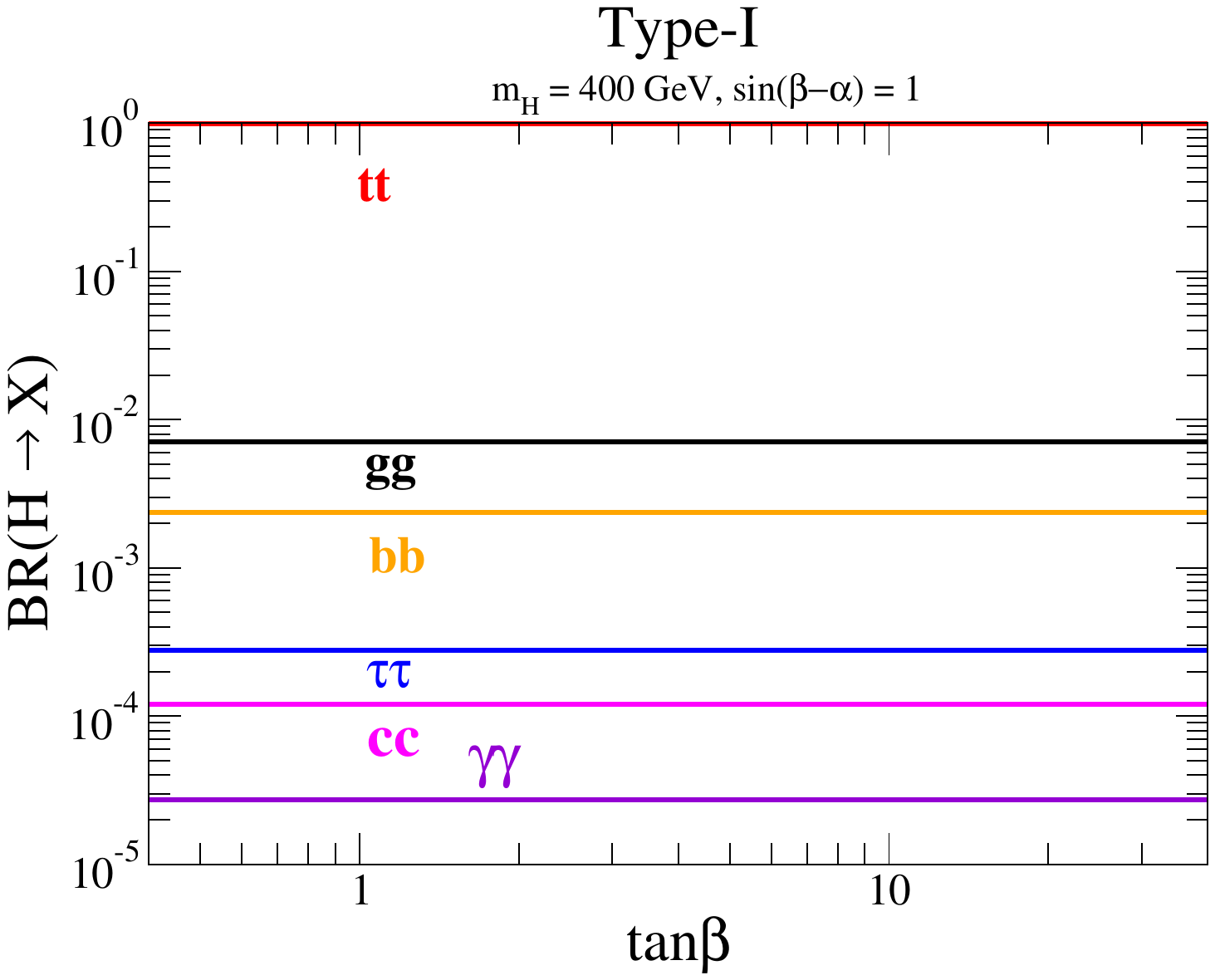}%\hspace{2mm}
 \includegraphics[width = 35mm]{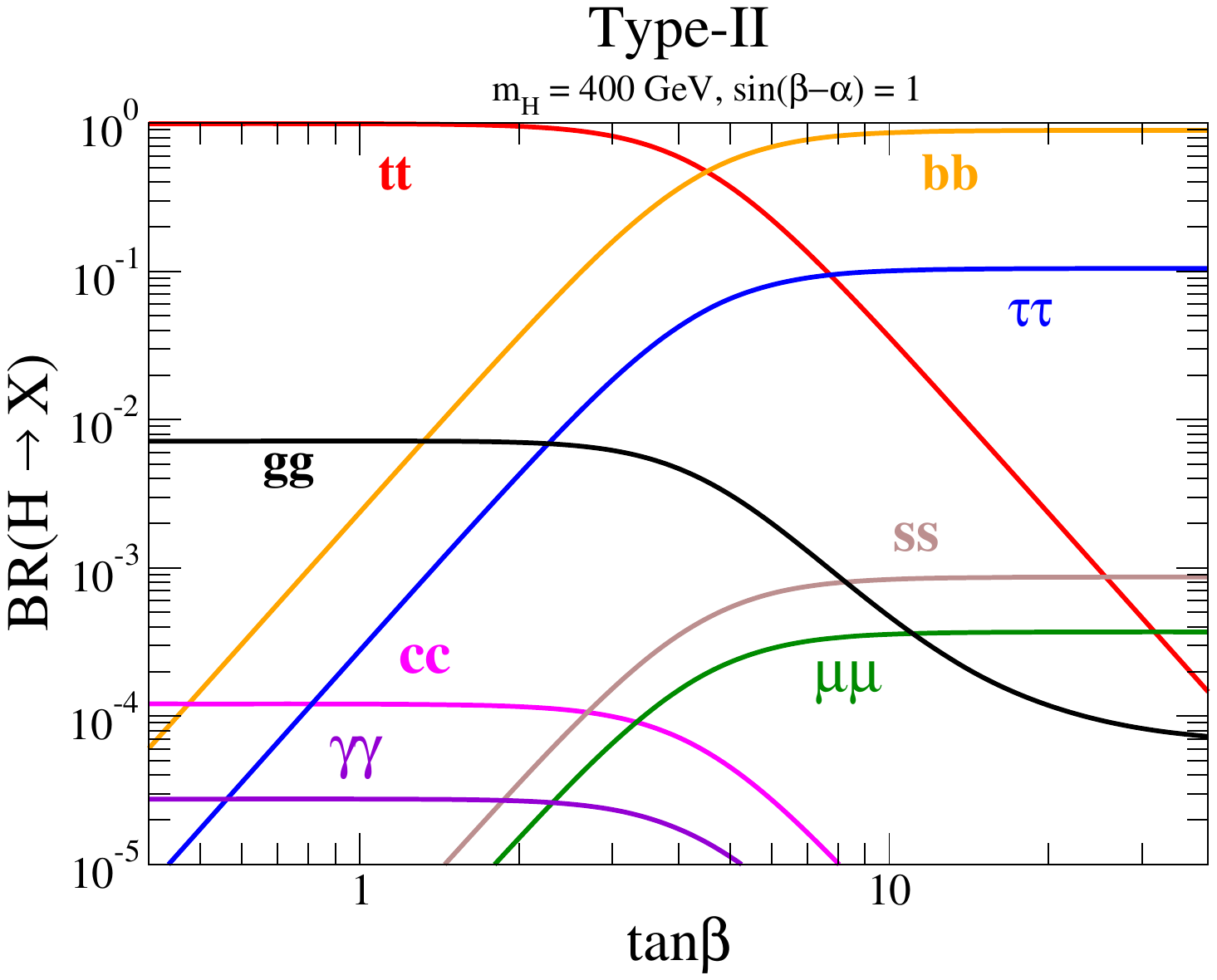}%\hspace{2mm}
 \includegraphics[width = 35mm]{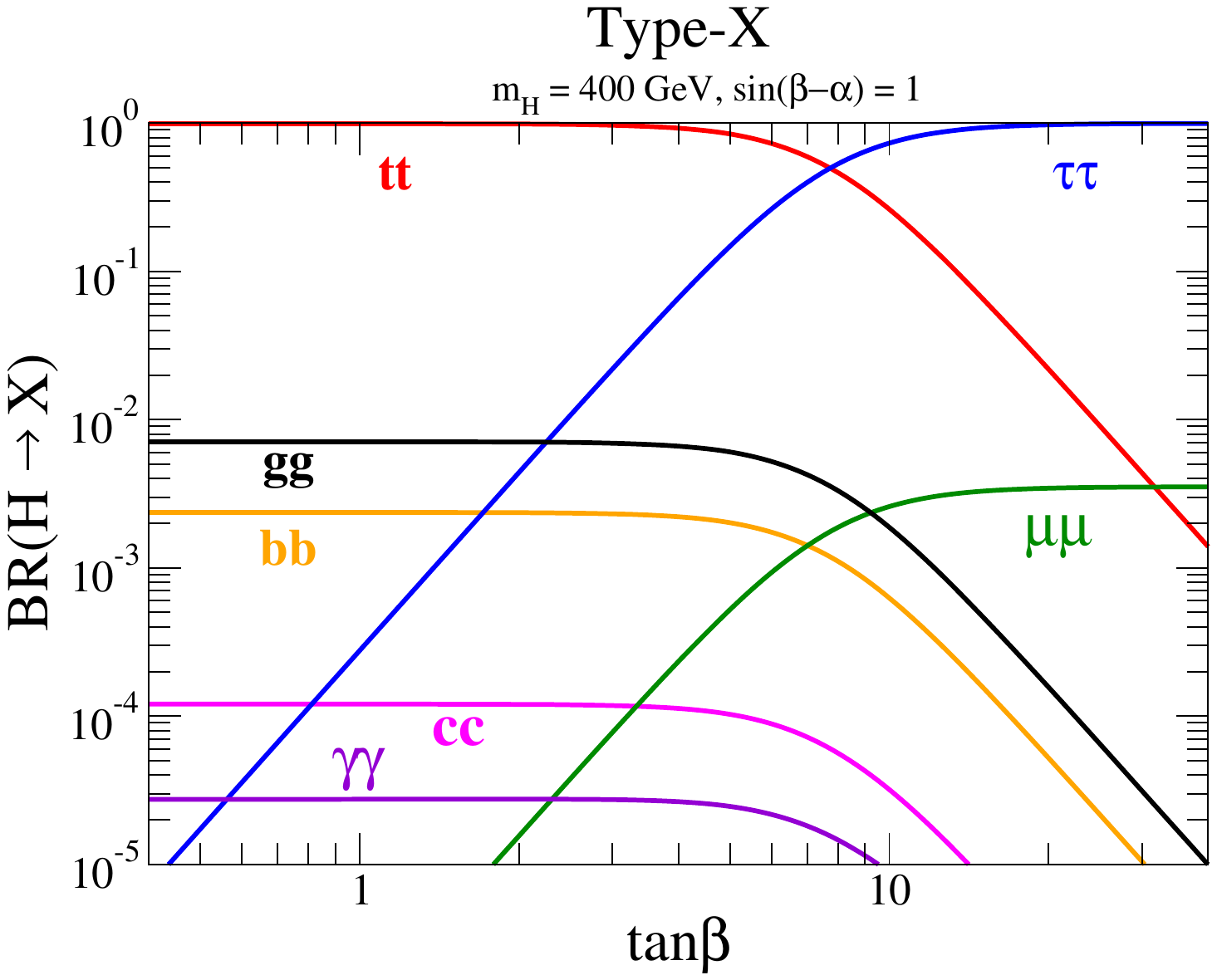}%\hspace{2mm}
 \includegraphics[width = 35mm]{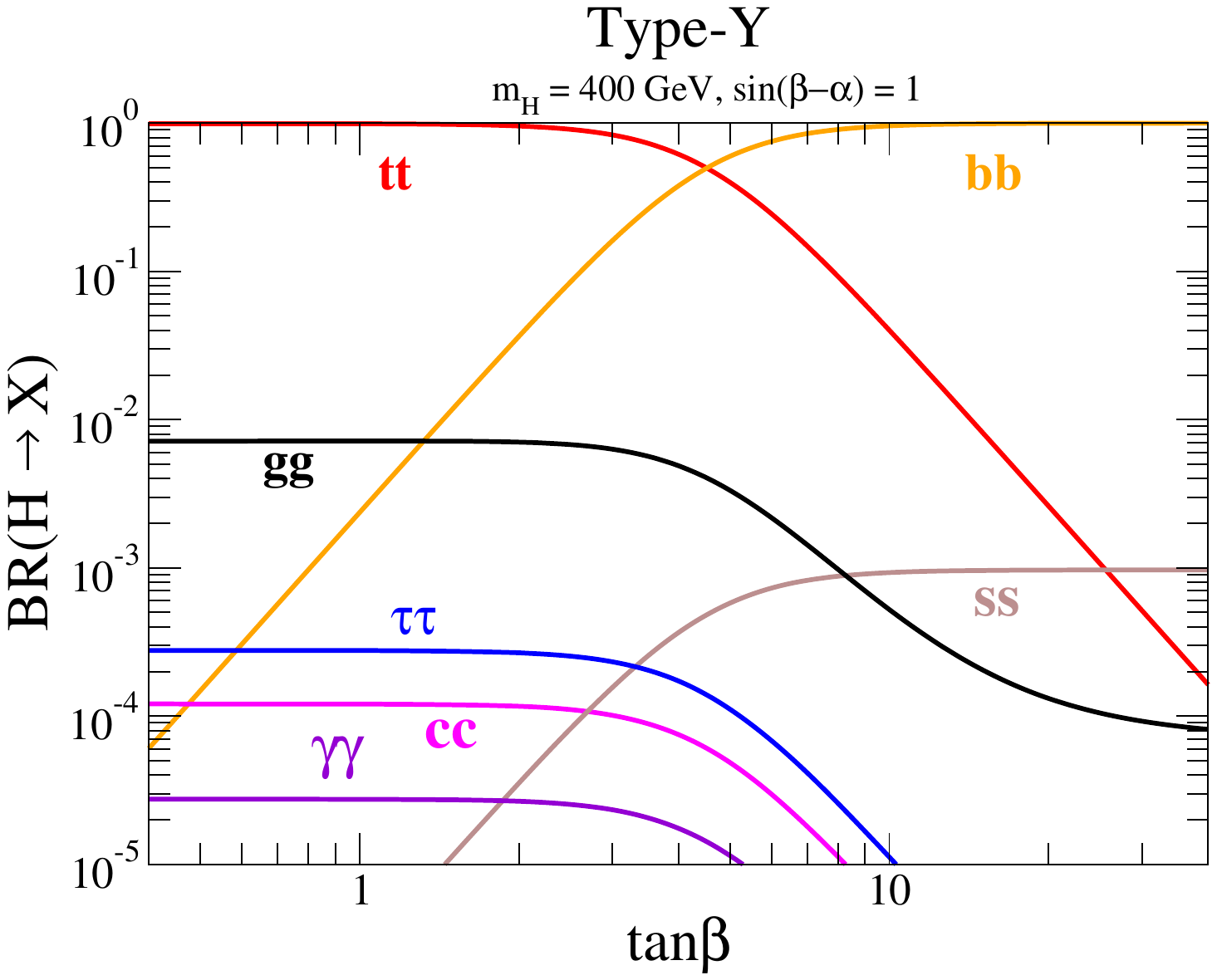}\\
\vspace{5mm}
  %\caption{XXX}
%\end{center}
%\end{figure}
%\begin{figure}[h]
%\begin{center}
 \includegraphics[width = 35mm]{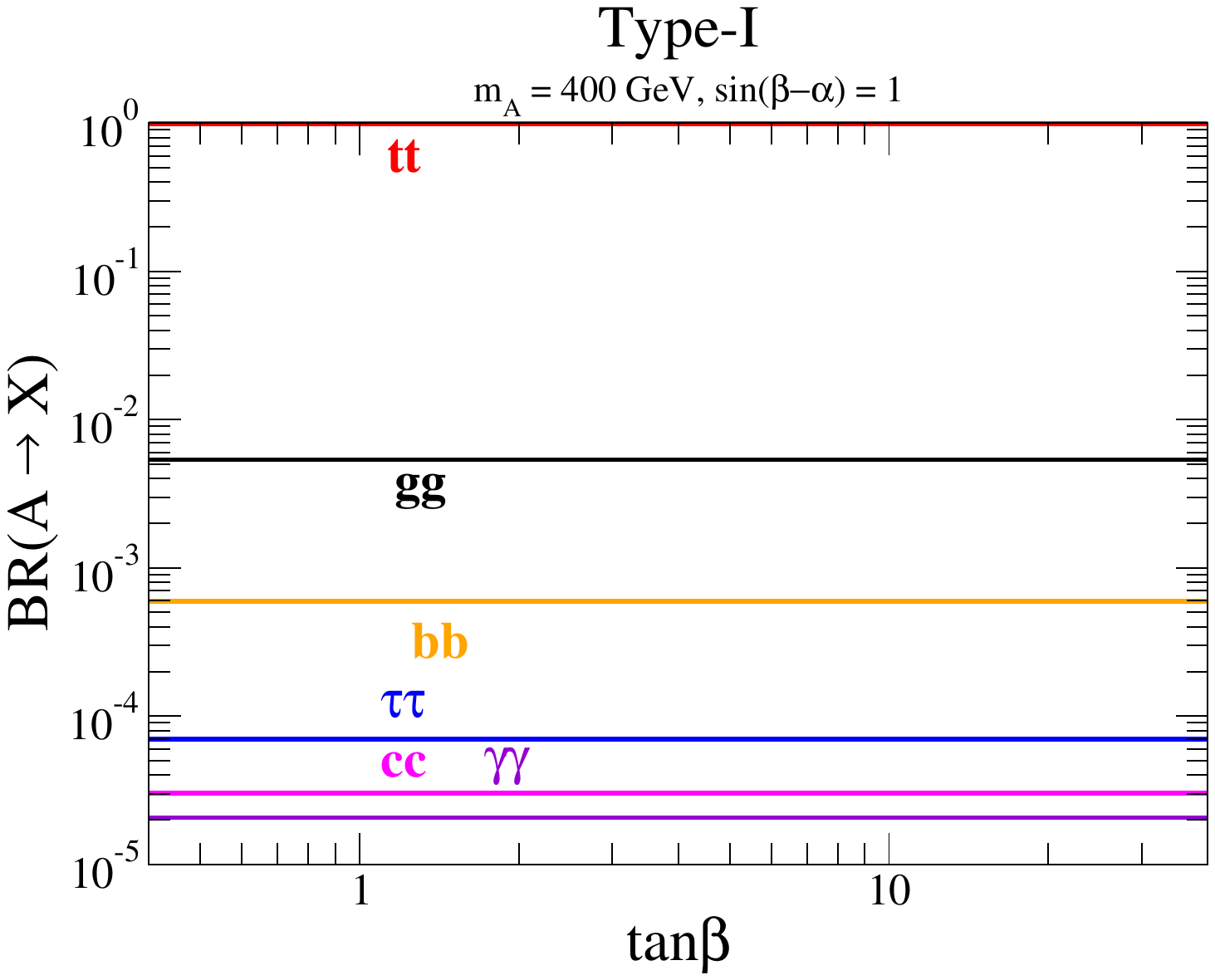}%\hspace{2mm}
 \includegraphics[width = 35mm]{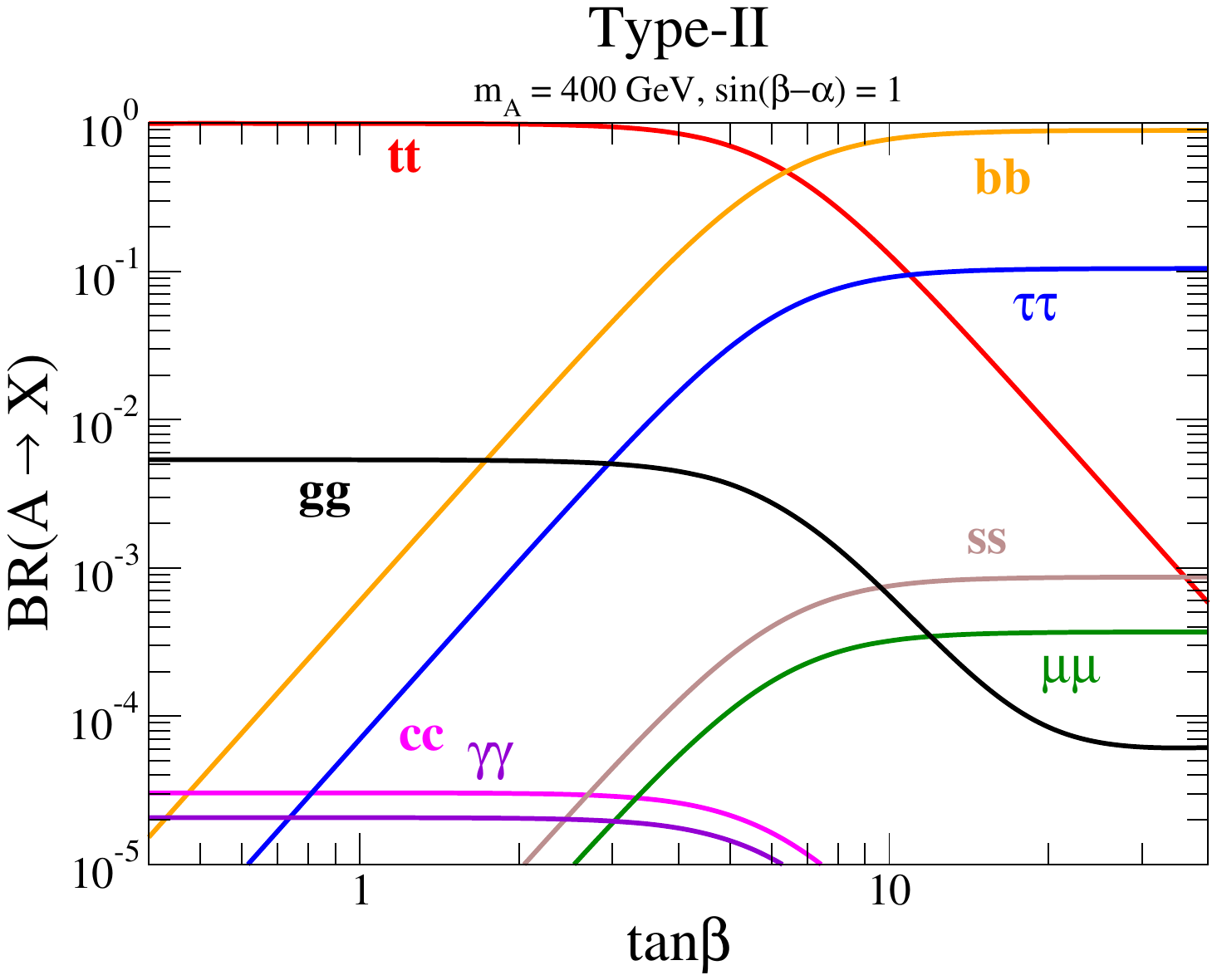}%\hspace{2mm}
 \includegraphics[width = 35mm]{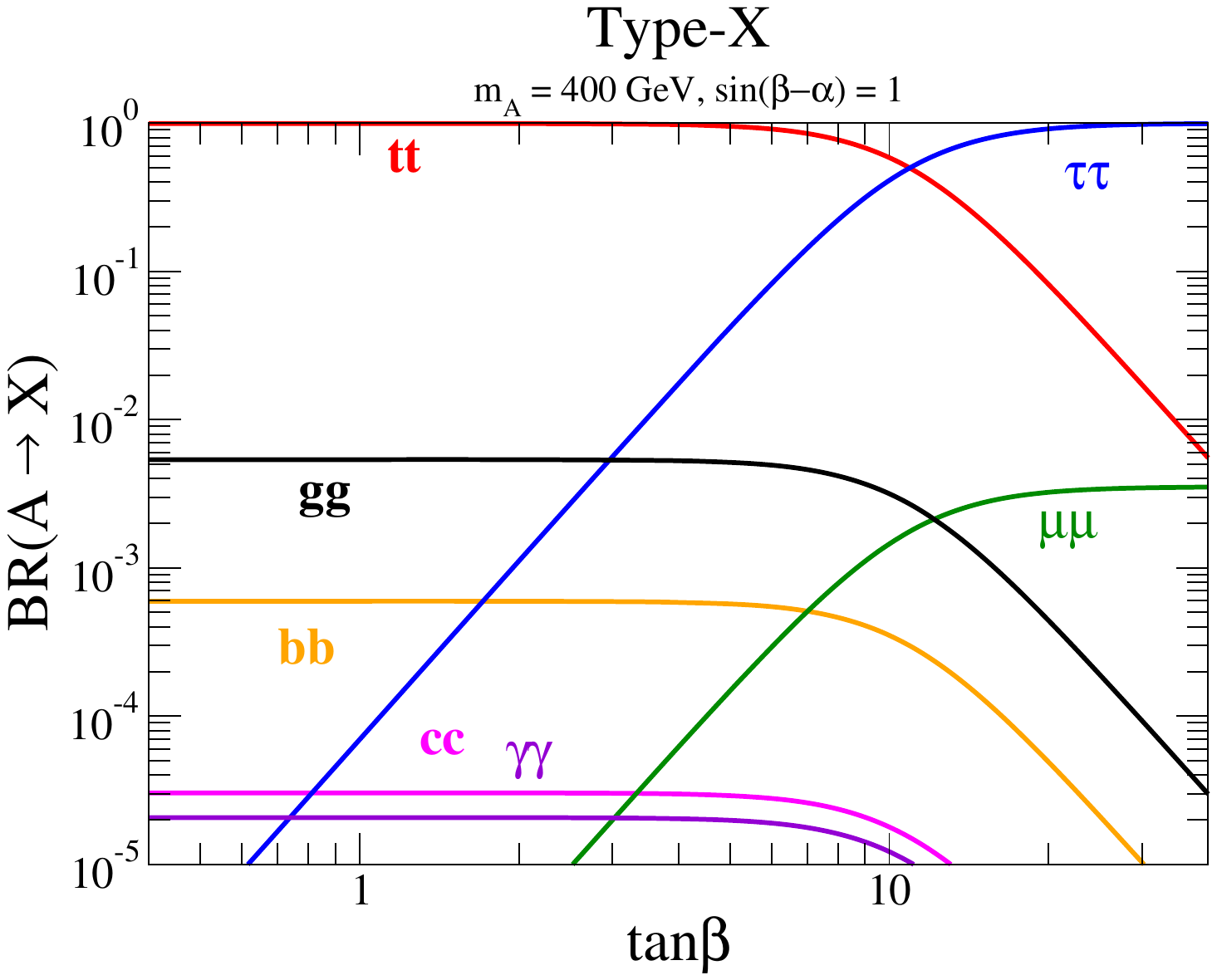}%\hspace{2mm}
 \includegraphics[width = 35mm]{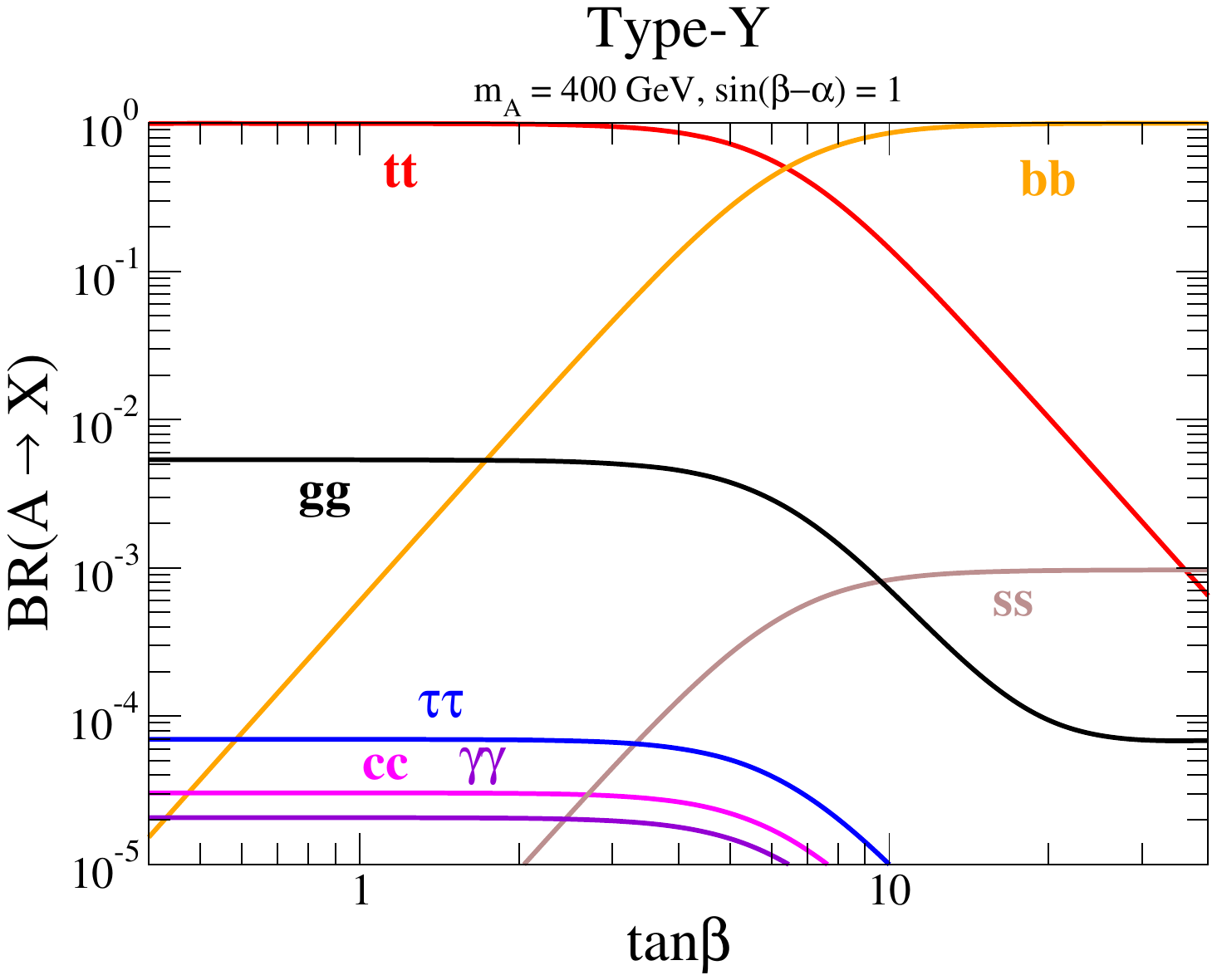}\\
\vspace{5mm}
  %\caption{XXX}
%\end{center}
%\end{figure}
%\begin{figure}[h]
%\begin{center}
 \includegraphics[width = 35mm]{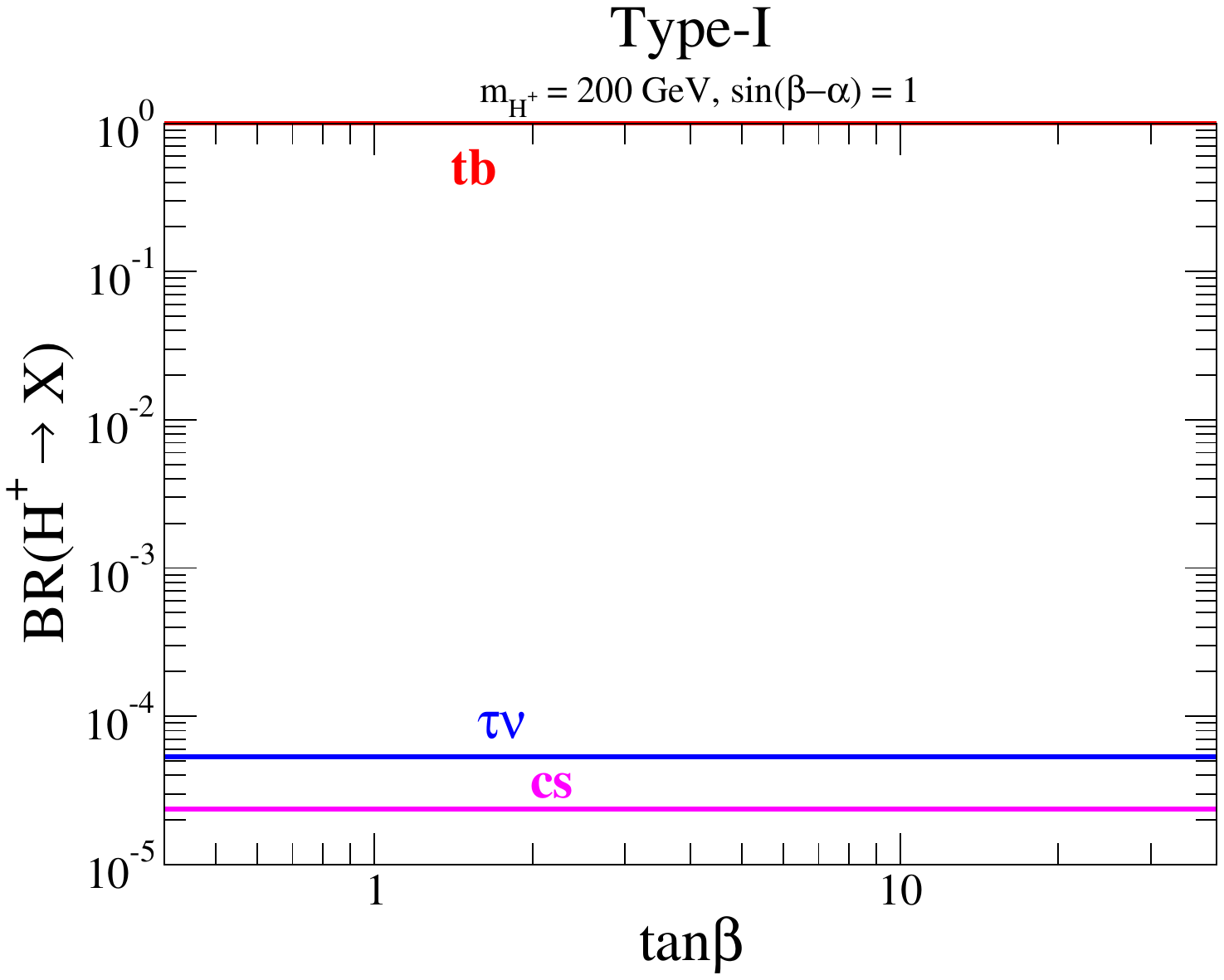}%\hspace{2mm}
 \includegraphics[width = 35mm]{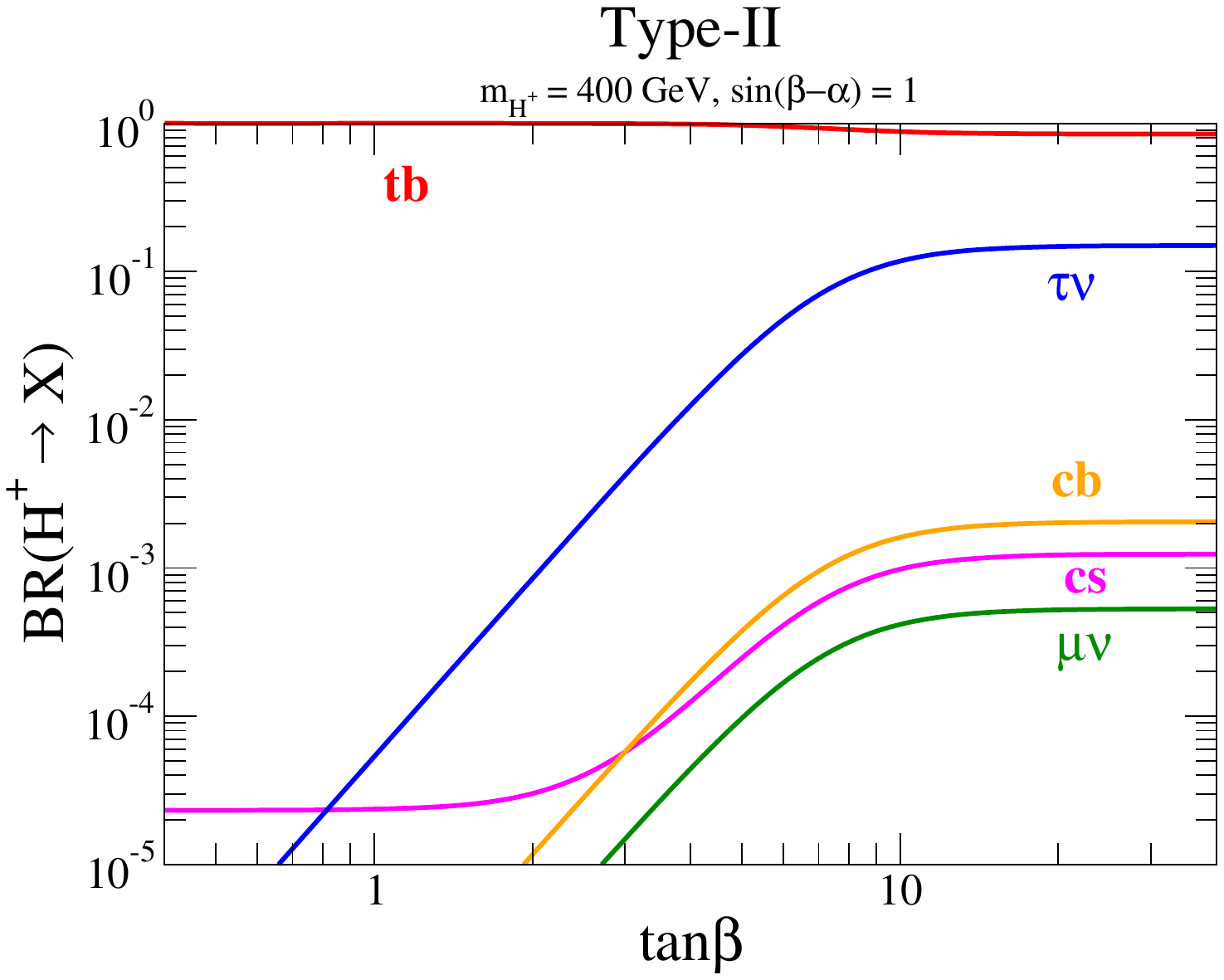}%\hspace{2mm}
 \includegraphics[width = 35mm]{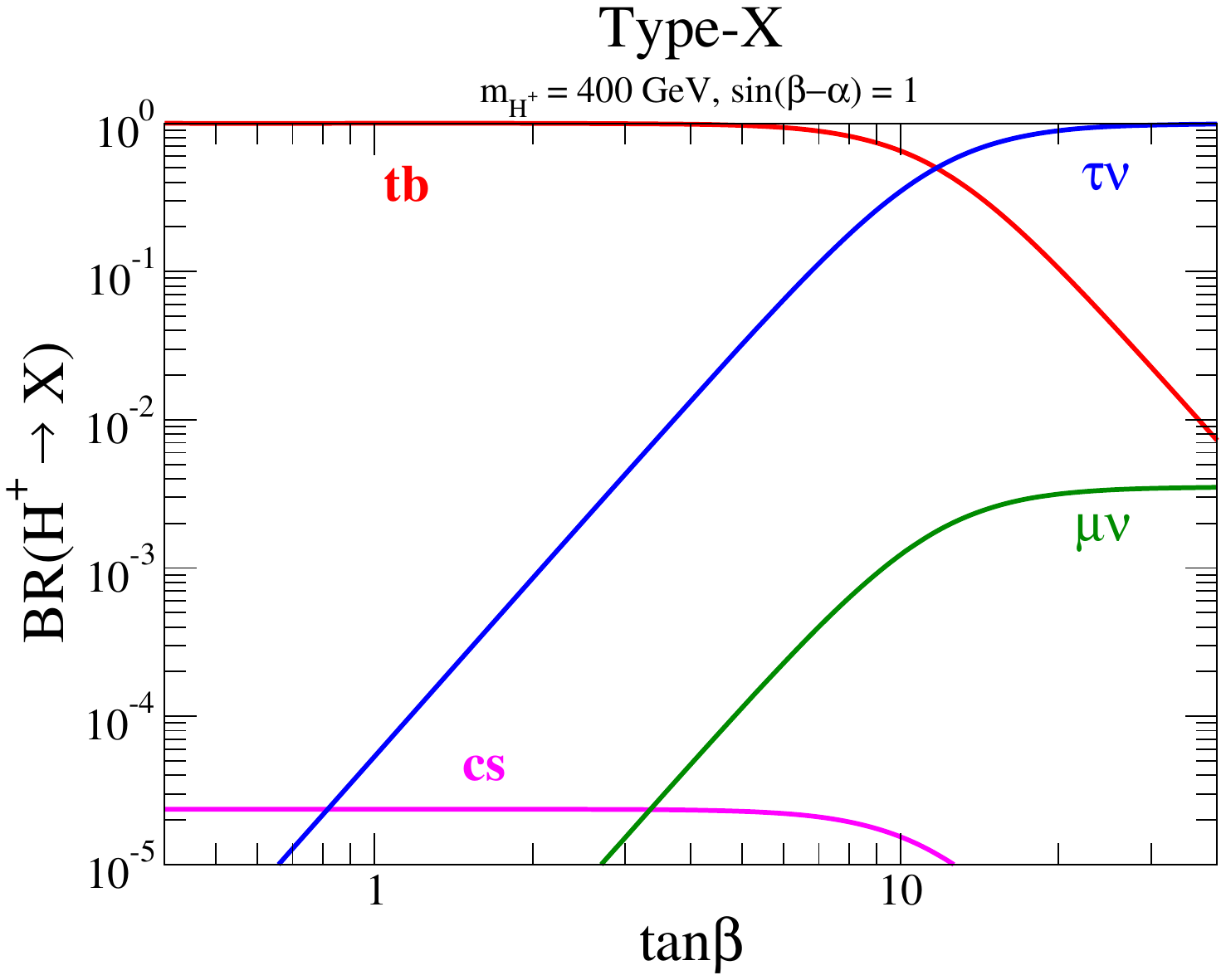}%\hspace{2mm}
 \includegraphics[width = 35mm]{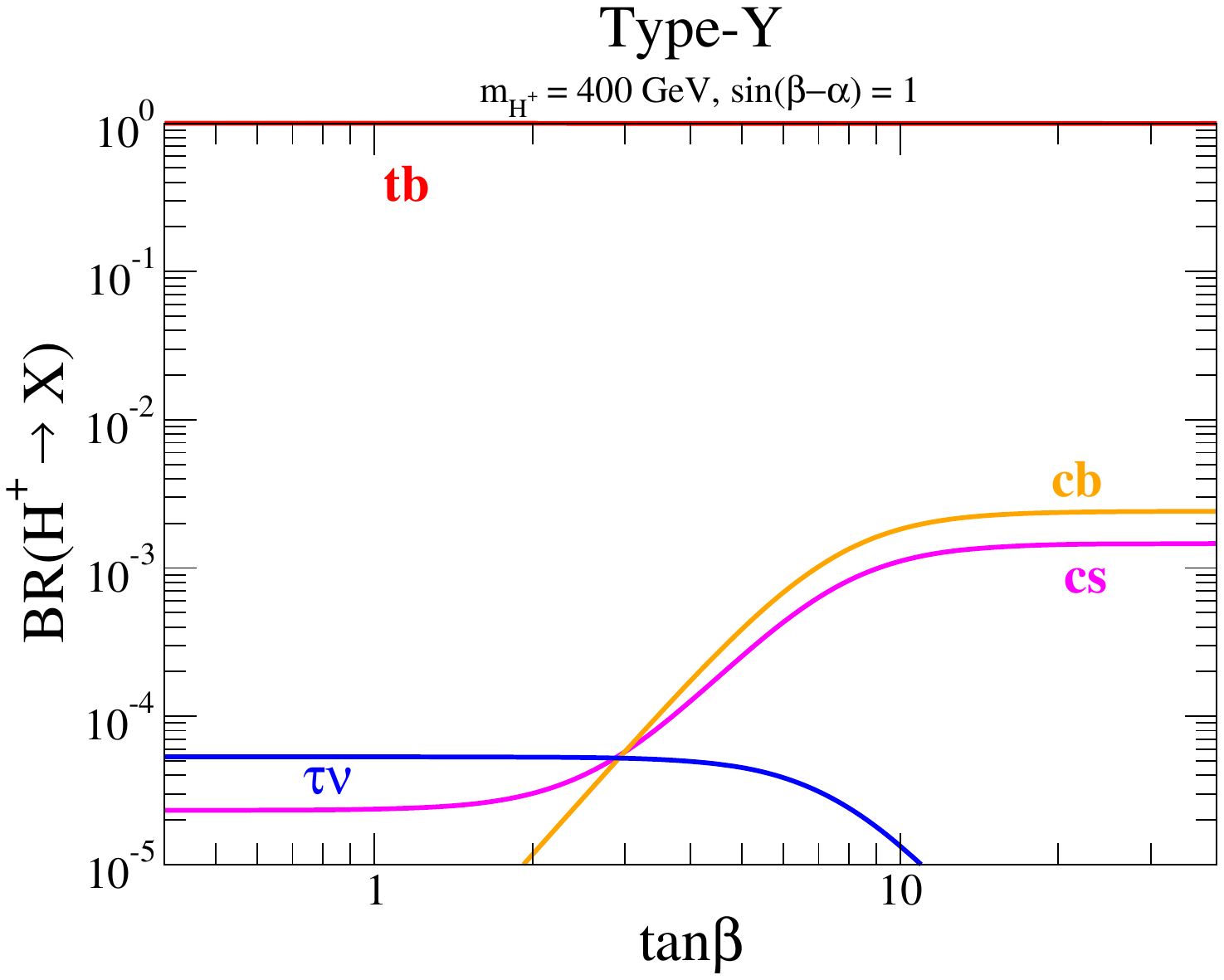}
 \caption{Decay branching ratios of $H$, $A$ and $H^\pm$
in the four different Types of 2HDM as a function of $\tan\beta$
for $m_H^{}=m_A^{}=m_{H^\pm}^{}=M=400$ GeV.
The SM-like limit $\sin(\beta-\alpha) =1$ is taken.\label{FIG:br_400}}
\end{center}
\end{figure}

In Figure~\ref{FIG:br_400}, figures of the decay
branching ratios of $H$, $A$ and $H^\pm$ similar to those in Fig.~\ref{FIG:br_200}
are shown for $\sin(\beta-\alpha)=1$ for $m_H^{}=m_A^{}=m_{H^\pm}^{}=M=400$ GeV.
The two-body decays $H/A \to t \bar t$ are now kinematically allowed in this case.
In general, the complexity of the $H$, $A$, $H^\pm$ decay schemes in
the various Types of Yukawa interactions
make it difficult to determine the underlying
model unless these scalars are created through a simple and well-characterized
pair-production reaction.  Thus, even if these scalars are discovered at the
LHC, it will be important to study them via the pair-production
process at the ILC.

% \begin{figure}
%\begin{center}
%\begin{minipage}{0.285\hsize}
%\includegraphics[width=4cm,angle=-90]{Chapter_Theory/figs/figs_br/BR_lh_1_sin96.pdf}
%\end{minipage}
%\begin{minipage}{0.23\hsize}
%\includegraphics[width=4cm,angle=-90]{Chapter_Theory/figs/figs_br/BR_lh_2_sin96.pdf}
%\end{minipage}
%\begin{minipage}{0.23\hsize}
%\includegraphics[width=4cm,angle=-90]{Chapter_Theory/figs/figs_br/BR_lh_X_sin96.pdf}
%\end{minipage}
%\begin{minipage}{0.23\hsize}
%\includegraphics[width=4cm,angle=-90]{Chapter_Theory/figs/figs_br/BR_lh_Y_sin96.pdf}
%\end{minipage}
%\hspace*{-10mm}
%\begin{minipage}{0.285\hsize}
%\includegraphics[width=4cm,angle=-90]{Chapter_Theory/figs/figs_br/BR_H_1_sin96.pdf}
%\end{minipage}
%\hspace{-6mm}
%\begin{minipage}{0.23\hsize}
%\includegraphics[width=4cm,angle=-90]{Chapter_Theory/figs/figs_br/BR_H_2_sin96.pdf}
%\end{minipage}
%\hspace{10mm}
%\begin{minipage}{0.23\hsize}
%\includegraphics[width=4cm,angle=-90]{Chapter_Theory/figs/figs_br/BR_H_X_sin96.pdf}
%\end{minipage}
%\begin{minipage}{0.23\hsize}
%\includegraphics[width=4cm,angle=-90]{Chapter_Theory/figs/figs_br/BR_H_Y_sin96.pdf}
%\end{minipage}
%
%\caption{Decay branching ratios of $h$ and $H$
%in the four different types of 2HDM as a function of $\tan\beta$
%for $m_h^{}=120$ GeV, $m_H^{}=150$ GeV, $M=148$ GeV and
%$\sin^2(\beta-\alpha) =0.96$.}
%\label{FIG:br_150'}
%\end{center}
%\end{figure}

\subsection{Constraints due to flavor physics}

\begin{figure}[b!]
\vspace{-0.8in}
\begin{minipage}{0.49\hsize}
\includegraphics[width=7.5cm,angle=0]{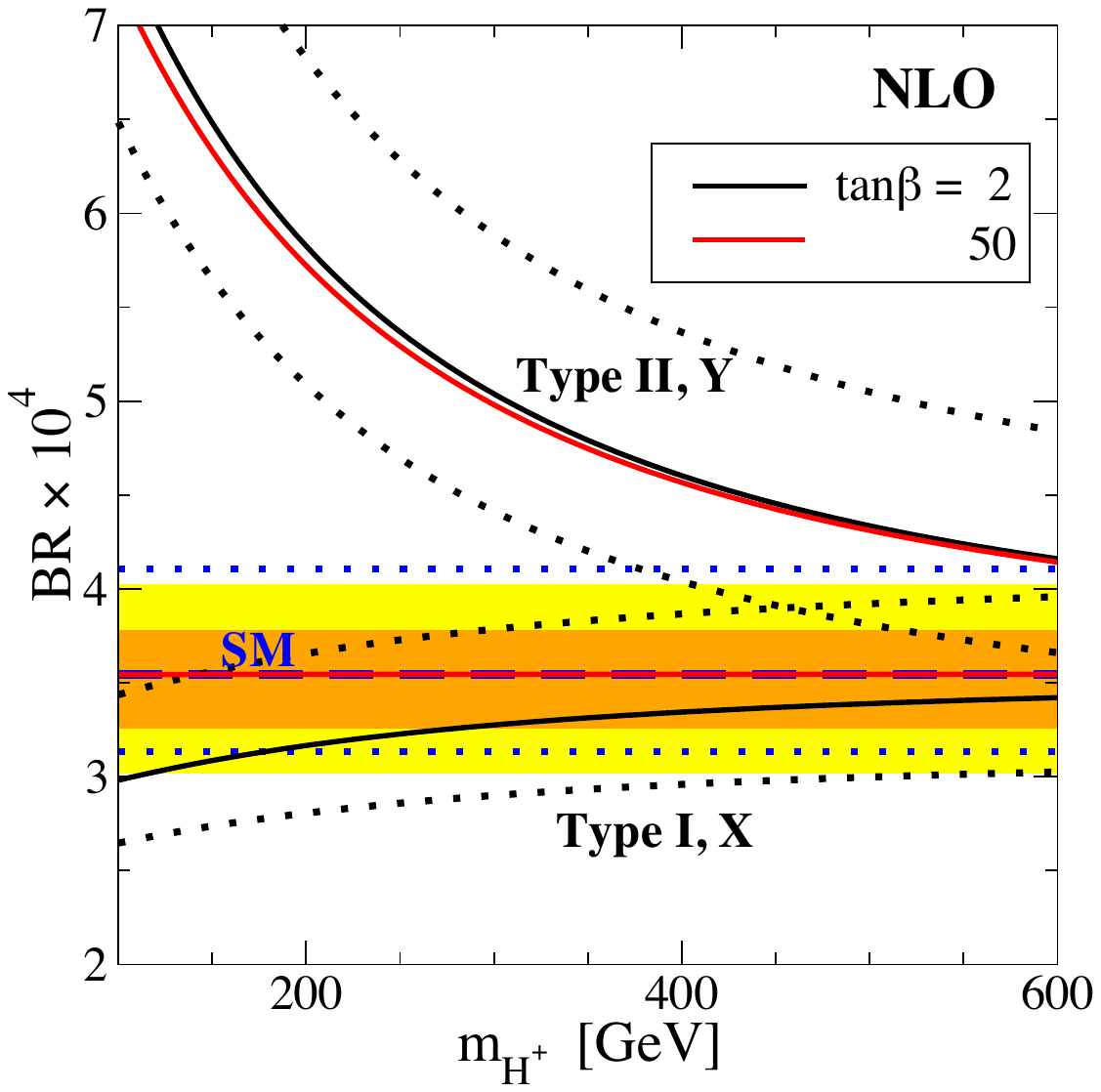}
\end{minipage}
\begin{minipage}{0.49\hsize}
\includegraphics[width=7.5cm,angle=0]{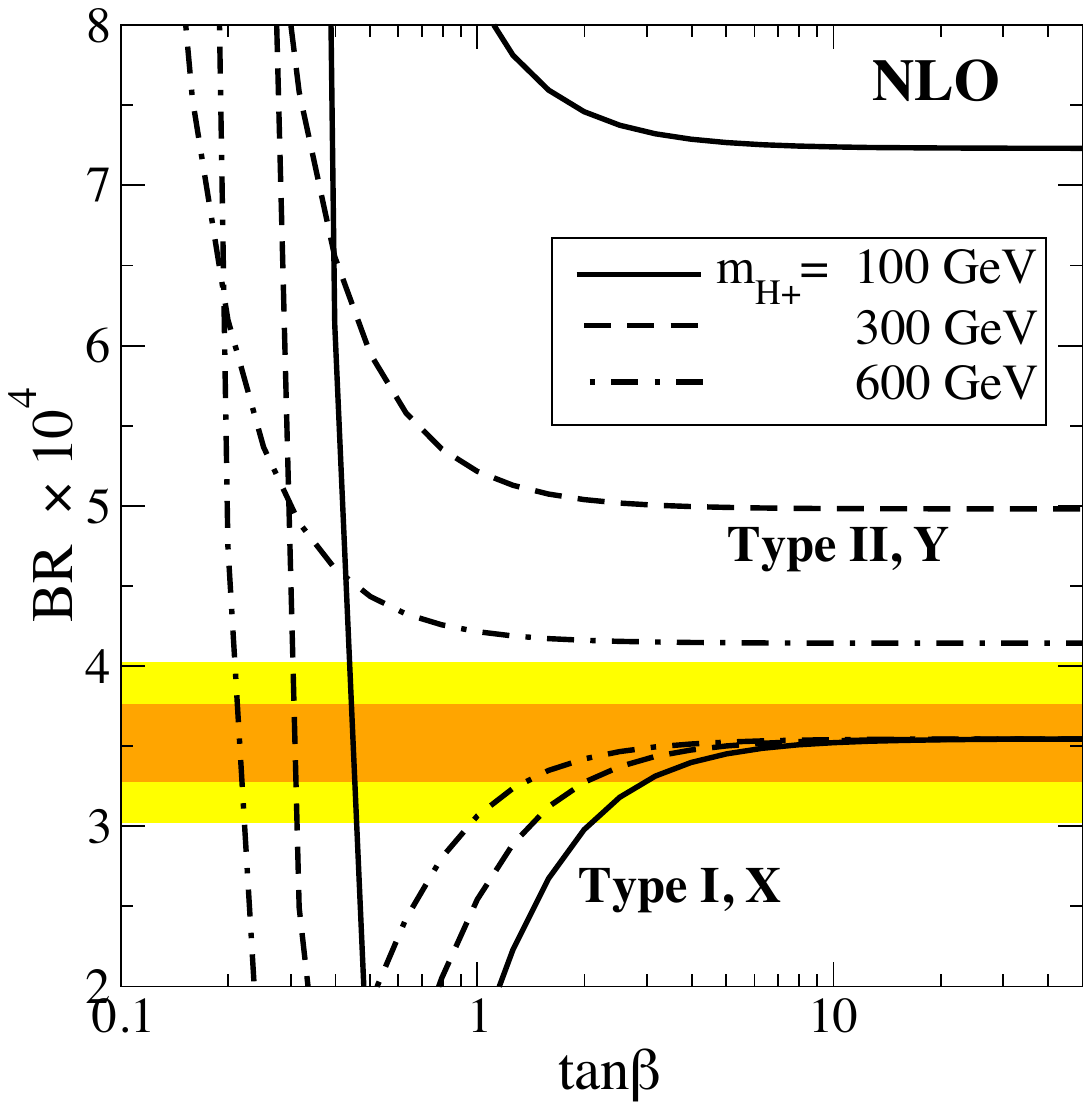}
\end{minipage}
\caption{Predictions of the decay branching ratio for $b\to s\gamma$
are shown at the NLO approximation as a function of $m_{H^\pm}^{}$ and $\tan\beta$.
The dark (light) shaded band represents $1\sigma$ $(2\sigma)$ allowed region
of current experimental data. In the left panel, solid (dashed) curves denote
the prediction for $\tan\beta=2$ $(50)$ in various 2HDMs. In the right panel, solid,
dashed and dot-dashed curves are those for $m_{H^\pm}^{}=100, 300$ and $600$ GeV,
respectively.}
\label{FIG:bsg}
\end{figure}

Indirect contributions of Higgs bosons to precisely measurable observables
can be used to constrain extended Higgs sectors.
In this section, we summarize the experimental bounds from flavor
experiments on the constrained 2HDMs introduced in
Section~\ref{specialforms}.   These bounds arise primarily due to
tree-level or loop diagrams that contain the charged Higgs boson.
The corresponding amplitudes involve the Yukawa interactions and hence
strongly depend on which Type of 2HDM is employed.

It is well known that the charged Higgs boson mass in the Type-II 2HDM
is stringently constrained by the precision measurements of the
radiative decay of $b\to s\gamma$~\cite{Hermann:2012fc}.
The process $b\to s\gamma$ receives contributions from the $W$ boson loop
and the charged Higgs boson loop in the 2HDM.
It is noteworthy that these two contributions
always constructively interfere in the Type-II (Type-Y) 2HDM,
whereas this is not the case in the Type-I (Type-X) 2HDM~\cite{Barger:1989fj,Aoki:2009ha,Su:2009fz,Logan:2009uf}.
In Fig.~\ref{FIG:bsg}~\cite{Aoki:2009ha}, we show the branching ratio of $B\to X_s\gamma$ for
each Type of 2HDM as a function of $m_{H^\pm}^{}$
(left-panel) and $\tan\beta$ (right-panel), which are evaluated at the next-to-leading
order (NLO) following the formulas in Ref.~\cite{Ciuchini:1997xe}.
The SM prediction at the NLO is also shown for comparison.
The theoretical uncertainty is about $15\%$
%\footnote{In Ref.~\cite{Ciuchini:1997xe},
%the theoretical uncertainty is smaller because the value for the error
%in $m_c^\text{pole}/m_b^\text{pole}$ is taken to be $7\%$, which gives main
%uncertainty in the branching ratio.}
in the branching ratio (as indicated by dotted curves in Fig.~\ref{FIG:bsg}),
which mainly comes from the pole mass of charm quark $m_c^\text{pole}=1.67\pm 0.07$
GeV ~\cite{Beringer:1900zz}.  (Note that Ref.~\cite{Ciuchini:1997xe} quotes and error in
$m_c^{\mathrm pole}$ of about $7\%$, which then leads to
a smaller theoretical uncertainty in the branching ratio for $b\to s\gamma$.)
The experimental bounds of the branching ratio are also indicated, where
the current world average value is given by
${\mathrm BR}(B\to X_s\gamma)=(3.52\pm 0.23\pm0.09)\times 10^{-4}$~\cite{Barberio:2008fa}.
One can see from Fig.~\ref{FIG:bsg} that the branching ratio in
the Type-I (Type-X) 2HDM lies within the 2 $\sigma$ experimental error
in all the regions of $m_{H^{\pm}}$ indicated for $\tan\beta\gtrsim 2$,
while that in the Type-II (Type-Y) 2HDM is far from the value indicated by
the data for a light charged Higgs boson region $(m_{H^\pm}^{}\lesssim 200$
GeV$)$.
In the right figure, a cancellation occurs in the Type-I (Type-X) 2HDM
since there are destructive interferences between the $W$ boson and
the $H^\pm$ contributions.
The results of these figures indicate that the $B\to X_s\gamma$
experimental results still permit a light charged Higgs boson
in the Type-I (Type-X) 2HDM.
We note that in the MSSM the chargino contribution can compensate
the charged Higgs boson contribution~\cite{Goto:1994ck,Ciuchini:1998xy}.
This cancellation weakens the limit on $m_{H^\pm}^{}$ from $b\to s\gamma$
in the Type-II 2HDM,
and allows a light charged Higgs boson
as in the Type-I (Type-X) 2HDM.
At the NNLO approximation, the branching ratio for $b\to s\gamma$
has been evaluated in the SM in
Refs.~\cite{Misiak:2006ab,Misiak:2006zs,Becher:2006pu}.
The predicted value at the NNLO approximation is less than that at the NLO approximation
over a wide range of renormalization scales.
The branching ratio for $b\to s\gamma$ in the Standard Model is $(3.15\pm 0.23)\times
10^{-4}$~\cite{Misiak:2006ab}, and a lower bound for $m_{H^\pm}^{}$, after adding the
NLO charged Higgs contribution, is found to be
$m_{H^\pm}^{} \gtrsim 380$~GeV ($95\%$ CL) in the Type-II (Type-Y)
2HDM~\cite{Hermann:2012fc}.  (Note that the calculation of
Refs.~\cite{Misiak:2006zs,Becher:2006pu} for
the NNLO branching ratio in the SM yields $(2.98\pm 0.26)\times
10^{-4}$, and the corresponding charged Higgs mass bound is somewhat relaxed.)
On the other hand, in the Type-I (Type-X) 2HDM, although the branching ratio
becomes smaller as compared to the NLO evaluation, no serious bound on
$m_{H^\pm}^{}$ can be found for $\tan\beta \gtrsim 2$.
Therefore, the charged Higgs boson mass is not expected to be strongly constrained in
the Type-I (Type-X) 2HDM even at the NNLO, and the conclusion that
the Type-I (Type-X) 2HDM
is favored for $m_{H^\pm}^{}\lesssim 200$ GeV based on the NLO
analysis should not be changed.

The decay $B\to\tau\nu$ has been examined in the Type-II 2HDM in Refs.~\cite{Akeroyd:2003zr,Krawczyk:2007ne}.
The data for ${\mathrm BR}(B^+\to\tau^+\nu_\tau)=(1.65\pm 0.34)\times 10^{-4}$
are obtained at the $B$ factories~\cite{Beringer:1900zz}.
The decay branching ratio can be written as~\cite{Krawczyk:2007ne}
\begin{align}
\frac{{\mathcal B}(B^+\to\tau^+\nu_\tau)_\text{2HDM}}{{\mathcal B}(B^+\to\tau^+\nu_\tau)_\text{SM}}
\simeq\left(1-\frac{m_B^2}{m_{H^\pm}^2}\xi_A^d\xi_A^e\right)^2,
\end{align}
where coefficients  $\xi_A^d$ and $\xi_A^e$ are defined in Table~\ref{yukawa_tab}.
In Fig.~\ref{FIG:mH+tanb}, the allowed region from the
experimental $B\to\tau\nu$ results
is shown in the Type-II 2HDM. The dark (light) shaded region denotes
the $2\sigma$ $(1\sigma)$ exclusion from the $B\to\tau\nu$ measurements.
The process is important only in the Type-II 2HDM at large values of $\tan\beta$.
The other Types of Yukawa interactions are not constrained by this process.

\begin{figure}[t!]
\begin{minipage}{0.49\hsize}
\includegraphics[width=7.5cm]{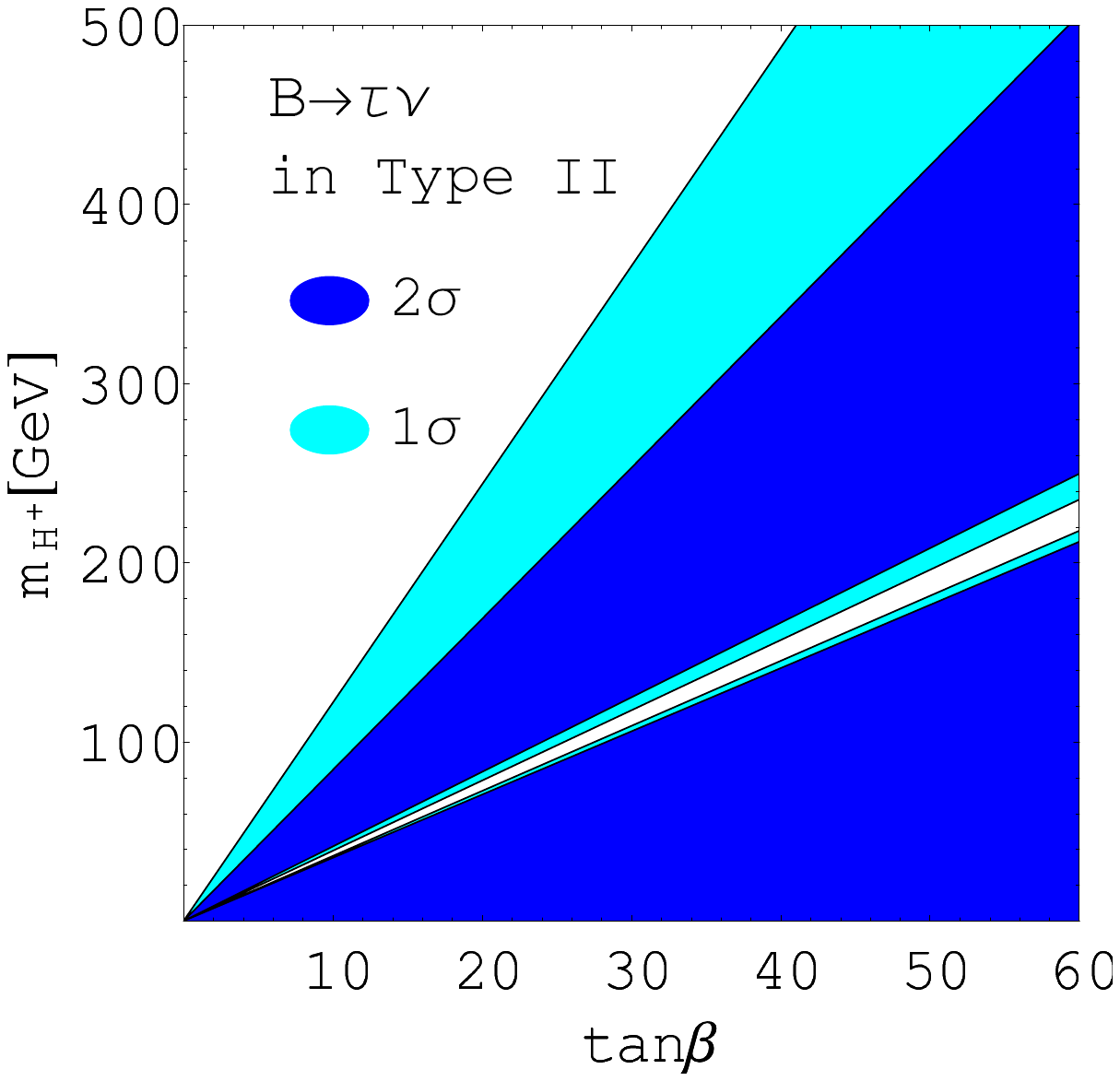}
\end{minipage}
\begin{minipage}{0.49\hsize}
\includegraphics[width=7.5cm]{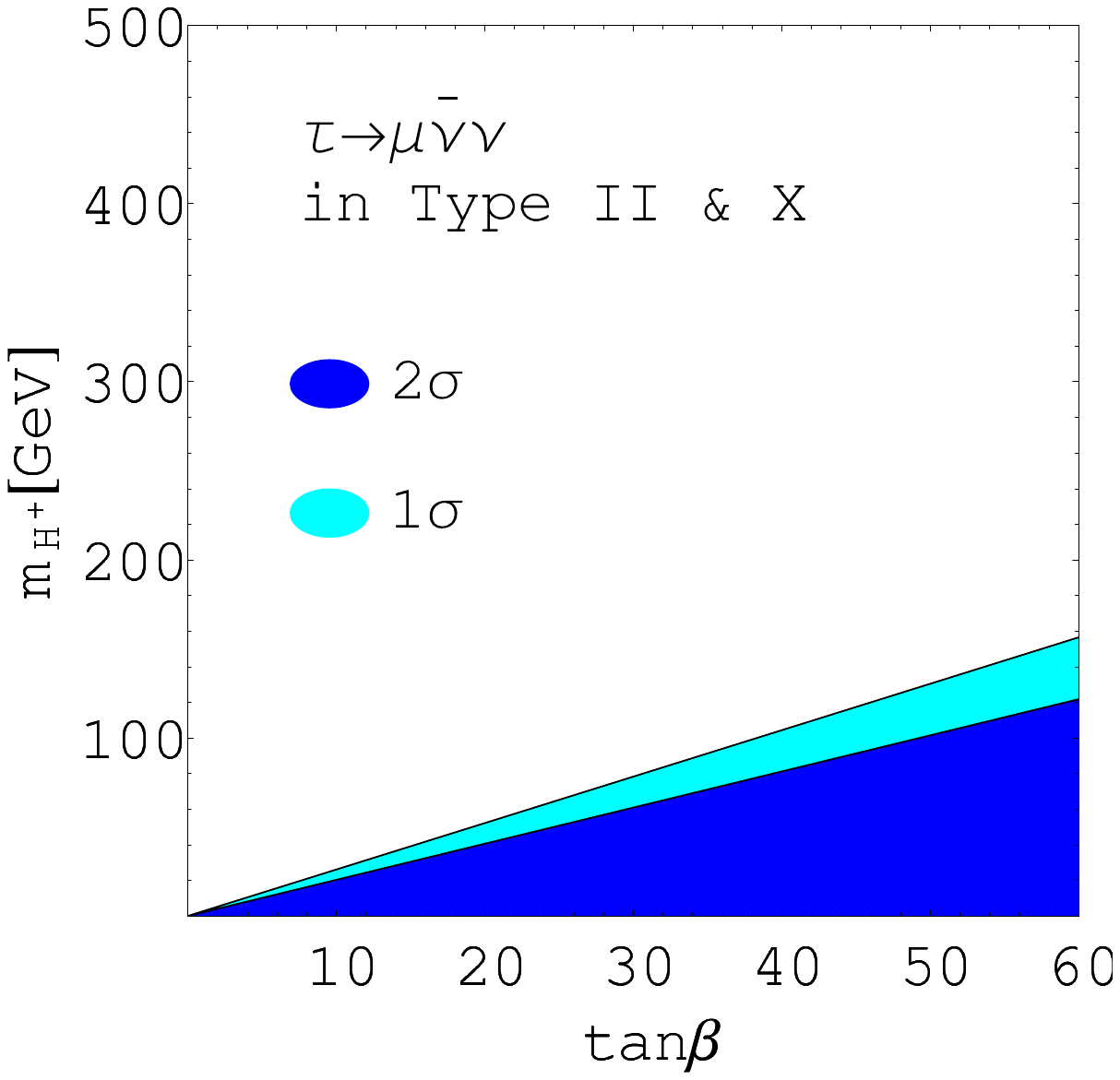}
\end{minipage}
\vspace{-0.4in}
\caption{Bounds from $B\to\tau\nu$ (left panel) and tau leptonic decay (right panel) on $m_{H^\pm}^{}$ as a function of $\tan\beta$ are shown. The dark (light) shaded region
corresponds to the $2\sigma$ $(1\sigma)$ exclusion of these experimental results.
In the Type-II 2HDM the wide parameter space is constrained by $B\to\tau\nu$,
while only the tau leptonic decays are important for the Type-X 2HDM.\label{FIG:mH+tanb}}
\end{figure}

The rate for the leptonic decay of the tau lepton, $\tau\to\mu\overline{\nu}\nu$,
can deviate from the SM value due to the presence of a light charged Higgs
boson~\cite{Krawczyk:2004na}.
The partial decay rate is approximately expressed as
\begin{align}
\frac{\Gamma_{\tau\to\mu\overline{\nu}\nu}^\text{2HDM}}{\Gamma_{\tau\to\mu\overline{\nu}\nu}^\text{SM}}
&\simeq 1-\frac{2m_\mu^2}{m_{H^\pm}^2}{\xi_A^e}^2 \kappa\left(\frac{m_\mu^2}{m_\tau^2}\right)+\frac{m_\mu^2m_\tau^2}{4m_{H^\pm}^4}{\xi_A^e}^4,
\end{align}
where the function $\kappa(x)$ is defined by
\begin{equation}
\kappa(x)=\frac{1+9x-9x^2-x^3+6x(1+x)\ln x}{1-8x+8x^3-x^4-12x^2\ln x}\,.
\end{equation}
%$\kappa(x)=g(x)/f(x)$ is defined by
%$f(x)=1-8x-12x^2\ln x+8x^3-x^4,$ and $g(x)=1+9x-9x^2-x^3+6x(1+x)\ln x$.
In the Type-II (Type-X) 2HDM, the leptonic Yukawa interaction
can be enhanced in the large $\tan\beta$ region. Hence, both model Types are weakly
constrained by tau decay data, as indicated in Fig.~\ref{FIG:mH+tanb}.

%++++++++++++++++++++++++++++++++++++++++++++++++++++++++++++++++++++++++++++++++++++++++++++++++
The precision measurement of the muon anomalous magnetic moment
can yield a mass bound on the Higgs boson in the SM~\cite{Jackiw:1972jz,Fujikawa:1972fe}.
This constraint can be applied to models with additional
interactions such as the 2HDM~\cite{Leveille:1977rc,Haber:1978jt,Krawczyk:1996sm}.
At the one-loop level, the 2HDM contribution is given by
\begin{align}
\delta a_\mu^{1-\text{loop}}
&\simeq \frac{G_Fm_\mu^4}{4\pi^2\sqrt2}
\left[\sum_{\phi^0=h,H}\frac{{\xi_{\phi^0}^e}^2}{m_{\phi^0}^2}
\left(\ln\frac{m_{\phi^0}^2}{m_\mu^2}-\frac76\right)
+\frac{{\xi_A^e}^2}{m_A^2}
\left(-\ln\frac{m_A^2}{m_\mu^2}+\frac{11}6\right)
-\frac{{\xi_A^e}^2}{6m_{H^\pm}^2}\right].
\end{align}
This process is also purely leptonic and only yields milder bounds
on the Higgs boson masses for very large $\tan\beta$ values
in the Type-II (Type-X) 2HDM.  No effective bound on the Type-I
(Type-Y) 2HDM is obtained.
It is also known that the two-loop (Barr-Zee type) diagram
can significantly affect $a_\mu$~\cite{Barr:1990vd}.
The contribution from this class of diagrams is only important
for large $\tan\beta$ values with smaller Higgs boson masses in the Type-II 2HDM.
For the other Types of 2HDM, a much less effective bound
on the parameter space is obtained.
%++++++++++++++++++++++++++++++++++++++++++++++++++++++++++++++++++++++++++++++++++++++++++++++++

\begin{figure}[t]
\centering
\includegraphics[width=7cm]{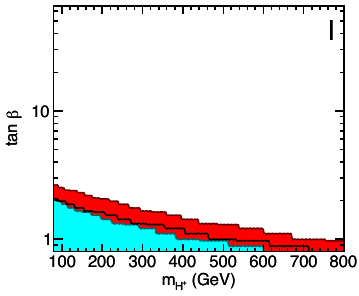}
\includegraphics[width=7cm]{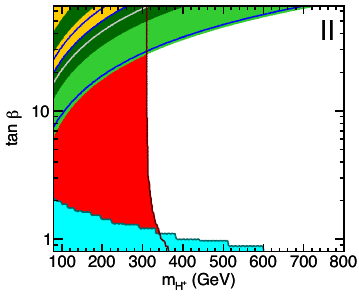}\\
\includegraphics[width=7cm]{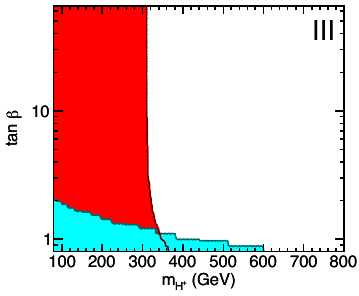}
\includegraphics[width=7cm]{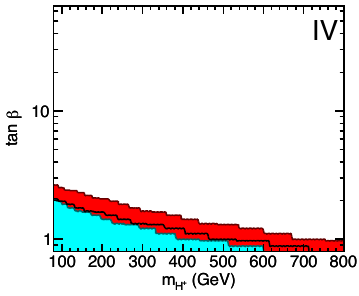}
\caption{
Excluded regions of the ($m_{H^+}\,,\,\tan\beta$) parameter space for
$Z_2$-symmetric 2HDM Types.  The Type Y and X models [cf.~Table~\ref{Tab:type}]
are denoted above by Type III and IV, respectively.
The color coding is as follows: $\mathrm{BR}(B\to X_s\gamma)$ (red), $\Delta_{0-}$ (black contour),
$\Delta M_{B_d}$ (cyan), $B_u\to\tau\nu_\tau$ (blue), $B\to D\tau\nu_\tau$ (yellow),
$K\to\mu\nu_\mu$ (gray contour), $D_s\to\tau\nu_\tau$ (light green),
and $D_s\to\mu\nu_\mu$ (dark green).
\label{fig:combined}
}
\end{figure}

The $B^0-\bar B^0$ mass differences, $\Delta M_{B_d}$ and $\Delta M_{B_s}$,
are sensitive to charged Higgs exchange via box-type diagrams in which top
quarks are also exchanged.  The data exclude relatively large top Yukawa couplings
that are proportional $m_t \cot\beta$ for smaller charged Higgs boson masses.
This constraint is common among the four Types of 2HDMs.
In light of the current data for $\Delta M_{B_d}$,
$\tan\beta < 1$ is ruled out for $m_{H^\pm} < 500$ GeV at 95 \% C.L.~\cite{Mahmoudi:2009zx}

All the important constraints on the parameter space for each Type of 2HDM
are summarized in Fig.~\ref{fig:combined}, where excluded regions from the data of
$\mathrm{BR}(B\to X_s\gamma)$, $\Delta_{0-}$, $\Delta M_{B_d}$, $B_u\to\tau\nu_\tau$,
$B\to D\tau\nu_\tau$, $K\to\mu\nu_\mu$, $D_s\to\tau\nu_\tau$, and $D_s\to\mu\nu_\mu$
are plotted in the ($m_{H^+}$\,,\,$\tan\beta$) plane~\cite{Mahmoudi:2009zx}.
Here, we have included among the list of flavor observables the degree
of isospin asymmetry in the exclusive decay mode $B\to K^*\gamma$,
defined as~\cite{Kagan:2001zk,Bosch:2001gv}
\begin{equation}
\Delta_{0-}\equiv\frac{\Gamma(\overline{B}\llsup{0}\to\overline{K}\llsup{*0})-
\Gamma(\overline{B}\llsup{\,-}\to\overline{K}\llsup{*\,-})}
{\Gamma(\overline{B}\llsup{0}\to\overline{K}\llsup{*0})+
\Gamma(\overline{B}\llsup{\,-}\to\overline{K}\llsup{*\,-})}\,.
\end{equation}
The exclusion of low $\tan\beta< 1$ in all four model Types for $m_{H^+}<500$~GeV,
arises as a result of three observables: $\mathrm{BR}(B\to X_s\gamma)$, $\Delta_{0-}$, and $\Delta M_{B_d}$.
The constraints at low $\tan\beta$ are similar between the model Types, since the couplings to the up-type quarks are universal. In the Type I 2HDM,
a value of $\tan\beta>1$ signals the decoupling of one Higgs doublet from the whole fermion sector.
In Type II and Type III (=Type Y), which share the same coupling pattern for the quarks, there
exists a $\tan\beta$-independent lower limit of $m_{H^+}\gtrsim 300$ GeV imposed by
$\mathrm{BR}(B\to X_s\gamma)$.  (This latter constraint is now
somewhat more stringent in light of Ref.~\cite{Hermann:2012fc}.)
No generic lower limit on $m_{H^+}$ is found in Type I and Type IV (=Type X) models.
Constraints for high $\tan\beta$ are only obtained in the Type II model.
This can be understood by noting that the leptonic and semi-leptonic observables
require $\tan\beta$-enhanced couplings $\lambda_{dd}\lambda_{\ell\ell}\sim \tan^2\beta\gg 1$
($d=d,s,b$) for the contributions to be interesting. In the Type III (=Type Y) and
and the Type IV (=Type X) 2HDMs, these couplings are instead always $\lambda_{dd}\lambda_{\ell\ell}=-1$,
while in Type I they are proportional to $\cot^2\beta$.

Finally, recently current data from BaBar of the $\bar B \to D \tau
\bar \nu$ and $\bar B \to D^\ast \tau \bar \nu$ slightly deviate from
the SM predictions by 2.0 $\sigma$ and 2.7 $\sigma$,
respectively~\cite{Lees:2012xj}.  Moreover, these data are also inconsistent with the
Type-I (X) and Type-II (Y) 2HDMs, since both decay rates, which depend
on the charged Higgs mass,  cannot be
explained simultaneously for the same value of $m_{H^\pm}$.
However, these data can be compatible in
the context of a more general 2HDM with unconstrained Higgs-quark
Yukawa interactions~\cite{Tanaka:2012nw}.  Meanwhile, there is no
confirmation yet of the BaBar results for $\bar B \to D \tau \bar\nu$
and $\bar B \to D^\ast \tau \bar \nu$ from the BELLE collaboration.
Thus, it is certainly premature to claim a definitive deviation from
the predictions of the Standard Model as well as all 2HDMs with Types
I, II, X, or Y Yukawa interactions.

\subsection{The inert 2HDM}
\label{inertsec}

The inert 2HDM is one of the simplest extensions of the
Standard Model~\cite{Barbieri:2006dq,Deshpande:1977rw}.
%In section \ref{specialforms}, we examined the constraints imposed on
%the 2HDM by various $\mathbb{Z}_2$ discrete symmetries in which $\Phi_1\to
%+\Phi_1$ and $\Phi_2\to -\Phi_2$.  As a consequence of this symmetry,
%$m_{12}^2=\lambda_6=\lambda_7=0$ in some basis which is rotated by an
%angle $\beta$ with respect to the Higgs basis.
The inert 2HDM is defined as a Type-I model
%(see Section~\ref{specialforms})
in which the
$\mathbb{Z}_2$ discrete symmetry is imposed in the Higgs basis.
The $\mathbb{Z}_2$ charge assignments for the inert 2HDM are given
in Table~\ref{tabinert}.  In particular, the Higgs basis field $H_2$,
which has no vev, is odd under the discrete symmetry.
As a result of this discrete symmetry, $Y_3=Z_6=Z_7=0$ and
there are no Yukawa couplings of $H_2$ to fermions.
%
%%%%%%%%%%%%%%%%%%%%%%%%%%%%%%%%%%%%%%%%%%%%%%%%%
\begin{table}[ht!]
 \begin{center}
 \caption{The $\mathbb{Z}_2$ charge assignments that define the Inert
   2HDM, where $H_1$ and $H_2$ are the Higgs doublet fields in the
   Higgs basis.}
\label{tabinert}
\begin{tabular}{|cl||c|c|c|c|c|c|}
\hline && $H_1$ & $H_2$ & $U_R^{}$ & $D_R^{}$ & $E_R^{}$ &
 $U_L$, $D_L$, $N_L$, $E_L$ \\  \hline
Inert 2HDM  && $+$ & $-$ & $+$ & $+$ & $+$ & $+$ \\
\hline
\end{tabular}
\end{center}
\end{table}
%%%%%%%%%%%%%%%%%%%%%%%%%%%%%%%%%%%%%%%%%%%%%

Since $Z_6=0$,
we are in the exact alignment limit, in which $h=\sqrt{2}\,\Re(H_1^0-v)$
is a mass-eigenstate whose couplings are equivalent to those of the
SM Higgs bosons.  We also identify
\beq
\Phi_2=\begin{pmatrix} H^+ \\ (H+iA)/\sqrt{2}\end{pmatrix}\,,
\eeq
where $H$, $A$ and $H^+$ are the other Higgs mass eigenstates.
The $\mathbb{Z}_2$ discrete symmetry is unbroken in the vacuum since
$\vev{H_2^0}=0$.  Thus, there are no couplings involving an odd
number of $H$, $A$ and $H^\pm$ fields.  In particular, the lightest
inert particle (LIP)  will be absolutely stable and is a potential dark matter
candidate~\cite{Barbieri:2006dq,LopezHonorez:2006gr,Gustafsson:2012aj,
Goudelis:2013uca}.
In addition, the inert 2HDM has rich phenomenological features.  For
example, the dark matter could play a critical role in the breaking of
the electroweak symmetry~\cite{Hambye:2007vf} and the triggering of the
electroweak phase transition~\cite{Chowdhury:2011ga}.  One can also
add a $\mathbb{Z}_2$-odd right-handed neutrino to the model thereby providing
a mechanism for generating the light neutrino masses at the one loop
level~\cite{Ma:2006km}.

The scalar potential for the inert 2HDM is given (in the Higgs basis)
by  \eq{higgsbasispot} with $Y_3=Z_6=Z_7=0$.   One can always rephase
$H_2\to e^{i\chi}H_2$ such that the one potentially complex parameter,
$Z_5$ is real.
Indeed, the sign of $Z_5$ is not physical, since the sign can be
flipped by redefining $H_2\to iH_2$.
Thus the Higgs sector of the inert 2HDM is CP-conserving
and depends on seven real
parameters $\{Y_1,Y_2,Z_1,Z_2,Z_3,Z_4,|Z_5|\}$.   The potential
minimum condition is given by $Y_1=-Z_1 v^2$.
Using \eqs{chhiggsmass}{mtwo}, it follows that the physical Higgs
masses are given by
\beqa
m^2_{h}&=&2Z_1 v^2\,,\\
m^2_{H,A}&=&Y_2+(Z_3+Z_4\pm |Z_5|)v^2\,,\\
m^2_{H^\pm}&=&Y_2+Z_3 v^2\,,
\eeqa
Here, we have adopted a notation in which $h$ corresponds to the
scalar whose properties are precisely those of the SM Higgs boson.  No
mass ordering of $h$ and $H$ is assumed.  Indeed,
\eqs{c2exact}{scexact} imply that either $\cos(\beta-\alpha)=0$ if
$h$ is identified as $h_1$ or $\sin(\beta-\alpha)=0$ if $h$ is identified
as~$h_2$.  In either case, this is the exact alignment limit with $h$
identified as the SM Higgs boson.

Moreover, we have used the
notation $H$ and $A$ above for the CP-even and odd bosons from the
inert sector.  However, an examination of the couplings of $H$ and $A$
implies only that $H$ and $A$ have CP-quantum numbers of opposite sign.
However, one cannot assign a unique CP-quantum number to $H$ and $A$
separately.  (For further details, see Ref.~\cite{Haber:2010bw}.)
The couplings of the inert scalars to the gauge bosons and Higgs
bosons are easily obtained by setting $\cos(\beta-\alpha)=0$
[$\sin(\beta-\alpha)=0$] if one identifies the SM Higgs boson with $h_1$ [$h_2$].

If we require that all scalar squared masses are positive, with $v^2=-Y_1/Z_1$,
then it follows that~\cite{Deshpande:1977rw}
\beq
Y_1<0\,,\qquad Z_1 Y_2>Z_3 Y_1\,,\qquad Z_1
Y_2>(Z_3+Z_4\pm|Z_5|)Y_1\,.
\eeq
If we also require that the scalar potential is bounded from below
(vacuum stability), then the following additional constraints must be
satisfied~\cite{Deshpande:1977rw}
\beq
Z_1>0\,,\qquad Z_2>0\,,\qquad Z_3>-(Z_1 Z_2)^{1/2}\,,\qquad Z_3+Z_4\pm |Z_5|
>-(Z_1 Z_2)^{1/2}\,.
\eeq
If one associates the dark matter with an electrically
neutral LIP then $m_{H^\pm}>m_{H,A}$, which yields~\cite{Ginzburg:2010wa}
\beq
Z_4<|Z_5|\,.
\eeq
Finally, one can impose the conditions of
perturbativity~\cite{Barbieri:2006dq} which can be used to
to restrict the magnitudes of the $Z_i$.

The seven parameters of the Higgs potential can be replaced by
the vev $v$, four masses of the Higgs boson and the inert scalars,
$(m_h,m_{H^\pm},m_{H},m_{A})$, the two of the scalar self-couplings
$Z_2$ and $Z_3$.   For example, we can use this set of input
parameters to compute $Y_1=-\half m_h^2$,
$Y_2=m^2_{H^\pm}-Z_3 v^2$, $Z_1 v^2=\half m_h^2$,
$Z_4 v^2=\half(m_H^2+m_A^2)-m_{H^\pm}^2$ and $|Z_5|v^2=\half|m_H^2-m_A^2|$.

Collider phenomenology of the inert scalars in the inert 2HDM has been studied in
Refs.~\cite{Barbieri:2006dq,Cao:2007rm,Lundstrom:2008ai,Goudelis:2013uca}.
In Ref.~\cite{Lundstrom:2008ai}, experimental bounds on the inert scalar
masses are obtained by using the experimental results at
the LEP collider~\cite{Acciarri:1999km,Abbiendi:1999ar,Abdallah:2003xe}.
At the LHC, even though the parameter regions where the inert scalars could be
discovered have been suggested~\cite{Dolle:2009ft,Miao:2010rg,Gustafsson:2012aj},
a detailed search for the inert scalars and a
determination of their masses and quantum numbers would be difficult.

The ILC phenomenology for the inert scalars has been considered in
Ref.~\cite{Aoki:2013lhm}.   Without loss of generality, we assume in what follows that $m_H<m_A$.
%to be the lighter of the two neutral inert scalars $H$ and $A$.
Four benchmark points for the mass spectrum of
inert scalars are listed in Table~\ref{tab:ilc}, which satisfy all available
theoretical and phenomenological constraints.
In the four benchmark points, the mass of $H$ is fixed to 65~GeV, so
that it does not permit the invisible decay of the SM Higgs boson,
$h\to HH$.
While a mass of $H$ up to $\sim80$~GeV is consistent with the dark matter relic
abundance analysis~\cite{LopezHonorez:2006gr,Gustafsson:2012aj,Goudelis:2013uca}, the
collider phenomenology does not change qualitatively by varying $m_H$
in this range.

\begin{table}[b!]
 \begin{tabular}{c||ccc|cc}
  & \multicolumn{3}{c|}{Inert scalar masses} &
  \multicolumn{2}{c}{ILC cross sections [$\sqrt{s}=250$~GeV (500~GeV)]} \\
  & $m_{H}$~[GeV] & $m_{A}$~[GeV] & $m_{H^\pm}$~[GeV] &
  $\sigma_{e^+e^-\to HA}$~[fb] & $\sigma_{e^+e^-\to H^+H^-}$~[fb] \\
  \hline
  (I)   & 65. & 73.  & 120. & 152. (47.) & 11. (79.) \\
  (II)  & 65. & 120. & 120. & 74. (41.)  & 11. (79.) \\
  (III) & 65. & 73.  & 160. & 152. (47.) & 0. (53.) \\
  (IV)  & 65. & 160. & 160. & 17. (35.)  & 0. (53.)
 \end{tabular}
 \caption{Masses of inert scalars and ILC cross sections
 for our four benchmark points.}\label{tab:ilc}
\end{table}

In Table~\ref{tab:ilc}, the cross sections of $HA$
production and $H^+H^-$ production at $\sqrt{s}=250$~GeV
and 500~GeV are shown.
The production cross sections of the inert
scalars are large enough to be tested  at the ILC.
The cross section of $HA$ production can take the largest value, i.e.\
186~fb at $\sqrt{s}=190$~GeV, 78~fb at $\sqrt{s}=280$~GeV, and 46~fb at
$\sqrt{s}=350$~GeV for cases (I, III), (II), and (IV), respectively.
The cross section of $H^+H^-$ production can take the largest value, i.e.\
96~fb at $\sqrt{s}=380$~GeV and 53~fb at $\sqrt{s}=500$~GeV for
cases (I, II) and (III, IV), respectively.
At $\sqrt{s}=1$~TeV, they are about 10~fb and 20~fb for $HA$
production and $H^+H^-$ production for all the four benchmark points, respectively.
For cases (II, IV), $H^{\pm}$ decays into $W^\pm H$ predominantly,
where we admit the $W$-boson to be off-shell if $m_{H^{\pm}}-m_{H}<m_W$.
For cases (I) and (III), $H^{\pm}\to W^\pm A$ decay would be
sizable as well, with the branching ratios about 32\% and 27\%,
respectively.
The decay of the $A$-boson is dominated by $A\to Z^{(*)}H$.

In the left panel of Fig.~\ref{fig:Mass},  the expected accuracy of
mass determination by the measurements of the four observables
for cases (I) and (II) at $\sqrt{s}=250$~GeV is shown~\cite{Aoki:2013lhm}.
The four bands are plotted in the ($m_H\,,\,m_{H^\pm}$) plane by
assuming that these four quantities are measured
in $\pm 2$~GeV accuracy (without any systematic shifts).
The accuracy of the $m_{H^\pm}$ ($m_{H}$) determination
would be $\pm 2$~GeV ($\pm 1$~GeV).
In the right panel of Fig.~\ref{fig:Mass}, the four bands are plotted in
the ($m_H\,,\,m_{H^\pm}$) plane by assuming that the four
observables are measured within the $\pm 2$~GeV accuracy.
By combining the four measurements with the uncertainty of $\pm 2$~GeV,
$m_{H^\pm}$ and $m_{H}$ can be determined in $\pm 1$~GeV accuracy.
\begin{figure}[t]
 \begin{center}
  \includegraphics[width=0.49\textwidth,clip]{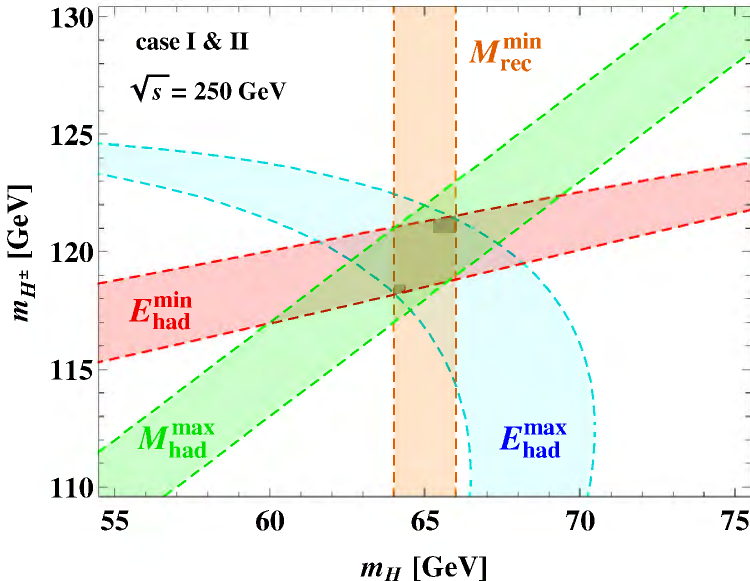}
  \includegraphics[width=0.49\textwidth,clip]{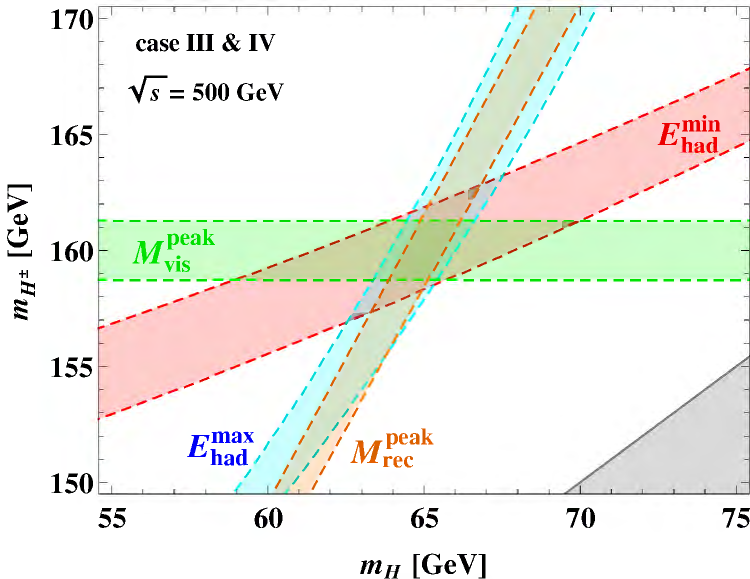}
  \caption{Determinations of $m_{H^\pm}$ and $m_{H}$ by the four
  observables are illustrated in the left [right] panel for the cases
  (I, II) [(III, IV)] at $\sqrt{s}=250$~GeV [500~GeV].
  Each observable is assumed to be measured in $\pm 2$~GeV accuracy.
  } \label{fig:Mass}
 \end{center}
\end{figure}
The determination of $m_{A}$ can also be achieved by combining the
observables in the process $e^+e^-\to HA$. However, at $\sqrt{s}=250$~GeV and
$\sqrt{s}=500$~GeV, since the two
constraints are very similar, these masses cannot be determined at one point.
In this case, one requires a fixed value of $m_{H}$ in the process
$e^+e^-\to H^+H^-$ as an input to determine $m_A$.
Then, the expected accuracy of the mass determination is
$\pm3$~GeV for the measurement of the observables in $\pm2$~GeV
accuracy.

The scenarios discussed above provide examples of parameter regions of the
inert 2HDM that cannot be detected at the LHC but can be probed in detail
at the ILC.

\subsection{The MSSM Higgs sector}\label{mssm}

In the minimal supersymmetric extension of the Standard Model (MSSM)
[see Ref.~\cite{HaberPDG} for a review and references],
all Standard Model particles are accompanied by an associated
supersymmetric partner whose spin differs by half a unit.  However,
adding a doublet hypercharge-one higgsino superpartner would yield
a gauge anomaly, rendering the theory mathematically inconsistent.
In the MSSM, this problem is overcome by considering the
supersymmetric extension of a two-Higgs doublet model, where the two
Higgs doublets possess hypercharges $\pm 1$, respectively.  As a
result, the corresponding two higgsino superpartners of opposite
hypercharge form a vector representation of spin-1/2 fermions, and the
gauge anomalies resulting from the higgsinos cancel exactly.

The Higgs sector of the MSSM is a 2HDM, whose Yukawa couplings and
scalar potential are constrained by supersymmetry (SUSY).
Instead of employing to hypercharge-one scalar doublets $\Phi_{1,2}$,
it is more convenient to introduce a $Y=-1$ doublet $H_d\equiv i\sigma_2 \Phi_1^*$
and a $Y=+1$ doublet
$H_u\equiv\Phi_2$:
\beq
H_d =\begin{pmatrix} H_d^1 \\ H_d^2 \end{pmatrix}=\begin{pmatrix}\Phi_1^{0\,*}\\
-\Phi_1^-\end{pmatrix}\,,\qquad\quad
H_u=\begin{pmatrix} H_u^1\\ H_u^2\end{pmatrix}=\begin{pmatrix}\Phi_2^+\\
\Phi_2^0\end{pmatrix}\,.
\eeq
The notation $H_{u,d}$ is motivated by the form of Higgs Yukawa Lagrangian,
\beq \label{mssmyuk}
\mathcal{L}_{\mathrm Yukawa}=-h_u^{ij}(\bar u_R^i u_L^j H_u^2-\bar u_R^i
d_L^j H_u^1)-h_d^{ij}(\bar d_R^i d_L^j H_d^1-\bar d_R^i u_L^j H_d^2)+{\mathrm h.c.}\,,
\eeq
which arises as a consequence of supersymmetry.
That is,  the neutral Higgs field $H_u^2$ couples exclusively to up-type quarks and
the neutral Higgs field $H_d^1$ couples exclusively to down-type
quarks.

In particular, the
so-called \textit{wrong Higgs interactions}~\cite{Haber:2007dj},
\beq \label{wronghiggs}
\mathcal{L}_{\mathrm wrong~Higgs}=-h_u^{\prime\,ij}(\bar u_R^i u_L^j H_d^{1\,*}+\bar u_R^i
d_L^j H_d^{2\,*})-h_d^{\prime,ij}(\bar d_R^i d_L^j H_u^{2\,*}
+\bar d_R^i u_L^j H_u^{1\,*})+{\mathrm h.c.}\,,
\eeq
are not supersymmetric (due to the appearance of the
complex-conjugated scalar fields in the terms exhibited explicitly
above).  Thus, the MSSM Higgs sector possesses Type-II Yukawa couplings
as a consequence of
supersymmetry (and not a $\mathbb{Z}_2$ discrete symmetry as discussed
in Section~\ref{specialforms}.)

The Higgs potential of the MSSM is:
\beqa
V&=&
\left(m_d^2+|\mu|^2\right) H_d^{i*}H_d^i
  + \left(m_u^2+|\mu|^2\right) H_u^{i*}H_u^i
  -b\left(\epsilon^{ij}H_d^iH_u^j+{\mathrm h.c.}\right) \nonumber \\
&&\qquad +\tfrac{1}{8}
\left(g^2 + g^{\prime\,2}\right) \left[H_d^{i*}H_d^i-H_u^{j*}H_u^j\right]^2
+\half g^2 |H_d^{i*}H_u^i|^2\,, \label{susypot}
\eeqa
where $\epsilon^{12}=-\epsilon^{21}=1$ and $\epsilon^{11}=\epsilon^{22}=0$,
and the sum over repeated indices is implicit.
In \eq{susypot}, $\mu$ is a supersymmetric Higgsino mass parameter and $m_d^2$,
$m_u^2$, $b$ are soft-supersymmetry-breaking squared-mass parameters.
The quartic
Higgs couplings are related to the SU(2) and U(1)$_{\mathrm Y}$ gauge couplings
as a consequence of SUSY.
%\vspace{-0.1in}

After minimizing the Higgs potential, the neutral components of the
Higgs fields (in an appropriately chosen phase convention)
acquire real positive vevs: $\vev{H_d^0}=v_d$ and $\vev{H_u^0}=v_u$,
where
$v^2\equiv v_d^2+v_u^2={2\mw^2/ g^2}=(174~{\mathrm GeV})^2$.
The ratio of the two vevs is
\beq
\tan\beta\equiv\frac{v_u}{v_d}\,,\qquad 0\leq\beta\leq\half\pi\,.
\eeq
In the Higgs basis, the phase of $H_2$ can be chosen such that
$Z_5$, $Z_6$ and $Z_7$ are real.  In particular,
\beqa
Z_1&=&Z_2=\quarter(g^2+g^{\prime\,2})\cos^2 2\beta\,,\qquad Z_3=Z_5+\quarter(g^2-g^{\prime\,2})\,,\qquad
Z_4=Z_5-\half g^2\,,\nonumber \\
Z_5&=&\quarter(g^2+g^{\prime\,2})\sin^2 2\beta\,,\qquad\qquad\quad
Z_7=-Z_6=\quarter(g^2+g^{\prime\,2})\sin 2\beta\cos 2\beta\,,\label{zsusy}
\eeqa
in the notation of Section~\ref{modelind}.
The existence of a Higgs basis where $Z_5$, $Z_6$ and $Z_7$ are
simultaneously real implies that the tree-level MSSM Higgs sector
is CP-conserving.  Thus the neutral Higgs mass-eigenstates
are states of definite CP.

The five physical Higgs particles
consist of a charged Higgs pair
\beq
H^\pm=H_d^\pm\sinb+ H_u^\pm\cosb\,,
\eeq
with squared mass given by $\mhpm^2 =\mha^2+\mw^2$,
one CP-odd scalar
\beq
A^0= \sqrt{2}\left({\mathrm Im\,}H_d^0\sinb+{\mathrm Im\,}H_u^0\cosb
\right)\,,
\eeq
with squared mass given by $m_A^2=2b/\sin 2\beta$, and two CP-even scalars
\beqa
h^0 &=&\sqrt{2}\bigl[ -({\mathrm Re\,}H_d^0-v_d)\sin\alpha+
({\mathrm Re\,}H_u^0-v_u)\cos\alpha\bigr]\,,\nonumber\\
H^0 &=& \sqrt{2}\bigl[({\mathrm Re\,}H_d^0-v_d)\cos\alpha+
({\mathrm Re\,}H_u^0-v_u)\sin\alpha\bigr]\,,\nonumber
\eeqa
where we have now labeled the Higgs fields according to their
electric charge.  The angle $\alpha$ arises when the CP-even Higgs
squared-mass matrix (in the $H_d^0$---$H_u^0$ basis) is
diagonalized to obtain the physical CP-even Higgs states.
Equivalently, one can perform the diagonalization of the CP-even
Higgs squared-mass matrix in the Higgs basis, in which case the
corresponding diagonalization angle is given by $\alpha-\beta$.
All Higgs masses and couplings can be expressed in terms of two
parameters usually chosen to be $\mha$ and $\tan\beta$.

%The charged Higgs mass is given by
%\beq
%\mhpm^2 =\mha^2+\mw^2\,.
%\eeq
The CP-even Higgs bosons $\hl$ and $\hh$ are eigenstates of the
squared-mass matrix, which in the Higgs basis is given by
\beq
\mathcal{M}_e^2 =
\begin{pmatrix} m^2_Z \cos^2 2\beta&
           -m^2_Z\sin 2\beta\cos 2\beta \\
  -m^2_Z\sin 2\beta\cos 2\beta&
  \mha^2+ m^2_Z \sin^2 2\beta \end{pmatrix}\,.
\eeq
The eigenvalues of $\mathcal{M}_e^2$ are
the squared-masses of the two CP-even Higgs scalars
\beq \label{mssmhm}
  m^2_{H,h} = \half \left( \mha^2 + m^2_Z \pm
                  \sqrt{(\mha^2+m^2_Z)^2 - 4m^2_Z \mha^2 \cos^2 2\beta}
                  \; \right)\,,
\eeq
and $\alpha$ is given by
\beq
\cos 2\alpha=-\cos 2\beta\frac{m_A^2-m_Z^2}{m_H^2-m_h^2}\,,\qquad\quad
\sin 2\alpha=-\sin 2\beta\frac{m_A^2+m_Z^2}{m_H^2-m_h^2}\,.
\eeq
Conventionally, one takes $0\leq\beta\leq\half\pi$ and
$-\half\pi\leq\alpha\leq 0$.  It follows that [cf.~\eqs{c2exact}{scexact}]:
\beqa
\cos^2(\beta-\alpha)&=&\frac{m_Z^2\cos^2 2\beta-m_h^2}{m_H^2-m_h^2}\,,
\label{cbma2}\\
\cos(\beta-\alpha)\sin(\beta-\alpha)&=&\frac{m_Z^2\sin 2\beta\cos 2\beta}{m_H^2-m_h^2}\,. \label{sbmacbma}
\eeqa
Note that $0\leq\beta-\alpha<\pi$ so that
$0\leq\sin(\beta-\alpha)\leq 1$ and the sign of $\cos(\beta-\alpha)$ is
given by the sign of $\sin 4\beta$.

The tree-level mass of the lightest CP-even Higgs
boson of the MSSM is bounded,
\beq \label{hbound}
\mhl\leq\mz |\cos 2\beta|\leq\mz\,.
\eeq
This inequality arises because all Higgs self-coupling
parameters of the MSSM are related to the squares of the electroweak
gauge couplings.  However, radiative corrections can boost the upper
bound of the lightest CP-even Higgs mass above its tree level bound of
$m_Z$.  The leading effects of the radiative corrections will be
discussed further below.

The tree-level couplings of the MSSM Higgs bosons are those of a
CP-conserving 2HDM with Type-II Higgs--fermion Yukawa couplings.
For example, the Higgs couplings to gauge boson pairs ($V=W$ or $Z$)
are given by
\beq
g\lsub{\hl VV}= g\lsub{V} m\lsub{V}\sinbma \,,\qquad\qquad
           g\lsub{\hh VV}= g\lsub{V} m\lsub{V}\cosbma\,,
\eeq
where
$g\lsub{V}\equiv \sqrt{2}m_V/v$.
There are no tree-level couplings of $\ha$ or $\hpm$ to $VV$.
The couplings of $V$ to two neutral Higgs bosons
(which must have opposite CP-quantum numbers) are denoted by
$g_{\phi\ha Z}(p_\phi-p_\ha)$, where $\phi=\hl$ or $\hh$ and the
momenta $p_\phi$ and $p_\ha$ point into the vertex, and
\beq
g\lsub{\hl\ha Z}={g\cosbma\over 2\cos\theta_W}\,,\qquad\qquad\quad
           g\lsub{\hh\ha Z}=-\,{g\sinbma\over 2\cos\theta_W}\,.
\eeq
The properties of the three-point and
four-point Higgs boson--vector boson couplings are conveniently summarized
in \eq{littletable} by listing the couplings that are proportional
to either $\sin(\beta-\alpha)$ or $\cos(\beta-\alpha)$ or are angle-independent.
Finally, the couplings of the MSSM Higgs bosons to quarks are given in
\eqs{t2}{chhiggsy2}
[with the corresponding coupling to leptons obtained by the
substitutions specified below \eq {chhiggsy2}].

%Especially noteworthy is the possible $\tan\beta$-enhancement
%of certain Higgs-fermion couplings.
%The general expectation in MSSM models is that $\tan\beta$
%lies in a range:
%$$
%1\lsim\tanb\lsim\frac{m_t}{m_b}\,.
%$$
%Near the upper limit of $\tan\beta$, we have roughly identical
%values for the top and bottom Yukawa couplings, $h_t\sim h_b$, since
%$$
%h_b = {\sqrt{2}\,m_b\over v_d}={\sqrt{2}\, m_b\over
%v\cos\beta}\,,\qquad\qquad
%h_t = {\sqrt{2}\,m_t\over v_u}={\sqrt{2}\, m_t\over v\sin\beta}\,.
%$$

The decoupling behavior of the MSSM Higgs sector is exhibited in
the limit of $\mha\gg\mz$, where the corresponding tree-level
squared-masses of the Higgs bosons are given by:
\beqa
\mhl^2 &\simeq &  \mz^2\cos^2 2\beta\,,\nonumber \\
\mhh^2 &\simeq &  \mha^2+\mz^2\sin^2 2\beta\,,\nonumber\\
\mhpm^2& = &  \mha^2+\mw^2\,.
\eeqa
Since $\sin(\beta-\alpha)\simeq 1$ and $m_H\simeq m_A$ in the decoupling limit,
\eq{sbmacbma} yields
\beq \label{cbmadecoup}
\cos(\beta-\alpha)={\mz^2\sin 4\beta\over 2\mha^2}+\mathcal{O}\left(\frac{m_Z^4}{m_A^4}\right)\,.
\eeq
Note that \eq{cbmadecoup} follows from the corresponding 2HDM result
given by \eq{z6decoupling} when $Z_6$ is given by its supersymmetric
value [cf.~\eq{zsusy}].

As expected, in the decoupling limit $\mha\simeq\mhh
\simeq\mhpm$, up to corrections of ${\cal O}(\mz^2/\mha)$, and
$\cos(\beta-\alpha)=0$ up to corrections of ${\cal O}(\mz^2/\mha^2)$.
%This is the decoupling limit, since at energy scales below
%approximately common mass of the heavy
%Higgs bosons $H^\pm$ $H^0$, $A^0$, the effective Higgs theory is
%precisely that of the SM.
In general, in the limit of $\cos(\beta-\alpha)\to 0$,
all the $h^0$ couplings to SM particles approach their SM limits.
In particular, if $\lambda_V$ is a Higgs coupling to vector bosons
and $\lambda_t$ [$\lambda_b$] are Higgs couplings to up-type
[down-type] fermions, then
\beqa
\frac{\lambda_{V}}{[\lambda_{V}]_{\mathrm SM}}&=&\sin(\beta-\alpha)=1+\mathcal{O}\left(\frac{m_Z^4\sin^2 4\beta}{m_A^4}\right)\,,\\
\frac{\lambda_{t}}{[\lambda_{t}]_{\mathrm
    SM}}&=&1+\mathcal{O}\left(\frac{m_Z^2\cos^2\beta\cos 2\beta}{m_A^2}\right)\,.\\
\frac{\lambda_{b}}{[\lambda_{b}]_{\mathrm
    SM}}&=&1+\mathcal{O}\left(\frac{m_Z^2\sin^2\beta\cos 2\beta}{m_A^2}\right)\,.\label{uf}
\eeqa
Note that the approach to decoupling is fastest in the case of the
$hVV$ couplings and slowest in the case of the $hbb$ couplings at
large $\tan\beta$ (where the corresponding trigonometric factor in
\eq{uf} is maximal).  This implies that at large $\tan\beta$,
a precision measurement of the
$h^0b\bar{b}$ coupling provides the greatest sensitivity to $m_A$~\cite{Carena:2001bg}.

When radiative corrections
are incorporated, additional parameters of the supersymmetric model
enter via virtual loops.  The impact of these corrections
can be significant~\cite{Haber:1990aw,Okada:1990vk,Ellis:1990nz}.
For example, the tree-level prediction for
the upper bound of the lightest $CP$-even Higgs mass,
given by \eq{hbound}\cite{Inoue:1982ej,Flores:1982pr,Gunion:1984yn}, can be
substantially modified when
radiative corrections are included.
The qualitative behavior of these radiative corrections can be most easily
seen in the large top squark mass limit.
Denoting the geometric mean of the two top squark squared masses by
$\msusyy\equiv\mstopa\mstopb$ and assuming that
$\mha>\mz$ and that the top squark mass
splitting is small compared to $M_S$,  the predicted upper bound for $m_h$
(which reaches its maximum at large $\tan\beta$ and $\mha\gg\mz$)
is approximately given by
\beq \label{eqmH1up0}
m_{h}^2\lsim
\mz^2\cos^2 2\beta+\frac{3g^2\mt^4}{8\pi^2\mw^2}
\left[\ln\left(\frac{\msusyy}{\mt^2}\right)+
\frac{X_t^2}{\msusyy}
\left(1-\frac{X_t^2}{12\msusyy}\right)\right],
\eeq
where $X_t\equiv A_t-\mu\cot\beta$ is the top squark mixing factor.

A more complete treatment of the radiative corrections~\cite{Degrassi:2002fi,Martin:2007pg,Kant:2010tf}\
shows that
\eq{eqmH1up0} somewhat overestimates the true upper bound of $\mhl$.
These more refined computations, which incorporate
renormalization group improvement and the
leading two-loop (and even some three-loop) contributions, yield $m_{h}\lsim 130$~GeV
(with an accuracy of a few GeV)
for $m_t=173$~GeV and $M_S\lsim 2$~TeV.  The observed Higgs
mass of 126 GeV is consistent with this bound.

Radiative corrections also can have an important impact on the
MSSM Higgs couplings.
Although radiatively-corrections to couplings generically tend to be at the
few-percent level, there is some potential for significant effects.
In certain cases, radiative corrections are enhanced for
values of $\tan\beta\gg 1$.
In addition,
CP-violating effects induced by complex SUSY-breaking parameters that
enter in loops, can yield new effects in the Higgs sector not present
at tree-level.

One leading correction of note is the renormalization of the mixing
angle $\alpha$.  This modifies the quantity $\cos(\beta-\alpha)$ which
governs the decoupling limit and
plays a critical role in the couplings of the Higgs bosons.  Employing
the same approximations used in obtaining \eq{eqmH1up0},
one finds that \eq{cbma2} is modified by replacing the
tree-level bound, $m_Z^2\cos^ 2\beta$, with the
radiatively-corrected bound given in  \eq{eqmH1up0}, and replacing
$m_h$ and $m_H$  with the corresponding loop-corrected masses.

However, this approximation can still miss some important loop
contributions that can drastically alter the tree-level results.
Let $M_{\mathrm SUSY}$ characterize the mass scale of SUSY-breaking
(or equivalently, the masses of the superpartners that
appear in the loops that contribute to the radiative corrections).  If
$M_{\mathrm SUSY}\gg m_A$, then one can integrate out the superpartners
that appear in the loops and obtain an
effective low-energy theory that is equivalent to the most general
2HDM.    In this effective 2HDM,
the supersymmetric value of $Z_6$ given in \eq{zsusy} is modified by
radiative corrections.  In certain special regions of the MSSM
parameter space, the radiative corrections are $\tan\beta$-enhanced
and can approximately cancel
out the tree-level value for a particular (large) value of
$\tan\beta$,
leaving $Z_6\simeq 0$.  This is the
alignment limit (which is not possible in the MSSM Higgs sector at
tree-level) and yields $\cos(\beta-\alpha)\simeq 0$ in light of
\eq{z6decoupling}.  In this case, the lightest CP-even Higgs
boson is very SM-like, whereas the other Higgs bosons of the model
need not be significantly heavier.

Moreover, the supersymmetric Yukawa couplings given
in \eq{mssmyuk} are modified by the radiative corrections,
$h_q\to h_q+\delta h_q$ ($q=u$, $d$).  In particular, since the
effective 2HDM below the SUSY-breaking scale
does not respect supersymmetry, the wrong-Higgs Yukawa
interactions given in \eq{wronghiggs} are also generated by the
radiative corrections; the corresponding Yukawa couplings will be
denoted by $h'_q\equiv \Delta h_q$.  Of particular interest are the
wrong Higgs Yukawa couplings to bottom quarks,
$$
 \Delta h _b \simeq h_b\left[\frac{2\alpha_s}{3\pi}\mu M_3
 \mathcal{I}(M_{\tilde b_1},M_{\tilde b_2},
 M_g) + \frac{h_t^2}{16\pi^2}\mu A_t \mathcal{I}
(M_{\tilde t_1}, M_{\tilde t_2}, \mu)\right]\,,
 $$
where, $M_3$ is the Majorana gluino mass,
$\mu$ is the supersymmetric Higgs-mass parameter, and $\widetilde b_{1,2}$
and $\widetilde t_{1,2}$ are the mass-eigenstate bottom squarks and top
squarks, respectively.   The loop integral is given by
\beq
\mathcal{I}(a,b,c) =
\frac{a^2b^2\ln(a^2/b^2) +   b^2c^2\ln(b^2/c^2)
+ c^2a^2\ln(c^2/a^2)}{(a^2-b^2)(b^2-c^2)(a^2-c^2)}\,.
\eeq
In the limit where at least one of the arguments of
$\mathcal{I}(a,b,c)$ is large,
\beq
\mathcal{I}(a,b,c)\sim 1/{\mathrm max}(a^2,b^2,c^2)\,.
\eeq
Thus, in the limit where $M_3\sim \mu\sim A_t\sim M_{\tilde b}\sim M_{\tilde t}
\sim M_{\mathrm SUSY}\gg m_Z$, the one-loop contributions to $\Delta h_b$ do
\textit{not} decouple.

The wrong-Higgs couplings yield
$\tan\beta$-enhanced modifications of some physical observables.
After expressing the Higgs doublet fields $H_d$ and $H_u$ in terms of
the physical Higgs mass-eigenstates, one can identify the
the $b$-quark mass,
\beq
m_b = h_bv \cos \beta \left(1 + \frac{\delta h_b}{h_b}
+ \frac{\Delta h_b \tan \beta}{h_b}\right)
\equiv h_bv\cos \beta (1 + \Delta_b)\,,
\eeq
which defines the quantity $\Delta_b$.
In the limit of large $\tan\beta$ the term proportional to $\delta
h_b$ can be neglected,
in which case,
\beq
\Delta_b\simeq \frac{\Delta h_b}{h_b}\tan\beta\,.
\eeq
Thus, $\Delta_b$ is $\tan\beta$--enhanced if $\tan\beta\gg 1$.
As previously noted, $\Delta_b$ survives in the limit of large $M_{\mathrm SUSY}$;
this effect does not decouple.

From the effective Yukawa Lagrangian, one can obtain the couplings of the physical Higgs
bosons to third generation fermions.
%Neglecting possible CP-violating effects,
%\beq \label{linthff}
%{\cal L}_{\mathrm int} =  -\sum_{q=t,b,\tau} \left[g_{\hl q\bar q}\hl q \bar{q} +
% g_{\hh q\bar q}\hh q \bar{q}-
% i g_{\ha q\bar q} A^0 \bar{q} \gamma_5 q\right] + \left[\bar b g_{H^- t\bar b}
%t H^- + {\mathrm h.c.}\right]\,.
%\eeq
For example, the one-loop corrections can generate measurable shifts
in the decay rate for $h^0\to b\bar b$,
\beq
g_{h^\circ b\bar b}= -\frac{gm_b}{2m_W}\frac{\sin\alpha}{\cos\beta}
\left[1+\frac{1}{1+\Delta_b}\left(\frac{\delta h_b}{h_b}-
\Delta_b\right)\left( 1 +\cot\alpha \cot\beta \right)\right]\,.
\eeq
At large $\tan\beta\sim 20$---50, $\Delta_b$ can be as large as 0.5 in magnitude and of
either sign, leading to a significant enhancement or suppression of the
Higgs decay rate to $b\bar b$.

If $m_A\sim\mathcal{O}(M_{\mathrm SUSY})$, then below the scale of
supersymmetry-breaking one must also integrate out the second Higgs
doublet (in the Higgs basis).  In this case, the
low-energy effective Higgs theory is a one-Higgs doublet model, and thus
$g_{h^0 b\bar b}$ must approach its SM value.  Indeed in this limit,
\beq
1+\cot\alpha\cot\beta=-\frac{2m_Z^2}{m_A^2}\cos
2\beta+\mathcal{O}\left(\frac{m_Z^4}{m_A^4}\right)\,.\nonumber
\eeq
Thus the previously non-decoupling SUSY radiative corrections
decouple for $m_A\gg m_Z$ as expected.

The one-loop corrected effective Higgs-fermion Lagrangian
can exhibit CP-violating effect due to possible CP-violating phases
in $\mu$, $A_t$ and $M_3$.  This leads to mixed-CP neutral Higgs
states and CP-violating couplings.  This is the so-called CPX scenario of the MSSM.
In the limit of $m_{H^\pm}\ll M_{\mathrm SUSY}$, the effective low-energy
theory is the most general CP-violating 2HDM.  Thus, the
model-independent treatment of the general 2HDM is applicable.
Further details on the CPX scenario can be found in Refs.~\cite{Carena:2000ks,Carena:2001fw,Carena:2002es,Accomando:2006ga}.

\section{Other extended Higgs sectors}

\subsection{Constraints from the tree-level rho parameter}

Apart from the part of the SM that is governed by the gauge
principle,  there is no theoretical principle for fixing the number of Higgs scalars.
Indeed, there are a variety of possibilities for non-minimal
Higgs sectors that are consistent with phenomenological requirements.
We have already treated in detail the two-Higgs doublet extension of
the SM in Section~\ref{tdhm}.   However, it is also possible
that the Higgs sector contains additional Higgs doublets and/or
one or more non-doublet representations.

A very strong constraint on exotic Higgs sectors derives from
the electroweak rho parameter whose experimental value is very close to unity.
The current value for the rho parameter is given by
$\rho=1.0008 ^{+0.0017}_{-0.0007}$~\cite{ErlerPDG}.
In the SM, the rho parameter is exactly unity at the tree level,
\begin{eqnarray}
 \rho = \frac{m_W^2}{m_Z^2 \cos^2\theta} =1.  \label{rho_sm}
\end{eqnarray}
Moreover, including radiative corrections, the
SM with the Higgs boson mass of 126 GeV yields a value of $\rho$ that
is consistent with the experimentally measure value~\cite{Baak:2012kk}.
Models with only Higgs doublets and singlets do not spoil the
tree-level prediction of $\rho=1$.  However,
the addition of scalars that transform under higher dimensional
representations generally violate the tree-level prediction of $\rho=1$.

In a general SU(2)$\times$U(1) model with $n$ scalar multiplets $\phi_i$ with isospin $T_i$ and
hypercharge $Y_i$, the rho parameter is given at the tree level by~\cite{Gunion:1989we}
\beq \label{rhogen}
\rho=  1+
\frac{\sum_{i} [4T_i(T_i+1)-3Y_i^2]|v_i|^2 c_i}
{\sum_{i} 2Y_i^2 |v_i|^2}\,,
\eeq
where
\beq
c_i=\begin{cases}  1, & (T,Y)\in \text{complex representation}, \\
                  \half, & (T,Y=0)\in \text{real representation}.
                  \end{cases}
\eeq
For a Higgs sector composed of complex ($c=1$) hypercharge-1 Higgs
doublets ($T=1/2$ and $Y=1$).  it follows that $\rho=1$, independently
of the value of the vacuum expectation value $v$.  One can also add
an arbitrary number of Higgs singlets ($T=Y=0$) without changing the
value of $\rho$.
To automatically have $\rho=1$ independently of the
Higgs vevs, one must satisfy the Diophantine equation~\cite{Tsao:1980em},
\beq \label{diaph}
(2T+1)^2-3Y^2=1\,,
\eeq
 for non-negative integer values of
$(2T,Y)$.  The smallest nontrivial solution beyond the complex
$Y=1$ Higgs doublet is a complex Higgs septet with $T=3$ and $Y=4$.

For extended Higgs sectors with multiplets that do not satisfy
\eq{diaph}, the tree-level value for $\rho$ will generally differ from
1.  In order to be consistent with the rho parameter data,
there are two possible strategies.  First, one can fine-tune the values of the
vevs $v_i$ such that $\rho=1$.  This may require some vevs to be
significantly smaller than 174~GeV, or it may require an unnatural
cancellation of terms in the numerator of the second term in \eq{rhogen}.
As an example of the first strategy, consider the effect of adding to
the Standard Model an extra hypercharge-two complex scalar triplet
field $\phi_2=\Delta$  ($T_2=1, Y_2= 2$) ,
which has been employed for generating the neutrino mass by the
so-called type-II seesaw mechanism~\cite{Mohapatra:1980yp}.  Denoting the
vev of the neutral component of the scalar triplet by $v_\Delta$,
\eq{rhogen} yields
\begin{eqnarray}
 \rho = \frac{1 + 2 v_\Delta^2/v^2}{1+4v_\Delta^2/v^2}.
\end{eqnarray}
Therefore, there is a strong upper bound for $v_\Delta$ ( $\lesssim$ a few GeV )
in light of the rho parameter data.

Second, one can make a clever
choice of Higgs multiplets such that the required cancellation of terms in the
numerator of the second term in \eq{rhogen} appears to be natural.  The simplest
example of this mechanism is the Georgi-Machacek model~\cite{Georgi:1985nv}, which consists
of the SM hypercharge-one complex scalar doublet $\Phi(T=\half,Y=1)$ ,
a complex hypercharge-two scalar triplet $\Delta(T=1,Y=2)$ and a real scalar triplet
$\xi(T=1,Y=0)$.  Suppose that the vevs of the neutral fields of
the two scalar triplets are equal, $v_\Delta=v_\xi$, which can be arranged with a
special choice of the scalar potential parameters corresponding to
an enhanced SU(2)$_L\times$SU(2)$_R$ global symmetry.  In this case,
it is easy to check that
\eq{rhogen} yields $\rho=1$ independently of the value of the vev of
the scalar doublet, $v_\Phi$,  and the common value of the triplet
vevs, $v_\Delta=v_\xi$.  Indeed
$v_\Delta/v_\Phi\sim 1$ is phenomenologically viable, which would lead to
a very different phenomenology than the simple doublet plus triplet
model considered above.  However, since the enhanced
SU(2)$_L\times$SU(2)$_R$ global symmetry of the scalar potential is
not respected by the hypercharge gauge interactions and the Yukawa
interactions, the condition that $v_\Delta=v_\xi$ is not stable under
radiative corrections and therefore must be considered as a tuning of parameters~\cite{Gunion:1990dt}.

%It is known that there is a variation model  with triplet fields proposed by by Georgi an%d Machacek,
%in which $\rho=1$ is satisfied~\cite{Georgi}. In this model, a complex triplet field $\D%elta$ ($T=1, Y=1$)  and
%a real triplet field $\xi$ ($T=1, Y=0$) are introduced, and the scalar potential is require%d
%to respect $SU(2)_L \times SU(2)_R$ global symmetry.
%In this model, because $\rho=1$ at the tree level, the constraint on the vacuum expect%ation
%value $v_\Delta$ ($=v_\xi$ due to the global symmetry)  is considerably relaxed and
%basically $v_\Delta/v_\phi \sim 1$ is possible.   Such a large value of vacuum expectat%ion
%values of triplet fields predicts much different collider signals.

\subsection{An upper bound for the Higgs coupling to vector boson pairs}
\label{Anupperbound}

Consider a CP-conserving
extended Higgs sector that has the
property that $\rho=1$ and no tree-level $ZW^\pm\phi^\mp$ couplings (where $\phi^\pm$ are physical charged scalars
that might appear in the scalar spectrum).   Then it follows that~\cite{Gunion:1990kf}
\beq
      \sum_i  g^2_{\phi_iVV}=g^2 m_W^2=g^2_{h_{\mathrm SM}VV}\,,\qquad\quad
m_W^2 g_{\phi_i ZZ}=m_Z^2 g_{\phi_i WW}\,,
\eeq
where the sum is taken over all neutral CP-even scalars $\phi_i$ and
$h_{\mathrm SM}$ is the Higgs boson of the SM.
In this case, it follows that $g_{\phi_i VV}\leq g_{h_{\mathrm SM}VV}$ for all $i$.
Models that contain only scalar singlets and doublets satisfy the requirements stated above and hence respect
the sum rule and the coupling relation given above.
However, it is possible to violate $g_{\phi_i VV}\leq g_{h_{\mathrm SM}VV}$
and $m_W^2 g_{\phi_i ZZ}=m_Z^2 g_{\phi_i WW}$ if tree-level $ZW^\pm\phi^\mp$
and/or $\phi^{++}W^-W^-$ couplings are present~\cite{Gunion:1990kf,Hisano:2013sn, Kanemura:2013mc}.  A more general sum rule is:
\beq
\sum_i g^2_{\phi_iVV}=g^2 m_W^2+\sum_k|g_{\phi^{++}_k W^- W^-}|^2\,.
\eeq
The Georgi-Machacek model provides an instructive example~\cite{Georgi:1985nv, Chanowitz:1985ug,Gunion:1989ci}.
This model
consists of a complex
Higgs doublet with $Y=1$, a complex Higgs triplet with $Y=2$ and a
real Higgs triplet with $Y=0$, with doublet vev $v_\Phi$ and triplet
vevs $v_\Delta=v_\xi$, such
that $v^2=v_\Phi^2+8v_\Delta^2$.

It is convenient to write~\cite{Gunion:1989ci}
$$
c_H\equiv\cos\theta_H=\frac{v_\Phi}{\sqrt{v_\Phi^2+8v_\Delta^2}}\,,
$$
and $s_H\equiv\sin\theta_H=(1-c_H^2)^{1/2}$.   Then, the following couplings are noteworthy:
\beqa
&&H_1^0 W^+ W^-:\quad g c_H m_W\,,\qquad\qquad\,\, \,   H_1^{\prime\,0}W^+ W^-:\quad \sqrt{8/3}gm_W s_H\,,\nn\\
&&H_5^0 W^+ W^-:\quad \sqrt{1/3}gm_W s_H\,,\qquad H_5^{++}W^-W^-:\quad \sqrt{2}gm_W s_H\,.\nn
\eeqa
$H_1^{\prime\,0}$ and $H_5^0$, $H_5^{++}$ have no coupling to fermions, whereas the $H_1^0$ coupling to fermions
is given by
$$
H_1^0 f\bar{f}:\quad \frac{gm_q}{2m_W c_H}\,.
$$
In general $H_1^0$ and $H_1^{\prime\,0}$ can mix.

In the absence of $H_1^0$--$H_1^{\prime\,0}$ mixing and $c_H=1$, we see that the couplings of $H_1^0$
match those of the SM.  In contrast, in the case of $s_H=\sqrt{3/8}$, the
$H_1^{\prime\,0}$ coupling to $W^+ W^-$ matches that of the SM.
Nevertheless, this does not saturate the $\Phi_iWW$ sum rule!  Moreover, it
is possible that the $H_1^{\prime\,0}W^+ W^-$ coupling is
\textit{larger} than $gm_W$, without violating the $\Phi_iWW$ sum rule.
Including $H_1^0$--$H_1^{\prime\,0}$ mixing allows for even more
baroque possibilities not possible in a multi-doublet extension of the
SM.  Deviations above the $h_{\mathrm SM}VV$ coupling by as much as $10\%$
or more are possible.
Thus we have demonstrated the possibility that
\begin{eqnarray}
g_{\phi VV}^2 > g_{h_{\mathrm SM} VV}^2\,, \label{eq:hVVexotic}
\end{eqnarray}
in Higgs sectors with exotic (larger than doublet) Higgs representations.

In Table~\ref{tab:hVV_exotic}, the deviations in the Higgs boson couplings from the SM values are
listed in various extended Higgs sectors (further details are given in Ref.~(\cite{Kanemura:2013mc}).
Moreover, except for the case of the 2HDM, the couplings of the SM like Higgs boson with
the weak gauge bosons shown in Table~\ref{tab:hVV_exotic} can be greater than 1.
%In particular, in the model by Georgi and Machacek,
%and that with a septet field, the coupling squared can deviate from the SM value by plu%s 10 \% or more.
At the ILC the $hVV$ couplings can be measured at the percent level.
Therefore, even if extra Higgs bosons are not discovered directly,
a Higgs sector with exotic multiplets can be distinguished
via the precision measurement of the $hVV$ coupling.
%with Eq.~(\ref{eq:hVVexotic}).
\begin{table}[t]
\begin{center}
{\renewcommand\arraystretch{1.3}
\begin{tabular}{|c||c|c|c|c|}\hline
Model & $\tan\beta$ &$\tan\beta'$& $c_{hWW}$ & $c_{hZZ}$  \\\hline\hline
$\phi_1+\phi_2$ (2HDM) &$v_{\phi_2}/v_{\phi_1}$&$v_{\phi_2}/v_{\phi_1}$ &$\sin(\beta-\alpha)$ & $\sin(\beta-\alpha)$   \\\hline
$\phi+\Delta$ (cHTM) &$\sqrt{2}v_\Delta/v_\phi$&$2v_\Delta/v_\phi$& $\cos\beta \cos\alpha + \sqrt{2}\sin\beta\sin\alpha$ & $\cos\beta' \cos\alpha + 2\sin\beta'\sin\alpha$ \\\hline
$\phi+\xi$ (rHTM) &$2v_\xi/v_\phi$&-& $\cos\beta \cos\alpha + 2\sin\beta\sin\alpha$ & $\cos\alpha$  \\\hline
$\phi+\Delta+\xi$ (GM model) &$2\sqrt{2}v_\Delta/v_\phi$& $2\sqrt{2}v_\Delta/v_\phi$ &$\cos\beta \cos\alpha +\frac{2\sqrt{6}}{3}\sin\beta \sin\alpha$ &$\cos\beta \cos\alpha +\frac{2\sqrt{6}}{3}\sin\beta \sin\alpha$ \\\hline
$\phi+\varphi_7$ &$4v_{\varphi_7}/v_\phi$& $4v_{\varphi_7}/v_\phi$ &$\cos\beta \cos\alpha +4\sin\beta \sin\alpha$ &$\cos\beta \cos\alpha +4\sin\beta \sin\alpha$ \\\hline
\end{tabular}}
\caption{The deviations in the Higgs boson couplings from the SM values in various extended Higgs sectors.
$\phi$, $\Delta$, $\xi$ and $\varphi_7$ are respectively denoted as Higgs fields with ($T,Y$)=($1/2,1$), ($1,2$), ($1,0$) and ($3,4$).
In the second and third column, $v_X$ is the vev of the Higgs field $X$.
The mixing angle $\alpha$ is defined for each extended Higgs sector in Ref.~\cite{Kanemura:2013mc}.
}
\label{tab:hVV_exotic}
\end{center}
\end{table}

\subsection{Adding Higgs singlets}

The introduction of an additional Higgs singlet field to the SM Higgs sector
does not affect $\rho=1$, and does not generate any flavor changing Higgs-mediated neutral current processes
as it does not couple to quarks, leptons and gauge bosons.
For example, such a singlet field has been introduced in new physics models with an
extra U(1) gauge symmetry, where the U(1) boson couples to B$-$L~\cite{Khalil:2006yi}.
A neutral singlet scalar field is also employed in the Next-to-Minimal
supersymmetric extension of the  Standard Model (NMSSM) along with the
second doublet field required in SUSY~\cite{Ellwanger:2009dp}.

The existence of a singlet field $\phi_2=S$ ($T_2=0$, $Y_2=2$) only change the Higgs boson couplings
via mixing of the singlet and doublet Higgs fields.
In the model with only one additional neutral singlet scalar field to the SM, $S$ and $\Phi$
can be parameterized as
\begin{eqnarray}
 \Phi=\left(\begin{array}{c}
 \omega^+ \\
 v + (\phi + i \chi)/\sqrt{2}\\ \end{array}
  \right) , \hspace{1cm} S=v_S + \phi_S
  \end{eqnarray}
where $v=174$ GeV, and $v_S$ is the vev of $S$. The two CP-even mass eigenstates $h$ and $H$  are
defined by
\begin{eqnarray}
   h = \phi \cos\theta - \phi_S \sin\theta, \hspace{1cm} H=\phi \sin\theta + \phi_S \cos\theta.
\end{eqnarray}
%In the model with one additional neutral complex singlet scalar field to the SM
%$\phi_2=\varphi$ ($T_2=0$, $Y_2=2$), $\varphi$ and $\Phi$ can be parameterized as
% \begin{eqnarray}
% \Phi=\left(\begin{array}{c}
% \omega^+ \\
% v + (\phi + i \chi)/\sqrt{2}\\ \end{array}
%  \right) , \hspace{1cm} S=v_\varphi + \frac{1}{\sqrt{2}}(\phi_\varphi + i \chi_\varphi)
%  \end{eqnarray}
%where $v=174$ GeV, and $v_\varphi$ is the vev of $\varphi$.
%The two CP-even mass eigenstates $h$ and $H$  are
%defined by
%\begin{eqnarray}
%   h = \phi \cos\theta - \phi_S \sin\theta, \hspace{1cm} H=\phi \sin\theta + \phi_S \cos\theta.
%\end{eqnarray}
In models with an extra U(1) gauge boson, this boson absorbs the CP-odd component $\chi_S$ via the Higgs mechanism.
The difference from the SM is just one additional CP-even scalar boson $H$.
There is no physical charged scalar state in this model.
All of the SM fields obtain mass from the vev, $v$, while the couplings of $h$ and $H$ are
obtained by the replacement of $\phi_{\mathrm SM } \to h \cos\theta + H \sin\theta$ in the Standard Model Lagrangian.

In the decoupling region $\theta \sim 0$, $h$ is the SM-like Higgs boson with its couplings reduced from their SM values
by $\cos\theta \sim 1 -\theta^2/2$. On the other hand, when $\tan\theta \sim {\cal O} (1)$, both $h$ and $H$ behave
as SM-like Higgs bosons, sharing the SM couplings to gauge bosons and fermions.  If $h$ and $H$ are almost degenerate
in mass, the two bosons might appear as a single SM Higgs boson in the LHC experiments.
At the ILC, by tagging the Higgs mass
in $e^+e^-\to Z +(h,H)$ by the invariant mass recoiling against the $Z$,
the two Higgs bosons could be better separated.
The ILC phenomenology of the Higgs sector in the minimal B-L model is surveyed in Ref.~\cite{Basso:2010si}.

%[Write More]

\subsection{Adding Higgs triplets}

Triplet Higgs fields are introduced in several new physics models.
An example of these models is the Higgs sector with the SM Higgs doublet $\Phi$ with
an additional triplet $\Delta$  with $T_2=1, Y_2= 2$ .  If  $\Delta$ carries the lepton number of 2, it can
couple to leptons by
\begin{eqnarray}
   {\cal L}_Y = h_{ij} \overline{L^{ic}} i\tau_2 \Delta L_L^j + {\mathrm h.c.}  \label{triplet_Yukawa}
   \end{eqnarray}
If $\Delta$ obtains the vev that is proportional to the explicit violation term
for the lepton number in the Higgs sector, then a neutrino mass matrix is generated,
\begin{eqnarray}
     {\cal M}_{ij}  = \sqrt{2} h_{ij}   v_\Delta.  \label{eq:neutrinomass}
\end{eqnarray}

The Higgs fields $\Phi$ and $\Delta$ are expressed in terms of component fields as
\begin{eqnarray}
 \Phi = \left( \begin{array}{c} \omega^+ \\ v_\Phi + (\phi + i \chi)/\sqrt{2} \\ \end{array}\right) ,
 \hspace{1cm}
 \Delta = \left(
 \begin{array}{cc}
\Delta^+/\sqrt{2} & \Delta^{++} \\
v_\Delta+(\delta + i \eta)/\sqrt{2} & - \Delta^+/\sqrt{2} \\
\end{array}
 \right),
\end{eqnarray}
 where $v_\Phi$ and $v_\Delta$ are vevs of $\Phi$ and $\Delta$. The physical scalar states are two CP-even ($h$ and $H$),
 a CP-odd ($A$), a singly charged pair ($H^\pm$) and a doubly charged pair ($H^{\pm\pm}$), which are
 derived from the component fields by diagonalizing the squared-mass matrices with the mixing angles $\alpha$, $\beta_0$ and $\beta_\pm$ as
 \begin{eqnarray}
&&  h = \phi \cos\alpha + \delta \sin\alpha, \hspace{4mm}\qquad H = - \phi \sin\alpha + \delta \cos\alpha,  \nonumber\\
 &&  A = -\chi \sin\beta_0 + \eta \cos\beta_0, \hspace{4mm}\,  H^\pm= - \phi^\pm \sin\beta_\pm + \Delta^\pm \cos\beta_\pm,
 \hspace{4mm}\, H^{++} = \Delta^{++}.
 \end{eqnarray}
In light of the constraint from the rho parameter, $v_\Delta \ll v$ must be taken, and then the masses of the scalar bosons
are given by
\begin{eqnarray}
  m_h^2 \simeq 2 \lambda_1 v^2,  \hspace{1cm} m^2_{H^{++}} - m_{H^+} \simeq m^2_{H^+} - m_A^2, \hspace{1cm} m_H^2 \simeq m_A^2,
\label{trip}
\end{eqnarray}
with $\alpha \ll 1$, $\beta_0 \ll 1$, and $\beta_\pm \ll 1$,
where $\lambda_1$ represents the quartic coupling constant of the doublet field.
Therefore,  $h$ behaves as the SM Higgs boson, and the others scalar states satisfy
the relations among the masses given in \eq{trip}.

The most interesting characteristic feature in this model is the existence of doubly charged Higgs bosons $H^{\pm\pm}$.
Its discovery would be a direct probe of the exotic Higgs sectors.
In general, doubly charged Higgs fields can arise from
the singlet with $Y=4$, the doublet with $Y=3$ and the triplet with $Y=2$.
In the model with an additional triplet field, the doubly charged Higgs bosons $H^{\pm\pm}$ can
decay into $\ell^\pm \ell^\pm$, $H^\pm W^\pm$ and $W^\pm W^\pm$
depending on the magnitude of $v_\Delta$~\cite{Perez:2008ha}.
In Fig.~\ref{FIG:BR1_HTM} , the branching ratios are shown as a function of
the vacuum expectation value of the triplet field, $v_\Delta$, for the cases with the mass difference $\Delta m=m_{H^{++}}-m_{H^+}=0$,
$10$ GeV and $30$ GeV~\cite{Aoki:2011pz}.
\begin{figure}[t]
\centering
\includegraphics[width=48mm]{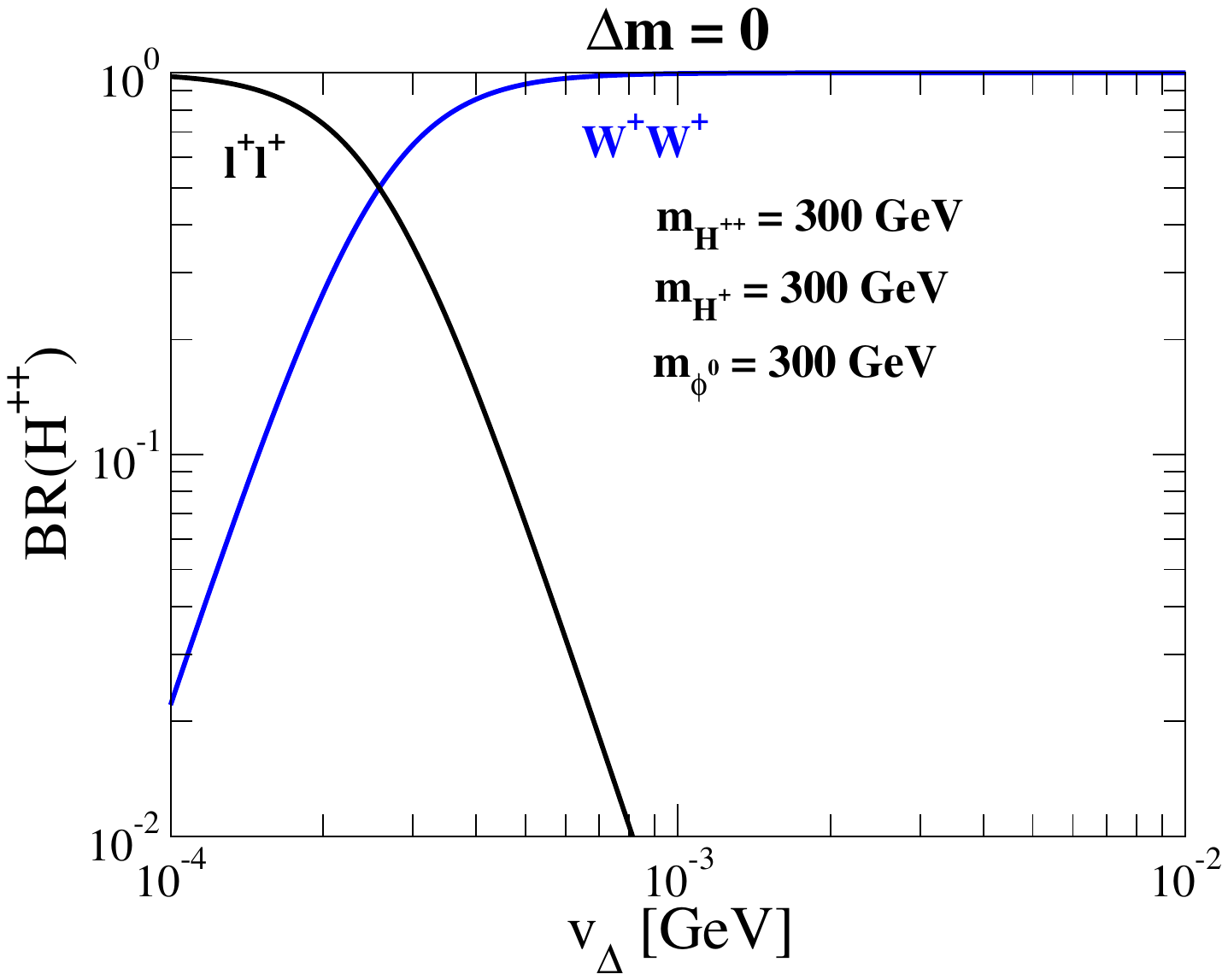}
\includegraphics[width=48mm]{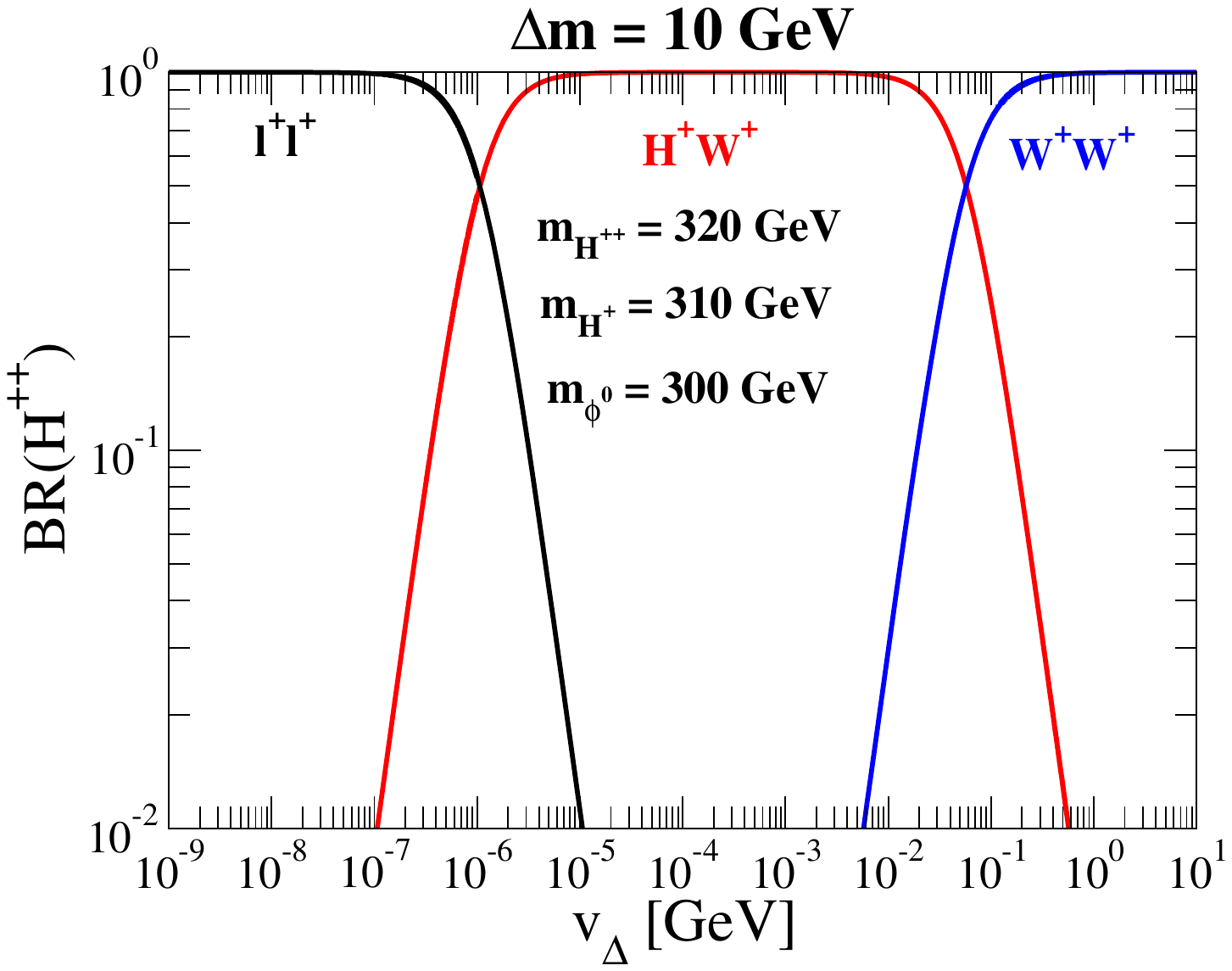}
\includegraphics[width=48mm]{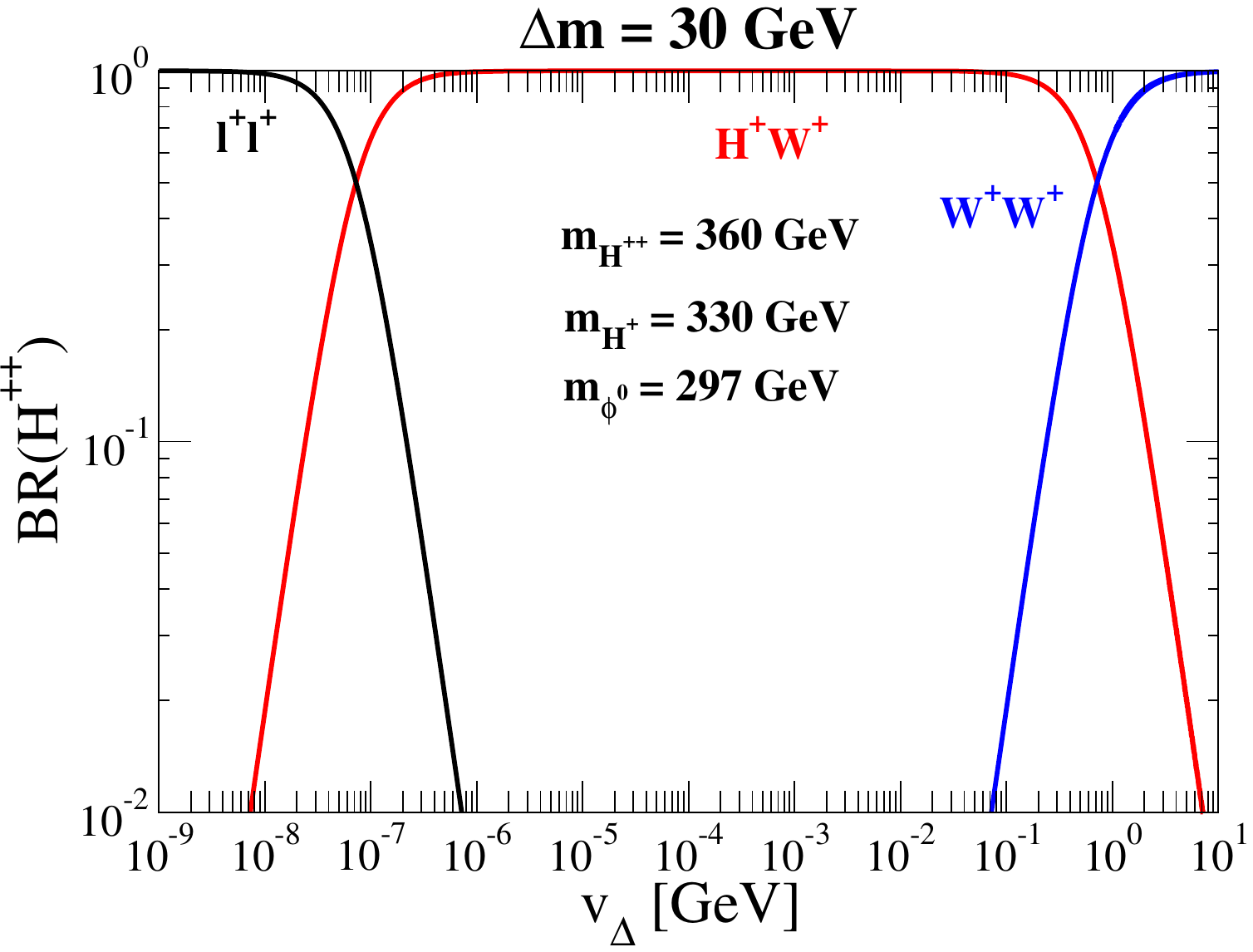}
\caption{Decay branching ratio of $H^{++}$ as a function of $v_\Delta$.
In the left figure, $m_{H^{++}}$ is
fixed to be 300 GeV, and $\Delta m$ is taken to be zero.
In the middle  figure, $m_{H^{++}}$ is
fixed to be 320 GeV, and $\Delta m$ is taken to be 10 GeV.
In the right figure, $m_{H^{++}}$ is
fixed to be 360 GeV, and $\Delta m$ is taken to be 30 GeV.}
\label{FIG:BR1_HTM}
\centering
\end{figure}
When $v_\Delta$ is smaller than $10^{-3}$ GeV, the dilepton decay $H^{\pm\pm} \to \ell^\pm \ell^\pm$ is dominant. The signal
directly shows the existence of the doubly charged scalar boson with lepton number 2, which can be a strong evidence for the
neutrino mass generation via Eq.~(\ref{eq:neutrinomass}).
At the LHC, the current search results for $H^{\pm\pm}$ using this decay mode gives the lower bound
on the mass of $m_{H^{++}} > 400$ GeV~\cite{ATLAS:2012hi,Chatrchyan:2012ya}.

\begin{figure}[b!]
\begin{center}
\includegraphics[width=90mm]{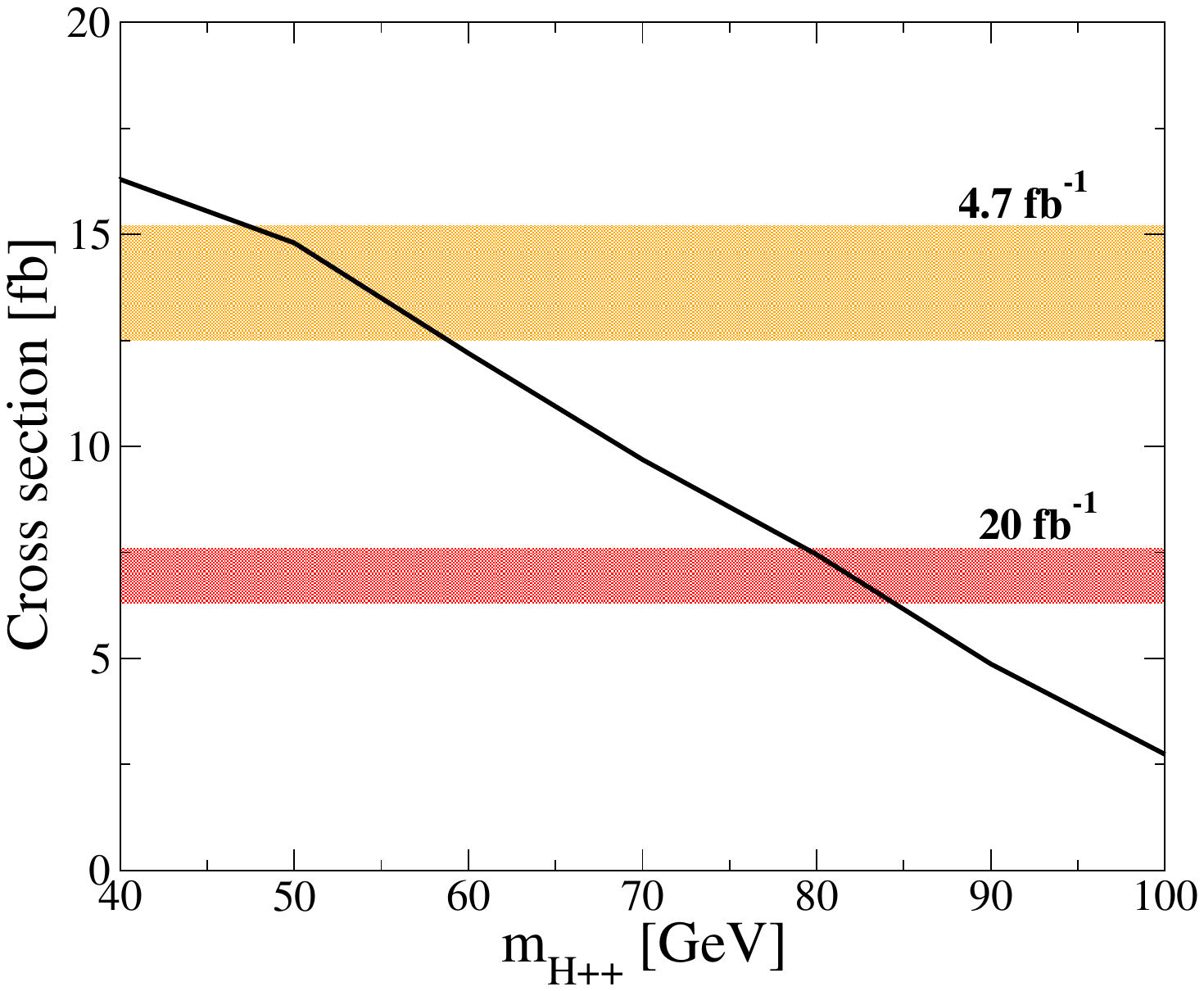}
\caption{The signal cross section after the $M_{\ell\ell}$ cut as a function of $m_{H^{++}}$ with the collision energy to be 7 TeV.
The light (dark) shaded band shows the 95\% CL (expected) upper limit for the cross section from the data for the
$\mu^+\mu^+$ channel with the integrate luminosity to be 4.7 fb$^{-1}$ (20 fb$^{-1}$).
%The width of the bands comes from the uncertainty of $\varepsilon_{\mathrm{fid}}$ for the $\mu\mu$ system between 59\% and 72\%~\cite{ATLAS}.
}
\label{FIG:diboson}
\end{center}
\end{figure}

In contrast,  when $v_\Delta$ is sufficiently larger than $10^{-3}$ GeV,  the diboson decay $H^{\pm\pm} \to W^\pm W^\pm$
becomes dominant. In this case, the signal can also be same sign four leptons, but its rate is reduced by
the branching ratios of leptonic decays of $W$s.
The current lower bound from this final state is only $m_{H^{++}} > 60$ GeV at 95\% CL
from data at the LHC with 5 fb$^{-1}$~\cite{Kanemura:2013vxa} as exhibited in Fig.~\ref{FIG:diboson}.
This bound is greatly relaxed as compared to the dilepton decay scenario.
 By the extrapolation of the data to 20~fb$^{-1}$ with the same collision energy,
the lower limit is estimated to be 85 GeV.

 If there is a mass difference between $H^{\pm\pm}$ and $H^\pm$, the parameter region where
 $H^{\pm\pm}$ can mainly decay into $H^\pm W^\pm$ appears. For example, for $v_\Delta=1$ GeV with a
 mass difference $\Delta m = m_{H^{++}} - m_{H^+} \sim 10$ GeV, the decay into $H^+W^+$ dominates for a wide range of $v_\Delta$.
 In this case, $H^{++}$ could be identified through its cascade decay.   If there is a mass difference with the opposite sign,
 $\Delta m \sim -10$ GeV for $v_\Delta \sim 1$ GeV,  $H^{\pm\pm} \to W^+W^+$ is dominant.

 There are wide regions of parameter space where the diboson decay is dominant. In this case, a relatively
 light $H^{\pm\pm}$ with a mass of few 100 GeV is expected to be still allowed even after
 the LHC run at 14 TeV with an integrated luminosity with 3000 fb$^{-1}$.
 The $H^{\pm\pm}$ can be pair produced at the ILC with $\sqrt{s} = 250$ GeV or $500$ GeV,
 and a doubly-charged Higgs signal can be easily detected.

\subsection{The NMSSM Higgs sector}

The minimal Higgs sector required for an anomaly-free supersymmetric
extension of the Standard Model consists of two Higgs doublets as
described in Section~\ref{mssm}.  But, the Higgs sector of the MSSM has
a number of troubling features.  First, the Higgs-Higgsino Lagrangian
contains a dimensionful parameter $\mu$, which in principle can be
arbitrarily large.  However, phenomenological considerations require
this parameter to be no larger than a typical supersymmetry-breaking
mass parameter, which should be of $\mathcal{O}(1~{\mathrm TeV})$ or less
in order to provide a natural explanation for the origin of the
scale of electroweak symmetry breaking (EWSB).  Second, the coefficients of
the quartic terms of the MSSM Higgs potential are fixed by the
SU(2)$\times$U(1) gauge couplings $g$ and $g'$.  This is the origin of
the tree-level bound $m_h\leq m_Z$, and implies that radiative
corrections due to loops of top-quarks must be large enough to explain
the observed mass of $m_h\simeq 126$~GeV.  This in turn requires
rather large top squark masses and/or mixing, which pushes at least
one of the top squark masses to values above 1~TeV.  Indeed, there is
already considerable tension in the MSSM between
achieving a large enough Higgs mass while maintaining a natural
explanation of the EWSB scale.

In the NMSSM, one add an additional complex singlet scalar field $S$ to
the MSSM Higgs sector.  A comprehensive review of the NMSSM
can be found in Refs.~\cite{Ellwanger:2009dp,Maniatis:2009re}.
The additional degrees of freedom of the NMSSM Higgs sector provide
an opportunity for ameliorating some of the troubling features of the
MSSM Higgs sector.  In the NMSSM, one can
set $\mu=0$ and generate an effective $\mu$ parameter dynamically that
is proportional to the vacuum expectation value of $S$.  Thus, a
phenomenologically acceptable NMSSM Higgs sector exist that contains
no new fundamental dimensionful parameters.  Second,
the NMSSM scalar potential contains a new quartic term proportional to
a dimensionless parameter $\lambda$ that is independent
of gauge couplings.  Thus, the mass of the observed Higgs boson now
depends on an unknown coupling, and it is significantly easier to
achieve the observed mass of $m_h\simeq 126$~GeV without extremely
large top squark masses and/or mixing.  As a result, the flexibility of the
additional degrees of freedom of the NMSSM can (somewhat) reduce the tension
with naturalness as compared with the MSSM.

In this section, we briefly review the structure of the Higgs sector
of the NMSSM.  We first consider the general NMSSM (sometimes called
the GNMSSM~\cite{Ross:2011xv}), in which all possible terms of dimension-four or less are
allowed in the Higgs Lagrangian.  The Higgs potential of the GNMSSM
is:
\beqa
V&=&(m_d^2+|\mu+\lambda S|^2)H_d^{i*}H_d^i+(m_u^2+|\mu+\lambda
S|^2)H_u^{i*}H_u^i-b(\epsilon^{ij}H_d^i H_u^j+{\mathrm h.c.})
\nonumber \\
&&\qquad +m_s^2|S|^2+(\xi_s S+\half b_s S^2+\tfrac{1}{3}\kappa A_\kappa S^3
-\lambda A_\lambda S\epsilon^{ij}H_d^i H_u^j+{\mathrm h.c.})\\
&&\qquad |\xi+\mu_s S +\kappa S^2-\lambda\epsilon^{ij}H_d^i H_u^j|^2
+\tfrac{1}{8}
\left(g^2 + g^{\prime\,2}\right) \left[H_d^{i*}H_d^i-H_u^{j*}H_u^j\right]^2
+\half g^2 |H_d^{i*}H_u^i|^2\,,\nonumber
\eeqa
where $\mu$ and $\mu_S$ are the supersymmetric Higgsino mass
parameters,
$\lambda$ and $\kappa$ are dimensionless supersymmetric scalar self-interaction
parameters, $m_d^2$, $m_u^2$, $b$, $b_s$ and $\xi$ are
soft-supersymmetry-breaking squared-mass parameters, $A_\kappa$ and
$A_\lambda$ are soft-supersymmetry-breaking mass parameters, and
$\xi_s$ is a soft-supersymmetry-breaking cubed-mass parameter.

To eliminate the supersymmetric preserving mass parameters, one can
impose a discrete $\mathbb{Z}_3$ symmetry, such that the scalar potential
is invariant under $\{H_d, H_u, S\}\to \omega\{H_d, H_u, S\}$ with
$\omega^3=1$ and $\omega\neq 1$.  In this case we have
$\mu=\mu_s=b=b_s=\xi=\xi_s=0$, and the scalar potential is specified in
terms of two supersymmetric dimensionless parameters, $\lambda$ and
$\kappa$, and five dimensionful supersymmetry-breaking parameters,
$m_d^2$, $m_u^2$, $m_s^2$, $A_\lambda$ and $A_\kappa$.  This model is
often referred to simply as the NMSSM (more accurately, it is also
called as the $\mathbb{Z}_3$-invariant NMSSM).

Unlike the MSSM, the tree-level (G)NMSSM allows for the possibility of
CP-violation in the Higgs sector.  For simplicity, we shall assume in
the following
that the (G)NMSSM scalar potential is CP-conserving, and take all
scalar potential parameters and vacuum expectation values to be real,
\beq
\langle{H_d^0}\rangle=v_d\,,\qquad\quad
 \langle{H_u^0}\rangle=v_u\,,\qquad\quad
 \langle{S}\rangle=v_s\,.
 \eeq
In this case, the scalar spectrum consists of three CP-even neutral
scalars, two CP-odd neutral scalars and a pair of charged Higgs
scalars $H^\pm$.  It is again convenient to go to the Higgs basis in
which linear combinations of the two doublet fields, denoted by $H_1$ and $H_2$, are defined such
that  $\langle{H_1^0}\rangle=v=174~{\mathrm GeV}$ and $\langle{H_2^0}\rangle=0$, where $v^2\equiv v_u^2+v_d^2$.
In this basis, the squared-mass matrix of the CP-even neutral Higgs bosons is given by
\beq \label{me}
\mathcal{M}_e^2 =
\begin{pmatrix} m^2_Z \cos^2 2\beta+\lambda^2 v^2 \sin^2 2\beta &
           -(m^2_Z-\lambda^2 v^2)\sin 2\beta\cos 2\beta  & M_{e13}^2
           \\
  -(m^2_Z-\lambda^2 v^2)\sin 2\beta\cos 2\beta&
  \mha^2+ m^2_Z \sin^2 2\beta & M_{e23}^2\\
  M_{e13}^2 & M_{e23}^2 & M_{e33}^2
   \end{pmatrix}\,,
\eeq
where
\beq
m_A^2\equiv \frac{2\bigl[b+\lambda(\mu_s v_s+\xi)+\lambda v_s(A_\lambda+\kappa v_s)\bigr]}{\sin 2\beta}\,,
\eeq
and $M_{e13}^2$, $M_{e23}^2$ and $M_{e33}^2$ can be expressed in terms
of the scalar potential parameters (explicit expressions can be found
in Ref.~\cite{Ross:2011xv}).  The parameter $m^2_A$ is no longer the
squared-mass of the CP-odd Higgs boson.  Rather, it is a diagonal
element of the CP-odd Higgs squared-mass matrix,
\beq
\mathcal{M}_o^2=\begin{pmatrix} m_A^2 & M_{o12}^2 \\ M_{o12}^2 & M_{o22}^2
\end{pmatrix}\,,
\eeq
where $M_{o12}^2$ and $M_{o22}^2$ can be expressed in terms of the
scalar potential parameters (explicit expressions can be found in
Ref.~\cite{Ross:2011xv}).
The squared-masses of the neutral Higgs bosons are obtained by
computing the eigenvalues of $\mathcal{M}^2_e$ and $\mathcal{M}_o^2$.
Finally, the charged Higgs mass is given by
\beq
m^2_{H^\pm}=m_W^2+m_A^2-\lambda^2 v^2\,.
\eeq

The phenomenology of the GNMSSM is richer than that of the MSSM due to
the additional Higgs states and the larger parameter space.  The
simplest scenario is an MSSM-like scenario in which either $M^2_{e33}$
and $M^2_{o22}$ are large
and/or $M_{e13}^2$, $M_{e23}^2$ and $M_{o12}^2$ are small.
In this case, the mixing
of the singlet and doublet Higgs components is suppressed and two of
the three CP-even Higgs bosons are governed by the $2\times 2$ block
obtained from the first two rows and columns of $\mathcal{M}^2_e$,
while the doublet-like CP-odd Higgs boson has mass $m_A$.
Nevertheless, the MSSM squared-mass relations are corrected by terms
of $\mathcal{O}(\lambda^2 v^2)$.  Using \eq{me}, it follows that
the CP-even Higgs squared-mass
inequality given by \eq{eqmH1up0} is modified to~\cite{Ellwanger:2006rm},
\beq
m_h^2\leq m_Z^2\cos^2 2\beta+\lambda^2 v^2\sin^2 2\beta+
\frac{3g^2\mt^4}{8\pi^2\mw^2}
\left[\ln\left(\frac{\msusyy}{\mt^2}\right)+
\frac{X_t^2}{\msusyy}
\left(1-\frac{X_t^2}{12\msusyy}\right)\right]\,,
\eeq
which includes the leading one-loop radiative correction.  Whereas the bound is
saturated at large $\tan\beta$ in the MSSM (where $\lambda=0$), we see
that in the NMSSM it is possible to have a SM-like Higgs boson with a
mass of 126 GeV for relatively modest values of $\tan\beta$ if
$\lambda$ is sufficiently large.  Indeed, in contrast to
the MSSM, one does not need as large a boost from the radiative corrections,
which means that lower top squark masses are allowed given the
observed Higgs mass (thereby lessening the tension with naturalness).
Typically, the region of interest corresponds to $\tan\beta\sim 2$ and
$\lambda\sim 0.7$.  For values of $\lambda>0.7$, the scalar
self-coupling running parameter $\lambda(Q)$ blows up at scales below
the Planck scale.  Although such a scenario is not consistent
with perturbative unification, it does lead to some interesting model
building opportunities for a highly natural Higgs boson with a mass of 126 GeV
over a wide range of parameters~\cite{Hall:2011aa}.

The scalar states that are dominantly singlet can only couple to gauge
bosons and fermions through the small doublet admixture in their wave
functions.  Thus, these states are very difficult to produce and
observe at the ILC.  One possible exception to this statement occurs
in the limit where both the mixing terms  $M_{e13}^2$, $M_{e23}^2$ and
$M_{o12}^2$ are small, and the diagonal elements of the CP-even and/or
CP-odd scalar squared-mass matrices are also small.  In this case, the
lightest scalar particles of the Higgs spectrum can be dominantly
singlet.  This leaves open the possibility that the decay channels
$h\to h_1 h_1$ and/or $h\to a_1 a_1$ are allowed, where $h$ is
identified as the observed SM-like Higgs boson and $h_1$ and $a_1$ are
the light dominantly singlet CP-even and CP-odd scalar states.
These light states would then decay dominantly
via their small scalar doublet admixtures into a pair of the heaviest
fermions that are kinematically allowed.
There are some experimental limits to this scenario due to
searches at LEP, Tevatron and LHC, but allowed regions of the (G)NMSSM
parameter space with light singlet-like scalars still persist.

Finally, it is possible that the mixing between singlet and doublet
components is not particularly small.  In this case, one can still
find parameter regimes (e.g. large $m_A$)
in which the lightest CP-even state is dominantly doublet and
SM-like (to be identified with the observed Higgs boson), while the
heavier states are admixtures of doublet and singlet states.
In this scenario, all Higgs states are in play and can be studied at
the ILC if their masses are less than half the center-of-mass energy.

%%%%%%%%% Part added by GW, Sept. 26
\section{Model-independent treatments of Higgs properties}

In the quest for identifying the underlying physics of electroweak
symmetry breaking it will be crucial to study the properties of the
observed signal at 126~GeV with high precision, taking into account also
the limits from Higgs searches in other regions of the parameter space.
Besides the interpretation of the experimental results in specific
models, it is also useful to express the properties of the Higgs sector
in terms of less model-dependent parameterizations. For the observed
signal at 126~GeV this refers in particular to its mass, spin, CP
properties and couplings.

While the mass can be determined in an essentially model-independent
way,
and the spin quantum number for a single resonance
can be obtained from confronting distinct
hypotheses for different spin states with the data (see below), the
determination of CP properties and couplings is more involved. An
observed state can in general consist of any admixture of CP-even and
CP-odd components. Testing the distinct hypotheses of a pure CP-even and
a pure CP-odd state can therefore only be a first step in determining
the CP properties of a new particle. While it is an obvious goal to
extract information on the couplings of the discovered state to other
particles, some care is necessary regarding the actual definition of
those couplings. It would be tempting to treat the
couplings in a certain model, for instance the SM, as independent
free parameters and perform a fit to the experimental data. This is not
possible, however, once (electroweak) higher-order corrections are
taken into account, since model-specific relations between the couplings
and the other parameters of the theory are required to ensure properties
like UV-finiteness and gauge cancellations.

Moreover, modifying a certain
coupling as compared to its form in the SM will in general change both
the overall coupling strength and the tensor structure of the coupling.
The latter implies a modification of
%a coupling will in general affect
the CP properties of the considered state. As a consequence, in
general the determination of couplings cannot be treated
separately from the determination of the CP properties.
%since a modification of the tensor structure of a coupling affects the CP
%properties.

Accordingly, in order to analyze the coupling properties of the
discovered state a well-defined framework is required where the
state-of-the art predictions within the SM
(or any other reference model), including all relevant higher-order
corrections, are supplemented by a parameterization of
possible deviations of the couplings from their reference values
(including also possible changes of the tensor structure). If one assumes that
effects of new physics occur only at rather high scales, such that the
contributions of heavy degrees of freedom
can be systematically integrated out, such
a framework can be formulated with the help of an effective Lagrangian.

%Such a framework
%can be formulated with the help of an effective Lagrangian where heavy
%fieds are systematically integrated out, under the assumption that the
%theory does not contain any additional light states that affect Higgs
%phenomenology.

\subsection{Effective Lagrangian treatments}
\label{sec:th_efflag}

Assuming that effects of new light particles in loops are absent,
physics beyond the Standard Model (BSM)
can be described via an effective Lagrangian in terms of the SM
fields. This approach has been pioneered in Ref.~\cite{Buchmuller:1985jz},
where a list of operators of dimensions 5 and 6 in the linear
parameterization of the Higgs sector with a Higgs doublet has been
provided. Those higher-dimensional operators arise from integrating out
the contributions of heavy degrees of freedom. Restricting to operators
of up to dimension 6 that are relevant for Higgs physics, such an
effective Lagrangian has the general form
\begin{equation}
{\cal L}_{\mathrm{eff}} = {\cal L}^{(4)}_{\mathrm{SM}}
+ \frac{1}{\Lambda^2}\sum_k \alpha_k {\cal O}_k,
\end{equation}
where  ${\cal L}^{(4)}_{\mathrm{SM}}$ is the SM Lagrangian,
${\cal O}_k \equiv {\cal O}^{d=6}_k$
denotes dimension-6 operators, $\alpha_k$ the corresponding
Wilson coefficients, and $\Lambda$ is the scale of new physics.

Taking into account all dimension-6 operators that are in accordance
with gauge invariance leads to a rather large number of operators.
A minimal complete basis can be constructed using the equations of
motions to eliminate redundant operators~\cite{Grzadkowski:2010es}.
Proposals that are suitable for the analysis of the upcoming data at the
LHC are currently under discussion, (e.g., see Ref.~\cite{Heinemeyer:2013tqa} and
references therein). For the analysis of the LHC results up to 2012 an
``interim framework'' has been adopted that is based on a simplified
approach using ``leading order inspired'' scale factors
$\kappa_i$~\cite{LHCHiggsCrossSectionWorkingGroup:2012nn,Heinemeyer:2013tqa}.
In particular, in order to make use of reinterpretations of searches
that have been performed within the context of the SM, in this approach
only overall changes in the coupling strengths are
considered, while effects that change kinematic distributions are not
taken into account.

%\subsection{Higher-dimensional operators}

\subsection{Simplified approach for the analysis of Higgs couplings}
\label{sec:th_kappas}

The searches for a SM-like Higgs carried out at the LHC so far have
mainly focused on the production processes gluon fusion, $gg \to H$,
weak-boson fusion, $qq^\prime \to qq^\prime H$, associated production
with $W$ or $Z$, $q \bar q \to WH / ZH$, and associated production with
a top-quark pair, $q \bar q / gg \to t \bar t H$. The searches were based on the decay channels $\gamma\gamma$, $ZZ^{(*)}$, $WW^{(*)}$,
$b \bar b$ and $\tau^+\tau^-$. The couplings involved in those processes
have been analyzed in an ``interim framework'' under the following simplifying
assumptions~\cite{LHCHiggsCrossSectionWorkingGroup:2012nn,Heinemeyer:2013tqa}:

\begin{itemize}

\item
The observed signal is assumed to correspond to a single narrow
resonance. The case of several, possibly overlapping, resonances
is not considered.

\item
The zero-width approximation is assumed for this state. This implies
that all channels can be decomposed into a production cross section
times a decay branching ratio.

\item
Only modifications of coupling strengths, i.e.\ of absolute values of
couplings, are considered. No modifications of the tensor structure as
compared to the SM case are taken into account. This means in
particular that the observed state is assumed to be a CP-even scalar.

\end{itemize}

In order to parameterize possible deviations from the SM predictions in
this framework scale factors $\kappa_i$ are introduced.
The scale factors $\kappa_i$ are defined in such a way that the cross
sections $\sigma_{ii}$ or the partial decay widths $\Gamma_{ii}$
associated with the SM particle $i$ scale
with the factor $\kappa_i^2$ when compared to the corresponding SM
prediction. For a process $ii \to H \to jj$ the application of the scale
factors results in the term $\kappa_i^2\kappa_j^2/\kappa_H^2$ relative
to the SM prediction, where $\kappa_H^2$ denotes the scale factor for
the total width of the observed signal.

By construction,
the SM predictions according to the current
state-of-the-art, i.e.\ including the available higher-order
corrections, are recovered if all $\kappa_i = 1$.
Since higher-order corrections in general do
not factorize with respect to the rescalings, the theoretical accuracy
degrades for $\kappa_i \neq 1$. This is a drawback of this simplified
framework in comparison to the effective Lagrangian approach discussed
above, where possible deviations from the SM predictions are
parameterized in a more systematic way.

For loop-induced processes such as $gg \to H$ and $H \to \gamma\gamma$
the scale factors $\kappa_g$ and $\kappa_\gamma$ are in general treated
as free parameters. If on the other hand
one assumes that there are no contributions of
BSM particles to the loops, those scale factors can be related to the
scale factors of the corresponding SM particles in the loop, e.g.\
$\kappa_\gamma =
\kappa_\gamma(\kappa_b, \kappa_t, \kappa_\tau, \kappa_W, m_H)$
in this approximation.

The total width $\Gamma_H$ is the sum of all Higgs partial widths. The
corresponding scale factor $\kappa_H^2$, i.e.\ $\Gamma_H = \kappa_H^2
\Gamma_H^\mathrm{SM}$, in general needs to be treated as
a free parameter. Under the assumption that no
additional BSM Higgs decay modes
(into either invisible or undetectable final states)
contribute to the total width and making additional approximations for
the scale factors of the currently undetectable decay modes into SM
particles, e.g.\ $\kappa_c = \kappa_t$, $\kappa_s = \kappa_b$ etc.,
the scale factor $\kappa_H^2$ can be related to the scale factors of the
partial widths of the different decay modes in the SM,
\begin{equation}
\kappa_H^2 = \kappa_H^2(\kappa_j, m_H) ,
\label{eq:kappaH}
\end{equation}
where $j = W, Z, b, \tau, \ldots$.

Within this interim framework, several benchmark parameterizations have
been considered. Since the available data do not permit a measurement of
the total width $\Gamma_H$, in general it is not possible to directly
determine scale factors $\kappa_i$, but one is limited to determining
ratios of scale factors of the form $\kappa_i\kappa_j/\kappa_H$.
If one assumes that no BSM Higgs decay modes contribute to the total width and
using the approximations mentioned above for the currently undetectable
SM modes, one can relate $\kappa_H$ to the other scale factors as given
in eq.~(\ref{eq:kappaH}), which makes an absolute determination of the
$\kappa_i$ possible (a milder assumption that also allows to constrain
the total width is to assume $\kappa_W \leq 1$ and $\kappa_Z \leq 1$~\cite{Duhrssen:2004cv,Heinemeyer:2013tqa}).

For the experimental analyses up to now often
benchmark parameterizations with two free parameters have been used. In
particular, a parameterization in terms of a common scale factor for the
couplings to fermions, $\kappa_F$, and a common scale factor for the
couplings to $W$ and $Z$, $\kappa_V$, has been considered, where
$\kappa_F = \kappa_t = \kappa_b = \kappa_\tau$ and
$\kappa_V = \kappa_W = \kappa_Z$. Besides assuming that all couplings to
fermions and the couplings to $W$ and $Z$ can be scaled with a universal
factor, in this case it is furthermore assumed that contributions of BSM
particles to the loop-induced processes are absent and that the
contributions to the total width can be approximated according to
eq.~(\ref{eq:kappaH}). The most general parameterization that has been
investigated up to now involves the parameters
$\kappa_V, \kappa_t, \kappa_b, \kappa_\tau, \kappa_g, \kappa_\gamma$~\cite{Chatrchyan:2013lba}.

%%%%%%%%% End of part added by GW, Sept. 26

%Suppose that  the signal corresponds to a CP-even scalar with the single
%narrow resonance and that only modifications from the SM predictions
%are the absolute values of the couplings rather than of the coupling strengths.
%Such a case leads to a simplified framework in terms of scale factors
%$\kappa_x$, which describe the modifications from the SM coupling
%constant of $hxx$ as $g_{hxx}=\kappa_x g_{hxx}({\mathrm SM})$.
%This intermediate framework has been adopted
%for the analyis of the data until 2012. A more complete framework that
%is based on an effective Lagrangian is currently being developed for the
%analysis of the data of the next LHC run.

\section{Alternative approaches to electroweak symmetry breaking
  dynamics}\label{alternate}

In the Standard Model, electroweak symmetry is broken by perturbative
scalar dynamics.  The scalar potential exhibits a minimum at a
non-zero value for the neutral component of a hypercharge one,
complex doublet of scalar fields.  The scalar fields are elementary
(not composite) degrees of freedom, at least at energy scales of order
1 TeV and below.  In all extensions of the Higgs sector discussed
previously in this document, the elementarity of the scalar fields is maintained and the
weakly-coupled nature of the scalar dynamics is preserved.

The Standard Model cannot be a fundamental theory of elementary
particle interactions to arbitrarily high energy scales.  At Planck
scale energies ($M_{\mathrm PL}\simeq 10^{19}~{\mathrm GeV}$), gravitational
phenomena at the microscopic scale can no longer be neglected.
Indeed, other new scales of fundamental physics may exist between the
scale of electroweak symmetry breaking (EWSB) of order $v=174$~GeV and the
Planck scale, e.g. the grand unification scale, the seesaw scale (that
governs right-handed neutrinos and is responsible for generating mass
for the light neutrinos) and the mass scale associated with dark matter.

In the Standard Model, the scale of EWSB is not protected by any known
symmetry.  Thus, it is deeply puzzling how the EWSB scale can be
stable with respect to the Planck scale (and other high energy scales
if they exist).  An equivalent statement is that there is no mechanism
in the Standard Model that can keep the mass of an elementary scalar
field much lighter than the highest fundamental mass scale of the
theory, $\Lambda$.  That is, the natural value for the squared-mass of
the scalar is~\cite{Weisskopf:1939zz}
\beq
m_h^2\sim \frac{g^2}{16\pi^2}\Lambda^2\,,
\eeq
where $g$ is the coupling of the scalar to other sectors of the theory.
That is, the scale of EWSB and the attendant Higgs mass is
extremely \textit{unnatural} if $\Lambda\gg \mathcal{O}(1~{\mathrm TeV})$.
Only if $\Lambda\sim 1$~TeV do we have a chance of providing a natural
mechanism for the EWSB dynamics~\cite{Susskind:1982mw}.

The quest for a natural theory of EWSB is one of the motivations for
TeV-scale supersymmetry~\cite{Maiani:1979cx,Witten:1981nf,Dimopoulos:1981zb,Sakai:1981gr,Kaul:1981wp,Kaul:1981hi}.
In this framework, elementary scalars are
related by supersymmetry to elementary fermionic superpartners.  The
fermion masses can be naturally small due to weakly broken chiral
symmetries, which in turn protects the masses of the scalar partners.
In theories of TeV-scale supersymmetry, we identify $\Lambda$ with the
scale of supersymmetry breaking.  Ideally, this scale should be no
larger than of $\mathcal{O}(1~{\mathrm TeV})$.
The fact that supersymmetry has not yet been discovered at the LHC
provides some tension for natural EWSB in the supersymmetric
framework.   A pedagogical review of alternative EWSB scenarios can be found in
Ref.~\cite{Bhattacharyya:2009gw}.
In this section, we shall briefly consider non-supersymmetric
approaches that could provide a natural explanation for EWSB dynamics.

One of the first leading contenders for a natural theory of EWSB dynamics
was technicolor.  (Reviews and references can be found in
Refs.~\cite{Farhi:1980xs,Kaul:1981uk,Hill:2002ap}.
In this approach, EWSB was generated by the
condensation of bilinears of new fermion fields.  No elementary scalar
fields were needed, and the naturalness problem associated with them
was avoided.  Unfortunately, this approach was ultimately
unsuccessful.  Apart from the fact that it was very difficult to
generate a realistic fermion mass spectrum (which required additional
new dynamics beyond the introduction of technicolor and the associated
techniquarks), the constraints of the precision electroweak observables
were extremely difficult to accommodate.  The discovery of Higgs boson
with Standard Model like properties may have provided the final nail in the
coffin (although not all technicolor proponents have conceded~\cite{Eichten:2012qb,Foadi:2012bb}).
Any theory of EWSB dynamics must explain the presence of a
weakly-coupled SM-like Higgs boson whose mass is considerably smaller
than $\mathcal{O}(1~{\mathrm TeV})$.

\subsection{The Higgs boson as a pseudo-Goldstone boson}

Apart from supersymmetry, there is a known mechanism that can yield
naturally light elementary scalars.  When a continuous global symmetry
is broken, one of the consequences is an exactly massless Goldstone
boson.  If this global symmetry is now explicitly broken, the would-be
Goldstone boson acquires a mass proportional to the strength of the
symmetry breaking.  This is a mechanism for producing naturally
light scalars---pseudo-Goldstone bosons whose masses are generated by
small symmetry-breaking effects~\cite{Georgi:1975tz}.

Thus, perhaps the Higgs boson is in fact a pseudo-Goldstone boson (PGB)
generated by strong dynamics associated with scale of new
physics $\Lambda$~\cite{Georgi:1984af,Dugan:1984hq}.
Even though the tree-level mass of the
PGB can be significantly smaller that $\Lambda$, one-loop corrections
to the PGB squared-mass due to the new physics at the scale $\Lambda$
will still be quadratically sensitive to $\Lambda$.  This would imply
that $\Lambda\sim\mathcal{O}(1~{\mathrm TeV})$, which is in conflict with
precision electroweak observables that do no show any sign of
strongly-coupled new physics effects at a mass scale of order 1 TeV.

By a clever construction, one can overcome this last objection by
arranging to have the quadratic sensitivity at one loop ameliorated by
a cancellation of contributions to the one-loop radiative
corrections. The quadratic sensitivity will persist at two-loops, but
the presence of the extra loop factor would imply that
$\Lambda\sim\mathcal{O}(10~{\mathrm TeV})$, which is no longer in conflict with
precision electroweak observables.  This is precisely the
mechanism employed by the little Higgs
Models~\cite{ArkaniHamed:2002pa}.    In this framework, the
Higgs boson is a PGB associated with the explicit breaking of some
global symmetry  But in this case, the global symmetry becomes exact
when two different interactions separately vanish (this phenomenon is
known as collective symmetry breaking).  That is, the
lightness of the Higgs boson mass is doubly protected.  Indeed, at one
loop the quadratic sensitivity of the Higgs squared-mass to $\Lambda$
vanishes due to the cancellation between Standard Model particles and
partner particles of the same spin (in contrast to supersymmetry where
the cancellation is due to partners that differ by a half a unit of
spin).  For example, the top quark must be accompanied by new
fermionic top partners whose masses should be of order 1 TeV.
Likewise, such models typically include additional gauge bosons, which
are partners of the $W^\pm$ and $Z$.

Numerous realizations of little Higgs models can be found in the
literature~\cite{Schmaltz:2005ky,Chen:2006dy,Perelstein:2005ka}.
The challenge of precision electroweak observables
is still present but can be overcome by introducing a discrete
$T$-parity~\cite{Cheng:2003ju,Cheng:2004yc}
(whose consequences are similar to that of $R$-parity in supersymmetry
models).  The presence of new physics (such as top partners and new
gauge bosons) can modify the properties of the 126 GeV Higgs boson and
provide additional constraints on model building.

An alternative approach for constructing Higgs bosons as PGBs arises
in composite models of Higgs bosons~\cite{Georgi:1984af,Dugan:1984hq}.
A pedagogical review of the recent progress in
developing realistic models of this type can be found in Ref.~\cite{Contino:2010rs}.
Such models often arise as low-energy effective theories of models constructed
in higher dimensions of spacetime, where the Higgs boson degree of freedom is
identified as the fifth component of a five-dimensional gauge field.
(A recent review of this mechanism, called
gauge-Higgs unification, can be found in Ref.~\cite{Serone:2009kf}.)
In this approach, $f\sim\mathcal{O}(1~{\mathrm TeV})$ characterizes
the scale of new strong interactions that produce the PGB when some
larger global symmetry (in which the Standard Model gauge group is
embedded) is broken.  The effective cutoff of the theory is
$\Lambda\sim 4\pi f\sim 10~{\mathrm TeV}$.  The natural value for the
Higgs mass is a few hundred GeV, so some amount of tuning is
required to identify the Higgs boson of these models with the 126 GeV
Higgs boson.  Typically, deviations from SM Higgs properties are
expected at the 10\% level or higher due to the composite nature of
the Higgs state.  Moreover, such approaches typically predict the
existence of other composite resonant states with masses below 1 TeV,
which can be searched for at the LHC~\cite{Pomarol:2012qf}.
These class of models are best studied using the
effective Lagrangian analysis of Section~\ref{sec:th_efflag}.

\subsection{The Higgs boson as a dilaton}

A massless scalar can also arise due to the spontaneous breaking of
conformal symmetry.  In this case, the corresponding massless
Goldstone boson is called a dilaton.  If there is a small explicit
breaking of conformal symmetry, the corresponding dilaton will be
light.  So perhaps the Higgs boson can be identified as the dilaton of
a broken conformal symmetry.  Indeed, if one sets the Higgs potential
of the Standard Model to zero, then the Standard Model is classically
scale invariant.  The Higgs field can take on any value (without
costing energy at the classical level).  If the Higgs field assumes a
nonzero value, then both the electroweak symmetry and the conformal
invariance is broken.  In this case, the Higgs boson is identified as
the dilaton of spontaneously broken conformal symmetry.  (In practice,
the resulting Higgs boson is not massless due to quantum effects that
break the conformal symmetry due to the conformal anomaly.)
Models of this type have been recently reviewed in Ref.~\cite{Terning:2013}.

It is not clear whether a theoretically consistent model of this type
exists.  One would have to demonstrate that given a model with
spontaneously broken conformal symmetry with a flat direction
for the dilaton, a suitable weak explicit breaking of that symmetry
can be introduced that picks out a unique vacuum and generates a small
mass for the dilaton.  Identifying the breaking scale $f$ with the
scale of EWSB $v$, it then follows that the leading couplings of the
dilaton to SM fermions is $m_f/v$ and to SM bosons is $m_b^2/v$ which
matches precisely with the couplings of the SM Higgs boson.   However,
there could be corrections to the one-loop couplings of the dilaton to
gluons and photons that depend in a model-independent way on the
details of the conformal sector, which would yield deviations from
the expected one-loop couplings of the SM Higgs boson~\cite{Bellazzini:2012vz}.
The most significant difference between a SM Higgs boson and a dilaton
Higgs boson would be found in the triple and quartic Higgs
self-couplings.  All these deviations can be parameterized via
the effective Lagrangian treatment of Section~\ref{sec:th_efflag}.

In models where $f\neq v$, the dilaton is a scalar distinct
from the Higgs boson.  This would yield a phenomenology quite
distinctive from that of a typical extended Higgs sector~\cite{Chacko:2012vm}.  However,
such an approach would not provide any fundamental understanding of
how the SM Higgs boson mass remains significantly smaller than the higher
energy scale that define this theory.

\section{Probing the properties of the signal at 126 GeV}

After the spectacular discovery of a signal at 126~GeV in the
Higgs searches at the LHC~\cite{Aad:2012tfa,Chatrchyan:2012ufa}, it is critically important to
determine the properties of the new state as comprehensively and as
accurately as possible. This information will provide crucial input for
identifying the nature of the
electroweak symmetry-breaking mechanism.

\subsection{Present status and prospects for the upcoming LHC runs}

We briefly discuss here the present status and the prospects for the
upcoming LHC runs. We keep this discussion at a rather qualitative level.
For a more quantitative treatment we refer to the latest results from
ATLAS and CMS (see in particular
\cite{Aad:2013xqa,Aad:2013wqa,Chatrchyan:2012jja,Chatrchyan:2013lba}
and references therein) and the
future projections that have been made under various assumptions.

The determination of the mass of the new particle is already at the
level of a precision measurement with the 2012 data, driven by the
$\gamma\gamma$ and $ZZ^* \to 4 \ell$ channels. The accuracy will further
improve with increasing statistics, requiring however a careful
treatment of systematic effects.

Concerning the spin of the discovered particle, the
observation in the $\gamma\gamma$ channel rules out the possibility of
a $J=1$ state as a consequence of the Landau--Yang theorem~\cite{Landau:1948kw,Yang:1950rg}. It should be
mentioned that there are two caveats to this argument. First,
the Landau--Yang theorem strictly applies to an
on shell resonance, so that the $J=1$ hypothesis can be excluded
only by making an additional small-width assumption~\cite{Ralston:2012ye}.
Second, the decay product could in principle
consist of two pairs of boosted
photons each misinterpreted as a single photon.
Nevertheless, assuming that the discovered state corresponds to a single
resonance rather than to several overlapping resonances corresponding to
different spin states, the spin of the discovered particle
can be determined by discriminating between
the distinct hypotheses for spin 0, (1), 2 states. Some care is
necessary in modeling possible incarnations of the spin~2 hypothesis in
order to make sure that the discriminating power of the analysis
actually refers to the spin properties rather than to some unphysical
behavior of the spin~2 implementation. The experimental results obtained
up to now are well
compatible with the spin~0 hypothesis~\cite{Chatrchyan:2012jja,Aad:2013xqa}, and there is growing evidence
against the alternative hypotheses.

The determination of the CP properties of the observed state is a much
more difficult task, since the observed state could in principle consist
of any admixture of CP-even and CP-odd components.
The analyses so far are mainly based on observables involving the coupling
of the new state to two gauge bosons, $HVV$, where $V = W, Z$,
in particular $H \to ZZ^* \to 4 \ell$.
The angular and kinematic distributions in these processes will only
provide sensitivity for a discrimination between CP-even and CP-odd
properties if a possible CP-odd component $A$ of the new state couples
with sufficient strength to $WW$ and $ZZ$. However,
in many models of physics beyond the
SM there is no lowest-order coupling between a pseudoscalar $A$ and a pair
of gauge bosons, so that the $AVV$ coupling is
strongly suppressed compared to the coupling of the CP-even component.
In this case, the angular and kinematic distributions will show
essentially no deviations from the expectations of a pure CP-even
state, even if
the state had a sizable CP-odd component. The difference between a pure
CP-even state and a state that is a
mixture of CP-even and CP-odd components would rather manifest itself
as a reduction of the total rate.  However, such a reduction in rate could
be caused by other effects (and there could even be a
compensation with other contributions leading to an enhancement of the
rate). The couplings of the Higgs boson to
fermions offer a more democratic test of its CP nature,
since in this case the CP-even and odd components can have the same
magnitude.

Using the results of the $H \to ZZ^* \to 4 \ell$ channel to discriminate
between the distinct hypotheses of a pure CP-even and
a pure CP-odd state has led to a growing evidence against the pure
CP-odd hypothesis. Furthermore, first results of more general analyses
that take into account the possibility of a CP-admixture have been
obtained. As explained above, in the channels involving the $HVV$
coupling the effects of even a large CP-admixture can be heavily
suppressed as a consequence of a small coupling of the CP-odd component
to gauge bosons.

Concerning the determination of the couplings and the total width of the
observed particle, a modification of a
coupling will give rise to a change in the tensor structure and thus in
the CP properties. The determination of coupling and CP properties are
therefore closely related. For the analysis of the data taken
at the LHC so far, an ``interim framework'' (described in Section~\ref{sec:th_kappas}) has been
introduced for determining coupling properties.  In this framework, it is assumed that
only the overall coupling strength gets modified while the tensor
structure of the different couplings is the same as in the SM.
In this way, results for scale factors $\kappa_i$ (or, with fewer
assumptions, ratios of scale factors) have been obtained
under certain assumptions, as discussed in Section~\ref{sec:th_kappas}. At
the present level of accuracy, these analyses do not show a significant
deviation from the SM predictions (the SM case corresponds to $\kappa_i = 1$
for all scale factors). Projections for future accuracies of the scale
factors $\kappa_i$ have also been discussed. The reported projections~\cite{CMSprojections}
should be interpreted with some care given the fact that one of the
goals of the analyses during the next run of the LHC will be to go
beyond the ``interim framework'' used for the definition of the
$\kappa_i$ in order to obtain more general results with less theoretical
assumptions (see Section~\ref{sec:th_efflag}).

The self-coupling $HHH$ will be very difficult to access at the LHC, even
with the integrated luminosities obtainable at the high-luminosity upgraded LHC. The
prospects are even
%much
worse for the quartic self-coupling $HHHH$.

The total decay width for a light Higgs boson with a mass in the
observed range is not expected to be directly observable at the LHC.  The
predicted total width of the Standard Model Higgs boson is about 4~MeV, which is
several orders of magnitude smaller than the LHC experimental mass
resolution. Furthermore, as all LHC channels rely on the identification
of Higgs decay products, the total Higgs width cannot be measured in
those analyses without additional assumptions. More sensitive
constraints on the total width than the ones limited by the experimental
mass resolution can be expected from the analysis of interference
effects between signal an background~\cite{Dixon:2003yb,Martin:2012xc,Martin:2013ula,Dixon:2013haa,Caola:2013yja}.
%(discuss impact of the Higgs width on interference between
%signal and background in the $\gamma\gamma$ channel).
The limited access to the total width at the LHC implies that
without further assumptions only ratios of couplings can be determined
rather than the absolute values of the couplings.

In summary, while the experimental information obtained so far
about the signal
at 126~GeV is compatible with the expectations for the
Higgs boson of the SM, a large variety of other interpretations of
the discovered particle is also possible, corresponding to very
different underlying physics.
Some scenarios of this kind have been discussed in
the previous sections.   These include
the possibility that the observed
state is composite or that it is an admixture or shares properties with
other scalar states of new physics.
Extended Higgs sectors are an simplest alternatives to the SM Higgs boson.
In this context the signal at 126~GeV can be
interpreted as the lightest state of an extended Higgs sector, but
interpretations involving at least one lighter Higgs state below
126~GeV, having significantly suppressed couplings to gauge bosons as
compared to the SM case, are also possible. The sensitivity for
discriminating among the different possible interpretations correlates
with the achievable precision in confronting the experimental results
with the theory predictions.

\subsection{Experimental precision required to discriminate between
different possible interpretations}

If the observed signal at about 126~GeV is interpreted as the lightest
state of an extended Higgs sector, this interpretation
typically refers to
the decoupling region of the respective model, where the lightest state
has SM-like properties, while the heavier Higgs states decouple from the
gauge bosons of the SM. A concrete example of this kind is the MSSM,
where solely from the measured mass value of about 126~GeV important
constraints can be inferred
if the signal is interpreted in terms of the light CP-even
Higgs boson $h$. Requiring the prediction for the mass
of the light CP-even Higgs boson, $m_h$, to be compatible
with the measured mass value of about 126~GeV leads to a lower bound of
about 200~GeV on the mass of the CP-odd Higgs boson, $m_A$, if the
masses of the superpartners are in the TeV range~\cite{Carena:2013qia}.
The value of $m_A$ is therefore much larger than
$m_Z$ in this case, which corresponds to the decoupling region of the
MSSM. This implies that the properties of the state at 126~GeV are
expected to be SM-like, and that one generically would not have expected any
deviations from SM-like properties in the LHC measurements of the new
resonance carried out so far.

In the actual decoupling limit, the couplings of the light Higgs state to
SM particles are exactly the same as for the SM Higgs.  Of course,
even if the couplings to SM particles were very close to the
SM values, there could still be deviations from the SM predictions
in the branching ratios and the total width if there is a
significant branching ratio into invisible BSM particles. The
deviations of the Higgs couplings from the SM limit depend on the mass scale
of the new physics.
In general 2HDM-type models (including the case of the MSSM) one
typically expects deviations from the SM predictions at the percent level for
BSM particles in the TeV range.  In this context, one expects
the largest deviations to occur in the couplings to fermions
that get their mass from the Higgs doublet with the smaller vacuum
expectation value.  For example, within the MSSM this refers in particular
to the couplings to $b \bar b$
and $\tau^+\tau^-$)~\cite{Carena:2001bg,Baer:2013cma}.
The couplings to $W$ and $Z$ are usually less
affected by deviations from the decoupling limit. This can be
illustrated in a general 2HDM by the feature that the deviation of the
$HVV$ coupling ($V = W, Z$) from its SM values behaves quadratically in an expansion of the
deviation term, while the couplings to fermions behave linearly, as discussed in
Section~\ref{sec:decouplalign}. The loop-induced couplings $H\gamma\gamma$
and $Hgg$ can be significantly affected by the presence of
relatively light BSM particles in the loops. See
Refs.~\cite{Baer:2013cma,Gupta:2012mi} for a discussion of other electroweak
symmetry breaking scenarios.

Examples of the coupling patterns in specific models are discussed in
the following section.

\subsection{Examples of analyses in different models}

%%%%%%%%  Coupling Deviation in 2HDM %%%%%%%%%%%%%%
%[Gauge coupling Yukawa coupling]

The decays of the Higgs bosons in the 2HDM depend on the Type of Yukawa
interactions. In the decoupling/alignment limit where $\sin(\beta-\alpha)=1$,
all the tree-level couplings of $h$ coincide with those in the SM, as discussed
in Section~\ref{sec:decouplalign}.
However at the loop level, the effects of the additional Higgs bosons $H$, $A$ and $H^\pm$
can generate deviations in the $h$ couplings from the SM predictions in the alignment
limit if the masses of the additional Higgs boson are not significantly heavier than
the mass of the external particles.
When $\sin(\beta-\alpha)$ is slightly smaller than 1,  the couplings of $h$
to various SM particles can differ from the SM predictions by mixing effects in addition to
the loop corrections due to extra fields.
The gauge couplings $hVV$
($VV=WW$ and $ZZ$) are modified by the factor $\sin(\beta-\alpha)$ relative to the SM values,
and Yukawa interactions of $h$ differ from the SM predictions by the factors given in
Table~\ref{yukawa_tab}.  The pattern of deviation in Yukawa couplings strongly
depends on the Type of Yukawa Interactions in the 2HDM.
Therefore, we can basically separate the type of an extended Higgs sector
by precision measurements of the couplings of $h$ at the ILC.

For example, we discuss here the deviations from the SM of the Yukawa couplings of $h$ in 2HDMs
with a softly broken $\mathbb{Z}_2$ discrete symmetry, which is imposed to avoid
tree-level Higgs-mediated flavor changing neutral currents.
The Yukawa interactions of the SM-like Higgs boson ($h$)  are given by
\begin{align}
{\mathcal L}_\text{yukawa}^\text{2HDM}
=&-\sum_f\frac{m_f}{v}\xi_h^f{\overline f}fh,
\end{align}
where the scaling factors $\xi^f_h$ are displayed in Table.\ref{yukawa_tab}.
%
%Note that the factors are conveniently rewritten as
%\begin{align}
%c_\alpha/s_\beta &= \sin(\beta-\alpha) +\cos(\beta-\alpha)/\tan\beta, \\
%-s_\alpha/c_\beta &= \sin(\beta-\alpha) -\cos(\beta-\alpha) \cdot \tan\beta.
%\end{align}
%
%\begin{table}[h]
%\begin{center}
%\begin{tabular}{|c|c|c|c|c|}
%\hline & Type I & Type II & Type X & Type Y\\ \hline \hline
%$\xi_h^u$ & $\frac{c_\alpha}{s_\beta}$ &
%$\frac{c_\alpha}{s_\beta}$
%& $\frac{c_\alpha}{s_\beta}$ & $\frac{c_\alpha}{s_\beta}$\\
%$\xi_h^d$ & $\frac{c_\alpha}{s_\beta}$ & $-\frac{s_\alpha}{c_\beta}$ &
%$\frac{c_\alpha}{s_\beta}$
%& $-\frac{s_\alpha}{c_\beta}$\\
%$\xi_h^\ell$ & $\frac{c_\alpha}{s_\beta}$ &
%$-\frac{s_\alpha}{c_\beta}$ &
%$-\frac{s_\alpha}{c_\beta}$ & $\frac{c_\alpha}{s_\beta}$\\
%\hline
%\end{tabular}
%\end{center}
%\caption{Scaling factors in 2HDMs} \label{Tab.ScalingFactor}
%\end{table}
%
The scaling parameters for the gauge couplings to $h$ are given by
$\kappa_V=\sin(\beta-\alpha)$, while those for the Yukawa interactions are
given by $\kappa_f=\xi_h^f$ for $f=u,d, \ell$.
The pattern in deviations for each coupling is different among the
various Types of Yukawa interactions.

In Fig.~\ref{FIG:lhc-ilc}, the scale factors $\kappa_f=\xi_h^f$
in the 2HDM with a softly broken $\mathbb{Z}_2$ symmetry are plotted on the  $\kappa_\ell$-$\kappa_d$ plane
and the $\kappa_\ell$-$\kappa_u$ plane as a function of
$\tan\beta$ and $\kappa_V^{}=\sin(\beta-\alpha)$ with $\cos(\beta-\alpha) \le 0$.
The points and the dashed curves denote changes of $\tan\beta$ by steps of one.
The scaling factor $\kappa_V$ for the Higgs-gauge-gauge couplings
is taken to be $\kappa_V^2 = 0. 99, 0.95$ and $0.90$.
For $\kappa_V^{}=1$, all the scaling factors for the couplings of $h$ to SM particles become unity.
In Fig.~\ref{FIG:lhc-ilc}, the current LHC constraints and the expected LHC and ILC
sensitivities for $\kappa_d$ and $\kappa_\ell$ at 68.27 \%  C.L. are also shown.
For the current LHC constraints (LHC30), we take the numbers
from the universal fit in Eq.~(18) of Ref.~\cite{Giardino:2013bma},
\begin{align}
\epsilon_b = -0.23 \pm0.31, \quad \epsilon_\tau = +0.00 \pm 0.19, \quad
\rho = \begin{pmatrix} 1 & 0.45 \\ 0.45 & 1 \end{pmatrix}
\end{align}
where $\kappa_x = 1 +\epsilon_x$.
Those including $\epsilon_t$ are not provided in Ref.~\cite{Giardino:2013bma},
because the uncertainties are much larger than unity.
For the future LHC sensitivities (LHC300 and LHC3000),
the expectation numbers are taken from the Scenario 1
in Table. 1 of Ref.~\cite{CMS:2012zoa}.
The central values and the correlations are assumed to be the same as in LHC30 (although in practice,
the correlations would change at a collision energy of $\sqrt{s}=14$ TeV).
The ILC sensitivities are based on the numbers in Table. 2.6 in Ref.~\cite{Baer:2013cma}.
The same central values and no correlation are assumed for the plots of ILC sensitivity curves.
Therefore, by precisely measuring the couplings of $h$ at the ILC, one can
detect deviations from the SM.  Those deviations can then be used to discriminate among the various types of extended Higgs sectors
by fingerprinting predictions in each model with the precision data of the $h$ coupling measurement.

\begin{figure}[t!]
\centering
\includegraphics[width=7cm]{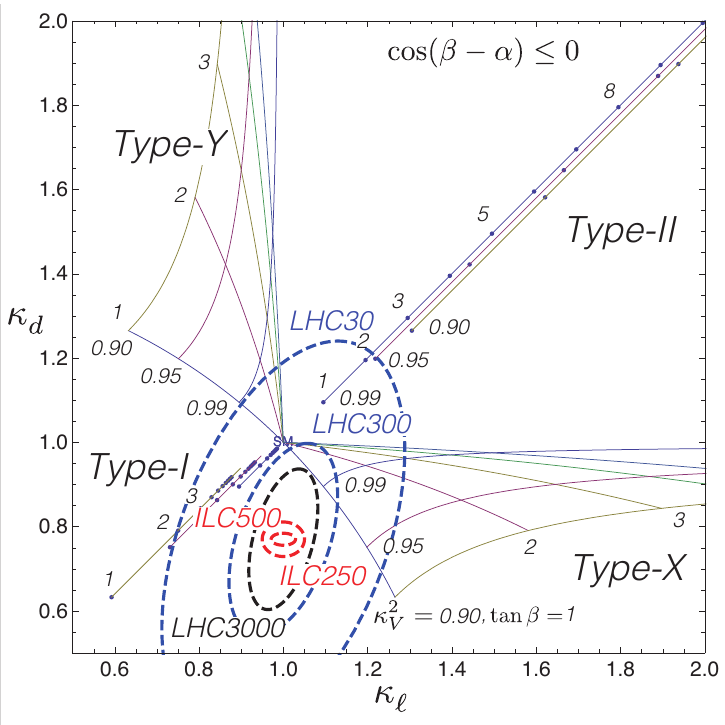}
\includegraphics[width=7cm]{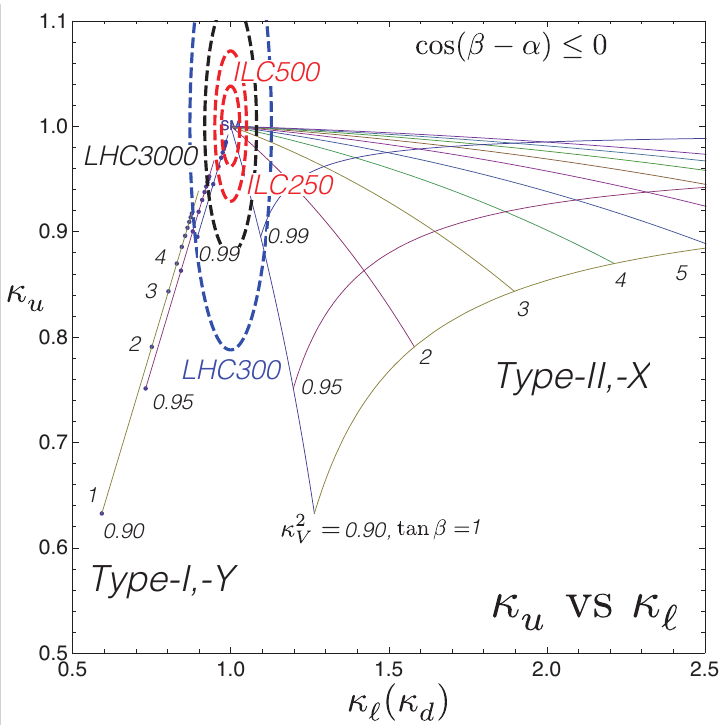}
\caption{
The deviation in $\kappa_f=\xi_h^f$ in the 2HDM with Type I, II, X and
Y Yukawa interactions are plotted
as a function of $\tan\beta=v_2/v_1$ and $\kappa_V^{}=\sin(\beta-\alpha)$
with $\cos(\beta-\alpha) \le 0$.
For the sake of clarity of illustration, several lines with $\kappa_x=\kappa_y$ are shifted
slightly so as to be separately visible.
The points and the dashed curves denote changes of $\tan\beta$ by one steps.
The scaling factor for the Higgs-gauge-gauge couplings is taken to be
$\kappa_V^2 = 0. 99, 0.95$ and $0.90$.
For $\kappa_V^{}=1$, all the scaling factors with SM particles become unity.
The current LHC constraints, expected LHC and ILC sensitivities on
(left) $\kappa_d$ and $\kappa_\ell$ and (right) $\kappa_u$ and $\kappa_\ell$
are added.\label{FIG:lhc-ilc}}
\end{figure}

%%%%%%% Exotic Higgs %%%%%%%%%%%

The behavior of the scaling factors depends on the structure of the extended Higgs sector.
For example, a model with mixing of the SM-like Higgs boson with a singlet Higgs field predicts a universal suppression
of the SM-like Higgs couplings, $\kappa_F^{} = \kappa_V^{} = \cos\alpha$,
where $\alpha$ is the mixing angle between the doublet field and the singlet field.
In contrast, $\kappa_F^{} \neq \kappa_V^{}$ in more complicated extended Higgs models
such as the 2HDM, the Georgi-Machacek model~\cite{Georgi:1985nv} and
doublet-septet model~\cite{Hisano:2013sn,Kanemura:2013mc}.
The scaling factors for these models are  summarized in
Table.~\ref{Tab.ScalingFactor2} .
%In these models, Yukawa interactions $hf\bar f$ are universally
%changed from the SM values.
Note that in exotic models with higher representation
scalar fields such as  the Georgi-Machacek model and doublet-septet model,
$\kappa_V$ can be greater than 1 as already discussed in Section~\ref{Anupperbound},
which is the clear signature of these exotic Higgs sectors.
\begin{table}[h!]
\begin{center}
\begin{tabular}{|c|c|c|c|c|}
\hline & Doublet-Singlet & 2HDM-I & Georgi-Machacek & Doublet-Septet\\ \hline \hline
$\tan\beta$ & --- & $v_2/v'_2$ & $v_2/(2\sqrt2\, v_3)$ & $v_2/(4\, v_7)$\\
$\xi_h^f$ & $c_\alpha$ & $\frac{c_\alpha}{s_\beta}$ & $\frac{c_\alpha}{s_\beta}$ & $\frac{c_\alpha}{s_\beta}$\\
$\xi_h^V$ & $c_\alpha$ & $s_{\beta-\alpha} (= s_\beta c_\alpha - c_\beta s_\alpha)$ & $s_\beta c_\alpha -\tfrac{2\sqrt6}3 c_\beta s_\alpha$ & $s_\beta c_\alpha-4 c_\beta s_\alpha$\\
\hline
\end{tabular}
\end{center}
\caption{Scaling factors in models with universal Yukawa coupling
constants.} \label{Tab.ScalingFactor2}
\end{table}

In Fig.~\ref{FIG:exotic}, the predictions for the scale factors of the universal Yukawa coupling $\kappa_F$ and
the gauge coupling $\kappa_V$ are plotted in the doublet-singlet model, the Type-I 2HDM, the
Georgi-Machacek model and the doublet-septet model for each set of $\tan\beta$ and $\alpha$.
The current LHC constraints, expected LHC and ILC
sensitivities for $\kappa_F$ and $\kappa_V$ at 68.27 \%  C.L. are also shown.
By precision measurements of $\kappa_V$ and $\kappa_F$ one can discriminate among exotic models.
The central values of the contours correspond to the the SM prediction.
For the contours for LHC~300 and LHC~3000, $\kappa_\tau$ is used for the scaling factor
of the Yukawa coupling, which exhibits the best sensitivity among the fermionic
channels.
For the contours for ILC250 and ILC500, the scaling factors are chosen as
$(\kappa_V, \kappa_F)=(\kappa_Z, \kappa_b)$ without making combinations.

\begin{figure}[t]
\centering
\includegraphics[width=9cm]{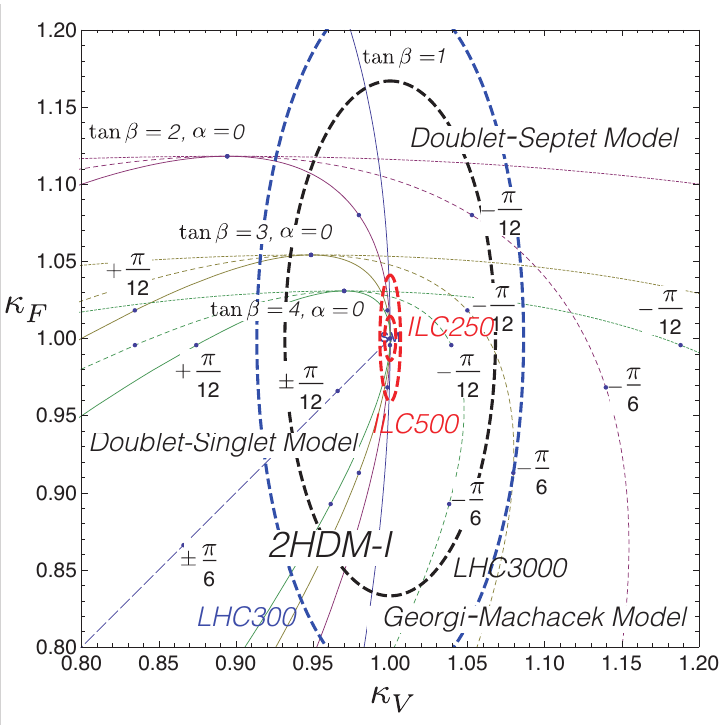}
\caption{The scaling factors in models with universal Yukawa couplings.}
\label{FIG:exotic}
\end{figure}

%ccccccccccccccc END cccccccccccccc
%%%%%%%%%%%%% The following subsection was Added in 5. August by Shinya
%\subsection{The mass scale of the second Higgs boson from the precision measurement of
%$\kappa_V$}

When $\kappa_V$ is slightly different from unity, we can obtain the upper bound on
the mass scale of the second Higgs boson.
Extended Higgs sectors usually contain additional  mass parameters  which are
irrelevant to electroweak symmetry breaking.
The mass of the second Higgs boson then is a free parameter
and can be taken to be very heavy, so that all the couplings of the lightest Higgs boson $h$
coincide with the SM value at tree level.
Although we cannot predict the mass of the second Higgs boson,
when  the coupling $hVV$ is slightly different from the SM prediction,
the upper bound on the heavy Higgs mass scale can be theoretically obtained as a function of $\kappa_V$
by using the properties of
the SM-like Higgs boson and the constraint from vacuum stability and perturbative unitarity.

In the case of the 2HDM with a softly-broken discrete $\mathbb{Z}_2$ symmetry,
the vacuum stability bound is given by~\cite{Deshpande:1977rw}
\begin{align}
\lambda_1>0\,,\qquad \lambda_2>0\,,\qquad \sqrt{\lambda_1\lambda_2}+\lambda_3+\text{min}\bigl\{0\,,\,\lambda_4+\lambda_5\,,\,\lambda_4-\lambda_5\bigr\} > 0.
\end{align}
The unitarity bounds are obtained by imposing the conditions,
$|x_i|<\half$,
where the $x_i$ are the eigenvalues of the $s$-wave amplitude matrix for the elastic scattering of two scalar states.
They are calculated in Ref.~\cite{Kanemura:1993hm},
\begin{align}
x_1^\pm &=  \frac{1}{16\pi}
\left[\tfrac{3}{2}(\lambda_1+\lambda_2)\pm\sqrt{\tfrac{9}{4}(\lambda_1-\lambda_2)^2+(2\lambda_3+\lambda_4)^2}\right],\\
x_2^\pm &=
\frac{1}{16\pi}\left[\tfrac{1}{2}(\lambda_1+\lambda_2)\pm\sqrt{\tfrac{1}{4}(\lambda_1-\lambda_2)^2+\lambda_4^2}\right],\\
x_3^\pm &= \frac{1}{16\pi}\left[
\tfrac{1}{2}(\lambda_1+\lambda_2)\pm\sqrt{\tfrac{1}{4}(\lambda_1-\lambda_2)^2+\lambda_5^2}
\right],\\
x_4 &= \frac{1}{16\pi}(\lambda_3+2\lambda_4-3\lambda_5),\qquad
x_5 = \frac{1}{16\pi}(\lambda_3-\lambda_5),\\
x_6 &= \frac{1}{16\pi}(\lambda_3+2\lambda_4+3\lambda_5),\qquad
x_7 = \frac{1}{16\pi}(\lambda_3+\lambda_5),\qquad
x_8 = \frac{1}{16\pi}(\lambda_3+\lambda_4).
\end{align}

\begin{figure}[t]
\centering
\includegraphics[width=8cm]{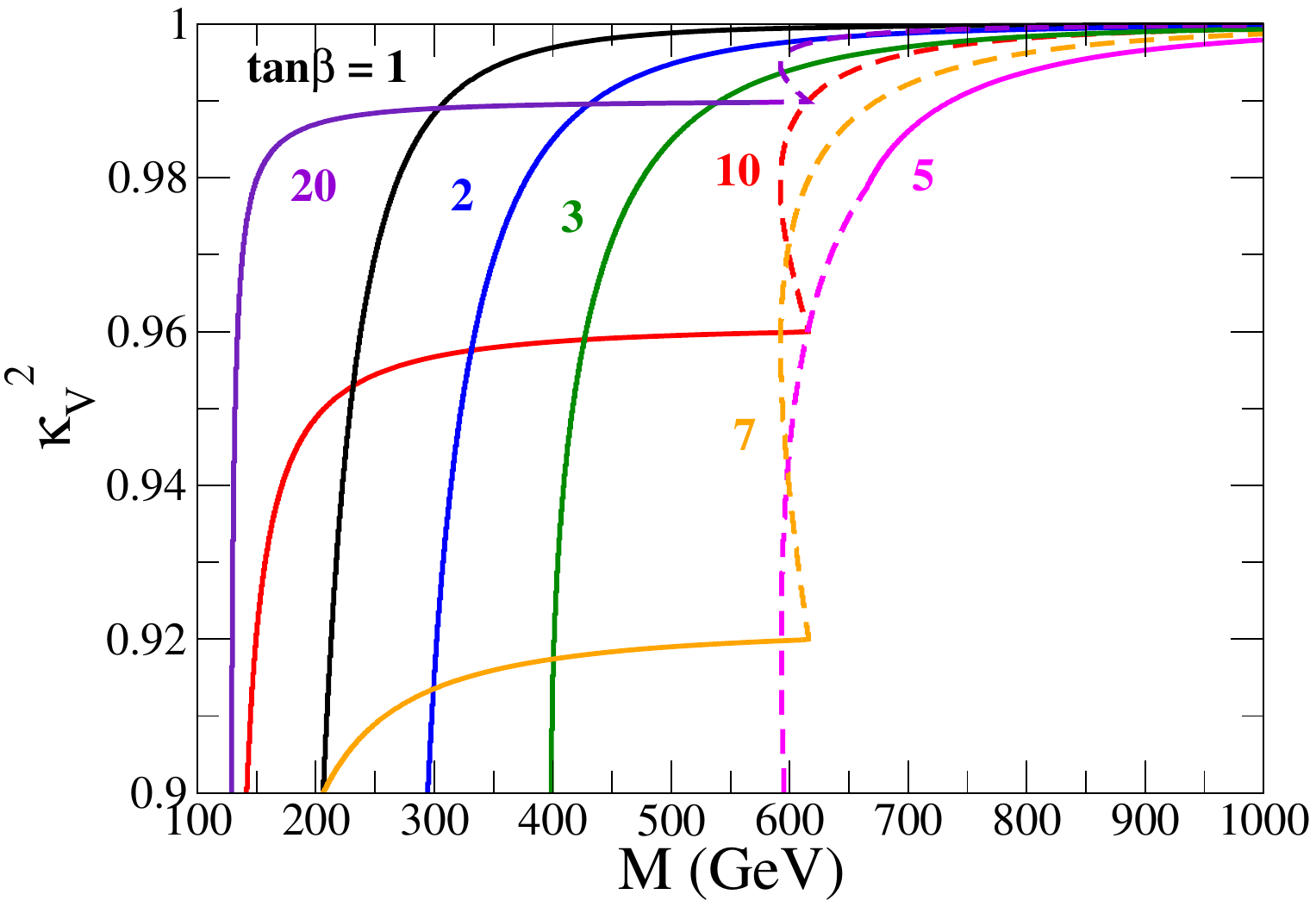}
\caption{Regions inside the curves are allowed by the constraints from unitarity and vacuum
stability in the ($M\,,\,\kappa_V^2$) plane
for each fixed value of $\tan\beta$.
We take $M=m_A=m_H=m_{H^+}$.
The solid and dashed curves correspond to the boundaries of the exclusion regions due to
vacuum stability and unitarity, respectively.
%The solid and dashed curves respectively show the
%changing points whether the region is allowed or excluded by the vacuum stability and unitarity. }
}
\label{FIG:kv_M}
\end{figure}
\begin{figure}[t]
\centering
\includegraphics[width=7cm]{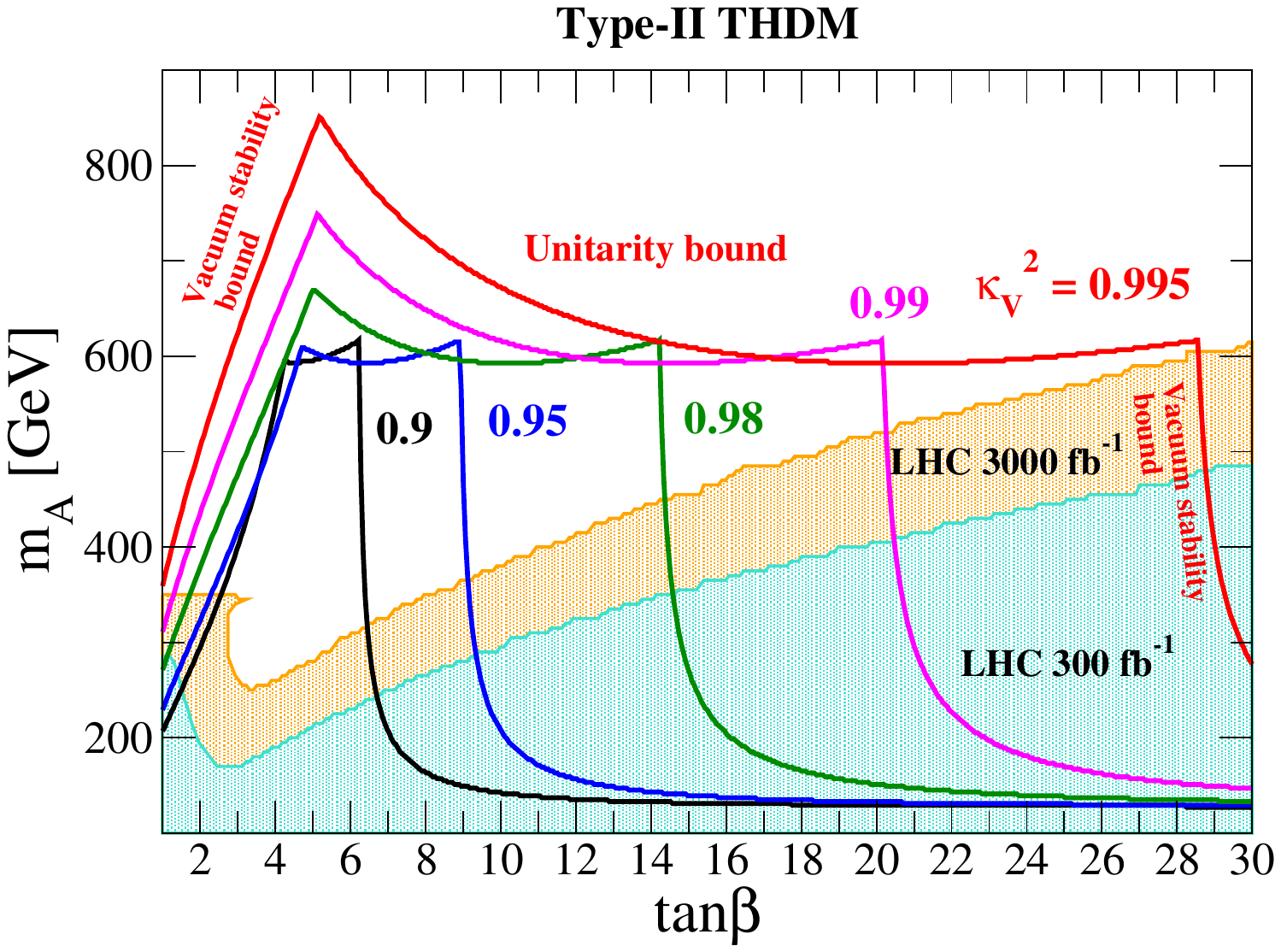}
\includegraphics[width=7cm]{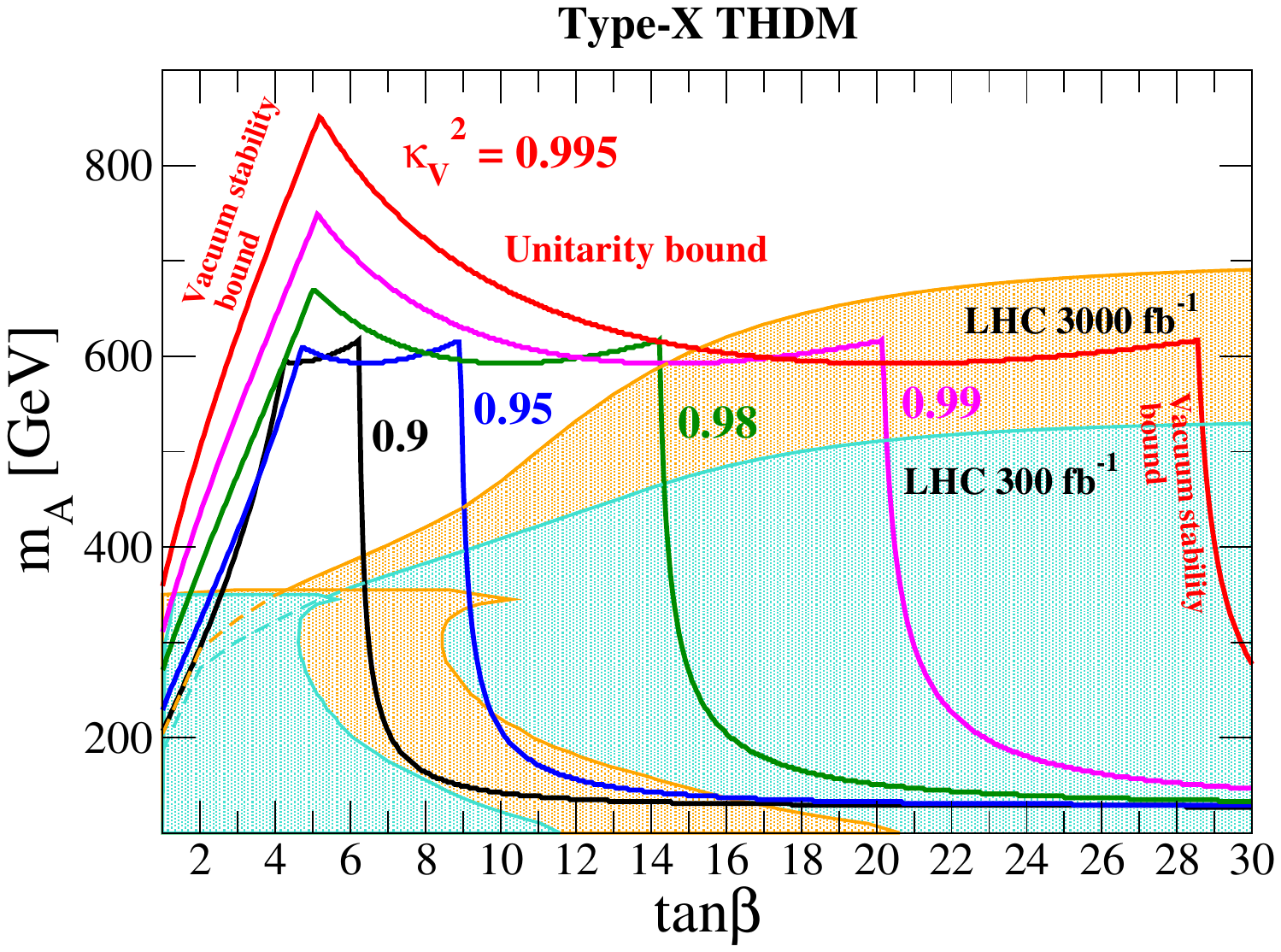}
\caption{
Regions below the curves are allowed by the constraints from unitarity and vacuum stability
in the ($\tan\beta\,,\,m_A$) plane for each fixed value of $\kappa_V^2$ for $M=m_A=m_H=m_{H^+}$
in the Type II and Type X 2HDMs.
Expected excluded areas of the parameter space are also shown by blue (orange) shaded regions
from the gluon fusion production and associate production of $A$ and $H$ with bottom quarks and
tau leptons at the LHC with the collision energy to be 14 TeV with an integrated luminosity of
300 fb$^{-1}$ (3000 fb$^{-1}$). }\label{FIG:tanb_M}
\end{figure}

In Fig.~\ref{FIG:kv_M}, an upper limit on the mass of the second lightest Higgs boson is shown
in the 2HDM with a softly broken $\mathbb{Z}_2$ discrete symmetry.
Regions on the left side of each curve are the allowed regions by the constraints from unitarity and vacuum
stability in the ($M\,,\,\kappa_V^2$) plane for each fixed value of $\tan\beta$.
We take $M=m_A=m_H=m_{H^+}$.  If the equality of the heavier Higgs masses is relaxed, then
the bound on the mass of the second lightest Higgs boson is typically stronger.
The solid and dashed curves correspond to the boundaries of the exclusion regions due to
vacuum stability and unitarity, respectively.

In Fig.~\ref{FIG:tanb_M}, the upper bound on the mass scale of the additional Higgs bosons
are shown as a function of $\tan\beta$ for each fixed value of
$\kappa_V^2=\sin^2(\beta-\alpha)$ under the constraints of perturbative unitarity and vacuum
stability~\cite{Kanemura:1993hm,Deshpande:1977rw}.
The expected discovery regions at LHC with 300 fb$^{-1}$ and 3000 fb$^{-1}$ are also
shown assuming a Type-II Yukawa interaction and a Type-X Yukawa interaction.
These discovery regions are obtained from the analysis of the tau lepton decay
of  $H$ and $A$ from gluon fusion production processes and
associate production processes with the bottom quarks and tau leptons,
\begin{align}
&gg \to \phi^0 \to \tau^+\tau^-,\\
&gg \to b\bar{b}\phi^0 \to b\bar{b}\tau^+\tau^-, \\
&gg \to \tau^+ \tau^-\phi^0 \to \tau^+\tau^- \tau^+ \tau^-,
\end{align}
where $\phi^0$ represents $H$ or $A$.
The cross section is obtained by rescaling the values of gluon fusion cross section
for $h_{\text{SM}}$ at 14 TeV from Ref.~\cite{Dittmaier:2011ti}, and the signal and background
analysis in the MSSM given in Ref.~\cite{Aad:2009wy} is used. The signal significance $\mathcal{S}$ is
computed by rescaling the results to the case of the 2HDMs, and the expected excluded
regions are obtained by requiring that $\mathcal{S} > 2$.
%
%If we relax the assumption of equal masses for the additional Higgs bosons, then the bounds are obtained for the
%mass scale of the second lightest Higgs boson.

For moderate values of $\tan\beta$, it
may not be possible to detect the second lightest Higgs boson of the 2HDM at the LHC.
In this case, it is important to determine the mass scale of the second Higgs boson
in an indirect way.  For example, it is possible to measure
$\kappa_V$ at the ILC with a precision at the one percent level or better.
If $\kappa_V$ is found to be slightly different from unity at the ILC at the percent level,
then the upper bound on the heavy Higgs mass scale can be obtained from perturbative unitarity.
If the deviation is a few percent, these upper bounds are above the discovery reach
at LHC with 300 fb$^{-1}$ in wide region of $\tan\beta$ in both Type-II and Type-X 2HDMs.
At the LHC with 3000 fb$^{-1}$, regions with relatively large $\tan\beta$ can be surveyed.
The ILC with a center-of-mass energy of 1~TeV can directly survey the extra Higgs bosons
with masses less than 500 GeV for relatively low $\tan\beta$ regions,
where the LHC cannot detect them.

\subsection{The $hhh$ coupling and electroweak baryogenesis}

How accurately we should measure the $hhh$ coupling?

\begin{figure}[b!]
\begin{center}
 \includegraphics[width=7cm]{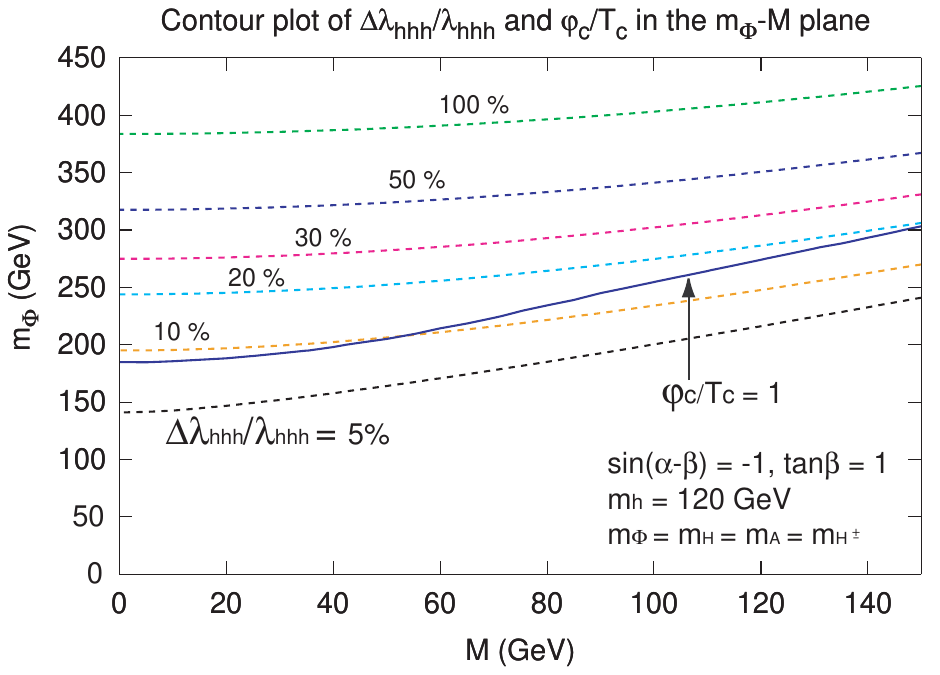}
  \caption{The region of strong first order phase transition
 ($\varphi_c/T_c>1$) required for successful electroweak baryogenesis
 and the contour plot of the deviation in the triple Higgs boson
 coupling from the SM prediction~\cite{Kanemura:2004ch}, where
 $m_\Phi$ represents the common mass of $H$, $A$ and $H^\pm$ and
 $M$ is the soft-breaking mass of the $\mathbb{Z}_2$ discrete symmetry in the Higgs
 potential. %Eq.~(\ref{eq:BigM}).
 }
  \label{fig:ewbg_hhh}
 \end{center}
\end{figure}

Given a sufficient accuracy of the $hhh$ coupling measurement,
one can test certain scenarios of electroweak baryogenesis.
The matter anti-matter asymmetry in our Universe cannot be explained within
the Standard Model of particle physics.  In particular,
the baryon-to-photon ratio is given by
$n_b/n_\gamma \simeq \text{(5.1---6.5)} \times 10^{-10}$
at 95 \% CL~\cite{FieldsPDG}, where $n_b$ is the difference in the number density
between baryons and anti-baryons and $n_\gamma$ is the photon number density.
In order to generate the baryon asymmetry from a baryon number symmetric
initial state, three conditions first given by Sakharov must be satisfied~\cite{Sakharov:1967dj}.
The electroweak gauge theory can satisfy these conditions
by employing sphaleron processes at high temperatures, C and CP violation in
the theory and the strongly first order phase transition
of the electroweak symmetry.
The mechanism of baryogenesis using such a scenario
is called electroweak baryogenesis~\cite{Kuzmin:1985mm,Cohen:1993nk,Morrissey:2012db},
which directly relates to the Higgs sector.
Electroweak baryogenesis is especially attractive because of its testability
at collider experiments.

In the SM, this scenario is already excluded by the data~\cite{Cohen:1993nk,Morrissey:2012db}.
The simplest viable model is the 2HDM~\cite{Fromme:2006cm},
which provides additional CP violating phases and a sufficiently strong first order
electroweak phase transition compatible with the 126 GeV SM-like
Higgs boson due to the loop effect of the extra Higgs bosons.
One of the interesting phenomenological predictions
for such a scenario is a large deviation in the
triple Higgs boson coupling~\cite{Grojean:2004xa,Kanemura:2004ch}.
The requirement of a sufficiently strong first order phase transition
results in a large deviation in the triple Higgs boson coupling
as seen in Fig.~\ref{fig:ewbg_hhh}.
This suggests that the electroweak baryogenesis scenario can be
tested by measuring the $hhh$ coupling with a 10\% accuracy.

An analysis of the first order phase transition has
also been performed in Ref.~\cite{Grojean:2004xa} using a simple Higgs
potential with higher order operators where similar
deviations in the $hhh$ coupling are predicted.
Moreover, the correlation between the condition of strong first order
phase transition and the deviation in the $hhh$ coupling from the SM prediction
can be seen in various extended Higgs models~\cite{Aoki:2008av,Kanemura:2012hr}.
Therefore, measuring the $hhh$ coupling accurately is a useful
probe of the class of models of electroweak baryogenesis.

The measurement of the $hhh$ coupling  for $m_h \simeq 126$ GeV
is very challenging at the LHC and even at the high luminosity upgrade of the LHC.
At the ILC, the $hhh$ coupling can be measured via
$e^+e^- \rightarrow Zhh$
and $e^+e^- \rightarrow hh \nu\bar\nu$.  As indicated in Chapter~\ref{sid:chapter_summary},
for the combined data taken at the ILC with $\sqrt{s}=250$ with 1150 fb$^{-1}$ and $500$ GeV with 1600 fb$^{-1}$,
the $hhh$ coupling can be measured with an accuracy of about 46\%.
By adding additional data from a run of $\sqrt{s}=1$ TeV with 2500 fb$^{-1}$,
one can determine the $hhh$ coupling to an accuracy of about 13\%.
Therefore, the scenario for electroweak baryogenesis would be
testable by measuring the triple Higgs boson coupling at the ILC.

\subsection{Value added by the ILC Higgs program post-LHC}

What will be the value added by the ILC Higgs program in
the context of the current and future results from the LHC?
We provide a qualitative assessment of this question in this section.
%will be the main topic of the following sections. We %therefore
%only briefly mention some qualitative features here.

The ILC will provide crucial information for identifying the underlying nature
of electroweak symmetry breaking. In particular,
high-precision measurements of the properties of
the signal observed at 126~GeV will be performed at the ILC.
For example, for the Higgs couplings to gauge bosons and
fermions, one typically expects an order of magnitude
improvement from the ILC measurements as compared to the
ultimate LHC precision. This expected accuracy provides a high
sensitivity for discriminating among possible realizations of
electroweak symmetry breaking, such as effects of an extended Higgs
sector, of additional states of new physics or deviations in the
couplings from the respective SM values that would occur in case the observed
signal is a composite state.

Besides those quantitative improvements, the ILC Higgs program will also
give rise to crucial qualitative improvements in studying the properties
of the observed signal. In particular, the
Higgsstrahlung process $e^+e^-\to ZH$ provides the unique opportunity
to make absolute measurements of Higgs couplings in a model-independent
way. The
clean experimental environment and the relatively low SM cross sections
for background processes allow  $e^+e^-\to ZH$
events to be selected based on the identification of two oppositely
charged leptons with invariant mass consistent with $m_Z$. The remainder
of the event, i.e.\ the Higgs decay, is
not considered in the event selection.

Because only the properties of the dilepton system are used in the
selection, this decay-mode independent measurement provides an absolute
determination of the Higgsstrahlung cross section.
Subsequently, by identifying the individual final states for different
Higgs and $Z$ decay modes, absolute measurements of the Higgs boson
branching fractions can be made.  Moreover,
the ILC provides a unique sensitivity to
invisible decay modes of the observed signal.
%Higgs boson.
%extending down to a branching
%ratio into invisible states as low as $1\%$.
If dark matter consists of a particle (or more than one) with
a mass that is less than half the mass of the observed
signal, there could be a significant branching ratio of the discovered
state at 126 GeV into a pair of dark matter particles.
If an invisible decay mode is detected, this could be the first
hint for the production of dark matter in collider experiments.
%This possibility can be studied in detail at the LC for all Higgs bosons
%within its kinematic reach.

Furthermore, the absolute measurements of the Higgs boson branching ratios
imply that the ILC can provide an absolute measurement of the
total width a model-independent way. This can be accomplished
using the relationship
between the total and partial decay widths, for example
\begin{equation}
    \Gamma_H = \frac{\Gamma(H \to W W^*)}{{\mathrm BR}(H \to W W^*)} ,
\end{equation}
where $\Gamma_H$ denotes the total width. The partial width
$\Gamma(H \to W W^*)$ can be determined from
the measurement of the $HWW$ coupling obtained from the
fusion process $e^+e^- \to H \nu\bar\nu$.
When combined with the direct measurement of ${\mathrm BR}(H \to W W^*)$,
the total Higgs width can be inferred.

The measurement of the Higgs trilinear self-coupling is of particular
importance, since it provides direct access to the form of the Higgs
potential that gives rise to electroweak symmetry breaking.
This measurement is therefore
crucial for experimentally establishing the scalar dynamics of electroweak symmetry breaking.
As mentioned above, the measurement of the Higgs trilinear self-coupling
will be extremely challenging at the LHC even with 3000\,fb$^{-1}$ of
data. This is due to the complexity of the final state and the
smallness of the cross sections. At the ILC the processes
$e^+e^- \to ZHH$ and $e^+e^- \to HH \nu\bar\nu$ provide sensitivity
to the trilinear self-coupling given sufficiently high luminosity.

Besides a high-precision determination of the properties of the observed
signal at 126~GeV, the ILC has also a high physics potential in the
direct search for additional states of an extended Higgs sector.
The search capacity of the ILC for the pair production of heavy Higgs states is
expected to be close to the kinematic limit of $\half\sqrt{s}$. An extended
Higgs sector could however also contain at least one state that is
{\em lighter\/} than 126~GeV with significantly
suppressed couplings to gauge bosons as compared to the case of a
SM-like Higgs.   The search for such a light Higgs state
in the mass range between 60~GeV and 100~GeV is very challenging
in the standard search channels at the LHC.  In contrast, at the
ILC there will be a high
sensitivity for probing scenarios of this kind.

%Additional Higgs states may be heavier but
%also lighter than the state at 126 GeV. The mass region below 100 GeV is
%essentially uncovered at the LHC so far.

\chapter{ILC Accelerator Parameters and Detector Concepts \label{sid:chapter_accelerator_detector}}
\section{ILC Accelerator Parameters} \label{sid:Accelerator_Detector:sec:accelparams}

\subsection{TDR Baseline ILC 250 - 500 GeV}

The International Linear Collider (ILC) is a high-luminosity linear electron-positron
collider based on \SI{1.3}{GHz} superconducting 
radio-frequency (SCRF) accelerating technology. 
Its center-of-mass-energy range is \SIrange{200}{500}{\GeV} (extendable
to \SI{1}{\TeV}). A schematic view of the accelerator complex,
indicating the location of the major sub-systems, is shown in \Fref{fig:tdres:ilcschematic}:

\thisfloatsetup{floatwidth=\textwidth}
\begin{figure}[htb]
  \includegraphics[trim=8 14 0 12,clip,width=\hsize]{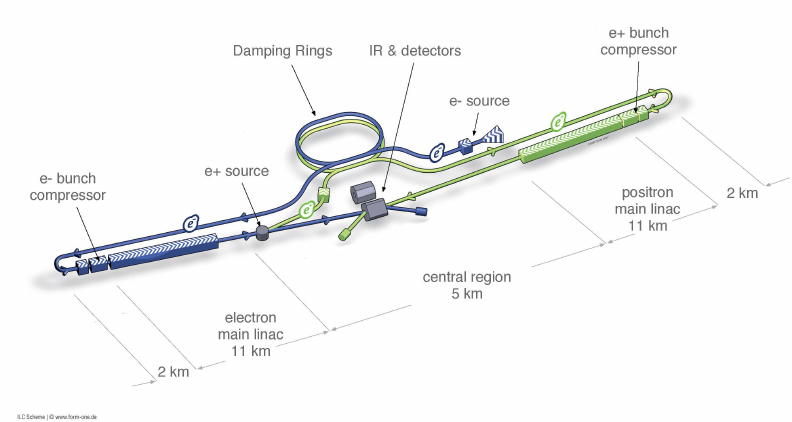}
  \caption[Schematic layout  of the ILC]
    {Schematic layout  of the ILC, indicating all the major subsystems (not to scale).}
  \label{fig:tdres:ilcschematic}
\end{figure}

\begin{itemize}

 \item a polarized electron source based on a photocathode DC gun;
 
 \item a polarized positron source in which positrons are obtained
 from electron-positron pairs by converting high-energy photons
 produced by passing the high-energy main electron beam through an
 undulator;

 \item \SI{5}{\GeV} electron and positron damping rings (DR) with a
 circumference of \SI{3.2}{\km}, housed in a common tunnel;

 \item beam transport from the damping rings to the main linacs,
 followed by a two-stage bunch-compressor system prior to injection
 into the main linac;

 \item two \SI{11}{\km} main linacs, utilizing \SI{1.3}{GHz} SCRF
 cavities operating at an average gradient
 of \SI{31.5}{\mega\volt/\meter}, with a pulse length
 of \SI{1.6}{\milli\second};

 \item two beam-delivery systems, each \SI{2.2}{\km} long, which bring
 the beams into collision with a \SI{14}{\milli\radian} crossing
 angle, at a single interaction point which can be occupied by two
 detectors in a so-called ``push-pull'' configuration.

 \end{itemize}

The total footprint of the ILC complex is $\sim$~\SI{31}{\km}
long. The electron source, positron source (including an independent
low-powered auxiliary source), and the electron and positron damping
rings are centrally located around the interaction region (IR) in the
Central Region. The damping-ring complex is displaced laterally to
avoid interference with the detector hall. The electron and positron
sources themselves are housed in the same (main accelerator) tunnels
as the beam-delivery systems, which reduces the overall cost and size
of the central-region underground construction.

The top-level parameters for the baseline operational range of
center-of-mass energies from 250 to \SI{1000}{\GeV} were set in close
discussion with the physics community that will exploit the ILC. The
baseline performance requirements thus obtained have been optimized
with respect to cost, physics performance and risk. All have been
either directly demonstrated, or represent justifiable extrapolations
from the current state of the art. Table~\ref{tab:tdres:prms} shows the parameters for
several center-of-mass energies, including possible upgrades and
staging.

The parameters in \Tref{tab:tdres:prms} represent relatively conservative operating
points resulting from optimization subject to the constraints imposed
by the various accelerator sub-systems. For example, the bunch charge,
bunch spacing and the total number of bunches in the damping rings are
limited by various instability thresholds (most notably the electron
cloud in the positron ring), realistic rise-times for the injection
and extraction kickers, and the desire to minimize the circumference
of the rings. Secondly, the maximum length of the beam pulse is
constrained to $\sim$~\SI{1.6}{ms}, which is routinely achieved in the available
\SI{1.3}{GHz} \SI{10}{MW} multi-beam klystrons and modulators. The beam current is
further constrained by the need to minimize the number of klystrons
(peak power) and higher-order modes (cryogenic load and beam
dynamics). Dynamic cryogenic load (refrigeration) is also a cost
driver, which limits the repetition rate of the machine. Thirdly, both
the electron and positron sources constrain the achievable beam
current and total charge: For the laser-driven photocathode polarized
electron source, the limits are set by the laser; for the
undulator-based positron source, the limits are set by the power
deposition in the photon target. The beam pulse length is further
constrained by the achievable performance of the warm RF capture
sections (both sources). Finally, at the interaction point,
single-bunch parameters are limited by the strong beam-beam effects
and requirements on both the beam-beam backgrounds and beam stability.

\thisfloatsetup{floatwidth=0.95\textwidth}
\begin{landscape}
\begin{table}[p] 
\caption{Summary table of the \SIrange{250}{500}{\GeV} baseline and luminosity and energy upgrade parameters. 
Also included is a possible 1st stage \SI{250}{\GeV} parameter set (half the original main linac length)}.
\setlength{\tabcolsep}{8pt}
\begin{tabular}{p{5.5cm} llcccccccccc }
\toprule
 &  &  & \multicolumn{3}{c}{Baseline \SI{500}{\GeV} Machine} &  & 1st Stage &  & L Upgrade &  & \multicolumn{2}{c}{$E\sub{CM}$ Upgrade} \\
\tcmidrule{4-6}
\tcmidrule{8-8}
\tcmidrule{10-10}
\tcmidrule{12-13}
 &  &  &  &  &  &  &  &  & & & A & B \\ 
Center-of-mass energy & $E\sub{CM}$  & \si{\GeV} & 250 & 350 & 500 &  & 250 &  & 500 &  & 1000 & 1000 \\
\midrule
Collision rate & $f\sub{rep}$ & \si{\Hz} & 5 & 5 & 5 &  & 5 &  & 5 &  & 4 & 4 \\
Electron linac rate & $f\sub{linac}$  & \si{\Hz} & 10 & 5 & 5 &  & 10 &  & 5 &  & 4 & 4 \\
Number of bunches & $n\sub b$ &  & 1312 & 1312 & 1312 &  & 1312 &  & 2625 &  & 2450 & 2450 \\
Bunch population & $N$  & $\times$\num{e10}  & 2.0 & 2.0 & 2.0 &  & 2.0 &  & 2.0 &  & 1.74 & 1.74 \\
Bunch separation & $\Delta t\sub b$ & \si{\ns} & 554 & 554 & 554 &  & 554 &  & 366 &  & 366 & 366 \\
Pulse current & $I\sub{beam}$  & \si{\mA} & 5.8 & 5.8 & 5.8 &  & 5.8 &  & 8.8 &  & 7.6 & 7.6 \\
 \\
Main linac average gradient & $G\sub a$ & \si{\MV\per\metre} & 14.7 & 21.4 & 31.5 & & 31.5 & & 31.5 & & 38.2 & 39.2 \\
Average total beam power & $P\sub{beam}$ & \si{\MW} & 5.9 & 7.3 & 10.5 &  & 5.9 &  & 21.0 &  & 27.2 & 27.2 \\
Estimated AC power & $P\sub{AC}$ & \si{\MW} & 122 & 121 & 163 &  & 129 &  & 204 &  & 300 & 300 \\
 \\
RMS bunch length & $\sigma\sub z$  & \si{\mm} & 0.3 & 0.3 & 0.3 &  & 0.3 &  & 0.3 &  & 0.250 & 0.225 \\
Electron RMS energy spread & $\Delta p/p$ & \% & 0.190 & 0.158 & 0.124 &  & 0.190 &  & 0.124 &  & 0.083 & 0.085 \\
Positron RMS energy spread & $\Delta p/p$ & \% & 0.152 & 0.100 & 0.070 &  & 0.152 &  & 0.070 &  & 0.043 & 0.047 \\
Electron polarization &$P\sub{-}$ & \% & 80 & 80 & 80 &  & 80 &  & 80 &  & 80 & 80 \\
Positron polarization & $P\sub{+}$ & \% & 30 & 30 & 30 &  & 30 &  & 30 &  & 20 & 20 \\
 \\
Horizontal emittance & $\gamma\epsilon\sub x$ & \si{\um} & 10 & 10 & 10 &  & 10 &  & 10 &  & 10 & 10 \\
Vertical emittance & $\gamma\epsilon\sub y$ & \si{\nm} & 35 & 35 & 35 &  & 35 &  & 35 &  & 30 & 30 \\
 \\
IP horizontal beta function & $\beta\sub x^{*}$ & \si{\mm} & 13.0 & 16.0 & 11.0 &  & 13.0 &  & 11.0 &  & 22.6 & 11.0 \\
IP vertical beta function & $\beta\sub y^{*}$ & \si{\mm} & 0.41 & 0.34 & 0.48 &  & 0.41 &  & 0.48 &  & 0.25 & 0.23 \\
 \\
IP RMS horizontal beam size & $\sigma\sub x^{*}$ & \si{\nm} & 729.0 & 683.5 & 474 &  & 729 &  & 474 &  & 481 & 335 \\
IP RMS vertical beam size & $\sigma\sub y^{*}$ & \si{\nm} & 7.7 & 5.9 & 5.9 &  & 7.7 &  & 5.9 &  & 2.8 & 2.7 \\
 \\
Luminosity & $L$ & $\times$\SI{e34}{\cm^{-2}\s^{-1}} & 0.75 & 1.0 & 1.8 &  & 0.75 &  & 3.6 &  & 3.6 & 4.9 \\
Fraction of luminosity in top 1\% & $L\sub{0.01}/L$  &  & 87.1\% & 77.4\% & 58.3\% &  & 87.1\% &  & 58.3\% &  & 59.2\% & 44.5\% \\
Average energy loss & $\delta\sub{BS}$ &  & 0.97\% & 1.9\% & 4.5\% &  & 0.97\% &  & 4.5\% &  & 5.6\% & 10.5\% \\
Number of pairs per bunch crossing & $N\sub{pairs}$  & $\times$\num{e3}  & 62.4 & 93.6 & 139.0 &  & 62.4 &  & 139.0 &  & 200.5 & 382.6 \\
Total pair energy per bunch crossing & $E\sub{pairs}$  & \si{\TeV} & 46.5 & 115.0 & 344.1 &  & 46.5 &  & 344.1 &  & 1338.0 & 3441.0 \\
\bottomrule
\end{tabular}
\label{tab:tdres:prms}
\end{table}
\end{landscape}

\subsection{Luminosity and Energy Upgrade Options}

The ILC TDR outlines two upgrades. One is the luminosity upgrade to double the average beam power by adding RF to the linacs.  The
second is  the energy 
upgrade to double the center of mass energy to 1 TeV by extending the main linacs. The latter will require substantial additional tunnel 
construction and the upgraded 1 TeV machine will consume more electrical power. The TDR also describes a possible first 
stage 250 GeV center of mass energy \"Higgs Factory\". These options are included in \Tref{tab:tdres:prms}.

Two additional options should be considered~\cite{Ross:2013aa}. The first is operation at 250 GeV center of mass energy following the baseline luminosity upgrade.  
The second is the potential for 
operation at 1.5 TeV center of mass energy. The latter is briefly mentioned in the ILC cover letter submission to the European Strategy 
Preparatory Group~\cite{Barish:2012es}
Here we only consider the luminosity upgrade at 250 GeV center of mass energy.
For operation at 250~GeV, a second step may be considered in which the collider is operated at 10 Hz, instead of 5 Hz, 
with an average beam power equivalent to that shown in the L Upgrade 500 column in \Tref{tab:tdres:prms}.  It is assumed in what follows that the full 
Baseline 500 and L Upgrade 500 have been completed. At that point, if the main linac gradient is reduced to half of nominal, the 
repetition rate can be doubled without substantially increasing the overall average power consumption (in a scheme quite similar 
to that adopted for the electron linac at center of mass energy below 300 GeV). Naturally, the average beam power is also the same 
as the L Upgrade 500 beam power. This second step scheme allows the ILC Light Higgs Factory luminosity to be increased by a factor 
four from 0.75 e34 cm-2s-1 to 3.0 e34  cm-2s-1.  \Tref{tab:loweupgrade}
and  \Fref{fig:loweupgrade} summarize 
the primary parameters for these three ILC operational modes.

\begin{table}
\caption{ILC Higgs factory operational modes}.
\begin{tabular}{llcccc }
\toprule
 &  &  &  1st Stage & Baseline ILC, after & High Rep Rate \\
 &  &  & Higgs Factory & Lumi Upgrade &  Operation \\
\tcmidrule{4-4}
\tcmidrule{5-5}
\tcmidrule{6-6}
 & & & & &  \\
Center-of-mass energy & $E\sub{CM}$  & \si{\GeV} & 250 & 250 & 250  \\
\midrule
Collision rate & $f\sub{rep}$ & \si{\Hz} & 5 & 5 & 10 \\
Electron linac rate & $f\sub{linac}$  & \si{\Hz} & 10 & 10 & 10 \\
Number of bunches & $n\sub b$ &  & 1312 & 2625 & 2625 \\
Pulse current & $I\sub{beam}$  & \si{\mA} & 5.8 & 8.75 & 8.75 \\
 \\
Average total beam power & $P\sub{beam}$ & \si{\MW} & 5.9 & 10.5 & 21 \\
Estimated AC power & $P\sub{AC}$ & \si{\MW} & 129 & 160 & 200 \\
 \\
Luminosity & $L$ & $\times$\SI{e34}{\cm^{-2}\s^{-1}} & 0.75 & 1.5 & 3.0 \\
\bottomrule
\end{tabular}
\label{tab:loweupgrade}
\end{table}

%%%%%%%%%%%%%%%%%%%%%%%%%%%%%%%%%%%%%%%%%%%%%%%%%%%%%%%%%%%%%%%%%
\begin{figure}
\begin{center}
\includegraphics[width=0.75\hsize]{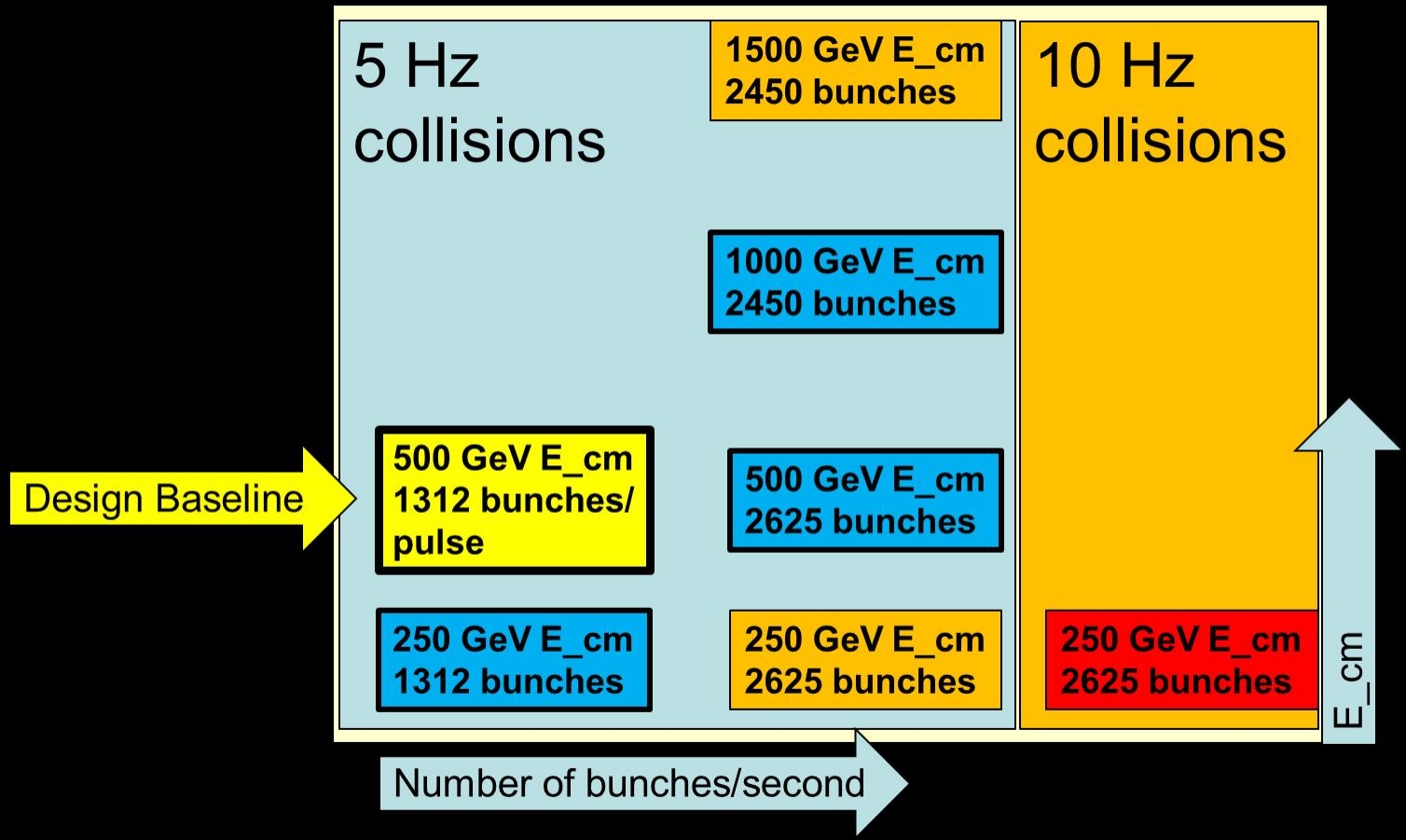}
\end{center}
\caption{ILC Stages and Upgrades. The baseline design (yellow) is fully optimized and represents 
the starting point for evaluating options. Three options 
(1st stage, L upgrade and TeV upgrade) are described in the TDR (blue). A further 3 options are mentioned here (orange and red). }
\label{fig:loweupgrade}
\end{figure}
%%%%%%%%%%%%%%%%%%%%%%%%%%%%%%%%%%%%%%%%%%%%%%%%%%%%%%%%%%%%%%%%%%%%%%%%%%%

The main impact of low energy ten Hz collision rate operation is on the injector systems: these must be able to cope with high repetition rate operation without 
any reduction in gradient, as is presently conceived only for the electron side (see~\cite{Adolphsen:2013jya}, Part II, Section 2.2.2, page 9). Furthermore, the positron source undulator 
must be able to produce adequate positrons using only the nominal 125 GeV luminosity-production electron beam. The latter may require further development of 
superconducting helical undulator technology as the helix pitch should be reduced from the present 12 cm (as demonstrated in Ref.~\cite{Adolphsen:2013jya}, Part I, Section 4.3.2, page 129)
to 0.9 cm without reducing the peak field. It is possible that a longer undulator with ILC TDR parameters would be adequate.  Alternatively, an intermediate 
solution could be considered with a reduced positron yield and possibly higher electron beam energy. For the latter additional electrical power would be 
required and the e+/e- beams might not have equal energy. This does not pose a problem for machine operation in principle but requires study. In addition, 
the positron injector system would be operated at 10 Hz, full gradient, requiring about two times more RF power and cryogenic capacity. Because of the 
low center of mass energy positron scheme, this aspect of electron injector operation is already accounted in the TDR.

\subsection{Gamma-Gamma Option}
High energy photon-photon collisions can be achieved by integrating high average power short-pulse lasers to the Linear Collider, enabling an expanded physics program for the facility including:
\begin{itemize}
\item Single Higgs production sensitive to charged particles of arbitrary mass
\item Greater reach for single production of supersymmetric Higgs bosons, H and A 
\item Probe of CP nature of the observed Higgs bosons through control of the polarization of the Compton photons that define the initial state
\item Anomalous couplings in single and double W boson production
\item Potential production of supersymmetric particles in electron-gamma collisions
\end{itemize}
The technology required to realize a photon linear collider continues to mature. Compton back-scattering technology is being developed worldwide for light source applications and high average power lasers continue to advance for Inertial Confinement Fusion. 

Compton scattering can transfer $\sim80\%$ of the incident electron energy to the backscattered photons when a 1 micron wavelength laser pulse is scattered from a 250~GeV electron beam. A laser pulse of 5 Joules, compressed to 1 ps width and focused to a diffraction limited spot can convert most of the incoming electrons in a bunch to high energy photons. An enormous amount of average laser power is required to provide 15,000 laser pulses per second to match the electron beam structure. Since most of the laser energy goes unused in the Compton process the required energy can be greatly reduced if the laser pulses can be recirculated.

A design of a recirculating cavity \cite{Will:2001ha} was created in 2001 which takes advantage of the long inter-bunch spacing in the superconducting machine to recirculate the laser pulses around the outside of the detector. Calculations showed that the required laser power could be reduced by a factor of 300 in this design. Recent studies have shown that a laser with sufficient phase stability to drive such a cavity is achievable with current technology. The available power saving for a recirculating system depends on the achievable cavity size that determines the number of times a laser pulse could be reused in a single electron bunch train.

Implementation of the photon collider option has several requirements for both the detector and the electron accelerator. Apertures must be opened in the forward part of the detector to allow the laser pulses to reach the Interaction Point and be focused a few millimeters before the electron beams collide. The electron beam will be left with an enormous energy spread after the Compton backscatter and a large crossing angle will be required in order to allow sufficient aperture for the spent beam to be extracted. Finally, the photon collider option will require its own beam dump design in order to handle the photon beam which will have about 50\% of the final beam energy.

Compton backscattering for the creation of MeV gamma-ray light sources is a world-wide activity. The basic techniques of bringing an electron beam and a laser pulse into collision is independent of the electron beam energy and these facilities are providing vital experience in the development of these techniques for the linear collider. These facilities are also developing the technology for recirculating laser pulses which will be critical to achieve a cost effective solution for the photon linear collider.
Current MeV gamma-ray sources include the ThomX\cite{Variola:2011zb} machine at LAL, the LUCX\cite{Fukuda:2010zzb} machine at KEK and the T-REX\cite{Hartemann:2010zz} machine at LLNL. The MightyLaser collaboration is developing a four mirror recirculating cavity for the demonstration of Compton backscattering at ATF\cite{Delerue:2011nk}.

While the photon linear collider has always been envisioned as a later stage to the basic linear collider program there may be advantages to considering it as a first stage. The photon collider requires an electron linear collider to drive it but it does not require positrons and it does not require flat electron beams at the Interaction Point in order to reduce the beamstrahlung. This opens up the possibility of creating a first stage linear collider without a positron source. The creation of a low-emittance RF electron gun. might also create the possibility of eliminating the damping rings in the first stage. Consideration of a dedicated photon collider Higgs factory as a first stage to the linear collider program is motivated by the discovery of a low mass Higgs boson at the LHC.

\subsection{Energy/Luminosity Running Scenarios}

It is of interest to consider the evolution of ILC  Higgs physics results over time 
given the ILC machine parameters defined in \Tref{tab:tdres:prms} and  \Tref{tab:loweupgrade}. 
 Taking eighteen years as a reasonable ILC lifetime, and using the concept of a Snowmass Year
where  an accelerator is assumed to run at its nominal luminosity for one-third of the time,
we assume that the ILC runs for a total of $18\times 10^{7}$ seconds at nominal luminosity
during its life.  Without optimization we make the simple assumption that we run 
for $3\times 10^7$~s at the baseline luminosity at each of the
center of mass energies 250, 500, and 1000 GeV, in that order.  Following those runs we go back and 
run for $3\times 10^7$~s at the upgraded luminosity at each of the three center of mass energies.

To avoid a proliferation of table entries, most results are only presented for 
the four different combinations of energy and luminosity listed in Table~\ref{tab:ecmlumruns}.  
Each scenario corresponds to the accumulated luminosity at different points in time.
In the summary chapter, however,  we present results for some alternative scenarios where, for example, 
runs at center of mass energies of 250 and 500 GeV take place at the upgraded luminosity before
any runs at 1000 GeV.

\begin{table}
 \begin{center}
 \begin{tabular}{lcccccccccc}
Nickname  & Ecm(1)    & Lumi(1)   &     +  & Ecm(2)         & Lumi(2)    &     +   & Ecm(3)         & Lumi(3)   & Runtime   & Wall Plug E         \cr              
          &         (GeV)  &          (fb$^{-1}$)     &  &       (GeV)  &          (fb$^{-1}$)      &  &       (GeV)  &          (fb$^{-1}$) & (yr) & (MW-yr)  \cr \hline
 ILC(250) &  250 & 250 & & & & & & & 1.1 & 130 \cr
 ILC(500) &  250 & 250 & & 500  & 500 & & &  & 2.0  & 270 \cr
 ILC(1000) &  250 & 250 & &  500  & 500 & &  1000 & 1000 & 2.9 & 540  \cr
 ILC(LumUp) &  250 & 1150 & &  500  & 1600 & &  1000 & 2500 & 5.8 & 1220 \cr
   \hline
   \end{tabular}
  \caption{Energy and luminosity scenarios assumed in this paper.}
\label{tab:ecmlumruns}
  \end{center}
\end{table}
%%%%%%%%%%%%%%%%%%%%%%%%%%%%%%%%%%%%%%%%%%%%%%%%%%%%%%%%%%%%%%%%%

\section{Detector Concepts} \label{sid:Accelerator_Detector:sec:detectorconcepts}
\subsection{ILD}

The ILD detector is a multi-purpose detector.
It has been designed for optimal particle-flow (PFA) performance and high precision 
vertexing and tracking performance. The tracking system consists of a high-precision pixel 
vertex detector, silicon trackers and a time-projection chamber.  The calorimeter system 
consists of highly segmented electro-magnetic calorimeter and hadron calorimeter.
They are placed inside a 3.5 Tesla solenoid magnet and achieves 
high precision measurements of particle flows, track momentum and vertexes. On the outside of the magnet coil, the iron 
return yoke is instrumented as a muon system and as a tail catcher calorimeter. 
The forward region is covered by 2 layers of pixel and 5 layers of silicon strip tracker. 
Calorimeter system covers down to 5 mrad from the outgoing beam except 
the hole for the in-coming beam due to 14 mrad crossing angle.
The quadrant view of the ILD detector is shown in Fig.\ref{fig:ILD-quad-view}.
Further detail of the detector will be found in the reference\cite{Behnke:2013lya}.
%% Half width
%\thisfloatsetup{floatwidth=\SfigwHalf,capposition=beside}
%\begin{figure}[h]
%    \centerline{\includegraphics[width=1.5\hsize]{Chapter_Accelerator_Detector/figs/ILD_quadrant_2.pdf}}
%    \caption{Quadrant view of the ILD detector concept.}
%    \label{fig:ILD-quad-view}
%\end{figure}

\begin{figure}[htb]
  \ffigbox{\CommonHeightRow{\begin{subfloatrow}[2]%
        \ffigbox[\FBwidth]{}{\includegraphics[height=\CommonHeight]{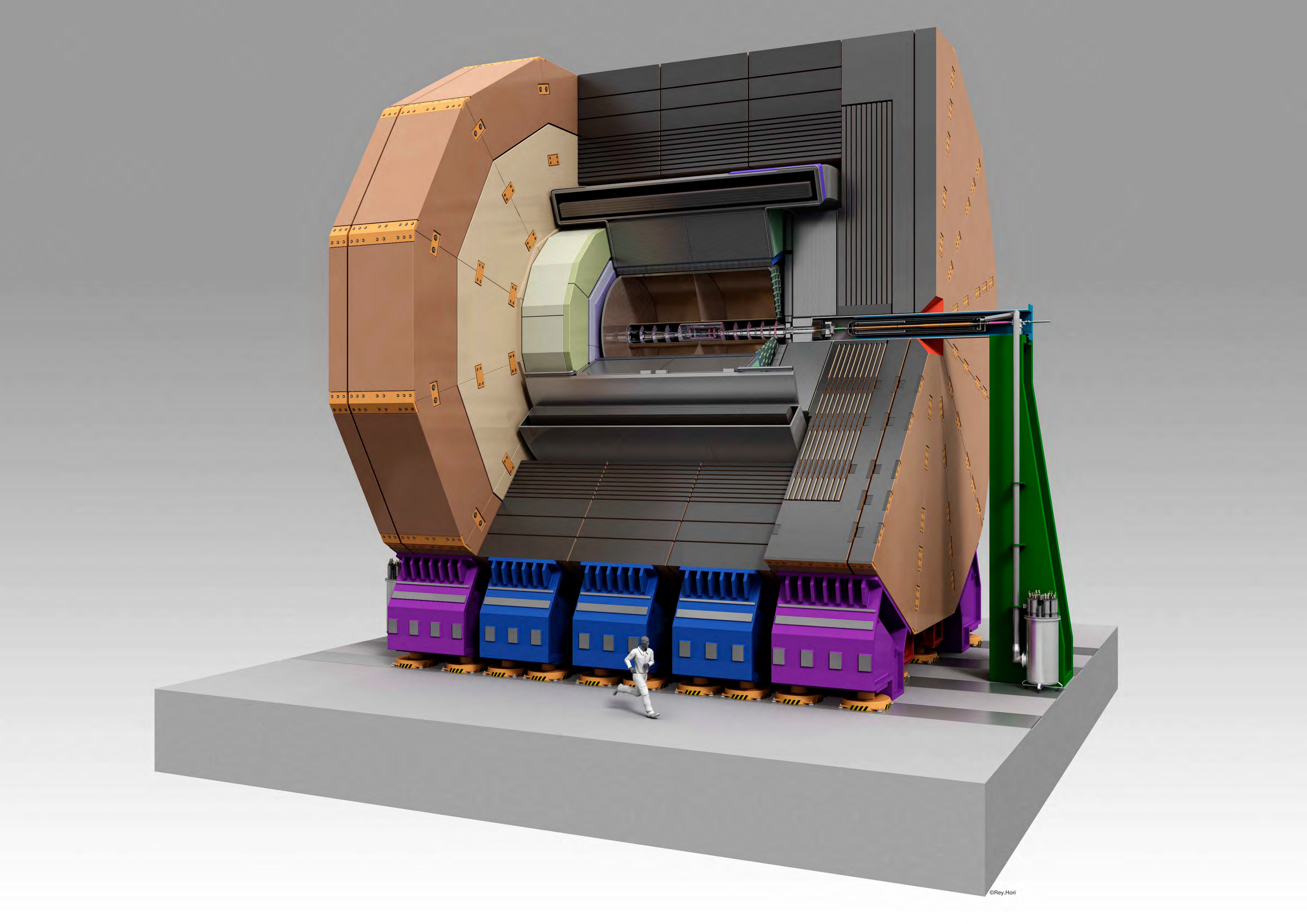}}
        \ffigbox[\FBwidth]{}{\includegraphics[trim=0 0 0 6,clip,height=\CommonHeight]{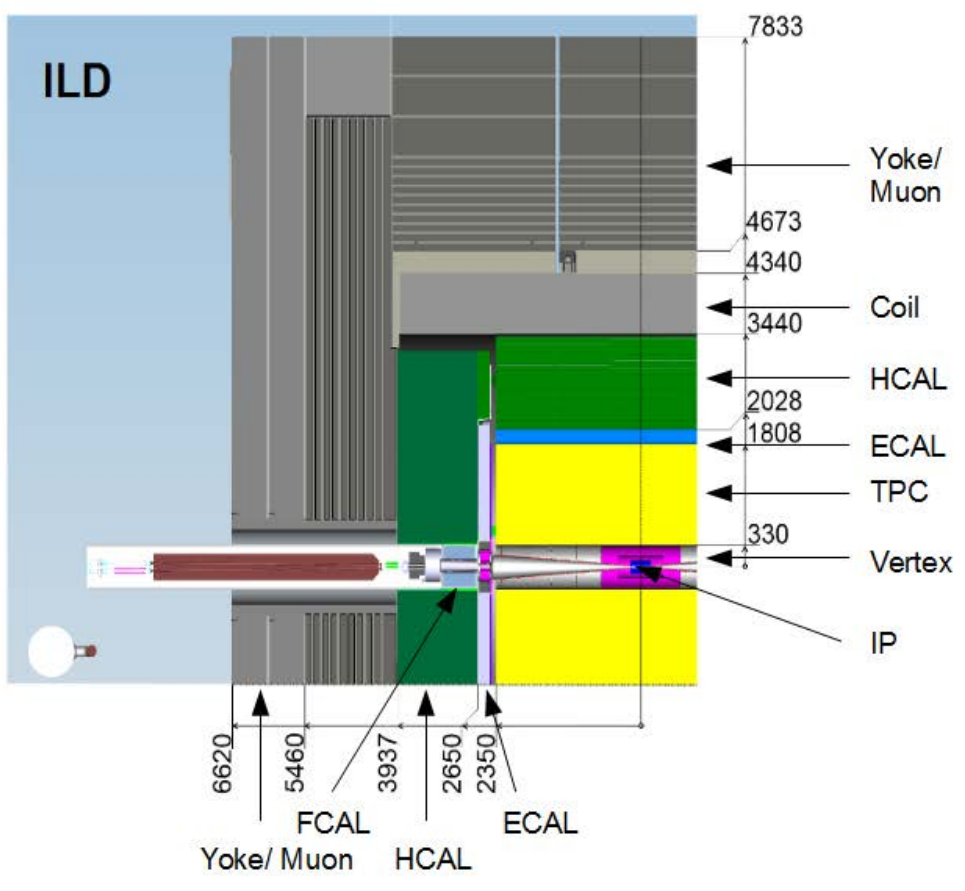}}
    \end{subfloatrow}}%
  }{%
    \caption{The ILD detector, showing (left) an isometric view on the platform, and (right) a quadrant view.}
    \label{fig:ILD-quad-view}
  }
\vspace{0.7cm}
\end{figure}

The ILC beam operates at 5 Hz, with 1 msec of beam collision period followed by 199 msec 
quiet period. This unique beam pulse structure allows data acquisition without a hard ware trigger and 
lower power consumption electronics system by adapting the pulsed operation of read out electronics.  
The requirement of the electronics system cooling is moderate and a thin detector system 
could be realized. 

Particle flow requires a thin tracker, to minimize interactions before the calorimeters and thick calorimeters 
to fully absorb the showers.  Thin vertex detector, as well as the small beam pipe radius, helps
a precise vertex reconstruction even for low momentum tracks.   
Figure.\ref{fig:ILD-detector-material} (left) shows the material in the detector in radiation lengths 
up to the end of the tracking system.  The amount of material up to the end of the tracking is 
mostly below 10\% for the full solid angle.  The right-hand plot shows the total interaction length 
including hadron calorimeter, showing a calorimeter coverage by 7 interaction length of coverage in almost all 
solid angle 
% Full width
\thisfloatsetup{floatwidth=\SfigwFull,capposition=beside}
\begin{figure}[h]
\begin{tabular}{cc}
\includegraphics[width=0.52\hsize,viewport={0 -10 600 500},clip]{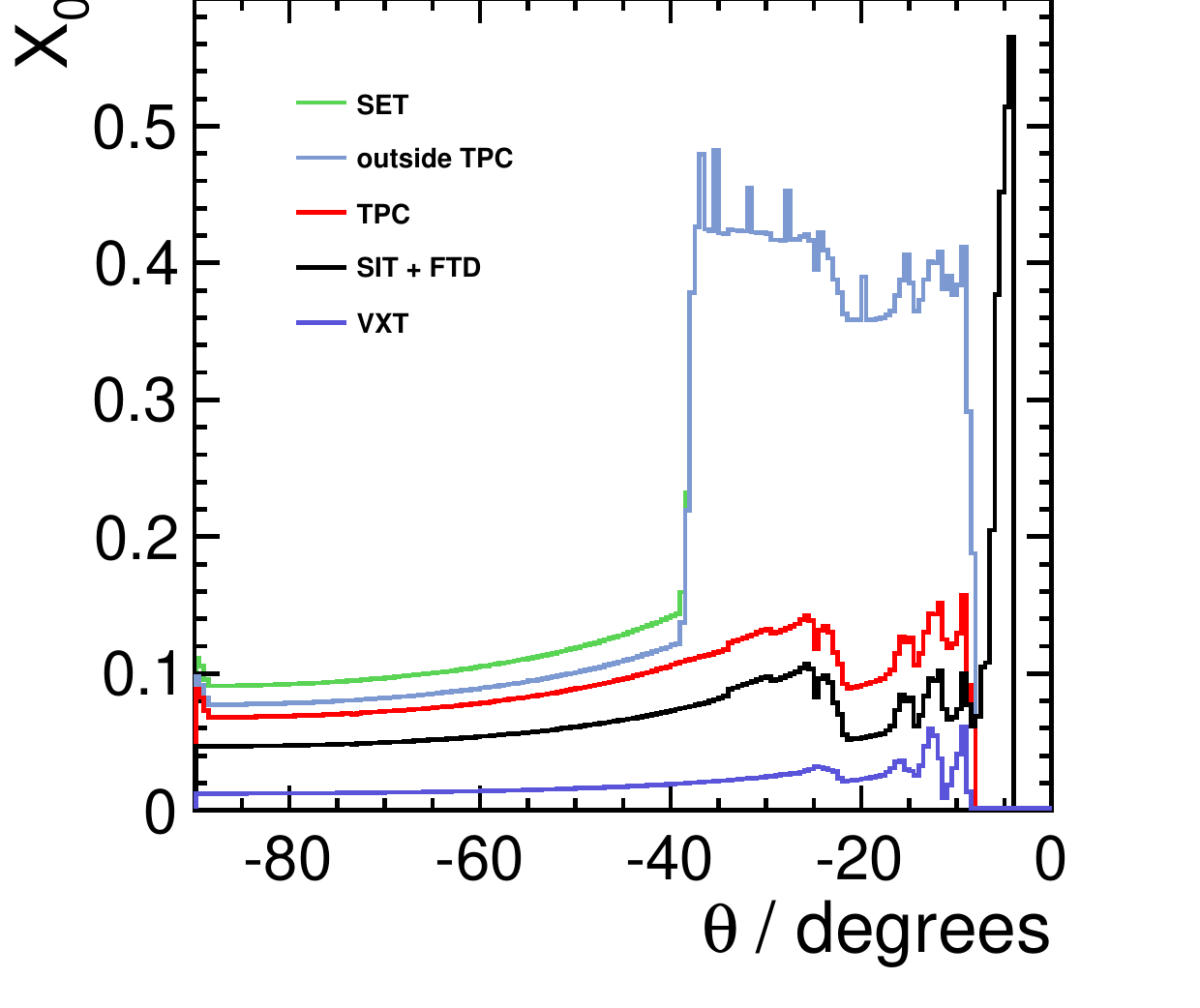} 
\includegraphics[width=0.5\hsize]{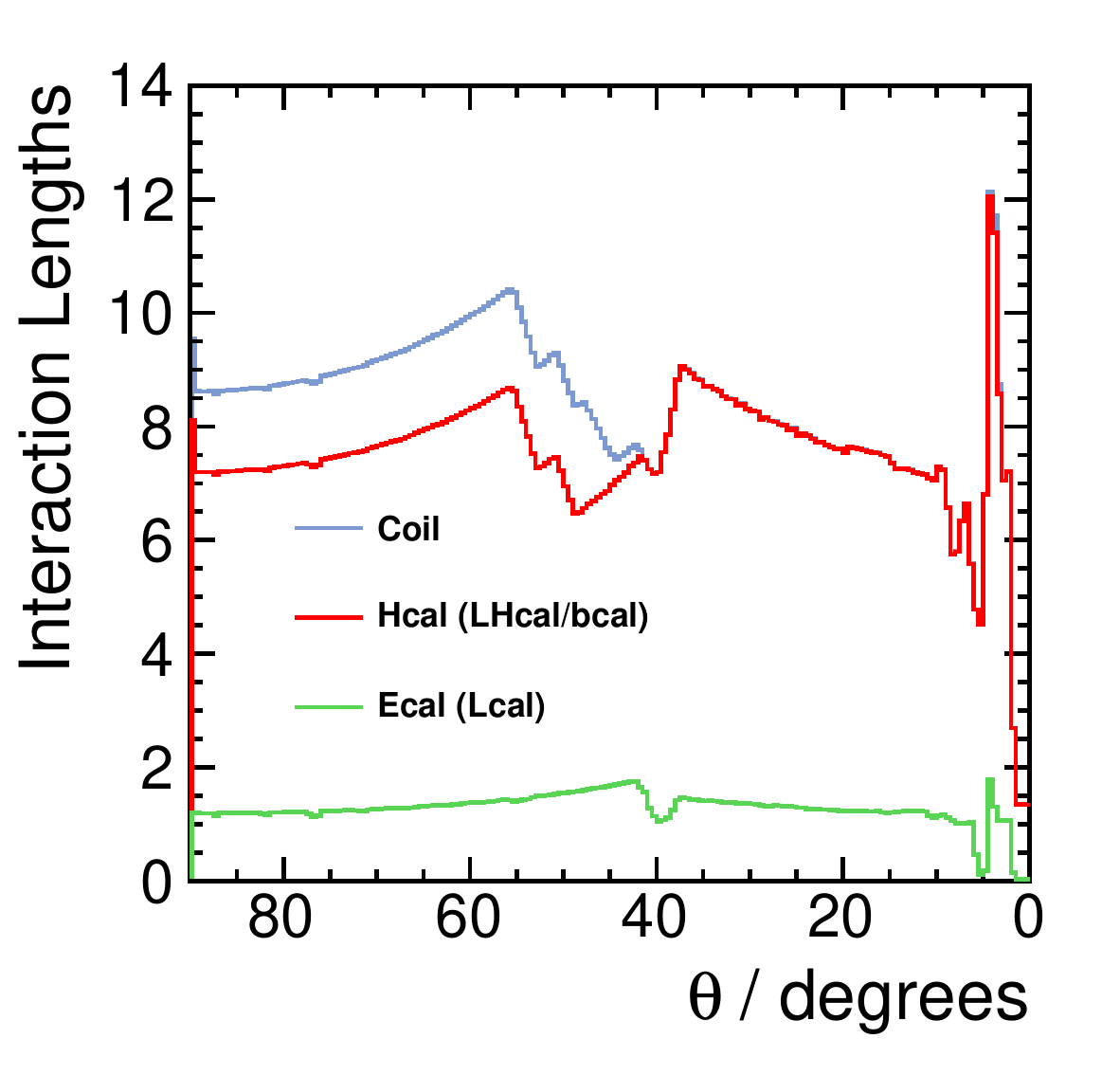}
\end{tabular}
    \caption{Left: Average total radiation length of the material in the tracking detectors 
   as a function of polar angle. Right: Total interaction length in the 
   detector up to the end of the calorimeter system and including the coil.}  
    \label{fig:ILD-detector-material}
\end{figure}

For the ILC TDR study, we have developed a realistic detector simulation model.
In the model, we have implemented materials for electronics, cooling system, 
and support structure based on the ILD baseline design in order to evaluate 
the detector performance as realistically as possible. 
The simulated data were analyzed with a realistic tracking software, (Marlin tracking packages\cite{Gaede:2006pj}), 
particle flow analysis(PandoraPFANew\cite{Marshall:2013bda}), flavor tagging analysis(LCFIPlus\cite{LCFIPlus}).  In physics event analysis, 
background events as low $P_T$ hadronic background due to 
collisions of bremsstrahlung or beamstrahlung photons and low energy 
electron/positron backgrounds hitting beam calorimeter were overlaid on signal events.

According to the performance study using 
$e^+e^-\rightarrow q\bar{q}$ events and single $\mu$ events, we have obtained
the jet energy resolution ( 90\% truncated RMS error ) of below 4% for nearly the full 
solid angle and the momentum resolution of $\sigma_{P_T}=2\times 10^{-5}$ GeV$^{-1}$ for 
high momentum tracks.  From the study using $e^+e^- \rightarrow t\bar{t}$ events, 
the average track reconstruction efficiency of 99.7\% for tracks greater than 
1 GeV across the entire polar angle range has been achieved.  For $e^+e^-\rightarrow q\bar{q}$ 
events at 91 GeV, $b$-quark($c$-quark) tagging purity at 60\% efficiency was about 100\% (60\%).

%The MC samples were generated by Whizard 1.95 and Physsim.  For 250 GeV study, the standard 
%model processes up to 4 fermion final states including those produced by beamstrahlung and 
%bremsstrahlung photons.  For 350 GeV study, the processes up to 6 fermion final states were considered.
%$t\bar{t}$ events were generated including the enhancement effect near threshold using Physsim generator. 

%The low Pt hadron events are copiously produced at ILC. They were overlaid to each events.
%0.22 events per BX at 250 GeV and 0.33 events per BX at 350 GeV

\subsection{SiD}

\sid is a general-purpose detector designed to perform precision measurements at
a Linear Collider\cite{Aihara:2009ad,Behnke:2013lya}. It satisfies the challenging detector
requirements for physics at the ILC. \sid is the result of many years
of creative design by physicists and engineers, backed up by a
substantial body of past and ongoing detector research and
development. While each component has benefitted from continual
development, the \sid design integrates these components into a
complete system for excellent measurements of jet energies, based on
the Particle Flow Algorithm (PFA) approach, as well as of charged
leptons, photons and missing energy.  The use of robust silicon vertexing and 
tracking makes \sid applicable
to a  wide range of energies from a Higgs factory to beyond \SI{1}{\TeV}. \sid has been designed in a cost-conscious manner, with the
compact design that minimizes the volumes of high-performing,
high-value, components, while maintaining critical levels of
performance. The restriction on dimensions is offset by the relatively
high central magnetic field from a superconducting solenoid.

\sid  is a compact detector based on a powerful silicon pixel vertex detector,
silicon tracking, silicon-tungsten electromagnetic calorimetry (ECAL)
and highly segmented hadronic calorimetry (HCAL). \sid also
incorporates a high-field solenoid, iron flux return, and a muon
identification system (see \Fref{sid:ConceptOverview:Ovw_1}). 

%The use of silicon
%sensors in the vertex, tracking and electromagnetic calorimetry, together with a
%high granularity gaseous hadron calorimeter, enables a unique integrated
%tracking system ideally suited to particle flow.

The choice of silicon detectors for tracking and vertexing ensures
that \sid is robust with respect to beam backgrounds or beam loss,
provides superior charged-particle momentum resolution, and eliminates
out-of-time tracks and backgrounds.  The main tracking detector and
calorimeters are ``live'' only during each single bunch crossing, so
beam-related backgrounds and low-\pT backgrounds from \gghadrons
processes will be reduced to the minimum possible levels. The \sid
calorimetry is optimized for excellent jet-energy measurement using
the PFA technique.  The complete tracking and calorimeter systems are
contained within a superconducting solenoid, which has
a \SI{5}{\tesla} field strength, enabling the overall compact
design. The coil is located within a layered iron structure that
returns the magnetic flux and is instrumented to allow the
identification of muons.

\begin{figure}[htb]
  \ffigbox{\CommonHeightRow{\begin{subfloatrow}[2]%
        \ffigbox[\FBwidth]{}{\includegraphics[height=\CommonHeight]{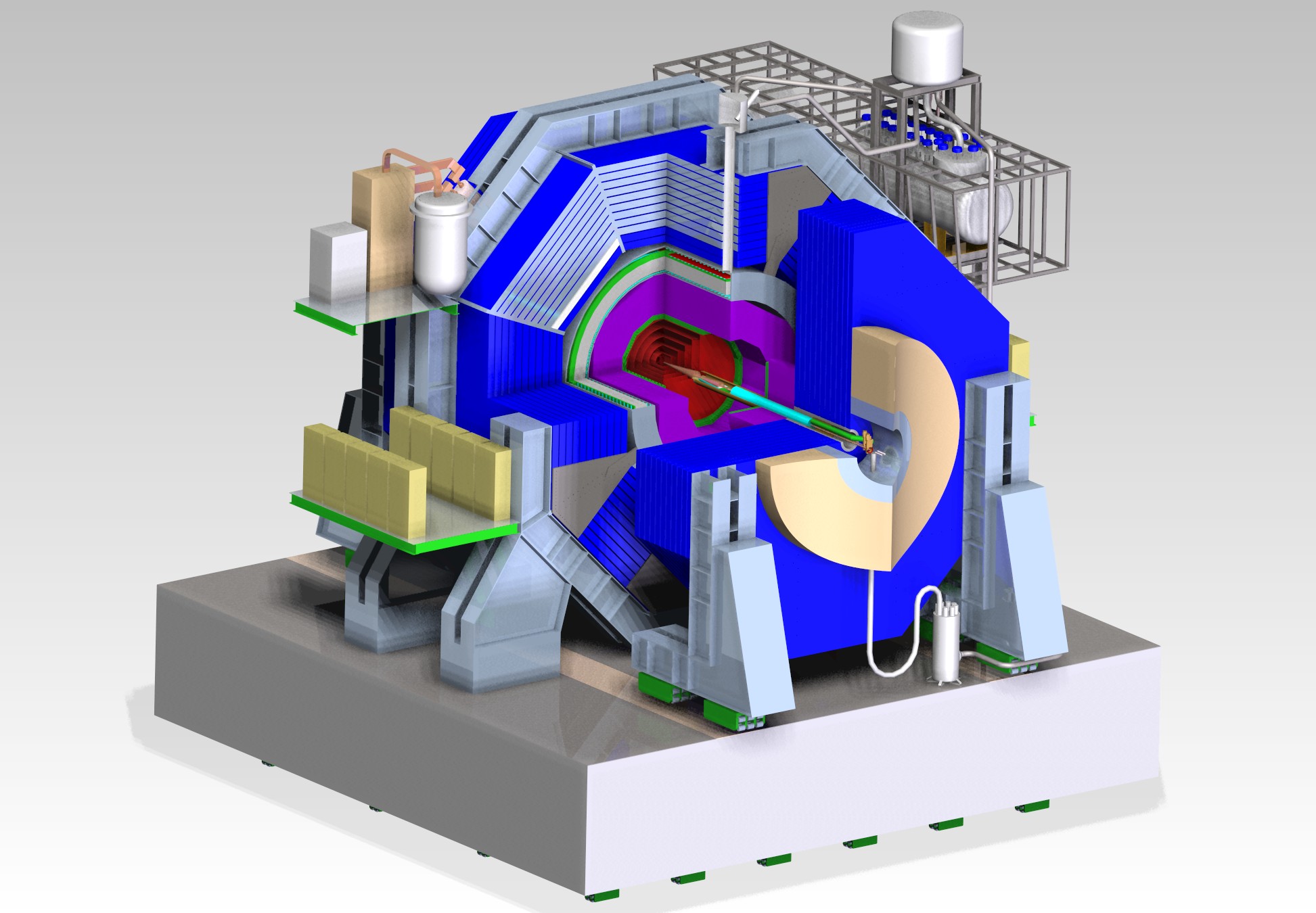}}
        \ffigbox[\FBwidth]{}{\includegraphics[trim=0 0 0 6,clip,height=\CommonHeight]{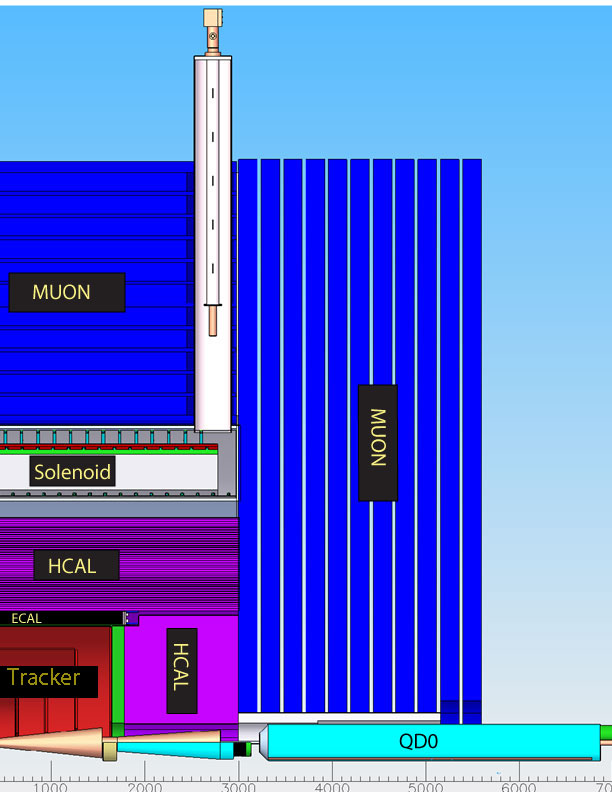}}
    \end{subfloatrow}}%
  }{%
    \caption{The \sid detector, showing (left) an isometric view on the platform, and (right) a quadrant section.
    Colour coding: tracking (red), ECAL (green), HCAL (violet) and the flux return
   (blue).}
    \label{sid:ConceptOverview:Ovw_1}
  }
\vspace{0.7cm}
\end{figure}

The tracking system is a key element as the particle-flow algorithm
requires excellent tracking with superb efficiency and good
two-particle separation. The requirements for precision measurements,
in particular in the Higgs sector, place high demands on the momentum
resolution at the level of $\delta (1/\pT) \sim$~2--\SI{5e-5}{(GeV/c)^{-1}}
and the material budget of the tracking system.
Highly efficient tracking is achieved using the pixel detector and
main tracker to recognize and measure prompt tracks.

The \sid vertex detector uses a barrel and disk layout. The barrel
section consists of five silicon pixel layers with a pixel size of
20$\times$\SI{20}{\square\micro\meter}. The forward and backward regions each have
four silicon pixel disks. In addition, there are three silicon pixel
disks at a larger distance from the interaction point to provide
uniform coverage for the transition region between the vertex detector
and the outer tracker. This configuration provides for very good
hermeticity with uniform coverage and guarantees excellent
charged-track pattern-recognition capability and impact-parameter
resolution over the full solid angle. The vertex detector design
relies on power pulsing during bunch trains to minimize heating and
uses forced air for its cooling.  The main tracker technology of
choice is silicon-strip sensors arrayed in five nested cylinders in
the central region with an outer cylinder radius of \SI{1.25}{\meter} and four
disks in each of the endcap regions. The geometry of the endcaps
minimizes the material budget to enhance forward tracking. The
detectors are single-sided silicon sensors with a readout pitch of
\SI{50}{\micro\meter}.

The choice of PFA  imposes a number of basic requirements on the calorimetry. 
The central calorimeter system must be contained within the solenoid in order 
to reliably associate tracks to energy deposits. The electromagnetic and
hadronic sections must have imaging capabilities that allow both efficient
track-following and correct assignment of energy clusters to tracks. These
requirements imply that the calorimeters must be finely segmented both
longitudinally and transversely. 

The combined ECAL and HCAL systems consist of a central barrel part and two
endcaps, nested inside the barrel. The entire barrel system is contained within
the volume of the cylindrical superconducting solenoid. The electromagnetic
calorimeter has silicon active layers between tungsten absorber layers. The
active layers use 3.5$\times$\SI{3.5}{mm^2} hexagonal silicon pixels, which provide excellent
spatial resolution. The structure has 30 layers in total, the first 20 layers
having a thinner absorber than the last ten layers. This configuration is a 
compromise between cost, electromagnetic shower radius, sampling frequency, and
shower containment. The total depth of the electromagnetic calorimeter is 26
radiation lengths (\xo) and one nuclear interaction length. The hadronic
calorimeter has a depth of 4.5 nuclear interaction lengths, consisting of
alternating steel plates and active layers. The baseline choice for the active
layers is the glass resistive-plate chamber with an individual readout segmentation of
10$\times$\SI{10}{mm^2}. 
Two special calorimeters are foreseen in the very forward region: LumiCal for
precise  measurement, and BeamCal for fast estimation, of the luminosity.

The \sid superconducting solenoid is based on the CMS solenoid
design philosophy and construction techniques, using a slightly modified CMS
conductor as its baseline design. Superconducting strand count in the coextruded
Rutherford cable was increased from 32 to 40 to accommodate the higher \SI{5}{\tesla}
central field. 
The flux-return yoke is instrumented with position sensitive detectors to serve
as both a muon filter and a tail catcher. The \sid Muon System
baseline design is based on scintillator technology, using extruded scintillator
readout with wavelength-shifting fiber and SiPMs. Simulation studies have shown
that nine or more layers of sensitive detectors yield adequate energy
measurements and good muon-detection efficiency and purity.

A large fraction of the software for the generation, simulation and
reconstruction is shared between the detector concepts. The \sid detector is
fully implemented and simulated using \slic, which is based on \geant. The background
originating from incoherent  pair interactions and from \gghadrons for one bunch
crossing is fully taken into account by the simulation.
The events are then passed through the reconstruction software suite, which
encompasses digitization, tracking, vertexing and the Pandora PFA algorithm.
The material budget of the simulated tracker and the simulated tracking
performance for single particles are shown in \Fref{sid:trackingplots}.

\begin{figure}[htb]
  \ffigbox{\CommonHeightRow{\begin{subfloatrow}[2]%
        \ffigbox[\FBwidth]{}{\includegraphics[height=\CommonHeight]{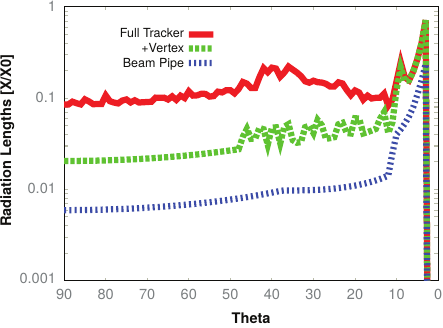}}
        \ffigbox[\FBwidth]{}{\includegraphics[height=\CommonHeight]{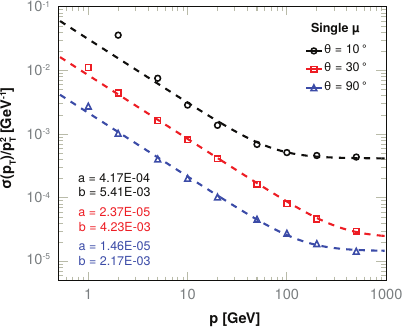}}
    \end{subfloatrow}}%
  }{%
    \caption{Left: \sid Tracker Material budget in terms of \xo. Right: the
   normalised transverse momentum resolution for single-muon events.}
    \label{sid:trackingplots}
  }
\vspace{1.0cm}
\end{figure}

\begin{figure}[htb]
  \ffigbox{\CommonHeightRow{\begin{subfloatrow}[2]%
        \ffigbox[\FBwidth]{}{\includegraphics[height=\CommonHeight]{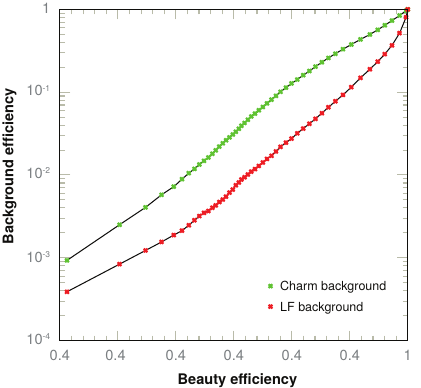}}
        \ffigbox[\FBwidth]{}{\includegraphics[height=\CommonHeight]{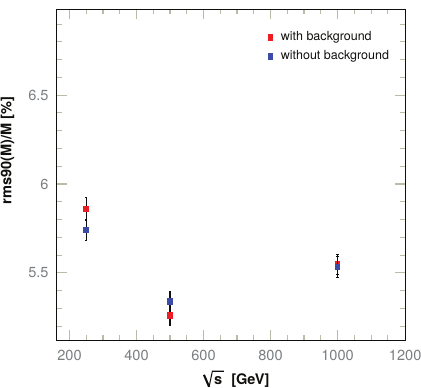}}
    \end{subfloatrow}}%
  }{%
    \caption{Left: Mis-identification efficiency of light quark (red points)
   and c quark events (green points) as b quark jets versus the b
   identification efficiency in di-jet events at \roots
   = \SI{91}{\GeV} including background from \gghadrons and incoherent
   pairs.  Right: Mass resolution of reconstructed $\PZ\PZ$ events
   with and without the backgrounds from
   \gghadrons and incoherent pairs at different values of \roots.}
    \label{sid:flavorpfaplots}
  }
\vspace{1.0cm}
\end{figure}

The material budget of the entire tracking system is less than 0.2 \xo down to
very low angles. The current design achieves an asymptotic momentum resolution 
of $\delta (1/\pT)  = $~\SI{1.46e-5}{(GeV/c)^{-1}} and an transverse 
impact parameter resolution better than \SI{2}{\micro\meter}.
The ability to tag bottom and charm decays with high purity has been a driving
factor in the design of the vertex detector. Figure~\ref{sid:flavorpfaplots}~(left)
illustrates the capability of the \sid to separate b-quarks also in
the presence of the full beam background.

Besides the detector performance, sophisticated reconstruction algorithms are
necessary to obtain a jet-energy resolution that allows the separation of hadronic W
and Z decays. To avoid a bias from possible tails, the rms$_{90}$ value is computed
to describe the energy or mass resolution of a particle-flow algorithm. It is
defined as the standard deviation of the distribution in the smallest range that
contains 90\% of the events. Figure~\ref{sid:flavorpfaplots}~(right) 
shows the mass resolution of
reconstructed \PZ bosons in \epem~$\rightarrow~\PZ\PZ$ events at different
collision energies, where one \PZ decays to neutrinos,  the other to two light
quarks that give rise to two jets.

\section{Systematic Errors} \label{sid:Accelerator_Detector:sec:systematicerrors}
   Most of the errors quoted in this document include statistical errors only.
For the three baseline luminosity scenarios this is an excellent
approximation of the total error.     For the luminosity upgrade scenario, 
however, some thought has to be given to systematic errors.   
\subsection{Flavor Tagging}

\subsubsection{Introduction}
We give a ballpark estimate of the systematic uncertainties
arising from $b$ tagging in the context of the Higgs branching ratio measurements. %
%\footnote{This comes from the request from M.~Peskin,
%who only gave us 7 days to produce this note.}
%
The strategy is to employ control samples to evaluate
the $b$ tagging efficiencies as well as
the fake rate due to non-$b$ jets (primarily $c$ jets) passing the
$b$ tag requirements.
For the former, we give an estimate using
a $b$ jet rich sample selecting the
$ZZ\rightarrow \ell\ell b\overline{b}$ process.
For the latter, we use the
$WW\rightarrow \ell\nu qq$ process
to obtain a control sample
containing very few $b$ jets in the event.
We then evaluate the impact on the uncertainties of
$BR(h\rightarrow b\overline{b})$ assuming
a center-of-mass energy of $\sqrt{s}=250$~GeV
and an integrated luminosity of $\mathcal{L}=250$~fb$^{-1}$ (nominal ILC case)
with an extrapolation to $\mathcal{L}=1150$~fb$^{-1}$ for the high luminosity ILC case.

For the $b$ tagging efficiency points,
we use the following two points in our estimates
$\epsilon=80\%$ and 50\%
with the $c$ and $uds$ fake rate summarized in
Tab.~\ref{tab:btag-wp}, which are read off from
Fig.~\ref{fig:btag}
which shows the signal and background efficiencies 
obtained using LCFIPlus.
\begin{table}[hbtp]
\centering
\caption{$b$ tagging working points and fake rate
  for $e^+e^-\rightarrow q\overline{q}$ samples at $\sqrt{s}=91.2$~GeV
  using LCFIPlus.}
\label{tab:btag-wp}
\begin{tabular}{cccc}
\hline
$b$ tag efficiency & $c$ fake rate & $q$ fake rate \\
\hline
80\% & 8\% & 0.8\% \\
50\% & 0.13\% & 0.05\% \\
\hline
\end{tabular}
\end{table}

\begin{figure}[hbtp]
\centering
\includegraphics[width=0.5\linewidth]{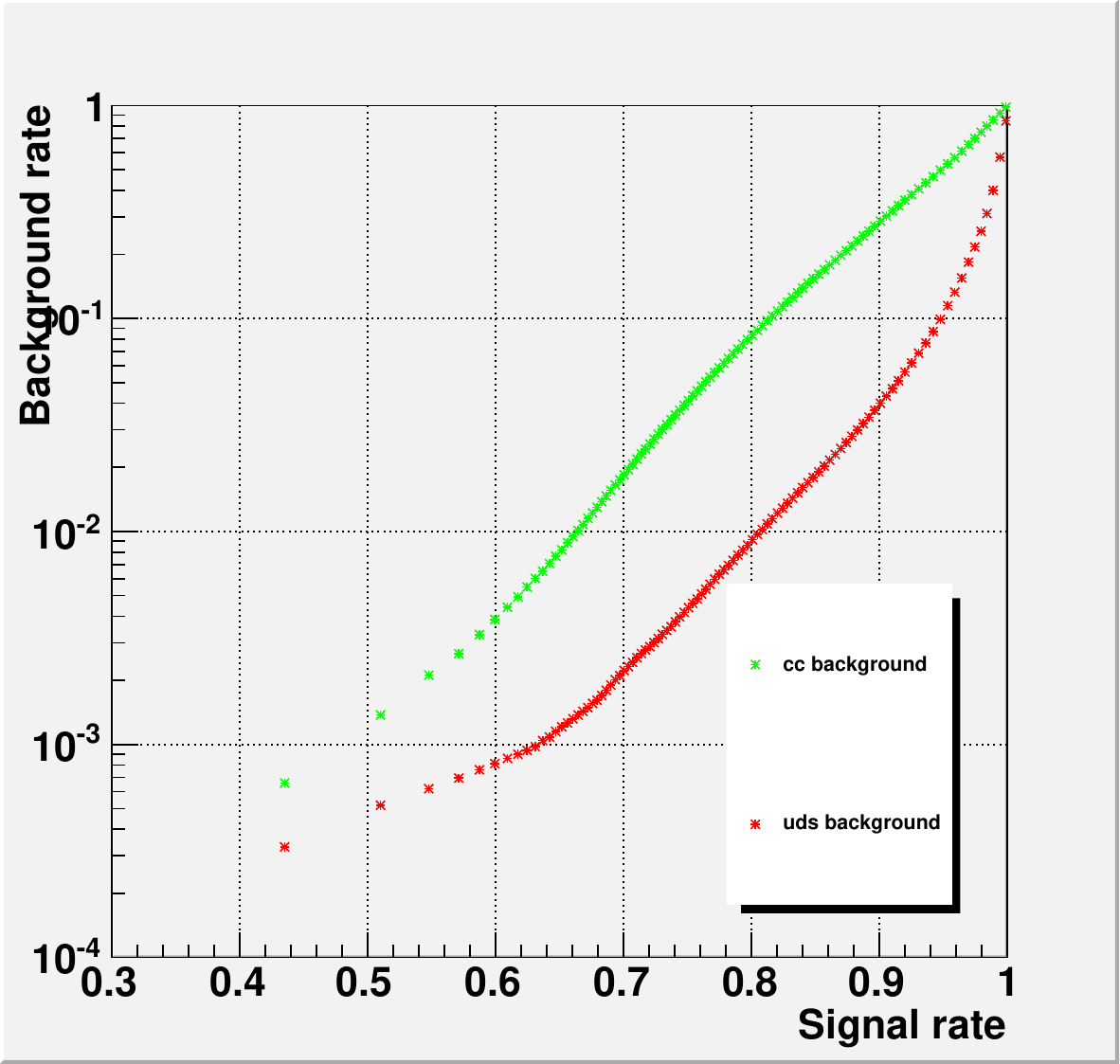}
\caption{$b$ tagging efficiencies versus background efficiencies
  for $e^+e^-\rightarrow q\overline{q}$ samples at $\sqrt{s}=91.2$~GeV
  using LCFIPlus.
}
\label{fig:btag}
\end{figure}

\subsubsection{Estimate of $b$ tag efficiency using
  $ZZ\rightarrow \ell^+\ell^- b\overline{b}$}
The signal efficiency is assumed to be 50\%,
by noting that the analysis will be very similar to the
$e^+e^-\rightarrow Zh\rightarrow \ell\ell qq$ analysis\cite{ild_loi_higgs_br}.
Background efficiency from the $WW$ process
is assumed to be 1\%, which should be a conservative estimate.
\begin{table}[hbtp]
\centering
\caption{Selection table for the $ZZ$ analysis.
The $b$ tag is applied to one of the two jets.}
\label{tab:zz-cut}
\begin{tabular}{cccc}
\hline
Process & Before selection & After selection &
Tag $b$ ($\epsilon=50\%$) \\
\hline
$ZZ\rightarrow\ell\ell bb$ & 30000 & 15000 & 7500 \\
$ZZ\rightarrow\ell\ell cc$ & 24000 & 12000 &   14 \\
$ZZ\rightarrow\ell\ell qq$ & 86000 & 43000 &   22 \\
$WW\rightarrow\ell\nu cs$  & $1.3\times10^6$ & 13000 &   13 \\
$WW\rightarrow\ell\nu ud$  & $1.3\times10^6$ & 13000 &    7 \\
\hline
\end{tabular}
\end{table}
The sample after the first $b$ tag is used as the control sample.

We apply the $b$ tagging with varying efficiencies to our control sample.
Since our sample achieves a purity of over $99\%$,
we can safely neglect the contribution of
fake $b$ jets in our estimate of $b$ tagging efficiencies.
We use the standard recipe for computing the
selection efficiencies and use the
uncertainty from the binomial distribution
$ \sqrt{ p(1-p)/N } $
where $N=7500$ and $p$ is chosen for the varying efficiencies.
The results are summarized in Tab.~\ref{tab:btag-result}.
\begin{table}[hbtp]
\centering
\caption{Expected $b$-tagging uncertainties at various selection efficiencies.}
\label{tab:btag-result}
\begin{tabular}{ccccc}
\hline
Efficiency & Uncertainty \\
\hline
80\% & 0.46\% \\
70\% & 0.53\% \\
60\% & 0.57\% \\
50\% & 0.58\% \\
  \hline
\end{tabular}
\end{table}
We therefore conclude that the uncertainty in the
$b$ tagging efficiency is around 0.3\%. Since the uncertainty
scales to the statistics of the control samples,
it goes down to around 0.15\% in the high luminosity ILC case.

\subsubsection{Estimate of $b$ tagging fake rate using
 the $WW\rightarrow \ell\nu qq$ process}

Here we assume a selection efficiency of 10\% for
$WW \rightarrow \ell\nu qq$ events.
The selection for this process could proceed by
selecting an isolated lepton with tight lepton identification criteria
and no more than one isolated lepton with loose lepton identification criteria.
We assume that the dominant background in this case will be
due to $ZZ\rightarrow \tau\tau bb$ events
where one $\tau$ will result in hadronic jets
while the other $\tau$ undergoes one-prong leptonic decay.

\begin{table}[hbtp]\centering
\caption{Summary of selection for the fake rate measurement.
  Here the $b$ tag selection is such that
    one of the two jets will pass the $b$ tag requirement
    at the specified efficiency.}
\label{tab:ww-cut}
\begin{tabular}{ccccc}
\hline
Process & Before selection & After selection &
$b$ tag ($\epsilon_b=80\%$) & $b$ tag ($\epsilon_b=50\%$) \\
\hline
$WW\rightarrow \ell\nu cs$
& $1.3\times 10^6$ & $1.3\times 10^5$ (10\%) &
11310 (8.7\%) & 234 (0.18\%) \\
$WW\rightarrow \ell\nu ud$
& $1.3\times 10^6$ & $1.3\times 10^5$ (10\%) &
2080 (1.6\%) & 130 (0.1\%) \\
$ZZ\rightarrow \tau\tau b\overline{b}$
& $8500$           & $85$ (1\%) &
82 (96\%) & 64 (75\%) \\
\hline
\end{tabular}
\end{table}

We give an estimate of the fake rate as follows.
The contribution of the fake rate from
$WW\rightarrow \ell\nu cs$ will be dominated by $c$ jets
as we can infer from Tab.~\ref{tab:btag-wp}.
As the uncertainty for this number,
we take the number of $WW\rightarrow \ell\nu ud$ events
as the full uncertainty, which should be a very conservative estimate.
This gives 20\% (100\%) as the relative uncertainty in the fake rate
for the case of $\epsilon_b=80\%$ ($50\%$).

\subsubsection{Estimate of branching ratio systematic uncertainty}

We now translate these results into the Higgs branching ratio analysis.
We take the nominal Higgs branching ratios of
$BR(h\rightarrow b\overline{b})=58\%$ and
$BR(h\rightarrow c\overline{c})=2.9\%$.
For the $b$ tagging working point of $\epsilon_b=80\%$,
the fake rate from $c$ jet is around $\epsilon_c=8\pm2\%$
including the uncertainty which was just estimated.
Applying this $b$ tag gives
$BR(h\rightarrow b\overline{b})\cdot\epsilon_b = 46.4 \pm 0.14\%$
and
$BR(h\rightarrow c\overline{c})\cdot\epsilon_c = 0.23 \pm 0.06\%$
where we took 0.3\% as the relative uncertainty in $\epsilon_b$
and 20\% for the uncertainty in $\epsilon_c$.
Similarly for the $\epsilon_b=50\%$ working point, we compute
$BR(h\rightarrow b\overline{b})\cdot\epsilon_b = 29.0 \pm 0.09\%$
and
$BR(h\rightarrow c\overline{c})\cdot\epsilon_c = 0.038 \pm 0.038\%$,
where we took 0.3\% as the relative uncertainty in $\epsilon_b$
and 100\% for the uncertainty in $\epsilon_c$.

We conclude that despite the large relative uncertainty in $c$ jet tagging,
the overall uncertainty is dominated by the uncertainty in $b$ jet tagging
due to the small $h\rightarrow c\overline{c}$ branching ratio.
It is estimated that the uncertainty in the $b$ tagging efficiency in the observable
$\sigma(e^+e^-\rightarrow Zh)\cdot BR(h\rightarrow b\overline{b})$ 
is at the 0.3\% level in the nominal ILC and at the 0.15\% level in the ILC luminosity upgrade case.
Prospects for improving these numbers include
refined selection of the control samples (before the first $b$ tagging)
and the addition of other $ZZ$ and $Z\gamma$ modes
which will require background estimates with an actual simulation analysis.
Moving up to $\sqrt{s}=350$~GeV provides additional clean control samples from 
fully leptonic top pair decays
$e^+e^-\rightarrow t\overline{t}\rightarrow b\ell\nu b\ell\nu$.

\subsubsection{Summary and prospects}
We put forth a ballpark argument for the $b$ tagging systematic uncertainty
in the context of the Higgs branching ratio measurement.
Our preliminary findings are that the dominant contribution comes from 
the uncertainty in the estimate of the $b$ efficiency,
which is at the level of 0.3\% (nominal ILC) / 0.15\% (high luminosity ILC) when applied to the
Higgs branching ratio measurement.
This number is expected to improve by including additional modes.
The contribution from the fake rate is found to be negligible.
It is highly desired to refine these estimates using
a proper simulation study including all background processes.

\subsection{Luminosity}

The number of Bhabha events
per bunch crossing for a detector with minimum and maximum polar angle coverage
$\theta_{min}$ and $\theta_{max}$ (in mrad) is:

$$ N = 0.5\mathrm{pb}\frac{L}{R}\int\limits_{\theta_{min}}^{\theta_{max}}\frac{d\mathrm{cos}\theta}{\mathrm{sin}^4\theta/2} \sim 6 \times 10^{-6} \left( %%@
\frac{1}{\theta_{min}^2}-\frac{1}{\theta_{max}^2}\right) $$

\noindent for \roots=0.5~TeV, L=2$\times10^{34}
\mathrm{cm}^{-2}\mathrm{s}^{-1}$, and bunch crossing rate R=$1.4\times10^4
\mathrm{s}^{-1}$. Our goal is to measure the luminosity normalization
with an accuracy of several $10^{-4}$ for \roots=0.5~TeV. 
To do
this one needs $\approx 10^8$ events collected over $\approx10^7$ s,
or about ten events per second. One can then calculate the absolute
luminosity with $\approx10\%$ statistical error every several
minutes during the run. 
With a bunch crossing rate of $1.4\times10^4
\mathrm{s}^{-1}$, we need $>10^{-3}$ events per bunch crossing. To
achieve this statistical accuracy, we start the fiducial region for
the precision luminosity measurement well away from the
beamstrahlung pair edge at $\theta$=20~mrad, with a fiducial
region beginning at $\theta_{min}$=46~mrad, which gives $\approx 2 \times 10^{-3}$
events per bunch crossing.

Since the Bhabha cross
section is $\sigma \sim 1/\theta^3$, the luminosity precision can be
expressed as

$$\frac{\Delta L}{L} = \frac{2\Delta\theta}{\theta_{min}},$$

\noindent
where $\Delta\theta$ is a systematic error (bias) in polar angle measurement
 and $\theta_{min} = 46$~mrad is the minimum polar angle of the fiducial region.
Because of the steep angular dependence, the precision of the minimum polar
angle measurement determines the luminosity precision.
To reach the luminosity precision goal of $10^{-3}$,
the polar angle must be measured with a precision
$\Delta\theta <$ 0.02~mrad and the radial positions of the sensors
must be controlled within 30~\micron relative to the IP.

\subsection{Polarization}

The primary polarization measurement comes from dedicated
Compton polarimeters detecting backscattered electrons and positrons.
A relative polarization error of 0.1\% is expected from  implementing polarimeters both upstream
and downstream of the Interaction Region.  In addition the polarization can be measured directly
with physics processes such as $e^+e^-\rightarrow W^+W^-$ .  Combining the two techniques we
assume a polarization systematic error on cross section times branching ratios of 
0.1\% and 0.05\% for the baseline and upgraded luminosities, respectively.

\subsection{Systematic Error Summary}

The systematic errors that are used throughout this paper are
summarized in \Tref{tab:syserrorsummary}.

\begin{table}
 \begin{center}
 \begin{tabular}{lcc}
\hline
  & Baseline   & LumUp \cr   \hline            
 luminosity &  0.1\% & 0.05\% \cr
 polarization &  0.1\% & 0.05\% \cr
 b-tag efficiency &   0.3\% & 0.15\% \cr
   \hline
   \end{tabular}
  \caption{Systematic errors assumed throughout the paper. }
\label{tab:syserrorsummary}
  \end{center}
\end{table}
%%%%%%%%%%%%%%%%%%%%%%%%%%%%%%%%%%%%%%%%%%%%%%%%%%%%%%%%%%%%%%%%%

  \label{sid:Accelerator_Detector:sec:systematicerrors:subsec:summary}

\chapter{Higgs Mass, ZH Cross Section, Spin and CP \label{sid:chapter_mass_spin_cp}}
\section{Higgs Mass and $\sigma(ZH)$ Measurements}
The Higgs mass and the total cross section for $e^+e^- \to Zh$ 
are measured simultaneously in the process
$e^+e^- \to Zh$, with $Z \to \mu^+\mu^-$, $Z \to e^+e^-$, and  $Z \to q\bar{q}$ decays. 
Here the shape of the distribution of the invariant mass recoiling against the
 reconstructed $Z$ provides a precise measurement of $m_h$, while the 
normalization of the distribution provides the total cross section $\sigma(ZH)$
 independently of the Higgs decay mode. In particular, the 
$\mu^+\mu^-X$ final state provides a particularly precise
 measurement as the $e^+e^-X$ channel suffers from larger
 experimental uncertainties due to bremsstrahlung.
 It should be noted that it is the capability to precisely
 reconstruct the recoil mass distribution from $Z \to \mu^+\mu^-$
  that defines the momentum resolution requirement for an ILC
  detector.   A measurement using $Z \to q\bar{q}$ decays appears to 
only be feasible at $\sqrt{s} \ge 350$~GeV.   A study of this channel
at  $\sqrt{s}=500$~GeV is presented here.

\subsection{$l^+l^-h$ at $\sqrt{s}=250$~GeV}

The reconstructed recoil mass distributions, 
calculated assuming the $Zh$ is produced with four-momentum 
$(\sqrt{s}, 0)$, are shown in Fig.\ref{mass:mass:fig:Mrecoil}. In the $e^+e^-X$ 
channel FSR and bremsstrahlung photons are identified and used
 in the calculation of the $e^+e^- (n\gamma)$ recoil mass. Fits to 
signal and background components are used to extract $m_h$ and $\sigma(ZH)$.
 Based on this model-independent analysis of Higgs production in
 the ILD detector, it is shown that $m_h$ can be determined with a
 statistical precision of $40$~MeV ($80$~MeV) from the $\mu^+\mu^-X$
  ($e^+e^-X$) channel. When the two channels are combined an
 uncertainty of $32$~MeV is obtained \cite{Abe:2010aa,Li:2012taa}. The
 corresponding
 model independent uncertainty on the Higgs production cross
 section is $2.6$\,\%. For a luminosity of 1150~fb$^{-1}$ at $\sqrt{s}$=250~GeV
(our scenario 4) the uncertainty on the Higgs mass and production cross
 section drop to  $15$~MeV and $1.2$\,\%, respectively.

Similar results were obtained from SiD
 \cite{Aihara:2009ad}. It should be emphasized that these measurements
 only used the information
 from the leptonic decay products of the $Z$ and are independent of
 the Higgs decay mode. As such this analysis technique could be
 applied 
even if the Higgs decayed invisibly and hence allows us to determine
 the absolute branching ratios including that of invisible Higgs
 decays. 

\begin{figure}
\includegraphics[width=0.49\hsize]{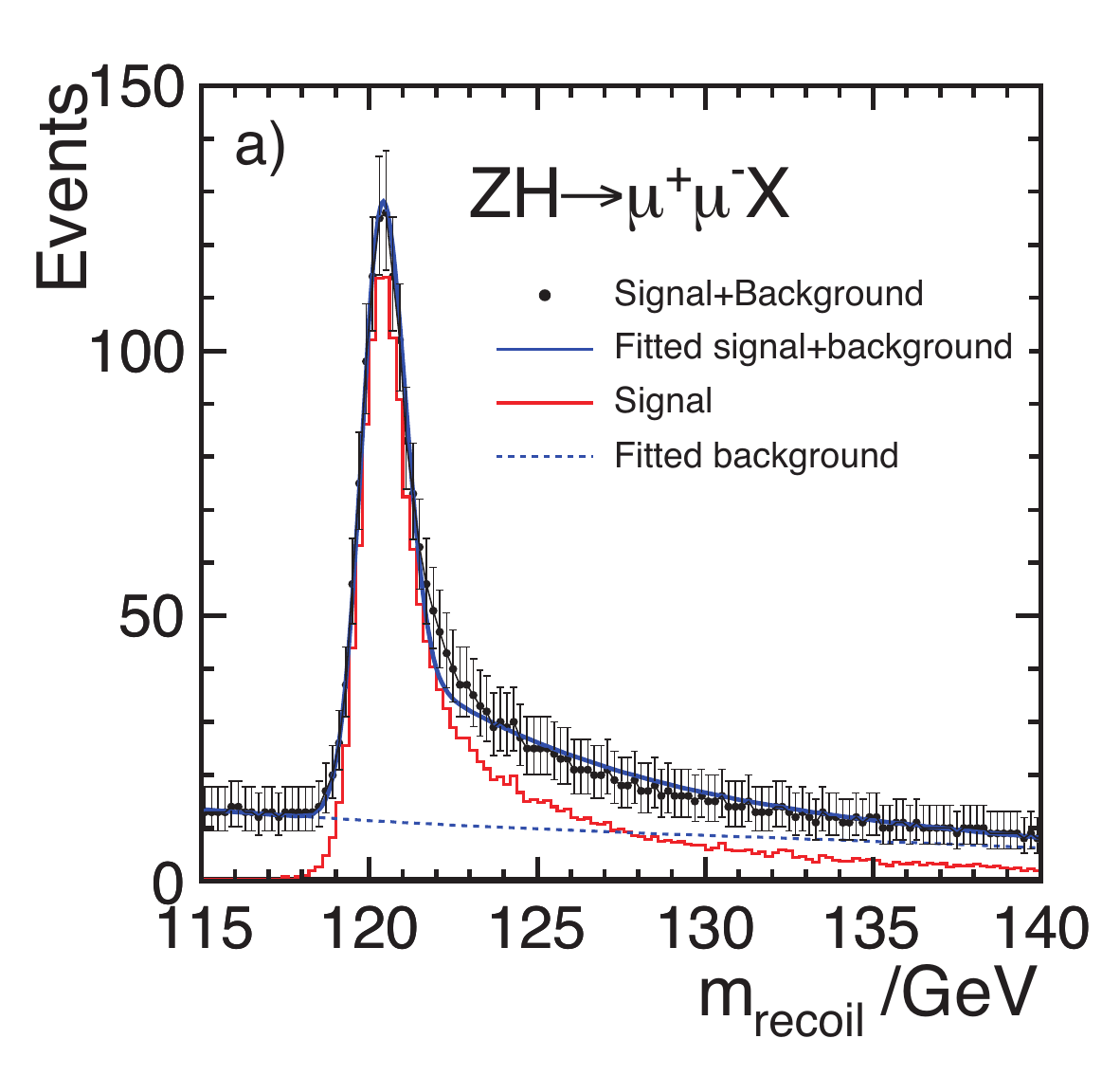} 
\includegraphics[width=0.49\hsize]{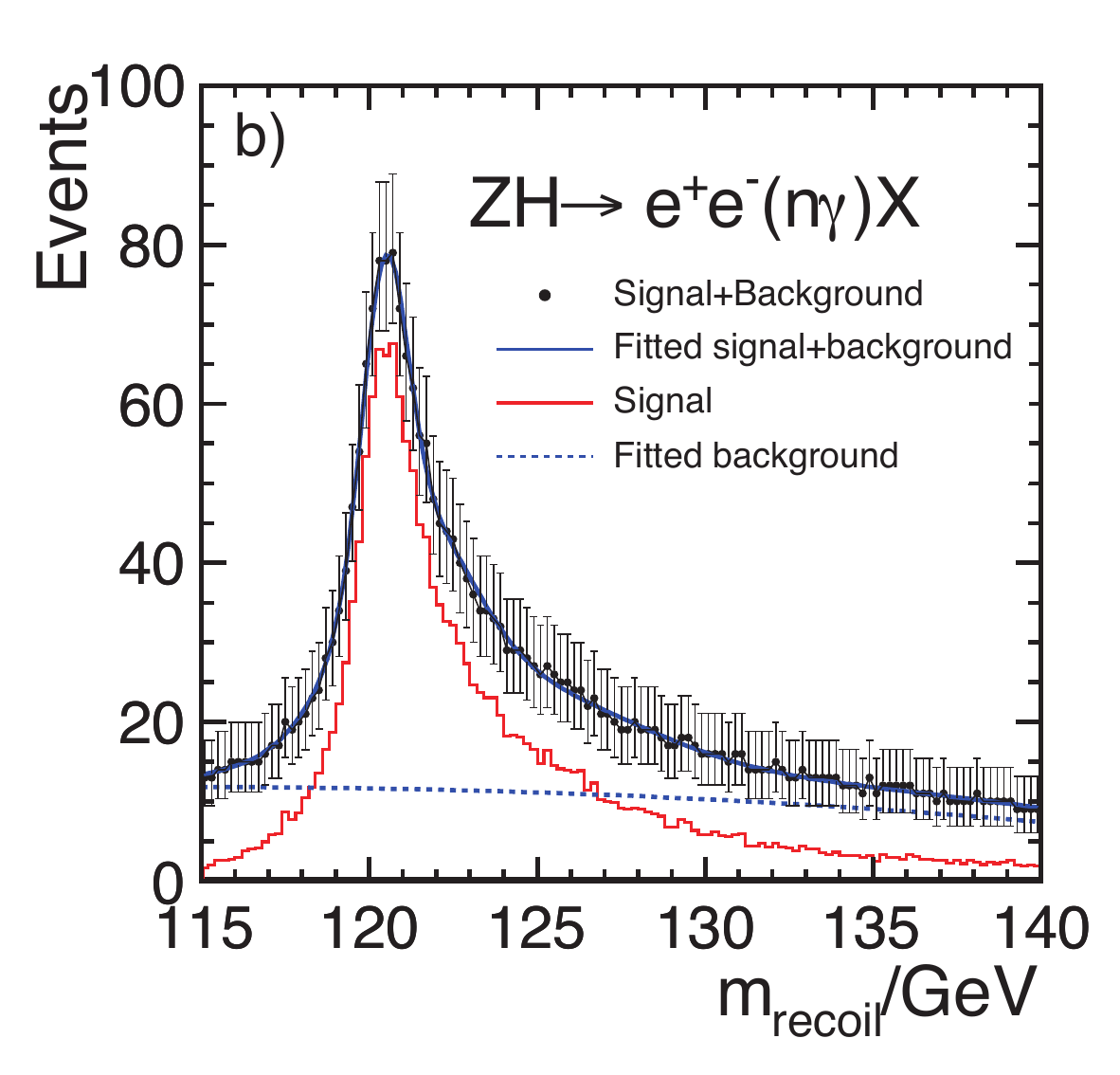}
\caption{Results of the model independent analysis of the Higgsstrahlung 
process $e^+e^- \to Zh$ at $\sqrt{s}=250$~GeV in which (a) $Z \to \mu^+\mu^-$  and 
(b) $Z \to e^+e^- (n\gamma)$. The results are shown for 
$P(e^+, e^-) = (+30 \%, -80 \%)$ beam polarization. \label{mass:mass:fig:Mrecoil}}
\end{figure}

%%%%%%%%%%%%%%%%%%%%%%%%%%%%%%%%%%%%%%%%%%%%%%%%%%%%%%%%%%%%%%
% llh at ecm=500 GeV
%%%%%%%%%%%%%%%%%%%%%%%%%%%%%%%%%%%%%%%%%%%%%%%%%%%%%%%%%%%%%%
\subsection{$l^+l^-h$ at $\sqrt{s}=500$~GeV}

\begin{figure}
\includegraphics[width=0.49\hsize, height=0.23\vsize, keepaspectratio=false]{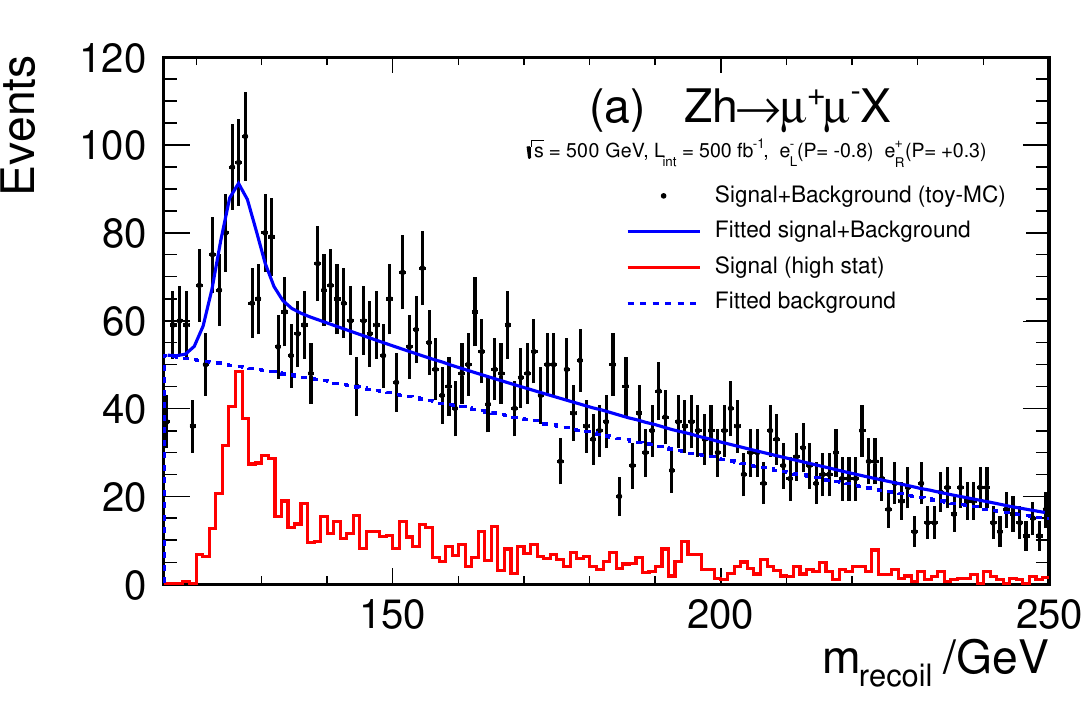} 
\includegraphics[width=0.49\hsize, height=0.23\vsize, keepaspectratio=false]{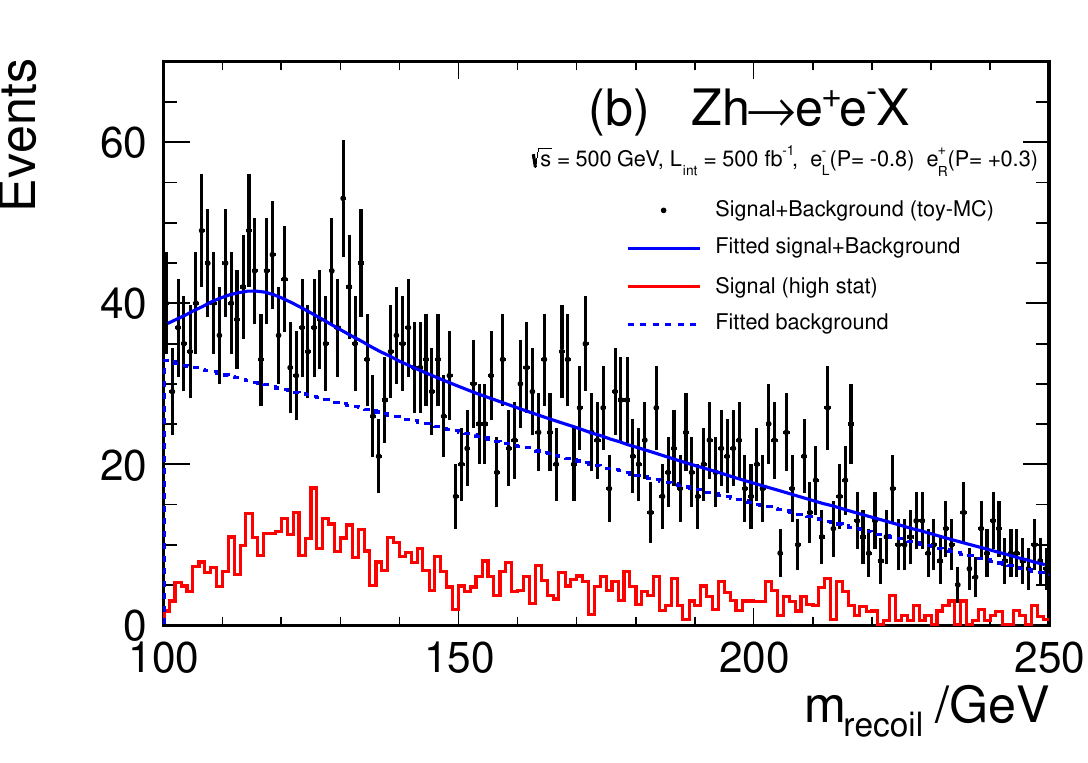}
\caption{Results of the model independent analysis of the Higgs-strahlung 
process $e^+e^- \to Zh$  at  $\sqrt{s}=500$~GeV in which (a) $Z \to \mu^+\mu^-$  and 
(b) $Z \to e^+e^- (n\gamma)$. The results are shown for 
$P(e^+, e^-) = (+30 \%, -80 \%)$ beam polarization. \label{fig:mumuheeh}}
\end{figure}

A Higgs recoil mass analysis has been done at $\sqrt{s} = 500$ GeV with ILD full detector simulation.
At $\sqrt{s} = 500$ GeV the $Zh$ cross section is about one-third compared to $\sqrt{s} = 250$ GeV.
Also there are numerous backgrounds from $t$-channel processes such as $ZZ$ at$\sqrt{s} = 500$ GeV.
Those aspects make the $Zh$ analysis at $\sqrt{s} = 500$ GeV less powerful than at $\sqrt{s} = 250$ GeV; 
however the result can be  combined with the $\sqrt{s} = 250$ result to improve the overall $Zh$ total cross section accuracy.

Firstly, lepton tagging is applied to both the muon and electron channels.
For the muon, the cluster energy is required to be smaller than half of the track energy.
For the electron, the ratio of the energy deposited in the ECAL to the total calorimeter energy must be greater than 90\%,
and the cluster energy is required to be between 70\% and 140\% with respect to the track energy.
For the electron channel, all neutral particles with $\cos\theta < 0.99$ with respect to the electron candidate
are added to the candidate electron to recover photons from final state radiation and bremsstrahlung.
If more than two lepton candidates are found, a pair giving the dilepton mass nearest to the $Z$ mass
is selected.

Cuts on the $Z$ mass, recoil mass, and di-lepton $p_T$ are applied.
Additional cuts are applied to the acoplanarity of the di-lepton system and the difference between the $p_T$ of the di-lepton and
the most energetic neutral particle.
Likelihood cuts are applied as the final cuts, with input variables of
di-lepton $p_T$, $Z$ mass, di-lepton $\cos\theta$ and acollinearity of di-lepton.

The resultant recoil mass distributions with fit are shown in Figure \ref{fig:mumuheeh}.
The cross section error is 6.49\% in $\mu\mu{}h$ channel and 7.10\% in $eeh$ channel.
The combined resolution for the $Zh\rightarrow\ell\ell{}h$ at $\sqrt{s} = 500$ GeV is 4.8\%.

%%%%%%%%%%%%%%%%%%%%%%%%%%%%%%%%%%%%%%%%%%%%%%%%%%%%%%%%%%%%%%
% Inclusive jet pair study
%%%%%%%%%%%%%%%%%%%%%%%%%%%%%%%%%%%%%%%%%%%%%%%%%%%%%%%%%%%%%%
\subsection{$q\bar{q}h$ at $\sqrt{s}=500$~GeV}

At $\sqrt{s}=500$ GeV, the total cross section for $e^{+}e^{-}\rightarrow q\bar{q}h$ is about 70~fb with 
an $e^{-}/e^{+}$ beam polarization of -80\%/+30\%.  About 35k such events are produced for 500 fb$^{-1}$ 
integrated luminosity. Unlike the situation at $\sqrt{s}=250$~GeV, the Z and h contain a significant boost at $\sqrt{s}=500$~GeV and 
the decay product jets  can be unambiguously  associated with the parent Z or h boson.  
The major background processes 
are 4-fermion $W^{+}W^{-}$, $Z^0Z^0$, and 2-fermion $q\bar{q}$.

  The energy of the  $Z$ from the $Zh$ process is more than 200 GeV, and  the two jets from the $Z$ 
overlap significantly.   Therefore, we reconstruct the hadronically decaying $Z$ as a single jet by the
$k_t$ jet algorithm with a jet radius of 1.2. From reconstructed jets, candidate jets are  preselected by
requiring that (1) the jet p$_{t}$ be greater than 50 GeV, (2) the jet mass be between 70 and 150 GeV/c$^{2}$, 
and (3) the jet energy be between 210 and 300 GeV.

With a fixed jet radius, both jet mass and jet energy are reduced if some decay products from the $Z$ are  outside
the jet radius. This effect was corrected by assuming a linear relationship between jet mass and jet energy;  in this way
 a better separation between $Z$ jets and non-$Z$ jets was achieved.  For the final selection 
we required that (1) the corrected jet mass be  between 87 and 105 GeV/c$^{2}$, (2) the maximum 
energy of a photon in the event be less than 100 GeV, (3) the number of particles in the jet be greater than 20, and
(4) the jet angle satisfy $|\cos\theta_{jet}|<0.7$.  

The recoil mass distribution of selected events is shown in Fig.~\ref{fig:incjet-recoil-mass-500}.
The figure shows the distribution after subtracting background events. The error bar and the central value of the histogram 
correspond to the actual event statistics. All standard model processes simulated for the ILC TDR
were considered as background. For the number of events with the recoil mass between 100 and 210 GeV/c$^{2}$, 
S/N is 11113/175437=0.063.  43\% of the backgrounds are due to 4-quark events through $ZZ$ and $WW$ processes. Other 4-fermion 
processes and 2-fermion hadron events constitute 26\% and 27\% of background events, respectively.  The relative cross section error for 
500 fb$^{-1}$ is 3.9\%~\cite{Miyamoto:2013zva}.

\begin{figure}
\begin{center}
\includegraphics[width=0.75\hsize]{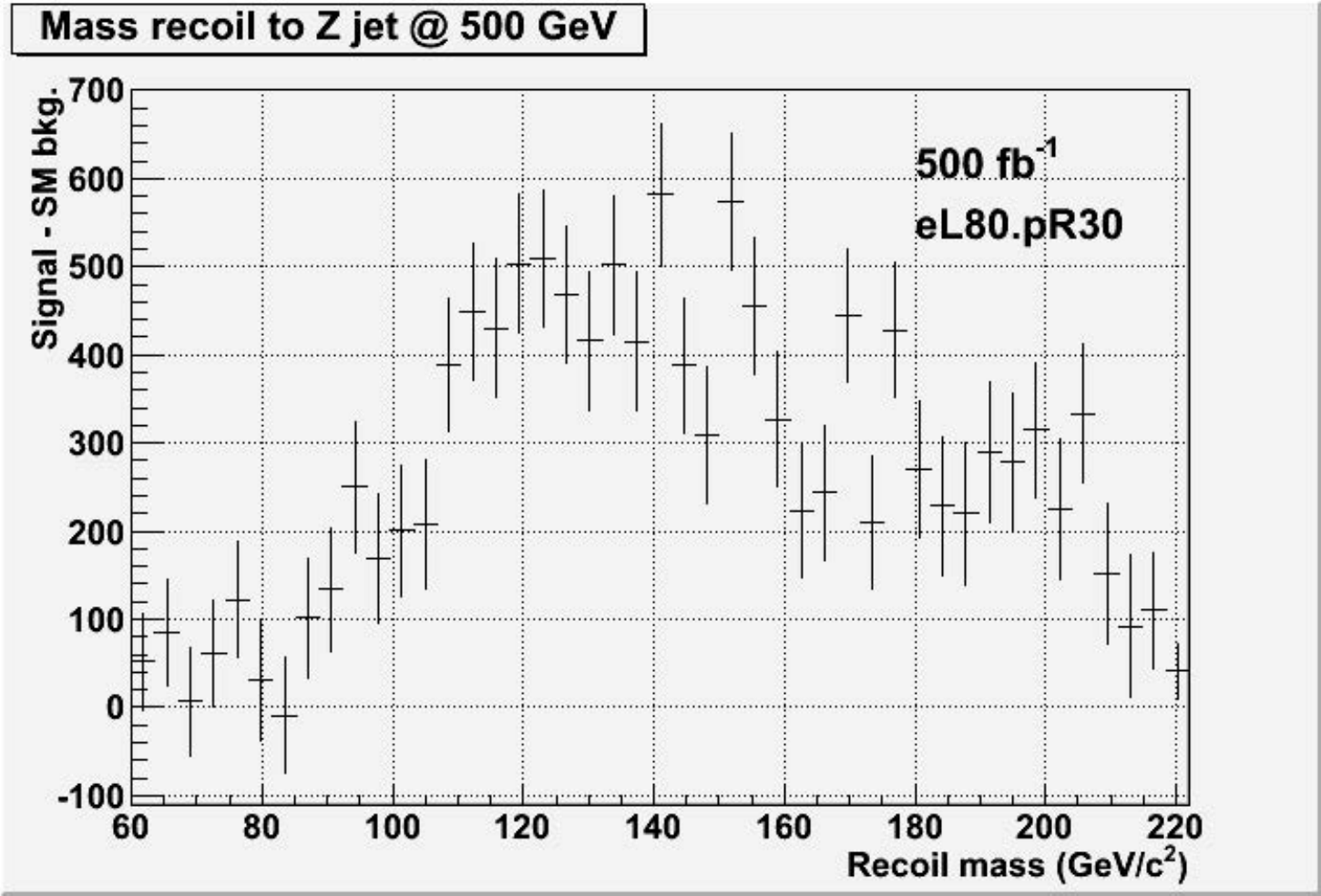}
\end{center}
\caption{The recoil mass of inclusive jet pair after subtracting background 
    processes.}
\label{fig:incjet-recoil-mass-500}
\end{figure}
%%%%%%%%%%%%%%%%%%%%%%%%%%%%%%%%%%%%%%%%%%%%%%%%%%%%%%%%%%%%%%%%%%%%%%%%%%%

\section{Higgs Spin Measurement}

The threshold behavior of the $Zh$ cross section has a characteristic
shape for each spin and each possible CP parity.  For spin 0,
the cross section rises as $\beta$ near the threshold for a CP even
state and as $\beta^3$ for a CP odd state.   For spin 2, 
 for the canonical 
form of the coupling to the energy-momentum tensor, the rise is also
$\beta^3$.
If the spin
 is higher than 2, the cross section will grow as a higher power of
 $\beta$. 
With a three-$20$\,fb$^{-1}$-point threshold scan of the $e^+e^- \to
Zh$ 
production cross section we can
separate these possibilities~\cite{Dova:2003py}
as
 shown in Fig.~\ref{fig:ZH:JPC}.  
 The discrimination of more 
general forms of the coupling is possible by the use of angular 
correlations in the boson decay; this is discussed in detail in
\cite{Miller:2001bi}.

At energies well above the $Zh$ threshold, the $Zh$ process will
 be dominated by longitudinal $Z$ production as implied by  the 
equivalence theorem. The reaction will then behave like a scalar pair 
production, showing the characteristic $\sim \sin^2\theta$ dependence
 if the $h$ particle's spin is zero. The measurement of the angular 
distribution will hence strongly corroborate that the $h$ is indeed a
scalar particle.

%%%%%%%%%%%%%%%%%%%%%%%%%%%%%%%%%%%%%%%%%%%%%%%%%%%%%%%%%%%%%%%%%
\begin{figure}
\begin{center}
\includegraphics[width=0.75\hsize]{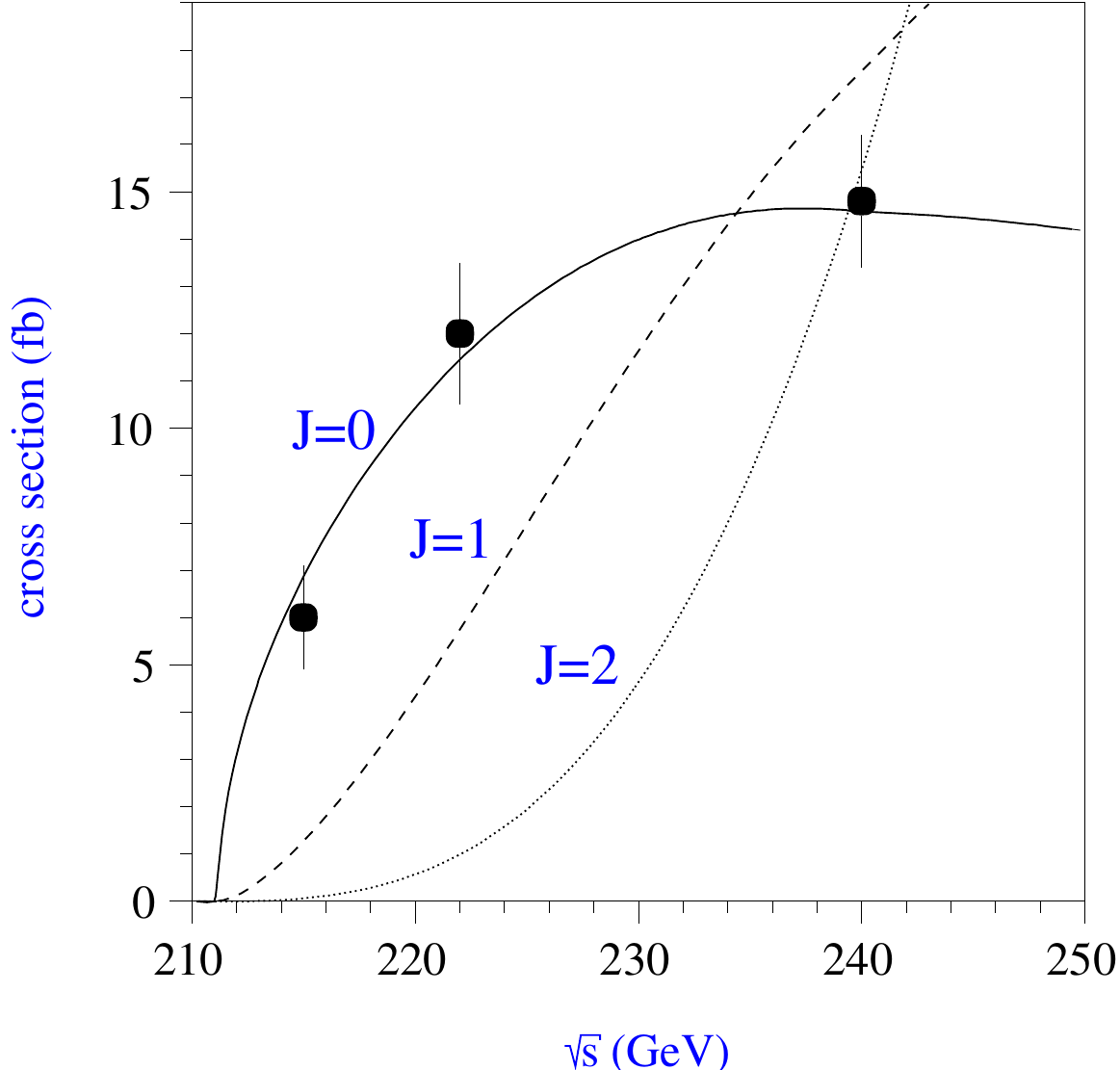}
\end{center}
\caption{ Threshold scan of the $e^+e^- \to Zh$ process for $m_h =
120$\,GeV, compared with theoretical predictions for  $J^{P}= 0^{+}$,
$1^{-}$, and $2^{+}$ \cite{Dova:2003py}.}
\label{fig:ZH:JPC}
\end{figure}
%%%%%%%%%%%%%%%%%%%%%%%%%%%%%%%%%%%%%%%%%%%%%%%%%%%%%%%%%%%%%%%%%%%%%%%%%%%
%%%%%%%%%%%%%%%%%%%%%%%%%%%%%%%%%%%%%%%%%%%%%%%%%%%%%%%%%%%%%%

%Here discuss new results from Katja Krueger presented at Seattle July 2013 meeting.

\section{Higgs Sector CP Measurements}
\subsection{Introduction }

The analytic power of the ILC is emphasized when we consider more 
detailed questions.
It is possible that  the $h$ is not a CP eigenstate but rather a
mixture of CP even and CP   odd components.   This occurs if there is
CP violation in the  Higgs sector.    It is known that CP violation
from the CKM matrix cannot explain the cosmological excess of baryons
over antibaryons; thus, a second source of CP violation in nature is
needed.   One possibility is that this  new CP violation comes from
the Higgs sector and gives rise to net baryon number at the
electroweak phase transitions, through mechanisms that we will discuss
in Section 9.1  of this report.   For these models, the $h$ mass
eigenstates can be mainly CP even but contain a small admixture of a
CP odd component.

\subsection{$e^+e^-\rightarrow ZH$ }

 A small CP odd contribution to the $hZZ$ coupling
can affect the threshold behavior.  The right-hand side of
Fig.~\ref{fig:ZH:CP}
shows the determination of this angle at a center of mass energy  of
350~GeV from the  value of the total cross section and from an 
appropriately defined optimal observable~\cite{Schumacher:2001ax}.

%%%%%%%%%%%%%%%%%%%%%%%%%%%%%%%%%%%%%%%%%%%%%%%%%%%%%%%%%%%%%%%%%
\begin{figure}
\begin{center}
\includegraphics[width=0.75\hsize]{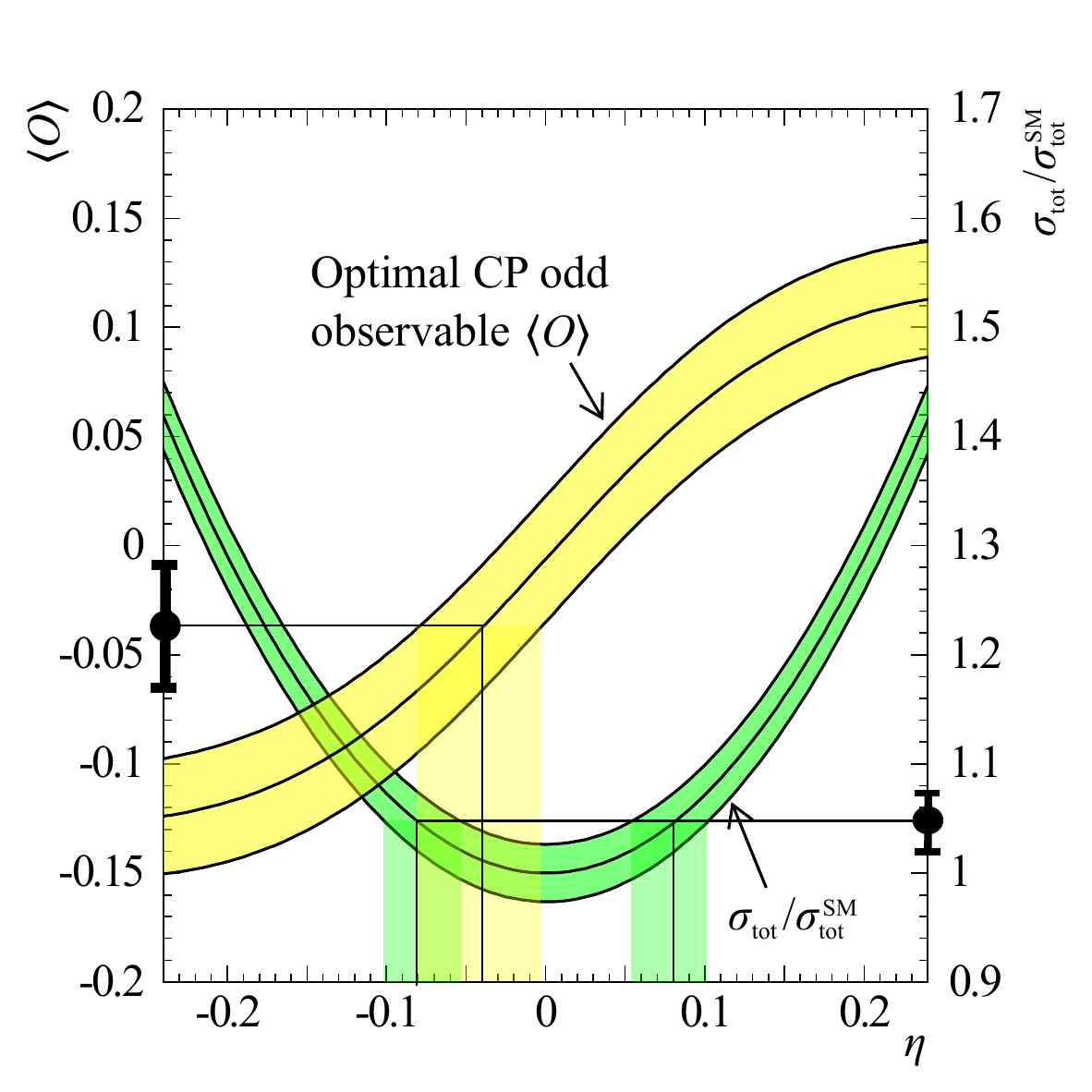}
\end{center}
\caption{  Determination of 
$CP$-mixing with $1$-$\sigma$ bands expected at
 $\sqrt{s}=350$\,GeV and $500$\,fb$^{-1}$ \cite{Schumacher:2001ax}. }
\label{fig:ZH:CP}
\end{figure}
%%%%%%%%%%%%%%%%%%%%%%%%%%%%%%%%%%%%%%%%%%%%%%%%%%%%%%%%%%%%%%%%%%%%%%%%%%%

%Here discuss new results from A. Whitbeck presented at July 2013 Seattle meeting \cite{Anderson:2013}.
A new result was presented during the Snowmass study \cite{Anderson:2013}
for the CP mixing that would appear in the $hZZ$ coupling in both the $pp$ and $e^+e^-$ colliders.
The analysis utilized a simplified detector simulation based on the smearing of parton-level information and simply assumed $30\,\%$ efficiency and $10\,\%$ background for the signal process: $e^+e^- \to Zh \to \mu^+\mu^- b \bar{b}$.
From the cross section and the observables concerning the production and decay angles of both the $Z$ and $h$ bosons, the analysis estimated the expected sensitivity to the effective fraction of events due to the  CP violating coupling, $f_{a3}$, which was then translated to that of the corresponding fraction of the anomalous contribution for the Higgs to two vector boson decays, $f_{a3}^{\mathrm dec}$, used in earlier CMS studies.
If the $Zh$ cross-section is first measured at the center of mass energy, $\sqrt{s}=250\,$GeV ($250\,$fb$^{-1}$), and then at $350$ ($350\,$fb$^{-1}$), $500$ ($500\,$fb$^{-1}$), and $1000\,$GeV ($1000\,$fb$^{-1}$), $f_{a3}$ can be measured to $0.035$, $0.041$, and $0.055$, respectively, which would translate to precision on $f_{a3}^{\mathrm dec}$ of $10^{-4}$, $4 \times 10^{-5}$, and $10^{-5}$, respectively.
However, the relative contributions of various possible anomalous couplings to the cross section might depend on the underlying dynamics that would appear as form factors in the anomalous couplings and would depend on the virtuality of $Z^*$.
At the ILC, the $q^2$ dependence can be separated by performing angular analyses separately at different energies since the virtuality of $Z^*$ is fixed at a fixed center-of-mass energy.
The expected precision of $f_{a3}$ is in the range of $0.03$ - $0.04$, being independent of the center-of-mass energy, and translates to $7 \times 10^{-4}$ to $8 \times 10^{-6}$, entering a region sensitive to a possible loop-induced CP-violating contribution.

\subsection{$H\rightarrow \tau^+\tau^-$ }

Tests of mixed CP property using the  $hZZ$ coupling may not be the
most  effective ones, since the CP odd $hZZ$ coupling is of higher
dimension and may be generated only through loops.  It is more
effective to use a coupling for which the CP even and CP odd
components are on the same footing.    An example is the $h$  coupling 
to $\tau^+\tau^-$, given by 
\beq 
   \Delta \L =     - {m_\tau\over v} h\  \bar \tau (\cos\alpha + i
   \sin\alpha \gamma^5) \tau 
\label{eq:taucouple}
\eeq{taucouple}
for a Higgs boson with a CP odd component.   The polarizations of the
final state $\tau$s can be determined from the kinematic distributions 
 of their
decay
products; the CP even and odd components interfere in these 
distributions~\cite{Kramer:1993jn}.   
In \cite{Desch:2003rw}, it
% KF 09/28/2013
%is estimated that the angle $\alpha$ can be determined at the ILC to
%an accuracy of 6$^\circ$.
is estimated that the angle $\alpha$ can be determined to
an accuracy of 6$^\circ$ with $1\,$ab$^{-1}$at $\sqrt{s}=350\,$GeV
in the case of maximal CP mixing, $\alpha=\pi/4$.
A similar study has been performed in \cite{Reinhard:2009} for a $120\,$GeV Higgs boson 
assuming the baseline ILC machine running at $\sqrt{s}=230\,$GeV for
an integrated luminosity of $500\,$fb$^{-1}$ with beam polarizations
of $(P_{e^-}, P_{e^+})=(-0,8,+0.3)$.
A full simulation for the $e^+e^- \to Zh \to \mu^+\mu^- \tau^+\tau^-$ mode 
in the study showed that with an inclusion of other $Z$ decay modes 
an expected statistical precision of $\Delta\alpha=0.135$ (i.e. $28\,\%$) could be
achieved for $\alpha=-\pi/8$ given the baseline integrated luminosity of $500\,$fb$^{-1}$.

% Say something here about the fact that maximal CP even-odd mixing is assumed in the Desch study.

\subsection{$e^+e^- \to t\bar{t}H$ }

 In the presence of CP violation, only the CP--even
component of the $HZZ$ coupling is projected out in Higgs decays to $ZZ$.  
 The $ZZ$  couplings of a pure
CP--odd $A$ state are zero at tree--level and are generated only through tiny
loop corrections.

The decays of the Higgs boson to fermions provide a more democratic probe of its CP
nature since, in this case, the CP--even and CP--odd components can have the
same magnitude. One therefore needs to look at channels where the Higgs boson is
produced and/or decays through these couplings.

A promising production mode for studying the Higgs CP properties is
 $\ee \to t\bar t H$.
The production of a spin 0 state with
arbitrary model-independent CP properties in association with a top
quark pair at the ILC was investigated in Ref.~\cite{BhupalDev:2007is,Godbole:2011hw}.
The CP properties of
the Higgs coupling to the top quarks were parametrized in a
model-independent way by a parameter $a$ for a CP-even Higgs, by a
parameter $b$ for a CP-odd Higgs and by simultaneously non-vanishing
$a$ and $b$ for a CP-mixed state:
\beq
C_{tt\Phi} = -i g_{ttH} (a+ib \gamma_5)\;.
\eeq
Notice that in the Standard Model, $a=1$ and $b=0$.

 These parameters were determined
by a measurement of the total cross section, the polarization
asymmetry of the top quark and the up-down asymmetry of the antitop
quark with respect to the top-electron plane. The former two
observables are CP-even and can be exploited to distinguish a CP-even
from a CP-odd Higgs boson.  Since the up-down asymmetry $A_\phi$ is
CP-odd, it can be exploited directly and unambiguously to test CP violation. 

The sensitivities to $a$ and $b$ were studied in each observable
separately before investigating the combination of all three observables. 
It was found that the total cross section is most sensitive to $a$ and to
some extent to $b$. The observables $P_t$ and $A_\phi$ do not exhibit
much sensitivity to $a$ and $b$, although polarization of the initial
$e^\pm$ beams slightly improves the sensitivity in case of $P_t$.  
The combination of all three observables, however, reduces
the error on $a$ for polarized $e^\pm$ beams as shown in Fig.~\ref{fig:tthCP}.
%%%%%%%%%%%%%%%%%%%%%%%%%%%%%%%%%%%%%%%%%%%%%%%%%%%%%%%%%%%%%%%%%
\begin{figure}
\begin{center}
\includegraphics[width=0.75\hsize]{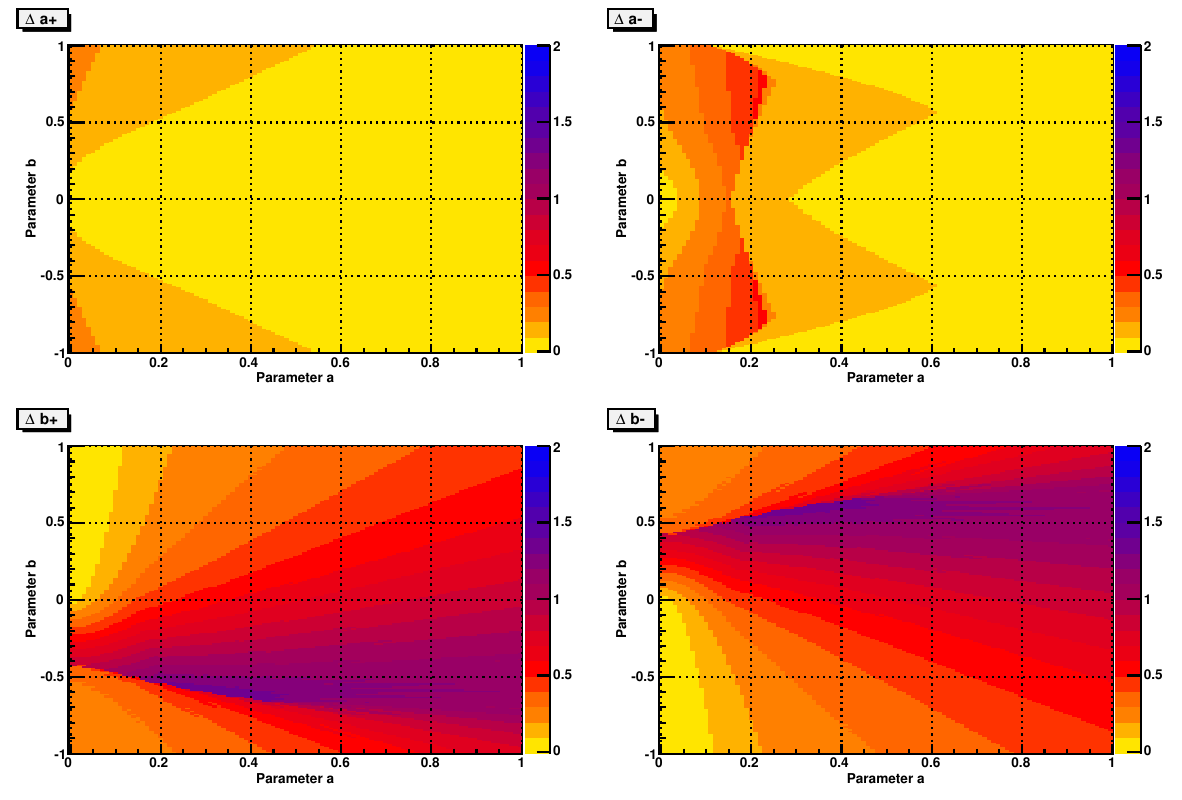}
\end{center}
\caption{ Errors $\Delta a^+$ (upper left) and
  $\Delta a^-$ (upper right)  on $a$ as well as $\Delta b^+$ (lower
  left) and $\Delta b^-$ (lower right) on $b$, by combining all 3
  observables $\sigma, P_t, A_\phi$, at $1 \sigma$ confidence level
  for $M_\Phi=120$ GeV and $\sqrt{s}=800$ GeV with ${\cal  L} = 500$
  fb$^{-1}$.  The electron and positron beams are polarized with $(P_{e^-}, P_{e^+})=(-0,8,+0.3)$.
  The colour code
  indicates the magnitude of the respective error.}
\label{fig:tthCP}
\end{figure}
%%%%%%%%%%%%%%%%%%%%%%%%%%%%%%%%%%%%%%%%%%%%%%%%%%%%%%%%%%%%%%%%%%%%%%%%%%%
If we assume that $a^2+b^2=1$ and parametrize $a$ and $b$ as $a = \cos\phi$ and $b=\sin\phi$,
as in eq. (\ref{eq:taucouple}) for $h \to \tau^+\tau^-$, then the cross section alone will be a measure of the mixing angle,
$\phi$.
Fig.\ref{fig:tth-cp}-(a) shows the $e^+e^- \to t\bar{t}h$ cross section as a function of $\sin^2\phi$ at three
different center of mass energies: $\sqrt{s}=500, 520,$ and $1000\,$GeV.
The cross section values are translated into the expected 1-$\sigma$ bounds and shown in Fig.\ref{fig:tth-cp}-(b) as a function of $\sin^2\phi$ for the three energies assuming $500\,$fb$^{-1}$ for $\sqrt{s}=500$ and $520\,$GeV, and the baseline $1\,$ab$^{-1}$ and the upgraded $2.5\,$ab$^{-1}$ at $\sqrt{s}=1\,$TeV \cite{Tanabe:2013tth-cp}.
The figure tells us that the contribution from the CP-odd component could be constrained to $\sim 5\%$ at 1-$\sigma$.
%%%%%%%%%%%%%%%%%%%%%%%%%%%%%%%%%%%%%%%%%%%%%%%%%%%%%%%%%%%%%%%%%
\begin{figure}
\begin{center}
\includegraphics[width=0.45\hsize]{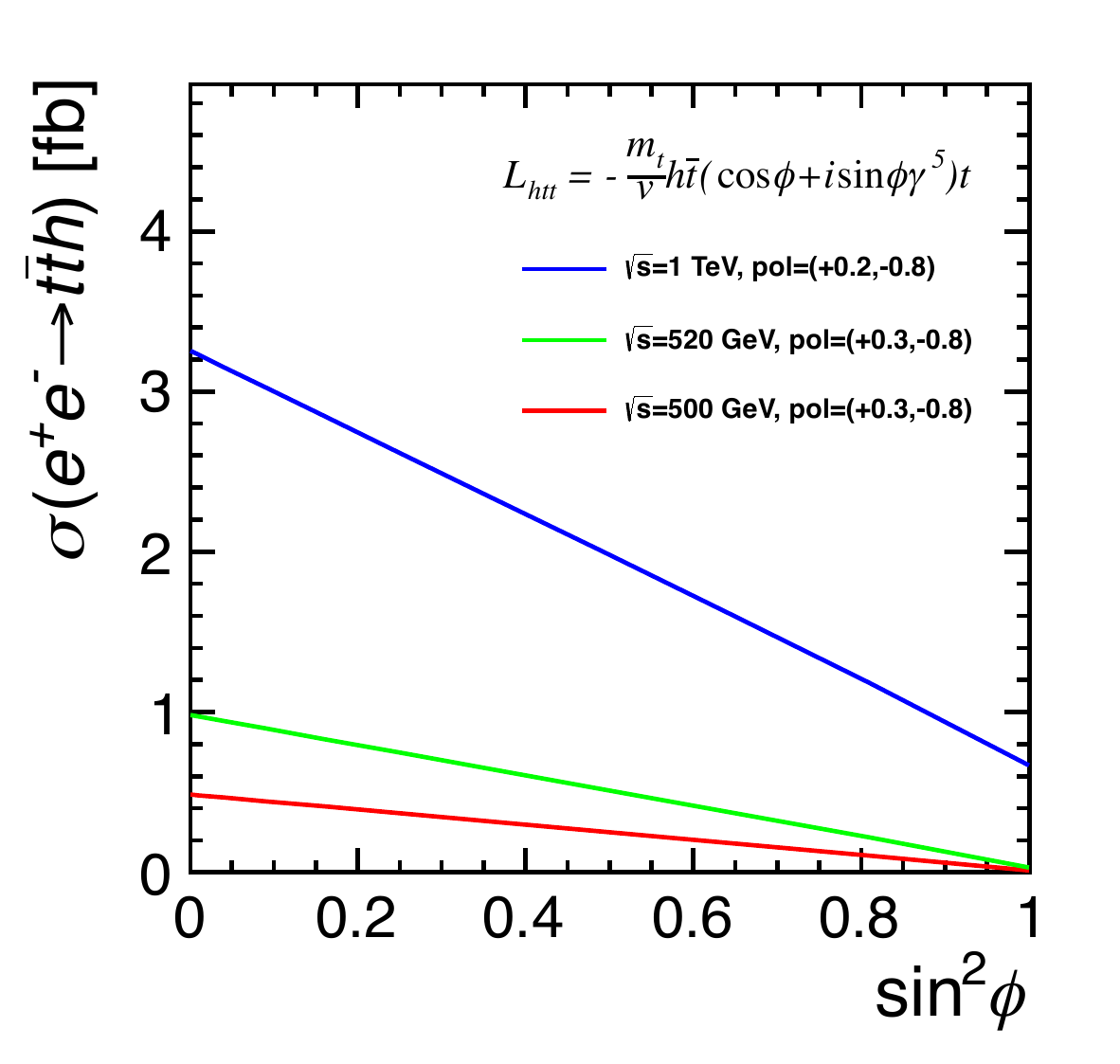}
\includegraphics[width=0.45\hsize]{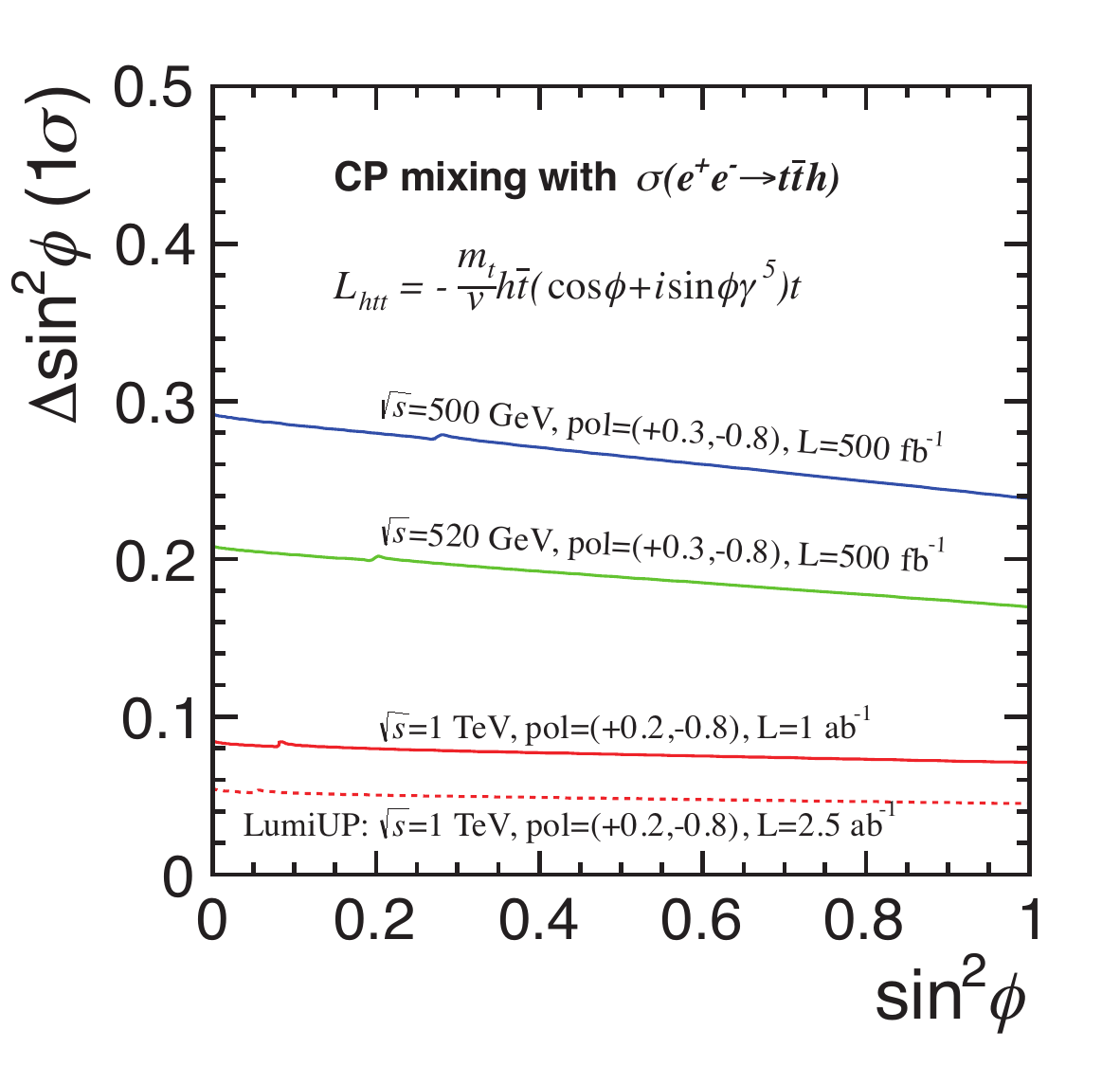}
\end{center}
\caption{(a) $e^+e^- \to t\bar{t}h$ cross section as a function of $\sin^2\phi$ at three
different center of mass energies: $\sqrt{s}=500, 520,$ and $1000\,$GeV,
(b) the expected 1-$\sigma$ bound on the CP-odd contribution, $\Delta\sin^2\phi$, as a function of $\sin^2\phi$ at the three energies. 
The corresponding beam polarizations and integrated luminosities are indicated in the figure.}
\label{fig:tth-cp}
\end{figure}
%%%%%%%%%%%%%%%%%%%%%%%%%%%%%%%%%%%%%%%%%%%%%%%%%%%%%%%%%%%%%%%%%%%%%%%%%%%

\chapter{Cross Section Times Branching Ratio Measurements I \label{sid:chapter_sigma_times_br_i}}
%\section{Introduction to Cross Section Times Branching Ratio Measurements at the ILC}

The measurement accuracies of the cross section times branching ratio ($\sigma\cdot BR$)
for Higgs decay to $b\bar{b}$, $c\bar{c}$, $gg$, $WW^*$ and $\tau^+\tau^-$ are 
described in this chapter.

%%%%%%%%%%%%%%%%%%%%%%%%%%%%%%%%%%%%%%%%%%%%%%%%%%%%%%%%%%%%%%%%%%%%%%%%%%%%%%%%%%%%%%%%%%%%
\section{$h\rightarrow b\bar{b},\ c\bar{c},\ gg$}

%%%%%%%%%%%%%%%%%%%%%%%%%%%%%%%%%%%%%%%%%%%%%%%%%%%%%%%%%%%%%%%%%%%%%%%%%%%%%%%%%%%%%%%%%%%%
\subsection{250 GeV and 350 GeV}

The measurement accuracies of the cross section times branching ratio $\Delta(\sigma\cdot BR)$ 
for Higgs decays to $b\bar{b}$, $c\bar{c}$ and gluons were studied in the ILD and SiD LOI's~\cite{Aihara:2009ad,Abe:2010aa} 
at $\sqrt{s}=250$ GeV assuming $m_h = 120$~GeV. A comprehensive study 
at 250 GeV and 350 GeV with $m_h = 120 $~GeV was reported in Ref.~\cite{Ono:2013higgsbr},
which is presented below. 

At these energies  the Higgsstrahlung process ($e^+e^-\rightarrow Zh)$ is the dominant contribution to 
the Higgs production.  Therefore, the event signatures of 4-jet($q\bar{q}h$) and 2-lepton ($e^+e^- or \mu^+\mu^-$)+2-jet ($\ell\bar{\ell}h$) 
were studied in addition to missing energy + 2-jet ($\nu\bar{\nu}h$) events.
In the case of the 4-jet analysis, the particles in the event were forcibly clustered to four-jets,
from which the dijet pairs for the $h$ and $Z$ candidates were selected 
as the pairs which minimized the dijet mass $\chi^2$ for $Z$ and $h$ bosons.
The background events were rejected by cuts on the number of tracks for each jet, 
the maximum scaled jet mass ($y_{max}$) needed to cluster as four jets, the thrust, 
the thrust angle and the Higgs production angle. The kinematical constraint fit was 
applied to the four jets to improve background rejection. Finally, the likelihood ratio (LR) was 
derived from the thrust, $\cos\theta_{thrust}$, the minimum angle between all the jets, 
the number of particles in the Higgs candidate jets, and the fitted $Z$ and Higgs masses.
The cut position to select 4-jet candidates was chosen to maximize the signal significance.
The background fractions after all cuts are 80\% $q\bar{q}q\bar{q}$ and 20\% $q\bar{q}$ at 250 GeV, and
60\% $q\bar{q}q\bar{q}$, 30\% $q\bar{q}$ and 10\% $t\bar{t}$ at 350 GeV

In the case of the 2-lepton + 2-jet mode an event must have an $e^+e^-$ or $\mu^+\mu^-$ pair with mass 
consistent with the $Z$, and the mass of everything else must be consistent with the $h$. Additionally, 
cuts on the production angle of the $Z$ and the lepton pair recoil mass were applied 
to improve the signal to noise ratio.

In the analysis of the missing energy+2-jet mode all visible objects were forced
into two jets, and the four vector sum of the two jets had to have a $P_T$ 
and  mass consistent with the Higgs. In contrast to the 1 TeV study, the recoil mass calculated from the two jets was
required to be consistent with the $Z$ mass because the Higgsstrahlung process is dominant at these energies; in addition 
this cut was effective in reducing backgrounds from non-Higgs four fermion processes. The likelihood ratio (LR)
was formed from the recoil mass, the number of particles, the jet momentum,  the jet pair mass and 
the minimum of the scaled jet mass for forced 2-jet clustering.  

With 250 fb$^{-1}$ at 250 GeV (250 fb$^{-1}$ at 350 GeV), the signal significance,  $S/\sqrt{S+B}$, is 
47.9 (66.4) for $\nu\bar{\nu}h$, 32.3(47.1) for $q\bar{q}h$, 22.4(16.7) for $e^+e^-h$ and 
28.2 (19.2) for $\mu\bar{\mu}h$.  

In order to evaluate the flavor content of the selected events, the flavor likeness of dijet events 
was calculated by LCFIPlus and fitted by a template consisting of $h\rightarrow b\bar{b}$, $h\rightarrow c\bar{c}$, $h\rightarrow gg$,
other Higgs decays and the standard model processes. Pseudo experiments
were performed with the fraction of $b\bar{b}$, $c\bar{c}$ and $gg$ as free parameters, and $\Delta \sigma\cdot BR$  was  determined by the widths of the fitted distribution.    The results are summarized in 
Table~\ref{table:bb-cc-gg-sigma-br}

\thisfloatsetup{floatwidth=\hsize,capposition=top}
\begin{table}[!h]
\caption[Sensitivity to $\sigma\cdot BR$ for $h\rightarrow b\bar{b}, c\bar{c}$, $gg$]
{
\label{table:bb-cc-gg-sigma-br}
Summary of the sensitivity to 
$\sigma\cdot BR$ for Higgs decay to $b\bar{b}$, $c\bar{c}$, $gg$
at 250 GeV with 250 fb$^{-1}$ and $P(e^{-})/P(e^{+}) = -80\%/+30\%$
and 
350 GeV with 250 fb$^{-1}$ and $P(e^{-})/P(e^{+}) = -80\%/+30\%$.
$m_h=120$~GeV was used for this analysis.
}
\begin{center}
\begin{tabular}{lllllll}
\toprule
 Energy & channel & missing+2-jet & 4-jet & $e^+e^-$+2-jet & $\mu^+\mu^-$+2-jet & Combined \\
\midrule
250 GeV &
${\Delta\sigma\cdot BR \over \sigma\cdot BR} (h\rightarrow b\bar{b})$ & 
1.7 & 1.5 & 3.8 & 3.3 & 1.0 \\
& ${\Delta\sigma\cdot BR \over \sigma\cdot BR} (h\rightarrow c\bar{c})$ &
11.2 & 10.2 & 26.8 & 22.6 & 6.9 \\ 
& ${\Delta\sigma\cdot BR \over \sigma\cdot BR} (h\rightarrow gg)$ & 
13.9 & 13.1 & 31.3 & 33.0 & 8.5 \\
\midrule
350 GeV &
${\Delta\sigma\cdot BR \over \sigma\cdot BR} (h\rightarrow b\bar{b})$ & 
1.4 & 1.5 & 5.3 & 5.1 & 1.0 \\
& ${\Delta\sigma\cdot BR \over \sigma\cdot BR} (h\rightarrow c\bar{c})$ &
8.6 & 10.1 & 30.5 & 30.9 & 6.2 \\
& ${\Delta\sigma\cdot BR \over \sigma\cdot BR} (h\rightarrow gg)$ & 
9.2 & 13.7 & 35.8 & 33.0 & 7.3 \\
\bottomrule
\end{tabular}
\end{center}
\end{table}

The $\sigma\cdot BR$ accuracies for a Higgs with 125~GeV mass were obtained by scaling the number 
of signal events according to the branching ratio while keeping the number of background events
the same.  From this extrapolation
${\Delta\sigma\cdot BR \over \sigma\cdot BR}$ for $h\rightarrow b\bar{b}, c\bar{c},$ and $gg$ 
were estimated to be 1.2\%, 8.3\% and 7.0\%, respectively, assuming 250~fb$^{-1}$ at $\sqrt{s}=250$~GeV\cite{Junping-keisuke-fitting}.

The $\nu_{e}\bar{\nu_{e}}h$ $WW-$fusion channel was studied in the $h\rightarrow b\bar{b}$ channel at $\sqrt{s}=250$~GeV
using the ILD full simulation\cite{durig-jenny-lcws2012}.
From a $\chi^{2}$ fit of the  missing mass distribution, the contributions from the
$WW$-fusion channel and the Higgsstrahlung channel were separated. A measurement 
accuracy of ${\Delta\sigma_{\nu\bar{\nu}h}\cdot BR \over  \sigma_{\nu\bar{\nu}h}\cdot BR}=0.11$ was obtained with 250 fb$^{-1}$ assuming 
$P(e^{-})/P(e^{+})=-80\%/+30\%$ and $m_h=126$~GeV.

%%%%%%%%%%%%%%%%%%%%%%%%%%%%%%%%%%%%%%%%%%%%%%%%%%%%%%%%%%%%%%%%%%%%%%%%%%%%%%%%%%%%%%%%%%%%
\subsection{500 GeV}

The Higgs decay to $b\bar{b}$ in the process $e^+e^-\rightarrow \nu\bar{\nu}h$ at 500 GeV was studied 
by ILD\cite{durig-jenny-lcws2012} using full simulation with $m_h=125$~GeV. 

In order to remove piled up background particles, the anti-$k_t$ jet algorithm was employed 
with the jet size parameter $R=1.5$. 
Events with an isolated muon or electron were removed, and events with 2 $b$-tagged jets were selected.
The visible energy and missing $P_T$ were required to be consistent with $\nu\bar{\nu}$ production, and
the recoil mass opposite the dijet was required to be greater than 172~GeV to reject
$Z\rightarrow \nu\bar{\nu}$ events. The dijet mass distribution for selected 
events is shown in Figure~\ref{fig:vvh-visible-dijet-mass-500GeV}.

The signal events were selected with 66\% efficiency and a signal-to-noise ratio of 3.7.   The main background 
was $e^+e^- \rightarrow \nu\bar{\nu}Z$, which is labelled as \texttt{4f\_sznu\_sl} in Figure~\ref{fig:vvh-visible-dijet-mass-500GeV}.
The signal significance for the $h\rightarrow b\bar{b}$ channel  was 150 and $\Delta(\sigma\cdot BR)/(\sigma\cdot BR)=0.667\%$.
% Half width
\thisfloatsetup{floatwidth=\SfigwFull,capposition=beside}
\begin{figure}[h]
  \centerline{\includegraphics[width=0.8\hsize]{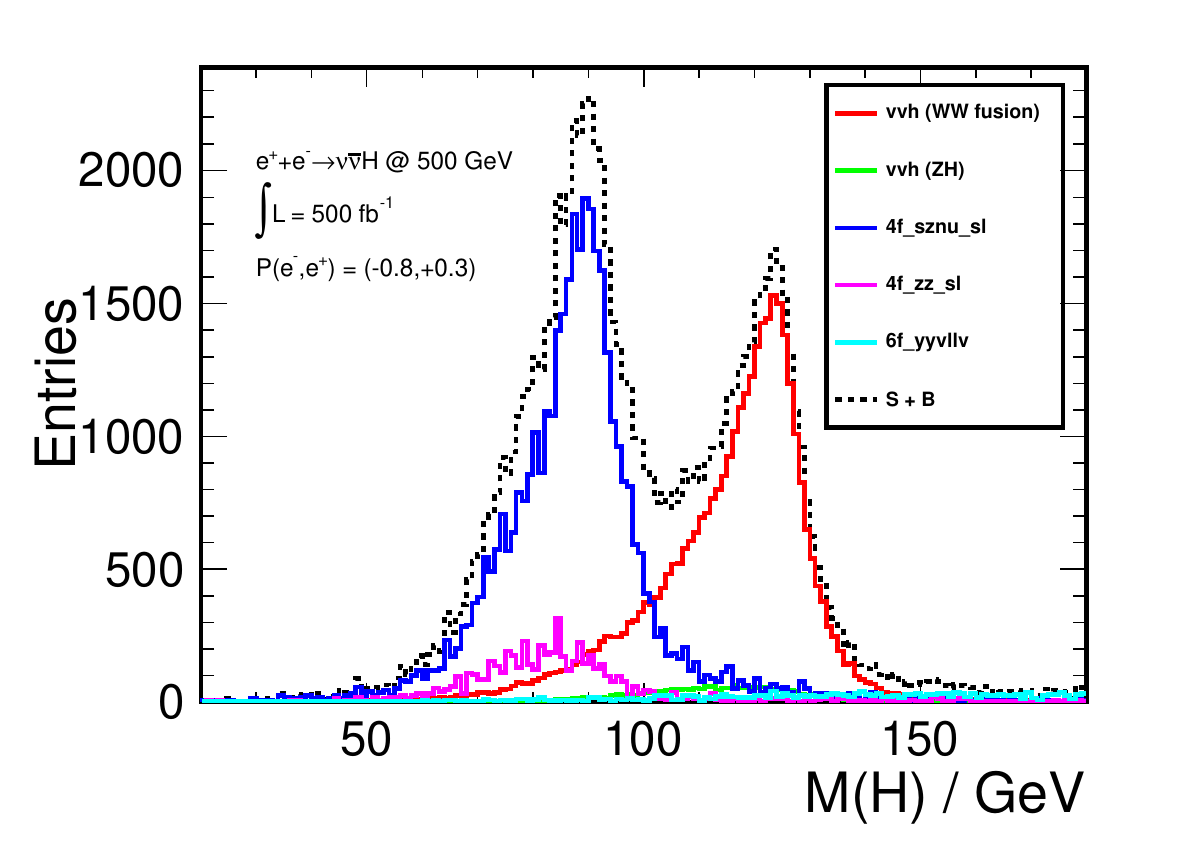}}
  \caption{The dijet mass distribution of $e^+e^-\rightarrow \nu\bar{\nu}h\rightarrow \nu\bar{\nu}b\bar{b}$ 
at $\sqrt{s}=500$~GeV assuming 500 fb$^{-1}$, $P(e^{-})/P(e^{+}) = -80\%/+30\%$, 
  and $m_h=125$~GeV. }
\label{fig:vvh-visible-dijet-mass-500GeV}
\end{figure}

%%%%%%%%%%%%%%%%%%%%%%%%%%%%%%%%%%%%%%%%%%%%%%%%%%%%%%%%%%%%%%%%%%%%%%%%%%%%%%%%%%%%%%%%%%%%
The decays $h\rightarrow c\bar{c}$ and $gg$ were studied at 500~GeV\cite{Ono:2013higgsdbd}  
using the ILD full simulation samples for the ILC TDR.
The analysis strategy is similar to the 1 TeV case in order to select  Higgs production via the 
$WW$ fusion process $e^+e^-\rightarrow \nu_{e}\bar{\nu_e}h$.  However, since 
no  low~$p_T$ $\gamma\gamma\rightarrow\ \mathrm{hadron}$ background was overlaid the 
Durham jet clustering algorithm\cite{Catani:1991} was applied instead of the 
$k_t$ algorithm.  Events with two jets were selected by cuts on $P_T$, $P_Z$, $P_{max}$, 
$N_{charged}$ with efficiencies of  57\%, 46\%, 65\% for $h\rightarrow b\bar{b}$, $c\bar{c}$ 
and $gg$, respectively. Among the background processes considered, $\nu\bar{\nu}q\bar{q}$ 
and $\nu\ell q\bar{q}$ were the largest.  Flavor composition was determined using
the template method described above. The following sensitivities were obtained assuming  500 fb$^{-1}$ and 
$P(e^{-})/P(e^{+})=-80\%/+30\%$:  
$\Delta(\sigma\cdot BR)/(\sigma\cdot BR)=0.6\%(b\bar{b})$, $5.2\%(c\bar{c}$) and $5.0\%(gg)$.
The result can be extrapolated to the case of $m_h=125$~GeV by scaling the signal yield by 
the total cross section and the branching ratio:
$\Delta(\sigma\cdot BR)/(\sigma\cdot BR)=0.66\%(b\bar{b})$, $6.2\%(c\bar{c}$) and $4.1\%(gg)$.
Note that this result for $h\rightarrow b\bar{b}$  
is consistent with the dedicated $h\rightarrow b\bar{b}$ study described earlier.

The results from the $Zh$ process were obtained by extrapolating the 250~GeV full simulation results. 
The number of signal and background events before template fitting were scaled according 
to the cross section, and then they were extrapolated according to the enhanced statistical significance from 
the template fitting. As a result, $\Delta\sigma \cdot BR \over \sigma \cdot BR$ 
with 500 fb$^{-1}$ at 500 GeV with $(-80\%, +30\%$) polarization was estimated 
to be 1.8\%, 13\%, and 11\% for $h\rightarrow b\bar{b}, c\bar{c},$ and $gg$ respectively.

%%%%%%%%%%%%%%%%%%%%%%%%%%%%%%%%%%%%%%%%%%%%%%%%%%%%%%%%%%%%%%%%%%%%%%%%%%%%%%%%%%%%%%%%%%%%
\subsection{1 TeV}
The Higgs decays to $b\bar{b}$, $c\bar{c}$, and $gg$ were studied at 1 TeV as one of the detector benchmark 
studies for the ILC TDR by the ILD and SiD concept groups. At this energy 
the Higgs is produced dominantly by the process $e^{+}e^{-} \rightarrow \nu\bar{\nu}h$.
Therefore, the event signature is a large missing  $P_T$ due to un-detected 
neutrinos and 2 jets from Higgs decays to $b\bar{b}, c\bar{c}$, and $gg$, with their 
invariant mass consistent with the Higgs.  To minimize the effect of the low $P_T$ 
hadron events, which are produced at an average rate of 4.1 events per bunch crossing at 1 TeV, 
both ILD and SiD employed the $k_t$ jet clustering algorithm with a size parameter, R, 
of 1.5 (1.1 in the case of ILD). 
%This algorithm helped to 
%reduce a contamination of non-Higgs particles in reconstructed Higgs signal.

After the jet clustering the candidate 2-jet events were selected by cuts on 
the visible $P_T$, visible energy, visible mass, the jet production angles,
and the number of tracks. In the case of the SiD analysis, these variables were used to form 
Fisher Discriminants implemented in TMVA together with the flavor tagging variables 
for $b$ jets and $c$ jets.  Fisher discriminants which maximized the significance
for each decay mode were used to obtain the final results.  The uncertainties
on the cross section times Higgs branching ratios were determined 
from the numbers of signal and background events passing each selection.
A typical Higgs mass distribution in the case of SiD is shown in Figure~\ref{fig:visible-mass-higgs}.
% Half width
\thisfloatsetup{floatwidth=\SfigwFull,capposition=beside}
\begin{figure}[h]
  \centerline{\includegraphics[width=0.8\hsize]{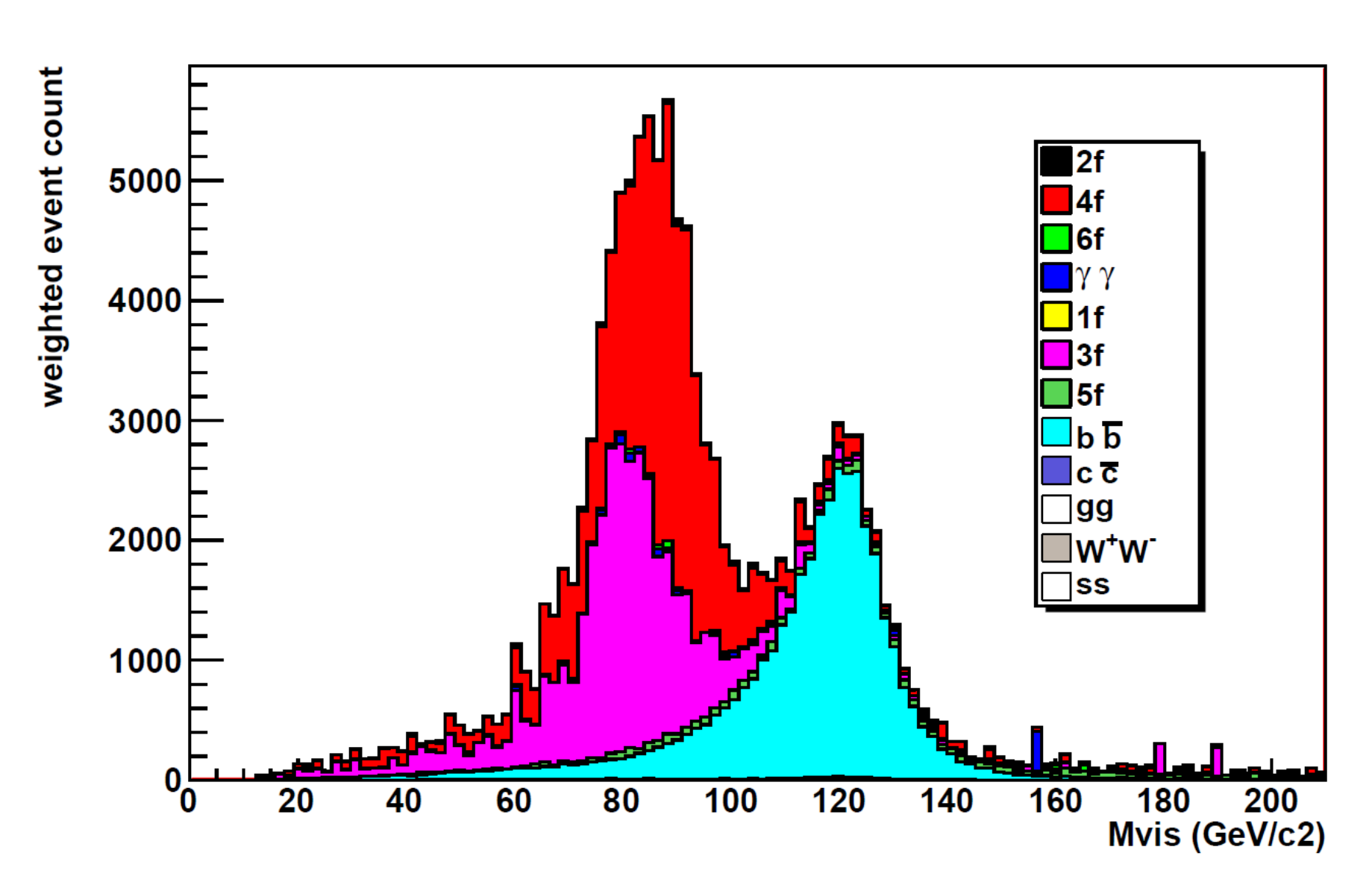}}
  \caption{The visible mass distribution for the $h\rightarrow b\bar{b}$ analysis
  without the visible mass cut for 500 fb$^{-1}$ and $P(e^{-})/P(e^{+}) = -80\%/+20\%$. }
\label{fig:visible-mass-higgs}
\end{figure}

In the case of the ILD analysis\cite{Ono:2013higgsdbd}, a flavor tagging template fitting 
was performed to extract $\sigma\cdot BR$ for the different channels.  The flavor templates of 
$h\rightarrow b\bar{b}$, $c\bar{c}$, $gg$, and background channels were obtained from the flavor 
tagging output of the LCFIPlus package.  
Taking into account the b-tagging efficiency systematic error of $0.3\%$, the accuracies for 1~ab$^{-1}$ and $P(e^{-})/P(e^{+}) = -80\%/+20\%$ beam polarization
were 0.49\%, 3.9\%, and 2.8\% for $h\rightarrow b\bar{b}$, $c\bar{c}$, 
and $gg$ respectively.   Following the publication of the ILC TDR, improvements to background rejection were developed~\cite{Junping-keisuke-fitting}, leading to relative 
$\sigma\cdot BR$ errors of 0.45\%, 3.1\%, and 2.3\% for $h\rightarrow b\bar{b}$, $c\bar{c}$, 
and $gg$ respectively.

% The ILD results were consistent with SiD except $c\bar{c}$ channel,
% where ILD got 3.9\%. 

%%%%%%%%%%%%%%%%%%%%%%%%%%%%%%%%%%%%%%%%%%%%%%%%%%%%%%%%%%%%%%%%%%%%%%%%%%%%%%%%%%%%%%%%%%%%%%%%%%%%%
\section{$h\rightarrow  WW^*$}
%%%%%%%%%%%%%%%%%%%%%%%%%%%%%%%%%%%%%%%%%%%%%%%%%%%%%%%%%%%%%%%%%%%%%%%%%%%%%%%%%%%%%%%%%%%%

%\subsection{250GeV}

%%%%%%%%%%%%%%%%%%%%%%%%%%%%%%%%%%%%%%%%%%%%%%%%%%%%%%%%%%%%%%%%%%%%%%%%%%%%%%%%%%%%%%%%%%%%%%%%%%%
\subsection{500 GeV}

%%%%%%%%%%%%%%%%%%%%%%%%%%%%%%%%%%%%%%%%%%%%%%%%%%%%%%%%%%%%%%%%%%%%%%%%%%%%%%%%%%%%%%%%%%%%%%%%%%%
The full simulation study of the process, $e^+e^- \rightarrow \nu\bar{\nu}H \rightarrow \nu\bar{\nu}WW^*$
was performed using the fully hadronic mode of $WW^*$.  In this case the event signature is 
4 jets with missing energy and missing momentum and mass consistent with Higgs.  The  2 jets from the $W^*$ are soft and the
piled up low $P_T$ particles due to $\gamma\gamma$ collisions have to be removed effectively. 
To this end, a multivariate analysis (MVA) to identify pile up particles was employed using $P_T$ and rapidity.  
In the case of charged particles, the closest approach to the interaction point along the beam axis ($Z_0$)
was also used to reduce background contamination. Figure~\ref{fig:Z0-and-MVA-for-BKG-regection} shows the boosted 
decision tree (BDT) MVA
response for neutral and charged particles.
% Half width
\thisfloatsetup{floatwidth=\SfigwFull,capposition=beside}
\begin{figure}[h]
  \begin{tabular}{cc} 
  \includegraphics[width=0.5\hsize]{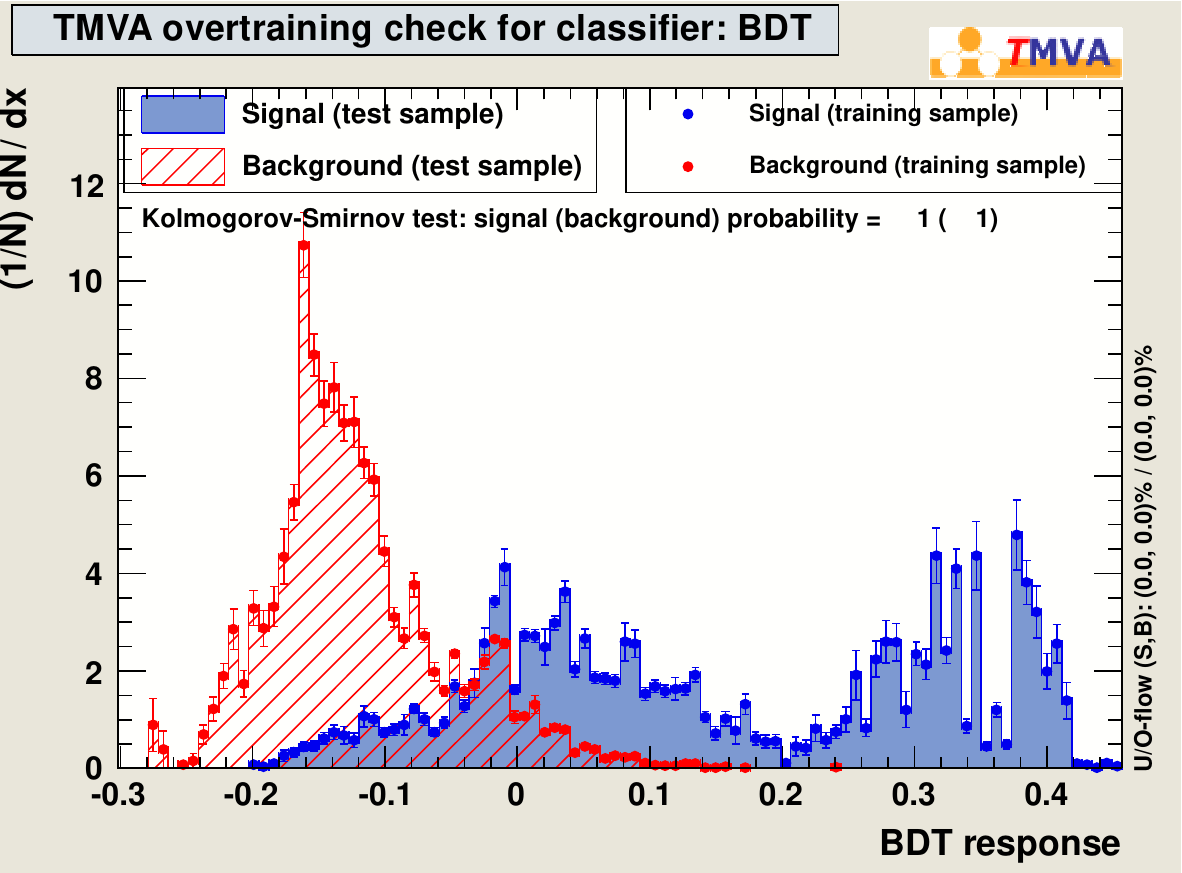}&
  \includegraphics[width=0.5\hsize]{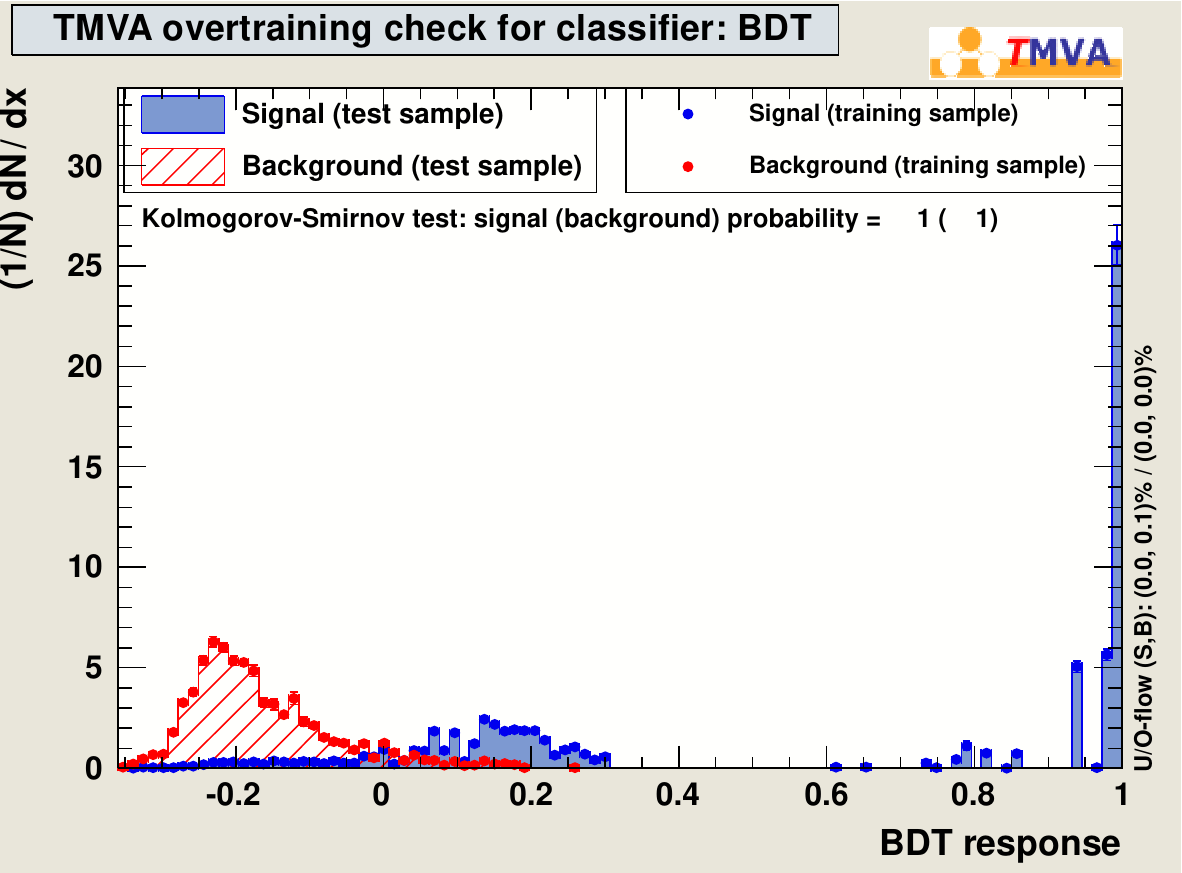}  \\
  \end{tabular}
  \caption{BDT response for particles from $\nu\bar{\nu}h\rightarrow WW^* \rightarrow q\bar{q}q\bar{q}$ 
  events and low $P_T$ hadron background events for neutral particles (left) and charged particles (right).}
\label{fig:Z0-and-MVA-for-BKG-regection}
\end{figure}

After rejecting background tracks by the MVA, 
events with isolated muons or electrons were removed 
and anti-$k_t$ jet clustering was employed to select 4-jet events.
Each jet was required to not be tagged as a $b$-jet, and one jet pair 
must have its mass consistent with the $W$ with the other jet pair mass
between 11 and 64~GeV. 
The 4-jet mass distibution for selected events is  shown in Figure~\ref{fig:vvh_500GeV_WW_4jetmass}.
% Half width
\thisfloatsetup{floatwidth=\SfigwFull,capposition=beside}
\begin{figure}[h]
  \centerline{
  \includegraphics[width=0.8\hsize]{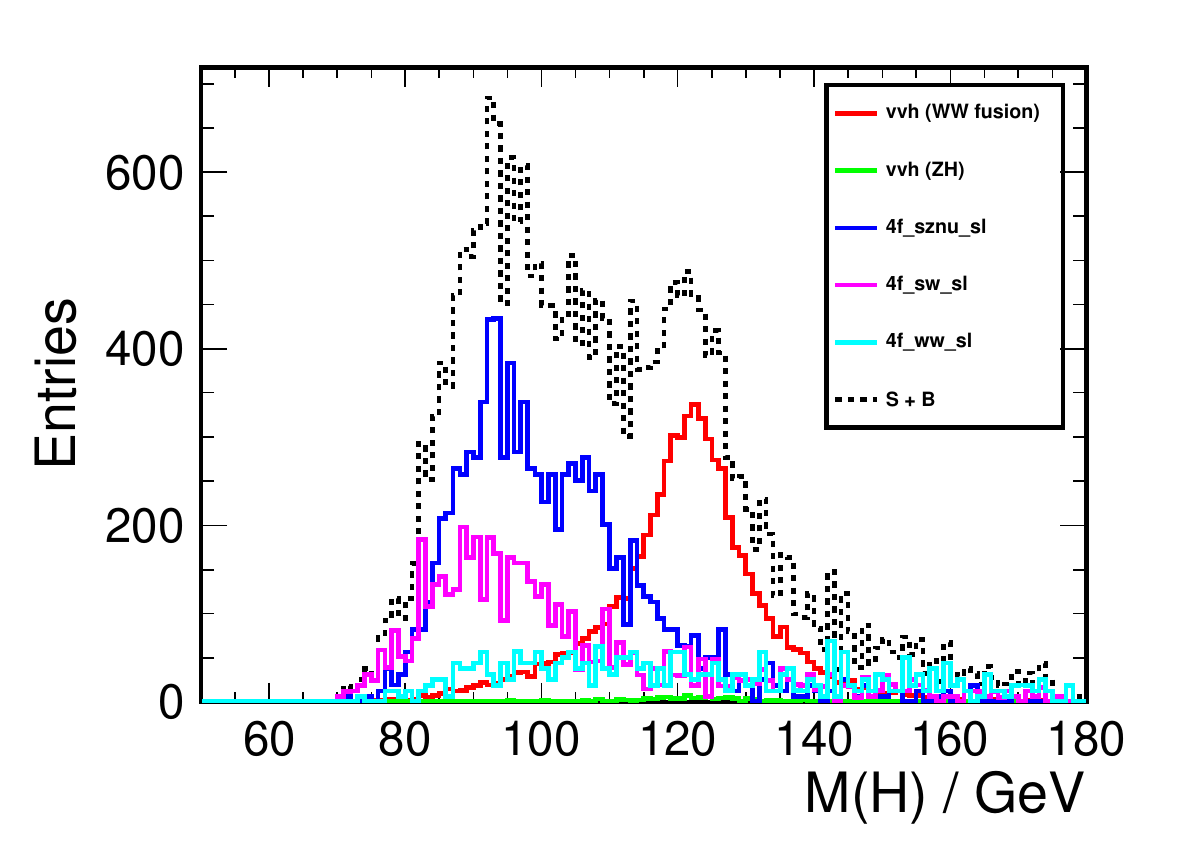}
  }
  \caption{4-jet mass distribution for selected events in the $h\rightarrow WW^*$ study.}
\label{fig:vvh_500GeV_WW_4jetmass}
\end{figure}

The signal selection efficiency was about 43\%.
The major backgrounds were other Higgs decays  and the
semi-leptonic channels for  $e^+e^-\rightarrow ZZ$ or $WW$. 
The signal-to-noise ratio was about 1.  For 500 fb$^{-1}$, the signal significance, $S/\sqrt{S+B}$,
was 35 and $\Delta(\sigma\cdot BR)/(\sigma\cdot BR)=2.8\%$.
When combined with an analysis of the semileptonic channel for $h\rightarrow WW*$~\cite{durig-jenny-lcws2012} the precision improves to 
$\Delta(\sigma\cdot BR)/(\sigma\cdot BR)=2.4\%$.

%%%%%%%%%%%%%%%%%%%%%%%%%%%%%%%%%%%%%%%%%%%%%%%%%%%%%%%%%%%%%%%%%%%%%%%%%%%%%%%%%%%%%%%%%%%%%%%%%%%
\subsection{1 TeV}
%%%%%%%%%%%%%%%%%%%%%%%%%%%%%%%%%%%%%%%%%%%%%%%%%%%%%%%%%%%%%%%%%%%%%%%%%%%%%%%%%%%%%%%%%%%%%%%%%%%
The decay $h\rightarrow WW^*$ was studied at 1 TeV by ILD and SiD for the ILC TDR using the fully hadronic 
decay mode of $WW^*$.   The signal final state is four jets consistent with $WW^*$,
with total mass consistent with the Higgs mass, and large missing energy 
and missing transverse momentum.  

In the ILD analysis  background from pile-up events was removed by 
employing the $k_t$ jet clustering algorithm with $R=0.9$ and $N_{jet}=4$. 
Further, the Durham algorithm was applied to force the remaining particles 
to be clustered into four jets, which were paired so that one dijet system 
had a mass consistent with the $W$, while the other had a mass between 15 and 60~GeV. 
To reduce background from $h\rightarrow b\bar{b}$ 
the $b$-likeness of each jet was required to be low.  The signal selection efficiency
was 12.4\%, and the remaining major backgrounds were 
4-fermions ($e^+e^-\rightarrow \nu\bar{\nu}q\bar{q}$), 
3-fermions ($e\gamma \rightarrow \nu q\bar{q}$) 
and other decay channels of the Higgs. 
The reconstructed Higgs mass distribution is shown in Figure~\ref{fig:massdistri-higgs-to-ww}.
%The precision of $\sigma\cdot BR$ was estimated from $\sqrt{S+B}/S$. 

With 1 ab$^{-1}$ luminosity and a beam 
polarization of $P(e^{-}) = -80\%$, $P(e^{+})  = +20\%$, ILD obtained 
${\Delta(\sigma\cdot BR) / (\sigma\cdot BR)} = 2.5\%$.  SiD obtained a similar result.
By including the semi-leptonic topology for $h\rightarrow WW^*$, and by using the particle-based MVA
technique to better reject pileup, the precision for $h\rightarrow WW^*$ improves to ${\Delta(\sigma\cdot BR) / (\sigma\cdot BR)} = 1.6\%$~\cite{Junping-keisuke-fitting}.

% Half width
\thisfloatsetup{floatwidth=\SfigwFull,capposition=beside}
\begin{figure}[h]
  \centerline{\includegraphics[width=0.8\hsize]{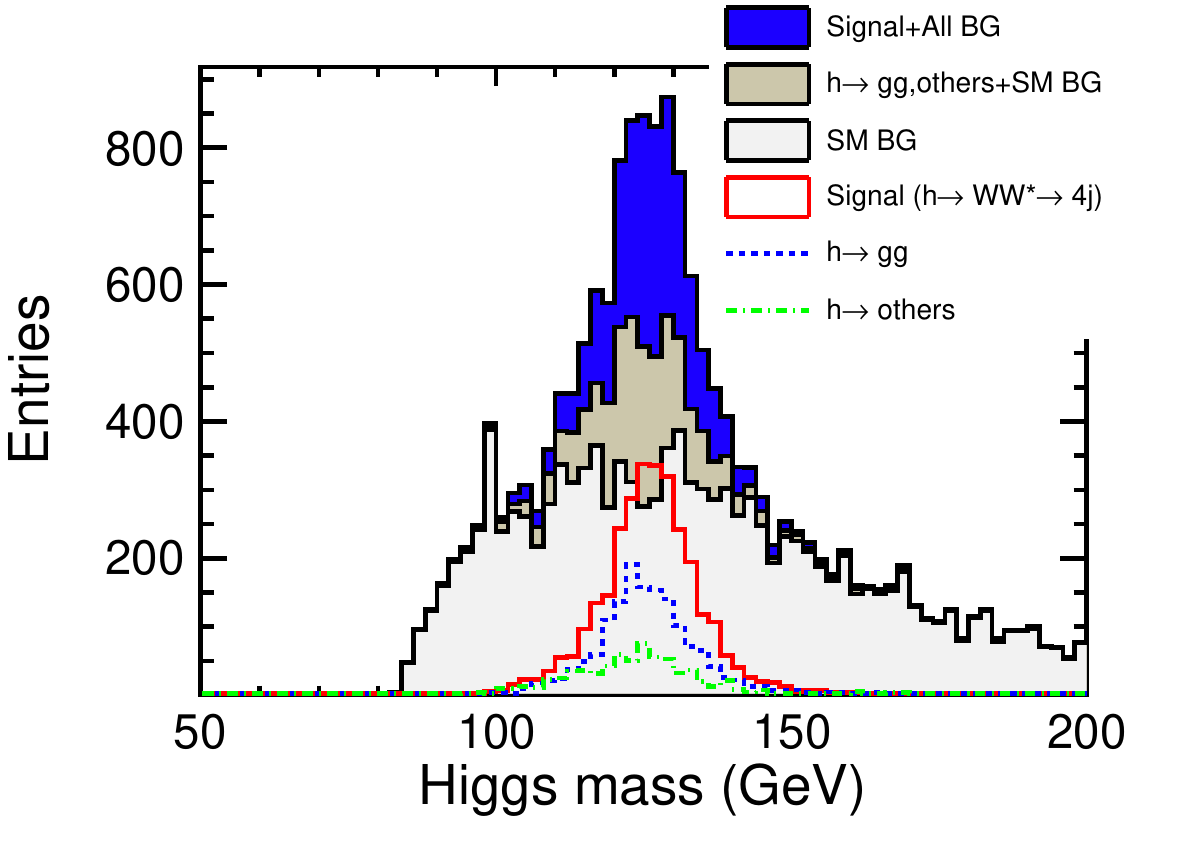}}
  \caption{ILD reconstructed Higgs mass distribution for the $h\rightarrow WW^*$ analysis in the fully hadronic decay channel.
}
\label{fig:massdistri-higgs-to-ww}
\end{figure}

%%%%%%%%%%%%%%%%%%%%%%%%%%%%%%%%%%%%%%%%%%%%%%%%%%%%%%%%%%%%%%%%%%%%%%%%%%%%%%%%%%%%%%%%%%%%%%%%%%%%%
\section{$h\rightarrow \tau^+\tau^- $}

\subsection{250 GeV}
The full simulation samples of the ILD LOI~\cite{Abe:2010aa} were used for the study of $h\rightarrow\tau^{+}\tau^{-}$
in Ref.~\cite{Kawada:2013tautau}. In this study the Higgsstrahlung process ($e^+e^- \rightarrow Zh$) at $\sqrt{s}=250$ GeV 
was considered using $m_h=120$~GeV and the $Z$ decay modes $Z\rightarrow l^+l^-$ ($l=e,\mu$) and $Z\rightarrow q\bar{q}$.

In the case of $Z\rightarrow l^+l^-$, events with an $l^+l^-$  mass consistent with the $Z$ 
were selected, where the  $l^+l^-$ tracks were required to come from the IP to reject such tracks from $\tau$ decay.
Particles other than those from the $Z$ were fed to a tau jet finder, which 
identified low mass ( $< 2 $~GeV ) jets using particles within 1 radian of an energetic track.
Signal events were required to have a $\tau^+$, a $\tau^-$ and an  $l^+l^-$ recoil mass close to the
Higgs mass.  The signal events were selected with an efficiency of $47\%$ and 
$62\%$ for the $e^+e^-$ and $\mu^+\mu^-$ channels, respectively.  The $S/N$ was 1.43 (1.44) for $e^+e^-$ $(\mu^+\mu^-)$,
and the signal significance was 8.0$\sigma$  (8.8$\sigma$) for the $e^+e^-$ ($\mu^+\mu^-$)
channel. 

In the case of $Z\rightarrow q\bar{q}$, a tau-jet was formed using 
an energetic track and all particles within 0.2 radians of the energetic track.   The mass of a tau-jet was required to be less than 2~GeV, 
and additional cuts on the $\tau$ energy and isolation were applied to 
reduce mis-identified quark jets.
% jets despite a loss of efficiency for $\tau$ decaying to 3-prong with neutral particles;
Low energy charged tracks found in  a jet were detached one by one until a unit charged jet 
with 1 or 3 prong charged multiplicity was obtained.    Once a  tau jet pair was found, the kinematics of the tau jet pair were
reconstructed assuming that the visible tau decay products and neutrinos were collinear, and that the missing 
momentum of the event was generated only by the neutrinos from the tau decays.  Following the reconstruction of the two $\tau$ jets,
$q\bar{q}$ jets were reconstructed by clustering the remaining particles with the Durham jet algorithm.
Variables such as jet mass, energy, and production angle were used together with particle multiplicities and the impact parameters of tracks from
$\tau$ jets to select  $Zh\rightarrow q\bar{q}\tau^{+}\tau^{-}$ 
events.  
%The reconstructed Higgs mass distribution is shown in 
%Figure~\ref{fig:h-to-tautau-collinear-mass}, 
%where Higgs signal is clearly seen above backgrounds, 
%main background being $e^+e^-\rightarrow q\bar{q}\tau^{+}\tau^{-}$ processes.  
The efficiency for signal selection was about 0.24 and the $S/N$ was 1.85. With an integrated luminosity of 250~fb$^{-1}$,
the signal significance was 25.8$\sigma$

%% Half width
%\thisfloatsetup{floatwidth=\SfigwHalf,capposition=beside}
%\begin{figure}[h]
%    \centerline{\includegraphics[width=1.5\hsize]{Chapter_Sigma_Times_BR_I/figs/higgs-tautau-collinear-mass.pdf}}
%    \caption{Collinear mass distribution in $h\rightarrow \tau^+\tau^-$ study.}
%    \label{fig:h-to-tautau-collinear-mass}
%\end{figure}

If we combine the results for  $Z\rightarrow l^+l^-$ and $Z\rightarrow q\bar{q}$, the
significance is 28.4, which corresponds to a measurement accuracy of $\Delta(\sigma\cdot BR)/(\sigma\cdot BR)=3.5\%$.
Table~\ref{tab:zh-to-tautau-kawada} shows the extrapolation of this result to the case of $m_h=125$~GeV, 
where it was assumed that the signal selection efficiency is unchanged.
%This result was extrapolated to the case of $m_h=125$~GeV by scaling the signal 
%yield by the cross section of $e^+e^-\rightarrow Zh$  and by using the branching ratio for $m_H=125$~GeV 
%and the same signal selection efficiency. The results for $m_h$=125~GeV are shown in .
\thisfloatsetup{floatwidth=\hsize,capposition=top}
\begin{table}[!h]
\caption{Relative error on $\sigma\cdot BR$ at $\sqrt{s}=250$~GeV for $h\rightarrow \tau^+\tau^-$ assuming $m_h=125$~GeV,
250 fb$^{-1}$ luminosity and  beam polarization $P(e^{-})=-80\%$ and $P(e^+)=+30\%$. The results were obtained by
scaling the errors for $m_h=120$~GeV.}
\label{tab:zh-to-tautau-kawada}
\begin{center}
\begin{tabular}{c c c | c c }
\toprule
$Z\rightarrow e^+e^-$ & $Z\rightarrow \mu^+\mu^-$ & $Z\rightarrow q\bar{q}$ & Combined &
${\Delta(\sigma\cdot BR)\over (\sigma\cdot BR) }$ \\
\midrule
$6.8\sigma$ & $7.4\sigma$ & $21.9\sigma$ & $24.1\sigma$ & $4.2\%$ \\
\bottomrule
\end{tabular}
\end{center}
\end{table}

%%%%%%%%%%%%%%%%%%%%%%%%%%%%%%%%%%%%%%%%%%%%%%%%%%%%%%%%%%%%%%%%%%%%%%%%%%%%%%%%%%%%%%%%%%%%%%%%%%
\subsection{500 GeV}

The decay $h\rightarrow \tau^+\tau^-$ was studied at $\sqrt{s}=500$~GeV using the ILD full simulation 
with $m_h = 125$~GeV~\cite{Kawada-Seattle-Higgs-tautau}.
% at Snowmass EF workshop at 
% Seattle.\cite{Kawada-Seattle-Higgs-tautau}.  
At this energy both Higgsstrahlung and $WW$ fusion 
processes contribute with comparable weight.

For the Higgsstrahlung process $e^+e^- \rightarrow Zh \rightarrow q\bar{q}h$, methods similar to those used at $\sqrt{s}=250$~GeV 
were employed. A signal efficiency of 21.0\% and a precision of ${\Delta(\sigma_{ZH}\cdot BR)\over \sigma_{ZH} \cdot BR} = 5.4\%$ were obtained
for -80\%/+30\% $e^{-}/e^{+}$ polarization and 500 fb$^{-1}$ luminosity.
Further improvement is expected by including the  $Z\rightarrow \ell\bar{\ell}$ mode.
% with 67.4\% purity was achieved and 
%was obtained preliminary. By including leptonic decay mode of $Z$, 
%${\Delta(\sigma_{ZH}\cdot BR)\over \sigma_{ZH} \cdot BR }= XXX\%$   %% FIX ME
%could be achieved with $P_{e^{-}}=-80\%$ and $P_{e^{+}}=+30\%$ and 500fb$^{-1}$.

In the $WW$ fusion case $e^+e^- \rightarrow \nu_e \bar{\nu}_e h\rightarrow  \nu_e \bar{\nu}_e \tau^+ \tau^-$, a jet with mass less than 2~GeV was considered 
a $\tau$ jet. The most energetic $\tau^{+}$ and $\tau^{-}$ were combined as the Higgs boson, and cuts
were applied to the tau pair mass and event missing energy.   A signal efficiency of 25\% and a precision of ${\Delta(\sigma_{\nu_e \bar{\nu}_e h}\cdot BR)\over \sigma_{\nu_e \bar{\nu}_e h} \cdot BR} = 9.0\%$ were obtained
for -80\%/+30\% $e^{-}/e^{+}$ polarization and 500 fb$^{-1}$ luminosity.

%@misc{Kawada-Seattle-Higgs-tautau,
%      author="S.~Kawada",
%      note={{Talk presented at Snowmass EF Workshop at Seattle, paper in preparation}}
%}

\chapter{Cross Section Times Branching Ratio Measurements II \label{sid:chapter_sigma_times_br_ii}}
\section{$h\rightarrow ZZ^*$}

A full simulation study of the decay $h\rightarrow ZZ^*$ has been performed using the process
$\epem \rightarrow  Zh \rightarrow ZZZ^*$
at \roots=250~GeV.  This decay has a SM branching ratio of 2.7\% given the Higgs mass of 125~GeV.
The final state is characterized by two on-shell Z bosons and one off-shell Z boson, leading to 
a variety of combinations of jets,  isolated leptons and missing energy. The analysis is
directed toward topologies where the $Z$ opposite the Higgs boson decays in any manner $Z\to q,l,\nu$,
while the Higgs decays without missing energy, $h\to ZZ^*\to q\bar{q}\, \  \mathrm{or}\, \ l^+l^-\, , \ l=e,\mu$.
The datasets used for this analysis are shown in Table~\ref{sid:benchmarking:tab:vvHdatasets}.

\begin{table}
\caption{\label{sid:benchmarking:tab:vvHdatasets} 
Simulated data samples used for the \nuenueH analysis.
}
\begin{center}
\begin{tabular}{c l l}\toprule
Process & $P(e^{-})/P(e^{+})$ & $N_{Events}$ \\ \midrule
 $f^+f^-h\,\ h \rightarrow  ZZ^*$ &+80\%/-30\%&120,012\\\midrule
All SM background mix&+80\%/-30\%&2,058,374\\
\bottomrule
\end{tabular}
\end{center}
\end{table}

%\begin{figure}[htbp]
%\includegraphics[width=0.85\textwidth]{vvhallbkgsig-allbkg.pdf}
%\caption{Visible mass distributions for the backgrounds and \zzH events.\label{sid:benchmarking:mvissig}}
%\end{figure}

\subsection{Event reconstruction for $ h \to  ZZ^*$}

Events are classified as 6-jet or 4-jet, depending on whether the visible
energy in the event is greater or less than 140~GeV;  here the term 
``jet'' can also refer to an isolated electron or muon.

In the low visible energy signal events 
we expect a 4-jet topology if the $Z$ and $Z^{*}$ decay visibly. One pair of jets
must have a mass consistent with the $Z$ mass. Events that have opposite signed electrons
or muons with a mass consistent with the $Z$ mass are unlikely to come from the
WW background. Because of large missing energy and momentum from the
invisible $Z$ decay, it is unlikely that the reconstructed $Z$ bosons are back-to-back
and so we cut on the angle between them. Cutting on the number of tracks helps
to remove much of the two-photon background.

The high visible energy signal events are largely true six jet events with all $Z$ bosons decaying
visibly. Backgrounds that come from ZZ and WW decays can be cut using the Durham jet clustering
y24 and y56 variables. All pairs of jets are tried for the pair most consistent with
the mass of the Z. Then from the remaining jets, the next pair most consistent with
the mass of the $Z$ is found and the remaining pair is taken as coming from the $Z^{*}$.
Each $Z$ is then paired with the $Z^{*}$ to see which one gives a mass most consistent
with coming from a 125 GeV Higgs. The analysis then proceeds similarly to the
4 jet analysis using this pair of  $ZZ^*$.

Before applying an MVA selection, preselection cuts are applied separately for
Evis$<$140 GeV and Evis$>$140 GeV. The preselection cuts exclude regions only where
there is almost no signal. Events are preselected based on the Higgs topology being studied using the
criteria shown in Table~\ref{sid:benchmarking:tab:hvvpresel}. 

\begin{table}[h!]
\caption{\label{sid:benchmarking:tab:hvvpresel} Overview of the preselections for the different 
Higgs decay modes. The cuts as well as the efficiencies for signal and background events are shown.}
\begin{center}
\begin{small}
\begin{tabular}{l l c c} \toprule
 Higgs decay & Preselection cuts & Signal eff. & Background eff. \\ \midrule
$ h \to  ZZ^*  (E_{vis}<140 GeV)$ & \parbox{4cm}{ 
$25 < p^{T}_{\text{vis}} < 70$~GeV \\ 
%$100 < E_{\text{vis}} < 400$~GeV \\
$95. < M^{\text{higgs}}_{\text{vis}} < 140.$~GeV \\
$|\cos(\theta_{\text{jet}})|< 0.90$ \\
$N_{\text{PFO}} > 5$ \\
$y_{\text{34}} > 0.$ \\
E$_{\text{Z}} > 120 GeV$
} &  & \\
\midrule
$ h \to  ZZ^*  (E_{vis}>140 GeV)$ & \parbox{4cm}{ 
%$25 < p^{T}_{\text{vis}} < 70$~GeV \\ 
%$100 < E_{\text{vis}} < 400$~GeV \\
$90. < M^{\text{higgs}}_{\text{vis}} < 160.$~GeV \\
$|\cos(\theta_{\text{jet}})|< 0.90$ \\
$N_{\text{PFO}} > 5$ \\
$y_{\text{34}} > 0.$ \\
E$_{\text{Z}} > 120 GeV$\\
$|$thrust$|$ < 0.98
} & & \\
\midrule
Both & \parbox{4cm}{
} & 77\%& $1.5 \times 10^{-2}$\\
\bottomrule
\end{tabular}
\end{small}
\end{center}
\end{table}

\newpage

\subsection{Multi-Variate Analysis}

After the preselection, multivariate methods, as implemented in TMVA, are then
used to maximize the significance (${S} / \sqrt{S+B}$) of the selection. They
are trained using 50\% of the signal and background events and done separately for the
different polarizations and integrated luminosities. The cuts on the
Fisher discriminant which maximize the significance for each decay mode are used
to obtain the final results. The input variables for the multivariate methods are:\\
\begin{itemize}
\item{visible mass of the event}
\item{the visible energy, mass and transverse momentum}
\item{B-Likeness from b-tag flavor tagging values}
\item{C-likeness from c-tag flavor tagging values}
\item{Number of High Energy Electrons}
\item{Higgs Mass = mass of the reconstructed  $ZZ^*$}
\item{reconstructed $Z$ energy}
\item{reconstructed $Z^{*}$ energy}
\item{cosine of the reconstructed $Z$ polar angle}
\item{cosine of the reconstructed $Z^{*}$ polar angle}
\item{reconstructed $Z$ mass}
\item{reconstructed $Z^{*}$ mass}
\item{the angle between the reconstructed $Z$ and $Z^{*}$ in the plane perpendicular to the beam axis.}
\item{the event thrust magnitude}
\item{number Charged Tracks}
\item{number of identified electrons}
\item{number of identified muons}
\item{Durham jet clustering y34 value}
\item{Durham jet clustering y56 value}
\item{lepton pair (PDG ID1 = -ID2) mass closest to m(Z)}
\item{jet pair mass closest to m(W)}
\item{sum of the absolute differences of the best W jet pair mass w.r.t. m(W)} 
\end{itemize}

The flavor tagging is used as implemented in the LCFIPlus package which uses boosted decision trees
on vertexing quantities to determine b-tag and c-tag probabilities for bottom and charm jets respectively.
It is trained using samples of four-jet events from $ZZ^{*} \to
\bbbar, \ccbar$~and~$\qqbar$ at  \roots = 350~GeV and the tagging is accordingly
applied to all signal and background samples.

\begin{figure}[htbp]
\includegraphics[width=0.90\textwidth]{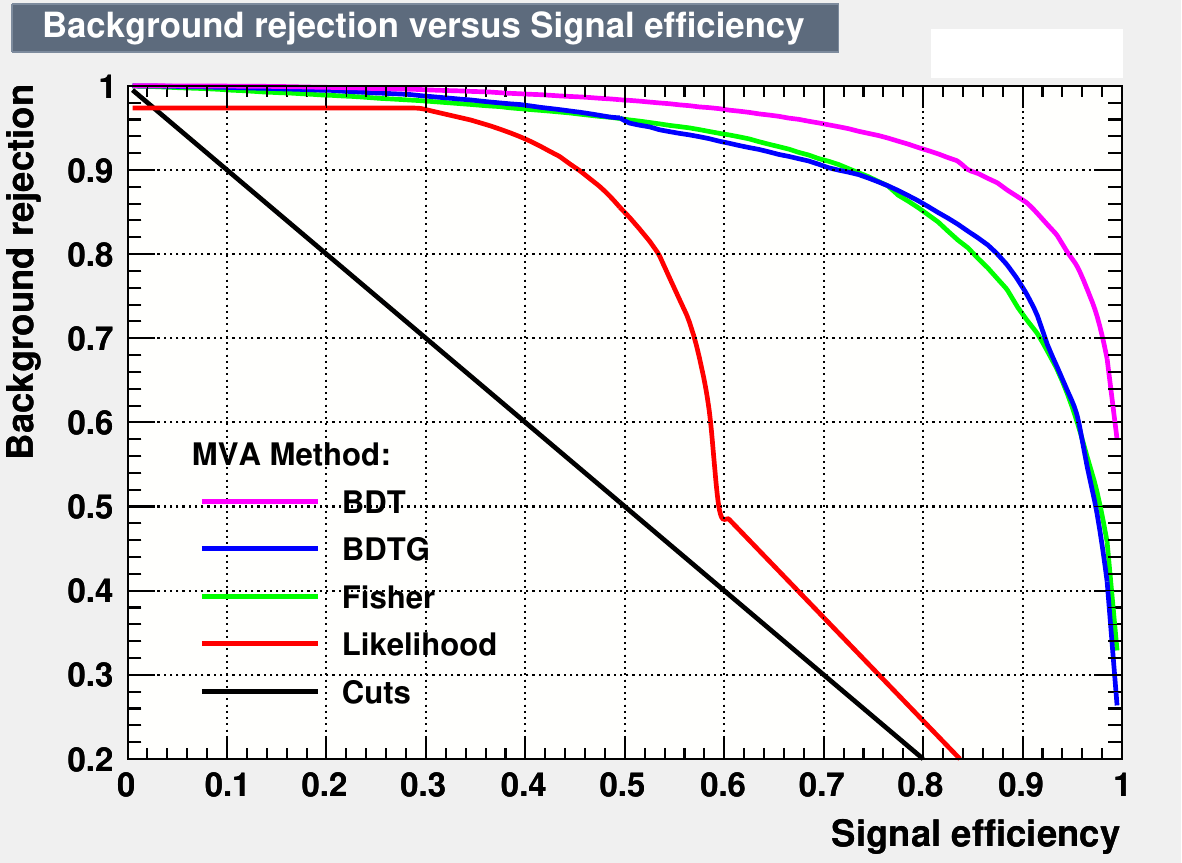}
\caption{Rejection of background vs. signal for selecting Higgs boson decays to 
 $ZZ^*$ from the BDT, BDTG, Fisher, Likelihood and CUTS multivariate methods.\label{sid:benchmarking:tmvaprobs}}
\end{figure}

The performance of the various MVA methods is shown in Figure~\ref{sid:benchmarking:tmvaprobs}. 
It is found that the BDT method significantly out performs the other methods.
Plots showing the efficiency and significance curves vs. cuts on the BDT 
output are shown in Figure~\ref{sid:benchmarking:vvHsig}.

\begin{figure}[htbp]
\includegraphics[width=0.90\textwidth]{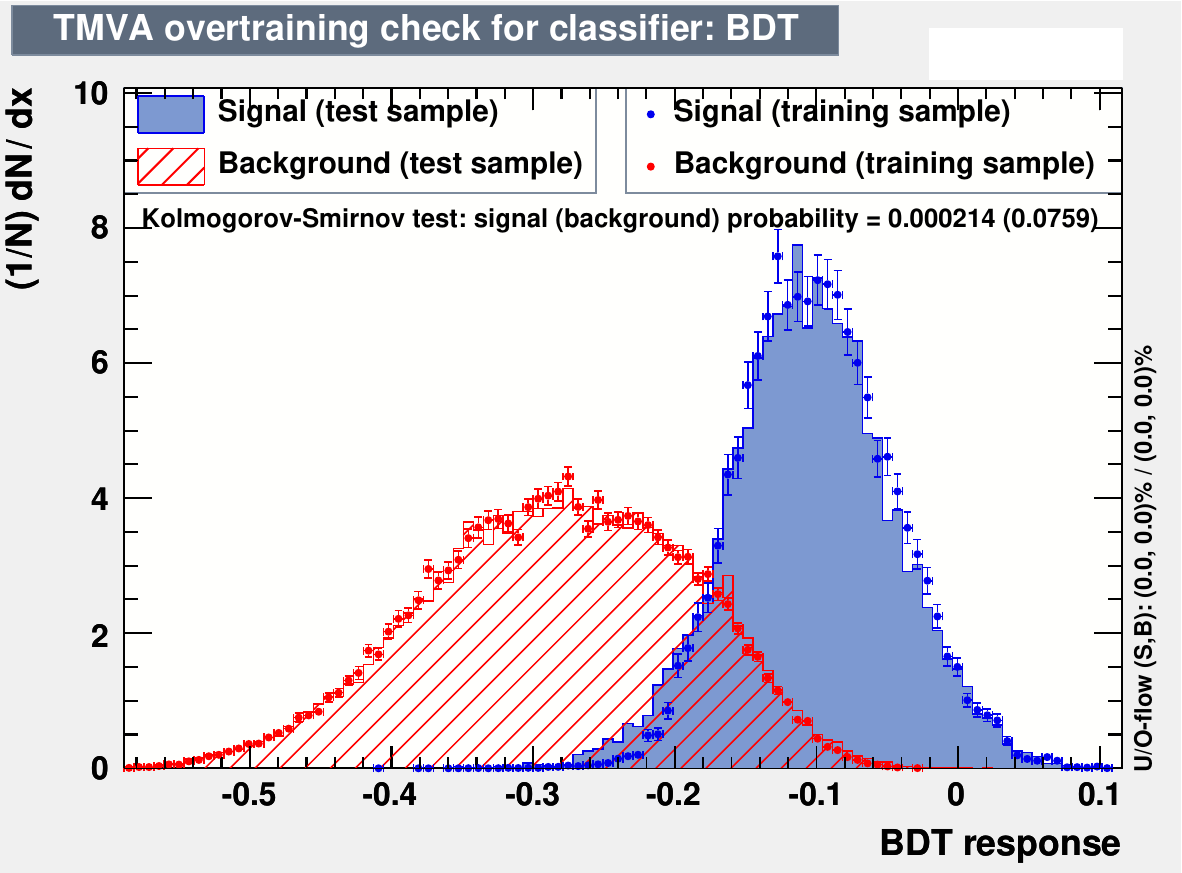}
\caption{
The multi-variate BDT output for the signal ($ h \to  ZZ^*$) and background for the training samples and test samples (points).\label{sid:benchmarking:vvHsig}}
\end{figure}

The composition of the samples of events passing all selections of the analysis are shown in Table~\ref{sid:benchmarking:tab:vvHcomp}
for the polarization P(\Pem) = +80\%, P(\Pep) = -30\% and 250~\fbinv. The fraction of events passing all selections is
10.8$\%$ for the signal and 0.0008$\%$ for the background. The significance of the signal after the preselection is 1.0. After applying the
cut on the BDT output, the significance is 5.6.

\begin{table}[h!]
\caption{\label{sid:benchmarking:tab:vvHcomp} 
Composition of the events passing all analysis selections for the polarizations P(\Pem) = +80\%, P(\Pep) = -30\% 
and an integrated luminosity of 250~\fbinv collected by SiD at a center of mass energy of 250~GeV.}
\begin{center}
\begin{tabular}{l r r r r}\toprule
 & $ h\rightarrow ZZ^*$  \\ 
 &  \multicolumn{1}{c}{(\%)}         \\
 \midrule
$\epem \to 2~\text{fermions}$ &  50 \\ 
$\epem \to 4~\text{fermions}$ & 462 \\ 
$\epem \to 6~\text{fermions}$ &   0 \\ 
$\gamgam \to X$               &   0 \\ 
$\gamma e^+ \to X$            &   0 \\ 
$e^- \gamma \to X$            &   0 \\ 
$qq h \to ZZ^{*}$           &  68 \\
$ee h,\mu\mu h \to ZZ^{*}$&  24 \\
$\tau\tau h \to ZZ^{*}$     &   3 \\
$\nu\nu h \to ZZ^{*}$       &  49 \\
\bottomrule
\end{tabular}
\end{center}
\end{table}

%\begin{figure}[htbp]
%\begin{center}
%\includegraphics[width=0.85\textwidth]{vvhallbkgsig-bbmva-linear.pdf}
%\caption{
%The visible mass distribution for the $h \rightarrow \bbbar$ selected events
%without the visible mass preselection cut for 250~\fbinv and the P(\Pem) = +80\%, P(\Pep) = -30\% polarization configuration.
%Note that the ``a'' in the legend represents $\gamma$.
%\label{sid:benchmarking:fig:mvisbb}}
%\end{center}
%\end{figure}
%
%The visible mass distribution for the $ h \rightarrow ZZ$ selected events
%with the visible mass preselection cut removed is shown in Figure~\ref{sid:benchmarking:fig:mvisbb} 
%for 250~\fbinv and the P(\Pem) = +80\%, P(\Pep) = -30\% polarization.

\subsection{Results for $ f^+f^-h \to  ZZ^*$}

The uncertainties on the cross sections times Higgs branching fractions, $\Delta ( \sigma \times BR )$, are determined from
the numbers of signal and background events passing each selection. For 250~\fbinv of \epem P(\Pem) = +80\%, P(\Pep) = -30\%
250 GeV collisions in the SiD detector this benchmark indicates that a precision of 
18\% can be obtained.

\section{$h\rightarrow  \gamma\gamma$}
        Fast Monte Carlo studies of $e^+e^- \to Zh \to f\bar{f}\gamma\gamma,\, \  f=q,\nu$ at $\sqrt{s}=250$~GeV~\cite{Boos:2000bz} and 
$e^+e^- \to \nu \bar{\nu}h \to \nu \bar{\nu}\gamma\gamma$ at $\sqrt{s}=1000$~GeV~\cite{Barklow:2003hz} have been supplemented recently 
with a full simulation study of $e^+e^- \to f\bar{f}h \to f\bar{f}\gamma\gamma,\, \   f=q,\nu$ at $\sqrt{s}=250$~GeV and $500$~GeV~\cite{Calancha:2013gm}.
These studies indicate that the ILC can measure $\sigma\cdot BR(h\to \gamma \gamma)$ with an accuracy of 34\% using $e^+e^- \to Zh$ 
  at $\sqrt{s}=250$~GeV  assuming 250~fb$^{-1}$. The process $e^+e^- \to \nu \bar{\nu}h \to \nu \bar{\nu}\gamma\gamma$  yields 
errors of 23\% for 500~fb$^{-1}$ at $\sqrt{s}=500$~GeV and 8.5\% for 1000~fb$^{-1}$ at $\sqrt{s}=1000$~GeV.

%\section{$h\rightarrow Z\gamma$}
\section{$h\rightarrow  \mu^+\mu^- $}

 The decay $h\rightarrow  \mu^+\mu^- $ has been studied using $e^+ e^- \rightarrow Zh \rightarrow  q\bar{q} \mu^+\mu^-$ at $\sqrt{s}=250$~GeV~\cite{Aihara:2009ad}
and $e^+ e^- \rightarrow  \nu \bar{\nu} h \rightarrow  \nu \bar{\nu} \mu^+\mu^- $ at $\sqrt{s}=1000$~GeV~\cite{Behnke:2013lya}.
This decay has a SM branching ratio of 0.02\%.  The very small event rate at the ILC can 
be compensated somewhat by the excellent $\delta (1/\pT) \sim$~2--\SI{5e-5}{(GeV/c)^{-1}} charged particle momentum resolution of the ILD and SiD detectors.

At $\sqrt{s}=250$~GeV the largest background is $e^+ e^- \rightarrow ZZ \to  q\bar{q} \mu^+\mu^-$. Following all cuts an error of 91\% for $\sigma\cdot BR(h\to \mu^+\mu^-$ 
was obtained in Ref.~\cite{Aihara:2009ad} for 250~fb$^{-1}$ assuming a 120~GeV Higgs mass.  
Scaling to a Higgs mass of 125~GeV this error becomes 100\%. 

Figure~\ref{fig:mumumass} shows the
reconstructed muon pair mass distributions for signal and background after all cuts at $\sqrt{s}=1000$~GeV.  At this center of mass energy the largest backgrounds 
following all cuts are $\epem\to \nu_e\bar{\nu}_e\mpmm$, $\epem\to W^+W^- \to \nu_\mu\bar{\nu}_\mu\mpmm $, and $\gamma\gamma\to W^+W^-\to  \nu_\mu\bar{\nu}_\mu\mpmm $.  With 1000~fb$^{-1}$
an error of 31\% was obtained in Ref.~\cite{Behnke:2013lya}.

%%%%%%%%%%%%%%%%%%%%%%%%%%%%%%%%%%%%%%%%%%%%%%%%%%%%%%%%%%%%%%%%%%%%%%%%%%%
\begin{figure}[t]
\begin{center}
      \includegraphics[width=0.47\linewidth]{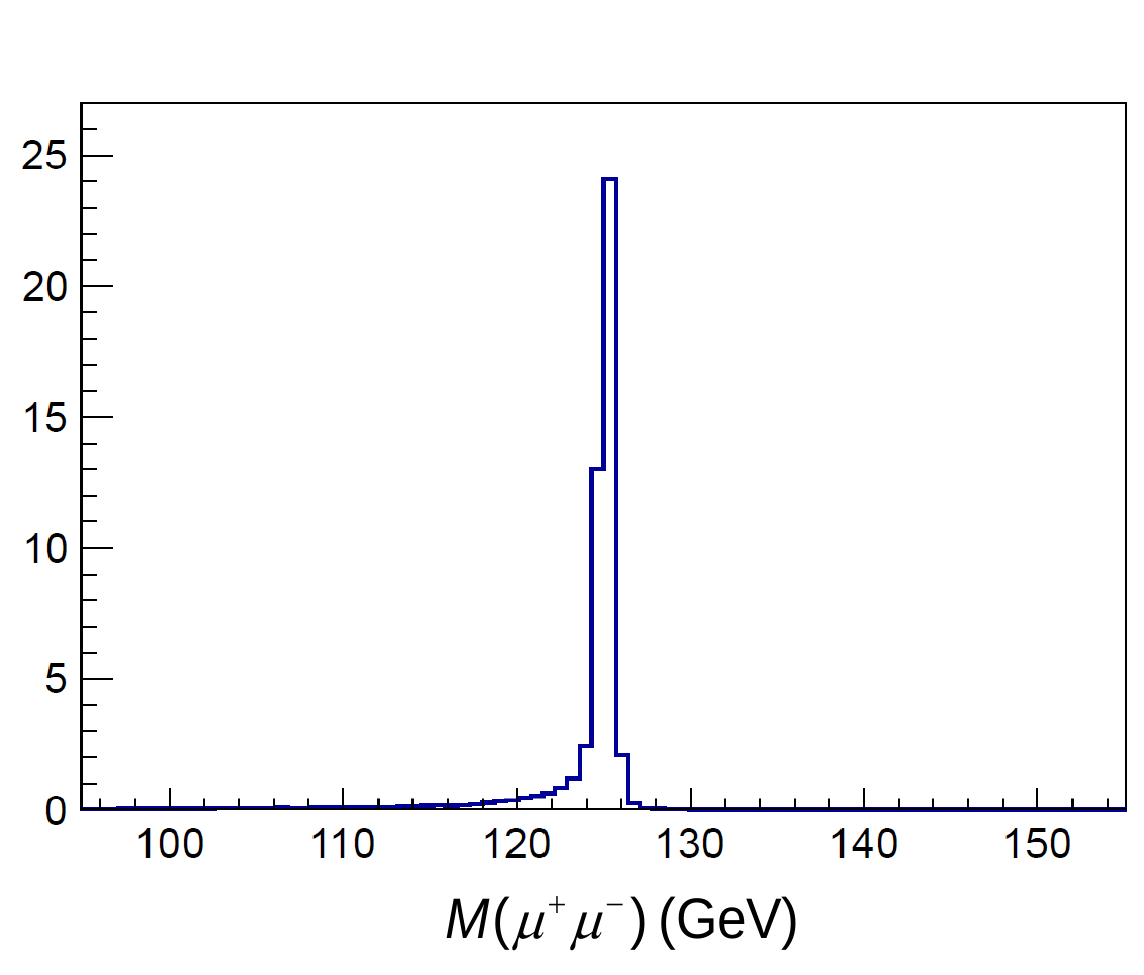}\        \includegraphics[width=0.47\linewidth]{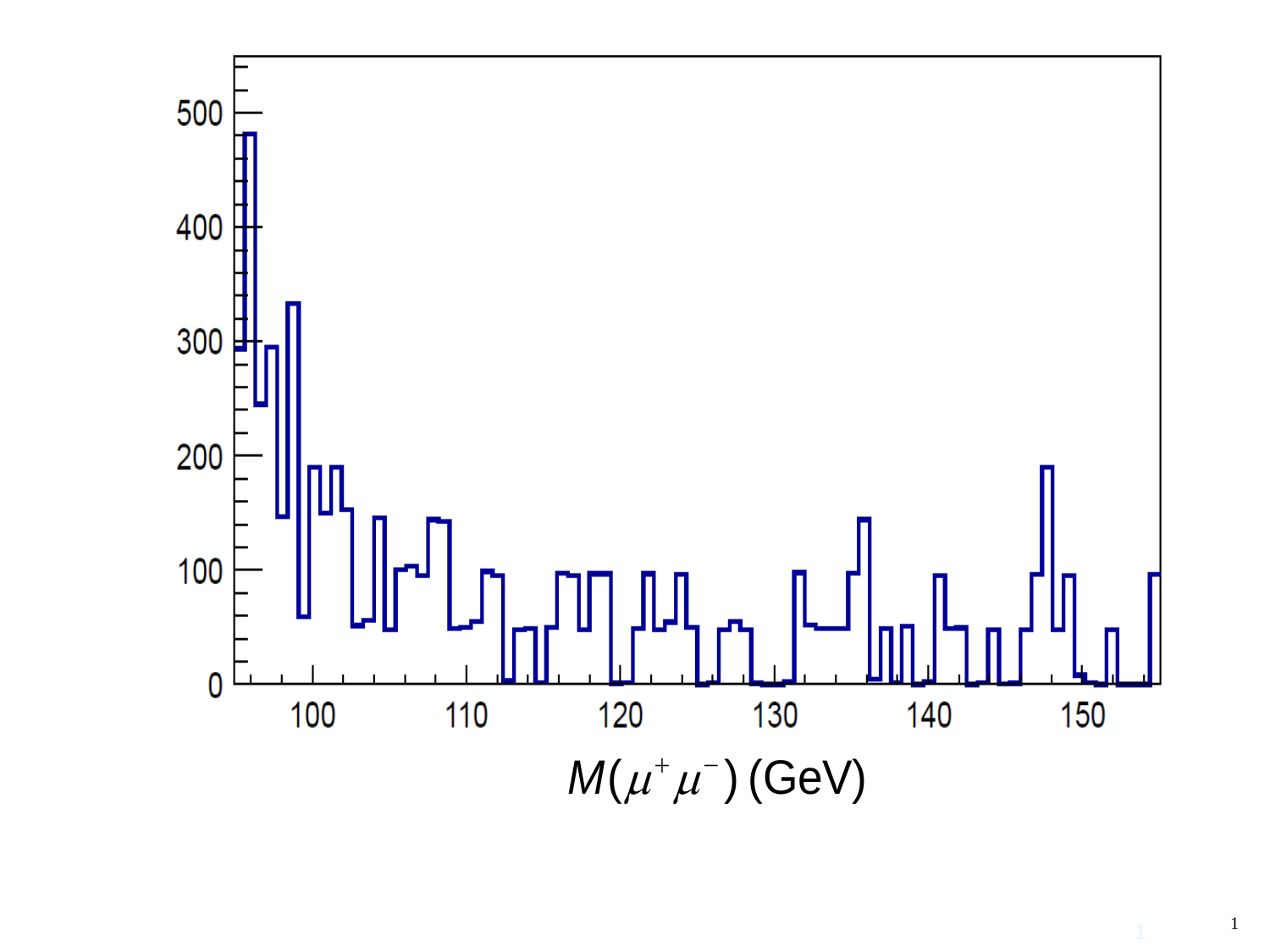}
\end{center}
\caption{ Muon pair mass for $e^+ e^- \rightarrow  \nu \bar{\nu} h \rightarrow  \nu \bar{\nu} \mu^+\mu^- $ at $\sqrt{s}=1000$~GeV (left) and for all 
Standard Model background (right) following all cuts. The plots are normalized to 1000~fb$^{-1}$ luminosity.
}
\label{fig:mumumass}
\end{figure}
%%%%%%%%%%%%%%%%%%%%%%%%%%%%%%%%%%%%%%%%%%%%%%%%%%%%%%%%%%%%%%%%%%%%%%%%%%%

%\subsection{$e^+ e^- \rightarrow Zh \rightarrow  f\bar{f} \mu^+\mu^-,\, \   f=q,\nu $ at $\sqrt{s}=250$~GeV}
%\subsection{$e^+ e^- \rightarrow  \nu \bar{\nu} h \rightarrow  \nu \bar{\nu} \mu^+\mu^- $ at $\sqrt{s}=1000$~GeV}

\section{Invisible Higgs Decays}
The $h$ decay to invisible final states, if any, can be measured by 
looking at the recoil mass under the condition that nothing observable 
is recoiling against the $Z$ boson. 
%%%
Higgs portal models predict such decays and provide a unique opportunity 
to access dark matter particles~\cite{Englert:2011yb}. 
% 
%.
%%%
The main background is $e^+e^- \to ZZ$ followed by one $Z$ decaying into 
a lepton pair or quark pair,  and the other into a neutrino pair. With an integrated 
luminosity of 250\,fb$^{-1}$ at $\sqrt{s} = 250$\,GeV, the ILC can set 
a 95\%~CL limit on the invisible branching ratio of
4.8\%
using the golden $Z \to \mu^+\mu^-$ mode alone~\cite{ref:2012onoc}.  Using 
other modes including $Z \to q\bar{q}$, we can improve this significantly to 
0.9\%~\cite{ref:2012yamamoto}.  Assuming a luminosity of 1150\,fb$^{-1}$ at $\sqrt{s} = 250$\,GeV
the 95\% CL limit is 0.4\%

\section{Top Yukawa Coupling Measurement}
   The 
cross section for the process $e^+e^- \to t\bar{t}h$ 
is 
significantly enhanced near the threshold due to the 
bound-state effects between $t$ and $\bar{t}$ 
\cite{Dittmaier:1998dz,Dawson:1998ej,Belanger:2003nm,Denner:2003zp,You:2003zq,Farrell:2005fk,Farrell:2006xe}.  
The effect is made 
obvious in the right-hand plot of Fig.~\ref{fig:sigtth}.  This enhancement
implies that 
the measurement of the top Yukawa coupling 
might be possible already at $\sqrt{s} = 500$~GeV~\cite{Juste:2006sv}. 
A serious simulation study at $\sqrt{s} = 500$~GeV 
was performed for the first time, with the QCD bound-state effects 
consistently taken into account for both signal and background cross sections,
in  \cite{Yonamine:2011jg}.

%%%%%%%%%%%%%%%%
\begin{figure}[t]
\begin{center}
      \includegraphics[width=0.47\linewidth]{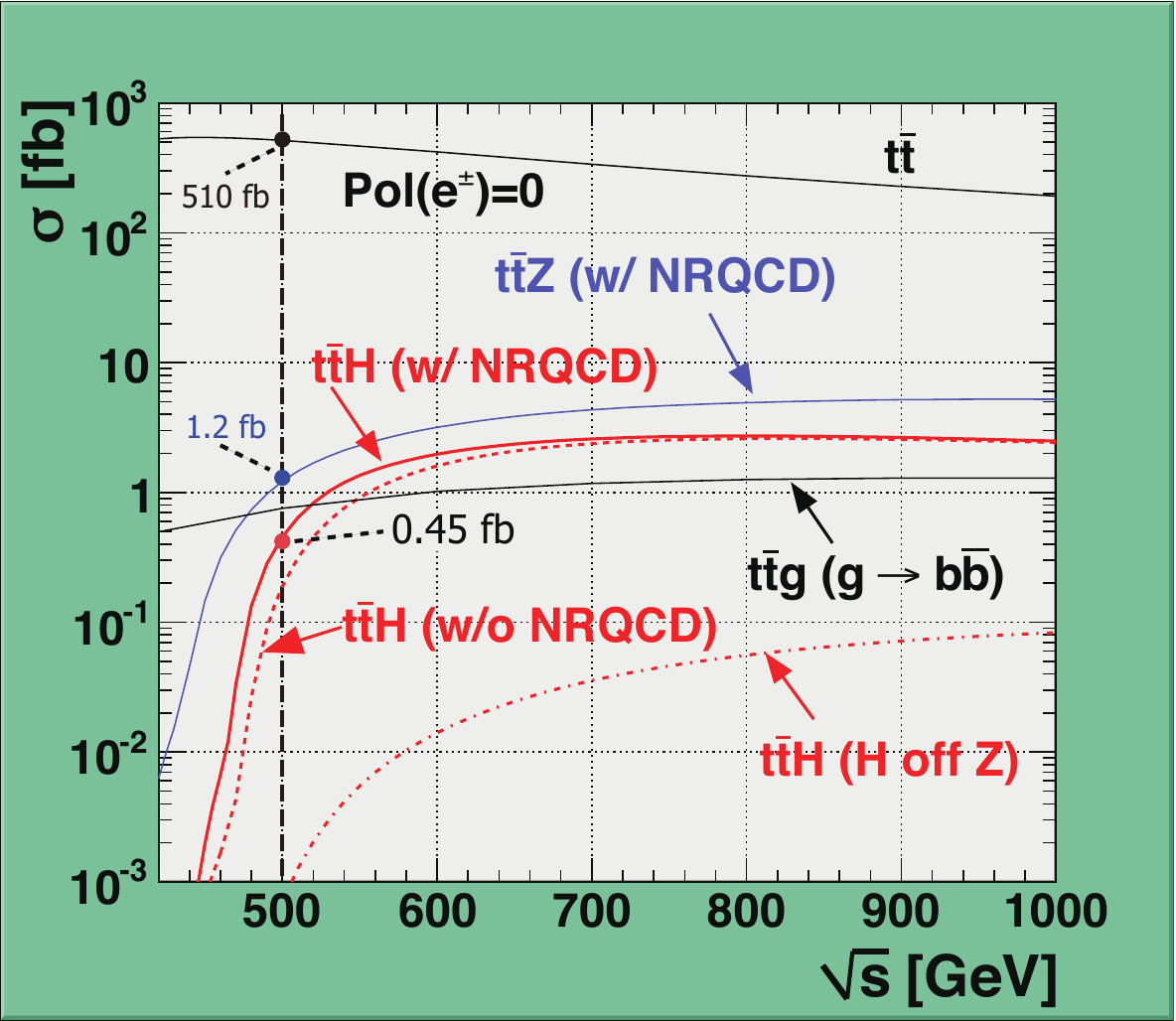}\  
      \includegraphics[width=0.47\linewidth]{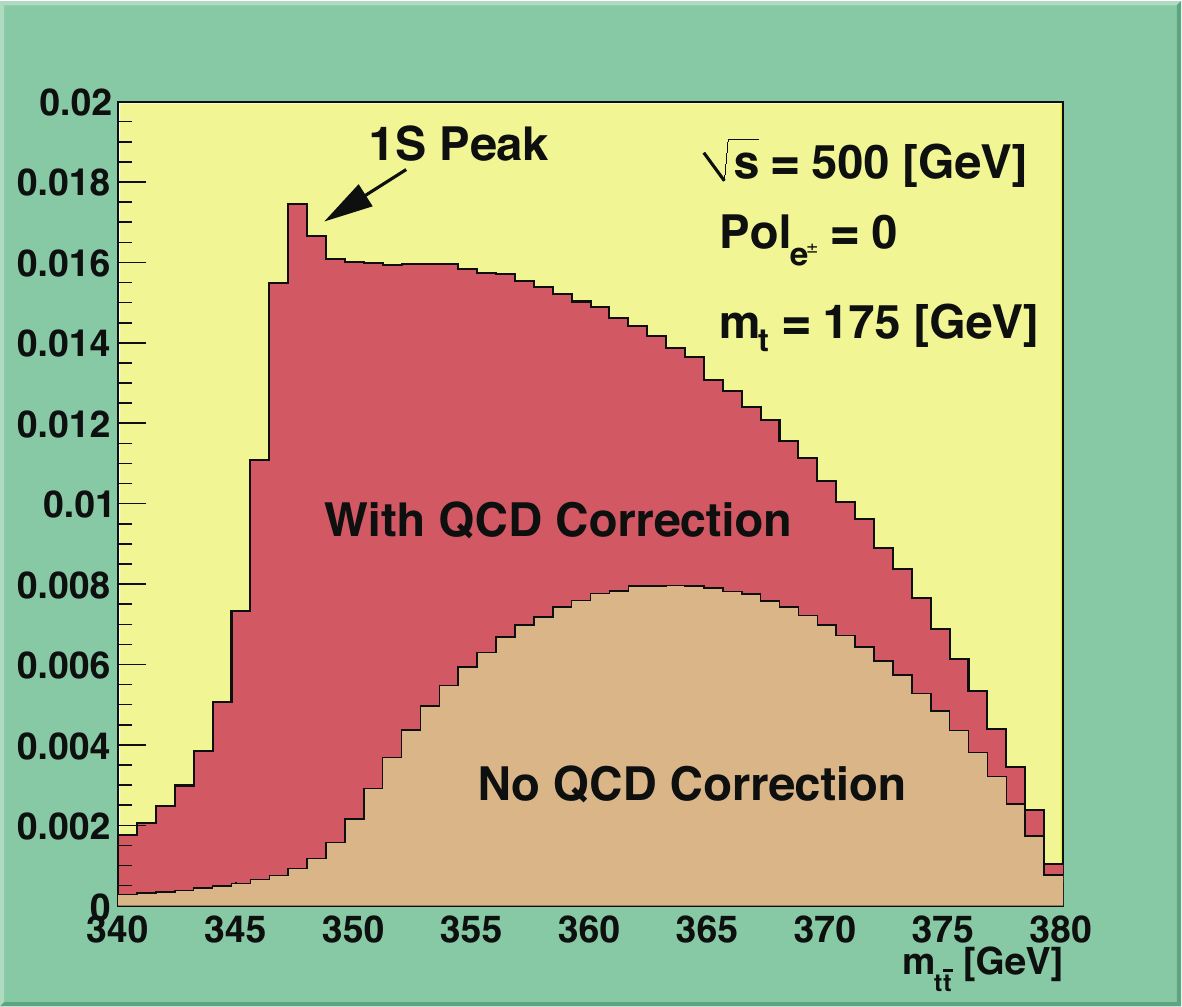}
\end{center}
\caption{
 Left:  Cross section for the $e^+e^- \to t\bar{t}h$ process 
as a function of $\sqrt{s}$, together with those 
of background processes, $e^+e^- \to t\bar{t}Z$, 
$\to t\bar{t}g^*$, and $\to t\bar{t}$.
 Right: The invariant mass distribution of the 
$t\bar{t}$ system from the $e^+e^- \to t\bar{t}h$ 
process with and without the non-relativistic QCD correction.
}
\label{fig:sigtth}
\end{figure}
%%%%%%%%%%%%%%%%%%%%%%%%%%%%%%%%%%%%%%%%%%%%%%%%%%%%%%%%%%%%%%%%%%%%%%%%%%%
%%%%%%%%%%%%%%%%%%%%%%%%%%%%%%%%%%%%%%%%%%%%%%%%%%%%%%%%%%%%%%%%%
\begin{figure}
\begin{center}
\includegraphics[width=0.25\hsize]{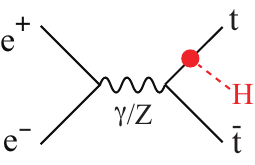}
\includegraphics[width=0.25\hsize]{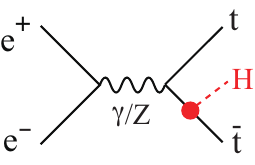}
\includegraphics[width=0.25\hsize]{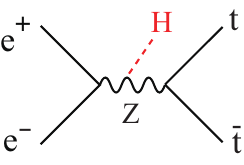}
\end{center}
\caption{
Three diagrams contributing to the $e^+e^- \to t\bar{t}h$ process. 
The $h$-off-$t$ or $\bar{t}$ diagrams, (a) and (b), contain the top 
Yukawa coupling while the $h$-off-$Z$ diagram (c) does not.}
\label{fig:tthdiagrams}
\end{figure}
%%%%%%%%%%%%%%%%%%%%%%%%%%%%%%%%%%%%%%%%%%%%%%%%%%%%%%%%%%%%%%%%%%%%%%%%%%%

The $e^+e^- \to t\bar{t}h$ reaction takes place through the three 
diagrams shown in Fig.~\ref{fig:tthdiagrams}.
As shown in Fig.~\ref{fig:sigtth} (left), the contribution from the 
irrelevant $h$-off-$Z$ diagram is negligible at $\sqrt{s} = 500$\,GeV, 
thereby allowing us to extract the top Yukawa coupling $g_t$ by just 
counting the number of signal events.
By combining the 8-jet and 6-jet-plus-lepton modes of $e^+e^- \to t\bar{t}h$ 
followed by $h \to b\bar{b}$, the analysis of \cite{Yonamine:2011jg}
showed that a measurement of the 
top Yukawa coupling to $\Delta g_t / g_t = 14.1\%$ is possible for 
$m_h=120$~GeV 
with polarized electron and positron beams of $(P_{e^-}, P_{e^+})=(-0,8,+0.3)$ 
and an integrated luminosity of $500$~fb$^{-1}$. This result obtained with a 
fast Monte Carlo simulation has just recently been corroborated by a full 
simulation~\cite{Tabassam:2012it}. 
When extrapolated to $m_h=125$~GeV, and taking into account a recent analysis improvement, 
the corresponding expected precision would be $\Delta g_t/g_t = 14.0\%$.

  It should be noted that a small increase in the center of mass 
energy beyond $\sqrt{s} = 500$\,GeV can increase the cross section for 
$e^+e^- \to t\bar{t}h$ significantly, as can  be seen in  Fig.~\ref{fig:sigtth}.   By increasing the center of mass
energy to  $\sqrt{s} = 520$\,GeV, for example, the cross section for  $e^+e^- \to t\bar{t}h$ can be doubled and hence the precision can be improved to $9.9\%$
assuming  $500$~fb$^{-1}$.

The $14$\% accuracy on the top quark Yukawa coupling expected at $\sqrt{s}=500$\,GeV can be significantly improved by the data taken at 1000\,GeV, 
thanks to the larger cross section and the less background from $e^+e^- \to t\bar{t}$. Fast simulations at 
$\sqrt{s}=800$\,GeV showed that we would be able to determine the top Yukawa coupling to $6$\% for 
$m_h=120$\,GeV, given an integrated luminosity of $1$\,ab$^{-1}$ and residual background uncertainty 
of $5$\%~\cite{Juste:1999af,Gay:2006vs}. 
As described in the Detector Volume of the ILC TDR~\cite{Behnke:2013lya}
full simulations just recently completed by SiD and ILD
show that the top Yukawa coupling can indeed be measured to a
statistical precision of $3.1$\% for $m_h = 125\,$GeV 
with~$1$\,ab$^{-1}$.

With luminosities of $1600$~fb$^{-1}$ at $500$~GeV and
$2500$~fb$^{-1}$ at $1000$~GeV, the statistical precision can be improved to $2.0\%$.

\section{Higgs Self Coupling Measurement}
%Explain why the measurement is difficult, small cross section, huge BGs, and
%in particular, emphasize the effect of the irreducible BG diagrams which forbid a simple-minded counting experiment as we assumed for the top Yukawa coupling measurement.
%Emphasize the energy dependence of the relative importance of the BG diagrams and discuss the possible weighting method to reduce the effect of the BG diagrams.
%Explain the current state-of-the-art and possible future improvements.

%%%%%%%%%%%%%%%%%%%%%%%%%%%%%%%%%%%%%%%%%%%%%%%%%%%%%%%%%%%%%%%%%
\thisfloatsetup{floatwidth=\SfigwFull,capposition=beside}
\begin{figure}[t]%
   \begin{subfigure}[b]{0.35\hsize-0.5\columnsep}%
 \includegraphics[width=\hsize]{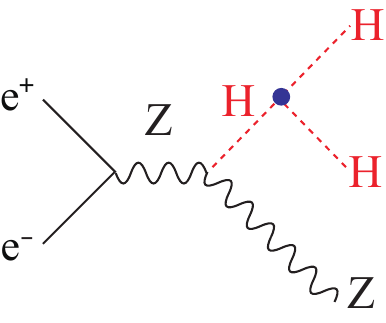}%
    \end{subfigure}%
    \hspace{\columnsep}%
    \begin{subfigure}[b]{0.38\hsize-0.5\columnsep}%
 \includegraphics[width=\hsize]{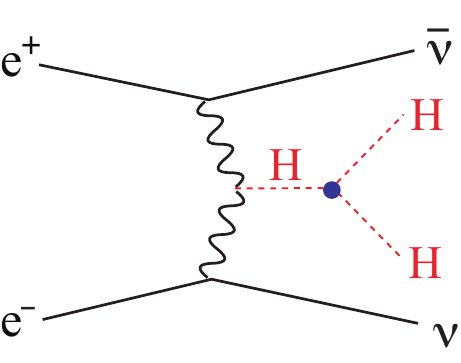}%     
    \end{subfigure}%
 \caption{
Relevant diagrams containing the triple Higgs coupling for the two
processes:
 $e^+e^-\rightarrow Zhh$ (left) and $e^+e^-\rightarrow\nu_e\overline{\nu}_e hh$.
}
\label{fig:HHHdiagrams}
\end{figure}

\thisfloatsetup{floatwidth=\SfigwFull,capposition=beside}
\begin{figure}[t]
 \includegraphics[width=0.8\hsize]{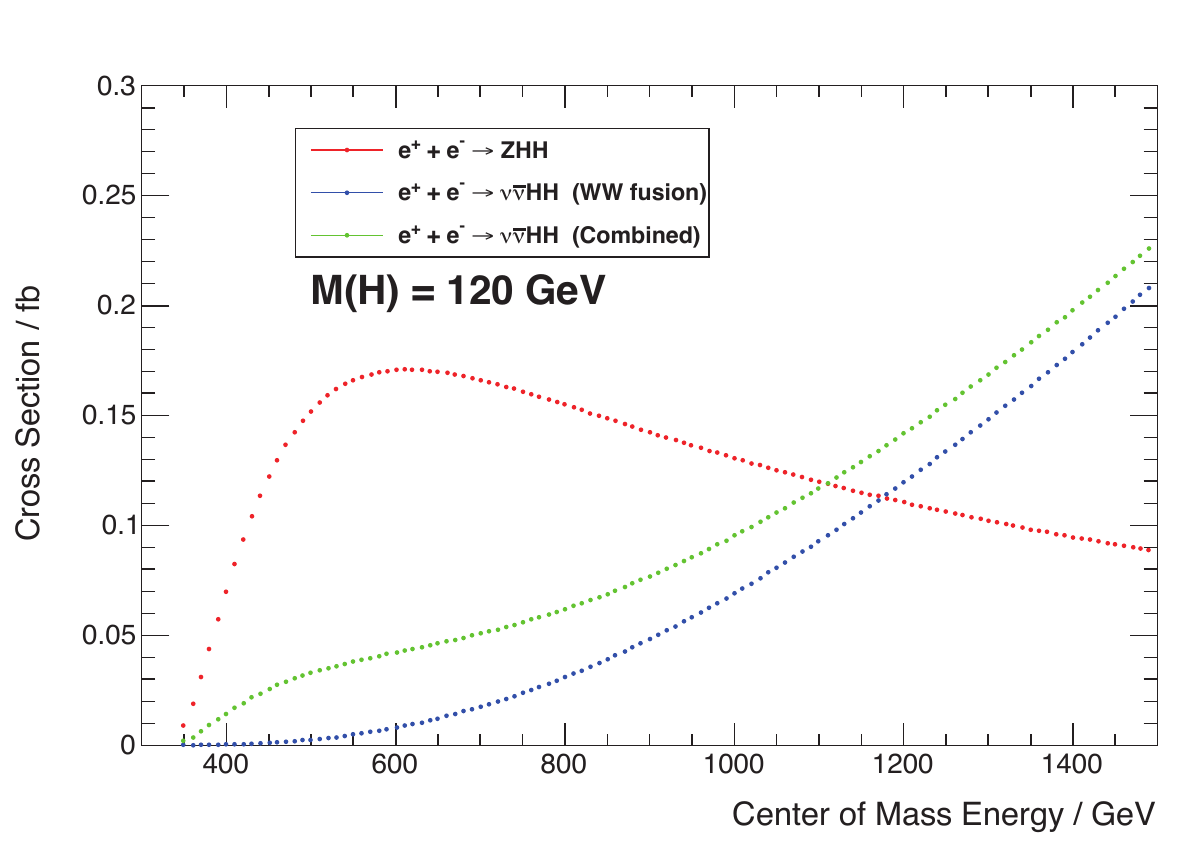}
\caption{
Cross sections for the two processes
 $e^+e^-\rightarrow Zhh$ (left) and $e^+e^-\rightarrow\nu_e\overline{\nu}_e hh$ 
as a function of $\sqrt{s}$ 
for $m_h=120$~GeV.}
\label{fig:sigHHH}
\end{figure}

%%%%%%%%%%%%%%%%%%%%%%%%%%%%%%%%%%%%%%

The triple Higgs boson coupling can be studied at the ILC through 
the processes $e^+e^-\rightarrow Zhh$ and 
$e^+e^-\rightarrow\nu_e\overline{\nu}_e hh$.  The relevant Feynman diagrams
are shown in Fig.~\ref{fig:HHHdiagrams}~\cite{Djouadi:1999gv}.
The cross sections for the two processes are plotted as a function of 
$\sqrt{s}$ for $m_h=120$\,GeV in Fig.~\ref{fig:sigHHH}.
The cross section reaches its maximum of about $0.18$\,fb at around 
$\sqrt{s}=500$\,GeV, which is dominated by the former process.

%%%%

%%%%%
 A full 
simulation study of the process $e^+e^- \to Zhh$ 
followed by $h \to b\bar{b}$ at $\sqrt{s}=500$~GeV has recently been carried out
using the ILD detector~\cite{Tian:2013hhbbbb}.
From the combined result of the three channels corresponding to 
different $Z$ decay modes, $Z \to l^+l^-$, $\nu\bar{\nu}$, and $q\bar{q}$, 
it was found that the process can be detected with an excess 
significance of $4.5$-$\sigma$ and the cross section can be 
measured to $\Delta \sigma / \sigma = 0.30$ for an integrated
 luminosity of $1.6$\,ab$^{-1}$ with beam polarization
 $(P_{e^-}, P_{e^+})=(-0,8,+0.3)$. Unlike the $e^+e^- \to t\bar{t}h$ case,
 however, the contribution from the background diagrams without the 
self-coupling is significant and the relative error on the 
self-coupling $\lambda$ is $\Delta \lambda / \lambda = 0.49$
 with a proper event weighting to enhance the contribution from
 the self-coupling diagram. 
 When extrapolated to $m_h=125$~GeV, taking into account a $20\%$ relative improvement
 expected from a recent preliminary full simulation result including $hh \to b\bar{b}WW^*$ mode,
 the precision would be improved to $46\%$.

At $\sqrt{s}=1000$\,GeV, the $e^+e^- \to \nu\bar{\nu}hh$ process will become significant~\cite{Yasui:2002se}. The cross 
section for this process is only  about $0.07$\,fb$^{-1}$, but the sensitivity to the self-coupling is 
potentially higher since the contribution from the background diagrams is smaller, leading to the relation
 $\Delta \lambda / \lambda \simeq 0.85 \times (\Delta \sigma_{\nu\bar{\nu}hh} / \sigma_{\nu\bar{\nu}hh})$, as 
compared to  $\Delta \lambda / \lambda \simeq 1.8 \times (\Delta \sigma_{Zhh} / \sigma_{Zhh})$ for the 
$e^+e^- \to Zhh$ process at 500\,GeV.
The measurement of the
self-coupling has been studied at 1~TeV with full simulation.
That analysis
is described in the Detector Volume of the ILC TDR~\cite{Behnke:2013lya}.  The
result, for  $2.5$\,ab$^{-1}$ with $(P_{e^-}, P_{e^+})=(-0.8,+0.2)$,
is $\Delta \lambda/\lambda \simeq 0.16$ for $m_h=125$~GeV.  This has recently been improved to $13\%$ 
with the inclusion of the $hh \to b\bar{b}WW^*$ mode~\cite{Kawada:2013hhbbww}.
Further improvements would be possible by adding more decay modes and/or improvements in jet clustering
\footnote{With perfect jet clustering we expect a $40\%$ relative improvement in the self-coupling precision.}.

%In addition to the fusion process, we also can use the $e^+e- \to Zhh$
%process 
%at $\sqrt{s}=1000$~GeV. 
%This process has somewhat less sensitivity, 
%$\Delta \lambda / \lambda \simeq 2.8 \times (\Delta \sigma_{Zhh} /
%\sigma_{Zhh})$.    The analysis gives  $\Delta
%\lambda/\lambda \simeq  0.53$.
%Combining all of the 
%measurements, 
%assuming integrated luminosities of $1600\,$fb$^{-1}$
% at $\sqrt{s}=500$\,GeV 
%and $2500\,$fb$^{-1}$ at $\sqrt{s}=1000$\,GeV with the left-handed beam combination: $(P_{e^-}, P_{e^+})=(-0.8,+0.2)$, 
%we expect that the Higgs self-coupling could be measured to $\Delta
%\lambda/\lambda \simeq  16.\%$.

\section{Cross Section Times Branching Ratio Summary}

The accuracies for all cross section and $\sigma\cdot BR$ measurements 
considered in this paper are summarized in  \Tref{tab:stimesbrbase} and \Tref{tab:stimesbrupgrade}.
 \Tref{tab:stimesbrbase} shows the accuracies assuming you run $3\times 10^7$~s at
the baseline differential luminosity for each of the center of mass energies 250, 500 and 1000~GeV. \Tref{tab:stimesbrupgrade}
gives the accuracies when you add the luminosities of   \Tref{tab:stimesbrbase} to  $3\times 10^7$~s
times the upgraded differential luminosities at each of the three center of mass energies.

\begin{table}[t]
 \begin{center}
 \begin{tabular}{|l|r|r|r|r|r|r|r|r|r|}
   \hline
  $\sqrt{s}$ and ${\cal L}$ & \multicolumn{2}{c|}{250\,fb$^{-1}$ at 250\,GeV} 
                                         & \multicolumn{4}{c|}{500\,fb$^{-1}$ at 500\,GeV} 
                                         & \multicolumn{3}{c|}{1\,ab$^{-1}$ at 1\,TeV }\\
  $(P_{e^-},P_{e^+})$ & \multicolumn{2}{c|}{(-0.8,+0.3)} 
                                      & \multicolumn{4}{c|}{(-0.8,+0.3)} 
                                      & \multicolumn{3}{c|}{(-0.8,+0.2)} \\
  \hline
         & $Zh$ & $\nu\bar{\nu}h$ 
             & $Zh$ & $\nu\bar{\nu}h$ & $t\bar{t}h$ & $Zhh$
             & $\nu\bar{\nu}h$ & $t\bar{t}h$  & $\nu\bar{\nu}hh$  \\
   \hline\hline
  $\Delta \sigma / \sigma$              & 2.6\%     &  -       & 3.0    &  -  &     &  42.7\%          &   & & 26.3\% \\ \hline
   BR(invis.)        & $ <$  0.9  \%   & -         & -             & -    & -   & &  \\ \hline \hline
   mode & \multicolumn{9}{c|}{$\Delta (\sigma \cdot BR) / (\sigma \cdot BR)$}  \\
  \hline
   $h \to b\bar{b}$              & 1.2\%     & 10.5\%     & 1.8\%     & 0.7\%   &  28\% &       & 0.5\%  &  6.0\% & \\
   $h \to c\bar{c}$              & 8.3\%     & -           & 13\%      & 6.2\%    & &    & 3.1\%   & & \\
   $h \to gg$                       & 7.0\%     & -          & 11\%       & 4.1\%   & &    & 2.3\%   & & \\
   $h \to WW^*$                 & 6.4\%      & -          & 9.2\%     & 2.4\%   & &     & 1.6\%   & & \\
   $h \to \tau^+\tau^-$        & 4.2\%     & -           & 5.4\%     & 9.0\%      & &   & 3.1\%   & & \\
   $h \to ZZ^*$                   & 18\%       & -          & 25\%      & 8.2\%   & &      & 4.1\%    & & \\
   $h \to \gamma\gamma$ & 34\% & -          & 34\% & 23\%    & &  &  8.5\%  & & \\
   $h \to \mu^+\mu^-$        & 100\%     & -         & -             & -        & &         & 31\%  & & \\
%
%   $h \to Z\gamma$ & ??\% & ??\% & ??\%  & ?\%   \\
   \hline
  \end{tabular}
  \caption{Expected accuracies for cross section and cross section times branching ratio
measurements for the $125\,$GeV $h$ boson assuming you run $3\times 10^7$~s at
the baseline differential luminosity for each center of mass energy. For invisible decays of the Higgs, 
the number quoted is the 95\% confidence upper limit on the branching ratio.
% Assumed integrated luminosities and beam polarizations are
% ${\cal L}=250\,{\mathrm fb}^{-1}$ and $(P_{e^-},P_{e^+})=(-0.8,+0.3)$ at $\sqrt{s}=250\,$GeV,
% ${\cal L}=500\,{\mathrm fb}^{-1}$ and $(P_{e^-},P_{e^+})=(-0.8,+0.3)$ at $\sqrt{s}=500\,$GeV,
% and
% ${\cal L}=1\,{\mathrm ab}^{-1}$ and $(P_{e^-},P_{e^+})=(-0.8,+0.2)$ at $\sqrt{s}=1\,$TeV,
}
\label{tab:stimesbrbase}
  \end{center}
\end{table}

\begin{table}[t]
 \begin{center}
 \begin{tabular}{|l|r|r|r|r|r|r|r|r|r|}
   \hline
  $\sqrt{s}$ and ${\cal L}$ & \multicolumn{2}{c|}{1150\,fb$^{-1}$ at 250\,GeV} 
                                         & \multicolumn{4}{c|}{1600\,fb$^{-1}$ at 500\,GeV} 
                                         & \multicolumn{3}{c|}{2.5\,ab$^{-1}$ at 1\,TeV }\\
  $(P_{e^-},P_{e^+})$ & \multicolumn{2}{c|}{(-0.8,+0.3)} 
                                      & \multicolumn{4}{c|}{(-0.8,+0.3)} 
                                      &  \multicolumn{3}{c|}{(-0.8,+0.2)} \\
  \hline
         & $Zh$ & $\nu\bar{\nu}h$ 
             & $Zh$ & $\nu\bar{\nu}h$ & $t\bar{t}h$ & $Zhh$
             & $\nu\bar{\nu}h$ & $t\bar{t}h$  & $\nu\bar{\nu}hh$  \\
   \hline\hline
  $\Delta \sigma / \sigma$              & 1.2\%     &  -       & 1.7     &  -  &     &  23.7\%          &   & & 16.7\% \\ \hline
   BR(invis.)        & $ <$  0.4  \%   & -         & -             & -   & &    & -    & & \\ \hline \hline
   mode & \multicolumn{9}{c|}{$\Delta (\sigma \cdot BR) / (\sigma \cdot BR)$}  \\
  \hline
   $h \to b\bar{b}$              & 0.6\%     & 4.9\%      & 1.0\%     & 0.4\% &  16\% &         & 0.3\%  & 3.8\% & \\
   $h \to c\bar{c}$              & 3.9\%     & -           & 7.2\%      & 3.5\%    & &    & 2.0\%   & & \\
   $h \to gg$                       & 3.3\%     & -          & 6.0\%       & 2.3\%  & &     & 1.4\%   & & \\
   $h \to WW^*$                 & 3.0\%      & -          & 5.1\%     & 1.3\%   & &     & 1.0\%   & & \\
   $h \to \tau^+\tau^-$        & 2.0\%     & -           & 3.0\%     & 5.0\%   & &      & 2.0\%   & & \\
   $h \to ZZ^*$                   & 8.4\%       & -          & 14\%      & 4.6\%   & &      & 2.6\%    & & \\
   $h \to \gamma\gamma$           & 16\%       & -          & 19\%       & 13\%    & &     &  5.4\%  & & \\
   $h \to \mu^+\mu^-$        & 46.6\%    & -         & -             & -        & &         & 20\%  & & \\ 
%
%   $h \to Z\gamma$ & ??\% & ??\% & ??\%  & ?\%   \\
   \hline
  \end{tabular}
  \caption{Expected accuracies for cross section and cross section times branching ratio
measurements for the $125\,$GeV $h$ boson assuming  you run $3\times 10^7$~s at
the sum of the baseline and upgrade differential luminosities for each center of mass energy.
For invisible decays of the Higgs, 
the number quoted is the 95\% confidence upper limit on the branching ratio.
% Assumed integrated luminosities and beam polarizations are
% ${\cal L}=250\,{\mathrm fb}^{-1}$ and $(P_{e^-},P_{e^+})=(-0.8,+0.3)$ at $\sqrt{s}=250\,$GeV,
% ${\cal L}=500\,{\mathrm fb}^{-1}$ and $(P_{e^-},P_{e^+})=(-0.8,+0.3)$ at $\sqrt{s}=500\,$GeV,
% and
% ${\cal L}=1\,{\mathrm ab}^{-1}$ and $(P_{e^-},P_{e^+})=(-0.8,+0.2)$ at $\sqrt{s}=1\,$TeV,
}
\label{tab:stimesbrupgrade}
  \end{center}
\end{table}

\chapter{Higgs  Couplings, Total Width and  Branching Ratios \label{sid:chapter_couplings}}
\section{Model Independent Determination of Higgs Couplings}
%Explain global fit including independent measurements such as $\sigma \times BR$s for
%both $\sigma_{Zh}$ (higgsstrahlung) and $\nu\bar{\nu}h$ ($W$-fusion), $\sigma_{Zh}$ and free parameters, 
%individual couplings ($g_{hAA}$) and the total width ($\Gamma_h$).
%Show results at each energy with data cumulatively combined.
%Emphasize the power of $W$-fusion process at higher energies, though it is not so trivial for $h \to \tau^+\tau^-$.
%Again point out that some most accurate measurements would be limited by the $\sigma_{Zh}$ error from the recoil mass measurement.
The sigma times branching ratio measurements in the previous chapters
%Tables~\ref{table:brall} and
%\ref{table:tthzhhvvhh}  
imply a very high level of precision for the
various Higgs boson couplings.  To quantify this  we perform a  global fit of the  Higgs boson couplings and total Higgs width  using
all the available cross section and cross section times branching ratio data.

Before discussing the global fit in detail, it would be helpful to show an 
example explaining how we get the absolute couplings and Higgs total width. 
Let's look at the following four independent measurements:
\beqa
Y_1 &=& \sigma_{ZH} = F_1\cdot g^2_{HZZ} \nonumber \\
Y_2 &=& \sigma_{ZH}\times {\mathrm Br}(H\rightarrow b\bar{b}) = F_2\cdot {g^2_{HZZ}g^2_{Hb\bar{b}} \over \Gamma_T} \nonumber \\
Y_3 &=& \sigma_{\nu\bar{\nu}H}\times {\mathrm Br}(H\rightarrow b\bar{b}) = F_3\cdot  {g^2_{HWW}g^2_{Hb\bar{b}} \over \Gamma_T} \nonumber \\
Y_4 &=& \sigma_{\nu\bar{\nu}H}\times  {\mathrm Br}(H\rightarrow WW^*) = F_4\cdot  {g^4_{HWW} \over \Gamma_T}\,, \nonumber
\eeqa
where $\Gamma_T$ is the Higgs total width, $g_{HZZ}$, $g_{HWW}$, and  $g_{Hb\bar{b}}$ are the couplings 
of the Higgs to $ZZ$, $WW$, and $b\bar{b}$, respectively, and $F_1$,  $F_2$,  $F_3$,  $F_4$ 
are calculable quantities.  It is straightforward to get the couplings with the 
following steps: 

\begin{enumerate}[label=\roman*.)]
\item from the measurement $Y_1$ we can get the coupling $g_{HZZ}$. 
\item from the ratio $Y_2/Y_3$ we can get the coupling ratio $g_{HZZ}/g_{HWW}$. 
\item with $g_{HZZ}$ and $g_{HZZ}/g_{HWW}$, we can get $g_{HWW}$. 
\item once we know $g_{HWW}$,  we can get the Higgs total width $\Gamma_T$ from the measurement $Y_4$
\item from the ratio $Y_3/Y_4$ we get the ratio $g_{Hbb}/g_{HWW}$, from which we obtain $g_{Hbb}$.
\end{enumerate}

This example already gave quite clear synergy between the two main Higgs production channels. The 
best energy
to investigate the Higgsstrahlung production $e^+e^-\rightarrow ZH$ is around 250 GeV, however the 
$e^+e^-\rightarrow \nu\bar{\nu}H$ at 250 GeV is very small. WW-fusion production will be fully open 
at 500 GeV with cross section one order of magnitude larger. This is one essential motivation to go to higher 
energy after running at 250 GeV.

We discuss in detail the model independent fit of the Higgs couplings for the ILC(1000) luminosity scenario.  For this scenario 
the 33 independent $\sigma \times {\mathrm Br}$ measurements in  \Tref{tab:stimesbrbase} are used as experimental input. The 
$\sigma \times {\mathrm Br}$ measurements are labelled with 
with $Y_i$, $i=1,2,...,33$.   The predicted values of these measurements as a function of the Higgs couplings are given by 
$Y^{'}_i=F_i\cdot {g^2_{HZZ}g^2_{HXX} \over \Gamma_T}$, 
or $Y^{'}_i=F_i\cdot {g^2_{HWW}g^2_{HXX} \over \Gamma_T}$, $Y^{'}_i=F_i\cdot {g^2_{Htt}g^2_{HXX} \over \Gamma_T}$,
where $XX$ means some specific decay particle from Higgs and $F_i$ is some factor corresponding to the decay.  In addition we have one absolute cross section 
measurement $Y_{34}=\sigma_{ZH}$ which can be predicted as 
$Y^{'}_{34}=F_{34}\cdot g^2_{HZZ}$. 
In  total we have 34 independent measurements and 10 fit parameters consisting of 9  
fundamental couplings $HZZ$, $HWW$,  $Hb\bar b$, $Hc \bar c$,
$Hgg$, $H\tau^+\tau^-$, $H\mu\mu$, $Htt$ and $H\gamma\gamma$,  and the Higgs total width $\Gamma_T$.

 The factors $F_i$ can be written 
\beq
  F_i=S_iG_i\ \ \ {\mathrm where}\ S_i=({\sigma_{ZH}\over g_Z^2})\,,\ ({\sigma_{\nu\bar\nu H}\over g_W^2})\,,\ {\mathrm or}\ ({\sigma_{t\bar{t}H}\over g_t^2})\,, \ {\mathrm and}\  G_i=({\Gamma_i\over g_i^2})\, .
\eeq
These are theoretical calculations with parametric and theoretical uncertainties.
Because the relevant quantities are ratios of cross sections and partial widths to couplings squared, the total theory errors for $S_i$, and particularly $G_i$, should be less than
the total theory errors for the corresponding cross sections and partial widths.  We believe that a total theory error of 0.5\% or less can be achieved for the $F_i$ parameters at the time of ILC running.
We quote coupling results assuming total theory errors of  $\Delta F_i/F_i=0.1\%$ and $\Delta F_i/F_i=0.5\%$.

The fitted couplings and width are obtained by minimizing the chi-square function $\chi^2$ defined by 
\beq
     \chi^2  = \sum^{34}_{i=1}({Y_i-Y^{'}_i \over \Delta Y_i})^2\,,
\eeq
where $\Delta Y_i$ is the square root of the sum in quadrature of the error on the measurement $Y_i$
and the total theory error for $Y^{'}_i$.
The results for theory errors of  $\Delta F_i/F_i=0.1\%$ and $\Delta F_i/F_i=0.5\%$ are summarized in \Tref{tab:modelindglobalfit0p1} and  \Tref{tab:modelindglobalfit0p5}, respectively.
%The results for the
%errors on Higgs couplings are shown in 
%Table~\ref{tab:globalfit}.  The
%four columns represent the errors from LHC (300 fb$^{-1}$, 1 detector)
%only, and then, cumulatively,
%ILC at 250~GeV, ILC at 500~GeV, and ILC at
%1000~GeV~\cite{Peskin:2012we}.

\begin{table}
 \begin{center}
 \begin{tabular}{lcccc}
Mode            &  ILC(250)        & ILC(500)           & ILC(1000)      & ILC(LumUp) \cr \hline
$\gamma\gamma$  &      18   \%      &    8.4  \%      &   4.0  \%       & 2.4  \%  \cr
$gg$            &   6.4  \%         &   2.3  \%       &   1.6  \%       & 0.9 \%  \cr
$WW$            &   4.8 \%         &  1.1   \%        &   1.1  \%       & 0.6 \%  \cr
$ZZ$            &      1.3 \%      & 1.0  \%          &   1.0  \%       & 0.5 \%  \cr
$t\bar t$       &     --          &  14   \%       &    3.1   \%        & 1.9 \%  \cr
$b\bar b$       &    5.3 \%         &  1.6   \%       &   1.3    \%     & 0.7 \%  \cr
$\tau^+\tau^-$   &     5.7 \%        &    2.3  \%      &   1.6  \%       & 0.9  \%  \cr
$c\bar c$       &     6.8   \%      &     2.8  \%     &   1.8    \%     & 1.0  \%  \cr
$\mu^+\mu^-$     &    91 \%              &     91 \%          &   16 \%         & 10   \%  \cr
$\Gamma_T(h)$    &    12   \%        &    4.9 \%       &   4.5   \%      & 2.3  \%  \cr
   \hline
   \end{tabular}
  \caption{Expected accuracies $\Delta g_i/g_i$ for Higgs boson couplings
for a completely model independent fit assuming theory errors of $\Delta F_i/F_i=0.1\%$
}
\label{tab:modelindglobalfit0p1}
  \end{center}
\end{table}

%%%%%%%%%%%%%%%%%%%%%%%%%%%%%%%%%%%%%%%%%%%%%%%%%%%%%%%%%%%%%%%%%
\begin{table}
 \begin{center}
 \begin{tabular}{lcccc}
Mode            &  ILC(250)        & ILC(500)           & ILC(1000)      & ILC(LumUp) \cr \hline
$\gamma\gamma$  &      18   \%      &    8.4  \%      &   4.0  \%       & 2.4  \%  \cr
$gg$            &   6.4  \%         &   2.3  \%       &   1.6  \%       & 0.9 \%  \cr
$WW$            &   4.9 \%         &  1.2   \%        &   1.1  \%       & 0.6 \%  \cr
$ZZ$            &      1.3 \%      & 1.0  \%          &   1.0  \%       & 0.5 \%  \cr
$t\bar t$       &     --          &  14   \%       &    3.2   \%        & 2.0 \%  \cr
$b\bar b$       &    5.3 \%         &  1.7   \%       &   1.3    \%     & 0.8 \%  \cr
$\tau^+\tau^-$   &     5.8 \%        &    2.4  \%      &   1.8  \%       & 1.0  \%  \cr
$c\bar c$       &     6.8   \%      &    2.8 \%     &   1.8    \%     & 1.1  \%  \cr
$\mu^+\mu^-$     &   91 \%              &     91 \%          &   16 \%         & 10   \%  \cr
$\Gamma_T(h)$    &    12   \%        &    5.0 \%       &   4.6   \%      & 2.5  \%  \cr
   \hline
   \end{tabular}
  \caption{Expxected accuracies $\Delta g_i/g_i$ for Higgs boson couplings
for a completely model independent fit assuming theory errors of $\Delta F_i/F_i=0.5\%$
}
\label{tab:modelindglobalfit0p5}
  \end{center}
\end{table}
%%%%%%%%%%%%%%%%%%%%%%%%%%%%%%%%%%%%%%%%%%%%%%%%%%%%%%%%%%%%%%%%%

\section{Model Independent Determination of Higgs Cross Sections and Higgs Branching Ratios}

Alternatively, in the $\chi^2$ of our global fit, we can define the fit parameters to be the three cross sections $\sigma_{ZH}$, $\sigma_{\nu\bar{\nu}H}$, $\sigma_{t\bar{t}H}$, and the eight branching ratios
${\mathrm Br}(H\rightarrow b\bar{b})$, ${\mathrm Br}(H\rightarrow c\bar{c})$, ${\mathrm Br}(H\rightarrow gg)$, ${\mathrm Br}(H\rightarrow WW^*)$, ${\mathrm Br}(H\rightarrow ZZ^*)$, ${\mathrm Br}(H\rightarrow \tau^+\tau^-)$, 
${\mathrm Br}(H\rightarrow \mu^+\mu^-)$, ${\mathrm Br}(H\rightarrow \gamma\gamma)$.  Taking again the ILC(1000) luminosity scenario as an example, we use 
the  34 independent cross section and cross section times branching ratio measurements from
 \Tref{tab:stimesbrbase} and appropriately redefined 
$Y^{'}_i$ functions to solve for the 11 parameters through the minimization of an alternate $\chi^2$ function.
The cross section and  branching ratio accuracies for all four of our energy and  luminosity scenarios
are summarized in \Tref{tab:brmodelindglobalfit}.

\begin{table}
 \begin{center}
 \begin{tabular}{|l|cccc|}
\hline
           &  ILC(250)        & ILC500           & ILC(1000)      & ILC(LumUp) \cr \hline
process & \multicolumn{4}{c|}{$\Delta \sigma /\sigma$} \\
\hline
$e^+e^-\rightarrow ZH$                  &      2.6 \%       & 2.0 \%           &  2.0 \%         & 1.0 \%  \cr
$e^+e^-\rightarrow \nu\bar{\nu}H$       &      11 \%      & 2.3 \%          &   2.2 \%       &  1.1 \%  \cr
$e^+e^-\rightarrow t\bar{t}H$           &       -        & 28 \%          &    6.3  \%       & 3.8 \%  \cr
\hline
mode & \multicolumn{4}{c|}{$\Delta \mathrm{Br} /\mathrm{Br}$} \\
\hline
$H\rightarrow ZZ$            &      19 \%      & 7.5  \%          &   4.2 \%       & 2.4 \%  \cr
$H\rightarrow WW$            &     6.9 \%         &  3.1 \%        &   2.5 \%       & 1.3 \%  \cr
$H\rightarrow b\bar b$       &     2.9  \%         &  2.2 \%       &   2.2 \%     & 1.1 \%  \cr
$H\rightarrow c\bar c$       &     8.7 \%      &      5.1 \%     &    3.4 \%     & 1.9 \%  \cr
$H\rightarrow gg$            &     7.5  \%         &   4.0 \%       &   2.9 \%       & 1.6 \%  \cr
$H\rightarrow \tau^+\tau^-$   &     4.9 \%        &    3.7 \%      &    3.0 \%       & 1.6 \%  \cr
$H\rightarrow \gamma\gamma$  &      34 \%      &    17 \%      &    7.9 \%       & 4.7 \%  \cr
$H\rightarrow \mu^+\mu^-$     &   100 \%            &     100 \%          &   31 \%         & 20 \%  \cr
\hline
   \end{tabular}
  \caption{Summary of expected accuracies for the three cross sections
and eight branching ratios obtained from an eleven parameter
global fit of all available data.}
\label{tab:brmodelindglobalfit}
  \end{center}
\end{table}

\section{Model-Dependent Coupling Parameterizations}
While the couplings of the Higgs boson  and the total
Higgs width can be determined at the ILC 
 without model assumptions, it is sometimes 
useful to extract couplings from ILC data within
the context of certain models.   Such
analyses makes it easier to compare the experimental
precision of the ILC with other facilities, such 
as the LHC, that cannot determine Higgs couplings 
in a model independent manner.  
\subsection{Benchmark Parameterizations of the LHC HXSWG}

The LHC Higgs Cross Section Working Group (HXSWG) has proposed 
a series of benchmark Higgs coupling parameterizations~\cite{LHCHiggsCrossSectionWorkingGroup:2012nn,Dittmaier:2011ti}.
We take as an example the parameterization with seven free parameters $\kappa_g,\kappa_{\gamma},\kappa_W,\kappa_Z,\kappa_b,\kappa_t,\kappa_{\tau}$
and a dependent parameter $\kappa_H(\kappa_i)$ described in Section~10.3.7 of Ref.~\cite{Dittmaier:2011ti}.
In this parameterization 2nd generation fermion Higgs couplings are related to 3rd generation couplings via $\kappa_c=\kappa_t$,
 $\kappa_\mu=\kappa_\tau$, etc., and the total Higgs width is assumed to be the sum of the partial widths for all Standard model decays. 
We implement these boundary conditions by adding two new  terms to our model independent chisquare function:
\beq
     \chi^2  = \sum^{i=33}_{i=1}({Y_i-Y^{'}_i \over \Delta Y_i})^2+({\xi_{ct}\over\Delta\xi_{ct}})^2+({\xi_{\Gamma}\over\Delta\xi_{\Gamma}})^2
\eeq
where

\beq
     \xi_{ct}=\kappa_c-\kappa_t={g_c \over g_c^{SM}}-{g_t \over g_t^{SM}}\,, \ \ \xi_\Gamma=\Gamma_T-\sum_{i=1}^{9}\Gamma_i\,,\ \ {\mathrm and} \ \ \Gamma_i=G_i\cdot g_i^2 \, .
\eeq
The error $\Delta\xi_{ct}$ is obtained by propagating the total theory errors on $g_c^{SM}$ and $g_t^{SM}$, while the error $\Delta\xi_\Gamma$ is obtained
by propagating the errors on $G_i$:
\beq
     \Delta\xi_\Gamma=\Gamma_{SM}\left[\sum_i ({\Delta G_i \over G_i})^2(BR_i)^2\right]^{\frac{1}{2}}  \approx \Gamma_{SM}{\Delta G \over G}\left[\sum_i (BR_i)^2\right]^{\frac{1}{2}}  \approx 0.63\ \Gamma_{SM}{\Delta G \over G} \, .
\eeq
   The results for the seven parameters in the HXSWG parameterization 
are shown in \Tref{tab:modeldep7parhxswgtheory0p1} and \Tref{tab:modeldep7parhxswgtheory0p5}
assuming all theory errors are given by $\Delta\xi_{ct}=\Delta G_i/G_i=\Delta F_i/F_i=0.1\%$ and $0.5\%$, respectively.

\begin{table}
 \begin{center}
 \begin{tabular}{lcccc}
Mode            &  ILC(250)        & ILC(500)           & ILC(1000)      & ILC(LumUp) \cr \hline
$\gamma\gamma$  &      17   \%      &    8.3  \%      &   3.8  \%       & 2.3  \%  \cr
$gg$            &   6.1  \%         &   2.0  \%       &   1.1  \%       & 0.7 \%  \cr
$WW$            &   4.7 \%         &  0.4   \%        &   0.3  \%       & 0.2 \%  \cr
$ZZ$            &     0.7 \%      &  0.5  \%          &  0.5  \%       & 0.3 \%  \cr
$t\bar t$       &    6.4 \%         &  2.5  \%       &   1.3    \%     & 0.9 \%  \cr
$b\bar b$       &    4.7 \%         &  1.0  \%       &   0.6    \%     & 0.4 \%  \cr
$\tau^+\tau^-$   &     5.2 \%        &    1.9  \%      &   1.3  \%       & 0.7  \%  \cr
$\Gamma_T(h)$    &    9.0   \%        &    1.7 \%       &   1.1   \%   & 0.8  \%  \cr  
   \hline
   \end{tabular}
  \caption{Expected accuracies $\Delta g_i/g_i$ for Higgs boson couplings and the total width $\Gamma_T(h)$ using the seven parameter
 HXSWG benchmark parameterization described in Section~10.3.7 of Ref.~\cite{Dittmaier:2011ti}
assuming all theory errors are given 0.1\%. 
}
\label{tab:modeldep7parhxswgtheory0p1}
  \end{center}
\end{table}
%%%%%%%%%%%%%%%%%%%%%%%%%%%%%%%%%%%%%%%%%%%%%%%%%%%%%%%%%%%%%%%%%

\begin{table}
 \begin{center}
 \begin{tabular}{lcccc}
Mode            &  ILC(250)        & ILC(500)           & ILC(1000)      & ILC(LumUp) \cr \hline
$\gamma\gamma$  &      17   \%      &    8.3  \%      &   3.8  \%       & 2.3  \%  \cr
$gg$            &   6.1  \%         &   2.0  \%       &   1.2  \%       & 0.7 \%  \cr
$WW$            &   4.7 \%         &  0.5   \%        &   0.3  \%       & 0.2 \%  \cr
$ZZ$            &     0.8 \%      &  0.5  \%          &  0.5  \%       & 0.3 \%  \cr
$t\bar t$       &    6.4 \%         &  2.6  \%       &   1.4    \%     & 0.9 \%  \cr
$b\bar b$       &    4.7 \%         &  1.0  \%       &   0.6   \%     & 0.4 \%  \cr
$\tau^+\tau^-$   &     5.2 \%        &    2.0  \%      &   1.3  \%       & 0.8  \%  \cr
$\Gamma_T(h)$    &    9.0   \%        &    1.8 \%       &   1.1   \%   & 0.9  \%  \cr  
   \hline
   \end{tabular}
  \caption{Expected accuracies $\Delta g_i/g_i$ for Higgs boson couplings  and the total width $\Gamma_T(h)$  using the seven parameter
 HXSWG benchmark parameterization described in Section~10.3.7 of Ref.~\cite{Dittmaier:2011ti}
and assuming  all theory errors are 0.5\%.
}
\label{tab:modeldep7parhxswgtheory0p5}
  \end{center}
\end{table}
%%%%%%%%%%%%%%%%%%%%%%%%%%%%%%%%%%%%%%%%%%%%%%%%%%%%%%%%%%%%%%%%%
\clearpage

\subsection{Higgs Couplings to $W$ and $Z$ Bounded by SM Couplings}

A different method to fit for Higgs couplings using LHC data is given in Ref.~\cite{Peskin:2012we},
where an effort is made to minimize the model dependence of the coupling fit.
Under rather general 
conditions~\cite{Gunion:1990kf}, each scalar with a vev makes a
 positive contribution to the masses of the $W$ and 
$Z$.  Since the  Higgs couplings to the $W$ and $Z$ also arise from the 
vev, this implies that the coupling of any single
Higgs field is bounded above by the coupling that would give the full 
mass of the vector bosons.   This implies
\beqpeskin
            g^2(hWW) \leq  g^2(hWW)|_{SM}  \quad \mbox{and }        
     g^2(hZZ) \leq  g^2(hZZ)|_{SM}
\eeqpeskin{upperbound}
Then the measurement of the $\sigma\cdot BR$  for a process such as 
$WW$ fusion to $h$ with decay to $WW^*$, which 
is proportional to   $g^4(hWW)/\Gamma_T$, puts an upper limit on 
$\Gamma_T$.    This constraint was first 
applied to Higgs coupling fitting by 
D\"uhrssen \textit{et al.}~\cite{Duhrssen:2004cv}.     In the 
literature, this constraint
is sometimes applied together with the relation
\beqpeskin
           g^2(hWW)/g^2(hZZ) = \cos^2\theta_w    \ .
\eeqpeskin{gsqrat}
The relation \leqn{gsqrat}, however, requires models in which 
the Higgs is a mixture of $SU(2)$ singlet and doublet
fields only, while \leqn{upperbound} is more general~\cite{Low:2010jp}.
An estimate of Higgs coupling errors from the LHC under the assumption of  Eqn.~\leqn{upperbound}
can be found in Ref.~\cite{Peskin:2012we}.

We have carried out a global 
fit to the ILC measurements under the constraint \leqn{upperbound}
  with  9 parameters representing 
independent Higgs boson couplings to $WW$, $ZZ$, $b\bar b$, 
$gg$, $\gamma\gamma$, $\tau^+\tau^-$, $c \bar c$, $t\bar t$, and the total Higgs width $\Gamma_T(h)$.  
The results for
the
errors on Higgs couplings are shown in Table~\ref{tab:globalfit}.  The 
four columns represent the combination of results from LHC (300 fb$^{-1}$, 1 detector)~\cite{Peskin:2012we} and  
our four ILC luminosity scenarios.

\begin{table}
 \begin{center}
 \begin{tabular}{lcccc}
      &  LHC(300 fb$^{-1}$) & LHC(300 fb$^{-1}$)    & LHC(300 fb$^{-1}$)   &  LHC(300 fb$^{-1}$)     \cr
Mode  &    +ILC(250)  & +ILC(500)  & +ILC(1000) & +ILC(LumUp) \cr \hline
$\gamma\gamma$  &   4.8  \%         &  4.2  \%       &    3.0  \%       & 2.0 \%  \cr
$gg$            &   3.8  \%         &  1.9  \%       &    1.1  \%       & 0.7  \%  \cr
$WW$            &   1.9  \%          &  0.2  \%      &    0.1 \%       &  0.1 \%  \cr
$ZZ$            &   0.4 \%         &  0.3  \%      &    0.3  \%      &  0.1 \%  \cr
$t\bar t$       &   12.0  \%        &  9.6   \%      &     2.9  \%       & 1.8 \%  \cr
$b\bar b$       &   2.8 \%          &  1.0  \%      &    0.6 \%       &  0.3 \%  \cr
$\tau^+\tau^-$   &   3.3 \%          &  1.8  \%       &    1.2  \%       & 0.7 \%  \cr
$c\bar c$       &   5.1   \%        &  2.6  \%       &     1.4  \%       &  0.8 \%  \cr
$\Gamma_T(h)$     &    4.7  \%      &  1.6 \%         &     0.9   \%      &  0.5 \%  \cr
   \hline
   \end{tabular}
  \caption{Expected accuracies for Higgs boson couplings under the 
assumption of Eqn.~\leqn{upperbound} and assuming LHC results with 300~fb$^{-1}$ are combined
with ILC results.
}
\label{tab:globalfit}
  \end{center}
\end{table}

%%%%%%%%%%%%%%%%%%%%%%%%%%%%%%%%%%%%%%%%%%%%%%%%%%%%%%%%%%%%%%%%%

%The result of this fit are shown graphically in Fig.~\ref{fig:PeskinHiggsILC}.
%
%
%
%\begin{figure}[t]
%\begin{center}
%     \includegraphics[width=6in]{Chapter_Couplings/figs/Higgsprofile-ILC-5.pdf}
%\caption{
%Estimate of the sensitivity of the ILC experiments to Higgs boson couplings 
%under the assumption of  Eqn.~\leqn{upperbound}  The plot
%  shows the  1~$\sigma$ confidence intervals
% as they emerge from the fit.  Deviation of the central
%  values from zero indicates a bias, which can be corrected for.  The 
%upper limit on the $WW$ and $ZZ$ couplings arises from the constraints
%\leqn{upperbound}.
% The bar for the invisible
%   channel gives the 1 $\sigma$ upper limit on the \textit{branching
%     ratio}.  The four sets of errors for each Higgs
%coupling represent the results for ILC energy and luminosity scenarios 1,2,3, and 4.   The 
%methodology leading to this figure is explained in Ref.~\cite{Peskin:2012we}.
%}
%\label{fig:PeskinHiggsILC}
%\end{center}
%\end{figure}

\section{Effective Higgs Operators}
\begin{figure}[t]
\begin{center}
\includegraphics[width=0.48\columnwidth]{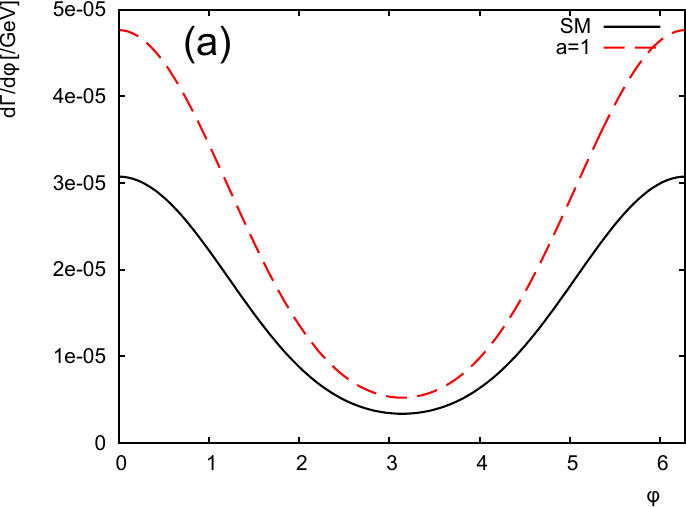}
\includegraphics[width=0.48\columnwidth]{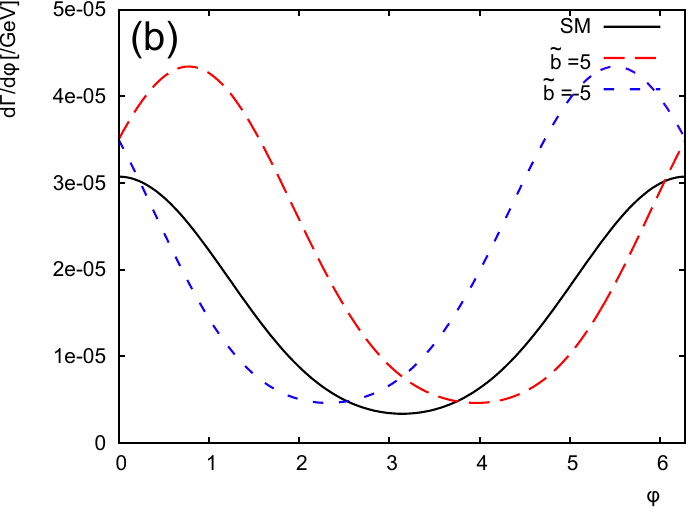}
\caption{
Distribution of the angle $\phi$ between two decay planes of $W$ and $W^*$ from the decay $H\to WW^\ast\to 4j$ with the inclusion of anomalous couplings \cite{Takubo:2010tc}.
(a) The SM curve along with that for $a=1$, $b=\tilde{b}=0$, $\Lambda=1$~TeV; the position of the minimum is the
same for both distributions.  (b) The SM result with the cases $\tilde{b}=\pm5$, $a=b=0$, $\Lambda=1$~TeV;
the position of the minimum is now shifted as discussed in the text.
From \cite{Takubo:2010tc}.
}
\label{fig:phi}
\end{center}
\end{figure}

 The $h \to WW^*$ decay provides an interesting
 opportunity to study its differential
 width and probe the Lorentz structure of the $hWW$ coupling through
 angular
 analyses of the decay products. The relevant part of the general
 interaction
 Lagrangian, which couples the Higgs boson to $W$ bosons in a both
 Lorentz- and gauge-symmetric fashion, can be parameterized as
\begin{equation}
\label{eq:Lhww}
\mathcal{L}_{\mathrm hWW}
= 2 m_W^2\left(\frac{1}{v}+\frac{a}{\Lambda}\right) h\ W_\mu^+ W^{-\mu}
+\frac{b}{\Lambda} h\ W^+_{\mu\nu} W^{-\mu\nu}
+\frac{\tilde b}{\Lambda} h\ \epsilon^{\mu\nu\sigma\tau} W^+_{\mu\nu} W^-_{\sigma\tau}\ ,
\end{equation}
where $W_{\mu\nu}^\pm$ is the usual gauge field strength tensor, 
$\epsilon^{\mu\nu\sigma\tau}$ is the Levi-Civita tensor, $v$ is the 
VEV of the Higgs field, and $\Lambda$ is a cutoff scale\footnote{
The Lagrangian (\ref{eq:Lhww}) is not by itself gauge invariant; to
restore explicit
 gauge invariance we must also include the corresponding anomalous
 couplings
 of the Higgs boson to $Z$ bosons and photons.
}. 
The real dimensionless coefficients, $a$, $b$, and $\tilde{b}$, are
all zero in the 
Standard Model and measure the anomaly in the $hWW$ coupling, which
arise
 from some new physics at the scale $\Lambda$. The coefficient $a$
 stands
 for the correction to the Standard Model coupling. 
 The 
coefficients $b$ and $\tilde{b}$ parametrize the
leading dimension-five non-renormalizable interactions 
and corresponding to 
$(\mathbold{E} \cdot \mathbold{E}
 - \mathbold{B} \cdot \mathbold{B})$-type $CP$-even and
 $(\mathbold{E} \cdot \mathbold{B})$-type $CP$-odd contributions.
%$(\mathbf{E} \cdot \mathbf{E}
% - \mathbf{B} \cdot \mathbf{B})$-type $CP$-even and
% $(\mathbf{E} \cdot \mathbf{B})$-type $CP$-odd contributions.
The $a$ coefficient, if nonzero, would modify just
 the normalization of the Standard Model coupling, while
 the $b$ and $\tilde{b}$ coefficients would change the angular
 correlations of the decay planes.  This effect is shown 
in Fig.~\ref{fig:phi}~\cite{Takubo:2010tc}. 
Nonzero $b$ and $\tilde{b}$ would also modify the momentum 
distribution of the $W$ boson in the Higgs rest frame. 
Simultaneous fits to $p_{W}$ and $\phi_{\mathrm{plane}}$ result in the
 contour plots in Figs.\ref{fig:cont_a-b} and \ref{fig:cont_a-bt}.

\thisfloatsetup{floatwidth=\SfigwFull,capposition=beside}
\begin{figure}[t]
 \includegraphics[width=\hsize]{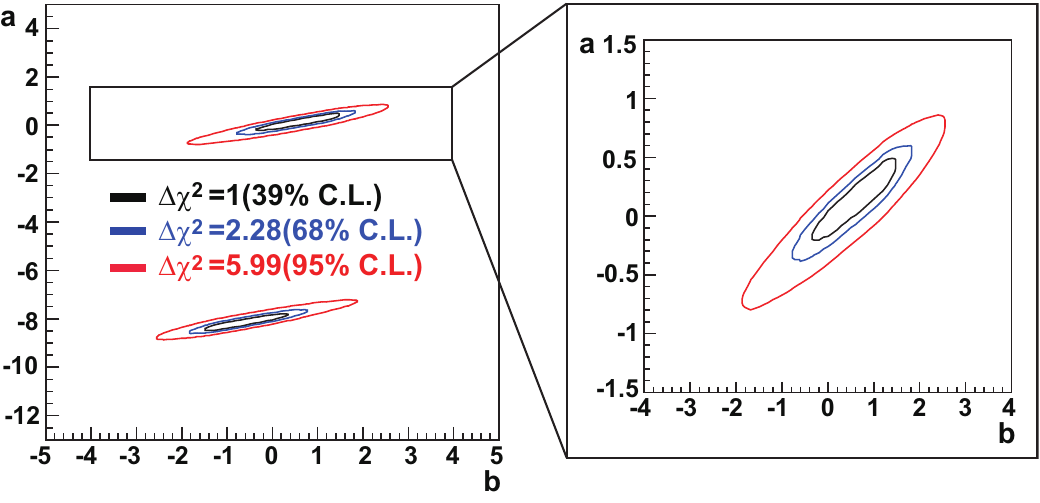}
\caption{Probability contours for $\Delta \chi^{2} =$ 1, 2.28, 
and 5.99 in the $a$-$b$ plane, which correspond to 39\%, 68\%, and 95\% C.L., respectively.}
\label{fig:cont_a-b}
\end{figure}

\thisfloatsetup{floatwidth=\SfigwFull,capposition=beside}
\begin{figure}[t]
 \includegraphics[width=\hsize]{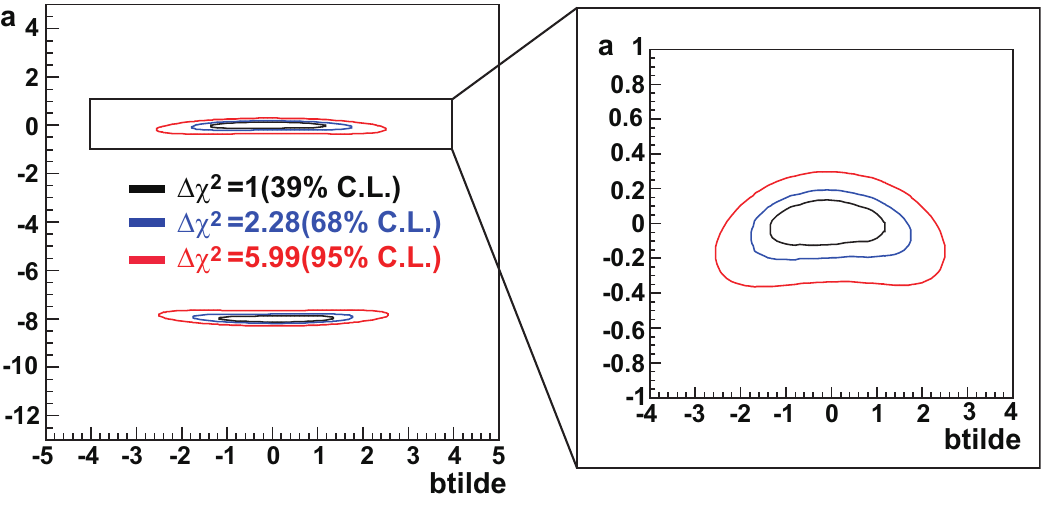}
 \caption{Contours similar to Fig.$\,$\ref{fig:cont_a-b} plotted in the $a$-$\tilde{b}$ plane.}
\label{fig:cont_a-bt}
\end{figure}
%%%%%%%%%%%%%%%%%%%%%%%%%%%%%%%%%%%%%%%%%%%%%%%%%%%%%%%%%%%%%%%%%%%

\chapter{Non-Minimal Higgs Models \label{sid:chapter_multiple_higgs}}
\section{Direct Production of Non-Minimal Higgs Bosons}

\def\hpm{H^\pm}
\def\mhh{m_{H}}
\def\mhl{m_{h}}
\def\mha{m_{A}}
\def\mhpm{m_{H^\pm}}
\def\hsm{h_{\mathrm SM}}
\def\mhsm{m_{\hsm}}
\def\go{G^0}
\def\hh{H}
\def\hl{h}
\def\ha{A}
\def\hp{H^+}
\def\hm{H^-}
\def\nn{\nonumber}
\def\abar{{\bar a}}
\def\bbar{{\bar b}}
\def\cbar{{\bar c}}
\def\dbar{{\bar d}}
\def\ebar{{\bar e}}
\def\fbar{{\bar f}}
\def\gbar{{\bar g}}
\def\hb{{\bar h}}
\def\qlo{Q^0_L}
\def\uro{U^0_R}
\def\dro{D^0_R}
\def\ql{Q_L}
\def\ur{U_R}
\def\dr{D_R}
\def\mud{M_U}
\def\mdd{M_D}
\def\eiuo{\eta_1^{U,0}}
\def\eiiuo{\eta_2^{U,0}}
\def\eido{\eta_1^{D,0}}
\def\eiido{\eta_2^{D,0}}
\def\eiu{\eta_1^U}
\def\eiiu{\eta_2^U}
\def\eid{\eta_1^D}
\def\eiid{\eta_2^D}
\def\eiuoi{\eta_i^{U,0}}
\def\eidoi{\eta_i^{D,0}}
\def\eiui{\eta_i^U}
\def\eidi{\eta_i^D}\def\lam{\lambda}
\def\eiuoa{\eta_a^{U,0}}
\def\eidoa{\eta_a^{D,0}}
\def\eiuoab{\eta_{\abar}^{U,0}}
\def\eidoab{\eta_{\abar}^{D,0}}
\def\eiua{\eta_a^U}
\def\eida{\eta_a^D}
\def\eiuab{\eta_{\abar}^U}
\def\eidab{\eta_{\abar}^D}
\def\kpu{\kappa^U}
\def\rhu{\rho^U}
\def\kpd{\kappa^D}
\def\rhd{\rho^D}
\def\rhds{(\rho^D)\lsup{*}}
\def\rhdt{(\rho^D)\lsup{T}}

%%%%%%%%%%%%%%%%%%%%%%%%

The discovery of additional Higgs bosons such as $H$, $A$, $H^\pm$ and
$H^{\pm\pm}$  would give direct evidence for extended Higgs sector.
As discussed in Section~\ref{specialforms} there are many possibilities for the 
decay branching ratios of these particles.
The ongoing searches at LHC 
rely on specific production and decay 
mechanisms that occupy only a part of the complete model parameter 
space.  At the ILC, the extended Higgs bosons are produced
in electroweak pair production through cross sections that depend only 
on the $SU(2)\times U(1)$ quantum numbers and the mixing angles. 
Thus, the reach of the ILC is typically limited to masses less than 
$\sqrt{s}/2$, but it is otherwise almost uniform over the parameter space.

 \subsection{Neutral Higgs pair production at ILC}

The signals from $HA$ production in the 
$bbbb$ and $bb\tau\tau$ channels, in the context of
the MSSM (Type-II 2HDM), was carried out in the studies 
of Ref.~\cite{Aguilar-Saavedra:2001rg,Desch:2004yb}.
A rather detailed detector simulation was performed in \cite{Desch:2004yb}, 
including  all the SM backgrounds 
at  $\sqrt{s}=500$, 800 and 1000 GeV.
Using a kinematical fit which imposes energy  momentum conservation and
under the assumed experimental conditions, a statistical accuracy on 
the Higgs boson mass
from 0.1 to 1 GeV is found to be achievable.
The topological cross section of $e^+e^- \to HA \to bbbb$
($e^+e^- \to HA \to \tau\tau bb$) could be determined with a relative
precision of 1.5\% to 7\% (4\% to 30\%).
The width of $H$ and $A$ could also be determined with an 
accuracy of 20\% to 40\%, depending
on the mass of the Higgs bosons.
Figure~\ref{FIG:4tau_dist} shows, on the left,
the  $\tau^+\tau^-$ invariant mass obtained by a kinematic
fit in $e^+e^- \to HA \to b\bar b \tau^+\tau^-$
 for $m_A=140$ GeV and $m_H =150$ GeV, for
  $\sqrt{s}=500$ GeV and 500 fb$^{-1}$~\cite{Desch:2004yb}.

The $\tau^+\tau^-\tau^+\tau^-$ and
$\mu^+\mu^-\tau^+\tau^-$ final states would be dominant for the 
type X (lepton specific) 2HDM.
When $\sqrt{s}=500$ GeV, assuming an integrated luminosity of
$500$ fb$^{-1}$, one expects to collect 16,000 (18,000) 
$\tau^+\tau^-\tau^+\tau^-$ events
in the type X (type II) 2HDM, and 110 (60) 
$\mu^+\mu^-\tau^+\tau^-$ events in the same models,
assuming  $m_H^{}=m_A^{}=m_{H^\pm}^{}=130$ GeV, $\sin(\beta-\alpha)=1$
and $\tan\beta=10$. These numbers do not change much for $\tan\beta\gtrsim 3$.
%$\tan\beta\gtrsim 3$.
It is important to recognize that the four-momenta
 of the $\tau$ leptons can be solved by a kinematic fit based on the
known center of mass energy and momentum, by applying
the collinear approximation to each set of $\tau$ lepton decay 
products~\cite{Schael:2006cr,Abdallah:2002qj}. 
Figure~\ref{FIG:4tau_dist} shows, on the right, 
the two dimensional invariant mass
distribution of the $\tau$ lepton pairs from the neutral 
Higgs boson decays as obtained with  
a simulation at 500~GeV in which the masses of the neutral Higgs bosons
are taken to be 130 GeV and 170 GeV~\cite{Tsumura}.

%%%%%%%%%%%%%%%%%%%%%%%%%%%%%%%%%%%%%%%%%%%%%%%%%%%%%%%%%
\begin{figure}[t]
\begin{center}
\includegraphics[width=0.42\hsize]{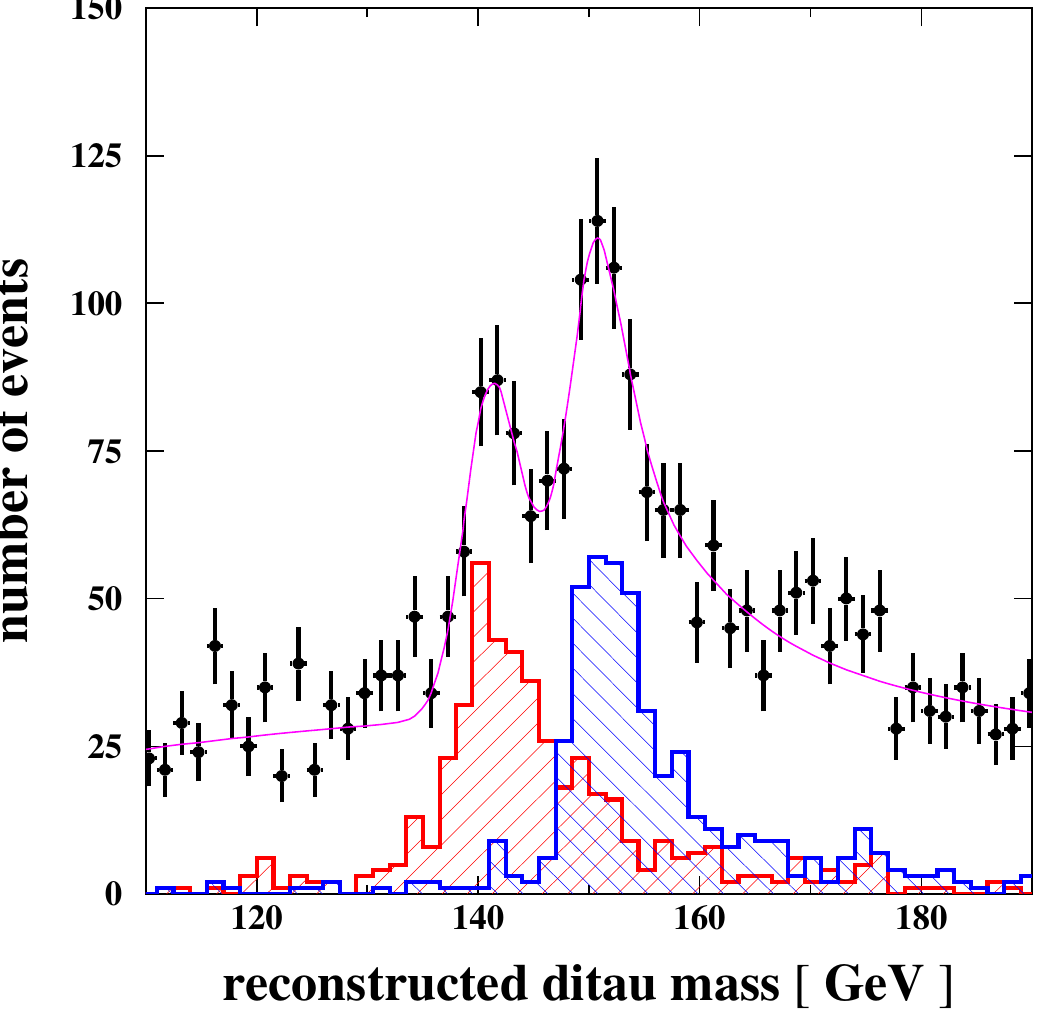}\
\includegraphics[width=0.46\hsize]{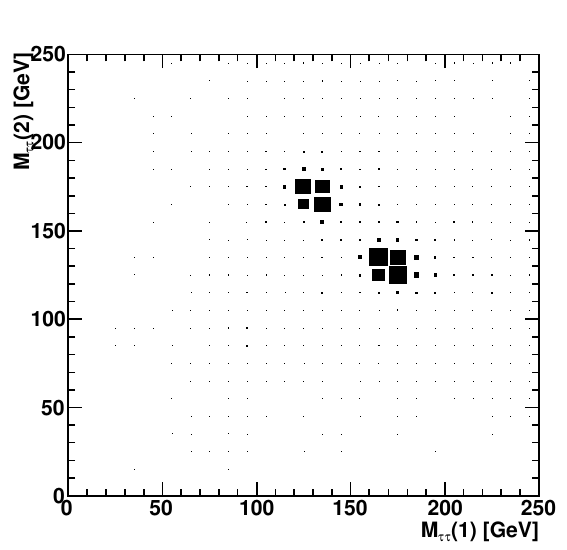}
\caption{Left:
Invariant mass reconstruction from the kinematical fit in the process
 $e^+e^- \to HA \to b\bar{ b} \tau^+\tau^-$ in the Type-II (MSSM like) 2HDM 
 for $m_A=140$ GeV and $m_H
 =150$ GeV at $\sqrt{s}=500$ GeV and 500 fb$^{-1}$~\cite{Desch:2004yb}
 Right:
Two dimensional distribution of ditau invariant mass
in $e^+e^-\to HA \to \tau^+\tau^- \tau^+\tau^-$ in the Type X (lepton
 specific) 2HDM for $m_A =170$ GeV and $m_H=130$ GeV 
for $\sqrt{s}=500$ GeV and 500 fb$^{-1}$~\cite{Tsumura}.
}
\label{FIG:4tau_dist}
\end{center}
\end{figure}
%%%%%%%%%%%%%%%%%%%%%%%%%%%%%%%%%%%%%%%%%%%%%%%%%%%%%%%

In an extended Higgs sector with singlets, it is very common to have lighter Higgs bosons
with suppressed couplings to the Z, but which can be seen at an $e^+e^-$ collider either through direct
production or by decays of the 125~GeV Higgs boson.   A specific  NMSSM example 
that has been studied is the cascade decay $h_1\rightarrow aa \rightarrow (\tau^+\tau^-)(\tau^+\tau^-)$
at the ILC\cite{Liu:2013gea}.  In addition to discovery, the masses can be measured to better than 1\%.

Although the associated Higgs production process $e^+e^- \to HA$ is
a promising one for testing the properties of the extended Higgs sectors, the
kinematic reach is restricted by $m_H + m_A < \sqrt{s}$ and is not
available beyond this limit.
Above the threshold of the $HA$ production, the
associated production processes
$t \bar t \Phi$, $b \bar b \Phi$ and $\tau^+\tau^- \Phi$
($\Phi=h, H, A$)  could be used~\cite{Djouadi:1992gp,Djouadi:1991tk}.
In particular, for $b \bar b \Phi$ and $\tau^+\tau^- \Phi$,
the mass reach is extended almost up to the collision energy.
The cross sections for these processes 
are proportional to the Yukawa interaction,
so they directly depend on the type of Yukawa coupling
in the 2HDM structure. In MSSM or the Type II 2HDM  (Type I 2HDM),
these processes are enhanced (suppressed) for large $\tan\beta$ values.
In Type X 2HDM, only the $\tau^+\tau^- H/A$ channels
could be significant while only $b \bar b H/A$ channels would be
important in Type I and Type Y 2HDMs. These reactions can then 
be used to discriminate
the type of the Yukawa interaction.

  \subsection{Charged Higgs boson  production}

At the ILC, charged Higgs bosons
 are produced in pairs in $e^+e^- \to H^+H^-$~\cite{Komamiya:1988rs}. The cross
section is a function only of $m_{H^\pm}$ and is independent of the
type of Yukawa interaction in the 2HDM. Therefore, as in the
case of the $HA$ production, the study of the final state channels
can be used to determine the type of Yukawa interaction.
When $m_{H^\pm} > m_t + m_b$, the main decay mode is $tb$ in Type I,
II and Y, while in Type X the main decay mode is $\tau\nu$ for
$\tan\beta >2$.
When $H^\pm$ cannot decay into $tb$, the main decay mode is
$\tau\nu$ except in Type Y for large $\tan\beta$ values.
For $m_{H^\pm} < m_t -m_b$, the charged Higgs boson can also be studied
via the decay of top quarks $t \to b H^\pm$ in 2HDMs except
in Type X 2HDM case with $\tan\beta >2$.

In the MSSM, a detailed simulation study  of this reaction
has been performed for the final state
$e^+e^- \to H^+H^- \to t \bar b \bar t b$ for $m_{H^\pm}=300$ GeV
at $\sqrt{s}=800$ GeV~\cite{Battaglia:2001be}. 
The final states is 4 $b$-jets with 4 non-$b$-tagged jets. Assuming
an integrated luminosity of 1 ab$^{-1}$, a mass resolution of
approximately 1.5\% can be achieved  (Figure~\ref{fig:tbtb} (left)).
The decay mode $tbtb$ can also
be used to determine $\tan\beta$, especially for relatively small
values, $\tan\beta <5$), where the production rate of the signal
strongly depends on this parameter.

%%%%%%%%%%%%%%%%%%%%%%%%%%%%%%%%%%%%%%%%%%%%%%%%%%%%%%%%
\begin{figure}[t]
\begin{center}
\includegraphics[width=0.45\hsize]{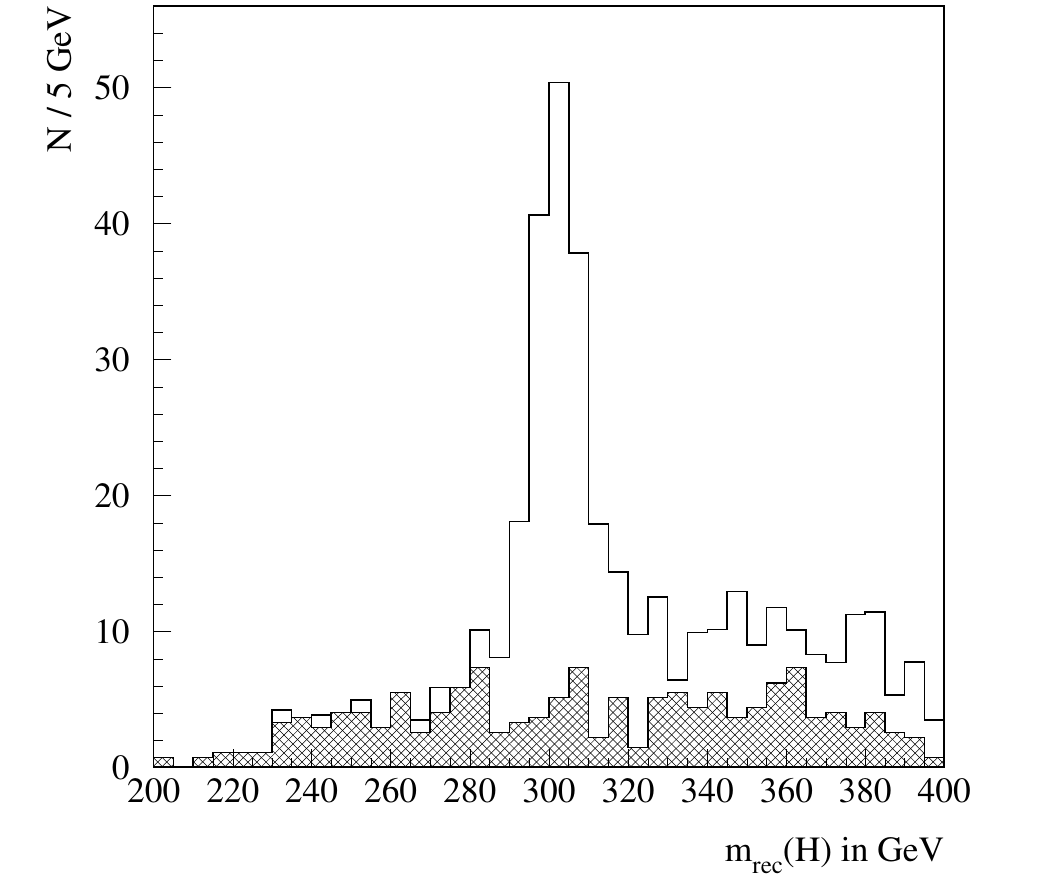}
\
 \includegraphics[width=0.53\hsize]{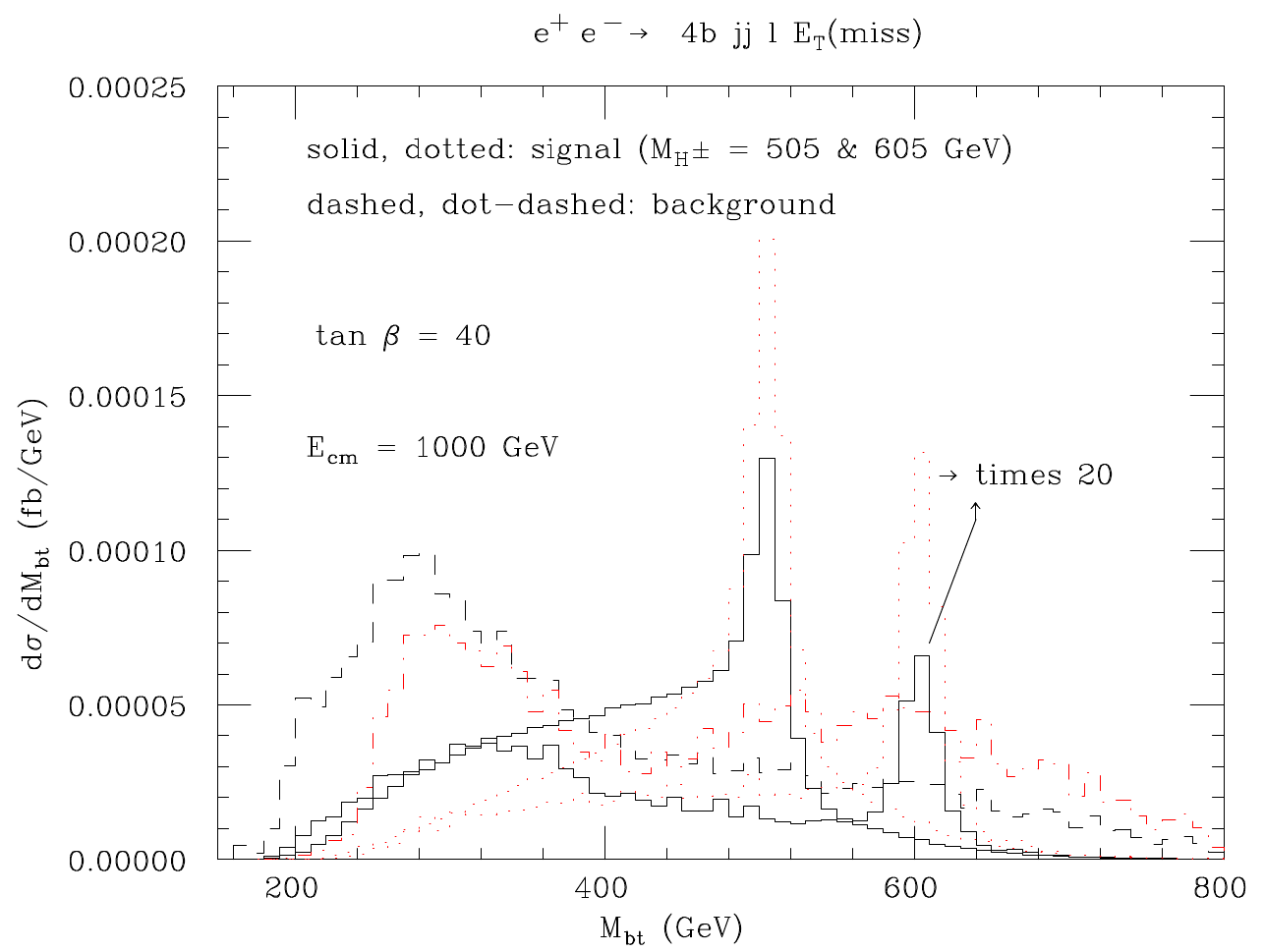}
\caption{
Left: Fitted charged Higgs boson mass in $H^+H^- \to (t \bar{ b})(\bar{ t }b)$
in the MSSM, with $m_{H^\pm}$ = 300 GeV, measured at the ILC
at CM energy 800 GeV with 1 ab$^{-1}$  of data.
The background is shown by the  dark histogram~\cite{Battaglia:2001be}.
Right: Differential distribution of  the reconstructed Higgs mass for the
 signal $e^+e^-\to b \bar{ t} H^+ + t \bar{b} H^-\to t \bar{ t} b \bar{ b}$
 and the background $e^+e^- \to t \bar{ t }g^\ast\to t \bar{ t} b \bar{ b}$
 in the MSSM  or the Type II 2HDM~\cite{Moretti:2003cd}.
}
\label{fig:tbtb}
\end{center}
\end{figure}
%%%%%%%%%%%%%%%%%%%%%%%%%%%%%%%%%%%%%%%%%%%%%%%%%

The pair production is kinematically limited to relatively light charged
Higgs bosons with $m_{H^\pm} < \sqrt{s}/2$.
When $m_{H^\pm} > \sqrt{s}/2$, one can make use of 
the single production processes 
 $e^+e^- \to t \bar b H^+$,
 $e^+e^- \to  \tau \bar \nu H^+$,
 $e^+e^- \to  W^- H^+$,
 $e^+e^- \to H^+ e^- \nu$ and
 their charge conjugates.
 The cross sections for the first two of these processes are
 directly proportional to the square of the Yukawa coupling constants. The
 others are one-loop induced.
 Apart from the pair production rate, these single production processes
 strongly depend on the type of Yukawa interaction in the 2HDM
 structure. 
 In general, their rates are small and quickly suppressed for larger
 values of $m_{H^\pm}$.
 They can be used only for limited parameter regions where $m_H^\pm$
 is just above the threshold for the
 pair production with very large or low $\tan\beta$
 values.

%%%%%%%%%%%%%%%%%%%%%%%%%%%%%%%%%%%%%%%%%%%%%%%%%%%%%%%%%%%%
\thisfloatsetup{floatwidth=\SfigwFull,capposition=beside}
\begin{figure}[t]
\includegraphics[width=0.7\hsize]{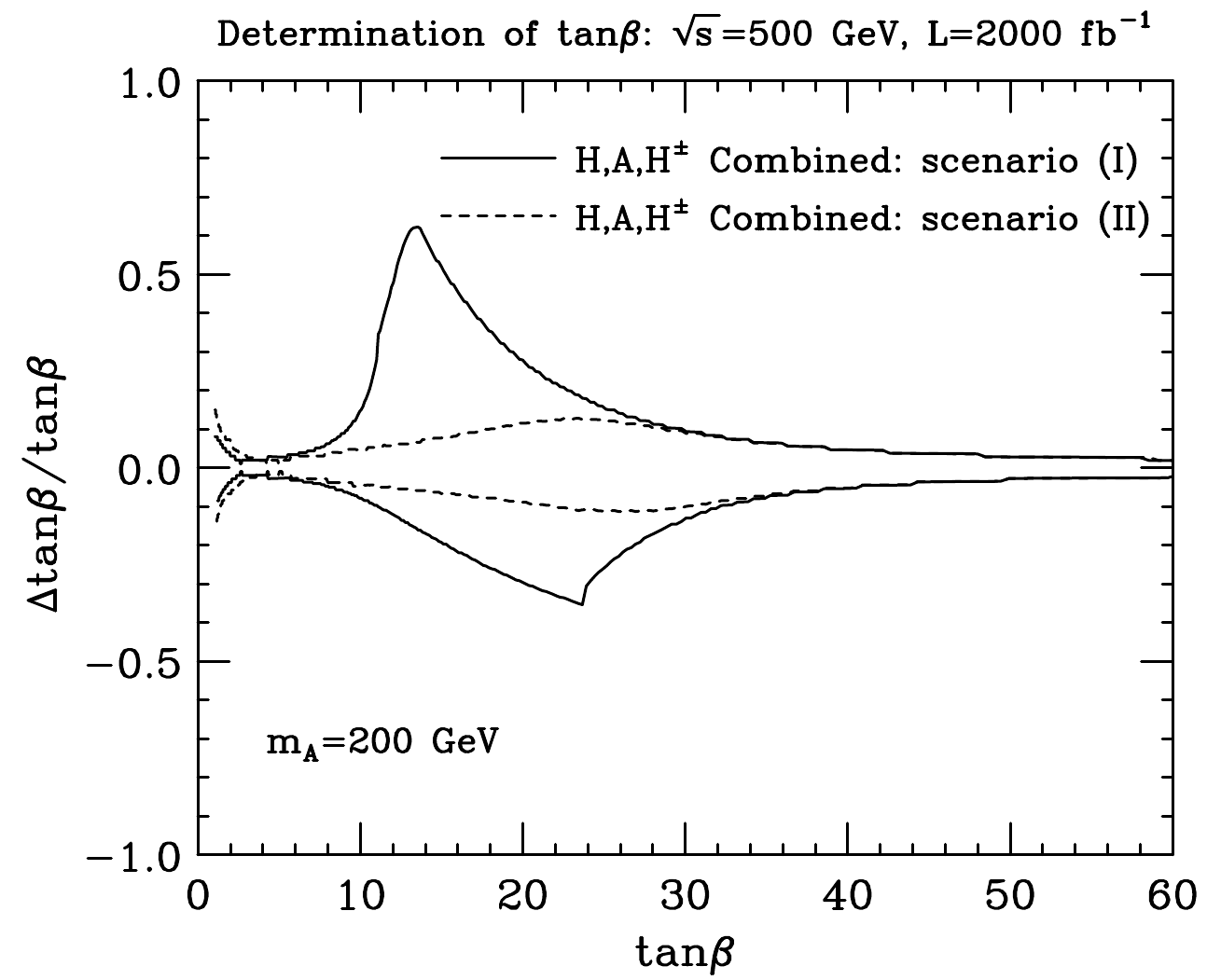} 
\caption{
Estimates of the 1 $\sigma$ statistical upper and lower bounds on $\tan\beta$
from ILC measurements, for an MSSM model with $m_{H^\pm} \sim m_A = 200$~GeV, 
assuming  $\sqrt{s}=500$ GeV and  2000 fb$^{-1}$ of data, from 
 \cite{Gunion:2002ip}.  The quantity plotted is the relative error,
$\Delta\tan\beta/\tan\beta$.} 
 \label{totalonly}
 \end{figure}
%%%%%%%%%%%%%%%%%%%%%%%%%%%%%%%%%%%%%%%%%%%%%%%%%%%%%%%%%%%%

 In Ref.~\cite{Moretti:2003cd}, a simulation study for the process
 $e^+e^- \to t \bar b H^- + b \bar t H^+ \to 4b + jj+\ell + p_T^{\mathrm
 miss}$ ($\ell=e$, $\mu$) has been done for $m_{H^\pm}$ just above
 the pair production threshold $m_{H^\pm} \simeq\sqrt{s}/2$. 
 It is  shown that this process provides a significant 
 signal of $H^\pm$ in a relatively small region just above
 $\sqrt{s}/2$, for very large or very small values of $\tan\beta$,
 assuming a high $b$-tagging efficiency.  The reconstructed $H^+$ mass
distribution is shown in the right-hand side of Fig.~\ref{fig:tbtb}.

\section{Measurements of $\tan\beta$ at the ILC}

 In multi-Higgs models, mixing angles between bosons with the same quantum numbers are
 important parameters.   In the CP-conserving two Higgs doublet model, there are two mixing angles
 $\alpha$ and $\beta$, where $\alpha$ is introduced to diagonalize the mass matrix of the CP-even scalar states,
 and $\tan\beta$ is defined as the ratio of vacuum expectation values of two Higgs doublets diagonalizing
 the charged and CP-odd scalar states.  All coupling constants associated with the Higgs bosons, 
 i.e. the couplings of $h$, $H$, $A$ and $H^\pm$ to gauge bosons, fermions and themselves,
  depend on these mixing angles.

 The information on $\sin(\beta-\alpha)$ ($\cos(\beta-\alpha)$) can be directly extracted
 from the precision measurement of the couplings of the SM-like boson $h$ (the extra Higgs boson $H$)
 to weak  gauge bosons, $hVV$ ($HVV$).
  At the LHC, the SM-like coupling $hVV$ ($VV=WW$ and $ZZ$) is being measured, and the current
  data indicates $\sin^2(\beta-\alpha) \simeq 1$ within the error of order 10-20\%.
  At the ILC,  the $hWW$ and $hZZ$ couplings can be measured precisely to  the percent level or better.
  %
  %Information on $\sin^2(\beta-\alpha) $ can be extracted via the precision measurement
  %of the $hVV$ coupling constants.

   When $\sin(\beta-\alpha)$ is precisely determined, all the Yukawa couplings $h f \overline f$ and $H f \overline f$
   are a function of $\tan\beta$, so that one can extract $\tan\beta$  by precise measurements
   of the Yukawa interactions.
   The $\tan\beta$ dependences in the Yukawa couplings are different for  each type of Yukawa interaction~\cite{Barger:1989fj,Aoki:2009ha,Su:2009fz,Logan:2009uf}.
   In the Type-II 2HDM, the $\tan\beta$ dependences are large for Yukawa interactions of $H$ and $A$ with down type
   fermions such as $H b \overline b$  $A b\overline b$, $H \tau^+\tau^-$ and $A \tau^+\tau^-$
   ($Y_{Hbb, Abb} \sim m_b\tan\beta$,  $Y_{H\tau\tau, A\tau\tau} \sim m_\tau\tan\beta$ ),
   while in the Type-X (lepton specific) 2HDM the Yukawa couplings of $H$ or $A$
   to charged leptons are sensitive to $\tan\beta$ ($Y_{H\tau\tau, A\tau\tau} \sim m_\tau \tan\beta$).

   In Fig.~\ref{FIG:Bbb} the branching ratios of $h\to b\overline b$, $H\to b \overline b$ and $A\to b\overline b$
   are shown as a function of $\tan\beta$ for a fixed value of $\sin^2(\beta-\alpha)=1$, $0.99$ and $0.98$
   in the Type-II 2HDM (MSSM)~\cite{Kanemura:2013eja}.  In Fig.~\ref{FIG:Btautau}, 
   similar figures for the branching ratios of $h\to\tau^+\tau^-$, $H\to \tau^+\tau^-$ 
   and $A\to\tau^+\tau^-$ are shown in the Type-X (lepton specific) 2HDM.
  \begin{figure}[t]
%----------------------------------------------------------------------------
 \centering
 \includegraphics[height=4.8cm]{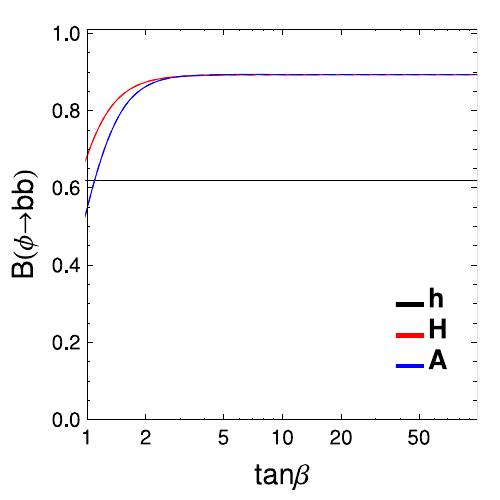}
 \includegraphics[height=4.8cm]{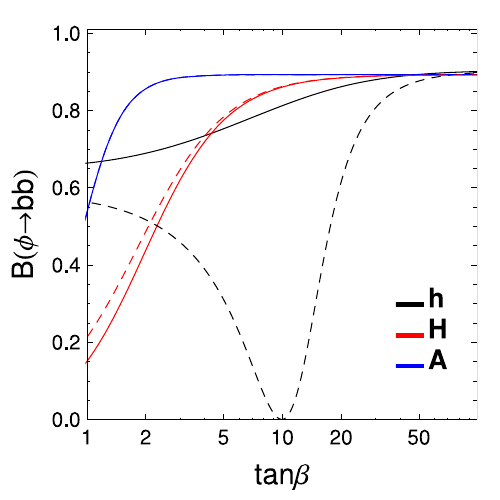}
 \includegraphics[height=4.8cm]{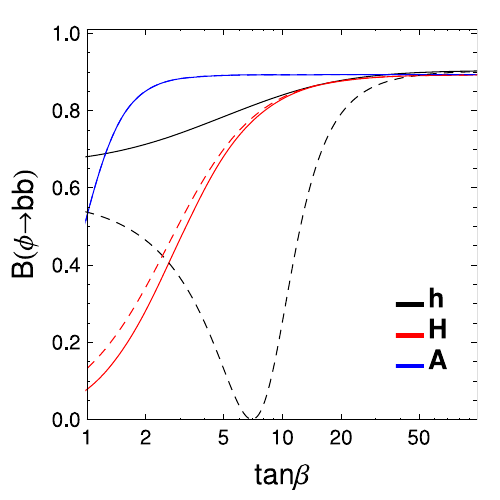}
 \caption{The decay branching ratios as a function of
 $\tan\beta$ for a fixed $\sin^2(\beta-\alpha)$ for $h\to b\bar
 b$ (black curves), $H\to b\bar b$ (red curves), and $A\to b\bar
 b$ (blue curves) in the Type-II 2HDM. From left to right, $\sin^2(\beta-\alpha)$ is
 taken to be $1$, $0.99$, and $0.98$.
 The solid (dashed) curves denote the case with $\cos(\beta-\alpha) \leq 0$
 ($\cos(\beta-\alpha) \geq 0$).
 }
 \label{FIG:Bbb}
%----------------------------------------------------------------------------
\end{figure}
\begin{figure}[tb]
%%----------------------------------------------------------------------------
 \centering
 \includegraphics[height=4.8cm]{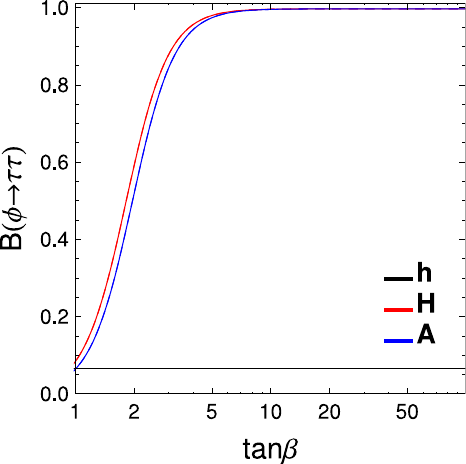}
 \includegraphics[height=4.8cm]{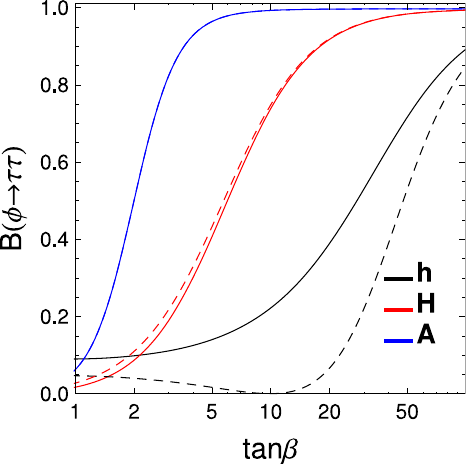}
 \includegraphics[height=4.8cm]{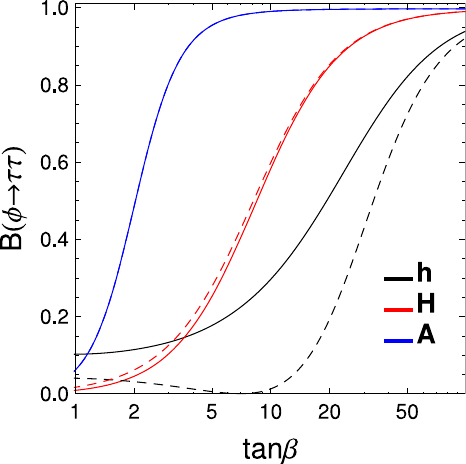}
 \caption{The decay branching ratios are shown as a function of
 $\tan\beta$ with a fixed value of $\sin^2(\beta-\alpha)$ for $h\to
 \tau\tau$ (black curves), for $H\to \tau\tau$ (red curves), and
 for $A\to \tau\tau$ (blue curves) in the Type-X 2HDM.
 From left to right, $\sin^2(\beta-\alpha)$ is taken to be $1$, $0.99$,
 and $0.98$.
 The solid (dashed) curves denote the case with $\cos(\beta-\alpha) \leq
 0$ ($\cos(\beta-\alpha) \geq 0$).
 }
 \label{FIG:Btautau}
%----------------------------------------------------------------------------
\end{figure}

 In Refs.~\cite{Barger:2000fi,Gunion:2002ip} methods using the production and decay of the $H$ and $A$ have been
 studied in the context of the MSSM.  Since the masses of the $H$ and $A$ can
 be measured by the invariant mass distributions in an appropriate decay mode in $e^+e^-\to HA$, the branching ratios can be predicted as a function of $\tan\beta$.
 Thus one can extract $\tan\beta$ by measuring the branching ratios of $H$ and $A$. Since the $\tan\beta$
 dependence of the branching ratio is large in small $\tan\beta$ regions, this method is useful for small $\tan\beta$.
 A second method~\cite{Barger:2000fi} is based on the measurement of the total decay widths of the $H$ and $A$.
 For large $\tan\beta$ values, the total decay widths are dominated by the $b\overline b$ and $\tau \tau$ decay modes in the Type-II and Type-X 2HDMs,
 respectively, whose partial widths are proportional to $(\tan\beta)^2$. Therefore, $\tan\beta$ can be extracted using this method
 in large $\tan\beta$ regions.

 In addition to these two methods, a new method using the precision measurement of the SM-like Higgs boson $h$ has been proposed
 in Ref.~\cite{Kanemura:2013eja}. This can be applied to the case where $\sin^2(\beta-\alpha) $ is smaller than unity through the $\tan\beta$ dependences
 in the Yukawa couplings for $h$.
   In the limit of $\sin^2(\beta-\alpha)=1$, the Yukawa couplings for the SM-like Higgs boson $h$ are identical to the SM ones,
   so that there is no $\tan\beta$ dependence in them.
    However, if $\sin^2(\beta-\alpha)$ turns out to be slightly smaller than unity in future precision measurements,
    then the Yukawa couplings for $h$ can also depend on $\tan\beta$ significantly. For example, for the Type-II 2HDM
    \begin{eqnarray}
       Y_{hb\overline b} &\sim& \sin(\beta-\alpha)  - \tan\beta \cos(\beta-\alpha) , \\
        Y_{h\tau\tau} &\sim& \sin(\beta-\alpha)  - \tan\beta \cos(\beta-\alpha),
    \end{eqnarray}
    and for the Type-X 2HDM
       \begin{eqnarray}
       Y_{hb\overline b} &\sim& \sin(\beta-\alpha)  + \cot\beta \cos(\beta-\alpha) , \\
        Y_{h\tau\tau} &\sim& \sin(\beta-\alpha)  - \tan\beta \cos(\beta-\alpha).
    \end{eqnarray}
 At the ILC, the main decay modes of $h$ can be measured precisely to the few percent level.
 The precision measurement of the decay of $h$ can be used to determine $\tan\beta$ for the case with $\sin(\beta-\alpha)<1$.

   In Fig.~\ref{FIG:2HDM-II}, the numerical results for the sensitivities of the
   $\tan\beta$ measurements are shown for the Type-II 2HDM~\cite{Kanemura:2013eja}.
   The production cross section and the number of the signal events are evaluated for
   $m_H=m_A=200$ GeV with $\sqrt{s} =500$ GeV and ${\cal L}_{\mathrm int}=250$ fb$^{-1}$. The acceptance ratio
   of the $4b$ final states in the $e^+e^-\rightarrow HA$ signal process is set to 50\%.
   The results for the three methods are shown. The results for 1 $\sigma$ (solid) and
   2 $\sigma$ (dashed) sensitivities for the branching ratios, the total width of $H$ and $A$, and
   the branching ratio of $h$ are plotted in the red, blue and black curves, respectively.
   The parameter $\sin^2(\beta-\alpha) $ is set to 1 (left), 0.99 (middle) and 0.98 (right)
   for $\cos(\beta-\alpha)<0$.

   In Fig.~\ref{FIG:2HDM-X}, the sensitivities to $\tan\beta$ are shown for the case of
   the Type-X 2HDM, where the channels $H\to \tau^+\tau^-$ and $A\to\tau^+\tau^-$
   are the main decay modes~\cite{Kanemura:2013eja}.
   With or without the assumption of $\sin^2(\beta-\alpha)=1$, the total width measurement of
   $H$ and $A$ is a useful probe for the large $\tan\beta$ regions.
   For the smaller $\tan\beta$ regions, the branching ratio measurement of $H$ and $A$
   can probe $\tan\beta$. For $\sin(\beta-\alpha)=0.99$ and $0.98$, the measurement of
   the branching ratio of $h\to\tau^+\tau^-$ can give good $\tan\beta$ sensitivity over a wide range of
   $\tan\beta$.

  Here, comments on the $\tan\beta$ measurements for the other 2HDM types are given.
 In the Type-I 2HDM, the Yukawa coupling constants are universally changed from those in the SM.
 In the SM-like limit, $\sin(\beta-\alpha)=1$, the Yukawa interactions for
 $H$ and $A$ become weak for $\tan\beta > 1$.
 As for the $\tan\beta$ measurement at the ILC, the method using the
 total width of $H$ and $A$ is useless, because the absolute value of the decay
 width is too small compared to the detector resolution.
 Without the SM-like limit, the branching ratio measurement of $H$ and $A$
 using the fermionic decay modes may be difficult, because the bosonic decay
 modes $H\to WW$ and $A\to Zh$ become important.
 Furthermore, the decays of $h$ are almost unchanged from the SM because
 there is  no $\tan\beta$ enhancement.
 Thus, the $\tan\beta$ determination in the Type-I 2HDM seems to be
 difficult even at the ILC.
 In the Type-Y 2HDM, the $\tan\beta$ sensitivity at the ILC would be
 similar to that of the Type-II 2HDM, because the Yukawa interactions of
 the neutral scalar bosons with the bottom quarks are enhanced by
 $\tan\beta$ in the same way.

 \begin{figure}[tb]
%----------------------------------------------------------------------------
 \centering
 \includegraphics[height=4.8cm]{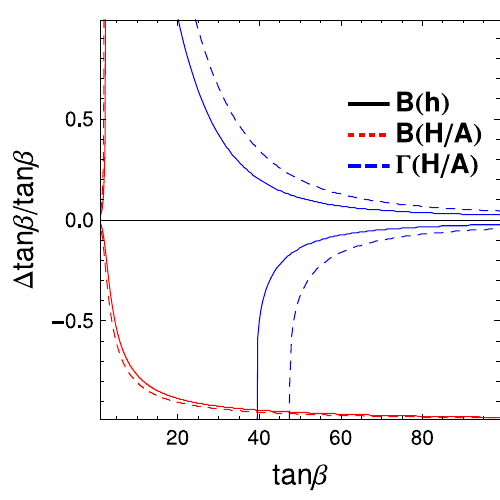}
 \includegraphics[height=4.8cm]{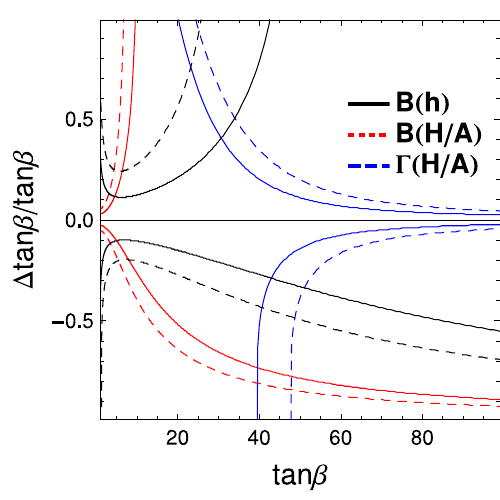}
 \includegraphics[height=4.8cm]{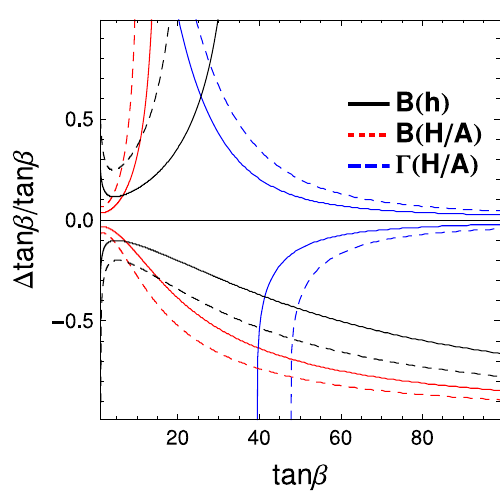}
 \caption{Sensitivities to the $\tan\beta$ measurement by the three
 methods in the Type-II 2HDM.
From left to right, $\sin^2(\beta-\alpha)$ is taken to be $1$, $0.99$,
 and $0.98$, with $\cos(\beta-\alpha) \le 0$.
Estimated $\Delta\tan\beta/\tan\beta$ using the branching ratio of
 $H/A\to b\bar b$ (red curves), the total width of $H/A$ (blue curves),
 and the branching ratio of $h\to b\bar b$ (black curves) are plotted as
 a function of $\tan\beta$.
The solid curves stand for $1\sigma$ sensitivities, and the dashed
 curves for $2\sigma$.
For $HA$ production, $m_H^{}=m_A^{}=200$ GeV with $\sqrt{s}=500$ GeV and
${\cal L}_{\mathrm int}=250$~fb$^{-1}$ are assumed.
For the $h\to b\bar b$ measurement, $\Delta{\cal
B}/{\cal B} = 1.3\%$ ($1\sigma$) and $2.7\%$ ($2\sigma$)
are used.
}~\label{FIG:2HDM-II}
\end{figure}

\begin{figure}[tb]
%----------------------------------------------------------------------------
 \centering
 \includegraphics[height=4.8cm]{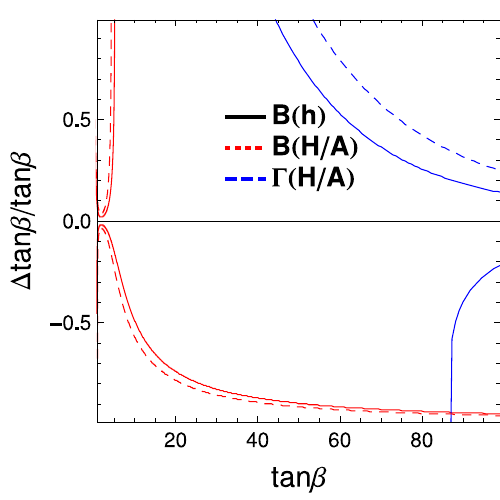}
 \includegraphics[height=4.8cm]{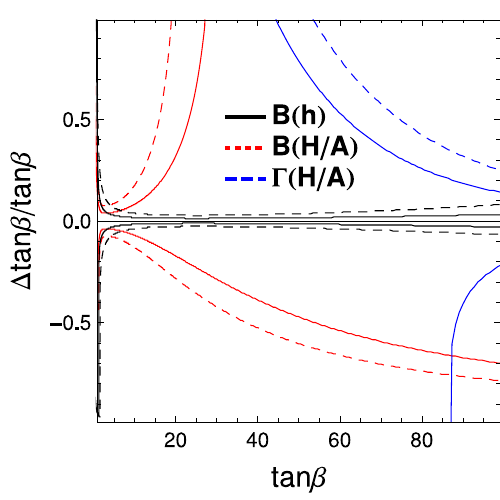}
 \includegraphics[height=4.8cm]{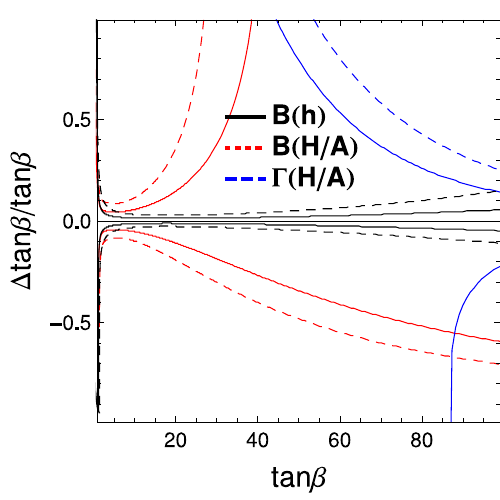}
 \caption{The same as FIG.~\ref{FIG:2HDM-II}, but $\tau\tau$ decay modes
 are used for the analysis in the Type-X 2HDM.
 From left to right, $\sin^2(\beta-\alpha)$ is taken to be $1$, $0.99$,
 and $0.98$, with $\cos(\beta-\alpha) \leq 0$.
 For ${\cal B}^h_{\tau\tau}$, $\Delta{\mathcal B}/{\mathcal B} = 2\%
 $ ($1\sigma$) and $5\%$ ($2\sigma$) are assumed.
 }
 \label{FIG:2HDM-X}
\end{figure}

\chapter{Gamma-Gamma and e-Gamma Option \label{sid:chapter_gamma_gamma}}

Higgs production in $\gamma\gamma$ collisions,
first studied in  \cite{Barklow:1990ah,Gunion:1992ce,Borden:1993cw}, 
 offers a unique 
capability to measure the two-photon width of the Higgs and
to determine its charge conjugation and parity (CP) 
composition through control of the photon
polarization. Both measurements have unique value in understanding
the nature of a Higgs boson eigenstate.  Photon-photon collisions
also offer one of the best means for producing a heavy Higgs
boson singly, implying significantly greater mass reach than
electron-positron production of a pair of Higgs bosons.

There are many important reasons for measuring 
the $\gamma\gamma$ coupling of a Higgs boson, generically denoted
$h$.  In the Standard Model, the coupling
of the Higgs boson, $h_{SM}$, to two photons 
receives contributions from loops containing any charged particle
whose mass arises in whole or part from the vacuum expectation
value (vev) of the neutral Higgs field.
In the limit of infinite mass for the charged particle in the loop, the
contribution asymptotes to a  value  that depends 
on the particle's spin (i.e., the contribution does not decouple).
Thus, a measurement of $\Gamma(h_{SM}\to\gamma\gamma)$ 
provides the possibility of revealing the presence of arbitrarily
heavy charged particles, since in the SM context all particles
acquire mass via the Higgs 
mechanism.\footnote{Loop contributions from charged particles that acquire
a large mass from some other mechanism, beyond
the SM context, will decouple as $({\mathrm mass})^{-2}$
and, if there is a SM-like Higgs boson $h$,
$\Gamma(h\to\gamma\gamma)$ will not be sensitive to their presence.}

Even if there are no new particles that acquire mass via the Higgs
mechanism, a precision measurement of $N(\gamma\gamma\to h\to X)$
for specific final states $X$ ($X=b\overline b,WW^*,\ldots$)
can allow one to distinguish between a $h$ that is part
of a larger Higgs sector and the SM $h_{SM}$. The ability to detect
deviations from SM expectations will be enhanced by combining this 
with other types of precision measurements for the SM-like Higgs boson. 
Observation of small deviations would be typical 
for an extended Higgs sector as one approaches the decoupling limit
in which all other Higgs bosons are fairly heavy, leaving
behind one SM-like light Higgs boson. In such models,
the observed small deviations could then be interpreted as implying the
presence of heavier Higgs bosons. 

The ability to detect $\gamma\gamma\to H^0, A^0$ will be of
greatest importance if the $H^0$ and $A^0$ cannot be detected either
at the LHC or in $e^+e^-$ collisions at the ILC.  In fact, there is
a very significant section of parameter space in the MSSM for which
this is the case. The $\gamma\gamma$ collider would also play a very important role
in exploring a non-supersymmetric general two-Higgs-doublet model
(2HDM) of which the MSSM Higgs sector is a special case.

Once one or several Higgs bosons have been detected, 
precision studies can be
performed. Primary on the list would be
the determination of the CP nature of any observed Higgs boson.
This and other types of measurements
become especially important if one is in the decoupling limit of a 2HDM.
The decoupling limit is defined by the situation in which 
there is a light SM-like Higgs boson, while the other
Higgs bosons ($H^0,A^0,H^\pm$) are heavy and quite degenerate.
In the MSSM context, such decoupling is automatic
in the limit of large $m_{A^0}$.
In this situation, a detailed scan to
separate the $H^0$ and $A^0$ would be very important and
entirely possible at the $\gamma\gamma$ collider. Further,
measurements of relative branching fractions for the $H^0$ and
$A^0$ to various possible final states would also be possible
and reveal much about the Higgs sector model. 
%In the MSSM context, the branching ratios for supersymmetric final states would be measurable; these are especially important for determining the basic supersymmetry breaking parameters \cite{Gunion:1995bh,Gunion:1997cc,Gunion:1996qd,Feng:1997xv,Muhlleitner:2001kw,Muhlleitner:2000jj}.

\section{Production Cross Sections and Luminosity Spectra}
The gamma-gamma option at the ILC opens a new opportunity for truly high energy two-photon physics that is not limited to the QCD studies performed by most $e^+e^-$ colliders. The production cross sections for charged particles are considerably larger in $\gamma\gamma$ collisions than in $e^+e^-$ enabling the study of new particles above threshold at a higher rate - e.g. $WW$ pair production at 500~GeV is a factor of 20 larger than in $e^+e^-$. This effect more than offsets that factor of $5-10$ lower $\gamma\gamma$ luminosity compared to the corresponding $e^+e^-$ collider. Similarly the cross sections for charged scalars, lepton and top pairs are a factor of $5-10$ higher at a photon collider compensating for the luminosity reduction.

The proposed technique for the gamma-gamma option consists of Compton backscattering a $\sim1$~MeV laser photons from the 125-500~GeV electron and position beams. The backscattered photon receives a large fraction of the incoming electron energy. This is described in detail in \cite{Asner:2001ia}. The maximum energy of the generated photons is given by $E^{max}_\gamma = xE_e/(1+x)$, where $E_e$ is the electron beam energy and $x = 4E_eE_L\cos^2(\theta/2)/m^2_ec^4$ with $E_L$ and $\theta$ the laser photon energy and angle between the electron and laser beam. The distance from the conversion point to the interaction point is in the range of a few millimeters to a few centimeters. The optimal values of $x$ are around 4.8, yielding $E^{max}\approx0.82 E_e$, which maximizes the spin-0 luminosity near $E_{\gamma\gamma}=0.8 E_{ee}$, for a particular configuration of beam and laser polarizations as shown in Figure~\ref{fig:gamgamlum_plot}. The fundamental laser wavelength is determined by available technology and are typically 1.054 $\mu m$. For machine energies of $\sqrt{s}$=250, 500, 1000 GeV the corresponding values of $x$ are 2.26, 4.52 and 9.03, respectively. The maximum $E_{\gamma\gamma}$ are 173 GeV, 409 GeV, and 900 GeV with the peak in the spin-0 luminosity somewhat lower. The optimal machine energy (using a 1.054 $\mu m$ laser) to study a 126~GeV Higgs-like particle is about 215~GeV with $x=1.94$. As mentioned above larger values of $x$ are desirable and can be obtained using non-linear optics to triple the laser frequency
\footnote{The efficiency
with which the standard $1.054~\mu$ laser beam is converted to $0.351~\mu$
is 70\%. Thus, roughly 40\% more laser power is required 
in order to retain the subpulse power.} 
In this case, the optimal machine energy to study a 126~GeV Higgs-like particle is $\sim$170 GeV and $x=4.55$ much closer to the optimal value.  

\thisfloatsetup{floatwidth=\SfigwFull,capposition=beside}
\begin{figure}
\includegraphics[width=0.99\hsize]{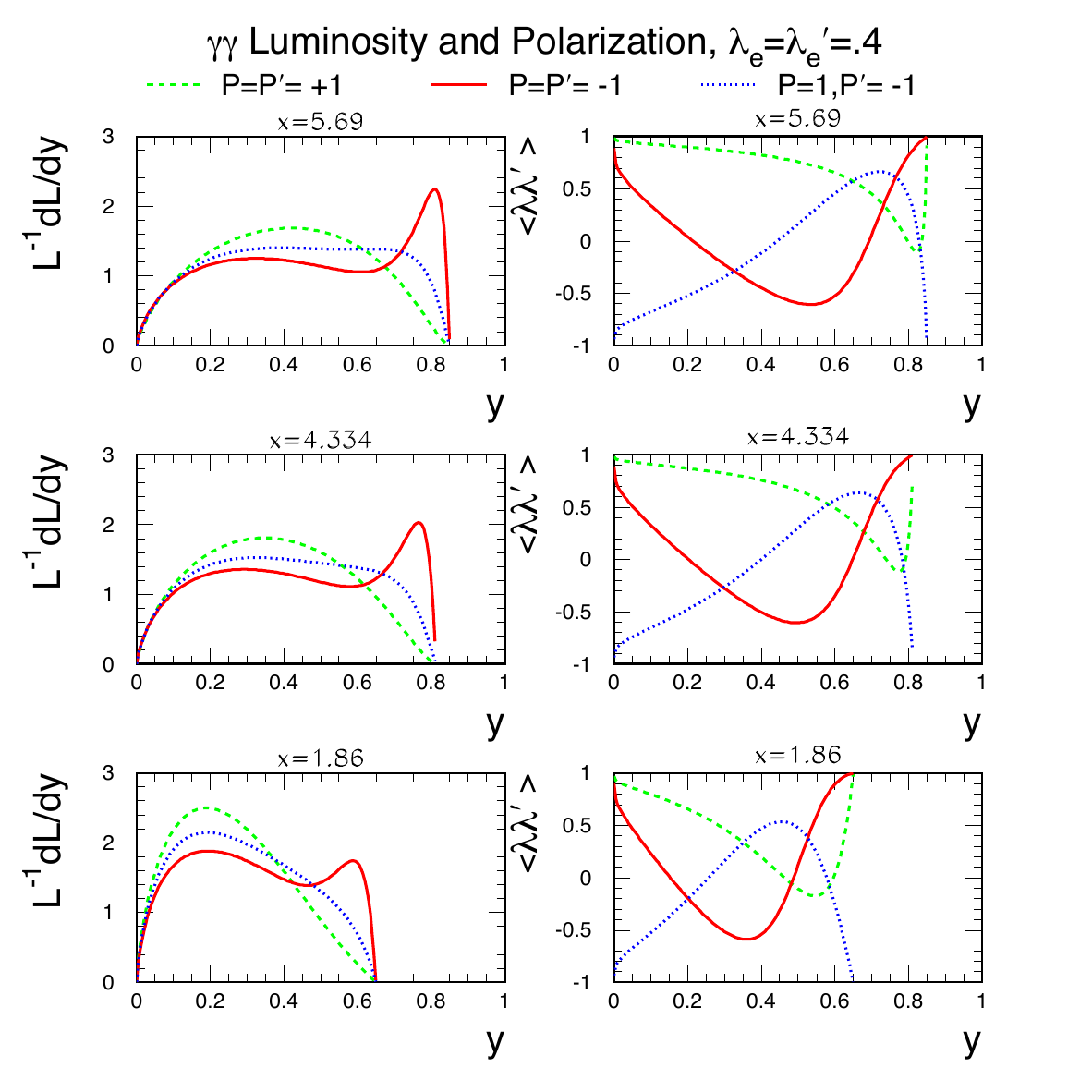}
\vspace*{0.1cm}
\caption[0]{The normalized differential luminosity
  ${1\over {\cal L}_{\gamma\gamma}}{d{\cal L}_{\gamma\gamma}\over dy}$ and the corresponding
  $\protect\vev{\lambda\lambda'}$ for $\lambda_e=\lambda'_e=.4$ (80\%
  polarization)
and three different choices of the initial laser photon polarizations
$P$ and $P'$. The distributions shown are for 
$\rho^2\ll 1$ \cite{Ginzburg:1981vm,Ginzburg:1982yr}. Results for  $x=5.69$,
$x=4.334$ and $x=1.86$ are compared.
}
\label{fig:gamgamlum_plot}
\end{figure}

\section{Higgs Studies}
A Standard Model-like Higgs boson $h$ arises in many models containing physics beyond
the SM. The $h\to\gamma\gamma$ coupling receives contributions from loops
containing any charged particle
whose mass, $M$, arises in whole or part from the vacuum expectation
value of the corresponding neutral Higgs field.
When the mass, $M$, derives in whole or part from the vacuum expectation
value ($v$) of the neutral Higgs
field associated with the $h$, 
then in the limit of $M\gg m_h$ for the particle in the loop, the
contribution asymptotes to a value that depends 
on the particle's spin (i.e., the contribution does not decouple).
As a result, a measurement of $\Gamma(h\to\gamma\gamma)$ 
provides the possibility of revealing the presence of 
heavy charged particles that acquire their mass via the Higgs mechanism.

In addition, we note that  $B(h\to X)$ 
is entirely determined by the spectrum
of particles with mass $<m_{h}/2$, and is not affected by heavy states
with $M>m_h/2$. Consequently,
measuring $N(\gamma\gamma\to h\to X)$ provides
an excellent probe of new heavy particles with mass deriving
from the Higgs mechanism.  

\thisfloatsetup{floatwidth=\SfigwFull,capposition=beside}
\begin{figure}
\includegraphics[width=0.99\hsize]{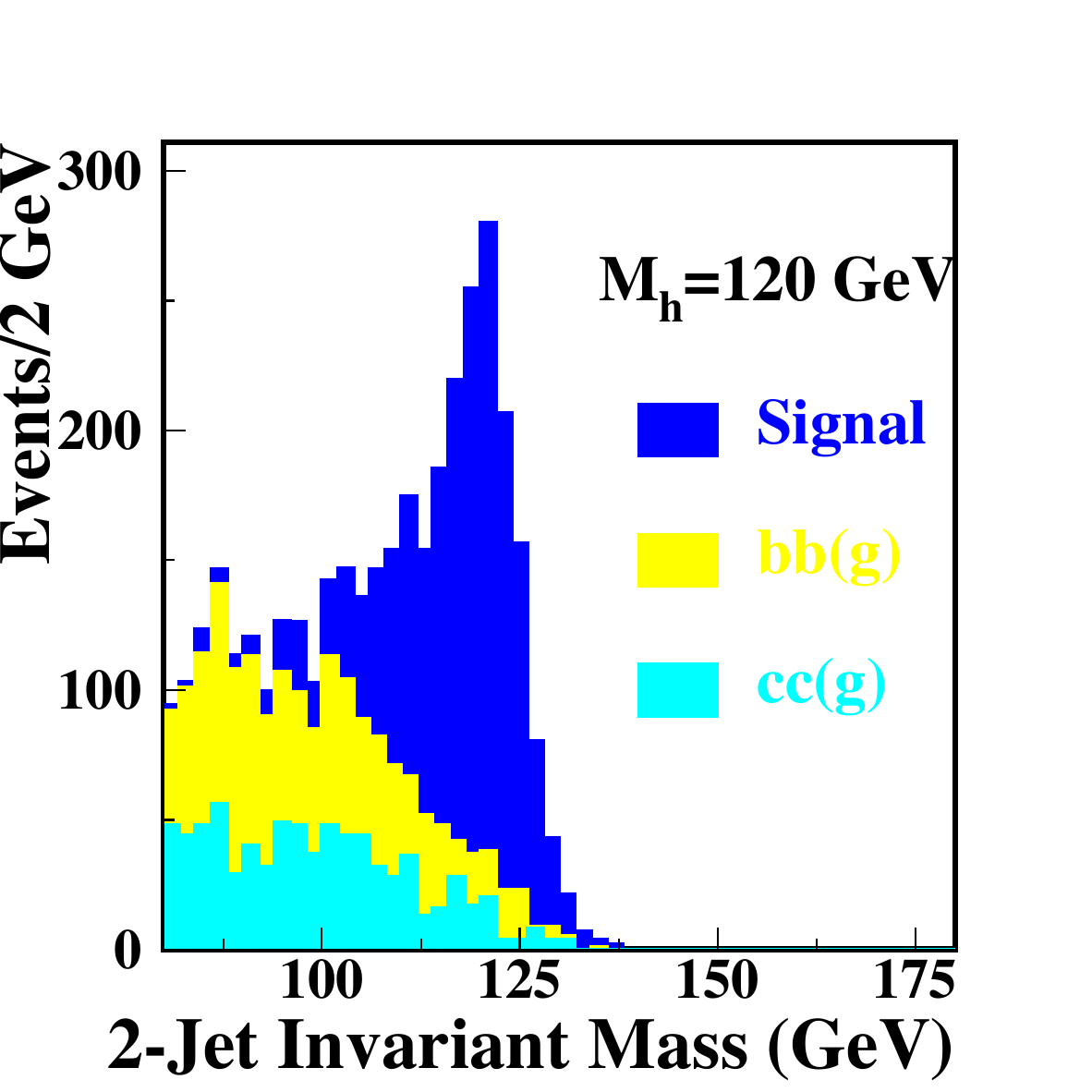}
\vspace*{0.1cm}
\caption{%Mass distributions for the 
Higgs signal and heavy quark backgrounds in units of events per 2 GeV
for a Higgs mass of 115~GeV and assuming a running year of $10^7$ sec
\cite{Asner:2001vh}.
}
\label{fig:higgs}
\end{figure}

%--------------------------------------
\subsection{$h_{SM}$ Mass Measurement}
A special feature of the $\gamma\gamma$ collider is the sharp edge of
the $\gamma\gamma$ luminosity function, as depicted in Fig.~\ref{fig:gamgamlum_plot}.
The position of this edge can be controlled by changing the electron beam
energy.  As it sweeps across the threshold for Higgs production, the 
number of, e.g., $b\overline b$ events will increase dramatically.  
%This phenomenon is already reflected in the sharpness of the excitation curves of Fig.~\ref{fig:excitation}.  
Since the position of this turn-on
depends on the Higgs mass, a threshold scan offers the possibility to
measure the Higgs mass kinematically, as developed in Ref.~\cite{Ohgaki:1999ez}.
%
%--------------------------------------

This possibility was studied in the context of
CLICHE~\cite{Asner:2001vh}, assuming that the Higgs mass is already
known to within a GeV or so. 
%Considered as a function of the $e^-e^-$ center-of-mass energy, as shown in
%Fig.~\ref{fig:excitation}, 
There is a point of optimum sensitivity to the
Higgs mass a few~GeV below the peak of the cross section. The raw number
of events at a single energy cannot be used to measure the
mass, however, because the $\gamma\gamma$ partial width cannot be assumed
known \textit{a priori}. There is another point, though, close to the maximum
of the cross section, at which there is no sensitivity to the Higgs mass,
and with maximum sensitivity to $\Gamma_{\gamma\gamma}$, allowing the
separation of these two quantities. These points are illustrated in
Fig.~\ref{fig:scan}.  Furthermore, the background can be estimated using
data obtained by running below the threshold.  To estimate the 
sensitivity of the yields to $m_H$, we work with a simple 
observable based on the ratio of background-subtracted yields
at peak and at threshold:
\begin{displaymath}
 Y = \frac{N_{\mathrm{peak}} - N_{\mathrm{below}}\cdot r_p}
          {N_{\mathrm{threshold}} - N_{\mathrm{below}}\cdot r_t}
\end{displaymath}
where $N$ is the number of events in a mass window logged at the peak,
on the threshold, and below threshold, and $r_p$ and $r_t$ are scale
factors to relate the background data taken below threshold to
the expectation at peak and at threshold.  We have propagated
statistical uncertainties, and, assuming one year of data on peak,
half a year on threshold and another half below threshold, we find
$\sigma_Y / Y = 0.088$.  This translates into an error
on the inferred Higgs mass of~100~MeV.  A more refined treatment should
improve this estimate somewhat. This estimate is 
obtained using the laser and beam energies proposed for CLIC~1 and the
analysis results are similar to those shown in in Fig.~\ref{fig:higgs}. It is still necessary to investigate how  
sensitive the luminosity function is to the shape of the luminosity curve.
It is not sensitive to the electron polarization precision.

\thisfloatsetup{floatwidth=\SfigwFull,capposition=beside}
\begin{figure}
\includegraphics[width=0.3\hsize]{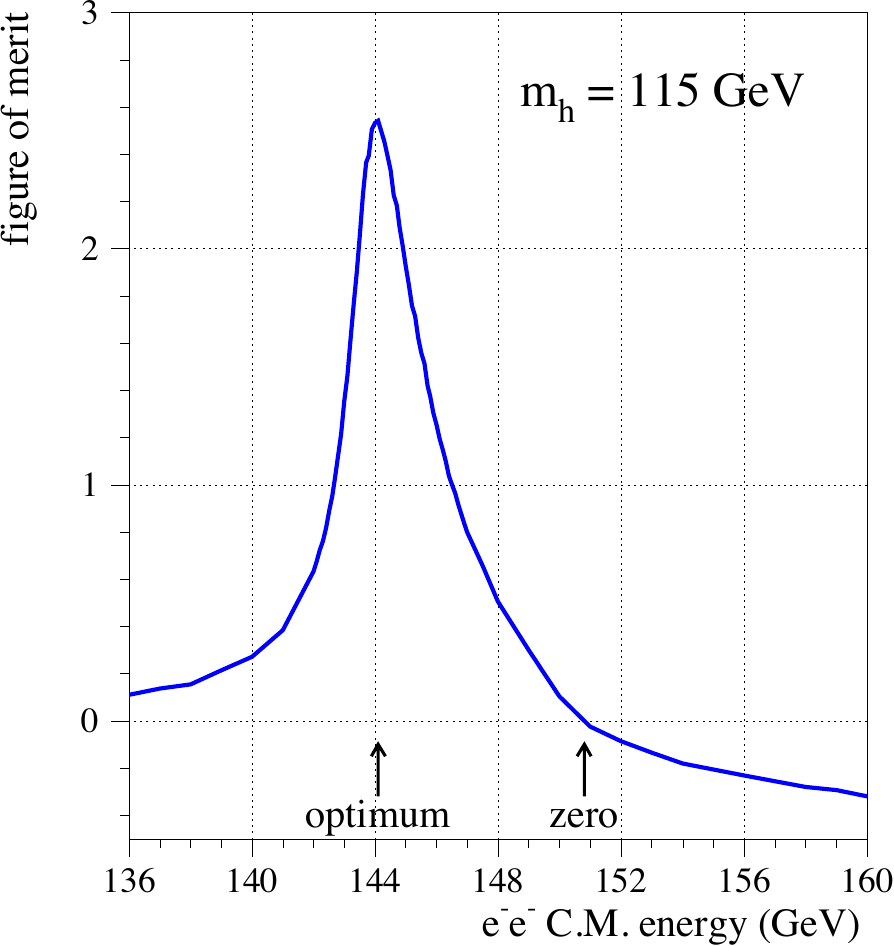}
\includegraphics[width=0.3\hsize]{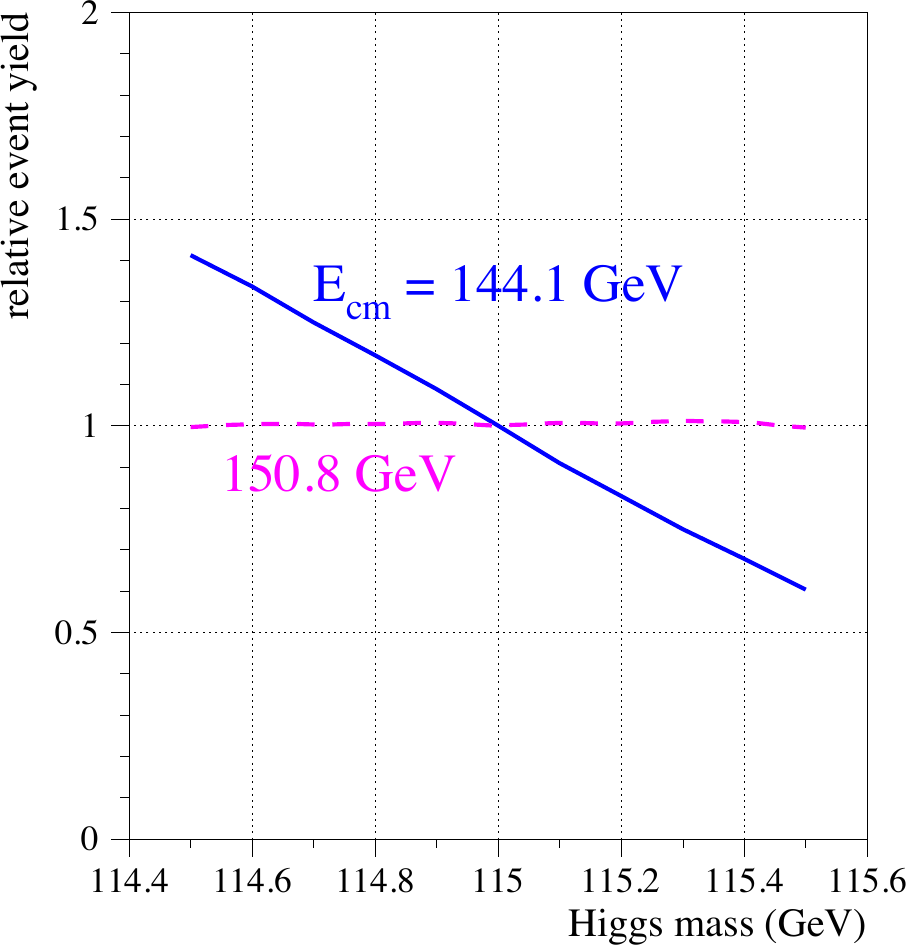}
\includegraphics[width=0.3\hsize]{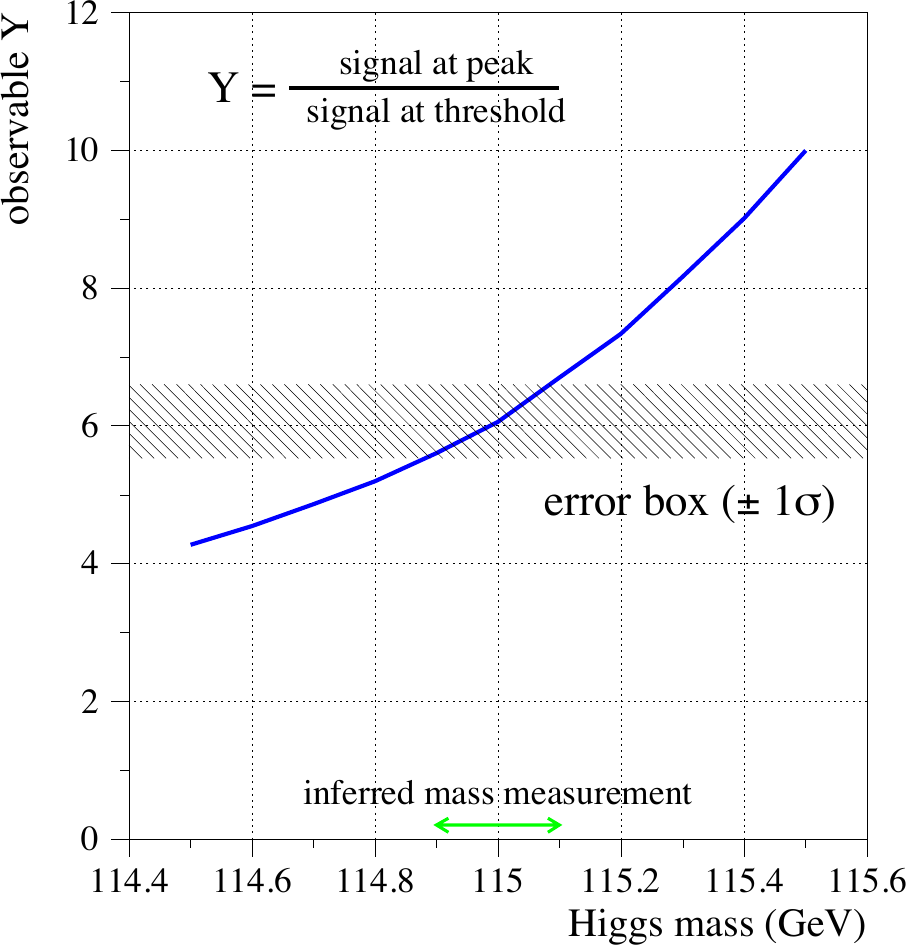}
\caption[0]{
(a) A figure of merit quantifying the measurement error on the mass as a function of the $e^-e^-$ center-of-mass energy. The optimum and zero 
sensitivity points are marked.
(b) Relative yield for a 115~GeV Higgs boson at the point of optimum
sensitivity and zero sensitivity to $m_H$.  
(c) Behavior of the observable~$Y$ as a function of $m_H$, 
and the projected error.
}
 \label{fig:scan}
\end{figure}

\subsection{Branching Fractions}

The precision to which the most important decay modes of a Standard Model Higgs boson can be measured at a gamma-gamma collider are presented in Table~\ref{tab:gghiggsbf}. One of the objectives of a gamma-gamma collider would be to test the Standard Model predictions for Higgs branching fractions and to use measurements of them to distinguish between the
Standard Model and its possible extensions, such as the minimal
supersymmetric extension of the Standard Model (MSSM) or a more general
two-Higgs-doublet model (2HDM).
\thisfloatsetup{floatwidth=\SfigwFull,capposition=beside}
%\thisfloatsetup{capposition=beside}
\begin{table}[htbp]
%\ttabbox
\begin{tabular}{cc}\toprule
 Measurement    &  Precision  \\ \midrule
 $\Gamma_{\gamma\gamma}\times B(h\to b \overline b)$ & 0.012 \\
 $\Gamma_{\gamma\gamma}\times B(h\to WW)$ & 0.035 \\
  $\Gamma_{\gamma\gamma}\times B(h\to \gamma\gamma)$ & 0.121 \\
   $\Gamma_{\gamma\gamma}\times B(h\to ZZ)$ & 0.064 \\
    $\Gamma_{\gamma\gamma}\times B(h\to \gamma Z)$ & 0.020 \\ \midrule
      $\Gamma_{\gamma\gamma}^*$  & 0.021 \\
       $\Gamma_{Total}^*$ & 0.13 \\ \midrule
        $m_{h_{SM}}$ $(h \to \gamma\gamma)$ & 61 MeV \\
         $CP$ Asymmetry $(h \to WW)$ & 0.035-0.040 \\ \bottomrule
*Taking $BR(h \to b \overline b)$ from $e^+e^-$ running at ILC &  \cr
\end{tabular}

\caption{Summary of Higgs Branching Fraction and other measurements in 3 years of design luminosity at a Higgs Factory. This study assumed a 120~GeV Standard Model-like Higgs Boson and accelerator parameters as described in \cite{Asner:2001vh}.}
\label{tab:gghiggsbf}
\end{table}

\subsubsection{$h \to b\overline b$}
If there are no new particles that acquire mass via the Higgs
mechanism, a precision measurement of $\Gamma({\widehat h}\to\gamma\gamma)$
can allow one to distinguish between a ${\widehat h}$ that is part
of a larger Higgs sector and the SM $h_{SM}$. Figure~\ref{fig:higgs}
shows the dijet invariant
mass distributions for the 
%$m_{h_{SM}}$~GeV 
Higgs signal and
for the $b\overline b(g)$ and $c\overline c(g)$ backgrounds,
%using the luminosity distribution
%of Fig.~\ref{fig:higgsspec}, 
after
all cuts.  

Due to the large branching ratio for $H\rightarrow\bar{b}b$ decay
for a Higgs mass $\sim 115$~GeV, this is the main channel for Higgs 
studies at CLICHE. This channel has received
the most attention and the studies are already quite
detailed~\cite{Ohgaki:1997jp,Asner:2001vh}. 
Our analysis includes perturbative QCD backgrounds,
including $\gamma \gamma \rightarrow {\bar b} b(g)$ and
$\gamma \gamma \rightarrow {\bar c} c(g)$.
The ${\bar q} q$ backgrounds are suppressed by choosing like polarizations
for the colliding photons, but this suppression is not so strong when the
final states contain additional gluons.

The mass resolution is around 6~GeV with a jet energy resolution of
$\sigma_E=0.6 \times \sqrt{E}$. The distribution in the dijet invariant 
mass,
$m_{jets}$, for a $m_H=115$~GeV Higgs found in this study 
with an integrated luminosity of 200~fb$^{-1}$ is shown in
Fig.~\ref{fig:higgs}. A clear signal peak can be seen above sharply falling
backgrounds. Including the three bins nearest to $m_{jets}\sim 115$~GeV,
we obtain 4952 signal events and 1100 background events. Thus, the
signal-to-background ratio is expected to be 4.5 after all cuts.
A feature which is not taken into account in these studies is the pile-up
of events from different bunch crossings.  Initial studies indicate that pile-up of order 10 bunch crossings degrades the Higgs signal only slightly.

This would yield a measurement of
$\Gamma(h_{SM}\to\gamma\gamma)B(h_{SM}\to b\overline b)$ with an accuracy of  
$\sqrt{S+B}/S\sim 1.2\%$ in 3 years of design luminosity at a Higgs Factory. This study assumed a 120~GeV Standard Model-like Higgs Boson and accelerator parameters as described in \cite{Asner:2001vh}.

\subsubsection{$h \to WW$}

Observation of this decay mode is extremely difficult at high-energy
$\gamma\gamma$ colliders, because of the large cross section for $W$~pair
production.  If the $\gamma\gamma$ center-of-mass energy is below the
$W^+W^-$ threshold, however, the continuum production of $W$ pairs is
greatly reduced, allowing the observation of resonant production through a
Higgs boson.  The sharp peak in the $\gamma\gamma$ luminosity function seen in
Fig.~\ref{fig:gamgamlum_plot} plays a key role here.
Figure~\ref{fig:wwcross}(a) compares the cross sections for the 
continuum $W$~pair production with the Higgs resonance curve. As shown,
the cross sections for $\sigma(\gamma\gamma\to W^+W^-)$ and 
${\cal B}r(h\to W^+W^-) \times \sigma(\gamma\gamma\to h)$ are comparable,
if  $E_{CM}(e^-e^-)=150$~GeV for a  $m_H=115$~GeV. 
One significant difference between the two type of events is
the energy distribution of the $W^+W^-$ pairs, as illustrated 
in Figure~\ref{fig:wwcross}(b).

Our study is concentrated on the hadronic decays of the $W$ pairs,
applying several kinematic cuts. One pair of jets must reconstruct to the
$W$~mass, while the other pair is required to saturate the remaining phase
space. This cuts allows us not only to reduce the  $W^+W^-$ pairs 
to those with energy similar to those  produced in Higgs events, but 
also to reject
 any possible $\gamma\gamma \to qq(g)$ background.  There must be at least
four jets in the event and  the jet reconstruction efficiency is assumed to 
be 100\%.  In contrast to the $h\to b\bar{b}$ analysis, here we are imposing 
a $y=0.003$ cut  in the Durham algorithm used in the jet reconstruction.
In addition, the transverse momentum is required  to be smaller than 0.1. 
After these cuts we have a 29\% reconstruction efficiency.
A comparison of the signal and the  background after cuts is given in
Fig.~\ref{fig:wwcross}(c), which corresponds to a signal-to-background ratio
of 1.3, and the statistical precision in the
signal rate measurement is expected to be 5\%.

The other event topologies (two leptons and missing energy, or one lepton,
missing energy and jets) remain to be studied.  Techniques similar to those
described in \cite{Dittmar:1996ss} may be used. We also believe that the
decay $H \rightarrow ZZ, Z\gamma$ might be interesting, despite their
relatively small branching ratios.

\thisfloatsetup{floatwidth=\SfigwFull,capposition=beside}
\begin{figure}
\includegraphics[width=0.49\hsize]{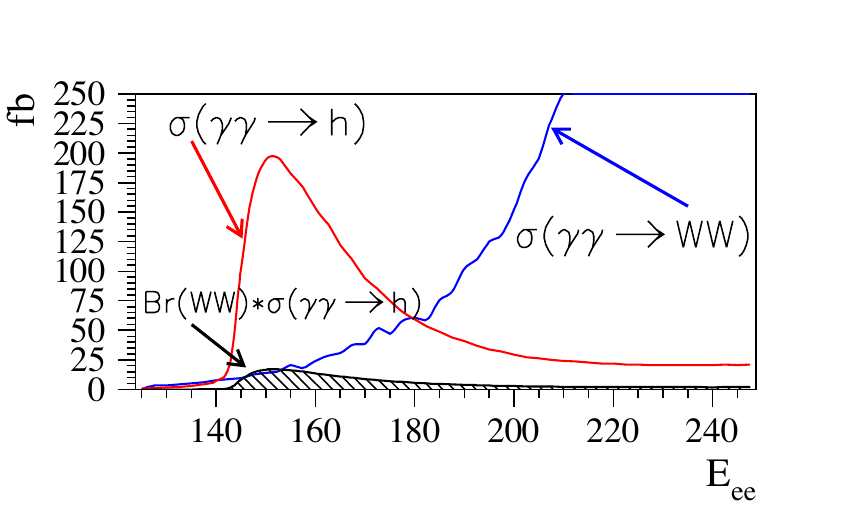}
\includegraphics[width=0.49\hsize]{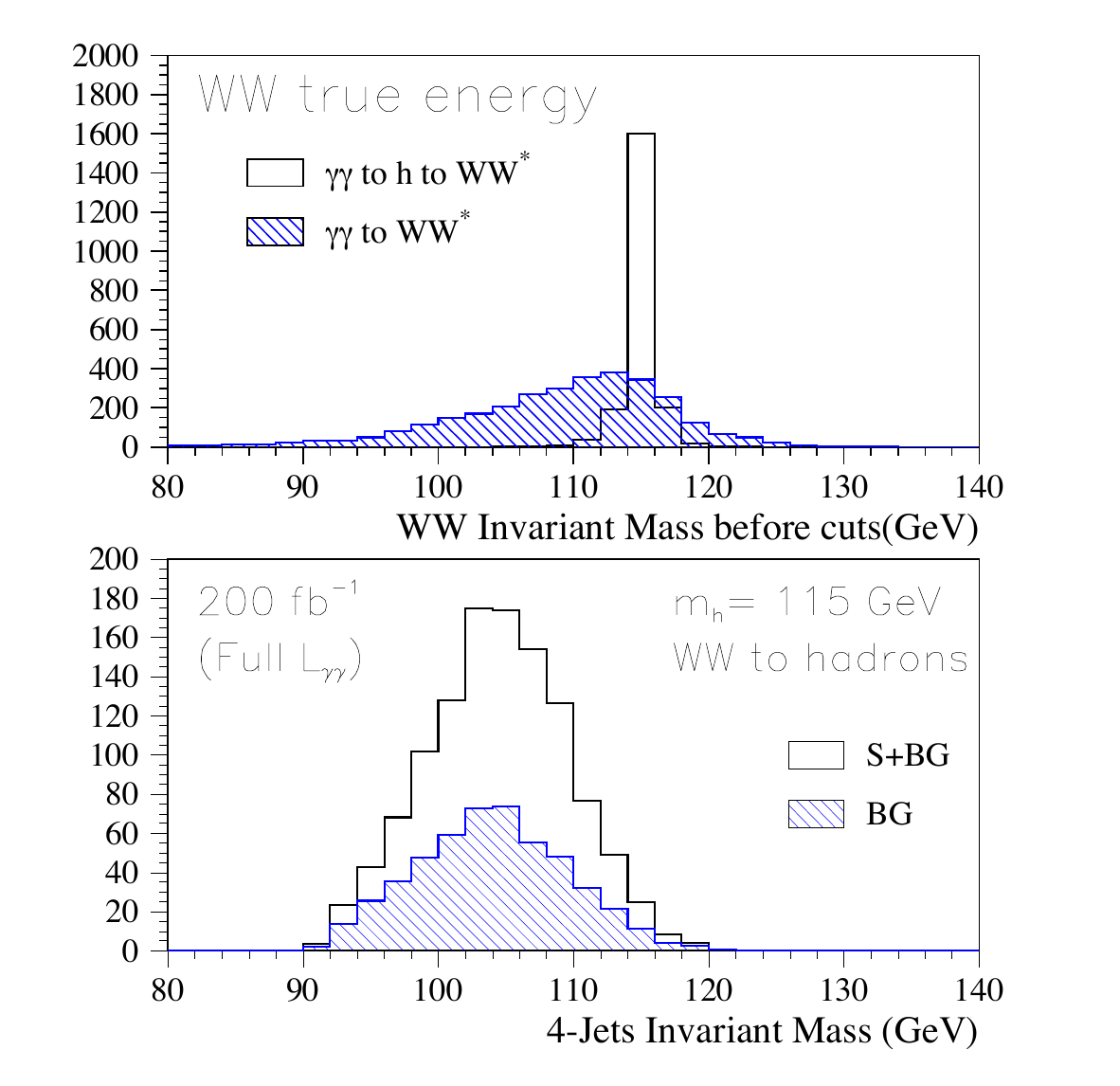}
\caption[0]{      
(a) Cross sections for $\gamma\gamma\rightarrow h$, 
$\gamma\gamma\rightarrow h \times {\cal B}r(h\to WW)$ for 
$m_H=115$~GeV and  $\gamma\gamma\rightarrow WW$
production. (b) Comparison of the  ideal invariant mass
of the $WW$ pairs  from signal and background events.
(c) Selection of the $WW$   decay mode of the Higgs boson for 
$m_H=115$~GeV, running at $E_{CM}(\gamma\gamma)=115$~GeV at CLICHE.
}
\label{fig:wwcross}
\end{figure}

\subsubsection{$h \to \gamma\gamma$}
In almost any phenomenological context, the decay $H \rightarrow
\gamma\gamma$ is a very rare one. However,
the number of Higgs events is large at a $\gamma\gamma$~collider, 
so an interesting number of $H \to \gamma\gamma$
events would be produced.  Furthermore, the backgrounds are expected
to be quite small, below 2~fb~\cite{Jikia:1993yc}, 
 since there is no tree-level coupling of photons,
and the box-mediated processes are peaked very sharply in the
forward direction. A  complete background study has not yet been made, but 
initial estimates indicate that a clear peak
in the $\gamma\gamma$~mass distribution should be observable, 
and we assume here that the background error would be negligible.

The number of events produced in this channel is proportional
to ${\Gamma_{\gamma\gamma}^2/\Gamma_{\mathrm{total}}}$. The quadratic
dependence is interesting, because if $\Gamma_{\mathrm{total}}$ could
be measured elsewhere, a small error on $\Gamma_{\gamma\gamma}$ would
be obtained. Similarly, if $\Gamma_{\gamma\gamma}$ is measured elsewhere, 
a small error $\Gamma_{\mathrm{total}}$  could be obtained.
In Fig.~\ref{fig:ggtogg}, we can see that a 10\% measurement of
${\Gamma_{\gamma\gamma}^2/\Gamma_{\mathrm{total}}}$ can be made with 
less than a year of data taking.

\thisfloatsetup{floatwidth=\SfigwFull,capposition=beside}
\begin{figure}
\includegraphics[width=0.99\hsize]{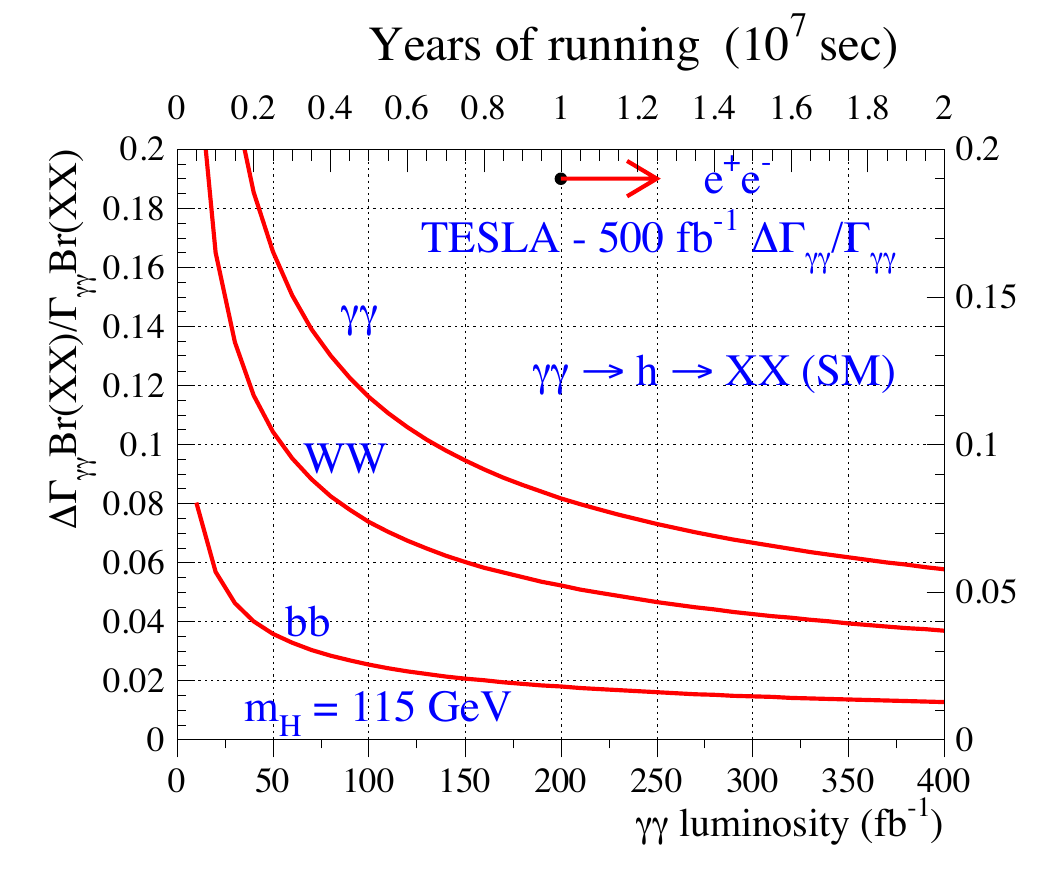}
\caption[0]{ 
The expected precision in the ${h \rightarrow \gamma\gamma}$ decay
width from direct measurements
of $h\rightarrow\gamma\gamma$ for $m_H = 115$~GeV. The precision is
less than in the equivalent measurement of $H\rightarrow WW, \bar{b}b$,
but this observable is unique to a low-energy $\gamma\gamma$ collider
like CLICHE. 
}
\label{fig:ggtogg}     
\end{figure}

The cleanliness of these events and good energy resolution in the 
electromagnetic calorimeter would allow for an independent measurement 
of the Higgs mass. Assuming that the calorimeter energy scales can
be sufficiently well calibrated, a resolution better than $100$~MeV
can be expected.

\subsection{Determining CP Nature of a Higgs Boson}
Precision studies of the Standard Model-like Higgs can be performed using the peaked luminosity spectrum (II) with $\sqrt s=m_{\mathrm Higgs}/y_{\mathrm peak}$.
These include: determination of CP properties; a detailed scan to
separate the $H^0$ and $A^0$ when in the decoupling limit of a 2HDM; branching ratios, and the ration of vacuum expectation values - $\tan \beta$.

Determination of the CP properties of
any spin-0 Higgs $\widehat h$ produced in $\gamma\gamma$
collisions is possible since $\gamma\gamma\to {\widehat h}$ must proceed at one
loop, whether ${\widehat h}$ is CP-even, CP-odd or a mixture.
As a result, the CP-even and CP-odd parts of ${\widehat h}$ have
$\gamma\gamma$ couplings
of similar size.   However, the structure of the couplings is very different:
\begin{equation}
{\cal A}_{CP=+}\propto \vec \epsilon_1\cdot\vec \epsilon_2\,,\quad
{\cal A}_{CP=-}\propto (\vec\epsilon_1\times\vec \epsilon_2)\cdot \hat p_{\mathrm beam}\,.
\end{equation}
By adjusting the orientation of the photon polarization vectors with
respect to one another, it is possible to determine the relative
amounts of CP-even and CP-odd content in the resonance ${\widehat h}$
\cite{Grzadkowski:1992sa}. 
If ${\widehat h}$ is a mixture, one can use helicity asymmetries for this purpose
 \cite{Grzadkowski:1992sa,Kramer:1993jn}.
However, if ${\widehat h}$ is either purely CP-even
or purely CP-odd, then one must employ transverse linear polarizations
\cite{Gunion:1994wy,Kramer:1993jn}. 

For a Higgs boson of pure CP, one finds that the Higgs cross section 
is proportional to
\begin{equation}
{d{\cal L}\over dE_{\gamma\gamma}}\left(1+\vev{\lambda\lambda'}+{\cal CP} \vev{\lambda_T\lambda_T'}
\cos2\delta \right)
\label{tranxsec}
\end{equation}
where ${\cal CP}=+1$ (${\cal CP}=-1$) 
for a pure CP-even (CP-odd) Higgs boson and
and $\delta$ is the angle between the transverse polarizations of
the laser photons. Thus, one measure of  
the CP nature of a Higgs is the asymmetry
for parallel vs. perpendicular orientation 
of the transverse linear polarizations of the initial laser beams.
In the absence of background, this would take the form
\begin{equation}
{\cal A}\equiv{N_{\parallel}-N_{\perp}\over N_{\parallel}+N_{\perp}}
={{{\cal L} CP}\vev{\lambda_T\lambda_T'}\over 1+\vev{\lambda\lambda'}} \,,
\label{asymzerob}
\end{equation}
which is positive (negative) for a CP-even (odd) state. 
The $b\overline b(g)$ and $c\overline c(g)$ backgrounds result 
in additional contributions 
to the  $N_{\parallel}+N_{\perp}$ denominator, which dilutes the
asymmetry. The backgrounds do not contribute to the numerator
for CP invariant cuts. Since, as described below, total
linear polarization for the laser beams translates into
only partial polarization for the back-scattered photons
which collide to form the Higgs boson, both $N_{\parallel}$
and $N_{\perp}$ will be non-zero for the signal.
The expected value of ${\cal A}$ must be carefully computed for a given model
and given cuts.
  
\thisfloatsetup{floatwidth=\SfigwFull,capposition=beside}
\begin{figure}
\includegraphics[width=0.99\hsize]{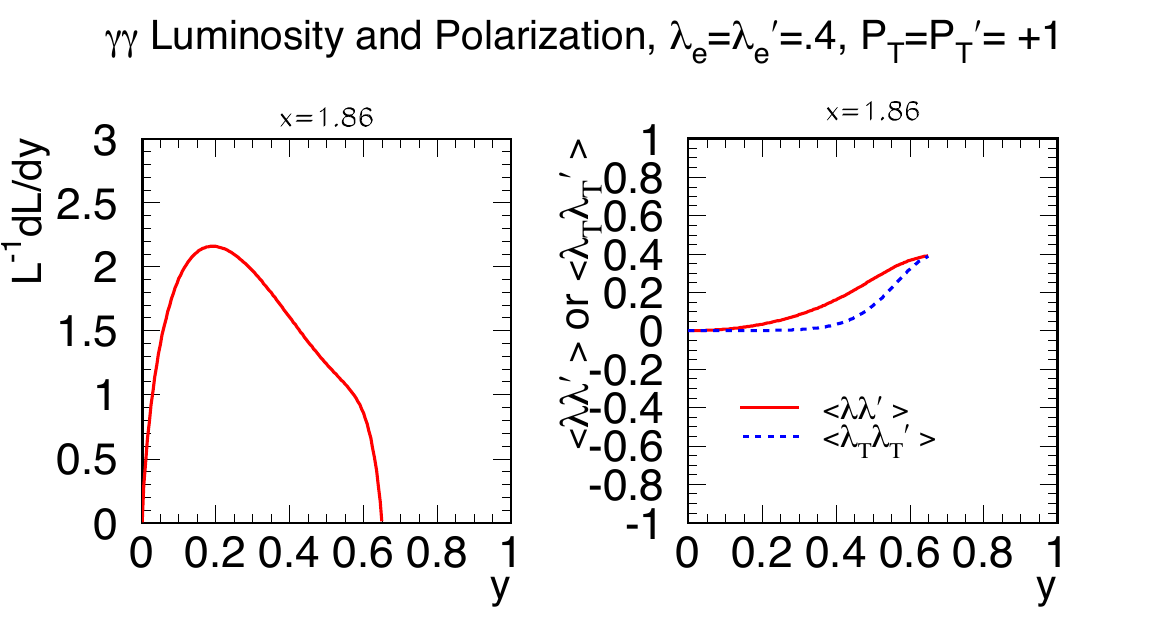}
\caption[0]{We plot the luminosities and corresponding $\vev{\lambda\lambda'}$
and $\vev{\lambda_T\lambda_T^\prime}$
for operation at $\sqrt s=206$~GeV and $x=1.86$,
assuming 100\% transverse polarization for the laser photons
and $\lambda_e=\lambda_e^\prime=0.4$. These plots are for the naive
non-CAIN distributions.
}
 \label{linlumnaive}
\end{figure}
\thisfloatsetup{floatwidth=\SfigwFull,capposition=beside}
\begin{figure}
\includegraphics[width=0.99\hsize]{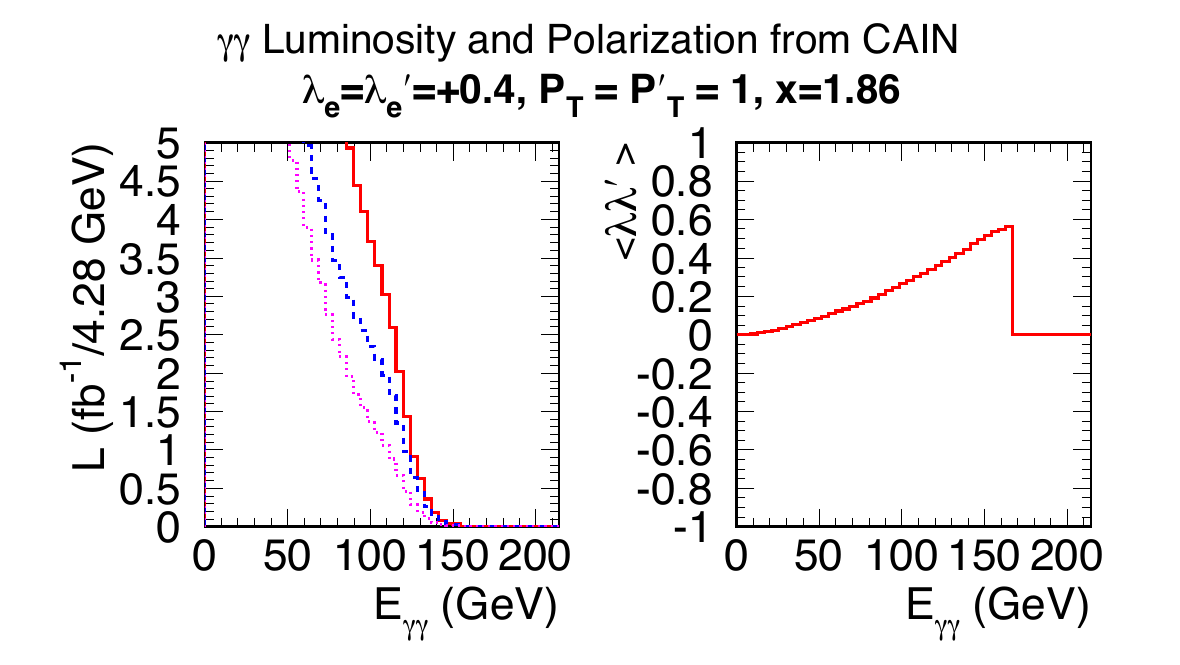}
%\vspace*{-2.7in}
\caption[0]{We plot the luminosity,
$L=d{\cal L}/dE_{\gamma\gamma}$, in units of ${\mathrm fb}^{-1}/4.28$~GeV 
and corresponding $\vev{\lambda\lambda'}$ predicted by CAIN
for operation at $\sqrt s=206$~GeV and $x=1.86$,
assuming 100\% transverse polarization for the laser photons
and $\lambda_e=\lambda_e^\prime=0.4$.
 The dashed (dotted) curve gives the component
of the total luminosity that derives from the $J_z=0$ ($J_z=2$) two-photon
configuration. The solid luminosity curve is the sum of these two
components and $\vev{\lambda\lambda'}=(L_{J_z=0}-L_{J_z=2})/(L_{J_z=0}+L_{J_z=2})$.
}
  \label{linlum}
\end{figure}

At the kinematic limit, $z=z_{\mathrm max}=x/(1+x)$, 
the ratio of $\lambda$ to $\lambda_T$ is given by
\begin{equation}
{\lambda\over \lambda_T}= \lambda_e x{2+x\over 1+x}\sim 1
\end{equation}
for $\lambda_e=0.4$ and $x=1.86$.
 Substantial luminosity and values of $\lambda_T$ close to the
maximum are achieved for moderately smaller $z$. 
%From (\ref{ptform}),
Operation at $x=1.86$ (corresponding to $\sqrt s=206$~GeV and laser
wave length of $\lambda\sim 1~\mu$) would allow 
$\lambda_T^{\mathrm max}\sim\lambda^{\mathrm max}\sim 0.6$.
Making these choices for both beams
is very nearly optimal for the CP study for the
following reasons. First, these choices will maximize 
${d{\cal L}\over dE_{\gamma\gamma}}\vev{\lambda_T\lambda_T'}$ 
at ${E_{\gamma\gamma}=120}$~GeV. As seen
from earlier equations, it is
the square root of the former quantity that
essentially determines the accuracy with which
the CP determination can be made. 
Second, $\lambda_e=\lambda_e'=0.4$ results in $\vev{\lambda\lambda'}>0$. This is 
desirable for suppressing the background. (If there
were no background, Eq.~(\ref{asymzerob}) implies
that the optimal choice would be to employ
$\lambda_e$ and $\lambda_e'$ such that $\vev{\lambda\lambda'}<0$.
However, in practice the background is very substantial and it
is very important to have $\vev{\lambda\lambda'}>0$ to suppress
it as much as possible.) 
In Fig.~\ref{linlumnaive}, we plot  
the naive luminosity distribution  
and associated values of $\vev{\lambda\lambda'}$ and $\vev{\lambda_T\lambda_T'}$
obtained for $\lambda_e=\lambda_e'=0.4$ and 100\% transverse polarization
for the laser beams.

As discussed in \cite{Gunion:1994wy}, the
asymmetry studies discussed below are not very sensitive to
the polarization of the colliding $e$ beams.
Thus, the studies could be performed in parasitic fashion during
$e^-e^+$ operation if the $e^+$ polarization is small. (As emphasized
earlier, substantial $e^+$ polarization would be needed for precision
studies of other $h_{SM}$ properties.)

\thisfloatsetup{floatwidth=\SfigwFull,capposition=beside}
\begin{figure}
\includegraphics[width=0.6\hsize]{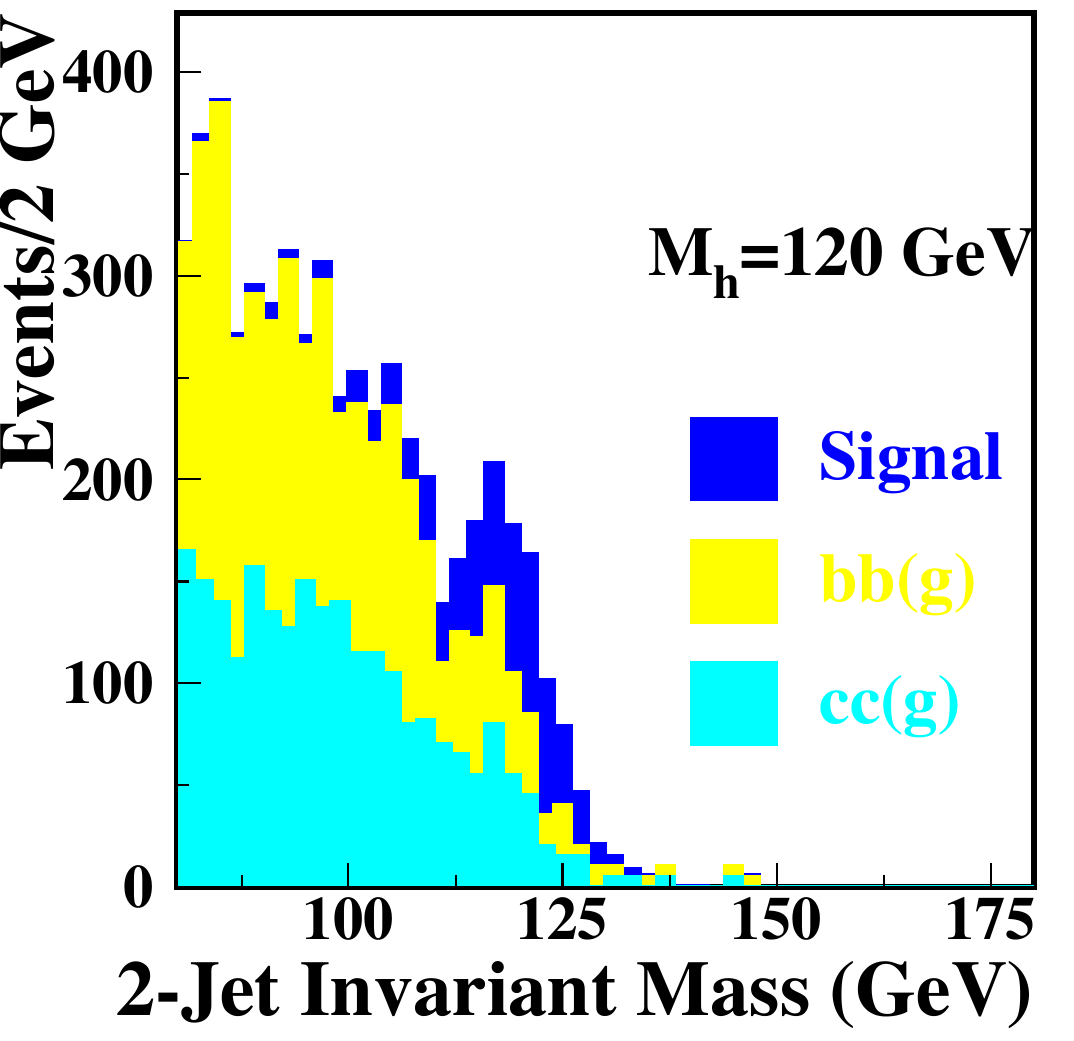}
%\vspace*{-2.7in}
\caption[0]{We plot the signal and $b\overline b$ and $c\overline c$
backgrounds for a SM Higgs boson with $m_{h_{SM}}=120$~GeV
assuming $\gamma\gamma$ operation at $\sqrt s=206$~GeV and $x=1.86$,
based on the luminosity and polarization distributions of Fig.~\ref{linlum}
for the case of linearly polarized laser photons.
The cross sections presented are those for $\delta=\pi/4$, \ie\
in the absence of any contribution from the transverse polarization
term in Eq.~(\ref{tranxsec}).
}
\label{linsignal}
\end{figure}

The luminosity distribution predicted by the CAIN
Monte Carlo for transversely polarized laser photons
and the corresponding result for $\vev{\lambda\lambda'}$ are plotted in
Fig.~\ref{linlum}. We note that
even though the luminosity spectrum is not peaked, it is very nearly the same 
at $E_{\gamma\gamma}=120$~GeV as in the circular polarization case.
As expected from our earlier discussion
of the naive luminosity distribution,
at $E_{\gamma\gamma}=120$~GeV we find  $\vev{\lambda\lambda'}\sim 
\vev{\lambda_T\lambda_T'}\sim 0.36$. Since CAIN includes multiple interactions
and non-linear Compton processes, the luminosity is actually
non-zero for $E_{\gamma\gamma}$ values above the naive kinematic limit of
$\sim 132$~GeV.  Both $\vev{\lambda\lambda'}$ and $\vev{\lambda_T\lambda_T'}$
continue to increase as one enters this region.  However, the luminosity
becomes so small that we cannot make effective use of this region
for this study. 
We employ these luminosity and polarization results 
in the vicinity of $E_{\gamma\gamma}=120$~GeV
in a full Monte Carlo for Higgs production and decay as outlined
earlier in the circular polarization case. All the same cuts
and procedures are employed.  

The resulting
signal and background rates for $\delta=\pi/4$
are presented in Fig.~\ref{linsignal}.
The width of the Higgs resonance peak is $5.0\pm 0.3$~GeV (using a Gaussian
fit), only slightly larger than in the circularly polarized case.
However, because of the shape of the luminosity distribution,
the backgrounds rise more rapidly for $m_{b\overline b}$
values below $120$~GeV than in the case of circularly polarized laser beams. 
Thus, it is best to use a slightly higher cut on the $m_{b\overline b}$
values in order to obtain the best statistical significance for the signal.
Ref.~\cite{Asner:2001ia} finds $\sim 360$
reconstructed two-jet signal events with $m_{b\overline b}\geq 114$~GeV
in one year of operation, with roughly 440 background events
in this same region. Under luminosity assumptions similar to those used in Table~\ref{tab:gghiggsbf}, this corresponds to a precision
of $\sqrt{S+B}/S\sim 0.032$ for the measurement
of $\Gamma(h_{SM}\to\gamma\gamma)B(h_{SM}\to b\overline b)$.
Not surprisingly, this is not as good as for the circularly polarized
setup, but it is still indicative of a very strong Higgs signal.
Turning to the CP determination, let us assume that 
we run 50\% in the parallel
polarization configuration and 50\% in the perpendicular
polarization configuration.
Then, because we have only 60\% linear
polarization for the colliding photons for $E_{\gamma\gamma} \sim 120$~GeV, 
$N_{\parallel}\sim 180[1+(0.6)^2]+273\sim 518$ and
$N_{\perp}\sim 180[1-(0.6)^2]+273=388$.
For these numbers, ${\cal A}=130/906\sim 0.14$.
The error in ${\cal A}$ (again with luminosity assumptions similar to those used in Table~\ref{tab:gghiggsbf})
is $\delta{\cal A}=\sqrt{N_{\parallel}N_{\perp}/N^3}\sim 0.007$
($N\equiv N_\parallel+N_\perp$), yielding 
${\delta{\cal A}\over{\cal A}}={\delta {\cal CP}\over {\cal CP}}\sim 0.05$.
This measurement would thus provide a 
fairly strong confirmation of the CP=+ nature of the $h_{SM}$
after three $10^7$ sec years devoted to this study.

%\section{Single H, A production}

%{\bf NEED TO SUMMARIZE from http://arxiv.org/abs/hep-ph/0110320}
%\subsection{Extending the Search for SUSY Particles}
%\subsection{Other BSM Models}

%{\bf - this stuff below although interesting probably not needed in ILC Higgs paper}
%\section{Other Standard Model Physics}
%\subsection{Top quark Production}
%\subsection{Triple Gauge Couplings}
%\section{Beyond the Standard Model}
%\subsection{Extending the Search for SUSY Particles}
%\subsection{Other BSM Models}

\section{Understanding gamma-gamma backgrounds at the ILC} %from 0111056
QCD aspects of gamma-gamma physics have been studied at electron-positron colliders over the last several decades years. At LEP, gamma-gamma collisions with $\sqrt{s}$ up to 140 GeV have been studied. 
Up to now, the photons have been produced via bremsstrahlung from the electron and positron beams, leading to soft energy spectra with only limited statistics at high $\sqrt{s}$, whereas the 
gamma-gamma option of the ILC will produce collisions in the high-energy part of the spectrum. A plethora of QCD physics topics in two-photon interactions can be addressed with a gamma-gamma collider, 
as discussed in \cite{Asner:2001vh}. These topics include total gamma-gamma to hadrons cross sections and studies of the (polarized) photon structure functions. Furthermore, good knowledge and understanding of 
two-photon processes will be essential for controlling physics background contributions to other processes and machine backgrounds at TeV and multi-TeV linear electron-positron colliders.

\section{Summary}
A gamma-gamma (and $e$-gamma) collider provide exciting physics opportunities that are  complementary to and thus strengthen the physics case for the ILC.
This section presented a summary of Higgs studies possible at a gamma-gamma collider. The broader physics program of a photon collider is summarized in Table~\ref{gammagammasumtab}.
\thisfloatsetup{floatwidth=\SfigwFull,capposition=beside}
\begin{figure}
\includegraphics[width=0.99\hsize]{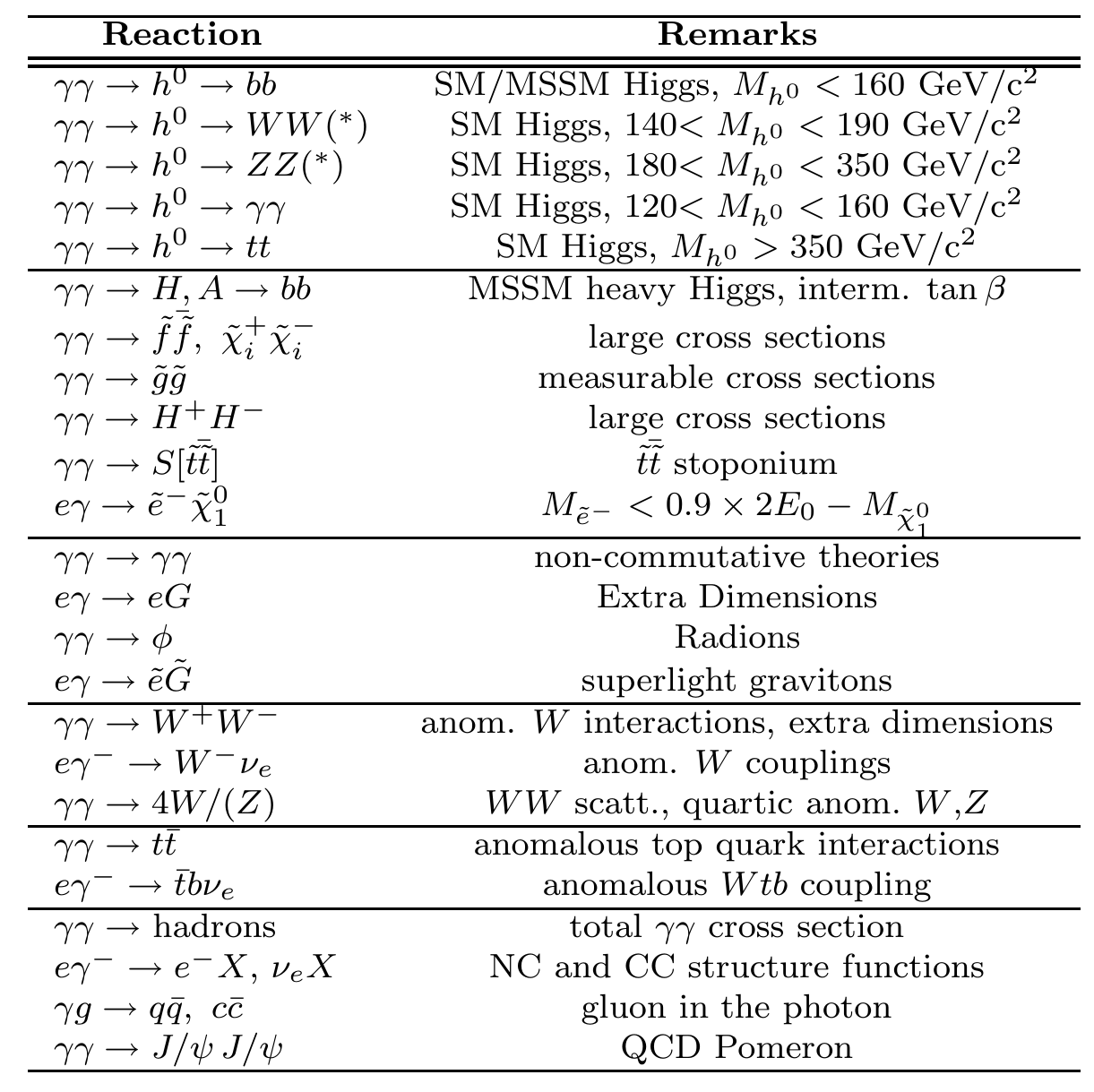}
%\vspace*{-2.7in}
\caption[0]{Summary of gamma-gamma collider golden modes}
\label{gammagammasumtab}
\end{figure}

\chapter{Summary \label{sid:chapter_summary}}

A summary of all model independent coupling precisions is given in 
\Tref{tab:summodelindglobalfit0p1}.
\begin{table}[h]
 \begin{center}
 \begin{tabular}{lcccc}
                &  ILC(250)        & ILC(500)           & ILC(1000)      & ILC(LumUp) \cr 
$\sqrt{s}$  (GeV)   &    250           &   250+500           & 250+500+1000    & 250+500+1000  \cr 
  L  (fb$^{-1}$)  &   250         &     250+500            & 250+500+1000   & 1150+1600+2500   \cr \hline
$\gamma\gamma$  &      18   \%      &    8.4  \%      &   4.0  \%       & 2.4  \%  \cr
$gg$            &   6.4  \%         &   2.3  \%       &   1.6  \%       & 0.9 \%  \cr
$WW$            &   4.8 \%         &  1.1   \%        &   1.1  \%       & 0.6 \%  \cr
$ZZ$            &      1.3 \%      & 1.0  \%          &   1.0  \%       & 0.5 \%  \cr
$t\bar t$        &      --           &   14    \%      &   3.1   \%      & 1.9  \%  \cr
$b\bar b$       &    5.3 \%         &  1.6   \%       &   1.3    \%     & 0.7 \%  \cr
$\tau^+\tau^-$   &     5.7 \%        &    2.3  \%      &   1.6  \%       & 0.9  \%  \cr
$c\bar c$       &     6.8   \%      &     2.8  \%     &   1.8    \%     & 1.0  \%  \cr
$\mu^+\mu^-$     &   91\%              &     91\%          &   16 \%         & 10   \%  \cr
$\Gamma_T(h)$    &    12   \%        &    4.9 \%       &   4.5   \%      & 2.3  \%  \cr
$hhh$             &      --          &     83 \%        &  21 \%          & 13 \%  \cr \hline
BR(invis.)       &  $ <$  0.9  \%   &  $ <$  0.9 \%  &  $ <$  0.9 \%  &  $ <$  0.4 \%  \cr
   \hline
   \end{tabular}
  \caption{Summary of expected accuracies  $\Delta g_i/g_i$ for model independent 
determinations of the Higgs boson couplings. The theory errors are $\Delta F_i/F_i=0.1\%$.  For the invisible
branching ratio, the numbers quoted are 95\% confidence upper limits.}
\label{tab:summodelindglobalfit0p1}
  \end{center}
\end{table}
%%%%%%%%%%%%%%%%%%%%%%%%%%%%%%%%%%%%%%%%%%%%%%%%%%%%%%%%%%%%%%%%%

For the purpose of comparing ILC coupling precisions with those of other facilities we present
the coupling errors in \Tref{tab:summatchlhctechnique0p1}.
\begin{table}[h]
 \begin{center}
 \begin{tabular}{lcccc}
                &  ILC(250)        & ILC(500)           & ILC(1000)      & ILC(LumUp) \cr \hline
$\sqrt{s}$  (GeV)   &    250           &   250+500           & 250+500+1000    & 250+500+1000  \cr 
  L  (fb$^{-1}$)  &   250         &     250+500            & 250+500+1000   & 1150+1600+2500   \cr \hline
$\gamma\gamma$  &      17   \%      &    8.3  \%      &   3.8  \%       & 2.3  \%  \cr
$gg$            &   6.1  \%         &   2.0  \%       &   1.1  \%       & 0.7 \%  \cr
$WW$            &   4.7 \%         &  0.4   \%        &   0.3  \%       & 0.2 \%  \cr
$ZZ$            &     0.7 \%      &  0.5  \%          &  0.5  \%       & 0.3 \%  \cr
$t\bar t$       &    6.4 \%         &  2.5  \%       &   1.3    \%     & 0.9 \%  \cr
$b\bar b$       &    4.7 \%         &  1.0  \%       &   0.6    \%     & 0.4 \%  \cr
$\tau^+\tau^-$   &     5.2 \%        &    1.9  \%      &   1.3  \%       & 0.7  \%  \cr 
$\Gamma_T(h)$    &    9.0   \%        &    1.7 \%       &   1.1   \%   & 0.8  \%  \cr  \hline
$\mu^+\mu^-$     &   91 \%              &     91 \%          &   16 \%    & 10   \%  \cr
$hhh$             &      --          &     83 \%        &  21 \%          & 13 \%  \cr
BR(invis.)       &  $ <$  0.9  \%   &  $ <$  0.9 \%  &  $ <$  0.9 \%  &  $ <$  0.4 \%  \cr \hline
$c\bar c$       &     6.8   \%      &     2.8 \%     &   1.8    \%     & 1.0  \%  \cr
   \hline
   \end{tabular}
  \caption{Summary of expected accuracies  $\Delta g_i/g_i$ of Higgs boson couplings 
using, for each coupling, the fitting technique that most closely matches
that used by LHC experiments.
For  $g_g,g_{\gamma},g_W,g_Z,g_b,g_t,g_{\tau},\Gamma_T(h)$ the seven parameter
 HXSWG benchmark parameterization described in Section~10.3.7 of Ref.~\cite{Dittmaier:2011ti} is used.
For the couplings $g_\mu$, $g_{hhh}$ and the limit on invisible branching ratio independent analyses are used.  
The charm coupling $g_c$ comes from our 10 parameter model independent fit.
All theory errors are $0.1\%$. 
  For the invisible
branching ratio, the numbers quoted are 95\% confidence upper limits.}
\label{tab:summatchlhctechnique0p1}
  \end{center}
\end{table}
%%%%%%%%%%%%%%%%%%%%%%%%%%%%%%%%%%%%%%%%%%%%%%%%%%%%%%%%%%%%%%%%%

In the energy and luminosity scenarios discussed in this paper it was assumed that
the luminosity upgrades at 250 and 500 GeV center of mass energy occurred after
the energy upgrade at 1000 GeV.  It is of interest to consider
a scenario where the 250 GeV and 500 GeV luminosity upgrade running occurs before the
energy upgrade to 1000 GeV.    This would correspond to the energies and luminosities in \Tref{tab:ecmlumrunsnotev}.

\begin{table}
 \begin{center}
 \begin{tabular}{lccccccc}
Nickname  & Ecm(1)    & Lumi(1)   &     +  & Ecm(2)         & Lumi(2)    & Runtime   & Wallplug E         \cr              
          &         (GeV)  &          (fb$^{-1}$)     &  &       (GeV)  &          (fb$^{-1}$)     & (yr) & (MW-yr)  \cr \hline
 ILC(250) &  250 & 250 & & & & 1.1 & 130 \cr
 ILC(500) &  250 & 250 & & 500  & 500 & 2.0  & 270 \cr
 ILC500(LumUp) &  250 & 1150 & &  500  & 1600 & 3.9 & 660 \cr
   \hline
   \end{tabular}
  \caption{Energy and luminosities assuming no running at 1 TeV center of mass energy.}
\label{tab:ecmlumrunsnotev}
  \end{center}
\end{table}
%%%%%%%%%%%%%%%%%%%%%%%%%%%%%%%%%%%%%%%%%%%%%%%%%%%%%%%%%%%%%%%%%

A summary of all model independent coupling precisions for the case where the 250 GeV and 500 GeV luminosity upgrade running occurs before the
energy upgrade to 1000 GeV is shown in \Tref{tab:summodelindglobalfit0p1notev}.
\begin{table}[h]
 \begin{center}
 \begin{tabular}{lccc}
                &  ILC(250)        & ILC(500)       &     ILC500(LumUp) \cr \hline
$\sqrt{s}$  (GeV)   &    250           &   250+500           & 250+500  \cr 
  L  (fb$^{-1}$)  &   250         &     250+500            & 1150+1600   \cr \hline
$\gamma\gamma$  &      18   \%      &    8.4  \%      & 4.5 \%   \cr
$gg$            &   6.4  \%         &   2.3  \%       & 1.2 \%  \cr
$WW$            &   4.8 \%         &  1.1   \%        & 0.6 \%  \cr
$ZZ$            &      1.3 \%      & 1.0  \%          & 0.5 \%  \cr
$t\bar t$        &      --           &   14    \%     & 7.8 \%   \cr
$b\bar b$       &    5.3 \%         &  1.6   \%       & 0.8 \% \cr
$\tau^+\tau^-$   &     5.7 \%        &    2.3  \%      & 1.2 \%    \cr
$c\bar c$       &     6.8   \%      &     2.8 \%     & 1.5 \%  \cr
$\mu^+\mu^-$     &   91 \%              &     91 \%          & 42 \%    \cr
$\Gamma_T(h)$    &    12   \%        &    4.9 \%       & 2.5 \%  \cr
$hhh$             &      --          &     83 \%        & 46 \% \cr \hline
BR(invis.)       &  $ <$  0.9  \%   &  $ <$  0.9 \%  &  $ <$  0.4 \% \cr 
   \hline
   \end{tabular}
  \caption{Summary of expected accuracies  $\Delta g_i/g_i$ for model independent 
determinations of the Higgs boson couplings. The theory errors are $\Delta F_i/F_i=0.1\%$.  For the invisible
branching ratio, the numbers quoted are 95\% confidence upper limits. 
}
\label{tab:summodelindglobalfit0p1notev}
  \end{center}
\end{table}
%%%%%%%%%%%%%%%%%%%%%%%%%%%%%%%%%%%%%%%%%%%%%%%%%%%%%%%%%%%%%%%%%

The facility comparison table in the case where the 250 GeV and 500 GeV luminosity upgrade running occurs before the
energy upgrade to 1000 GeV is shown in \Tref{tab:summatchlhctechnique0p1notev}.

\begin{table}[h]
 \begin{center}
 \begin{tabular}{lccc}
                &  ILC(250)        & ILC(500)       &     ILC500(LumUp) \cr \hline
$\sqrt{s}$  (GeV)   &    250           &   250+500           & 250+500  \cr 
  L  (fb$^{-1}$)  &   250         &     250+500            & 1150+1600   \cr \hline
$\gamma\gamma$  &      17   \%      &    8.3  \%      & 4.4 \%  \cr
$gg$            &   6.1  \%         &   2.0  \%       & 1.1 \%  \cr
$WW$            &   4.7 \%         &  0.4   \%        & 0.3 \%  \cr
$ZZ$            &     0.7 \%      &  0.5  \%          & 0.3 \%   \cr
$t\bar t$       &    6.4 \%         &  2.5  \%       & 1.4 \%  \cr
$b\bar b$       &    4.7 \%         &  1.0  \%       & 0.6 \%   \cr
$\tau^+\tau^-$   &     5.2 \%        &    1.9  \%      & 1.0 \%    \cr
$\Gamma_T(h)$    &    9.0   \%        &    1.7 \%       & 1.0 \%  \cr  \hline
$\mu^+\mu^-$     &   91 \%              &     91 \%          & 42 \%   \cr
$hhh$             &      --          &     83 \%        & 46 \%  \cr 
BR(invis.)       &  $ <$  0.9  \%   &  $ <$  0.9 \%  &  $ <$  0.4 \%  \cr \hline
$c\bar c$       &     6.8   \%      &     2.8 \%     & 1.5 \%  \cr
   \hline
   \end{tabular}
  \caption{Summary of expected accuracies  $\Delta g_i/g_i$ of Higgs boson couplings 
using, for each coupling, the fitting technique that most closely matches
that used by LHC experiments.
For  $g_g,g_{\gamma},g_W,g_Z,g_b,g_t,g_{\tau},\Gamma_T(h)$ the seven parameter
 HXSWG benchmark parameterization described in Section~10.3.7 of Ref.~\cite{Dittmaier:2011ti} is used.
For the couplings $g_\mu$, $g_{hhh}$ and the limit on invisible branching ratio independent analyses are used.  
The charm coupling $g_c$  comes from our 10 parameter model independent fit.
All theory errors are $0.1\%$. 
  For the invisible
branching ratio, the numbers quoted are 95\% confidence upper limits.  
}
\label{tab:summatchlhctechnique0p1notev}
  \end{center}
\end{table}
%%%%%%%%%%%%%%%%%%%%%%%%%%%%%%%%%%%%%%%%%%%%%%%%%%%%%%%%%%%%%%%%%

A comparison of  model independent coupling precisions with and without 1 TeV running
is shown in \Tref{tab:summodelindglobalfit0p1notevvstev}.
\begin{table}[h]
 \begin{center}
 \begin{tabular}{lcc}
                &     ILC500(LumUp) & ILC(LumUp) \cr \hline
$\sqrt{s}$  (GeV)   & 250+500  & 250+500+1000    \cr 
  L  (fb$^{-1}$)  & 1150+1600  & 1150+1600+2500     \cr \hline
$\gamma\gamma$  & 4.5 \%    & 2.4  \%  \cr
$gg$            & 1.2 \%   & 0.9 \%  \cr 
$WW$           & 0.6 \%  & 0.6 \%  \cr
$ZZ$            & 0.5\%  & 0.5 \%  \cr
$t\bar t$        & 7.8 \%   & 1.9  \%  \cr
$b\bar b$       & 0.8 \% & 0.7 \%  \cr
$\tau^+\tau^-$   & 1.2 \%    & 0.9  \%  \cr
$c\bar c$       & 1.5 \%  & 1.0  \%  \cr
$\mu^+\mu^-$     & 42 \%    & 10   \%  \cr
$\Gamma_T(h)$    & 2.5 \%  & 2.3  \%  \cr
$hhh$            & 46 \%  & 13 \%  \cr \hline
BR(invis.)      &  $ <$  0.4 \% &  $ <$  0.4 \%  \cr 
   \hline
   \end{tabular}
  \caption{Summary of expected accuracies  $\Delta g_i/g_i$ for model independent 
determinations of the Higgs boson couplings. The theory errors are $\Delta F_i/F_i=0.1\%$.  For the invisible
branching ratio, the numbers quoted are 95\% confidence upper limits. 
}
\label{tab:summodelindglobalfit0p1notevvstev}
  \end{center}
\end{table}
%%%%%%%%%%%%%%%%%%%%%%%%%%%%%%%%%%%%%%%%%%%%%%%%%%%%%%%%%%%%%%%%%

The facility comparison table with and without 1 TeV running
is shown in \Tref{tab:summatchlhctechnique0p1notevvstev}.

\begin{table}[h]
 \begin{center}
 \begin{tabular}{lcc}
                &     ILC500(LumUp) & ILC(LumUp) \cr \hline
$\sqrt{s}$  (GeV)   & 250+500  & 250+500+1000    \cr 
  L  (fb$^{-1}$)  & 1150+1600  & 1150+1600+2500     \cr \hline
$\gamma\gamma$  & 4.4 \%   & 2.3  \%  \cr
$gg$            & 1.1 \%  & 0.7 \%  \cr
$WW$            & 0.3 \%  & 0.2 \%  \cr
$ZZ$            & 0.3 \%   & 0.3 \%  \cr
$t\bar t$       & 1.4 \%  & 0.9 \%  \cr
$b\bar b$       & 0.6 \%   & 0.4 \%  \cr
$\tau^+\tau^-$   & 1.0 \%    & 0.7  \%  \cr
$\Gamma_T(h)$    & 1.0 \%  & 0.8  \%  \cr  \hline
$\mu^+\mu^-$     & 42 \%    & 10   \%  \cr
$hhh$            & 46 \%  & 13 \%  \cr 
BR(invis.)      &  $ <$  0.4 \%  &  $ <$  0.4 \%  \cr \hline
$c\bar c$       & 1.5 \%  & 1.0  \%  \cr
   \hline
   \end{tabular}
  \caption{Summary of expected accuracies  $\Delta g_i/g_i$ of Higgs boson couplings 
using, for each coupling, the fitting technique that most closely matches
that used by LHC experiments.
For  $g_g,g_{\gamma},g_W,g_Z,g_b,g_t,g_{\tau},\Gamma_T(h)$ the seven parameter
 HXSWG benchmark parameterization described in Section~10.3.7 of Ref.~\cite{Dittmaier:2011ti} is used.
For the couplings $g_\mu$, $g_{hhh}$ and the limit on invisible branching ratio independent analyses are used.  
The charm coupling $g_c$  comes from our 10 parameter model independent fit.
All theory errors are $0.1\%$. 
  For the invisible
branching ratio, the numbers quoted are 95\% confidence upper limits.  
}
\label{tab:summatchlhctechnique0p1notevvstev}
  \end{center}
\end{table}
%%%%%%%%%%%%%%%%%%%%%%%%%%%%%%%%%%%%%%%%%%%%%%%%%%%%%%%%%%%%%%%%%

%\input{Chapter_Benchmarking/benchmarking.tex}
\cleardoublepage
\phantomsection % fix hyperlinking
\addcontentsline{toc}{chapter}{Bibliography}
\fancyhead[LE]{\textit{\nouppercase{Bibliography}}}
\fancyhead[RO]{\textit{\nouppercase{Bibliography}}}
\bibliographystyle{kp}
\bibliography{Bibliography/ilchiggsbibliography}

\begingroup\raggedright\begin{thebibliography}{297}
\expandafter\ifx\csname natexlab\endcsname\relax\def\natexlab#1{#1}\fi

\bibitem[Behnke et~al.(2013)Behnke, Brau, Foster, Fuster, Harrison,
  et~al.]{Behnke:2013xla}
T.~Behnke, J.~E. Brau, B.~Foster, J.~Fuster, M.~Harrison, {\em et~al.}, ``{The
  International Linear Collider Technical Design Report - Volume 1: Executive
  Summary}'',
 \href{http://xxx.lanl.gov/abs/1306.6327}{{\ttfamily arXiv:1306.6327}}.
%%CITATION = ARXIV:1306.6327;%%.

\bibitem[Baer et~al.(2013)Baer, Barklow, Fujii, Gao, Hoang,
  et~al.]{Baer:2013cma}
H.~Baer, T.~Barklow, K.~Fujii, Y.~Gao, A.~Hoang, {\em et~al.}, ``{The
  International Linear Collider Technical Design Report - Volume 2: Physics}'',
 \href{http://xxx.lanl.gov/abs/1306.6352}{{\ttfamily arXiv:1306.6352}}.
%%CITATION = ARXIV:1306.6352;%%.

\bibitem[Adolphsen et~al.(2013{\natexlab{a}})Adolphsen, Barone, Barish,
  Buesser, Burrows, et~al.]{Adolphsen:2013jya}
C.~Adolphsen, M.~Barone, B.~Barish, K.~Buesser, P.~Burrows, {\em et~al.},
  ``{The International Linear Collider Technical Design Report - Volume 3.I:
  Accelerator R \& D in the Technical Design Phase}'',
 \href{http://xxx.lanl.gov/abs/1306.6353}{{\ttfamily arXiv:1306.6353}}.
%%CITATION = ARXIV:1306.6353;%%.

\bibitem[Adolphsen et~al.(2013{\natexlab{b}})Adolphsen, Barone, Barish,
  Buesser, Burrows, et~al.]{Adolphsen:2013kya}
C.~Adolphsen, M.~Barone, B.~Barish, K.~Buesser, P.~Burrows, {\em et~al.},
  ``{The International Linear Collider Technical Design Report - Volume 3.II:
  Accelerator Baseline Design}'',
 \href{http://xxx.lanl.gov/abs/1306.6328}{{\ttfamily arXiv:1306.6328}}.
%%CITATION = ARXIV:1306.6328;%%.

\bibitem[Behnke et~al.(2013)Behnke, Brau, Burrows, Fuster, Peskin,
  et~al.]{Behnke:2013lya}
T.~Behnke, J.~E. Brau, P.~N. Burrows, J.~Fuster, M.~Peskin, {\em et~al.},
  ``{The International Linear Collider Technical Design Report - Volume 4:
  Detectors}'',
 \href{http://xxx.lanl.gov/abs/1306.6329}{{\ttfamily arXiv:1306.6329}}.
%%CITATION = ARXIV:1306.6329;%%.

\bibitem[Heuer(2003)]{Heuer:2003nn}
R.~Heuer, ``{Parameters for the Linear Collider}'',
2003.
%%CITATION = INSPIRE-640245;%%.

\bibitem[Weinberg(1995)]{Weinberg:1995mt}
S.~Weinberg, ``{The Quantum Theory of Fields. Vol. 1: Foundations}'',
1995.
%%CITATION = INSPIRE-406190;%%.

\bibitem[Weinberg(1996)]{Weinberg:1996kr}
S.~Weinberg, ``{The Quantum Theory of Fields. Vol. 2: Modern applications}'',
1996.
%%CITATION = INSPIRE-430948;%%.

\bibitem[Englert and Brout(1964)]{Englert:1964et}
F.~Englert and R.~Brout, ``{Broken Symmetry and the Mass of Gauge Vector
  Mesons}'', {\em Phys.Rev.Lett.} {\bfseries 13} (1964)
321--323.
%%CITATION = PRLTA,13,321;%%.

\bibitem[Higgs(1964{\natexlab{a}})]{Higgs:1964ia}
P.~W. Higgs, ``{Broken symmetries, massless particles and gauge fields}'', {\em
  Phys.Lett.} {\bfseries 12} (1964){\natexlab{a}}
132--133.
%%CITATION = PHLTA,12,132;%%.

\bibitem[Higgs(1964{\natexlab{b}})]{Higgs:1964pj}
P.~W. Higgs, ``{Broken Symmetries and the Masses of Gauge Bosons}'', {\em
  Phys.Rev.Lett.} {\bfseries 13} (1964){\natexlab{b}}
508--509.
%%CITATION = PRLTA,13,508;%%.

\bibitem[Guralnik et~al.(1964)Guralnik, Hagen, and Kibble]{Guralnik:1964eu}
G.~Guralnik, C.~Hagen, and T.~Kibble, ``{Global Conservation Laws and Massless
  Particles}'', {\em Phys.Rev.Lett.} {\bfseries 13} (1964)
585--587.
%%CITATION = PRLTA,13,585;%%.

\bibitem[Guralnik and Hagen(1965)]{Guralnik:1965uza}
G.~S. Guralnik and C.~R. Hagen, ``{Massless particles and the goldstone
  theorem}'', {\em Nuovo Cim.} {\bfseries 43} (1965), no.~1,
1--32.
%%CITATION = NUCIA,43,1;%%.

\bibitem[Kibble(1967)]{Kibble:1967sv}
T.~Kibble, ``{Symmetry breaking in nonAbelian gauge theories}'', {\em
  Phys.Rev.} {\bfseries 155} (1967)
1554--1561.
%%CITATION = PHRVA,155,1554;%%.

\bibitem[Weinberg(1967)]{Weinberg:1967tq}
S.~Weinberg, ``{A Model of Leptons}'', {\em Phys.Rev.Lett.} {\bfseries 19}
  (1967)
1264--1266.
%%CITATION = PRLTA,19,1264;%%.

\bibitem[Glashow(1961)]{Glashow:1961tr}
S.~Glashow, ``{Partial Symmetries of Weak Interactions}'', {\em Nucl.Phys.}
  {\bfseries 22} (1961)
579--588.
%%CITATION = NUPHA,22,579;%%.

\bibitem[Salam(1968)]{Salam:1968rm}
A.~Salam, ``{Weak and Electromagnetic Interactions}'', {\em Conf.Proc.}
  {\bfseries C680519} (1968)
367--377.
%%CITATION = CONFP,C680519,367;%%.

\bibitem[King(1995)]{King:1994yr}
S.~F. King, ``{Dynamical electroweak symmetry breaking}'', {\em
  Rept.Prog.Phys.} {\bfseries 58} (1995) 263--310,
 \href{http://xxx.lanl.gov/abs/hep-ph/9406401}{{\ttfamily
  arXiv:hep-ph/9406401}}.
%%CITATION = HEP-PH/9406401;%%.

\bibitem[Aad et~al.(2012)]{Aad:2012tfa}
{\bfseries ATLAS Collaboration} Collaboration, G.~Aad {\em et~al.},
  ``{Observation of a new particle in the search for the Standard Model Higgs
  boson with the ATLAS detector at the LHC}'', {\em Phys.Lett.} {\bfseries
  B716} (2012) 1--29,
 \href{http://xxx.lanl.gov/abs/1207.7214}{{\ttfamily arXiv:1207.7214}}.
%%CITATION = ARXIV:1207.7214;%%.

\bibitem[Chatrchyan et~al.(2012)]{Chatrchyan:2012ufa}
{\bfseries CMS Collaboration} Collaboration, S.~Chatrchyan {\em et~al.},
  ``{Observation of a new boson at a mass of 125 GeV with the CMS experiment at
  the LHC}'', {\em Phys.Lett.} {\bfseries B716} (2012) 30--61,
 \href{http://xxx.lanl.gov/abs/1207.7235}{{\ttfamily arXiv:1207.7235}}.
%%CITATION = ARXIV:1207.7235;%%.

\bibitem[Aad et~al.(2013{\natexlab{a}})]{Aad:2013wqa}
{\bfseries ATLAS Collaboration} Collaboration, G.~Aad {\em et~al.},
  ``{Measurements of Higgs boson production and couplings in diboson final
  states with the ATLAS detector at the LHC}'', {\em Phys.Lett.} {\bfseries B}
  (2013){\natexlab{a}}
 \href{http://xxx.lanl.gov/abs/1307.1427}{{\ttfamily arXiv:1307.1427}}.
%%CITATION = ARXIV:1307.1427;%%.

\bibitem[Aad et~al.(2013{\natexlab{b}})]{Aad:2013xqa}
{\bfseries ATLAS Collaboration} Collaboration, G.~Aad {\em et~al.}, ``{Evidence
  for the spin-0 nature of the Higgs boson using ATLAS data}'', {\em
  Phys.Lett.} {\bfseries B726} (2013){\natexlab{b}} 120--144,
 \href{http://xxx.lanl.gov/abs/1307.1432}{{\ttfamily arXiv:1307.1432}}.
%%CITATION = ARXIV:1307.1432;%%.

\bibitem[Chatrchyan et~al.(2013{\natexlab{a}})]{Chatrchyan:2013lba}
{\bfseries CMS Collaboration} Collaboration, S.~Chatrchyan {\em et~al.},
  ``{Observation of a new boson with mass near 125 GeV in pp collisions at
  $\sqrt{s}$ = 7 and 8 TeV}'', {\em JHEP} {\bfseries 1306} (2013){\natexlab{a}}
  081,
 \href{http://xxx.lanl.gov/abs/1303.4571}{{\ttfamily arXiv:1303.4571}}.
%%CITATION = ARXIV:1303.4571;%%.

\bibitem[Chatrchyan et~al.(2013{\natexlab{b}})]{Chatrchyan:2012jja}
{\bfseries CMS Collaboration} Collaboration, S.~Chatrchyan {\em et~al.},
  ``{Study of the Mass and Spin-Parity of the Higgs Boson Candidate Via Its
  Decays to $Z$ Boson Pairs}'', {\em Phys.Rev.Lett.} {\bfseries 110}
  (2013){\natexlab{b}} 081803,
 \href{http://xxx.lanl.gov/abs/1212.6639}{{\ttfamily arXiv:1212.6639}}.
%%CITATION = ARXIV:1212.6639;%%.

\bibitem[Llewellyn~Smith(1973)]{LlewellynSmith:1973ey}
C.~Llewellyn~Smith, ``{High-Energy Behavior and Gauge Symmetry}'', {\em
  Phys.Lett.} {\bfseries B46} (1973)
233--236.
%%CITATION = PHLTA,B46,233;%%.

\bibitem[Cornwall et~al.(1973)Cornwall, Levin, and
  Tiktopoulos]{Cornwall:1973tb}
J.~M. Cornwall, D.~N. Levin, and G.~Tiktopoulos, ``{Uniqueness of spontaneously
  broken gauge theories}'', {\em Phys.Rev.Lett.} {\bfseries 30} (1973)
1268--1270.
%%CITATION = PRLTA,30,1268;%%.

\bibitem[Cornwall et~al.(1974)Cornwall, Levin, and
  Tiktopoulos]{Cornwall:1974km}
J.~M. Cornwall, D.~N. Levin, and G.~Tiktopoulos, ``{Derivation of Gauge
  Invariance from High-Energy Unitarity Bounds on the $S$ Matrix}'', {\em
  Phys.Rev.} {\bfseries D10} (1974)
1145.
%%CITATION = PHRVA,D10,1145;%%.

\bibitem[Gunion et~al.(1991)Gunion, Haber, and Wudka]{Gunion:1990kf}
J.~Gunion, H.~Haber, and J.~Wudka, ``{Sum rules for Higgs bosons}'', {\em
  Phys.Rev.} {\bfseries D43} (1991)
904--912.
%%CITATION = PHRVA,D43,904;%%.

\bibitem[Erler and Langacker(????)]{ErlerPDG}
J.~Erler and P.~Langacker, ``{Electroweak model and constraints on new physics,
  in Ref.~\cite{Beringer:1900zz}, pp.~136--156}'',.

\bibitem[Baak et~al.(2012)Baak, Goebel, Haller, Hoecker, Kennedy, Kogler,
  Moenig, Schott, and Stelzer]{Baak:2012kk}
M.~Baak, M.~Goebel, J.~Haller, A.~Hoecker, D.~Kennedy, R.~Kogler, K.~Moenig,
  M.~Schott, and T.~Stelzer, ``{The Electroweak Fit of the Standard Model after
  the Discovery of a New Boson at the LHC}'', {\em Eur.Phys.J.} {\bfseries C72}
  (2012) 2205,
 \href{http://xxx.lanl.gov/abs/1209.2716}{{\ttfamily arXiv:1209.2716}}.
%%CITATION = ARXIV:1209.2716;%%.

\bibitem[Gunion et~al.(2000)Gunion, Haber, Kane, and Dawson]{Gunion:1989we}
J.~F. Gunion, H.~E. Haber, G.~L. Kane, and S.~Dawson, ``{The Higgs Hunter's
  Guide}'', {\em Front.Phys.} {\bfseries 80} (2000)
1--448.
%%CITATION = FRPHA,80,1;%%.

\bibitem[Abers and Lee(1973)]{Abers:1973qs}
E.~Abers and B.~Lee, ``{Gauge Theories}'', {\em Phys.Rept.} {\bfseries 9}
  (1973)
1--141.
%%CITATION = PRPLC,9,1;%%.

\bibitem[Minkowski(1977)]{Minkowski:1977sc}
P.~Minkowski, ``{$\mu\to e\gamma$ at a Rate of One Out of 1-Billion Muon
  Decays?}'', {\em Phys.Lett.} {\bfseries B67} (1977)
421.
%%CITATION = PHLTA,B67,421;%%.

\bibitem[Gell-Mann et~al.(1979)Gell-Mann, Ramond, and Slansky]{GellMann:1980vs}
M.~Gell-Mann, P.~Ramond, and R.~Slansky, ``{Complex Spinors and Unified
  Theories}'', {\em Conf.Proc.} {\bfseries C790927} (1979) 315--321,
 \href{http://xxx.lanl.gov/abs/1306.4669}{{\ttfamily arXiv:1306.4669}}.
%%CITATION = ARXIV:1306.4669;%%.

\bibitem[Yanagida(1980)]{Yanagida:1980xy}
T.~Yanagida, ``{Horizontal Symmetry and Masses of Neutrinos}'', {\em
  Prog.Theor.Phys.} {\bfseries 64} (1980)
1103.
%%CITATION = PTPKA,64,1103;%%.

\bibitem[Mohapatra and Senjanovic(1980)]{Mohapatra:1979ia}
R.~N. Mohapatra and G.~Senjanovic, ``{Neutrino Mass and Spontaneous Parity
  Violation}'', {\em Phys.Rev.Lett.} {\bfseries 44} (1980)
912.
%%CITATION = PRLTA,44,912;%%.

\bibitem[Mohapatra and Senjanovic(1981)]{Mohapatra:1980yp}
R.~N. Mohapatra and G.~Senjanovic, ``{Neutrino Masses and Mixings in Gauge
  Models with Spontaneous Parity Violation}'', {\em Phys.Rev.} {\bfseries D23}
  (1981)
165.
%%CITATION = PHRVA,D23,165;%%.

\bibitem[Ellis et~al.(1976)Ellis, Gaillard, and Nanopoulos]{Ellis:1975ap}
J.~R. Ellis, M.~K. Gaillard, and D.~V. Nanopoulos, ``{A Phenomenological
  Profile of the Higgs Boson}'', {\em Nucl.Phys.} {\bfseries B106} (1976)
292.
%%CITATION = NUPHA,B106,292;%%.

\bibitem[Shifman et~al.(1979)Shifman, Vainshtein, Voloshin, and
  Zakharov]{Shifman:1979eb}
M.~A. Shifman, A.~Vainshtein, M.~Voloshin, and V.~I. Zakharov, ``{Low-Energy
  Theorems for Higgs Boson Couplings to Photons}'', {\em Sov.J.Nucl.Phys.}
  {\bfseries 30} (1979)
711--716.
%%CITATION = SJNCA,30,711;%%.

\bibitem[Carena et~al.(2012)Carena, Low, and Wagner]{Carena:2012xa}
M.~Carena, I.~Low, and C.~E. Wagner, ``{Implications of a Modified Higgs to
  Diphoton Decay Width}'', {\em JHEP} {\bfseries 1208} (2012) 060,
 \href{http://xxx.lanl.gov/abs/1206.1082}{{\ttfamily arXiv:1206.1082}}.
%%CITATION = ARXIV:1206.1082;%%.

\bibitem[Appelquist and Carazzone(1975)]{Appelquist:1974tg}
T.~Appelquist and J.~Carazzone, ``{Infrared Singularities and Massive
  Fields}'', {\em Phys.Rev.} {\bfseries D11} (1975)
2856.
%%CITATION = PHRVA,D11,2856;%%.

\bibitem[Sirlin and Zucchini(1986)]{Sirlin:1985ux}
A.~Sirlin and R.~Zucchini, ``{Dependence of the Quartic Coupling
  $\bar{h}_{\overline{\mathrm MS}}$ on $m_H$ and the Possible Onset of New Physics
  in the Higgs Sector of the Standard Model}'', {\em Nucl.Phys.} {\bfseries
  B266} (1986)
389.
%%CITATION = NUPHA,B266,389;%%.

\bibitem[Kanemura et~al.(2004)Kanemura, Okada, Senaha, and
  Yuan]{Kanemura:2004mg}
S.~Kanemura, Y.~Okada, E.~Senaha, and C.-P. Yuan, ``{Higgs coupling constants
  as a probe of new physics}'', {\em Phys.Rev.} {\bfseries D70} (2004) 115002,
 \href{http://xxx.lanl.gov/abs/hep-ph/0408364}{{\ttfamily
  arXiv:hep-ph/0408364}}.
%%CITATION = HEP-PH/0408364;%%.

\bibitem[Cohen et~al.(1993)Cohen, Kaplan, and Nelson]{Cohen:1993nk}
A.~G. Cohen, D.~Kaplan, and A.~Nelson, ``{Progress in electroweak
  baryogenesis}'', {\em Ann.Rev.Nucl.Part.Sci.} {\bfseries 43} (1993) 27--70,
 \href{http://xxx.lanl.gov/abs/hep-ph/9302210}{{\ttfamily
  arXiv:hep-ph/9302210}}.
%%CITATION = HEP-PH/9302210;%%.

\bibitem[Morrissey and Ramsey-Musolf(2012)]{Morrissey:2012db}
D.~E. Morrissey and M.~J. Ramsey-Musolf, ``{Electroweak baryogenesis}'', {\em
  New J.Phys.} {\bfseries 14} (2012) 125003,
 \href{http://xxx.lanl.gov/abs/1206.2942}{{\ttfamily arXiv:1206.2942}}.
%%CITATION = ARXIV:1206.2942;%%.

\bibitem[Kanemura et~al.(2005)Kanemura, Okada, and Senaha]{Kanemura:2004ch}
S.~Kanemura, Y.~Okada, and E.~Senaha, ``{Electroweak baryogenesis and quantum
  corrections to the triple Higgs boson coupling}'', {\em Phys.Lett.}
  {\bfseries B606} (2005) 361--366,
 \href{http://xxx.lanl.gov/abs/hep-ph/0411354}{{\ttfamily
  arXiv:hep-ph/0411354}}.
%%CITATION = HEP-PH/0411354;%%.

\bibitem[Heinemeyer et~al.(2013)]{Heinemeyer:2013tqa}
{\bfseries LHC Higgs Cross Section Working Group} Collaboration, S.~Heinemeyer
  {\em et~al.}, ``{Handbook of LHC Higgs Cross Sections: 3. Higgs
  Properties}'',
 \href{http://xxx.lanl.gov/abs/1307.1347}{{\ttfamily arXiv:1307.1347}}.
%%CITATION = ARXIV:1307.1347;%%.

\bibitem[Cabibbo et~al.(1979)Cabibbo, Maiani, Parisi, and
  Petronzio]{Cabibbo:1979ay}
N.~Cabibbo, L.~Maiani, G.~Parisi, and R.~Petronzio, ``{Bounds on the Fermions
  and Higgs Boson Masses in Grand Unified Theories}'', {\em Nucl.Phys.}
  {\bfseries B158} (1979)
295--305.
%%CITATION = NUPHA,B158,295;%%.

\bibitem[Degrassi et~al.(2012)Degrassi, Di~Vita, Elias-Miro, Espinosa, Giudice,
  Isidori, and Strumia]{Degrassi:2012ry}
G.~Degrassi, S.~Di~Vita, J.~Elias-Miro, J.~R. Espinosa, G.~F. Giudice,
  G.~Isidori, and A.~Strumia, ``{Higgs mass and vacuum stability in the
  Standard Model at NNLO}'', {\em JHEP} {\bfseries 1208} (2012) 098,
 \href{http://xxx.lanl.gov/abs/1205.6497}{{\ttfamily arXiv:1205.6497}}.
%%CITATION = ARXIV:1205.6497;%%.

\bibitem[Sher(1989)]{Sher:1988mj}
M.~Sher, ``{Electroweak Higgs Potentials and Vacuum Stability}'', {\em
  Phys.Rept.} {\bfseries 179} (1989)
273--418.
%%CITATION = PRPLC,179,273;%%.

\bibitem[Kanemura et~al.(1999)Kanemura, Kasai, and Okada]{Kanemura:1999xf}
S.~Kanemura, T.~Kasai, and Y.~Okada, ``{Mass bounds of the lightest CP even
  Higgs boson in the two Higgs doublet model}'', {\em Phys.Lett.} {\bfseries
  B471} (1999) 182--190,
 \href{http://xxx.lanl.gov/abs/hep-ph/9903289}{{\ttfamily
  arXiv:hep-ph/9903289}}.
%%CITATION = HEP-PH/9903289;%%.

\bibitem[Branco et~al.(2012)Branco, Ferreira, Lavoura, Rebelo, Sher, and
  Silva]{Branco:2011iw}
G.~Branco, P.~Ferreira, L.~Lavoura, M.~Rebelo, M.~Sher, and J.~P. Silva,
  ``{Theory and phenomenology of two-Higgs-doublet models}'', {\em Phys.Rept.}
  {\bfseries 516} (2012) 1--102,
 \href{http://xxx.lanl.gov/abs/1106.0034}{{\ttfamily arXiv:1106.0034}}.
%%CITATION = ARXIV:1106.0034;%%.

\bibitem[Lee(1973)]{Lee:1973iz}
T.~Lee, ``{A Theory of Spontaneous T Violation}'', {\em Phys.Rev.} {\bfseries
  D8} (1973)
1226--1239.
%%CITATION = PHRVA,D8,1226;%%.

\bibitem[Glashow and Weinberg(1977)]{Glashow:1976nt}
S.~L. Glashow and S.~Weinberg, ``{Natural Conservation Laws for Neutral
  Currents}'', {\em Phys.Rev.} {\bfseries D15} (1977)
1958.
%%CITATION = PHRVA,D15,1958;%%.

\bibitem[Paschos(1977)]{Paschos:1976ay}
E.~Paschos, ``{Diagonal Neutral Currents}'', {\em Phys.Rev.} {\bfseries D15}
  (1977)
1966.
%%CITATION = PHRVA,D15,1966;%%.

\bibitem[Davidson and Haber(2005)]{Davidson:2005cw}
S.~Davidson and H.~E. Haber, ``{Basis-independent methods for the
  two-Higgs-doublet model}'', {\em Phys.Rev.} {\bfseries D72} (2005) 035004,
 \href{http://xxx.lanl.gov/abs/hep-ph/0504050}{{\ttfamily
  arXiv:hep-ph/0504050}}.
%%CITATION = HEP-PH/0504050;%%.

\bibitem[Branco et~al.(1999)Branco, Lavoura, and Silva]{Branco:1999fs}
G.~C. Branco, L.~Lavoura, and J.~P. Silva, ``{CP Violation}'', {\em
  Int.Ser.Monogr.Phys.} {\bfseries 103} (1999)
1--536.
%%CITATION = IMPHA,103,1;%%.

\bibitem[Haber and O'Neil(2006)]{Haber:2006ue}
H.~E. Haber and D.~O'Neil, ``{Basis-independent methods for the
  two-Higgs-doublet model. II. The Significance of tan beta}'', {\em Phys.Rev.}
  {\bfseries D74} (2006) 015018,
 \href{http://xxx.lanl.gov/abs/hep-ph/0602242}{{\ttfamily
  arXiv:hep-ph/0602242}}.
%%CITATION = HEP-PH/0602242;%%.

\bibitem[Haber and O'Neil(2011)]{Haber:2010bw}
H.~E. Haber and D.~O'Neil, ``{Basis-independent methods for the
  two-Higgs-doublet model III: The CP-conserving limit, custodial symmetry, and
  the oblique parameters S, T, U}'', {\em Phys.Rev.} {\bfseries D83} (2011)
  055017,
 \href{http://xxx.lanl.gov/abs/1011.6188}{{\ttfamily arXiv:1011.6188}}.
%%CITATION = ARXIV:1011.6188;%%.

\bibitem[Peskin and Takeuchi(1992)]{Peskin:1991sw}
M.~E. Peskin and T.~Takeuchi, ``{Estimation of oblique electroweak
  corrections}'', {\em Phys.Rev.} {\bfseries D46} (1992)
381--409.
%%CITATION = PHRVA,D46,381;%%.

\bibitem[Veltman(1977)]{Veltman:1977kh}
M.~Veltman, ``{Limit on Mass Differences in the Weinberg Model}'', {\em
  Nucl.Phys.} {\bfseries B123} (1977)
89.
%%CITATION = NUPHA,B123,89;%%.

\bibitem[Froggatt et~al.(1992{\natexlab{a}})Froggatt, Moorhouse, and
  Knowles]{Froggatt:1991qw}
C.~Froggatt, R.~Moorhouse, and I.~Knowles, ``{Leading radiative corrections in
  two scalar doublet models}'', {\em Phys.Rev.} {\bfseries D45}
  (1992){\natexlab{a}}
2471--2481.
%%CITATION = PHRVA,D45,2471;%%.

\bibitem[Froggatt et~al.(1992{\natexlab{b}})Froggatt, Moorhouse, and
  Knowles]{Froggatt:1992wt}
C.~Froggatt, R.~Moorhouse, and I.~Knowles, ``{Two scalar doublet models with
  softly broken symmetries}'', {\em Nucl.Phys.} {\bfseries B386}
  (1992){\natexlab{b}}
63--114.
%%CITATION = NUPHA,B386,63;%%.

\bibitem[Grimus et~al.(2008)Grimus, Lavoura, Ogreid, and Osland]{Grimus:2008nb}
W.~Grimus, L.~Lavoura, O.~Ogreid, and P.~Osland, ``{The Oblique parameters in
  multi-Higgs-doublet models}'', {\em Nucl.Phys.} {\bfseries B801} (2008)
  81--96,
 \href{http://xxx.lanl.gov/abs/0802.4353}{{\ttfamily arXiv:0802.4353}}.
%%CITATION = ARXIV:0802.4353;%%.

\bibitem[Lee et~al.(1977)Lee, Quigg, and Thacker]{Lee:1977eg}
B.~W. Lee, C.~Quigg, and H.~Thacker, ``{Weak Interactions at Very
  High-Energies: The Role of the Higgs Boson Mass}'', {\em Phys.Rev.}
  {\bfseries D16} (1977)
1519.
%%CITATION = PHRVA,D16,1519;%%.

\bibitem[Weldon(1984)]{Weldon:1984wt}
H.~A. Weldon, ``{The Effects of Multiple Higgs Bosons on Tree Unitarity}'',
  {\em Phys.Rev.} {\bfseries D30} (1984)
1547.
%%CITATION = PHRVA,D30,1547;%%.

\bibitem[Haber and Nir(1990)]{Haber:1989xc}
H.~E. Haber and Y.~Nir, ``{Multiscalar Models With a High-energy Scale}'', {\em
  Nucl.Phys.} {\bfseries B335} (1990)
363.
%%CITATION = NUPHA,B335,363;%%.

\bibitem[Gunion and Haber(2003)]{Gunion:2002zf}
J.~F. Gunion and H.~E. Haber, ``{The CP conserving two Higgs doublet model: The
  Approach to the decoupling limit}'', {\em Phys.Rev.} {\bfseries D67} (2003)
  075019,
 \href{http://xxx.lanl.gov/abs/hep-ph/0207010}{{\ttfamily
  arXiv:hep-ph/0207010}}.
%%CITATION = HEP-PH/0207010;%%.

\bibitem[Craig et~al.(2013)Craig, Galloway, and Thomas]{Craig:2013hca}
N.~Craig, J.~Galloway, and S.~Thomas, ``{Searching for Signs of the Second
  Higgs Doublet}'',
 \href{http://xxx.lanl.gov/abs/1305.2424}{{\ttfamily arXiv:1305.2424}}.
%%CITATION = ARXIV:1305.2424;%%.

\bibitem[Haber(2013)]{Haber:2013}
H.~E. Haber, ``{The decoupling and alignment limits of the Two-Higgs Doublet
  Model}'', {\em In preparation}, 2013.

\bibitem[Djouadi et~al.(1996)Djouadi, Haber, and Zerwas]{Djouadi:1996ah}
A.~Djouadi, H.~Haber, and P.~Zerwas, ``{Multiple production of MSSM neutral
  Higgs bosons at high-energy $e^+ e^-$ colliders}'', {\em Phys.Lett.}
  {\bfseries B375} (1996) 203--212,
 \href{http://xxx.lanl.gov/abs/hep-ph/9602234}{{\ttfamily
  arXiv:hep-ph/9602234}}.
%%CITATION = HEP-PH/9602234;%%.

\bibitem[Djouadi et~al.(1999{\natexlab{a}})Djouadi, Kilian, Muhlleitner, and
  Zerwas]{Djouadi:1999gv}
A.~Djouadi, W.~Kilian, M.~Muhlleitner, and P.~Zerwas, ``{Testing Higgs
  selfcouplings at $e^+ e^-$ linear colliders}'', {\em Eur.Phys.J.} {\bfseries
  C10} (1999){\natexlab{a}} 27--43,
 \href{http://xxx.lanl.gov/abs/hep-ph/9903229}{{\ttfamily
  arXiv:hep-ph/9903229}}.
%%CITATION = HEP-PH/9903229;%%.

\bibitem[Djouadi et~al.(1999{\natexlab{b}})Djouadi, Kilian, Muhlleitner, and
  Zerwas]{Djouadi:1999rca}
A.~Djouadi, W.~Kilian, M.~Muhlleitner, and P.~Zerwas, ``{Production of neutral
  Higgs boson pairs at LHC}'', {\em Eur.Phys.J.} {\bfseries C10}
  (1999){\natexlab{b}} 45--49,
 \href{http://xxx.lanl.gov/abs/hep-ph/9904287}{{\ttfamily
  arXiv:hep-ph/9904287}}.
%%CITATION = HEP-PH/9904287;%%.

\bibitem[Ferrera et~al.(2008)Ferrera, Guasch, Lopez-Val, and
  Sola]{Ferrera:2007sp}
G.~Ferrera, J.~Guasch, D.~Lopez-Val, and J.~Sola, ``{Triple Higgs boson
  production in the Linear Collider}'', {\em Phys.Lett.} {\bfseries B659}
  (2008) 297--307,
 \href{http://xxx.lanl.gov/abs/0707.3162}{{\ttfamily arXiv:0707.3162}}.
%%CITATION = ARXIV:0707.3162;%%.

\bibitem[Hodgkinson et~al.(2009)Hodgkinson, Lopez-Val, and
  Sola]{Hodgkinson:2009uj}
R.~N. Hodgkinson, D.~Lopez-Val, and J.~Sola, ``{Higgs boson pair production
  through gauge boson fusion at linear colliders within the general 2HDM}'',
  {\em Phys.Lett.} {\bfseries B673} (2009) 47--56,
 \href{http://xxx.lanl.gov/abs/0901.2257}{{\ttfamily arXiv:0901.2257}}.
%%CITATION = ARXIV:0901.2257;%%.

\bibitem[Lopez-Val and Sola(2010)]{LopezVal:2009qy}
D.~Lopez-Val and J.~Sola, ``{Neutral Higgs-pair production at Linear Colliders
  within the general 2HDM: Quantum effects and triple Higgs boson
  self-interactions}'', {\em Phys.Rev.} {\bfseries D81} (2010) 033003,
 \href{http://xxx.lanl.gov/abs/0908.2898}{{\ttfamily arXiv:0908.2898}}.
%%CITATION = ARXIV:0908.2898;%%.

\bibitem[Glashow et~al.(1970)Glashow, Iliopoulos, and Maiani]{Glashow:1970gm}
S.~Glashow, J.~Iliopoulos, and L.~Maiani, ``{Weak Interactions with
  Lepton-Hadron Symmetry}'', {\em Phys.Rev.} {\bfseries D2} (1970)
1285--1292.
%%CITATION = PHRVA,D2,1285;%%.

\bibitem[Haber et~al.(1979)Haber, Kane, and Sterling]{Haber:1978jt}
H.~Haber, G.~L. Kane, and T.~Sterling, ``{The Fermion Mass Scale and Possible
  Effects of Higgs Bosons on Experimental Observables}'', {\em Nucl.Phys.}
  {\bfseries B161} (1979)
493.
%%CITATION = NUPHA,B161,493;%%.

\bibitem[Donoghue and Li(1979)]{Donoghue:1978cj}
J.~F. Donoghue and L.~F. Li, ``{Properties of Charged Higgs Bosons}'', {\em
  Phys.Rev.} {\bfseries D19} (1979)
945.
%%CITATION = PHRVA,D19,945;%%.

\bibitem[Hall and Wise(1981)]{Hall:1981bc}
L.~J. Hall and M.~B. Wise, ``{Flavor changing Higgs boson couplings}'', {\em
  Nucl.Phys.} {\bfseries B187} (1981)
397.
%%CITATION = NUPHA,B187,397;%%.

\bibitem[Barger et~al.(1990)Barger, Hewett, and Phillips]{Barger:1989fj}
V.~D. Barger, J.~Hewett, and R.~Phillips, ``{New Constraints on the Charged
  Higgs Sector in Two Higgs Doublet Models}'', {\em Phys.Rev.} {\bfseries D41}
  (1990)
3421--3441.
%%CITATION = PHRVA,D41,3421;%%.

\bibitem[Aoki et~al.(2009)Aoki, Kanemura, Tsumura, and Yagyu]{Aoki:2009ha}
M.~Aoki, S.~Kanemura, K.~Tsumura, and K.~Yagyu, ``{Models of Yukawa interaction
  in the two Higgs doublet model, and their collider phenomenology}'', {\em
  Phys.Rev.} {\bfseries D80} (2009) 015017,
 \href{http://xxx.lanl.gov/abs/0902.4665}{{\ttfamily arXiv:0902.4665}}.
%%CITATION = ARXIV:0902.4665;%%.

\bibitem[Su and Thomas(2009)]{Su:2009fz}
S.~Su and B.~Thomas, ``{The LHC Discovery Potential of a Leptophilic Higgs}'',
  {\em Phys.Rev.} {\bfseries D79} (2009) 095014,
 \href{http://xxx.lanl.gov/abs/0903.0667}{{\ttfamily arXiv:0903.0667}}.
%%CITATION = ARXIV:0903.0667;%%.

\bibitem[Logan and MacLennan(2009)]{Logan:2009uf}
H.~E. Logan and D.~MacLennan, ``{Charged Higgs phenomenology in the
  lepton-specific two Higgs doublet model}'', {\em Phys.Rev.} {\bfseries D79}
  (2009) 115022,
 \href{http://xxx.lanl.gov/abs/0903.2246}{{\ttfamily arXiv:0903.2246}}.
%%CITATION = ARXIV:0903.2246;%%.

\bibitem[Hermann et~al.(2012)Hermann, Misiak, and Steinhauser]{Hermann:2012fc}
T.~Hermann, M.~Misiak, and M.~Steinhauser, ``{$\overline{B}\to X_s \gamma$ in
  the Two Higgs Doublet Model up to Next-to-Next-to-Leading Order in QCD}'',
  {\em JHEP} {\bfseries 1211} (2012) 036,
 \href{http://xxx.lanl.gov/abs/1208.2788}{{\ttfamily arXiv:1208.2788}}.
%%CITATION = ARXIV:1208.2788;%%.

\bibitem[Ciuchini et~al.(1998)Ciuchini, Degrassi, Gambino, and
  Giudice]{Ciuchini:1997xe}
M.~Ciuchini, G.~Degrassi, P.~Gambino, and G.~Giudice, ``{Next-to-leading QCD
  corrections to $B\to X_s\gamma$: Standard model and two Higgs doublet
  model}'', {\em Nucl.Phys.} {\bfseries B527} (1998) 21--43,
 \href{http://xxx.lanl.gov/abs/hep-ph/9710335}{{\ttfamily
  arXiv:hep-ph/9710335}}.
%%CITATION = HEP-PH/9710335;%%.

\bibitem[Beringer et~al.(2012)]{Beringer:1900zz}
{\bfseries Particle Data Group} Collaboration, J.~Beringer {\em et~al.},
  ``{Review of Particle Physics (RPP)}'', {\em Phys.Rev.} {\bfseries D86}
  (2012)
010001.
%%CITATION = PHRVA,D86,010001;%%.

\bibitem[Barberio et~al.(2008)]{Barberio:2008fa}
{\bfseries Heavy Flavor Averaging Group} Collaboration, E.~Barberio {\em
  et~al.}, ``{Averages of $b$-hadron and $c$-hadron properties at the end of
  2007}'',
 \href{http://xxx.lanl.gov/abs/0808.1297}{{\ttfamily arXiv:0808.1297}}.
%%CITATION = ARXIV:0808.1297;%%.

\bibitem[Goto and Okada(1995)]{Goto:1994ck}
T.~Goto and Y.~Okada, ``{Charged Higgs mass bound from the $b \to s\gamma$
  process in the minimal supergravity model}'', {\em Prog.Theor.Phys.}
  {\bfseries 94} (1995) 407--416,
 \href{http://xxx.lanl.gov/abs/hep-ph/9412225}{{\ttfamily
  arXiv:hep-ph/9412225}}.
%%CITATION = HEP-PH/9412225;%%.

\bibitem[Ciuchini et~al.(1998)Ciuchini, Degrassi, Gambino, and
  Giudice]{Ciuchini:1998xy}
M.~Ciuchini, G.~Degrassi, P.~Gambino, and G.~Giudice, ``{Next-to-leading QCD
  corrections to $B\to X_s\gamma$ in supersymmetry}'', {\em Nucl.Phys.}
  {\bfseries B534} (1998) 3--20,
 \href{http://xxx.lanl.gov/abs/hep-ph/9806308}{{\ttfamily
  arXiv:hep-ph/9806308}}.
%%CITATION = HEP-PH/9806308;%%.

\bibitem[Misiak and Steinhauser(2007)]{Misiak:2006ab}
M.~Misiak and M.~Steinhauser, ``{NNLO QCD corrections to the $\overline{B}\to
  X_s\gamma$ matrix elements using interpolation in $m_c$}'', {\em Nucl.Phys.}
  {\bfseries B764} (2007) 62--82,
 \href{http://xxx.lanl.gov/abs/hep-ph/0609241}{{\ttfamily
  arXiv:hep-ph/0609241}}.
%%CITATION = HEP-PH/0609241;%%.

\bibitem[Misiak et~al.(2007)Misiak, Asatrian, Bieri, Czakon, Czarnecki,
  et~al.]{Misiak:2006zs}
M.~Misiak, H.~Asatrian, K.~Bieri, M.~Czakon, A.~Czarnecki, {\em et~al.},
  ``{Estimate of $B(\overline{B}\to X_s\gamma$) at
  $\mathcal{O}(\alpha_s^2)$}'', {\em Phys.Rev.Lett.} {\bfseries 98} (2007)
  022002,
 \href{http://xxx.lanl.gov/abs/hep-ph/0609232}{{\ttfamily
  arXiv:hep-ph/0609232}}.
%%CITATION = HEP-PH/0609232;%%.

\bibitem[Becher and Neubert(2007)]{Becher:2006pu}
T.~Becher and M.~Neubert, ``{Analysis of Br$(\overline{B}\to X_s\gamma)$ at
  NNLO with a cut on photon energy}'', {\em Phys.Rev.Lett.} {\bfseries 98}
  (2007) 022003,
 \href{http://xxx.lanl.gov/abs/hep-ph/0610067}{{\ttfamily
  arXiv:hep-ph/0610067}}.
%%CITATION = HEP-PH/0610067;%%.

\bibitem[Akeroyd and Recksiegel(2003)]{Akeroyd:2003zr}
A.~Akeroyd and S.~Recksiegel, ``{The Effect of $H^\pm$ on $B^\pm\to \tau^\pm
  \nu_\tau$ and $B^\pm\to \mu^\pm \nu_\mu$}'', {\em J.Phys.} {\bfseries G29}
  (2003) 2311--2317,
 \href{http://xxx.lanl.gov/abs/hep-ph/0306037}{{\ttfamily
  arXiv:hep-ph/0306037}}.
%%CITATION = HEP-PH/0306037;%%.

\bibitem[Krawczyk and Sokolowska(2007)]{Krawczyk:2007ne}
M.~Krawczyk and D.~Sokolowska, ``{The Charged Higgs boson mass in the 2HDM:
  Decoupling and CP violation}'', {\em eConf} {\bfseries C0705302} (2007)
  HIG09,
 \href{http://xxx.lanl.gov/abs/0711.4900}{{\ttfamily arXiv:0711.4900}}.
%%CITATION = ARXIV:0711.4900;%%.

\bibitem[Krawczyk and Temes(2005)]{Krawczyk:2004na}
M.~Krawczyk and D.~Temes, ``{2HDM(II) radiative corrections in leptonic tau
  decays}'', {\em Eur.Phys.J.} {\bfseries C44} (2005) 435--446,
 \href{http://xxx.lanl.gov/abs/hep-ph/0410248}{{\ttfamily
  arXiv:hep-ph/0410248}}.
%%CITATION = HEP-PH/0410248;%%.

\bibitem[Jackiw and Weinberg(1972)]{Jackiw:1972jz}
R.~Jackiw and S.~Weinberg, ``{Weak interaction corrections to the muon magnetic
  moment and to muonic atom energy levels}'', {\em Phys.Rev.} {\bfseries D5}
  (1972)
2396--2398.
%%CITATION = PHRVA,D5,2396;%%.

\bibitem[Fujikawa et~al.(1972)Fujikawa, Lee, and Sanda]{Fujikawa:1972fe}
K.~Fujikawa, B.~Lee, and A.~Sanda, ``{Generalized Renormalizable Gauge
  Formulation of Spontaneously Broken Gauge Theories}'', {\em Phys.Rev.}
  {\bfseries D6} (1972)
2923--2943.
%%CITATION = PHRVA,D6,2923;%%.

\bibitem[Leveille(1978)]{Leveille:1977rc}
J.~P. Leveille, ``{The Second Order Weak Correction to $(g-2)_\mu$ in Arbitrary
  Gauge Models}'', {\em Nucl.Phys.} {\bfseries B137} (1978)
63.
%%CITATION = NUPHA,B137,63;%%.

\bibitem[Krawczyk and Zochowski(1997)]{Krawczyk:1996sm}
M.~Krawczyk and J.~Zochowski, ``{Constraining 2HDM by present and future muon
  $g-2$ data}'', {\em Phys.Rev.} {\bfseries D55} (1997) 6968--6974,
 \href{http://xxx.lanl.gov/abs/hep-ph/9608321}{{\ttfamily
  arXiv:hep-ph/9608321}}.
%%CITATION = HEP-PH/9608321;%%.

\bibitem[Barr and Zee(1990)]{Barr:1990vd}
S.~M. Barr and A.~Zee, ``{Electric Dipole Moment of the Electron and of the
  Neutron}'', {\em Phys.Rev.Lett.} {\bfseries 65} (1990)
21--24.
%%CITATION = PRLTA,65,21;%%.

\bibitem[Mahmoudi and St\r{a}l(2010)]{Mahmoudi:2009zx}
F.~Mahmoudi and O.~St\r{a}l, ``{Flavor constraints on the two-Higgs-doublet
  model with general Yukawa couplings}'', {\em Phys.Rev.} {\bfseries D81}
  (2010) 035016,
 \href{http://xxx.lanl.gov/abs/0907.1791}{{\ttfamily arXiv:0907.1791}}.
%%CITATION = ARXIV:0907.1791;%%.

\bibitem[Kagan and Neubert(2002)]{Kagan:2001zk}
A.~L. Kagan and M.~Neubert, ``{Isospin breaking in $B\to K^* \gamma$ decays}'',
  {\em Phys.Lett.} {\bfseries B539} (2002) 227--234,
 \href{http://xxx.lanl.gov/abs/hep-ph/0110078}{{\ttfamily
  arXiv:hep-ph/0110078}}.
%%CITATION = HEP-PH/0110078;%%.

\bibitem[Bosch and Buchalla(2002)]{Bosch:2001gv}
S.~W. Bosch and G.~Buchalla, ``{The Radiative decays $B\to V \gamma$ at
  next-to-leading order in QCD}'', {\em Nucl.Phys.} {\bfseries B621} (2002)
  459--478,
 \href{http://xxx.lanl.gov/abs/hep-ph/0106081}{{\ttfamily
  arXiv:hep-ph/0106081}}.
%%CITATION = HEP-PH/0106081;%%.

\bibitem[Lees et~al.(2012)]{Lees:2012xj}
{\bfseries BaBar Collaboration} Collaboration, J.~Lees {\em et~al.},
  ``{Evidence for an excess of $\overline{B} \to D^{(*)} \tau^-\bar{\nu}_\tau$
  decays}'', {\em Phys.Rev.Lett.} {\bfseries 109} (2012) 101802,
 \href{http://xxx.lanl.gov/abs/1205.5442}{{\ttfamily arXiv:1205.5442}}.
%%CITATION = ARXIV:1205.5442;%%.

\bibitem[Tanaka and Watanabe(2013)]{Tanaka:2012nw}
M.~Tanaka and R.~Watanabe, ``{New physics in the weak interaction of
  $\overline{B}\to D^{(*)}\tau\bar\nu$}'', {\em Phys.Rev.} {\bfseries D87}
  (2013), no.~3, 034028,
 \href{http://xxx.lanl.gov/abs/1212.1878}{{\ttfamily arXiv:1212.1878}}.
%%CITATION = ARXIV:1212.1878;%%.

\bibitem[Barbieri et~al.(2006)Barbieri, Hall, and Rychkov]{Barbieri:2006dq}
R.~Barbieri, L.~J. Hall, and V.~S. Rychkov, ``{Improved naturalness with a
  heavy Higgs: An Alternative road to LHC physics}'', {\em Phys.Rev.}
  {\bfseries D74} (2006) 015007,
 \href{http://xxx.lanl.gov/abs/hep-ph/0603188}{{\ttfamily
  arXiv:hep-ph/0603188}}.
%%CITATION = HEP-PH/0603188;%%.

\bibitem[Deshpande and Ma(1978)]{Deshpande:1977rw}
N.~G. Deshpande and E.~Ma, ``{Pattern of Symmetry Breaking with Two Higgs
  Doublets}'', {\em Phys.Rev.} {\bfseries D18} (1978)
2574.
%%CITATION = PHRVA,D18,2574;%%.

\bibitem[Lopez~Honorez et~al.(2007)Lopez~Honorez, Nezri, Oliver, and
  Tytgat]{LopezHonorez:2006gr}
L.~Lopez~Honorez, E.~Nezri, J.~F. Oliver, and M.~H. Tytgat, ``{The Inert
  Doublet Model: An Archetype for Dark Matter}'', {\em JCAP} {\bfseries 0702}
  (2007) 028,
 \href{http://xxx.lanl.gov/abs/hep-ph/0612275}{{\ttfamily
  arXiv:hep-ph/0612275}}.
%%CITATION = HEP-PH/0612275;%%.

\bibitem[Gustafsson et~al.(2012)Gustafsson, Rydbeck, Lopez-Honorez, and
  Lundstrom]{Gustafsson:2012aj}
M.~Gustafsson, S.~Rydbeck, L.~Lopez-Honorez, and E.~Lundstrom, ``{Status of the
  Inert Doublet Model and the Role of multileptons at the LHC}'', {\em
  Phys.Rev.} {\bfseries D86} (2012) 075019,
 \href{http://xxx.lanl.gov/abs/1206.6316}{{\ttfamily arXiv:1206.6316}}.
%%CITATION = ARXIV:1206.6316;%%.

\bibitem[Goudelis et~al.(2013)Goudelis, Herrmann, and
  St\r{a}l]{Goudelis:2013uca}
A.~Goudelis, B.~Herrmann, and O.~St\r{a}l, ``{Dark matter in the Inert Doublet
  Model after the discovery of a Higgs-like boson at the LHC}'',
 \href{http://xxx.lanl.gov/abs/1303.3010}{{\ttfamily arXiv:1303.3010}}.
%%CITATION = ARXIV:1303.3010;%%.

\bibitem[Hambye and Tytgat(2008)]{Hambye:2007vf}
T.~Hambye and M.~H. Tytgat, ``{Electroweak symmetry breaking induced by dark
  matter}'', {\em Phys.Lett.} {\bfseries B659} (2008) 651--655,
 \href{http://xxx.lanl.gov/abs/0707.0633}{{\ttfamily arXiv:0707.0633}}.
%%CITATION = ARXIV:0707.0633;%%.

\bibitem[Chowdhury et~al.(2012)Chowdhury, Nemevsek, Senjanovic, and
  Zhang]{Chowdhury:2011ga}
T.~A. Chowdhury, M.~Nemevsek, G.~Senjanovic, and Y.~Zhang, ``{Dark Matter as
  the Trigger of Strong Electroweak Phase Transition}'', {\em JCAP} {\bfseries
  1202} (2012) 029,
 \href{http://xxx.lanl.gov/abs/1110.5334}{{\ttfamily arXiv:1110.5334}}.
%%CITATION = ARXIV:1110.5334;%%.

\bibitem[Ma(2006)]{Ma:2006km}
E.~Ma, ``{Verifiable radiative seesaw mechanism of neutrino mass and dark
  matter}'', {\em Phys.Rev.} {\bfseries D73} (2006) 077301,
 \href{http://xxx.lanl.gov/abs/hep-ph/0601225}{{\ttfamily
  arXiv:hep-ph/0601225}}.
%%CITATION = HEP-PH/0601225;%%.

\bibitem[Ginzburg et~al.(2010)Ginzburg, Kanishev, Krawczyk, and
  Sokolowska]{Ginzburg:2010wa}
I.~Ginzburg, K.~Kanishev, M.~Krawczyk, and D.~Sokolowska, ``{Evolution of
  Universe to the present inert phase}'', {\em Phys.Rev.} {\bfseries D82}
  (2010) 123533,
 \href{http://xxx.lanl.gov/abs/1009.4593}{{\ttfamily arXiv:1009.4593}}.
%%CITATION = ARXIV:1009.4593;%%.

\bibitem[Cao et~al.(2007)Cao, Ma, and Rajasekaran]{Cao:2007rm}
Q.-H. Cao, E.~Ma, and G.~Rajasekaran, ``{Observing the Dark Scalar Doublet and
  its Impact on the Standard-Model Higgs Boson at Colliders}'', {\em Phys.Rev.}
  {\bfseries D76} (2007) 095011,
 \href{http://xxx.lanl.gov/abs/0708.2939}{{\ttfamily arXiv:0708.2939}}.
%%CITATION = ARXIV:0708.2939;%%.

\bibitem[Lundstrom et~al.(2009)Lundstrom, Gustafsson, and
  Edsjo]{Lundstrom:2008ai}
E.~Lundstrom, M.~Gustafsson, and J.~Edsjo, ``{The Inert Doublet Model and LEP
  II Limits}'', {\em Phys.Rev.} {\bfseries D79} (2009) 035013,
 \href{http://xxx.lanl.gov/abs/0810.3924}{{\ttfamily arXiv:0810.3924}}.
%%CITATION = ARXIV:0810.3924;%%.

\bibitem[Acciarri et~al.(2000)]{Acciarri:1999km}
{\bfseries L3 Collaboration} Collaboration, M.~Acciarri {\em et~al.}, ``{Search
  for charginos and neutralinos in $e^{+} e^{-}$ collisions at $\sqrt{s}$ =
  189-GeV}'', {\em Phys.Lett.} {\bfseries B472} (2000) 420--433,
 \href{http://xxx.lanl.gov/abs/hep-ex/9910007}{{\ttfamily
  arXiv:hep-ex/9910007}}.
%%CITATION = HEP-EX/9910007;%%.

\bibitem[Abbiendi et~al.(2000)]{Abbiendi:1999ar}
{\bfseries OPAL Collaboration} Collaboration, G.~Abbiendi {\em et~al.},
  ``{Search for chargino and neutralino production at $\sqrt{s} = 189$~GeV at
  LEP}'', {\em Eur.Phys.J.} {\bfseries C14} (2000) 187--198,
 \href{http://xxx.lanl.gov/abs/hep-ex/9909051}{{\ttfamily
  arXiv:hep-ex/9909051}}.
%%CITATION = HEP-EX/9909051;%%.

\bibitem[Abdallah et~al.(2003)]{Abdallah:2003xe}
{\bfseries DELPHI Collaboration} Collaboration, J.~Abdallah {\em et~al.},
  ``{Searches for supersymmetric particles in $e^+ e^-$ collisions up to
  208-GeV and interpretation of the results within the MSSM}'', {\em
  Eur.Phys.J.} {\bfseries C31} (2003) 421--479,
 \href{http://xxx.lanl.gov/abs/hep-ex/0311019}{{\ttfamily
  arXiv:hep-ex/0311019}}.
%%CITATION = HEP-EX/0311019;%%.

\bibitem[Dolle et~al.(2010)Dolle, Miao, Su, and Thomas]{Dolle:2009ft}
E.~Dolle, X.~Miao, S.~Su, and B.~Thomas, ``{Dilepton Signals in the Inert
  Doublet Model}'', {\em Phys.Rev.} {\bfseries D81} (2010) 035003,
 \href{http://xxx.lanl.gov/abs/0909.3094}{{\ttfamily arXiv:0909.3094}}.
%%CITATION = ARXIV:0909.3094;%%.

\bibitem[Miao et~al.(2010)Miao, Su, and Thomas]{Miao:2010rg}
X.~Miao, S.~Su, and B.~Thomas, ``{Trilepton Signals in the Inert Doublet
  Model}'', {\em Phys.Rev.} {\bfseries D82} (2010) 035009,
 \href{http://xxx.lanl.gov/abs/1005.0090}{{\ttfamily arXiv:1005.0090}}.
%%CITATION = ARXIV:1005.0090;%%.

\bibitem[Aoki et~al.(2013)Aoki, Kanemura, and Yokoya]{Aoki:2013lhm}
M.~Aoki, S.~Kanemura, and H.~Yokoya, ``{Reconstruction of Inert Doublet Scalars
  at the International Linear Collider}'', {\em Phys.Lett.} {\bfseries B725}
  (2013) 302--309,
 \href{http://xxx.lanl.gov/abs/1303.6191}{{\ttfamily arXiv:1303.6191}}.
%%CITATION = ARXIV:1303.6191;%%.

\bibitem[Haber(????)]{HaberPDG}
H.~E. Haber, ``{Supersymmetry, Part I (Theory), in Ref.~\cite{Beringer:1900zz},
  pp.~1420--1437}'',.

\bibitem[Haber and Mason(2008)]{Haber:2007dj}
H.~E. Haber and J.~D. Mason, ``{Hard supersymmetry-breaking 'wrong-Higgs'
  couplings of the MSSM}'', {\em Phys.Rev.} {\bfseries D77} (2008) 115011,
 \href{http://xxx.lanl.gov/abs/0711.2890}{{\ttfamily arXiv:0711.2890}}.
%%CITATION = ARXIV:0711.2890;%%.

\bibitem[Carena et~al.(2002)Carena, Haber, Logan, and Mrenna]{Carena:2001bg}
M.~S. Carena, H.~E. Haber, H.~E. Logan, and S.~Mrenna, ``{Distinguishing a MSSM
  Higgs boson from the SM Higgs boson at a linear collider}'', {\em Phys.Rev.}
  {\bfseries D65} (2002) 055005,
 \href{http://xxx.lanl.gov/abs/hep-ph/0106116}{{\ttfamily
  arXiv:hep-ph/0106116}}.
%%CITATION = HEP-PH/0106116;%%.

\bibitem[Haber and Hempfling(1991)]{Haber:1990aw}
H.~E. Haber and R.~Hempfling, ``{Can the mass of the lightest Higgs boson of
  the minimal supersymmetric model be larger than $m_Z$?}'', {\em
  Phys.Rev.Lett.} {\bfseries 66} (1991)
1815--1818.
%%CITATION = PRLTA,66,1815;%%.

\bibitem[Okada et~al.(1991)Okada, Yamaguchi, and Yanagida]{Okada:1990vk}
Y.~Okada, M.~Yamaguchi, and T.~Yanagida, ``{Upper bound of the lightest Higgs
  boson mass in the minimal supersymmetric standard model}'', {\em
  Prog.Theor.Phys.} {\bfseries 85} (1991)
1--6.
%%CITATION = PTPKA,85,1;%%.

\bibitem[Ellis et~al.(1991)Ellis, Ridolfi, and Zwirner]{Ellis:1990nz}
J.~R. Ellis, G.~Ridolfi, and F.~Zwirner, ``{Radiative corrections to the masses
  of supersymmetric Higgs bosons}'', {\em Phys.Lett.} {\bfseries B257} (1991)
83--91.
%%CITATION = PHLTA,B257,83;%%.

\bibitem[Inoue et~al.(1982)Inoue, Kakuto, Komatsu, and Takeshita]{Inoue:1982ej}
K.~Inoue, A.~Kakuto, H.~Komatsu, and S.~Takeshita, ``{Low-Energy Parameters and
  Particle Masses in a Supersymmetric Grand Unified Model}'', {\em
  Prog.Theor.Phys.} {\bfseries 67} (1982)
1889.
%%CITATION = PTPKA,67,1889;%%.

\bibitem[Flores and Sher(1983)]{Flores:1982pr}
R.~A. Flores and M.~Sher, ``{Higgs Masses in the Standard, Multi-Higgs and
  Supersymmetric Models}'', {\em Annals Phys.} {\bfseries 148} (1983)
95.
%%CITATION = APNYA,148,95;%%.

\bibitem[Gunion and Haber(1986)]{Gunion:1984yn}
J.~Gunion and H.~E. Haber, ``{Higgs Bosons in Supersymmetric Models. 1.}'',
  {\em Nucl.Phys.} {\bfseries B272} (1986)
1.
%%CITATION = NUPHA,B272,1;%%.

\bibitem[Degrassi et~al.(2003)Degrassi, Heinemeyer, Hollik, Slavich, and
  Weiglein]{Degrassi:2002fi}
G.~Degrassi, S.~Heinemeyer, W.~Hollik, P.~Slavich, and G.~Weiglein, ``{Towards
  high precision predictions for the MSSM Higgs sector}'', {\em Eur.Phys.J.}
  {\bfseries C28} (2003) 133--143,
 \href{http://xxx.lanl.gov/abs/hep-ph/0212020}{{\ttfamily
  arXiv:hep-ph/0212020}}.
%%CITATION = HEP-PH/0212020;%%.

\bibitem[Martin(2007)]{Martin:2007pg}
S.~P. Martin, ``{Three-loop corrections to the lightest Higgs scalar boson mass
  in supersymmetry}'', {\em Phys.Rev.} {\bfseries D75} (2007) 055005,
 \href{http://xxx.lanl.gov/abs/hep-ph/0701051}{{\ttfamily
  arXiv:hep-ph/0701051}}.
%%CITATION = HEP-PH/0701051;%%.

\bibitem[Kant et~al.(2010)Kant, Harlander, Mihaila, and
  Steinhauser]{Kant:2010tf}
P.~Kant, R.~Harlander, L.~Mihaila, and M.~Steinhauser, ``{Light MSSM Higgs
  boson mass to three-loop accuracy}'', {\em JHEP} {\bfseries 1008} (2010) 104,
 \href{http://xxx.lanl.gov/abs/1005.5709}{{\ttfamily arXiv:1005.5709}}.
%%CITATION = ARXIV:1005.5709;%%.

\bibitem[Carena et~al.(2000)Carena, Ellis, Pilaftsis, and
  Wagner]{Carena:2000ks}
M.~S. Carena, J.~R. Ellis, A.~Pilaftsis, and C.~Wagner, ``{CP violating MSSM
  Higgs bosons in the light of LEP-2}'', {\em Phys.Lett.} {\bfseries B495}
  (2000) 155--163,
 \href{http://xxx.lanl.gov/abs/hep-ph/0009212}{{\ttfamily
  arXiv:hep-ph/0009212}}.
%%CITATION = HEP-PH/0009212;%%.

\bibitem[Carena et~al.(2002)Carena, Ellis, Pilaftsis, and
  Wagner]{Carena:2001fw}
M.~S. Carena, J.~R. Ellis, A.~Pilaftsis, and C.~Wagner, ``{Higgs boson pole
  masses in the MSSM with explicit CP violation}'', {\em Nucl.Phys.} {\bfseries
  B625} (2002) 345--371,
 \href{http://xxx.lanl.gov/abs/hep-ph/0111245}{{\ttfamily
  arXiv:hep-ph/0111245}}.
%%CITATION = HEP-PH/0111245;%%.

\bibitem[Carena and Haber(2003)]{Carena:2002es}
M.~S. Carena and H.~E. Haber, ``{Higgs boson theory and phenomenology}'', {\em
  Prog.Part.Nucl.Phys.} {\bfseries 50} (2003) 63--152,
 \href{http://xxx.lanl.gov/abs/hep-ph/0208209}{{\ttfamily
  arXiv:hep-ph/0208209}}.
%%CITATION = HEP-PH/0208209;%%.

\bibitem[Accomando et~al.(2006)Accomando, Akeroyd, Akhmetzyanova, Albert,
  Alves, et~al.]{Accomando:2006ga}
E.~Accomando, A.~Akeroyd, E.~Akhmetzyanova, J.~Albert, A.~Alves, {\em et~al.},
  ``{Workshop on CP Studies and Non-Standard Higgs Physics}'',
 \href{http://xxx.lanl.gov/abs/hep-ph/0608079}{{\ttfamily
  arXiv:hep-ph/0608079}}.
%%CITATION = HEP-PH/0608079;%%.

\bibitem[Tsao(1980)]{Tsao:1980em}
H.-S. Tsao, ``{Higgs Boson Quantum Numbers and the Pell Equation}'', {\em in
  Proceedings of the 1980 Guangzhou Conference on Theoretical Particle Physics}
  {\bfseries edited by H.~Ning and T.~Hung-yuan, (Science Press, Beijing,
  1980)} (1980)
1240.
%%CITATION = COO-2232B-199 ETC.;%%.

\bibitem[Georgi and Machacek(1985)]{Georgi:1985nv}
H.~Georgi and M.~Machacek, ``{Doubly charged Higgs bosons}'', {\em Nucl.Phys.}
  {\bfseries B262} (1985)
463.
%%CITATION = NUPHA,B262,463;%%.

\bibitem[Gunion et~al.(1991)Gunion, Vega, and Wudka]{Gunion:1990dt}
J.~Gunion, R.~Vega, and J.~Wudka, ``{Naturalness problems for $\rho = 1$ and
  other large one loop effects for a standard model Higgs sector containing
  triplet fields}'', {\em Phys.Rev.} {\bfseries D43} (1991)
2322--2336.
%%CITATION = PHRVA,D43,2322;%%.

\bibitem[Hisano and Tsumura(2013)]{Hisano:2013sn}
J.~Hisano and K.~Tsumura, ``{The Higgs boson mixes with an SU(2) septet
  representation}'',
 \href{http://xxx.lanl.gov/abs/1301.6455}{{\ttfamily arXiv:1301.6455}}.
%%CITATION = ARXIV:1301.6455;%%.

\bibitem[Kanemura et~al.(2013)Kanemura, Kikuchi, and Yagyu]{Kanemura:2013mc}
S.~Kanemura, M.~Kikuchi, and K.~Yagyu, ``{Probing exotic Higgs sectors from the
  precise measurement of Higgs boson couplings}'',
 \href{http://xxx.lanl.gov/abs/1301.7303}{{\ttfamily arXiv:1301.7303}}.
%%CITATION = ARXIV:1301.7303;%%.

\bibitem[Chanowitz and Golden(1985)]{Chanowitz:1985ug}
M.~S. Chanowitz and M.~Golden, ``{Higgs boson triplets with
  $m_W=m_Z\cos\theta_W$}'', {\em Phys.Lett.} {\bfseries B165} (1985)
105.
%%CITATION = PHLTA,B165,105;%%.

\bibitem[Gunion et~al.(1990)Gunion, Vega, and Wudka]{Gunion:1989ci}
J.~Gunion, R.~Vega, and J.~Wudka, ``{Higgs triplets in the standard model}'',
  {\em Phys.Rev.} {\bfseries D42} (1990)
1673--1691.
%%CITATION = PHRVA,D42,1673;%%.

\bibitem[Khalil(2008)]{Khalil:2006yi}
S.~Khalil, ``{Low scale B$-$L extension of the Standard Model at the LHC}'',
  {\em J.Phys.} {\bfseries G35} (2008) 055001,
 \href{http://xxx.lanl.gov/abs/hep-ph/0611205}{{\ttfamily
  arXiv:hep-ph/0611205}}.
%%CITATION = HEP-PH/0611205;%%.

\bibitem[Ellwanger et~al.(2010)Ellwanger, Hugonie, and
  Teixeira]{Ellwanger:2009dp}
U.~Ellwanger, C.~Hugonie, and A.~M. Teixeira, ``{The Next-to-Minimal
  Supersymmetric Standard Model}'', {\em Phys.Rept.} {\bfseries 496} (2010)
  1--77,
 \href{http://xxx.lanl.gov/abs/0910.1785}{{\ttfamily arXiv:0910.1785}}.
%%CITATION = ARXIV:0910.1785;%%.

\bibitem[Basso et~al.(2011)Basso, Moretti, and Pruna]{Basso:2010si}
L.~Basso, S.~Moretti, and G.~M. Pruna, ``{The Higgs sector of the minimal B$-$L
  model at future Linear Colliders}'', {\em Eur.Phys.J.} {\bfseries C71} (2011)
  1724,
 \href{http://xxx.lanl.gov/abs/1012.0167}{{\ttfamily arXiv:1012.0167}}.
%%CITATION = ARXIV:1012.0167;%%.

\bibitem[Fileviez~Perez et~al.(2008)Fileviez~Perez, Han, Huang, Li, and
  Wang]{Perez:2008ha}
P.~Fileviez~Perez, T.~Han, G.-y. Huang, T.~Li, and K.~Wang, ``{Neutrino Masses
  and the CERN LHC: Testing Type II Seesaw}'', {\em Phys.Rev.} {\bfseries D78}
  (2008) 015018,
 \href{http://xxx.lanl.gov/abs/0805.3536}{{\ttfamily arXiv:0805.3536}}.
%%CITATION = ARXIV:0805.3536;%%.

\bibitem[Aoki et~al.(2012)Aoki, Kanemura, and Yagyu]{Aoki:2011pz}
M.~Aoki, S.~Kanemura, and K.~Yagyu, ``{Testing the Higgs triplet model with the
  mass difference at the LHC}'', {\em Phys.Rev.} {\bfseries D85} (2012) 055007,
 \href{http://xxx.lanl.gov/abs/1110.4625}{{\ttfamily arXiv:1110.4625}}.
%%CITATION = ARXIV:1110.4625;%%.

\bibitem[Aad et~al.(2012)]{ATLAS:2012hi}
{\bfseries ATLAS Collaboration} Collaboration, G.~Aad {\em et~al.}, ``{Search
  for doubly-charged Higgs bosons in like-sign dilepton final states at
  $\sqrt{s}=7$ TeV with the ATLAS detector}'', {\em Eur.Phys.J.} {\bfseries
  C72} (2012) 2244,
 \href{http://xxx.lanl.gov/abs/1210.5070}{{\ttfamily arXiv:1210.5070}}.
%%CITATION = ARXIV:1210.5070;%%.

\bibitem[Chatrchyan et~al.(2012)]{Chatrchyan:2012ya}
{\bfseries CMS Collaboration} Collaboration, S.~Chatrchyan {\em et~al.}, ``{A
  search for a doubly-charged Higgs boson in $pp$ collisions at $\sqrt{s}=7$
  TeV}'', {\em Eur.Phys.J.} {\bfseries C72} (2012) 2189,
 \href{http://xxx.lanl.gov/abs/1207.2666}{{\ttfamily arXiv:1207.2666}}.
%%CITATION = ARXIV:1207.2666;%%.

\bibitem[Kanemura et~al.(2013)Kanemura, Yagyu, and Yokoya]{Kanemura:2013vxa}
S.~Kanemura, K.~Yagyu, and H.~Yokoya, ``{First constraint on the mass of
  doubly-charged Higgs bosons in the same-sign diboson decay scenario at the
  LHC}'',
 \href{http://xxx.lanl.gov/abs/1305.2383}{{\ttfamily arXiv:1305.2383}}.
%%CITATION = ARXIV:1305.2383;%%.

\bibitem[Maniatis(2010)]{Maniatis:2009re}
M.~Maniatis, ``{The Next-to-Minimal Supersymmetric extension of the Standard
  Model reviewed}'', {\em Int.J.Mod.Phys.} {\bfseries A25} (2010) 3505--3602,
 \href{http://xxx.lanl.gov/abs/0906.0777}{{\ttfamily arXiv:0906.0777}}.
%%CITATION = ARXIV:0906.0777;%%.

\bibitem[Ross and Schmidt-Hoberg(2012)]{Ross:2011xv}
G.~G. Ross and K.~Schmidt-Hoberg, ``{The Fine-Tuning of the Generalised
  NMSSM}'', {\em Nucl.Phys.} {\bfseries B862} (2012) 710--719,
 \href{http://xxx.lanl.gov/abs/1108.1284}{{\ttfamily arXiv:1108.1284}}.
%%CITATION = ARXIV:1108.1284;%%.

\bibitem[Ellwanger and Hugonie(2007)]{Ellwanger:2006rm}
U.~Ellwanger and C.~Hugonie, ``{The Upper bound on the lightest Higgs mass in
  the NMSSM revisited}'', {\em Mod.Phys.Lett.} {\bfseries A22} (2007)
  1581--1590,
 \href{http://xxx.lanl.gov/abs/hep-ph/0612133}{{\ttfamily
  arXiv:hep-ph/0612133}}.
%%CITATION = HEP-PH/0612133;%%.

\bibitem[Hall et~al.(2012)Hall, Pinner, and Ruderman]{Hall:2011aa}
L.~J. Hall, D.~Pinner, and J.~T. Ruderman, ``{A Natural SUSY Higgs Near 126
  GeV}'', {\em JHEP} {\bfseries 1204} (2012) 131,
 \href{http://xxx.lanl.gov/abs/1112.2703}{{\ttfamily arXiv:1112.2703}}.
%%CITATION = ARXIV:1112.2703;%%.

\bibitem[Buchmuller and Wyler(1986)]{Buchmuller:1985jz}
W.~Buchmuller and D.~Wyler, ``{Effective Lagrangian Analysis of New
  Interactions and Flavor Conservation}'', {\em Nucl.Phys.} {\bfseries B268}
  (1986)
621.
%%CITATION = NUPHA,B268,621;%%.

\bibitem[Grzadkowski et~al.(2010)Grzadkowski, Iskrzynski, Misiak, and
  Rosiek]{Grzadkowski:2010es}
B.~Grzadkowski, M.~Iskrzynski, M.~Misiak, and J.~Rosiek, ``{Dimension-Six Terms
  in the Standard Model Lagrangian}'', {\em JHEP} {\bfseries 1010} (2010) 085,
 \href{http://xxx.lanl.gov/abs/1008.4884}{{\ttfamily arXiv:1008.4884}}.
%%CITATION = ARXIV:1008.4884;%%.

\bibitem[David et~al.(2012)]{LHCHiggsCrossSectionWorkingGroup:2012nn}
{\bfseries LHC Higgs Cross Section Working Group} Collaboration, A.~David {\em
  et~al.}, ``{LHC HXSWG interim recommendations to explore the coupling
  structure of a Higgs-like particle}'',
 \href{http://xxx.lanl.gov/abs/1209.0040}{{\ttfamily arXiv:1209.0040}}.
%%CITATION = ARXIV:1209.0040;%%.

\bibitem[Duhrssen et~al.(2004)Duhrssen, Heinemeyer, Logan, Rainwater, Weiglein,
  et~al.]{Duhrssen:2004cv}
M.~Duhrssen, S.~Heinemeyer, H.~Logan, D.~Rainwater, G.~Weiglein, {\em et~al.},
  ``{Extracting Higgs boson couplings from CERN LHC data}'', {\em Phys.Rev.}
  {\bfseries D70} (2004) 113009,
 \href{http://xxx.lanl.gov/abs/hep-ph/0406323}{{\ttfamily
  arXiv:hep-ph/0406323}}.
%%CITATION = HEP-PH/0406323;%%.

\bibitem[Weisskopf(1939)]{Weisskopf:1939zz}
V.~Weisskopf, ``{On the Self-Energy and the Electromagnetic Field of the
  Electron}'', {\em Phys.Rev.} {\bfseries 56} (1939)
72--85.
%%CITATION = PHRVA,56,72;%%.

\bibitem[Susskind(1984)]{Susskind:1982mw}
L.~Susskind, ``{The Gauge Hierarchy Problem, Technicolor, Supersymmetry, and
  all that}'', {\em Phys.Rept.} {\bfseries 104} (1984)
181--193.
%%CITATION = PRPLC,104,181;%%.

\bibitem[Maiani(1979)]{Maiani:1979cx}
L.~Maiani, ``{All You Need to Know about the Higgs Boson}'', {\em Conf.Proc.}
  {\bfseries C7909031} (1979)
1--52.
%%CITATION = INSPIRE-148587;%%.

\bibitem[Witten(1981)]{Witten:1981nf}
E.~Witten, ``{Dynamical Breaking of Supersymmetry}'', {\em Nucl.Phys.}
  {\bfseries B188} (1981)
513.
%%CITATION = NUPHA,B188,513;%%.

\bibitem[Dimopoulos and Georgi(1981)]{Dimopoulos:1981zb}
S.~Dimopoulos and H.~Georgi, ``{Softly Broken Supersymmetry and SU(5)}'', {\em
  Nucl.Phys.} {\bfseries B193} (1981)
150.
%%CITATION = NUPHA,B193,150;%%.

\bibitem[Sakai(1981)]{Sakai:1981gr}
N.~Sakai, ``{Naturalness in Supersymmetric Guts}'', {\em Z.Phys.} {\bfseries
  C11} (1981)
153.
%%CITATION = ZEPYA,C11,153;%%.

\bibitem[Kaul(1982)]{Kaul:1981wp}
R.~K. Kaul, ``{Gauge Hierarchy in a Supersymmetric Model}'', {\em Phys.Lett.}
  {\bfseries B109} (1982)
19.
%%CITATION = PHLTA,B109,19;%%.

\bibitem[Kaul and Majumdar(1982)]{Kaul:1981hi}
R.~K. Kaul and P.~Majumdar, ``{Cancellation of Quadratically Divergent Mass
  Corrections in Globally Supersymmetric Spontaneously Broken Gauge
  Theories}'', {\em Nucl.Phys.} {\bfseries B199} (1982)
36.
%%CITATION = NUPHA,B199,36;%%.

\bibitem[Bhattacharyya(2011)]{Bhattacharyya:2009gw}
G.~Bhattacharyya, ``{A Pedagogical Review of Electroweak Symmetry Breaking
  Scenarios}'', {\em Rept.Prog.Phys.} {\bfseries 74} (2011) 026201,
 \href{http://xxx.lanl.gov/abs/0910.5095}{{\ttfamily arXiv:0910.5095}}.
%%CITATION = ARXIV:0910.5095;%%.

\bibitem[Farhi and Susskind(1981)]{Farhi:1980xs}
E.~Farhi and L.~Susskind, ``{Technicolor}'', {\em Phys.Rept.} {\bfseries 74}
  (1981)
277.
%%CITATION = PRPLC,74,277;%%.

\bibitem[Kaul(1983)]{Kaul:1981uk}
R.~K. Kaul, ``{Technicolor}'', {\em Rev.Mod.Phys.} {\bfseries 55} (1983)
449.
%%CITATION = RMPHA,55,449;%%.

\bibitem[Hill and Simmons(2003)]{Hill:2002ap}
C.~T. Hill and E.~H. Simmons, ``{Strong dynamics and electroweak symmetry
  breaking}'', {\em Phys.Rept.} {\bfseries 381} (2003) 235--402,
 \href{http://xxx.lanl.gov/abs/hep-ph/0203079}{{\ttfamily
  arXiv:hep-ph/0203079}}.
%%CITATION = HEP-PH/0203079;%%.

\bibitem[Eichten et~al.(2012)Eichten, Lane, and Martin]{Eichten:2012qb}
E.~Eichten, K.~Lane, and A.~Martin, ``{A Higgs Impostor in Low-Scale
  Technicolor}'',
 \href{http://xxx.lanl.gov/abs/1210.5462}{{\ttfamily arXiv:1210.5462}}.
%%CITATION = ARXIV:1210.5462;%%.

\bibitem[Foadi et~al.(2013)Foadi, Frandsen, and Sannino]{Foadi:2012bb}
R.~Foadi, M.~T. Frandsen, and F.~Sannino, ``{125 GeV Higgs from a not so light
  Technicolor Scalar}'', {\em Phys.Rev.} {\bfseries D87} (2013) 095001,
 \href{http://xxx.lanl.gov/abs/1211.1083}{{\ttfamily arXiv:1211.1083}}.
%%CITATION = ARXIV:1211.1083;%%.

\bibitem[Georgi and Pais(1975)]{Georgi:1975tz}
H.~Georgi and A.~Pais, ``{Vacuum Symmetry and the PseudoGoldstone
  Phenomenon}'', {\em Phys.Rev.} {\bfseries D12} (1975)
508.
%%CITATION = PHRVA,D12,508;%%.

\bibitem[Georgi and Kaplan(1984)]{Georgi:1984af}
H.~Georgi and D.~B. Kaplan, ``{Composite Higgs and Custodial SU(2)}'', {\em
  Phys.Lett.} {\bfseries B145} (1984)
216.
%%CITATION = PHLTA,B145,216;%%.

\bibitem[Dugan et~al.(1985)Dugan, Georgi, and Kaplan]{Dugan:1984hq}
M.~J. Dugan, H.~Georgi, and D.~B. Kaplan, ``{Anatomy of a Composite Higgs
  Model}'', {\em Nucl.Phys.} {\bfseries B254} (1985)
299.
%%CITATION = NUPHA,B254,299;%%.

\bibitem[Arkani-Hamed et~al.(2002)Arkani-Hamed, Cohen, Gregoire, and
  Wacker]{ArkaniHamed:2002pa}
N.~Arkani-Hamed, A.~G. Cohen, T.~Gregoire, and J.~G. Wacker, ``{Phenomenology
  of electroweak symmetry breaking from theory space}'', {\em JHEP} {\bfseries
  0208} (2002) 020,
 \href{http://xxx.lanl.gov/abs/hep-ph/0202089}{{\ttfamily
  arXiv:hep-ph/0202089}}.
%%CITATION = HEP-PH/0202089;%%.

\bibitem[Schmaltz and Tucker-Smith(2005)]{Schmaltz:2005ky}
M.~Schmaltz and D.~Tucker-Smith, ``{Little Higgs review}'', {\em
  Ann.Rev.Nucl.Part.Sci.} {\bfseries 55} (2005) 229--270,
 \href{http://xxx.lanl.gov/abs/hep-ph/0502182}{{\ttfamily
  arXiv:hep-ph/0502182}}.
%%CITATION = HEP-PH/0502182;%%.

\bibitem[Chen(2006)]{Chen:2006dy}
M.-C. Chen, ``{Models of little Higgs and electroweak precision tests}'', {\em
  Mod.Phys.Lett.} {\bfseries A21} (2006) 621--638,
 \href{http://xxx.lanl.gov/abs/hep-ph/0601126}{{\ttfamily
  arXiv:hep-ph/0601126}}.
%%CITATION = HEP-PH/0601126;%%.

\bibitem[Perelstein(2007)]{Perelstein:2005ka}
M.~Perelstein, ``{Little Higgs models and their phenomenology}'', {\em
  Prog.Part.Nucl.Phys.} {\bfseries 58} (2007) 247--291,
 \href{http://xxx.lanl.gov/abs/hep-ph/0512128}{{\ttfamily
  arXiv:hep-ph/0512128}}.
%%CITATION = HEP-PH/0512128;%%.

\bibitem[Cheng and Low(2003)]{Cheng:2003ju}
H.-C. Cheng and I.~Low, ``{TeV symmetry and the little hierarchy problem}'',
  {\em JHEP} {\bfseries 0309} (2003) 051,
 \href{http://xxx.lanl.gov/abs/hep-ph/0308199}{{\ttfamily
  arXiv:hep-ph/0308199}}.
%%CITATION = HEP-PH/0308199;%%.

\bibitem[Cheng and Low(2004)]{Cheng:2004yc}
H.-C. Cheng and I.~Low, ``{Little hierarchy, little Higgses, and a little
  symmetry}'', {\em JHEP} {\bfseries 0408} (2004) 061,
 \href{http://xxx.lanl.gov/abs/hep-ph/0405243}{{\ttfamily
  arXiv:hep-ph/0405243}}.
%%CITATION = HEP-PH/0405243;%%.

\bibitem[Contino(2010)]{Contino:2010rs}
R.~Contino, ``{The Higgs as a Composite Nambu-Goldstone Boson}'',
 \href{http://xxx.lanl.gov/abs/1005.4269}{{\ttfamily arXiv:1005.4269}}.
%%CITATION = ARXIV:1005.4269;%%.

\bibitem[Serone(2010)]{Serone:2009kf}
M.~Serone, ``{Holographic Methods and Gauge-Higgs Unification in Flat Extra
  Dimensions}'', {\em New J.Phys.} {\bfseries 12} (2010) 075013,
 \href{http://xxx.lanl.gov/abs/0909.5619}{{\ttfamily arXiv:0909.5619}}.
%%CITATION = ARXIV:0909.5619;%%.

\bibitem[Pomarol and Riva(2012)]{Pomarol:2012qf}
A.~Pomarol and F.~Riva, ``{The Composite Higgs and Light Resonance
  Connection}'', {\em JHEP} {\bfseries 1208} (2012) 135,
 \href{http://xxx.lanl.gov/abs/1205.6434}{{\ttfamily arXiv:1205.6434}}.
%%CITATION = ARXIV:1205.6434;%%.

\bibitem[Cs\'aki et~al.(2013)Cs\'aki, Grojean, and Terning]{Terning:2013}
C.~Cs\'aki, C.~Grojean, and J.~Terning, ``{Alternatives to an Elementary
  Higgs}'', {\em In preparation}, 2013.

\bibitem[Bellazzini et~al.(2013)Bellazzini, Csaki, Hubisz, Serra, and
  Terning]{Bellazzini:2012vz}
B.~Bellazzini, C.~Csaki, J.~Hubisz, J.~Serra, and J.~Terning, ``{A Higgslike
  Dilaton}'', {\em Eur.Phys.J.} {\bfseries C73} (2013) 2333,
 \href{http://xxx.lanl.gov/abs/1209.3299}{{\ttfamily arXiv:1209.3299}}.
%%CITATION = ARXIV:1209.3299;%%.

\bibitem[Chacko et~al.(2013)Chacko, Franceschini, and Mishra]{Chacko:2012vm}
Z.~Chacko, R.~Franceschini, and R.~K. Mishra, ``{Resonance at 125 GeV: Higgs or
  Dilaton/Radion?}'', {\em JHEP} {\bfseries 1304} (2013) 015,
 \href{http://xxx.lanl.gov/abs/1209.3259}{{\ttfamily arXiv:1209.3259}}.
%%CITATION = ARXIV:1209.3259;%%.

\bibitem[Landau(1948)]{Landau:1948kw}
L.~Landau, ``{On the angular momentum of a two-photon system}'', {\em
  Dokl.Akad.Nauk Ser.Fiz.} {\bfseries 60} (1948)
207--209.
%%CITATION = DANKA,60,207;%%.

\bibitem[Yang(1950)]{Yang:1950rg}
C.-N. Yang, ``{Selection Rules for the Dematerialization of a Particle Into Two
  Photons}'', {\em Phys.Rev.} {\bfseries 77} (1950)
242--245.
%%CITATION = PHRVA,77,242;%%.

\bibitem[Ralston(2012)]{Ralston:2012ye}
J.~P. Ralston, ``{The Need to Fairly Confront Spin-1 for the New Higgs-like
  Particle}'',
 \href{http://xxx.lanl.gov/abs/1211.2288}{{\ttfamily arXiv:1211.2288}}.
%%CITATION = ARXIV:1211.2288;%%.

\bibitem[CMS(????)]{CMSprojections}
Projections of Higgs Boson measurements with 30 fb$^{-1}$ at 8 TeV and 300
  fb$^{-1}$ at 14 TeV,
  \href{https://twiki.cern.ch/twiki/bin/view/CMSPublic/HigProjectionEsg2012TWiki}.

\bibitem[Dixon and Siu(2003)]{Dixon:2003yb}
L.~J. Dixon and M.~S. Siu, ``{Resonance continuum interference in the diphoton
  Higgs signal at the LHC}'', {\em Phys.Rev.Lett.} {\bfseries 90} (2003)
  252001,
 \href{http://xxx.lanl.gov/abs/hep-ph/0302233}{{\ttfamily
  arXiv:hep-ph/0302233}}.
%%CITATION = HEP-PH/0302233;%%.

\bibitem[Martin(2012)]{Martin:2012xc}
S.~P. Martin, ``{Shift in the LHC Higgs diphoton mass peak from interference
  with background}'', {\em Phys.Rev.} {\bfseries D86} (2012) 073016,
 \href{http://xxx.lanl.gov/abs/1208.1533}{{\ttfamily arXiv:1208.1533}}.
%%CITATION = ARXIV:1208.1533;%%.

\bibitem[Martin(2013)]{Martin:2013ula}
S.~P. Martin, ``{Interference of Higgs diphoton signal and background in
  production with a jet at the LHC}'', {\em Phys.Rev.} {\bfseries D88} (2013)
  013004,
 \href{http://xxx.lanl.gov/abs/1303.3342}{{\ttfamily arXiv:1303.3342}}.
%%CITATION = ARXIV:1303.3342;%%.

\bibitem[Dixon and Li(2013)]{Dixon:2013haa}
L.~J. Dixon and Y.~Li, ``{Bounding the Higgs Boson Width Through
  Interferometry}'',
 \href{http://xxx.lanl.gov/abs/1305.3854}{{\ttfamily arXiv:1305.3854}}.
%%CITATION = ARXIV:1305.3854;%%.

\bibitem[Caola and Melnikov(2013)]{Caola:2013yja}
F.~Caola and K.~Melnikov, ``{Constraining the Higgs boson width with $ZZ$
  production at the LHC}'', {\em Phys.Rev.} {\bfseries D88} (2013) 054024,
 \href{http://xxx.lanl.gov/abs/1307.4935}{{\ttfamily arXiv:1307.4935}}.
%%CITATION = ARXIV:1307.4935;%%.

\bibitem[Carena et~al.(2013)Carena, Heinemeyer, St\r{a}l, Wagner, and
  Weiglein]{Carena:2013qia}
M.~Carena, S.~Heinemeyer, O.~St\r{a}l, C.~Wagner, and G.~Weiglein, ``{MSSM
  Higgs Boson Searches at the LHC: Benchmark Scenarios after the Discovery of a
  Higgs-like Particle}'',
 \href{http://xxx.lanl.gov/abs/1302.7033}{{\ttfamily arXiv:1302.7033}}.
%%CITATION = ARXIV:1302.7033;%%.

\bibitem[Gupta et~al.(2012)Gupta, Rzehak, and Wells]{Gupta:2012mi}
R.~S. Gupta, H.~Rzehak, and J.~D. Wells, ``{How well do we need to measure
  Higgs boson couplings?}'', {\em Phys.Rev.} {\bfseries D86} (2012) 095001,
 \href{http://xxx.lanl.gov/abs/1206.3560}{{\ttfamily arXiv:1206.3560}}.
%%CITATION = ARXIV:1206.3560;%%.

\bibitem[Giardino et~al.(2013)Giardino, Kannike, Masina, Raidal, and
  Strumia]{Giardino:2013bma}
P.~P. Giardino, K.~Kannike, I.~Masina, M.~Raidal, and A.~Strumia, ``{The
  universal Higgs fit}'',
 \href{http://xxx.lanl.gov/abs/1303.3570}{{\ttfamily arXiv:1303.3570}}.
%%CITATION = ARXIV:1303.3570;%%.

\bibitem[Collaboration(2012)]{CMS:2012zoa}
{\bfseries CMS Collaboration} Collaboration, T.~C. Collaboration, ``{Jet Energy
  Scale performance in 2011}'',
2012.
%%CITATION = CMS-DP-2012-006 ETC.;%%.

\bibitem[Kanemura et~al.(1993)Kanemura, Kubota, and Takasugi]{Kanemura:1993hm}
S.~Kanemura, T.~Kubota, and E.~Takasugi, ``{Lee-Quigg-Thacker bounds for Higgs
  boson masses in a two doublet model}'', {\em Phys.Lett.} {\bfseries B313}
  (1993) 155--160,
 \href{http://xxx.lanl.gov/abs/hep-ph/9303263}{{\ttfamily
  arXiv:hep-ph/9303263}}.
%%CITATION = HEP-PH/9303263;%%.

\bibitem[Dittmaier et~al.(2011)Dittmaier, Mariotti, Passarino, Tanaka,
  et~al.]{Dittmaier:2011ti}
{\bfseries LHC Higgs Cross Section Working Group} Collaboration, S.~Dittmaier,
  C.~Mariotti, G.~Passarino, R.~Tanaka, {\em et~al.}, ``{Handbook of LHC Higgs
  Cross Sections: 1. Inclusive Observables}'',
 \href{http://xxx.lanl.gov/abs/1101.0593}{{\ttfamily arXiv:1101.0593}}.
%%CITATION = ARXIV:1101.0593;%%.

\bibitem[Aad et~al.(2009)]{Aad:2009wy}
{\bfseries ATLAS Collaboration} Collaboration, G.~Aad {\em et~al.}, ``{Expected
  Performance of the ATLAS Experiment---Detector, Trigger and Physics}'',
 \href{http://xxx.lanl.gov/abs/0901.0512}{{\ttfamily arXiv:0901.0512}}.
%%CITATION = ARXIV:0901.0512;%%.

\bibitem[Fields and Sarkar(????)]{FieldsPDG}
B.~Fields and S.~Sarkar, ``{Big-Bang Nucleosynthesis, in
  Ref.~\cite{Beringer:1900zz}, pp.~275--279}'',.

\bibitem[Sakharov(1967)]{Sakharov:1967dj}
A.~Sakharov, ``{Violation of CP Invariance, C Asymmetry, and Baryon Asymmetry
  of the Universe}'', {\em Pisma Zh.Eksp.Teor.Fiz.} {\bfseries 5} (1967)
32--35.
%%CITATION = ZFPRA,5,32;%%.

\bibitem[Kuzmin et~al.(1985)Kuzmin, Rubakov, and Shaposhnikov]{Kuzmin:1985mm}
V.~Kuzmin, V.~Rubakov, and M.~Shaposhnikov, ``{On the Anomalous Electroweak
  Baryon Number Nonconservation in the Early Universe}'', {\em Phys.Lett.}
  {\bfseries B155} (1985)
36.
%%CITATION = PHLTA,B155,36;%%.

\bibitem[Fromme et~al.(2006)Fromme, Huber, and Seniuch]{Fromme:2006cm}
L.~Fromme, S.~J. Huber, and M.~Seniuch, ``{Baryogenesis in the two-Higgs
  doublet model}'', {\em JHEP} {\bfseries 0611} (2006) 038,
 \href{http://xxx.lanl.gov/abs/hep-ph/0605242}{{\ttfamily
  arXiv:hep-ph/0605242}}.
%%CITATION = HEP-PH/0605242;%%.

\bibitem[Grojean et~al.(2005)Grojean, Servant, and Wells]{Grojean:2004xa}
C.~Grojean, G.~Servant, and J.~D. Wells, ``{First-order electroweak phase
  transition in the standard model with a low cutoff}'', {\em Phys.Rev.}
  {\bfseries D71} (2005) 036001,
 \href{http://xxx.lanl.gov/abs/hep-ph/0407019}{{\ttfamily
  arXiv:hep-ph/0407019}}.
%%CITATION = HEP-PH/0407019;%%.

\bibitem[Aoki et~al.(2009)Aoki, Kanemura, and Seto]{Aoki:2008av}
M.~Aoki, S.~Kanemura, and O.~Seto, ``{Neutrino mass, dark matter and baryon
  asymmetry via TeV-scale physics without fine-tuning}'', {\em Phys.Rev.Lett.}
  {\bfseries 102} (2009) 051805,
 \href{http://xxx.lanl.gov/abs/0807.0361}{{\ttfamily arXiv:0807.0361}}.
%%CITATION = ARXIV:0807.0361;%%.

\bibitem[Kanemura et~al.(2013)Kanemura, Senaha, Shindou, and
  Yamada]{Kanemura:2012hr}
S.~Kanemura, E.~Senaha, T.~Shindou, and T.~Yamada, ``{Electroweak phase
  transition and Higgs boson couplings in the model based on supersymmetric
  strong dynamics}'', {\em JHEP} {\bfseries 1305} (2013) 066,
 \href{http://xxx.lanl.gov/abs/1211.5883}{{\ttfamily arXiv:1211.5883}}.
%%CITATION = ARXIV:1211.5883;%%.

\bibitem[Ross and Walker(2013)]{Ross:2013aa}
M.~Ross and N.~Walker, ``{Private communication}'', 2013.

\bibitem[Barish et~al.(2012)]{Barish:2012es}
B.~Barish {\em et~al.}, ``{International Linear Collider Submission to European
  Strategy Discussion}'', 2012.

\bibitem[Will et~al.(2001)Will, Quast, Redlin, and Sandner]{Will:2001ha}
I.~Will, T.~Quast, H.~Redlin, and W.~Sandner, ``{A laser system for the TESLA
  photon collider based on an external ring resonator}'', {\em
  Nucl.Instrum.Meth.} {\bfseries A472} (2001)
79--85.
%%CITATION = NUIMA,A472,79;%%.

\bibitem[Variola(2011)]{Variola:2011zb}
A.~Variola, ``{The ThomX Project}'', {\em Conf.Proc.} {\bfseries C110904}
  (2011)
1903--1905.
%%CITATION = CONFP,C110904,1903;%%.

\bibitem[Fukuda et~al.(2010)Fukuda, Araki, Aryshev, Honda, Terunuma,
  et~al.]{Fukuda:2010zzb}
M.~Fukuda, S.~Araki, A.~Aryshev, Y.~Honda, N.~Terunuma, {\em et~al.}, ``{Status
  and Future Plan of the Accelerator for Laser Undulator Compact X-ray Source
  LUCX}'', {\em Conf.Proc.} {\bfseries C100523} (2010)
TUPD089.
%%CITATION = CONFP,C100523,TUPD089;%%.

\bibitem[Hartemann et~al.(2010)Hartemann, Albert, Anderson, Barty, Bayramian,
  et~al.]{Hartemann:2010zz}
F.~Hartemann, F.~Albert, S.~Anderson, C.~Barty, A.~Bayramian, {\em et~al.},
  ``{Overview of Mono-energetic Gamma-ray Sources and Applications}'', {\em
  Conf.Proc.} {\bfseries C100523} (2010)
TUPD098.
%%CITATION = CONFP,C100523,TUPD098;%%.

\bibitem[Delerue et~al.(2011)Delerue, Bonis, Chaikovska, Chiche, Cizeron,
  et~al.]{Delerue:2011nk}
N.~Delerue, J.~Bonis, I.~Chaikovska, R.~Chiche, R.~Cizeron, {\em et~al.},
  ``{High flux polarized gamma rays production: first measurements with a
  four-mirror cavity at the ATF}'', {\em Conf.Proc.} {\bfseries C110904} (2011)
  1446--1448,
 \href{http://xxx.lanl.gov/abs/1110.3243}{{\ttfamily arXiv:1110.3243}}.
%%CITATION = ARXIV:1110.3243;%%.

\bibitem[Gaede(2006)]{Gaede:2006pj}
F.~Gaede, ``{Marlin and LCCD: Software tools for the ILC}'', {\em
  Nucl.Instrum.Meth.} {\bfseries A559} (2006)
177--180.
%%CITATION = NUIMA,A559,177;%%.

\bibitem[Marshall and Thomson(2013)]{Marshall:2013bda}
J.~Marshall and M.~Thomson, ``{The Pandora Particle Flow Algorithm}'',
 \href{http://xxx.lanl.gov/abs/1308.4537}{{\ttfamily arXiv:1308.4537}}.
%%CITATION = ARXIV:1308.4537;%%.

\bibitem[LCF(????)]{LCFIPlus}
LCFIPlus,
  \href{https://confluence.slac.stanford.edu/display/ilc/LCFIPlus}{https://confluence.slac.stanford.edu/display/ilc/LCFIPlus}.

\bibitem[Aihara et~al.(2009)Aihara, Burrows, Oreglia, et~al.]{Aihara:2009ad}
E.~Aihara, H., E.~Burrows, P., E.~Oreglia, M., {\em et~al.}, ``{SiD Letter of
  Intent}'',  \href{http://xxx.lanl.gov/abs/0911.0006}{{\ttfamily
  arXiv:0911.0006}},
{\href{http://arxiv.org/pdf/0911.0006v1}{arXiv:0911.0006}, SLAC-R-944}.
%%CITATION = ARXIV:0911.0006;%%.

\bibitem[ild(????)]{ild_loi_higgs_br}
Higgs Hadronic Branching Ratios in the $ZH\rightarrow llqq$ channel,
  \href{http://ilcild.org/documents/ild-letter-of-intent/ild-loi-material/higgs\_branching\_ratios.pdf/at\_download/file}{http://ilcild.org/documents/ild-letter-of-intent/ild-loi-material/higgs\_branching\_ratios.pdf/at\_download/file}.

\bibitem[Abe et~al.(2010)]{Abe:2010aa}
{\bfseries ILD Concept Group---Linear Collider Collaboration} Collaboration,
  T.~Abe {\em et~al.}, ``{The International Large Detector: Letter of
  Intent}'',
 \href{http://xxx.lanl.gov/abs/1006.3396}{{\ttfamily arXiv:1006.3396}}.
%%CITATION = ARXIV:1006.3396;%%.

\bibitem[Li et~al.(2012)]{Li:2012taa}
{\bfseries ILD Design Study Group} Collaboration, H.~Li {\em et~al.}, ``{$HZ$
  Recoil Mass and Cross Section Analysis in ILD}'',
 \href{http://xxx.lanl.gov/abs/1202.1439}{{\ttfamily arXiv:1202.1439}}.
%%CITATION = ARXIV:1202.1439;%%.

\bibitem[Miyamoto(2013)]{Miyamoto:2013zva}
A.~Miyamoto, ``{A measurement of the total cross section of $\sigma_{Zh}$ at a
  future $e^{+}e^{-}$ collider using the hadronic decay mode of $Z$}'',
 \href{http://xxx.lanl.gov/abs/1311.2248}{{\ttfamily arXiv:1311.2248}}.
%%CITATION = ARXIV:1311.2248;%%.

\bibitem[Dova et~al.(2003)Dova, Garcia-Abia, and Lohmann]{Dova:2003py}
M.~Dova, P.~Garcia-Abia, and W.~Lohmann, ``{Determination of the Higgs boson
  spin with a linear $e^+ e^-$ collider}'',
 \href{http://xxx.lanl.gov/abs/hep-ph/0302113}{{\ttfamily
  arXiv:hep-ph/0302113}}.
%%CITATION = HEP-PH/0302113;%%.

\bibitem[Miller et~al.(2001)Miller, Choi, Eberle, Muhlleitner, and
  Zerwas]{Miller:2001bi}
.~Miller, D.J., S.~Choi, B.~Eberle, M.~Muhlleitner, and P.~Zerwas, ``{Measuring
  the spin of the Higgs boson}'', {\em Phys.Lett.} {\bfseries B505} (2001)
  149--154,
 \href{http://xxx.lanl.gov/abs/hep-ph/0102023}{{\ttfamily
  arXiv:hep-ph/0102023}}.
%%CITATION = HEP-PH/0102023;%%.

\bibitem[Schumacher(2001)]{Schumacher:2001ax}
M.~Schumacher, ``{Determination of the CP quantum numbers of the Higgs boson
  and test of CP invariance in the Higgsstrahlung process at a future $e^+ e^-$
  linear collider}'',
2001.
%%CITATION = LC-PHSM-2001-003 ETC.;%%.

\bibitem[Anderson et~al.(2013)Anderson, S., Caola, Gao, Gritsan, Martin,
  Melnikov, Schulze, Tran, Whitbeck, and Zhou]{Anderson:2013}
I.~Anderson, B.~S., F.~Caola, Y.~Gao, V.~Gritsan, C.~Martin, K.~Melnikov,
  M.~Schulze, N.~Tran, A.~Whitbeck, and Y.~Zhou, ``{Constraining anomalous
  $HVV$ interactions at proton and lepton colliders}'',
 \href{http://xxx.lanl.gov/abs/hep-ph/1309.4819}{{\ttfamily
  arXiv:hep-ph/1309.4819}}.
%%CITATION = HEP-PH/13094189;%%.

\bibitem[Kramer et~al.(1994)Kramer, Kuhn, Stong, and Zerwas]{Kramer:1993jn}
M.~Kramer, J.~H. Kuhn, M.~Stong, and P.~Zerwas, ``{Prospects of measuring the
  parity of Higgs particles}'', {\em Z.Phys.} {\bfseries C64} (1994) 21--30,
 \href{http://xxx.lanl.gov/abs/hep-ph/9404280}{{\ttfamily
  arXiv:hep-ph/9404280}}.
%%CITATION = HEP-PH/9404280;%%.

\bibitem[Desch et~al.(2004)Desch, Imhof, Was, and Worek]{Desch:2003rw}
K.~Desch, A.~Imhof, Z.~Was, and M.~Worek, ``{Probing the CP nature of the Higgs
  boson at linear colliders with tau spin correlations: The case of mixed
  scalar--pseudoscalar couplings}'', {\em Phys.Lett.} {\bfseries B579} (2004)
  157--164,
 \href{http://xxx.lanl.gov/abs/hep-ph/0307331}{{\ttfamily
  arXiv:hep-ph/0307331}}.
%%CITATION = HEP-PH/0307331;%%.

\bibitem[Reinhard(2009)]{Reinhard:2009}
M.~Reinhard, ``{CP violation in the Higgs sector with a next-generation
  detector at the ILC}'', PhD thesis, 2009.

\bibitem[Bhupal~Dev et~al.(2008)Bhupal~Dev, Djouadi, Godbole, Muhlleitner, and
  Rindani]{BhupalDev:2007is}
P.~Bhupal~Dev, A.~Djouadi, R.~Godbole, M.~Muhlleitner, and S.~Rindani,
  ``{Determining the CP properties of the Higgs boson}'', {\em Phys.Rev.Lett.}
  {\bfseries 100} (2008) 051801,
 \href{http://xxx.lanl.gov/abs/0707.2878}{{\ttfamily arXiv:0707.2878}}.
%%CITATION = ARXIV:0707.2878;%%.

\bibitem[Godbole et~al.(2011)Godbole, Hangst, Muhlleitner, Rindani, and
  Sharma]{Godbole:2011hw}
R.~Godbole, C.~Hangst, M.~Muhlleitner, S.~Rindani, and P.~Sharma,
  ``{Model-independent analysis of Higgs spin and CP properties in the process
  $e^+ e^- \to t \bar t \Phi$}'', {\em Eur.Phys.J.} {\bfseries C71} (2011)
  1681,
 \href{http://xxx.lanl.gov/abs/1103.5404}{{\ttfamily arXiv:1103.5404}}.
%%CITATION = ARXIV:1103.5404;%%.

\bibitem[Tanabe(2013)]{Tanabe:2013tth-cp}
T.~Tanabe, ``{Private communication}'', 2013.

\bibitem[{Ono, Hiroaki and Akiya Miyamoto}(2013)]{Ono:2013higgsbr}
{Ono, Hiroaki and Akiya Miyamoto}, ``{A study of measurement precision of the
  Higgs boson branching ratios at the Internationa Linear Collider}'', {\em
  Euro.Phys.J.} {\bfseries C73} (2013) 2343.

\bibitem[{Junping Tian and Keisuke Fujii}(2013)]{Junping-keisuke-fitting}
{Junping Tian and Keisuke Fujii}, ``{Summary of Higgs coupling measurement with
  staged running of ILC and 250 GeV, 500 GeV and 1 TeV}'', {\em
  LC-REP-2013-021}, 2013.

\bibitem[{Junping Tian, Claude Duerig, Keisuke Fujii, and Jenny
  List}(2013)]{durig-jenny-lcws2012}
{Junping Tian, Claude Duerig, Keisuke Fujii, and Jenny List}, ``{Determination
  of the Higgs total width with $WW$-fusion production at ILC up to 500 GeV}'',
  {\em LC-REP-2013-022}, 2013.

\bibitem[{Hiroaki Ono}(2013)]{Ono:2013higgsdbd}
{Hiroaki Ono}, ``{ Higgs banching ratio study for DBD detector benchmarking in
  ILD}'', {\em LC-REP-2013-005}, 2013.

\bibitem[Catani et~al.(1991)Catani, Dokshitzer, Olsson, Turnock, and
  Webber]{Catani:1991}
S.~Catani, Y.~L. Dokshitzer, M.~Olsson, G.~Turnock, and B.~Webber, ``{New
  clustering algorithm for multi-jet cross-sections in $e^+ e^-$
  annihilation}'', {\em Phys.Lett.} {\bfseries B269} (1991)
432--438.
%%CITATION = PHLTA,B269,432;%%.

\bibitem[{Shin-ichi Kawada, Keisuke Fujii, Taikan Suehara, Tohru Takahashi,
  Tomohiko Tanabe}(2013)]{Kawada:2013tautau}
{Shin-ichi Kawada, Keisuke Fujii, Taikan Suehara, Tohru Takahashi, Tomohiko
  Tanabe}, ``{ $H\rightarrow \tau^+\tau^{-}$ branching ratio study at
  $\sqrt{s}=250$ GeV at the ILC with the ILD detector}'', {\em
  LC-REP-2013-001}, 2013.

\bibitem[{Shin-ichi Kawada, Keisuke Fujii, Taikan Suehara, Tohru Takahashi, and
  Tomohiko Tanabe}(2013)]{Kawada-Seattle-Higgs-tautau}
{Shin-ichi Kawada, Keisuke Fujii, Taikan Suehara, Tohru Takahashi, and Tomohiko
  Tanabe}, ``{Evaluation of measurement accuracy of $h\rightarrow
  \tau^+\tau^{-}$ branching ratio at the ILC with the $\sqrt{s}=250$ GeV and
  500 GeV}'',  \href{http://xxx.lanl.gov/abs/1308.5489}{{\ttfamily
  arXiv:1308.5489}}.

\bibitem[Boos et~al.(2001)Boos, Brient, Reid, Schreiber, and
  Shanidze]{Boos:2000bz}
E.~Boos, J.~Brient, D.~Reid, H.~Schreiber, and R.~Shanidze, ``{Measuring the
  Higgs branching fraction into two photons at future linear $e^+ e^-$
  colliders}'', {\em Eur.Phys.J.} {\bfseries C19} (2001) 455--461,
 \href{http://xxx.lanl.gov/abs/hep-ph/0011366}{{\ttfamily
  arXiv:hep-ph/0011366}}.
%%CITATION = HEP-PH/0011366;%%.

\bibitem[Barklow(2003)]{Barklow:2003hz}
T.~L. Barklow, ``{Higgs coupling measurements at a 1 TeV linear collider}'',
 \href{http://xxx.lanl.gov/abs/hep-ph/0312268}{{\ttfamily
  arXiv:hep-ph/0312268}}.
%%CITATION = HEP-PH/0312268;%%.

\bibitem[Calancha(????)]{Calancha:2013gm}
C.~Calancha, ``{Presentation at the 32nd General Meeting of the ILC Physics
  Subgroup, June 2013}''.
  \url{http://ilcphys.kek.jp/meeting/physics/archives/2013-06-22/calancha_2013_06_22_generalMeeting.pdf}.

\bibitem[Englert et~al.(2011)Englert, Plehn, Zerwas, and
  Zerwas]{Englert:2011yb}
C.~Englert, T.~Plehn, D.~Zerwas, and P.~M. Zerwas, ``{Exploring the Higgs
  portal}'', {\em Phys.Lett.} {\bfseries B703} (2011) 298--305,
 \href{http://xxx.lanl.gov/abs/1106.3097}{{\ttfamily arXiv:1106.3097}}.
%%CITATION = ARXIV:1106.3097;%%.

\bibitem[Ono(????)]{ref:2012onoc}
H.~Ono, ``{ presentation at the Asian Physics and Software Meeting, June,
  2012.}''.

\bibitem[Yamamoto(????)]{ref:2012yamamoto}
A.~Yamamoto, ``{ presentation at the Asian Physics and Software Meeting, June,
  2012.}''.

\bibitem[Dittmaier et~al.(1998)Dittmaier, Kramer, Liao, Spira, and
  Zerwas]{Dittmaier:1998dz}
S.~Dittmaier, M.~Kramer, Y.~Liao, M.~Spira, and P.~Zerwas, ``{Higgs radiation
  off top quarks in $e^+ e^-$ collisions}'', {\em Phys.Lett.} {\bfseries B441}
  (1998) 383--388,
 \href{http://xxx.lanl.gov/abs/hep-ph/9808433}{{\ttfamily
  arXiv:hep-ph/9808433}}.
%%CITATION = HEP-PH/9808433;%%.

\bibitem[Dawson and Reina(1999)]{Dawson:1998ej}
S.~Dawson and L.~Reina, ``{QCD corrections to associated Higgs boson heavy
  quark production}'', {\em Phys.Rev.} {\bfseries D59} (1999) 054012,
 \href{http://xxx.lanl.gov/abs/hep-ph/9808443}{{\ttfamily
  arXiv:hep-ph/9808443}}.
%%CITATION = HEP-PH/9808443;%%.

\bibitem[Belanger et~al.(2003)Belanger, Boudjema, Fujimoto, Ishikawa, Kaneko,
  et~al.]{Belanger:2003nm}
G.~Belanger, F.~Boudjema, J.~Fujimoto, T.~Ishikawa, T.~Kaneko, {\em et~al.},
  ``{Full $\mathcal{O}(\alpha)$ electroweak and $\mathcal{O}(\alpha_s)$
  corrections to $e^+ e^- \to t\bar{t}H$}'', {\em Phys.Lett.} {\bfseries B571}
  (2003) 163--172,
 \href{http://xxx.lanl.gov/abs/hep-ph/0307029}{{\ttfamily
  arXiv:hep-ph/0307029}}.
%%CITATION = HEP-PH/0307029;%%.

\bibitem[Denner et~al.(2004)Denner, Dittmaier, Roth, and Weber]{Denner:2003zp}
A.~Denner, S.~Dittmaier, M.~Roth, and M.~Weber, ``{Radiative corrections to
  Higgs boson production in association with top quark pairs at $e^+ e^-$
  colliders}'', {\em Nucl.Phys.} {\bfseries B680} (2004) 85--116,
 \href{http://xxx.lanl.gov/abs/hep-ph/0309274}{{\ttfamily
  arXiv:hep-ph/0309274}}.
%%CITATION = HEP-PH/0309274;%%.

\bibitem[You et~al.(2003)You, Ma, Chen, Zhang, Yan-Bin, et~al.]{You:2003zq}
Y.~You, W.-G. Ma, H.~Chen, R.-Y. Zhang, S.~Yan-Bin, {\em et~al.},
  ``{Electroweak radiative corrections to $e^+ e^-\to t\bar{t}h$ at linear
  colliders}'', {\em Phys.Lett.} {\bfseries B571} (2003) 85--91,
 \href{http://xxx.lanl.gov/abs/hep-ph/0306036}{{\ttfamily
  arXiv:hep-ph/0306036}}.
%%CITATION = HEP-PH/0306036;%%.

\bibitem[Farrell and Hoang(2005)]{Farrell:2005fk}
C.~Farrell and A.~H. Hoang, ``{The Large Higgs energy region in Higgs
  associated top pair production at the linear collider}'', {\em Phys.Rev.}
  {\bfseries D72} (2005) 014007,
 \href{http://xxx.lanl.gov/abs/hep-ph/0504220}{{\ttfamily
  arXiv:hep-ph/0504220}}.
%%CITATION = HEP-PH/0504220;%%.

\bibitem[Farrell and Hoang(2006)]{Farrell:2006xe}
C.~Farrell and A.~H. Hoang, ``{Next-to-leading-logarithmic QCD corrections to
  the cross- section $\sigma(e^+ e^- \to t\bar{t}H)$ at 500~GeV}'', {\em
  Phys.Rev.} {\bfseries D74} (2006) 014008,
 \href{http://xxx.lanl.gov/abs/hep-ph/0604166}{{\ttfamily
  arXiv:hep-ph/0604166}}.
%%CITATION = HEP-PH/0604166;%%.

\bibitem[Juste et~al.(2006)Juste, Kiyo, Petriello, Teubner, Agashe,
  et~al.]{Juste:2006sv}
A.~Juste, Y.~Kiyo, F.~Petriello, T.~Teubner, K.~Agashe, {\em et~al.}, ``{Report
  of the 2005 Snowmass top/QCD working group}'',
 \href{http://xxx.lanl.gov/abs/hep-ph/0601112}{{\ttfamily
  arXiv:hep-ph/0601112}}.
%%CITATION = HEP-PH/0601112;%%.

\bibitem[Yonamine et~al.(2011)Yonamine, Ikematsu, Tanabe, Fujii, Kiyo,
  et~al.]{Yonamine:2011jg}
R.~Yonamine, K.~Ikematsu, T.~Tanabe, K.~Fujii, Y.~Kiyo, {\em et~al.},
  ``{Measuring the top Yukawa coupling at the ILC at $\sqrt{s}=500$ GeV}'',
  {\em Phys.Rev.} {\bfseries D84} (2011) 014033,
 \href{http://xxx.lanl.gov/abs/1104.5132}{{\ttfamily arXiv:1104.5132}}.
%%CITATION = ARXIV:1104.5132;%%.

\bibitem[Tabassam and Martin(2012)]{Tabassam:2012it}
H.~Tabassam and V.~Martin, ``{Top Higgs Yukawa Coupling Analysis from $e^+ e^-
  \to t\bar{t}H \to bW^+\, \bar{b}W^- \,b\bar{b}$}'',
 \href{http://xxx.lanl.gov/abs/1202.6013}{{\ttfamily arXiv:1202.6013}}.
%%CITATION = ARXIV:1202.6013;%%.

\bibitem[Juste and Merino(1999)]{Juste:1999af}
A.~Juste and G.~Merino, ``{Top Higgs-Yukawa coupling measurement at a linear
  $e^+ e^-$ collider}'',
 \href{http://xxx.lanl.gov/abs/hep-ph/9910301}{{\ttfamily
  arXiv:hep-ph/9910301}}.
%%CITATION = HEP-PH/9910301;%%.

\bibitem[Gay(2007)]{Gay:2006vs}
A.~Gay, ``{Measurement of the top-Higgs Yukawa coupling at a Linear $e^+ e^-$
  Collider}'', {\em Eur.Phys.J.} {\bfseries C49} (2007) 489--497,
 \href{http://xxx.lanl.gov/abs/hep-ph/0604034}{{\ttfamily
  arXiv:hep-ph/0604034}}.
%%CITATION = HEP-PH/0604034;%%.

\bibitem[{Junping Tian and Keisuke Fujii}(2013)]{Tian:2013hhbbbb}
{Junping Tian and Keisuke Fujii}, ``{Study of Higgs self-coupling at the ILC
  based on the full detector simulation at $\sqrt{s}= 500$~GeV and $\sqrt{s} =
  1$~TeV}'', {\em LC-REP-2013-003}, 2013.

\bibitem[Yasui et~al.(2002)Yasui, Kanemura, Kiyoura, Odagiri, Okada,
  et~al.]{Yasui:2002se}
Y.~Yasui, S.~Kanemura, S.~Kiyoura, K.~Odagiri, Y.~Okada, {\em et~al.},
  ``{Measurement of the Higgs self-coupling at JLC}'',
 \href{http://xxx.lanl.gov/abs/hep-ph/0211047}{{\ttfamily
  arXiv:hep-ph/0211047}}.
%%CITATION = HEP-PH/0211047;%%.

\bibitem[{Masakazu Kurata, Tomohiko Tanabe, Junping Tian, Keisuke Fujii, and
  Taikan Suehara}(2013)]{Kawada:2013hhbbww}
{Masakazu Kurata, Tomohiko Tanabe, Junping Tian, Keisuke Fujii, and Taikan
  Suehara}, ``{Study of Higgs self-coupling at the ILC Using $HH\to
  b\bar{b}\,WW^*$}'', {\em LC-REP-2013-025}, 2013.

\bibitem[Peskin(2012)]{Peskin:2012we}
M.~E. Peskin, ``{Comparison of LHC and ILC Capabilities for Higgs Boson
  Coupling Measurements}'',
 \href{http://xxx.lanl.gov/abs/1207.2516}{{\ttfamily arXiv:1207.2516}}.
%%CITATION = ARXIV:1207.2516;%%.

\bibitem[Low and Lykken(2010)]{Low:2010jp}
I.~Low and J.~Lykken, ``{Revealing the electroweak properties of a new scalar
  resonance}'', {\em JHEP} {\bfseries 1010} (2010) 053,
 \href{http://xxx.lanl.gov/abs/1005.0872}{{\ttfamily arXiv:1005.0872}}.
%%CITATION = ARXIV:1005.0872;%%.

\bibitem[Takubo et~al.(2013)Takubo, Hodgkinson, Ikematsu, Fujii, Okada,
  et~al.]{Takubo:2010tc}
Y.~Takubo, R.~N. Hodgkinson, K.~Ikematsu, K.~Fujii, N.~Okada, {\em et~al.},
  ``{Measuring anomalous couplings in $H\to WW^*$ decays at the International
  Linear Collider}'', {\em Phys.Rev.} {\bfseries D88} (2013) 013010,
 \href{http://xxx.lanl.gov/abs/1011.5805}{{\ttfamily arXiv:1011.5805}}.
%%CITATION = ARXIV:1011.5805;%%.

\bibitem[Aguilar-Saavedra et~al.(2001)]{Aguilar-Saavedra:2001rg}
{\bfseries ECFA/DESY LC Physics Working Group} Collaboration, J.~A.
  Aguilar-Saavedra {\em et~al.}, ``{TESLA Technical Design Report Part III:
  Physics at an $e^+e^-$ Linear Collider}'',
  \href{http://xxx.lanl.gov/abs/hep-ph/0106315}{{\ttfamily hep-ph/0106315}}.

\bibitem[Desch et~al.(2004)Desch, Klimkovich, Kuhl, and
  Raspereza]{Desch:2004yb}
K.~Desch, T.~Klimkovich, T.~Kuhl, and A.~Raspereza, ``{Study of Higgs boson
  pair production at linear collider}'',
 \href{http://xxx.lanl.gov/abs/hep-ph/0406229}{{\ttfamily
  arXiv:hep-ph/0406229}}.
%%CITATION = HEP-PH/0406229;%%.

\bibitem[Schael et~al.(2006)]{Schael:2006cr}
{\bfseries ALEPH Collaboration, DELPHI Collaboration, L3 Collaboration, OPAL
  Collaboration, LEP Working Group for Higgs Boson Searches} Collaboration,
  S.~Schael {\em et~al.}, ``{Search for neutral MSSM Higgs bosons at LEP}'',
  {\em Eur.Phys.J.} {\bfseries C47} (2006) 547--587,
 \href{http://xxx.lanl.gov/abs/hep-ex/0602042}{{\ttfamily
  arXiv:hep-ex/0602042}}.
%%CITATION = HEP-EX/0602042;%%.

\bibitem[Abdallah et~al.(2003)]{Abdallah:2002qj}
{\bfseries DELPHI Collaboration} Collaboration, J.~Abdallah {\em et~al.},
  ``{Search for doubly charged Higgs bosons at LEP-2}'', {\em Phys.Lett.}
  {\bfseries B552} (2003) 127--137,
 \href{http://xxx.lanl.gov/abs/hep-ex/0303026}{{\ttfamily
  arXiv:hep-ex/0303026}}.
%%CITATION = HEP-EX/0303026;%%.

\bibitem[Tsumura(????)]{Tsumura}
K.~Tsumura, ``{presentation at the KILC12 workshop (2012)}'',.

\bibitem[Liu and Potter(2013)]{Liu:2013gea}
T.~Liu and C.~Potter, ``{Exotic Higgs Decay $h\to aa$ at the International
  Linear Collider: a Snowmass White Paper}'',
 \href{http://xxx.lanl.gov/abs/1309.0021}{{\ttfamily arXiv:1309.0021}}.
%%CITATION = ARXIV:1309.0021;%%.

\bibitem[Djouadi et~al.(1992{\natexlab{a}})Djouadi, Kalinowski, and
  Zerwas]{Djouadi:1992gp}
A.~Djouadi, J.~Kalinowski, and P.~Zerwas, ``{Measuring the $H t\bar{t}$
  coupling in $e^+ e^-$ collisions}'', {\em Mod.Phys.Lett.} {\bfseries A7}
  (1992){\natexlab{a}}
1765--1769.
%%CITATION = MPLAE,A7,1765;%%.

\bibitem[Djouadi et~al.(1992{\natexlab{b}})Djouadi, Kalinowski, and
  Zerwas]{Djouadi:1991tk}
A.~Djouadi, J.~Kalinowski, and P.~Zerwas, ``{Higgs radiation off top quarks in
  high-energy $e^+ e^-$ colliders}'', {\em Z.Phys.} {\bfseries C54}
  (1992){\natexlab{b}}
255--262.
%%CITATION = ZEPYA,C54,255;%%.

\bibitem[Komamiya(1988)]{Komamiya:1988rs}
S.~Komamiya, ``{Searching for charged Higgs bosons at $\sim 1/2$---$1$~TeV
  $e^+e^-$ colliders}'', {\em Phys.Rev.} {\bfseries D38} (1988)
2158.
%%CITATION = PHRVA,D38,2158;%%.

\bibitem[Battaglia et~al.(2001)Battaglia, Ferrari, Kiiskinen, and
  Maki]{Battaglia:2001be}
M.~Battaglia, A.~Ferrari, A.~Kiiskinen, and T.~Maki, ``{Pair production of
  charged Higgs bosons at future linear $e^+ e^-$ colliders}'', {\em eConf}
  {\bfseries C010630} (2001) E3017,
 \href{http://xxx.lanl.gov/abs/hep-ex/0112015}{{\ttfamily
  arXiv:hep-ex/0112015}}.
%%CITATION = HEP-EX/0112015;%%.

\bibitem[Moretti(2004)]{Moretti:2003cd}
S.~Moretti, ``{Detection of heavy charged Higgs bosons in $e^+ e^-\to
  t\bar{b}H^-$ production at future linear colliders}'', {\em Eur.Phys.J.}
  {\bfseries C34} (2004) 157--163,
 \href{http://xxx.lanl.gov/abs/hep-ph/0306297}{{\ttfamily
  arXiv:hep-ph/0306297}}.
%%CITATION = HEP-PH/0306297;%%.

\bibitem[Gunion et~al.(2003)Gunion, Han, Jiang, and Sopczak]{Gunion:2002ip}
J.~F. Gunion, T.~Han, J.~Jiang, and A.~Sopczak, ``{Determining tan beta with
  neutral and charged Higgs bosons at a future $e^+ e^-$ linear collider}'',
  {\em Phys.Lett.} {\bfseries B565} (2003) 42--60,
 \href{http://xxx.lanl.gov/abs/hep-ph/0212151}{{\ttfamily
  arXiv:hep-ph/0212151}}.
%%CITATION = HEP-PH/0212151;%%.

\bibitem[Kanemura et~al.(2013)Kanemura, Tsumura, and Yokoya]{Kanemura:2013eja}
S.~Kanemura, K.~Tsumura, and H.~Yokoya, ``{Determination of $\tan\beta$ from
  the Higgs boson decay at linear colliders}'',
 \href{http://xxx.lanl.gov/abs/1305.5424}{{\ttfamily arXiv:1305.5424}}.
%%CITATION = ARXIV:1305.5424;%%.

\bibitem[Barger et~al.(2001)Barger, Han, and Jiang]{Barger:2000fi}
V.~D. Barger, T.~Han, and J.~Jiang, ``{$\tan\beta$ determination from heavy
  Higgs boson production at linear colliders}'', {\em Phys.Rev.} {\bfseries
  D63} (2001) 075002,
 \href{http://xxx.lanl.gov/abs/hep-ph/0006223}{{\ttfamily
  arXiv:hep-ph/0006223}}.
%%CITATION = HEP-PH/0006223;%%.

\bibitem[Barklow(1990)]{Barklow:1990ah}
T.~Barklow, ``{Particle physics research at a 500 GeV $e^+ e^-$ linear
  collider}'', {\em Conf.Proc.} {\bfseries C9006252} (1990)
440--450.
%%CITATION = CONFP,C9006252,440;%%.

\bibitem[Gunion and Haber(1993)]{Gunion:1992ce}
J.~F. Gunion and H.~E. Haber, ``{Higgs boson production in the photon-photon
  collider mode of a high-energy $e^+ e^-$ linear collider}'', {\em Phys.Rev.}
  {\bfseries D48} (1993)
5109--5120.
%%CITATION = PHRVA,D48,5109;%%.

\bibitem[Borden et~al.(1993)Borden, Bauer, and Caldwell]{Borden:1993cw}
D.~L. Borden, D.~A. Bauer, and D.~O. Caldwell, ``{Higgs boson production at a
  photon linear collider}'', {\em Phys.Rev.} {\bfseries D48} (1993)
4018--4028.
%%CITATION = PHRVA,D48,4018;%%.

\bibitem[Asner et~al.(2003)Asner, Gronberg, and Gunion]{Asner:2001ia}
D.~M. Asner, J.~B. Gronberg, and J.~F. Gunion, ``{Detecting and studying Higgs
  bosons at a photon-photon collider}'', {\em Phys.Rev.} {\bfseries D67} (2003)
  035009,
 \href{http://xxx.lanl.gov/abs/hep-ph/0110320}{{\ttfamily
  arXiv:hep-ph/0110320}}.
%%CITATION = HEP-PH/0110320;%%.

\bibitem[Ginzburg et~al.(1983)Ginzburg, Kotkin, Serbo, and
  Telnov]{Ginzburg:1981vm}
I.~Ginzburg, G.~Kotkin, V.~Serbo, and V.~I. Telnov, ``{Colliding $\gamma e$ and
  $\gamma \gamma$ Beams Based on the Single Pass Accelerators (of VLEPP
  Type)}'', {\em Nucl.Instrum.Meth.} {\bfseries 205} (1983)
47--68.
%%CITATION = NUIMA,205,47;%%.

\bibitem[Ginzburg et~al.(1984)Ginzburg, Kotkin, Panfil, Serbo, and
  Telnov]{Ginzburg:1982yr}
I.~Ginzburg, G.~Kotkin, S.~Panfil, V.~Serbo, and V.~I. Telnov, ``{Colliding
  $\gamma e$ and $\gamma \gamma$ Beams Based on the Single Pass $e^+ e^-$
  Accelerators. 2. Polarization Effects. Monochromatization Improvement}'',
  {\em Nucl.Instrum.Meth.} {\bfseries A219} (1984)
5--24.
%%CITATION = NUIMA,A219,5;%%.

\bibitem[Asner et~al.(2003)Asner, Burkhardt, De~Roeck, Ellis, Gronberg,
  et~al.]{Asner:2001vh}
D.~Asner, H.~Burkhardt, A.~De~Roeck, J.~Ellis, J.~Gronberg, {\em et~al.},
  ``{Higgs physics with a $\gamma \gamma$ collider based on CLIC I}'', {\em
  Eur.Phys.J.} {\bfseries C28} (2003) 27--44,
 \href{http://xxx.lanl.gov/abs/hep-ex/0111056}{{\ttfamily
  arXiv:hep-ex/0111056}}.
%%CITATION = HEP-EX/0111056;%%.

\bibitem[Ohgaki(2000)]{Ohgaki:1999ez}
T.~Ohgaki, ``{Precision mass determination of the Higgs boson at photon-photon
  colliders}'', {\em Int.J.Mod.Phys.} {\bfseries A15} (2000) 2605--2612,
 \href{http://xxx.lanl.gov/abs/hep-ph/0002083}{{\ttfamily
  arXiv:hep-ph/0002083}}.
%%CITATION = HEP-PH/0002083;%%.

\bibitem[Ohgaki et~al.(1997)Ohgaki, Takahashi, and Watanabe]{Ohgaki:1997jp}
T.~Ohgaki, T.~Takahashi, and I.~Watanabe, ``{Measuring the two photon decay
  width of intermediate mass Higgs at a photon-photon collider}'', {\em
  Phys.Rev.} {\bfseries D56} (1997) 1723--1729,
 \href{http://xxx.lanl.gov/abs/hep-ph/9703301}{{\ttfamily
  arXiv:hep-ph/9703301}}.
%%CITATION = HEP-PH/9703301;%%.

\bibitem[Dittmar and Dreiner(1997)]{Dittmar:1996ss}
M.~Dittmar and H.~K. Dreiner, ``{How to find a Higgs boson with a mass between
  155 GeV---180 GeV at the LHC}'', {\em Phys.Rev.} {\bfseries D55} (1997)
  167--172,
 \href{http://xxx.lanl.gov/abs/hep-ph/9608317}{{\ttfamily
  arXiv:hep-ph/9608317}}.
%%CITATION = HEP-PH/9608317;%%.

\bibitem[Jikia and Tkabladze(1993)]{Jikia:1993yc}
G.~Jikia and A.~Tkabladze, ``{Neutral gauge boson pair production at the photon
  linear collider: $\gamma \gamma \to Z Z$, $\gamma Z$, $\gamma \gamma$}'',
  {\em Proceedings of the 2nd International Workshop on Physics and Experiments
  with Linear $e^+ e^-$ Colliders, 26-30 Apr 1993. Waikoloa, Hawaii, edited by
  F.A. Harris, S.L. Olsen, S. Pakvasa, X. Tata. (World Scientific, River Edge,
  NJ)}, 1993
558--562.
%%CITATION = INSPIRE-369468;%%.

\bibitem[Grzadkowski and Gunion(1992)]{Grzadkowski:1992sa}
B.~Grzadkowski and J.~Gunion, ``{Using back scattered laser beams to detect CP
  violation in the neutral Higgs sector}'', {\em Phys.Lett.} {\bfseries B294}
  (1992) 361--368,
 \href{http://xxx.lanl.gov/abs/hep-ph/9206262}{{\ttfamily
  arXiv:hep-ph/9206262}}.
%%CITATION = HEP-PH/9206262;%%.

\bibitem[Gunion and Kelly(1994)]{Gunion:1994wy}
J.~Gunion and J.~Kelly, ``{Determining the CP eigenvalues of the neutral Higgs
  bosons of the minimal supersymmetric model in $\gamma \gamma$ collisions}'',
  {\em Phys.Lett.} {\bfseries B333} (1994) 110--117,
 \href{http://xxx.lanl.gov/abs/hep-ph/9404343}{{\ttfamily
  arXiv:hep-ph/9404343}}.
%%CITATION = HEP-PH/9404343;%%.

\end{thebibliography}\endgroup

\end{document}